\documentclass[singlecolumn,prd,amssymb,amsmath,preprintnumbers,aps,12pt,nofootinbib,notitlepage,superscriptaddress]{revtex4-2}
\usepackage{lineno}
\usepackage{bm,color,xcolor}
\usepackage{slashed} 
\usepackage{graphicx}
\usepackage{booktabs}
\usepackage{makecell}
\usepackage{float}
\usepackage{overpic}
\usepackage{subfig}
\usepackage{multirow}
\usepackage{comment}
\usepackage{mathtools}
\usepackage{appendix}
\usepackage[colorlinks=true
,urlcolor=blue
,anchorcolor=blue
,citecolor=blue 
,filecolor=blue
,linkcolor=blue
,menucolor=blue
,linktocpage=true
,pdfproducer=medialab
,pdfa=true
]{hyperref}
\usepackage{lineno}
\usepackage{feynmp-auto}
\usepackage{upgreek}
\usepackage[normalem]{ulem}
\usepackage{siunitx}
\usepackage{adjustbox}

\DeclareMathOperator{\Br}{Br}

\newcommand{\beq}{\begin{equation}}
\newcommand{\eeq}{\end{equation}}
\newcommand{\bea}{\begin{eqnarray}}
\newcommand{\eea}{\end{eqnarray}}

\newcommand{\st}{{\tilde t}}

\newcommand{\ave}[1]{\left\langle #1 \right\rangle}

\newcommand{\neutralino}[1]{\tilde{\chi}_{#1}^0}

\newcommand{\iab}{\text{~ab}^{-1}}

\newcommand{\mM}{\mathcal{M}}
\newcommand{\mF}{\mathcal{F}}

\newcommand{\eqal}[1]{\begin{align}#1\end{align}}

\DeclareUnicodeCharacter{202F}{~}
\DeclareUnicodeCharacter{2212}{\ensuremath{-}}
\usepackage{url}
\begin{document}

 
\begin{flushright}
\end{flushright}

\title{\bf \Large 
New Physics Search at the CEPC: a General Perspective
}


\author{Xiaocong Ai} 
\affiliation{School of Physics, Zhengzhou University, Zhengzhou 450001, China}

\author{Stefan Antusch}
\affiliation{Department of Physics, University of Basel, CH-4056 Basel, Switzerland}

\author{Peter Athron}
\affiliation{Department of Physics and Institute of Theoretical Physics, Nanjing Normal University, Nanjing 210023, China}

\author{Yunxiang Bai} 
\affiliation{Institute of High Energy Physics, Chinese Academy of Sciences, Beijing 100049, China}
\affiliation{TIANFU Cosmic Ray Research Center, Chengdu, Sichuan, China}

\author{Shou-Shan Bao} 
\affiliation{Institute of Frontier and Interdisciplinary Science, Shandong University, Qingdao 266237, China}
\affiliation{Key Laboratory of Particle Physics and Particle Irradiation (MOE), Shandong University, Qingdao 266237, China}

\author{Daniele Barducci}
\affiliation{Dipartimento di Fisica ``Enrico Fermi", Universit\'a di Pisa, Largo Bruno Pontecorvo 3, I-56127 Pisa, Italy}
\affiliation{INFN, Sezione di Pisa, Largo Bruno Pontecorvo 3, I-56127 Pisa, Italy}

\author{Xiao-Jun Bi} 
\affiliation{Institute of High Energy Physics, Chinese Academy of Sciences, Beijing 100049, China}
\affiliation{University of Chinese Academy of Sciences (UCAS), Beijing 100049, China}

\author{Tianji Cai} 
\affiliation{School of Physical Science and Engineering, Tongji University, Shanghai 200092, China}
\affiliation{State Key Laboratory of Autonomous Intelligent Unmanned Systems, and MOE Frontiers Science Center for Intelligent Autonomous Systems, Tongji University, Shanghai 200092, China}

\author{Lorenzo Calibbi}
\affiliation{School of Physics, Nankai University, Tianjin 300071, China}

\author{Junsong Cang} 
\affiliation{Theoretical and Scientiﬁc Data Science, Scuola Internazionale Superiore di Studi Avanzati (SISSA), Via Bonomea 265, 34136 Trieste, Italy}

% ...

\maketitle
\clearpage

\section*{Editors}

\begin{flushleft}
{
\setlength{\baselineskip}{0.9\baselineskip}
\textbf{\uppercase\expandafter{\romannumeral 2}. Introduction}\\
Yu Gao, \href{mailto:gaoyu@ihep.ac.cn}{gaoyu@ihep.ac.cn}\\
\vspace{1.1em}
\textbf{\uppercase\expandafter{\romannumeral 3}. Description of CEPC facility}\\
Manqi Ruan, \href{mailto:manqi.ruan@ihep.ac.cn}{manqi.ruan@ihep.ac.cn}\\ 
\vspace{1.1em}
\textbf{\uppercase\expandafter{\romannumeral 4}. Exotic Higgs potential and Exotic Higgs/Z/top decays}\\
Yaquan Fang, \href{mailto:fangyq@ihep.ac.cn}{fangyq@ihep.ac.cn}\\
Zhao Li, \href{mailto:zhaoli@ihep.ac.cn}{zhaoli@ihep.ac.cn}\\ 
Zhen Liu, \href{mailto:zliuphys@umn.edu}{zliuphys@umn.edu} \\
\vspace{1.1em}
\textbf{\uppercase\expandafter{\romannumeral 5}. Electroweak phase transition and gravitational waves}\\
Ke-pan Xie, \href{mailto:kepan.xie@unl.edu}{kepan.xie@unl.edu}\\
Fa Peng Huang, \href{mailto:huangfp8@sysu.edu.cn}{huangfp8@sysu.edu.cn}\\
Sai Wang, \href{mailto:wangsai@ihep.ac.cn}{wangsai@ihep.ac.cn}\\
Michael Ramsey-Musolf, \href{mailto:mjrm@sjtu.edu.cn}{mjrm@sjtu.edu.cn}\\
Bruce Mellado Garcia, \href{mailto:Bruce.Mellado.Garcia@cern.ch}{Bruce.Mellado.Garcia@cern.ch}\\ 
\vspace{1.1em}
\textbf{\uppercase\expandafter{\romannumeral 6}. Dark Matter and Dark Sector}\\
Jia Liu, \href{mailto:jialiu@pku.edu.cn}{jialiu@pku.edu.cn}\\
Xiao-Ping Wang, \href{mailto:hcwangxiaoping@buaa.edu.cn}{hcwangxiaoping@buaa.edu.cn}\\
Yongchao Zhang, \href{mailto:zhangyongchao@seu.edu.cn}{zhangyongchao@seu.edu.cn}\\
P. S. Bhupal Dev, \href{mailto:bdev@wustl.edu}{bdev@wustl.edu}\\
Peiwen Wu, \href{mailto:pwwu@seu.edu.cn}{pwwu@seu.edu.cn}\\ 
\vspace{1.1em}
\textbf{\uppercase\expandafter{\romannumeral 7}. Long-lived Particle Searches}\\
Xiang Chen, \href{mailto:brianchenfff@sjtu.edu.cn}{brianchenfff@sjtu.edu.cn}\\
Liang Li, \href{mailto:liangliphy@sjtu.edu.cn}{liangliphy@sjtu.edu.cn}\\
Ying-nan Mao, \href{mailto:ynmao@whut.edu.cn}{ynmao@whut.edu.cn}\\
Kechen Wang, \href{mailto:kechen.wang@whut.edu.cn}{kechen.wang@whut.edu.cn}\\
Zeren Simon Wang, \href{mailto:wzs@hfut.edu.cn}{wzs@hfut.edu.cn}\\
Peiwen Wu, \href{mailto:pwwu@seu.edu.cn}{pwwu@seu.edu.cn}\\ 
\vspace{1.1em}
\textbf{\uppercase\expandafter{\romannumeral 8}. Supersymmetry}\\
Tianjun Li, \href{mailto:tli@itp.ac.cn}{tli@itp.ac.cn}\\
Lei Wu, \href{mailto:leiwu@njnu.edu.cn}{leiwu@njnu.edu.cn}\\
Xuai Zhuang, \href{mailto:zhuangxa@ihep.ac.cn}{zhuangxa@ihep.ac.cn}\\
Da Xu, \href{mailto:xuda@ihep.ac.cn}{xuda@ihep.ac.cn}\\ 
\vspace{1.1em}
\textbf{\uppercase\expandafter{\romannumeral 9}. Flavor Portal New Physics}\\
Lingfeng Li, \href{mailto:l.f.li165@gmail.com}{l.f.li165@gmail.com}\\
Xin-Qiang Li, \href{mailto:xqli@mail.ccnu.edu.cn}{xqli@mail.ccnu.edu.cn}\\ 
\vspace{1.1em}
\textbf{\uppercase\expandafter{\romannumeral 10}. Neutrino Physics}\\
Bhupal Dev, \href{mailto:bdev@wustl.edu}{bdev@wustl.edu}\\
Wei Liu, \href{mailto:wei.liu@njust.edu.cn}{wei.liu@njust.edu.cn}\\
Yongchao Zhang, \href{mailto:zhangyongchao@seu.edu.cn}{zhangyongchao@seu.edu.cn}\\ 
\vspace{1.1em}
\textbf{\uppercase\expandafter{\romannumeral 11}. More Exotics}\\
Yu Gao, \href{mailto:gaoyu@ihep.ac.cn}{gaoyu@ihep.ac.cn}\\
Zuowei Liu, \href{mailto:zuoweiliu@nju.edu.cn}{zuoweiliu@nju.edu.cn}\\ 
\vspace{1.1em}
\textbf{\uppercase\expandafter{\romannumeral 12}. Global Fits}\\
Jiayin Gu, \href{mailto:jiayin_gu@fudan.edu.cn}{jiayin\_gu@fudan.edu.cn}\\
Yang Zhang, \href{mailto:zhangyang2025@htu.edu.cn}{zhangyang2025@htu.edu.cn}\\
Yong Du, \href{mailto:yongdu5@impcas.ac.cn}{yongdu5@impcas.ac.cn}\\
Tao Han, \href{mailto:than@pitt.edu}{than@pitt.edu}\\
Shufang Su, \href{mailto:shufang@arizona.edu}{shufang@arizona.edu}\\
Wei Su, \href{mailto:suwei26@mail.sysu.edu.cn}{suwei26@mail.sysu.edu.cn}\\
Yongcheng Wu, \href{mailto:ycwu@njnu.edu.cn}{ycwu@njnu.edu.cn}\\ 
\vspace{1.1em}
\textbf{\uppercase\expandafter{\romannumeral 13}. Conclusion}\\
Jia Liu, \href{mailto:jialiu@pku.edu.cn}{jialiu@pku.edu.cn}\\ 
Zhen Liu, \href{mailto:zliuphys@umn.edu}{zliuphys@umn.edu}\\
Manqi Ruan, \href{mailto:manqi.ruan@ihep.ac.cn}{manqi.ruan@ihep.ac.cn}\\ 
\vspace{1.1em}
\textbf{Glossary}\\
Xuai Zhuang, \href{mailto:zhuangxa@ihep.ac.cn}{zhuangxa@ihep.ac.cn}\\
}
\end{flushleft}

\clearpage

\newpage

\tableofcontents
\clearpage

\clearpage
\section{Executive Summary}

A next generation, high-intensity electron-positron collider “Higgs factory”, such as the Circular Electron-Positron Collider (CEPC), is among the highest priority for the global high energy collider physics community. The CEPC can provide unprecedented opportunities for making fundamental discoveries and providing decisive insights in the quest for a “New Standard Model (SM)” of nature’s fundamental interactions. The CEPC could:
\begin{itemize} 
    \item Identify the origin of matter, especially the mechanism related to the first-order phase transition in the early Universe, which could produce a detectable gravitational wave signal. 
    \item Discover dark matter, particularly dark matter particles with a mass between one tenth and 100 times the proton mass. 
    \item Observe an array of new physics smoking guns, with sensitivities orders of magnitude better than those of existing facilities. 
\end{itemize}

The SM of Particle Physics is a triumph of the past half a century, as it predicts and interprets almost all the phenomena observed in experiments from the highest energies with colliders to low energy “tabletop” studies. On the other hand, deep mysteries exist concerning the most fundamental interactions of matter and the space-time fabric of the Universe, including the nature of dark matter, the origin of “visible” matter, the vast hierarchy of elementary particle masses, the quantum nature of gravity, and the mechanism of inflation. These mysteries challenge us to look for “new physics” beyond the SM and General Relativity. Indeed, physicists believe that the SM is simply a low-energy effective theory that reflects aspects of the more profound theory that answers the aforementioned mysteries. Uncovering this ``New SM", the profound theory who supports the SM is the primary mission for particle physics in the post-Higgs boson era. 

The Higgs boson can play a crucial role in addressing open questions in the SM. It is connected to the origin of both visible and dark matter of the Universe, the origin of neutrino masses, the stability of the Universe, and the self-consistency of the particle physics theory at quantum level. Studying the Higgs boson with the highest possible precision is therefore a promising path toward deeper insights.
The combination of experimentally producing a vast number of Higgs bosons in a “clean” environment of electron-positron collisions, and theoretically interpreting these measurements at high confidence, makes the CEPC-like Higgs factory a cornerstone of the global particle physics vision. The CEPC will also produce a high statistics sample of the $Z$ and $W$ bosons, and potentially large statistics of top quarks, further enhancing the prospects for discovery. Coupled with advances in the precision of theoretical computations, the CEPC will provide a uniquely powerful lens in the search for the New SM in multiple avenues.

Among the most important, yet unexplored, arena is the way that the Higgs field contributes to the energy of the Universe. Crucial physics information resides in the “Higgs potential”, whose simple form in the SM is largely untested to date. The electroweak phase transition is an important yet unexplored milestone in the early universe. The shape of the Higgs potential determines the dynamics of that era. New Physics that significantly alters the Higgs potential could lead to “smoking gun” phenomena in electron-positron collisions, which provide crucial information about the electroweak phase transition.
In this vein, a CEPC-like Higgs factory could also provide critical information about the generation of the cosmic matter-antimatter imbalance and the nature of dark matter. The origin of “visible” matter---which makes up stars, planets, and human life itself---remains a long-standing mystery. Among the proposed theories addressing this question, one stands out as particularly promising for discovery at next-generation collider experiments: ``electroweak baryogenesis", which requires a first-order electroweak phase transition occurring about 10 picoseconds after the Big Bang. This phenomenon implies significant modifications to the SM Higgs potential, which could be probed through precision measurements of Higgs boson at the CEPC. Electroweak baryogenesis, therefore, falls directly within the central scientific goals of the CEPC, which can provide a decisive test with high discovery potential. In fact, the occurrence of a first-order electroweak phase transition in the early Universe means significant change to the thermal history of the Universe predicted by the SM. The CEPC has the potential to cover most of the theoretical phase space predicted by relevant new physics models.
Making this scenario more appealing, a first-order electroweak phase transition could also generate detectable gravitational wave signals. CEPC measurements will coincide with those expected from the next generation of gravitational wave detectors such as LISA, Taiji, and TianQin, offering a powerful synergy between terrestrial and astrophysical probes.

Discovering the identity and characteristics of dark matter is an equally compelling challenge. Even the basic property of the dark matter mass remains unknown. The CEPC has a strong comparative advantage in detecting a relatively light particle dark matter candidate with a mass between one tenth and 100 times the proton mass. If there exists a new “dark force” between dark matter particles, studies have also demonstrated that a CEPC-like collider is particularly powerful in testing scenarios where the dark force carrier is relatively light. In this regard, the CEPC is highly complementary to many low-background, deep underground dark matter direct detection experiments.

The CEPC could directly detect a variety of phenomena predicted by a wide range of new physics models.
Thanks to the clean collision environments, significant yields of massive SM particles, and advanced detector-reconstruction technologies, the CEPC has unprecedented sensitivities to a suite of new physics signatures, including long-lived particles; exotic decays of the Higgs and $Z$ bosons; rare or SM–forbidden decays of heavy quarks and leptons; exotic mono-photon events; and many others.
Recent studies suggest that the CEPC's sensitivities to direct new physics signals could exceed that of existing facilities by several orders of magnitude, making it highly complementary to the LHC and other facilities.
In this way, the CEPC provides excellent discovery potential for heavy neutrino partners and new light particles such as axion-like particles and dark photons. Moreover, it could offer crucial insights into the underlying principles of new physics and point toward a more complete fundamental theory, such as supersymmetry.

A CEPC-like Higgs factory will provide tremendous discovery potential for new physics and decisive insights needed to resolve long-standing puzzles, including matter generation, the nature of dark matter, etc. This discovery power is rooted in the huge amount of “clean” data: five orders of magnitude more $Z$ bosons than LEP, the last generation $Z$ factory, and one million pristine Higgs boson events. To fully realize and further enhance the discovery power of the CEPC, the following studies are critical: ongoing theoretical development, including both new physics model building and interpretation framework;  a new generation of high-precision calculations; timely completion of innovative detector design and reconstruction algorithm development; exploration of synergies between different facilities.
Last but not least, rapid advancements in emerging technologies, such as the artificial intelligence and quantum technologies, will undoubtedly further amplify the discovery capabilities of a Higgs factory.

\newpage

\clearpage
\section{Introduction}
\label{sec:introduction}

\begin{figure}[!htb]
\includegraphics[width=0.9\textwidth]{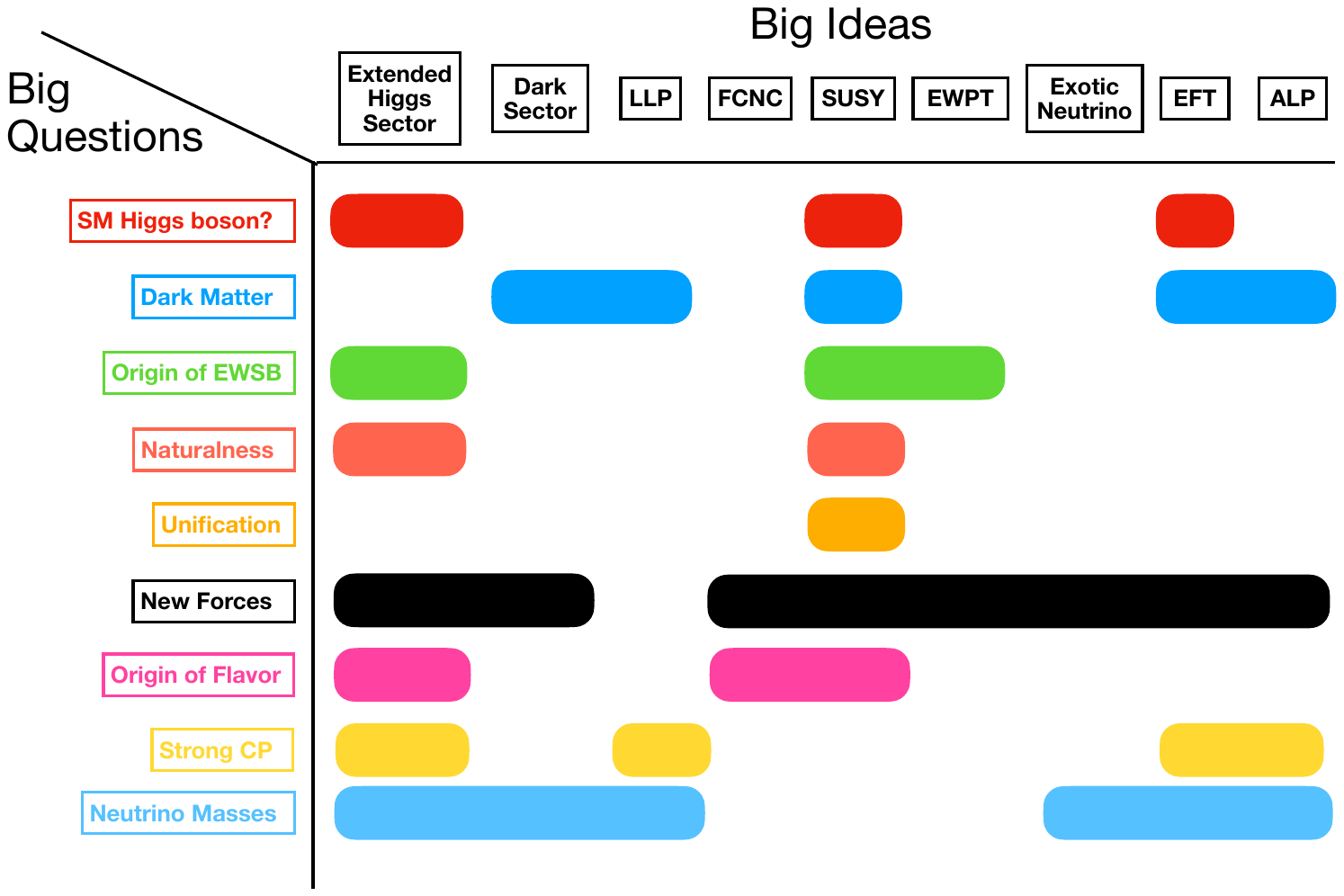}
\caption{\label{BigQuestions}Big questions and big ideas of the BSM landscape.}
\end{figure}

The Higgs boson is central to many mysteries of the Standard Model (SM) of particle physics, and a key to discovering new phenomena near the electroweak scale. Important questions include the scale of the electroweak unification, the nature of the electroweak phase transition, the flavor structure of fermions and so on. The Higgs field is also deeply connected with many fundamental phenomena beyond the Standard Model, such as the asymmetry of matter and anti-matter in the Universe, the presence of dark matter and dark energy, and even the mechanism for cosmic inflation. 
Each of the new phenomena may guide us to certain new mechanisms beyond the SM. For instance, as shown in Fig.~\ref{BigQuestions} along with other big topics, the investigation on the electroweak symmetry breaking and the origin of elementary mass may reveal a more complicated Higgs potential with additional bosons, and much more aggressive alternatives to the SM.
The 2012 discovery of the Higgs boson completed the last piece in the jigsaw puzzle of the SM's fundamental particle spectrum, and it offers a potent probe for these above-mentioned mysteries and phenomena. 
A Higgs factory that can measure the properties of the Higgs boson to an unprecedented precision 
is vital for this exploration. 

The LHC has so far served as a powerful Higgs factory. 
The high luminosity run of the LHC (HL-LHC) is projected to ultimately produce 100 million Higgs bosons.
However, proton-proton collision has large backgrounds and theoretical/systematic uncertainties, and
the typical accuracy of the Higgs property at the HL-LHC is expected to be limited to a few percent~\cite{Dawson:2022zbb}.

In comparison to the LHC, an electron-positron collider will have significant advantages in Higgs boson measurements. 
An electron-positron collider is by nature free of QCD backgrounds. The ratio of the Higgs signal versus the SM background is about 7-8 orders of magnitude higher than that at the HL-LHC. 
An electron-positron collider can produce precise and manipulable initial states that help to determine the Higgs boson's decay width and its couplings. 
Several future electron-positron colliders have been proposed, including the International Linear Collider (ILC) \cite{ILCInternationalDevelopmentTeam:2022izu}, the Compact Linear Collider (CLIC) \cite{CLIC:2018fvx}, the Future Circular Collider (FCC) \cite{FCC:2018evy,FCC:2025lpp,FCC:2025uan}, and
the Circular Electron Positron Collider (CEPC) \cite{CEPC-SPPCStudyGroup:2015csa,CEPC-SPPCStudyGroup:2015esa}. At the same time, a number of alternative possibilities are under consideration, e.g. a 125 GeV Muon collider~\cite{deBlas:2022aow}, C3~\cite{Bai:2021rdg}, ReliC~\cite{Litvinenko:2022qbd}, CERC~\cite{Litvinenko:2022mrt}, etc. 

The CEPC is proposed by the High Energy Physics community immediately after the Higgs boson was discovered. The CEPC working group was initiated in September of 2013.
In the year 2015, the CEPC working group presented the pre-CDR \cite{CEPC-SPPCStudyGroup:2015csa,CEPC-SPPCStudyGroup:2015esa} as the first milestone of CEPC study. Intensive R\&D and physics study in the following years led to the 2018 delivery of the CEPC Conceptual Design Report (CDR)~\cite{CEPCStudyGroup:2018ghi}, reporting no show-stoppers identified for this gigantic machine. In 2023, the Accelerator Technical Design Report (TDR) was released~\cite{CEPCStudyGroup:2023quu}, demonstrating that technology is ready for construction.

The CEPC is designed to host a main circular ring with a total circumference of 100 kilometers. 
The facility is designed to operate at several benchmark center-of-mass energies: $E_{\rm CM} = 91.2$ GeV as a $Z$ factory, $E_{\rm CM} \simeq 160$ GeV for $W$ boson pair production threshold and $E_{\rm CM} = 240$ GeV as a Higgs factory. The CEPC center-of-mass energy is capable of upgrading to 360 GeV, enabling $t\bar{t}$ pair production. 
Considering future upgrades, the CEPC underground tunnel is designed to have a large diameter such that it could host both CEPC and the future super proton-proton collider (SPPC) at the same time~\cite{Tang:2022fzs}.

In the TDR design~\cite{CEPCStudyGroup:2023quu}, the CEPC envisioned the collider to operate with two main detectors.
The nominal operation plan focuses on the Higgs operation, which lasts for 10 years; 
and it also includes 2 years of data taking at the $Z$ pole and 1 year for the W threshold scan. 
In its nominal operation plan with 50 MW synchrotron radiation power per beam, the CEPC is expected to deliver the total integrated luminosity of 100, 6.9, and 20 ab$^{-1}$ for $Z$-pole, $WW$, and Higgs factory runs, respectively.
The CEPC will produce approximately four trillion $Z$ bosons, nearly one billion $W$ bosons (mostly produced at Higgs operation), and over four million Higgs bosons.
After the high energy upgrade, the CEPC will operate for at least five years at a 360 GeV center-of-mass energy with 1 ab$^{-1}$ integrated luminosity. About 500 thousand $t\bar{t}$ events and 150 thousand inclusive Higgs events will be produced during this run,
see Table~\ref{tab:Yield_T1}.

\begin{table}[t]
	\centering
  	\resizebox{0.8 \textwidth}{!}{
	\begin{tabular}[t]{ccccc}
		\toprule[1pt]
		Operation mode & Z factory  & WW threshold  & Higgs factory & $t\bar{t}$ \\
		\midrule
		$\sqrt{s}$ (GeV) & 91.2 & 160 & 240 & 360 \\
		Run time (year) & 2 & 1 & 10 & 5 \\
		\makecell{Instantaneous luminosity \\ ($10^{34} \text{cm}^{-2}\text{s}^{-1}$, per IP)} & 191.7 & 26.6 & 8.3 & 0.83 \\
		\makecell{Integrated luminosity \\ (ab$^{-1}$, 2~IPs)} & 100 & 6 & 20 & 1 \\
		Event yields & $4.1 \times 10^{12}$ & $2 \times 10^{8}$ & $4.3 \times 10^{6}$ & $0.6 \times 10^{6}$ \\
		
		\bottomrule[1pt]
	\end{tabular}
	}
	\caption{Nominal CEPC operation scheme, and the physics yield, of four different modes. See \cite{Gao:2022lew} for details. }
	\label{tab:Yield_T1}
\end{table}

Physics study groups have continued to explore a wide range of topics to identify the CEPC's scientific capabilities. High priority searches include Higgs boson precision measurements, precise electroweak measurements, flavor physics, QCD-related measurements, and new physics. Physics studies identified a handful of critical detector requirements, quantified its impact on different physics benchmarks, and brought up clear performance goals for the CEPC detector design. To name a few, such requirements include the separation of final state particles, precise reconstruction of energy/momentum for different species of final state particles, the identification of physics objects in high-multiplicity events, good calibration of beam energy and instant luminosity, and so on. 
With these demands from physics study, the CEPC detector group dedicated a series of detector R$\&$D programs to the optimization between achieving physics requirements and leveraging the latest detector technology. 

\begin{figure}
    \centering
    \includegraphics[width=0.7\linewidth]{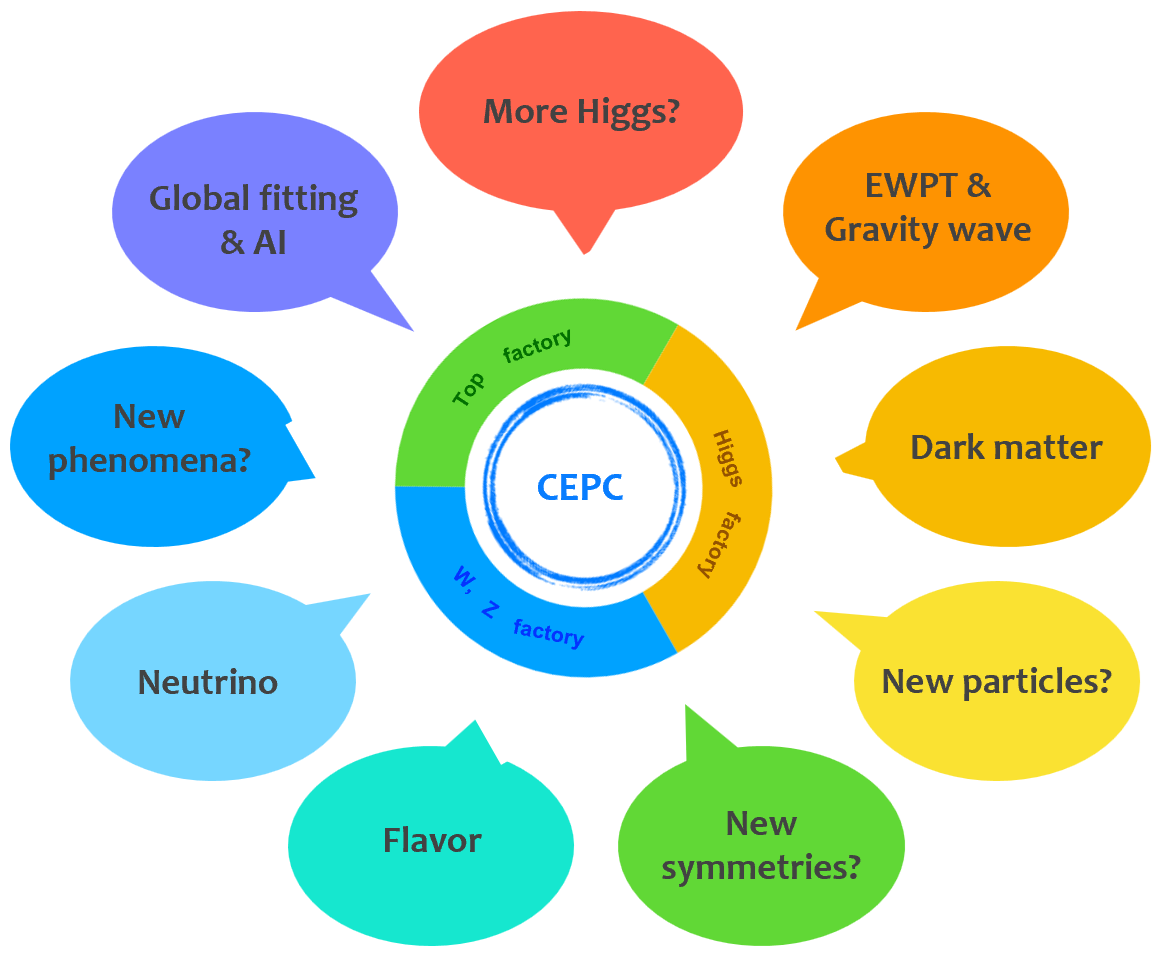}
    \caption{A cartoon of new physics program at the CEPC.}
    \label{fig:CEPC_BSM}
\end{figure}

The main body of this white paper is dedicated to the new physics potential at the CEPC. As precision measurement of the SM scalar sector is central among the CEPC's scientific goals, this white paper enlists major new physics interests that are deeply related to the electroweak scalar potential: precision measurements of the Higgs boson's couplings and decay branching ratios, new particles, novel phenomena, and many others. Out of the extremely broad kaleidoscope of new physics theory and phenomenology, we will focus on strategies that exploit the CEPC's advantage in its high precision with particle identity, low detection threshold, clean background, etc. A number of imminent search topics with strong interest are illustrated in Fig.~\ref{fig:CEPC_BSM}, in which the Higgs (and the $Z$) boson production channel play a major role. Detailed reviews will be presented in dedicated sections of the white paper . 

Aiming at SM electroweak test as a central task, the CEPC is designed with world-leading performance in the precision measurement of the electroweak scalar potential. Beyond the Standard Model searches can benefit immensely from the CEPC's precision in measuring the decay of the Higgs boson, the $Z$ boson and the top quark. Section~\ref{sec:Higgsexotic} provides a comprehensive review of exotic $h/Z/t$ decay sensitivities at the CEPC and the expected sensitivity on beyond-SM scalar potentials. Supersymmetry, dark sector motivated signals, and the decay channels yielding long-lived particles are emphasized as benchmark scenarios of particular interest. Cross-check measurements on potential Higgs-like hints from existing LHC results are also discussed.

Encouraged by progress in gravitational wave observations, high-energy phase transitions during the Early Universe gained much popularity in recent years. The nature of the electroweak phase transition sensitively depends on the details of the scalar potential, and the future colliders test of the CEPC offers a unique cross-check on our Universe's early phase-transition history. One of the essential questions to answer is whether the transition is a cross-over like in the vanilla SM, or it may undergo a more violent process, especially type-I transition with bubble formation and contribute non-trivially to the cosmic baryon asymmetry, stochastic gravitational wave background, etc. It is shown that the CEPC's precision can test scalar potentials that generate visible signals for future gravitational wave experiments. In Section~\ref{sec:EwptGw} we list the crucial measurements for electroweak phase transition, including the $Zh$ production cross-section, the Higgs boson decay width, and its branching ratios, which are highly sensitive to new physics corrections.


Dark sector and dark matter can be revealed via searches involving missing momentum, and dark sector models with leptonic and electroweak connections to the SM are of particular interest. The CEPC is designed with state-of-art detectors and will offer excellent reconstruction of missing energy/momentum. In Section~\ref{sec:DMDS}, we present the recent progress on selected dark matter/sector models and their studies for the designed high luminosity at Z-pole, Higgs-factory, and 360 GeV runs of the CEPC. The projected sensitivities have been provided for a number of such models, e.g. lepton-portal dark sectors, neutrino and electromagnetically interacting dark particles, leptophilic dark matter, visible in Z-decays, etc.


Long-lived particles (LLPs) have risen to a heated collider search target recently. The LLP scenario typically features a massive BSM particle with a much prolonged lifetime due to its near-degenerate mass to another particle, or highly suppressed couplings, and it can leave a novel signal inside the collider's detectors. In Section~\ref{sec:LLP}, we first briefly review the computational methodology for LLP production, and then we proceed to discuss the projected sensitivities with the main detector, proposed far detectors, and the beam dump. LLPs are predicted in many BSM theories, and LLP searches during Higgs and Z boson decay are important benchmark scenarios at the CEPC. Dedicated sensitivity studies are presented for supersymmetry, vector-like lepton extended models, axion-like particles, extra neutral gauge bosons, etc.

Supersymmetry provides an elegant framework to answer the SM's gauge hierarchy problem, and it has been extensively searched for by existing colliders. The CEPC offers search windows on many SUSY scenarios that are difficult for higher energy hadronic collisions. In Section~\ref{sec:SUSY}, we present the studies on the CEPC's direct search sensitivity on less massive SUSY electroweakinos and charged sleptons, where robust improvements on existing limits can be expected. It is also shown that SUSY-induced exotic channels with large missing energy and a diphoton signal yield strong limits on multi-TeV selectrons, much heavier than the CEPC center-of-mass energy. Relatively light sleptons are of strong interest due to their potential role in explaining the recent muon $g-2$ excess, and their synergy with the CEPC sensitivity is discussed in detail. The projected limits from ILC/CLIC SUSY search are also included for comparison.

The $Z$-pole run of CEPC can produce large samples of flavored hadrons and leptons, such as $B,D$ mesons and $\tau$ leptons, which offer a powerful measurement of new physics with flavored couplings. A more comprehensive review of CEPC's flavor physics is presented in a dedicated white paper. In Section~\ref{sec:FlavorPortal}, we summarize the CEPC's potential in new physics search via flavor portal. Important BSM search aspects include potential corrections to the Cabibbo–Kobayashi–Maskawa matrix, testing the presence of flavor-changing neutral current, violation of lepton flavor universality, etc. Precision testing on the flavor symmetries of the SM can test new physics effects from a high energy scale. We include recent studies on charged lepton flavor violation, $b,c$-hadron decays, and the search for light BSM degree of freedom during flavor transitions.

Neutrino oscillation provides a clear indication of physics beyond the Standard Model, such as the seesaw mechanism. The precision of the CEPC will offer powerful tests for the underlying neutrino models. Section~\ref{sec:neutrino} discusses the CEPC's advantage in neutrino-related searches and the potential connection to leptogenesis. Relevant scenarios include the heavy neutrino search at the CEPC's main/far detectors and the beam dump, promptly decaying heavy neutrinos with high lepton multiplicity and visible lepton-number violation, active-sterile neutrino transition and non-standard effective neutrino interactions.

In addition, the CEPC features a unique opportunity for even more exotic physics searches. In particular, the low hadronic background and high sensitivity to soft leptons, photons, and jets empower careful investigation of exotic physical processes with high lepton multiplicity, or those that require good lepton reconstruction, energy resolution, and flavor recognition. Section~\ref{sec:exotica} discusses how exotic models involving lepton and photon interactions can benefit from such capability. For instance, the characteristic Chern-Simons term involving photons and axion-like particles can be probed to high precision at the CEPC. At Tera-$Z$ and Higgs factory runs, the rare decay of the Higgs boson and electroweak gauge boson can be of particular interest. The high design luminosity at these runs offers a powerful probe into right-handed neutrino, extended scalars, exotic lepton models, etc. Noticeably the high sensitivity with leptons also empowers precision tests on the SM lepton interactions, such as the dipole moment of $\mu$ and $\tau$ leptons, effective non-standard neutrino interactions. These measurements provide a complementary test of any underlying BSM theory, generating large corrections to lepton form factors. In addition, strong interest in quantum entanglement has appeared in collider physics. Section~\ref{sec:exotica} also includes newly completed analyses on its application in leptonic Higgs boson decays. 

Recent developments in global fitting techniques, like GAMBIT, etc., greatly strengthen our ability to navigate through vast new physics model spaces. Global fits maximize data's theoretical output by efficiently analyzing and comparing a large number of different models. This powerful computation capability can quickly identify models or parameter spaces with the highest priority, making robust references for theory interpretation. In Section~\ref{sec:global git}, we include recent global-fitting analyses for well-motivated SMEFT, 2HDM and SUSY models, based on the CEPC's design specifics at $Z$-pole, Higgs factory, and $t\bar{t}$ runs.

\medskip

\clearpage
\section{Description of CEPC facility} 
\label{sec:detector}

The CEPC accelerator is designed to provide an instantaneous luminosity in the range of 5--192 ($\times 10^{34}\,{\rm cm}^{-2}{\rm s}^{-1}$) at center-of-mass energies from 91 to 240 GeV~\cite{CEPCStudyGroup:2023quu}.
According to its nominal operation plan, which includes 10 years of data taking at the Higgs factory mode, 2 years at the $Z$ pole, and 1 year for the $WW$ threshold scan, the CEPC is expected to deliver approximately 4 million Higgs boson events, 4 trillion (Tera) $Z$ boson events, and 1 billion (Giga) $W$ boson events.
The massive production of these elementary particles not only enables high-precision measurements of SM parameters but also provides fertile ground for uncovering potential deviations induced by new physics.
The new physics search at the CEPC primarily relies on three portals: the Higgs portal, the $Z$ portal, and direct searches for BSM signatures.
To effectively handle the high event rates and complex event topology characteristic of these physics processes, the detector must meet the following key performance requirements:

\begin{itemize}
    \item \textbf{High stability and calibration}
    
    The CEPC operates in a high-luminosity environment, producing a vast number of events that demand exceptional detector stability. Consistent performance under such conditions is crucial for minimizing systematic uncertainties. To ensure high data quality, precise monitoring and calibration systems are required so that rare signals—such as deviations in Higgs branching ratios or rare Z decays—are not obscured by noise or beam-induced backgrounds.
    
    \item \textbf{Large acceptance}
    
    The acceptance encompasses not only the solid angle coverage, but also the energy and momentum thresholds for reconstructing final state particles. In considering the high event rate, the acceptance shall also be extended to the time dimension.
    In general, a coverage up to $|\cos\theta| < 0.99$ is required to ensure capture of particles over a wide angular range, which is critical for reconstructing full event topologies in Higgs, $Z$, and BSM processes. 
    Low energy and momentum thresholds of $\mathcal{O}(100)$~MeV are essential for detecting soft particles, such as photons and pions from heavy hadron decays or exotic BSM signatures. 
     
    \item \textbf{Excellent intrinsic subdetector resolution}

    The intrinsic resolution of each subdetector is crucial for all physics measurements, particularly for new physics, as it underpins the precision and accuracy required to detect and analyze novel phenomena.
    Typically, the intrinsic momentum resolution of the tracker should reach 0.1\% level in the barrel region. The intrinsic energy resolutions of the ECAL and HCAL are expected to be better than $3\%/\sqrt{E\text{(GeV)}}$ and $40\%/\sqrt{E\text{(GeV)}}$, respectively.
    Moreover, to efficiently reconstruct decay vertices of $\tau$ lepton and heavy-flavor hadrons, the vertex position resolution is expected to be better than 5~$\upmu$m, with the vertex detector placed sufficiently close to the interaction point~\cite{Vcb_Hao_Talk}.

    \item \textbf{Better Boson Mass Resolution (BMR)}

    BMR refers to the relative mass resolution of hadronically decayed massive bosons (e.g. Higgs, $Z$, and $W$ bosons)~\cite{CEPCStudyGroup:2018ghi}. 
    It is a key metric to quantify the overall reconstruction performance of hadronic systems.
    For new physics searches through the Higgs and $Z$ portals, BMR is particularly important, as the majority of Higgs and $Z$ bosons decay hadronically.
    It is also directly relevant to the reconstruction of missing energy and momentum, which is critical for dark matter searches.
    Generally, BMR $<$ 4\% is required in Higgs measurements, such as the Higgs width determination via $e^+e^-\to\nu\bar{\nu}H(\to b\bar{b})$~\cite{Zhao:2018jiq}, the measurement of $H\to \tau^+\tau^-$ via $Z(\to q\bar{q})H(\to \tau^+\tau^-)$~\cite{Yu:2020bxh}, and the study of Higgs invisible decays via $Z(\to q\bar{q})H(\to \rm invisible)$~\cite{CEPCStudyGroup:2018ghi}.
    While achieving a BMR better than 3\% is highly beneficial for precision Higgs studies, it also significantly enhances sensitivity to new physics signals involving boosted hadronic final states or missing energy signatures.
    
    \item \textbf{Excellent particle flow reconstruction and particle identification}

    New physics measurements often involve complex final states with overlapping jets, boosted objects, soft decay products, and significant missing energy. To effectively disentangle these signatures from overwhelming SM backgrounds, excellent particle flow reconstruction~\cite{Marshall:2013bda,Ruan:2013rkk} and particle identification (PID) are indispensable. High-fidelity reconstruction of individual particles enables precise jet substructure analysis, accurate missing energy measurement, and efficient lepton isolation---all of which are critical in processes such as dark matter searches, exotic Higgs decays, and LLP detection.
    
    In addition, excellent pattern recognition capabilities are essential for handling dense event topologies, particularly in scenarios with high-multiplicity final states or collimated decay products. 
    Robust pattern recognition ensures reliable track-cluster association, secondary vertex reconstruction, and object separation, directly impacting the sensitivity to rare or unconventional signatures predicted by many BSM models.

    \item \textbf{Jet origin identification (JOI)}
    
    JOI is the procedure to determine the type of quark or gluon from which a jet originates. Typically, 11 types are considered: $b$, $\bar{b}$, $c$, $\bar{c}$, $s$, $\bar{s}$, $u$, $\bar{u}$, $d$, $\bar{d}$, $g$. As a natural extension of jet flavor tagging, quark-gluon jet discrimination, and jet charge measurements, JOI provides a powerful tool for new physics searches. 
    Many BSM models predict exotic decays or final states that preferentially produce jets from specific quark flavors or exhibit flavor asymmetries. By identifying the initiating parton flavor on a jet-by-jet basis, JOI enables the construction of new discriminating observables that are sensitive to such effects.
    In addition, JOI can help reduce SM backgrounds, particularly in multijet final states, where the jet flavor composition often differs significantly between signal and background processes.

\end{itemize}

Since the proposal of the CEPC, multiple detector concepts have been proposed and optimized to address these requirements, demonstrating state-of-the-art performance. These developments have progressed alongside advances in detector technology and reconstruction algorithms. 
We refer to three benchmark detector concepts, which are used in the simulations in this white paper, providing reference performance for relevant physics potential studies.

\begin{figure}[t]
    \centering
    \includegraphics[scale=.45]{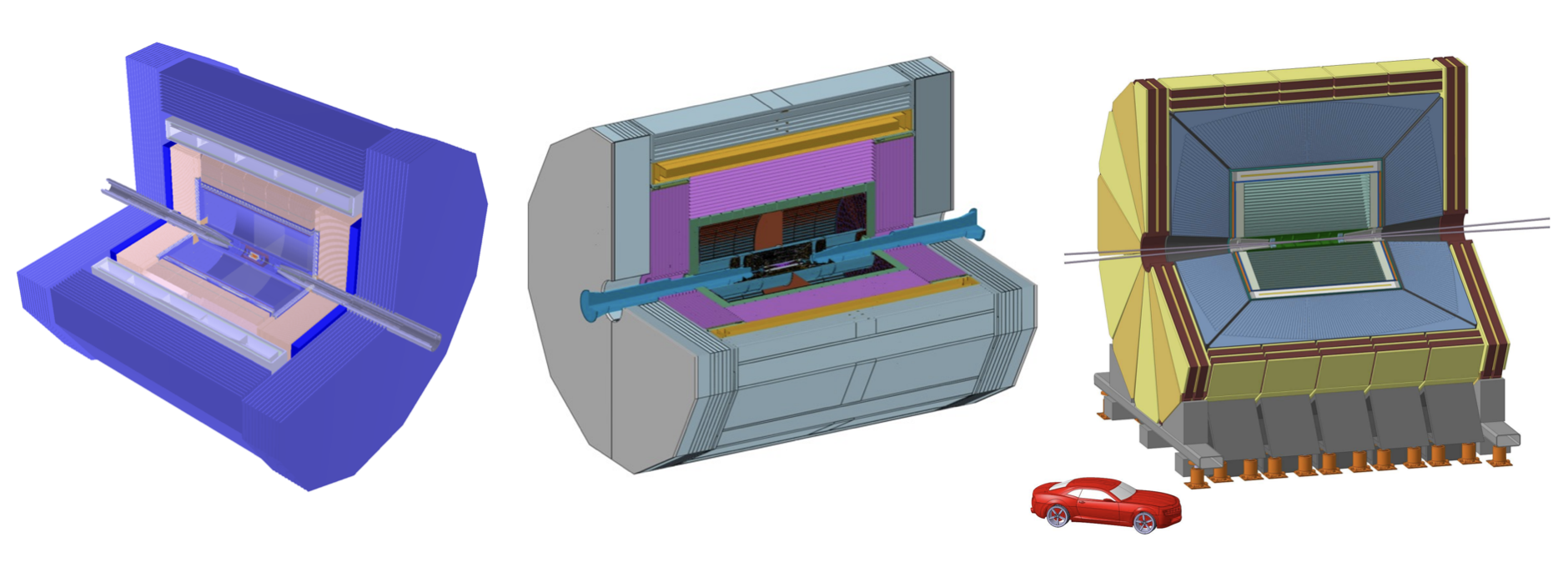}
    \caption{Schematic layouts of \textbf{LEFT:} CEPC CDR detector~\cite{CEPCStudyGroup:2018ghi}, \textbf{MIDDLE:} Ref-TDR detector~\cite{refTDR_Mingshui_talk}, and \textbf{RIGHT:} IDEA detector~\cite{IDEA_Snowmass_2021}.}
    \label{fig:cepc_detectors}
\end{figure}

Let's start with the CEPC CDR detector~\cite{CEPCStudyGroup:2018ghi}, which follows the particle flow principle, as shown in the left panel of Fig.~\ref{fig:cepc_detectors}.
It features a high-precision tracking system, a high-granularity calorimeter system, and a high magnetic field of 3 Tesla.
By virtue of the particle-flow-oriented design, the CDR detector delivers excellent tracking efficiency, lepton identification, and precise hadronic reconstruction, providing a solid basis for new physics studies.
Specifically, the tracking system demonstrates an efficiency close to 90\% and a relative momentum resolution approaching $\mathcal{O}(10^{-3})$ for tracks with momenta above 1 GeV in the barrel region.
The photon energy resolution reaches $17\%/\sqrt{E\text{(GeV)}} \oplus 1\%$, achieved by a sampling Si-W ECAL, which features the high granularity critical for particle flow reconstruction.
For PID, the CDR detector offers a $K/\pi$ separation better than 2\,$\sigma$ in the momentum range up to 20 GeV by combining time-of-flight (TOF) and $dE/dx$ information. 
In inclusive $Z\to q\bar{q}$ events, the $K^\pm$ identification efficiency and purity both exceed 95\%~\cite{Zhu:2022hyy}.
For hadronic systems, a BMR of 3.8\% is achieved for hadronically decayed $W$, $Z$, and Higgs bosons. This allows $W$/$Z$ separation beyond 2\,$\sigma$ in hadronic final states and improves missing energy/momentum resolution, which is crucial for new physics searches.

After the release of the CEPC CDR, intensive detector R\&D efforts have continued to address the CEPC physics requirements, leading to the development of the Ref-TDR detector~\cite{refTDR_Mingshui_talk, refTDR_Jianchun_talk}, shown in the middle panel of Fig.~\ref{fig:cepc_detectors}.
It demonstrates significant improvements in electromagnetic (EM) and hadronic energy resolutions, PID performance, and vertexing. 
The Ref-TDR detector employs a particle-flow-compatible homogeneous crystal ECAL to enhance the EM resolution to $1.3\%/\sqrt{E\text{(GeV)}} \oplus 0.7\%$.
A high-density glass-scintillator HCAL is utilized to improve the hadronic energy resolution to $30\%/\sqrt{E} \oplus 6.5\%$, nearly a factor of two better than the CDR performance~\cite{Hu:2023dbm}.
The Ref-TDR detector also features a pixelated TPC that provides precise $dE/dx$~\cite{Zhu:2022hyy,She:2023puo} and $dN/dx$~\cite{Cuna:2021sho} measurements, both of which are essential for PID. 
Furthermore, the outermost silicon tracker integrates TOF capability with a resolution of 50 ps per MIP, further enhancing PID performance.
The vertex detector adopts stitching technology~\cite{4th_vtx_talk} to significantly reduce the material budget. 
Table~\ref{tab:detector_parameter} summarizes some key parameters and performance of the Ref-TDR detector.

\begin{table}[t]
\begin{center}
\resizebox{.45\textwidth}{!}{\begin{tabular}{c|c}
    \toprule[1pt]
    Subdetector  &  Parameter and Performance \\
    \hline
    Vertex  & \makecell{Inner radius of 11 mm\\ Material budget of 0.77\% X$_{0}$} \\
    \hline
    TPC  &  \makecell{0.5 mm $\times$ 0.5 mm readout\\ $dN/dx$ resolution 3\%} \\
    \hline
    TOF  &  $\sigma_{T}$ = 50 ps per MIP  \\
    \hline
    ECAL &  $\sigma_{E}/E$ = $1.3\%/\sqrt{E} \oplus 0.7\%$  \\
    \hline
    HCAL &  $\sigma_{E}/E$ = $30\%/\sqrt{E} \oplus 6.5\%$  \\
    \hline
    BMR & 3.87\% \\
    \bottomrule[1pt]
\end{tabular}}
\end{center}
\caption{Key parameters and performance of Ref-TDR detector~\cite{refTDR_Mingshui_talk}.}
\label{tab:detector_parameter}
\end{table}

An alternative detector concept known as IDEA~\cite{IDEA_Snowmass_2021}, with schematic shown in the right panel of Fig.~\ref{fig:cepc_detectors}, is also utilized in some physics potential studies. In comparison to the CDR and Ref-TDR detectors, the IDEA detector incorporates a dual readout calorimeter system to attain superior energy resolution for both EM and hadronic showers. Moreover, the IDEA detector operates with a reduced magnetic field of 2 Tesla while compensating for this reduction by offering a larger tracking volume.

Besides the aforementioned detectors, several far detectors are proposed to search for LLP. Compared to the main detector near the collision point, far detectors can significantly enhance the ability to detect LLP, particularly when the decay length of the LLP reaches on the order of hundreds of meters. Examples of such detector proposals, such as FADEPC and LAYCAST, are discussed in detail in section~\ref{sec:LLP}.
Far detectors are expected to work synergistically with the main detector, greatly extending the CEPC's capability for exploring new physics.

In addition to the detector technology, reconstruction algorithms have progressed significantly. Leveraging advanced machine learning algorithms, new concepts such as one-to-one (1-1) correspondence reconstruction~\cite{Wang:2024eji} and jet origin identification (JOI)~\cite{Liang:2023wpt} have been proposed and greatly enhanced reconstruction performance.

\begin{figure}[!t]
    \centering
    \adjustbox{valign=c}{\includegraphics[scale=0.405]{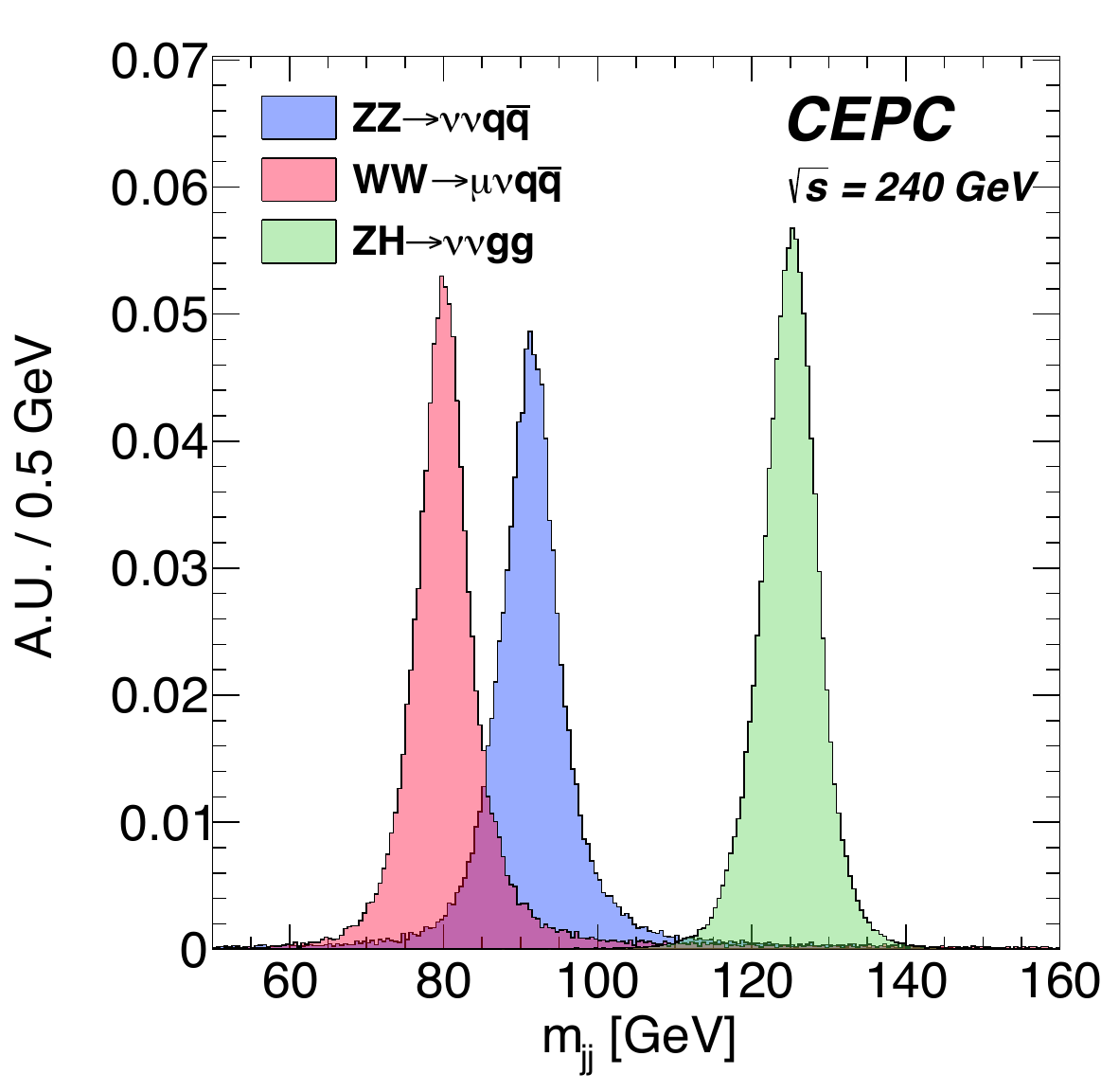}}
    \adjustbox{valign=c}{\includegraphics[scale=0.33]{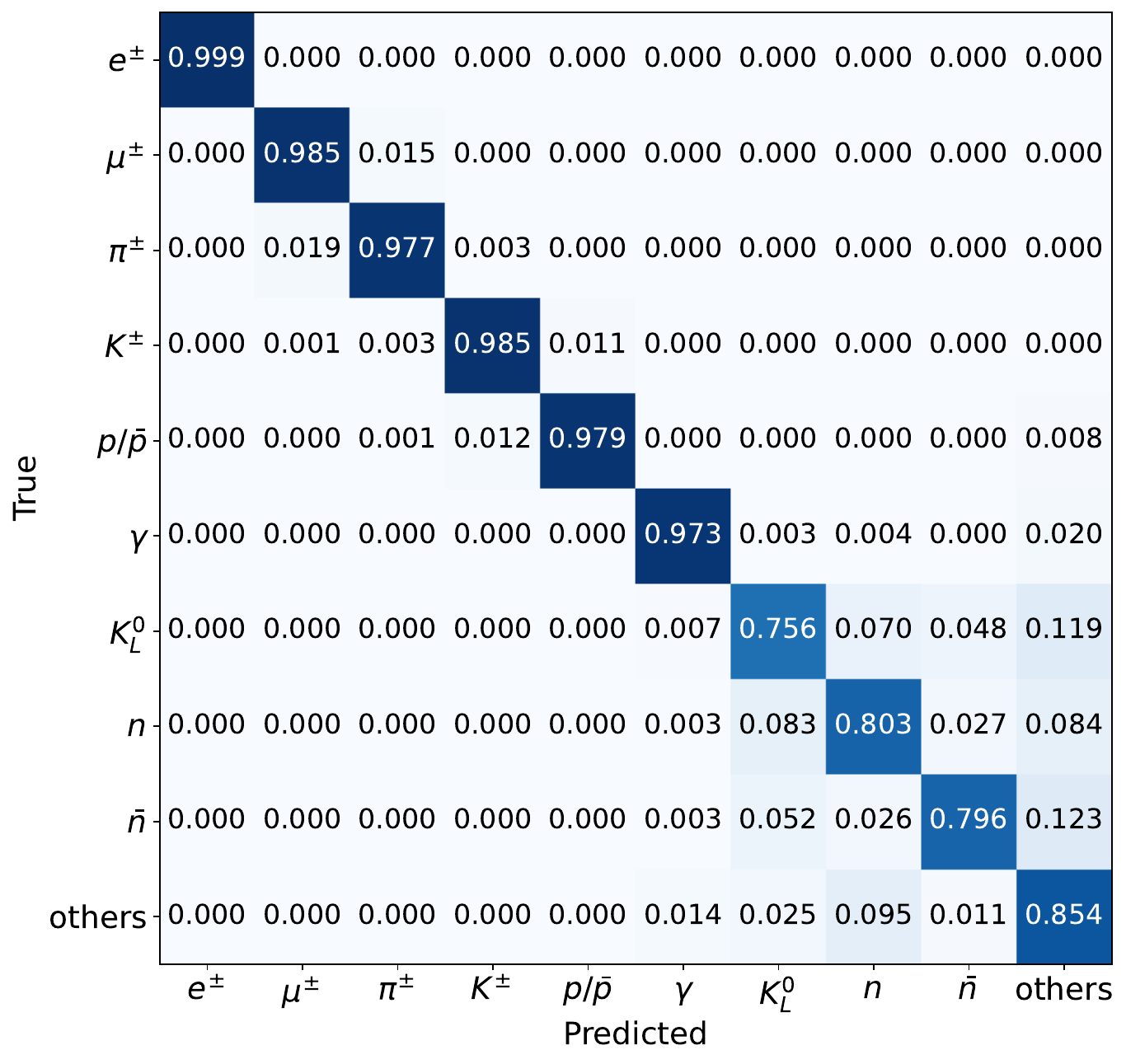}}
    \caption{
    \textbf{LEFT:} Invariant mass distributions of hadronically decayed Higgs, $W$, and $Z$ bosons derived by 1-1 correspondence reconstruction.
    \textbf{RIGHT:} Confusion matrix of well-reconstructed particles identification.
    Both plots are taken from Ref.~\cite{Wang:2024eji}.
    }
    \label{fig:1-1}
\end{figure}

1-1 correspondence reconstruction, as an ultimate goal and a natural extension of particle flow, aims to establish a 1-1 correspondence between visible and reconstructed particles, offering a high-quality and holistic description of physics events.
However, due to limitations in pattern recognition, current particle flow reconstruction still suffers from significant confusion effects---including fake particles, failures in track-cluster matching, and particle loss caused by shower overlaps---which severely violate the 1-1 correspondence relationship.
By leveraging the powerful pattern recognition capability of machine learning and a novel detector design featuring a 5-dimensional (5D) calorimeter which provides extra time information, 1-1 correspondence reconstruction has been realized at the full simulation level in the CEPC environment~\cite{Wang:2024eji}.
In the benchmark process of Higgs decaying to di-jets, over 90\% of visible energy can be successfully mapped to well-reconstructed particles that not only maintain a one-to-one correspondence relationship but are also associated with the correct combination of cluster and track, achieving a BMR less than 3\%, as shown in the left panel of Fig.~\ref{fig:1-1}.
Performing simultaneous identification on these well-reconstructed particles, efficiencies of 97\% to nearly 100\% for charged particles ($e^{\pm}$, $\mu^{\pm}$, $\pi^{\pm}$, $K^{\pm}$, $p/\bar{p}$) and photons ($\gamma$), and 75\% to 80\% for neutral hadrons ($K_L^0$, $n$, $\bar{n}$) are observed.
For physics measurements of Higgs to invisible and exotic decays---golden channels for probing new physics---1-1 correspondence could enhance discovery power by 10\% to up to a factor of two.

\begin{figure}[t]
    \centering
    \includegraphics[width=0.45\textwidth]{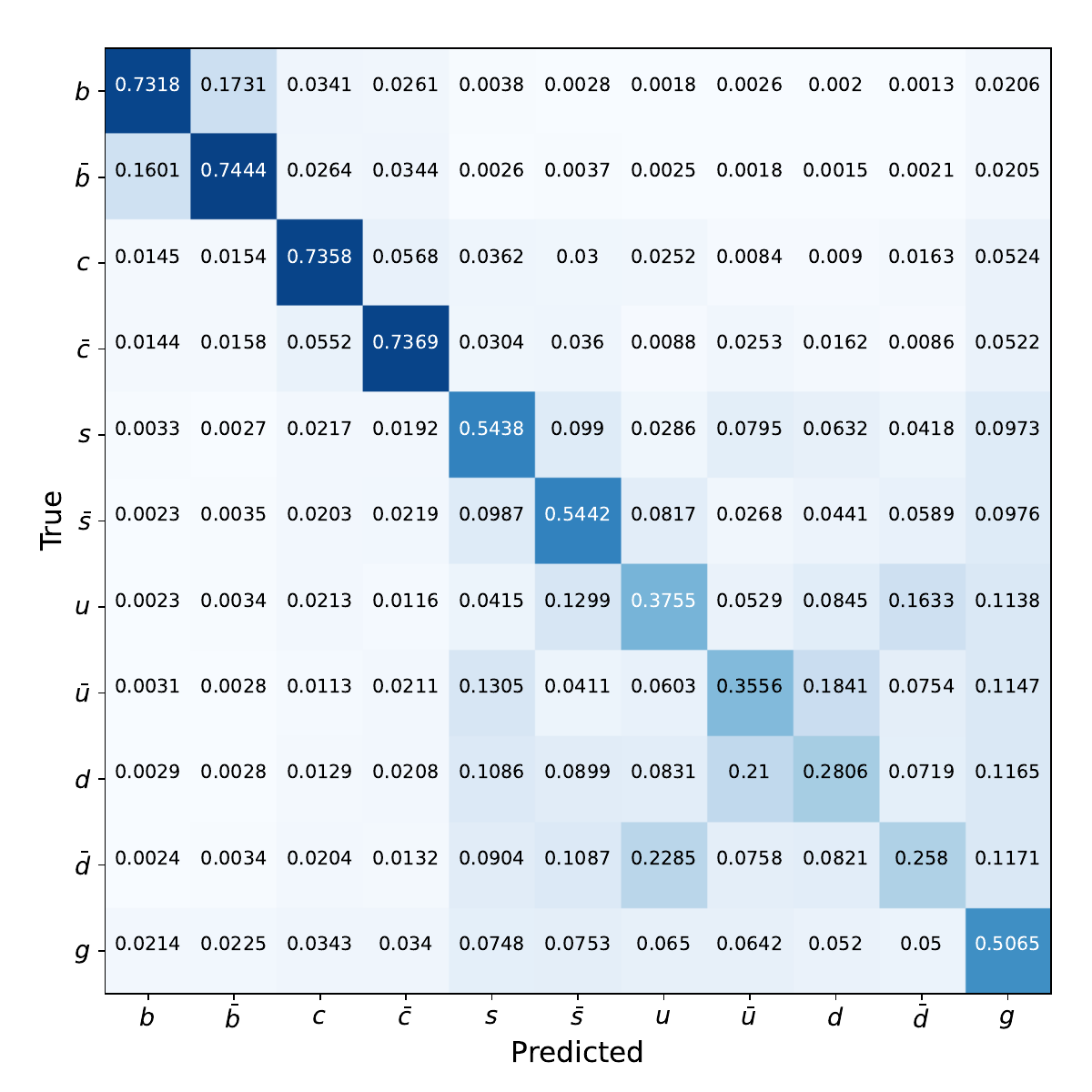}
    \includegraphics[width=0.45\textwidth]{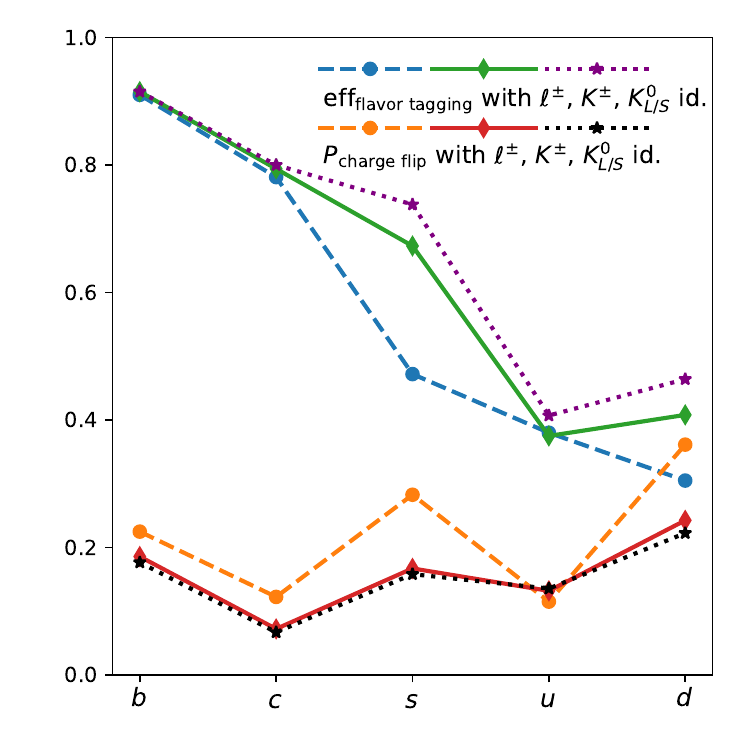}
    \caption{Jet origin identification performance of full simulated Higgs/$Z$ to di-jet processes with CEPC conceptual detector~\cite{Liang:2023wpt}. \textbf{LEFT:} The confusion matrix $M_{11}$ with perfect identification of leptons and charged hadrons. \textbf{RIGHT:} Jet flavor tagging efficiency and charge flip rates for quark jets with different scenarios of particle identification: with only lepton identification, plus identification of charged hadrons, plus identification of neutral kaons.}
    \label{fig:JOI}
\end{figure}

Similarly benefiting from machine learning, the concept of JOI is realized at the full simulation level~\cite{Liang:2023wpt} using the CEPC CDR detector and advanced algorithms, including the Arbor~\cite{Ruan:2013rkk} particle flow reconstruction and the ParticleNet~\cite{Qu:2019gqs} deep learning architecture.
JOI enables the simultaneous identification of $b$, $c$, and $s$ quarks with efficiencies ranging from 70\% to 90\%, and $u$, $d$ quarks with efficiencies around 40\%. Charge misidentification between quarks and antiquarks is controlled at the 10--20\% level, with an ideal lepton and charged hadron identification, as illustrated in Fig.~\ref{fig:JOI}.
This capability significantly extends the flavor sensitivity of hadronic final states and brings substantial improvements to measurements involving rare and exotic Higgs decays---key channels for exploring BSM physics. By precisely identifying the flavor composition of jets, JOI provides powerful handles to suppress backgrounds and isolate potential signals of new physics.

\clearpage
\section{Exotic Higgs potential and Exotic Higgs/Z/top decays}
\label{sec:Higgsexotic}
\subsection{Introduction}
The novel physics phenomena may manifest through the exotic decay channels of the Higgs boson, $Z$ boson and top quark. Especially,  it may reveal the structure of the Higgs portal, generic Higgs potential or new effective operators. The research effort devoted to investigating Higgs/$Z$/top exotic decays will effectively improve the precision of the coupling measurements for relevant particles and will constitute an indispensable element of the scientific agenda for the prospective Higgs factories. These capacities could be significantly enhanced with the introduction of cutting-edge machine learning technologies~\cite{Li_2022,Liang:2023wpt}. In the following sections, we will review some representative investigations for the relevant exotic decays of Higgs, $Z$ boson and top quark. 

\subsection{Model-independent Sensitivity to Exotic Higgs decays}

\begin{figure}[!tbp]
\centering
\includegraphics[scale=0.55,clip]{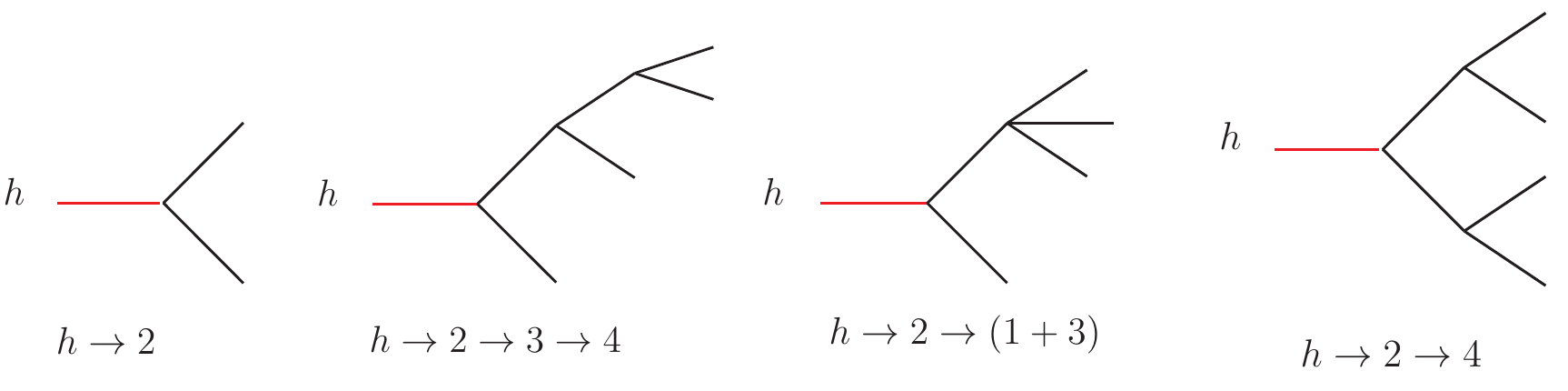}
\caption{\label{fig:topo}Representative topologies of the Higgs exotic decays~\cite{Liu:2016zki}. }
\end{figure}

A comprehensive assessment of the sensitivity of lepton colliders to exotic decay channels of the Higgs boson into various final states was presented in Ref.~\cite{Liu:2016zki}, with a particular emphasis on the channels that face considerable challenges at hadron Colliders. The findings of the investigation indicate that lepton colliders exhibit notable sensitivity potential in the discussed decay channels. The present analysis is concentrated on the two-body Higgs decays into BSM particles, denoted as $X_i$, through the decay process $h\to X_1 X_2$. These particles are permitted to undergo further decays, potentially resulting in four-body final states at most. The cascade decay modes are systematically categorized into four distinct cases, as illustrated in Fig.~\ref{fig:topo}. A broad class of the underlying BSM theoretical frameworks, including but not limited to singlet scalar extensions, two-Higgs-doublet models, SUSY models, various Higgs portals, and gauge extensions of the SM, as referenced in~\cite{Curtin:2013fra,Liu:2016zki,deFlorian:2016spz,Cepeda:2021rql}, provide the theoretical motivation for the exploration of these exotic decay channels.

\begin{figure}[!bp] 
\centering
\includegraphics[width=1.0\textwidth,height=0.3\textheight]{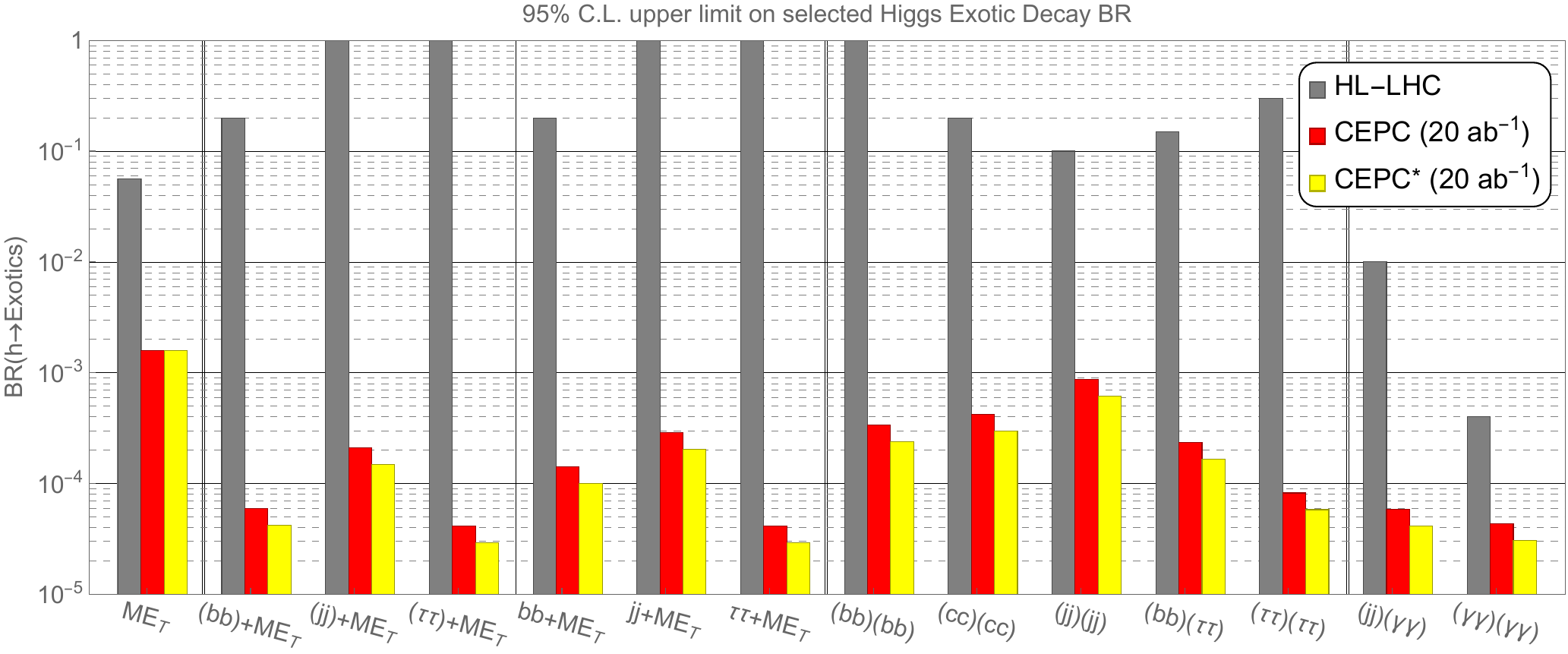}
\caption{The 95\% C.L. upper limit on selected Higgs exotic decay branching fractions at HL-LHC and CEPC, based on Ref~\cite{Liu:2016zki}.}\label{fig:ExoticHiggssummary}
\end{figure}

At the CEPC $240$~GeV, the predominant mechanism for the Higgs boson production is the associated production with a $Z$ boson. 
The decay of the $Z$ boson into detectable final states facilitates the identification of the Higgs boson via the recoil mass method. Implementing a selection criterion centered around the peak of the recoil mass significantly reduces background processes from SM processes. Several channels have been investigated in Ref.~\cite{Liu:2016zki}. The results of which are shown in Fig.~\ref{fig:ExoticHiggssummary}. The analysis provides the projected exclusion limits at the 95\% C.L. for the CEPC with an integrated luminosity of 20~ab$^{-1}$. Additionally, the forecasted sensitivities for the LHC represented by gray bars are included. The projections for the LHC are based on the most current sensitivity estimates. However, several of these projections are either non-existent or notably conservative. More contemporaneous investigations, such as those presented in Ref.~\cite{Shelton:2021xwo} concerning the decay $h\to 4\tau$, and Ref.~\cite{Kato:2022} about the decay $h\to 4b$, have shown the consistency of these sensitivity projections.

\begin{table}[!ht]
\setlength{\tabcolsep}{20pt}
\begin{center}
    \begin{tabular}{lcccc}
\hline
\multirow{2}{*}{Decay channel} & \multicolumn{4}{c}{Upper limit on the branching ratio at 95\% C.L.}            \\ \cline{2-5} 
                               & HL-LHC            & CEPC              & ILC               & FCC-ee            \\ \cline{1-5}
ME$_T$                         & 0.056             & 0.0028            & 0.0025            & 0.005             \\
$(b\bar{b})+$ME$_T$            & 0.2               & $1\times 10^{-4}$ & $2\times 10^{-4}$ & $5\times 10^{-5}$ \\
$(jj)+$ME$_T$                  & -                 & $5\times 10^{-4}$ & $5\times 10^{-4}$ & $2\times 10^{-4}$ \\
$(\tau\tau)+$ME$_T$            & 1                 & $8\times 10^{-4}$ & $1\times 10^{-3}$ & $3\times 10^{-4}$ \\
\hline
$b\bar{b}+$ME$_T$              & 0.2               & $3\times 10^{-4}$ & $4\times 10^{-4}$ & $1\times 10^{-4}$ \\
$jj+$ME$_T$                    & -                 & $5\times 10^{-4}$ & $7\times 10^{-4}$ & $2\times 10^{-4}$ \\
$\tau\tau+$ME$_T$              & -                 & $8\times 10^{-4}$ & $1\times 10^{-3}$ & $3\times 10^{-4}$ \\
\hline
$(b\bar{b})(b\bar{b})$         & 0.2               & $4\times 10^{-4}$ & $9\times 10^{-4}$ & $3\times 10^{-4}$ \\
$(c\bar{c})(c\bar{c})$         & 0.2               & $8\times 10^{-4}$ & $1\times 10^{-3}$ & $3\times 10^{-4}$ \\
$(jj)(jj)$                     & 0.1               & $1\times 10^{-3}$ & $2\times 10^{-3}$ & $7\times 10^{-4}$ \\
$(b\bar{b})(\tau\tau)$         & 0.15              & $4\times 10^{-4}$ & $6\times 10^{-4}$ & $2\times 10^{-4}$ \\
$(\tau\tau)(\tau\tau)$         & $0.2\sim 0.4$     & $1\times 10^{-4}$ & $2\times 10^{-4}$ & $5\times 10^{-5}$ \\
$(jj)(\gamma\gamma)$           & 0.01              & $1\times 10^{-4}$ & $2\times 10^{-4}$ & $3\times 10^{-5}$ \\
$(\gamma\gamma)(\gamma\gamma)$ & $4\times 10^{-4}$ & $1\times 10^{-4}$ & $1\times 10^{-4}$ & $3\times 10^{-5}$ \\
\hline
\end{tabular}
\end{center}
\caption{The projection of upper limits at 95\% C.L. on selected exotic decay branching ratios for various channels at the HL-LHC is compared to those at various lepton colliders \cite{Liu:2016zki}, as shown in Fig.~\ref{fig:ExoticHiggssummary}. }\label{table:summarybar}
\end{table}

The LHC is expected to impose stringent constraints on a large number of decay channels that involve muons, electrons, and photons. In the context of the more formidable channels that are dependent on the detection of jets, heavy quarks, and tau leptons, the prospective enhancements in sensitivity compared to current LHC projections span from one to four orders of magnitude. This substantial improvement is attributed to the reduced QCD background and the effectiveness of Higgs boson identification through the recoil mass method that is expected to be employed at forthcoming lepton collider facilities. 
Specifically, for exotic decays of the Higgs boson without missing energy, the anticipated improvements in detection sensitivity range between two to three orders of magnitude. An exception is noted for the $(\gamma\gamma)(\gamma\gamma)$ decay channel, which is projected to enhance merely one order of magnitude. This particular channel at the LHC benefits from the ability to reconstruct the Higgs boson mass from the final state particles, which significantly aids in the discrimination between signal and background. Moreover, decay channels involving electrons, muons, and photons, which are considered to be relatively clean signatures at the LHC, stand to gain from the higher statistics that will be provided by the HL-LHC, thereby utilizing the higher event counts to improve statistical precision. Table~\ref{table:summarybar} summarises the upper limit on the branching ratios for different channels for the HL-LHC, CEPC, ILC and FCC-ee. 

\subsection{Exotic Higgs potential}
The exotic Higgs potential is often realized through the extension of the SM Higgs potential, leading to the SM-$s$ model. This model is a minimal theoretical extension that introduces a new scalar particle, typically denoted as $s$, which can mix with the Higgs boson. The Lagrangian for this extension is given by:

\begin{equation}
\mathcal{L} = \mathcal{L}_\text{kin} + \frac{\mu_s^2}{2} S^2 - \frac{\lambda_s}{4!} S^4 - \frac{\kappa}{2} S^2 |H|^2 + \mu^2 |H|^2 - \lambda |H|^4,
\end{equation}
where $ \mathcal{L}_\text{kin} $ represents the kinetic terms, 
$ \mu_s^2 $ and $ \lambda_s $ are parameters associated with the new scalar $ s $, 
$ \kappa $ describes the mixing between $ s $ and the Higgs boson $ H $, 
and $ \mu^2 $ and $ \lambda $ are the mass and self-coupling parameters of the Higgs boson, respectively.

The SM-$s$ model \cite{Curtin:2013fra} is motivated by theoretical considerations such as naturalness, 
which tackles the hierarchy problem between the Higgs mass and the Planck scale; 
potential interactions with dark matter, linking the visible 
and dark sectors of the universe; and the electroweak phase transition, 
which could influence the early universe's dynamics and baryon asymmetry. 

The new scalar $ s $ can lead to exotic Higgs decays, such as $ h \to ss $, 
which are not present in the SM. These decays are of particular interest 
because they could provide evidence for physics beyond the SM 
and help resolve some of the current puzzles in particle physics.

\begin{figure}[!tp]
     \centering
     \includegraphics[width =0.65\textwidth,height=0.4\textheight]{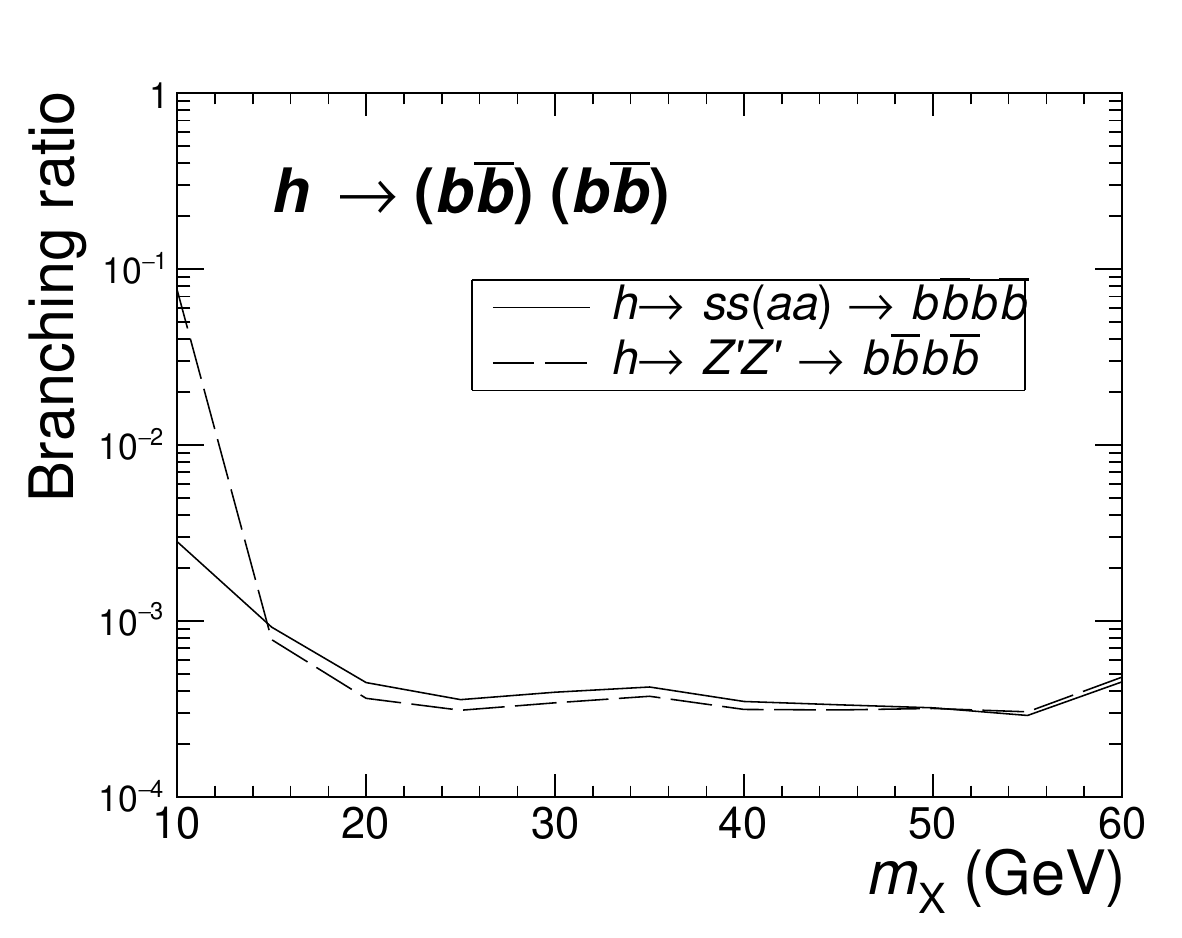}
     \caption{The 95\% C.L. exclusive bound on $\Br(h \to (\bar bb)(\bar bb)) $, based on Ref.~\cite{Liu:2016zki}.}\label{bbbb}
 \end{figure}

At the LHC, detecting the decay $ h \to ss $ is challenging due to significant background processes. 
However, the projected sensitivity of the HL-LHC for the $bb\tau\tau$ final state is $ \Br(h \to ss) < 2.8 \times 10^{-2} $ 
according to CMS projections \cite{CMS:2019rsy}. 
At the future electron-positron colliders, the sensitivity for the $bbbb$ final state is much more promising \cite{Liu:2016zki}.
With at least three b-tagged jets required in the final state,  
the b-tagging efficiency can be conservatively chosen to be 80\%, 
and the charm mis-tagging rate and the light flavor mis-tagging rate can be set to be 9\% and 1\%, respectively.
As shown in Fig.~\ref{bbbb}, the future lepton collider with 5 ab$^{-1}$ integrated luminosity can exclude branching 
fractions of $h \to ss \to (\bar bb)(\bar bb)$ down to $3\times 10^{-4} \sim 4\times 10^{-4}$, in the range of mediator mass $20$~GeV $<m_s< 60$~GeV.

The enhanced sensitivity at electron-positron colliders is attributed to their cleaner experimental environment 
and the precision can be significantly improved with high statistics in the next-generation facilities. 
This makes them ideal for studying rare processes and searching for new physics, such as the exotic Higgs decays discussed here.

\subsection{Higgs exotic decays in supersymmetry}
Numerous supersymmetric extensions of the Standard Model have the potential to produce the exotic decay channels of the Higgs boson. 
In recent years, the next-to-minimal supersymmetric standard model (NMSSM) 
and its slightly modified version, semi-constrained NMSSM (scNMSSM) have attracted significant attention for providing theoretically compelling explanations of Higgs boson phenomena. 
In particular, a comprehensive analysis has been undertaken to execute two primary investigations 
on the exotic decays of the Higgs boson at the CEPC.

First, there has been a concerted effort to clarify the nature of the invisible decay channels of the 125 GeV Higgs boson 
into potential dark matter (DM) candidates, specifically into pairs of the lightest neutralinos 
$ h \rightarrow \tilde{\chi}^0_1 \tilde{\chi}^0_1 $ \cite{Wang:2020tap,Wang:2020dtb,Wang:2020xta}. 
Theoretical predictions suggest the existence of four distinct funnel-annihilation processes 
for the lightest supersymmetric particle (LSP) $\tilde{\chi}^0_1$, 
corresponding to intermediate states $h_2$, $Z$, $h_1$, and $a_1$. 
The composition of the LSP is constrained to either a singlino-dominant or a higgsino-dominant form. 
In the scenario where the LSP is predominantly singlino, it is supposed to achieve the observed dark matter relic density. 
As demonstrated in Ref.~\cite{Wang:2020dtb}, the branching fraction of the Higgs boson's invisible decay could be as low as approximately $10^{-5}$. While this fraction exceeds the capability of the HL-LHC, it can be reached through measurements at the CEPC.
Meanwhile, a higgsino-dominant LSP is predicted to yield an insufficient relic density 
and is associated with a substantial branching fraction for the Higgs invisible decay, 
leaving it in the investigation capability of the CEPC.

\begin{table}[!htb]
  \centering
  \setlength{\tabcolsep}{3.5mm}
    \begin{tabular}{ccccc}
    \hline
    \multirow{2}*{Decay Mode}&
    \multicolumn{4}{c}{Future colliders}\cr
    \cline{2-5}
    &HL-LHC&CEPC&FCC-$ee$&ILC\cr
    \hline
\hline
($b\bar{b}$)($b\bar{b}$)&$650\text{fb}^{-1}$(@II)&$0.42\text{fb}^{-1}$(@III)&$0.41\text{fb}^{-1}$(@III)&$0.31\text{fb}^{-1}$(@II)\cr
($jj$)($jj$)&-&$21\text{fb}^{-1}$(@II)&$18\text{fb}^{-1}$(@II)&$25\text{fb}^{-1}$(@II)\cr
($\tau^+\tau^-$)($\tau^+\tau^-$)&-&$0.26\text{fb}^{-1}$(@III)&$0.22\text{fb}^{-1}$(@III)&$0.31\text{fb}^{-1}$(@III)\cr
($b\bar{b}$)($\tau^+\tau^-$)&$1500\text{fb}^{-1}$(@II)&$4.6\text{fb}^{-1}$(@II)&$3.6\text{fb}^{-1}$(@II)&$4.4\text{fb}^{-1}$(@II)\cr
($\mu^+\mu^-$)($\tau^+\tau^-$)&$1000\text{fb}^{-1}$(@II)&-&-&-\cr
\hline
    \end{tabular}
      \caption{The minimum integrated luminosity for discovering the exotic Higgs decay at the future colliders \cite{Liu:2016zki}, where the ``@I, II, III'' means the three different scenarios. Scenario I: $h_2$ is SM-like Higgs, and the light scalar $a_1$ is CP-odd; Scenario II: $h_1$ is SM-like Higgs, and the light scalar $a_1$ is CP-odd; Scenario III: $h_2$ is SM-like Higgs, and the light scalar $h_1$ is CP-even.}
  \label{t1}
\end{table}

Second, investigations have been directed towards the decay of the Higgs boson into lighter CP-odd or CP-even Higgs states, 
with particular attention given to the processes $ h_2 \rightarrow a_1 a_1 $, $ h_1 \rightarrow a_1 a_1 $, 
and $ h_2 \rightarrow h_1 h_1 $ manifesting in final states comprising four bottom quarks ($4b$), four jets ($4j$), 
a pair of bottom quarks and a pair of tau leptons ($2b2\tau$), 
and four tau leptons ($4\tau$) respectively \cite{Ma:2020mjz}. 
Three distinct scenarios are compared and their relative sensitivities are evaluated 
in the context of detecting these exotic Higgs decay modes at the HL-LHC as well as at forthcoming lepton colliders, including the CEPC. 
The predominant mechanism for the production of the SM-like Higgs boson is identified as the $ \rm Zh $ channel. 
Empirical findings suggest that the most efficient strategy for the detection of these exotic decays at the CEPC 
would be via the $4\tau$ channel, with the requisite minimum integrated luminosity for potential discoveries 
being as modest as 0.26 $ \text{fb}^{-1} $. 
Table \ref{t1} presents the minimum required integrated luminosities for the detection of these exotic Higgs decays 
across various experimental setups, including the HL-LHC, CEPC, FCC-ee, and ILC. 
The analysis indicates that the luminosity threshold requisite for discovery at the CEPC 
is comparable to that of the FCC-ee, highlighting the competitive potential 
of these facilities in the search for new physics phenomena.

\subsection{ Exotic Decays via Dark Sector}
\label{sec:exotic-decay-in-DS}

\subsubsection{Higgs Exotic Decays via Dark Sector}
Within the framework of the lepton portal dark matter model, 
the relic abundance is determined by the fermion portal coupling (($\mathcal{L}_\chi$)) involving the Majorana fermion DM candidate $\chi$, 
the singlet charged scalar mediator $S^\pm$, and the SM right-handed lepton \cite{Bai:2013iqa}. In Ref.~\cite{Liu:2021mhn}, the Lagrangian is further expanded with another Scalar portal interaction ($\mathcal{L}_S$) as
\begin{align}
    \mathcal{L}_\chi =&~ \frac12\bar\chi i\slashed{\partial}\chi-\frac12m_\chi\bar\chi\chi+ y_\ell\left(\bar{\chi}_L S^\dagger \ell_R+{\rm h.c.}\right), \label{eq:fermion-portal-DM}\\
\mathcal{L}_S =&~ \left(D^\mu S \right)^\dagger D_\mu S 
- \left( \mu_H^2|H|^2+\mu_S ^2| S |^2+\lambda_H|H|^4+\lambda_S | S |^4+2\lambda_{H S }|H|^2| S |^2 \right) \,.
\label{eq:scalar-portal-coupling}
\end{align}
The incorporation of the scalar portal interaction not only leads to a significant enhancement 
of the detection capabilities at the LHC through the $gg\to h^\star \to S^+S^-$ process 
but also induces novel signal channels, including exotic decay processes and the deviations in Higgs boson couplings. 
These aspects yield promising prospects for the exploration and detailed examination of the model.

The experimental results from the LEP have excluded the existence of a charged scalar $S^\pm$ with a mass $m_S$ below 100 GeV, thus negating the possibility of the on-shell decay $h\to S^+ S^-$. Nevertheless, in the event that the mass of the dark matter candidate $\chi$, $m_\chi$, is less than half the mass of the Higgs boson, $m_h/2 \approx 62.5~\text{GeV}$, the portal coupling $\lambda_{HS}$ may activate the exotic decay processes such as the three- or four-body decays $h\to S^\pm \ell^\mp \chi $ or $h\to \ell^\pm\chi\ell^\mp\chi$. These decays are mediated by either one or two virtual $S^\pm$, depending on whether $m_S$ is less than $m_h$. The decay rates for these processes are directly proportional to  $y_\ell^2\lambda_{HS}^2$ or $y_\ell^4\lambda_{HS}^2$. This relationship presents a novel method for probing the parameters $\lambda_{HS}$ and $y_\ell$. By calibrating $y_\ell$ to the theoretical value $y_\ell^{th}$, which is derived from the necessary relic abundance, one may deduce constraints on the $\lambda_{HS}$ coupling for a given set of mass parameters ($m_S$,$m_\chi$).

\begin{figure}[t]
\centering	
\includegraphics[width=0.49 \columnwidth]{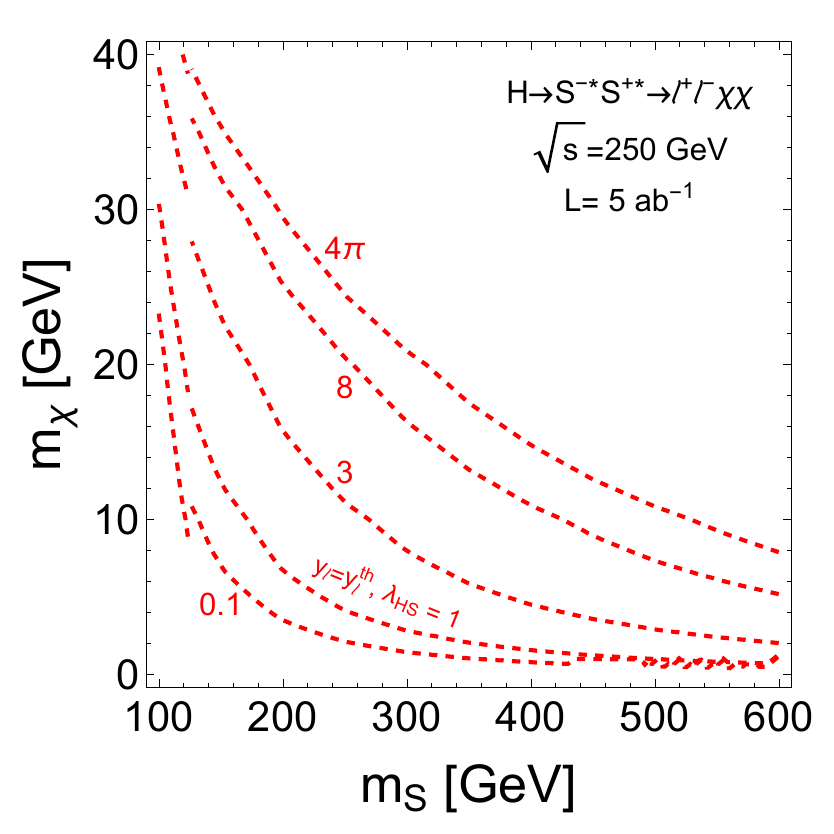}
\includegraphics[width=0.49 \columnwidth]{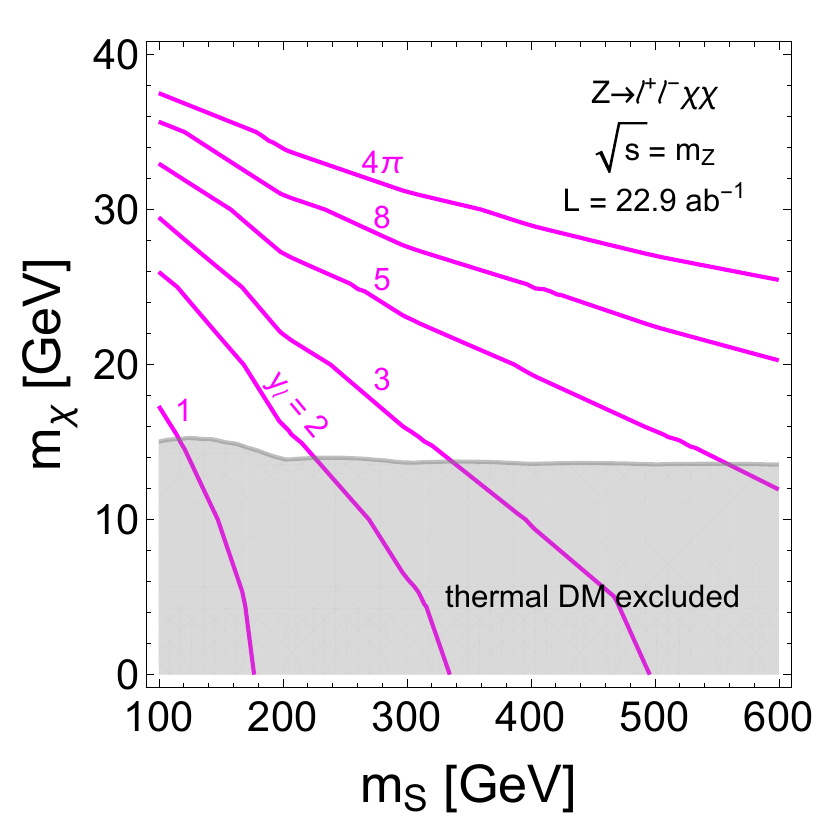}
\caption{\textit{Left}: The Higgs exotic decay $h \to S^{+(*)} S^{-(*)} \to \ell^+\chi \ell^-\chi$, mediated by off-shell charged scalars $S^{\pm}$, probes the Higgs portal coupling $\lambda_{HS}$ (Eq.~\ref{eq:scalar-portal-coupling}) and the fermion portal coupling $y_\ell$ (Eq.~\ref{eq:fermion-portal-DM}). Since the cross section for the DM annihilation process $\bar{\chi}\chi \to \ell^+\ell^-$ scales as $y_\ell^4$, we fix $y_\ell = y_\ell^{\rm th}$ to satisfy the thermal DM relic abundance for each point in the $(m_S, m_\chi )$ 2D plane. The dashed red lines correspond to fixed values of $\lambda_{HS}$, with the lower-left corner of each line excluded by the Higgs exotic decay searches $h \to S^{\pm(*)} S^{\mp(*)} \to \ell^+\chi \ell'^-\chi$.  
\textit{Right}: The exotic $Z$ boson decay $Z \to \ell^+\chi \ell^-\chi$, dominantly mediated by an off-shell $S^{\pm}$ attached to the external charged lepton leg, probes the fermion portal coupling $y_\ell$ (Eq.~\ref{eq:fermion-portal-DM}). The magenta contours show the $95\%$ C.L. limits on $y_\ell$ for each point in the $(m_S, m_\chi)$ plane, derived from exotic $Z \to \ell^+\ell^- + \slashed{E}$ searches at a Tera-$Z$ factory. For each mass configuration, the corresponding $y_\ell$ is used to compute the thermal DM relic density, with the gray shaded region excluded by relic abundance constraints. Figures taken from Ref.~\cite{Liu:2021mhn}.
}
\label{fig:Scalar-Fermion-portal-constraints}
\end{figure}

For a certain integrated luminosity, it is feasible to derive constraints on the branching ratio $Br(h\to S^{\pm(*)}S^{\mp(*)}\to \ell^+\chi\ell^{\prime -}\chi)$. These constraints can subsequently be converted into upper bounds for $\lambda_{HS}$, relying on the determination of $y_\ell$ through the conditions imposed by the DM relic abundance, as illustrated in the left panel of Fig.~\ref{fig:Scalar-Fermion-portal-constraints}. An apparent discontinuity in the depicted curves around $m_S=125~\text{GeV}$ corresponds to a transition in the available phase space, shifting from a three-body to a four-body decay mechanism. In conclusion, the future CEPC is anticipated to impose stringent constraints on the interaction strengths for scenarios with a relatively light DM candidate, $m_\chi \lesssim 30 ~\text{GeV}$, and a mediator scalar mass $m_S$ within the sub-TeV scale.

Another example is the Higgs decay into a dark shower, i.e., a shower of dark-sector particles, which can be bosons or fermions, for example, composite neutrinos~\cite{Chacko:2020zze}. These can either decay promptly or be long-lived and their decay back to visible SM particles can be either hadronic or leptonic. The process is motivated by generic considerations of hidden sector strong dynamics. It also appears in the discussion of neutral naturalness~\cite{Craig:2015pha}.  Current studies have been focusing on the Higgs decays into a pair of twin glueballs~\cite{Curtin:2015fna,Liu:2018wte,Alipour-fard:2018mre,Liu:2020vur,Carena:2022yvx,Wang:2022dkz}, but this is only a subclass of the generic Higgs decays into these final states. This dark shower channel is also motivated by the class of models with a large number of light scalars~\cite{Jung:2021tym}, e.g., Naturalness~\cite{Arkani-Hamed:2016rle}, electroweak scale as a trigger~\cite{Arkani-Hamed:2020yna}, and delayed or non-restored electroweak  symmetry~\cite{Meade:2018saz,Baldes:2018nel,Glioti:2018roy,Matsedonskyi:2020mlz}.

\subsubsection{Z Exotic Decays via Dark Sector}
Beside the exotic decays of the 125GeV Higgs boson, the exotic decay of the $Z$ boson constitutes an additional prospect for the investigation beyond the SM physics \cite{Liu:2017zdh}. To refine the precision in measuring SM parameters, the forthcoming CEPC will include the operation at the $Z$ resonance \cite{CEPCStudyGroup:2018ghi, Abada:2019lih}, expected to produce 4 Tera Z bosons.

Notably, within the context of the model under consideration~\cite{Liu:2021mhn}, there exists an exotic decay mode $Z\to\ell^+\chi\ell'^-\chi$, which results in a final state characterized by a pair of leptons accompanied by missing energy. This decay channel is mediated by two distinct types of Feynman diagrams: the first involves a pair of virtual $S^{\pm}$ particles via the $ZS^+S^-$ vertex, while the second proceeds via a single virtual $S^{\pm}$ that emerges from the $Z\ell^+\ell^-$ vertex, with the scalar $S$ coupling to one of the leptons. Given the assumption that the charged scalar $S^\pm$ possesses a mass exceeding 100 GeV, thereby surpassing the mass of any other particle involved, the decay width is predominantly governed by the second diagram type. This decay width exhibits a dependence that is proportional to $y_\ell^4 m_S^{-4}$.

\begin{figure}[!tbp]
  \centering
   \includegraphics[width =1.0\textwidth,height=0.35\textheight]{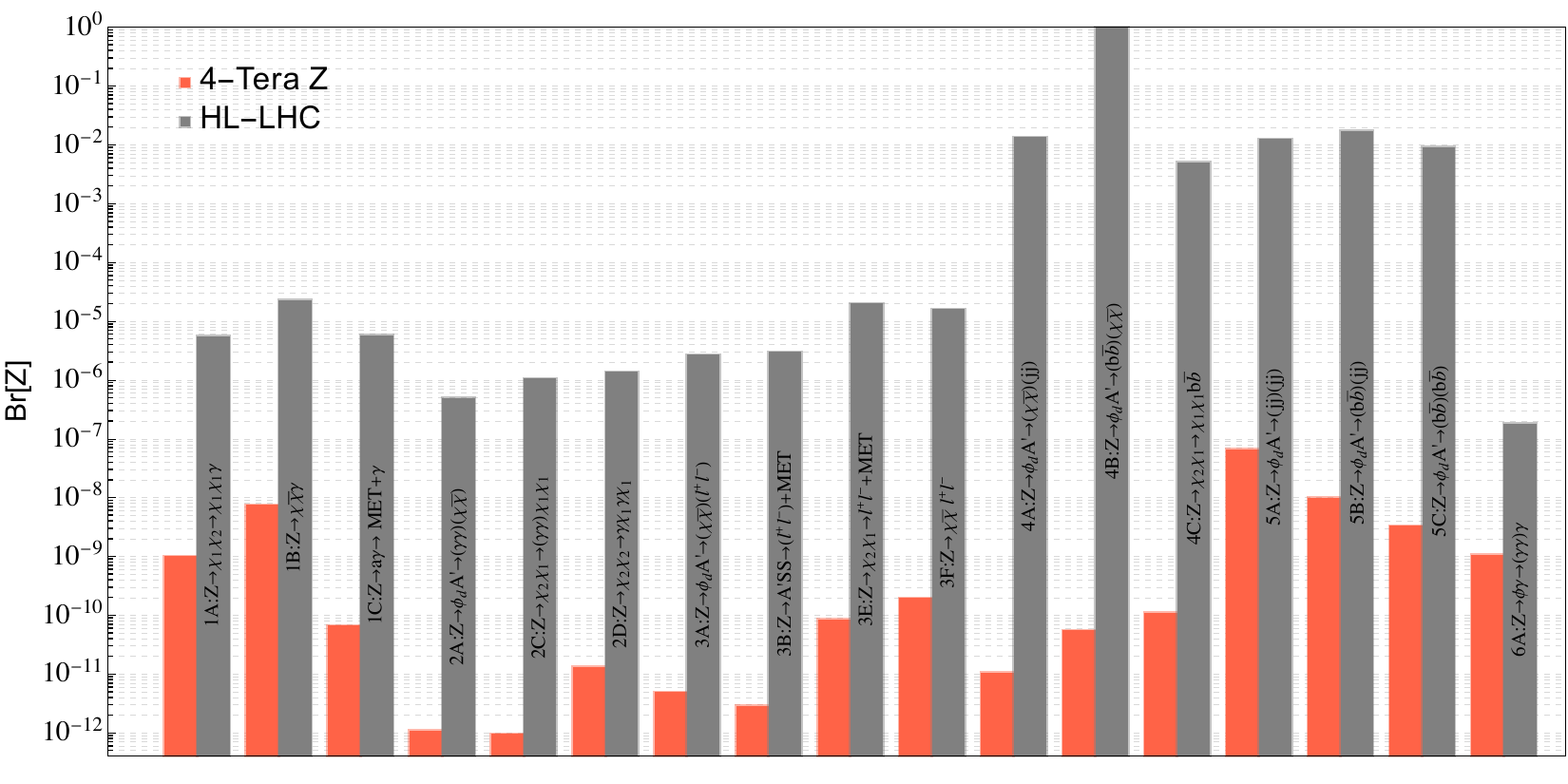}
   \caption{The reach for the branching ratio of various exotic $Z$ decay modes at the future $Z$-factories (rescaled to four Tera $Z$) and the HL-LHC at 13 TeV with $\mathcal{L} = 3 ~\rm{ab^{-1}}$~\cite{Liu:2017zdh}. The sensitivities, in general, generally also depend on model parameters, such as the masses of the mediator and dark matter. In this figure, we take the best case for each category.}\label{fig:summary-Z}
\end{figure}

In the right panel of Fig.~\ref{fig:Scalar-Fermion-portal-constraints}, the $95\%$ C.L. constraint on the branching ratio for the exotic decay mode is approximately $\Br(Z\to e^-e^+ \chi\chi) \lesssim 10^{-9}$ for the Tera $Z$ scenario. However, this upper bound is sensitive to the mass parameters $m_\chi$ and $m_S$. In the right panel of Fig.~\ref{fig:Scalar-Fermion-portal-constraints}, the $95\%$ C.L. upper limit on $y_\ell$ is represented by the region above the magenta contours. This constraint is compared with the requirement for $y_\ell^{\rm th}$ derived from thermal relic considerations. It becomes apparent that, under the Tera $Z$ framework, DM candidates with a mass of $m_\chi \lesssim 13 $ GeV are excluded by the search for the exotic decay $Z\to e^-e^+ \chi\chi$, as theoretically $y_\ell^{\rm th}$ exceeds the derived limit from the $Z$ exotic decay. The exclusion zone is illustrated in gray and denoted as "thermal DM excluded". This constraint serves as a supplementary bound for scenarios with large scalar mass $m_S$, compared with the constraints imposed by the LHC, which are not predicated on the on-shell production of $S$. Moreover, considering that both the decay width for the exotic decay and the DM annihilation cross-section are proportional to $y_\ell^4 m_S^{-4}$, the exclusion boundary can be extended horizontally to substantially high values of $m_S$. Consequently, this provides a robust constraint for the DM model.

Besides the specific exotic $Z$ decay channel above, Ref.~\cite{Liu:2017zdh} has studied a broad range of dark sector models and model-independent exotic Z decay channels at future $e^+ e^-$ colliders with the Giga Z and Tera Z options. Four general categories of dark sector models have been included: Higgs portal dark matter, vector portal dark matter, inelastic dark matter and axion-like particles (ALPs). Focusing on channels motivated by the dark sector models, an independent model study of 
the sensitivities of $Z$-factories is also carried out. The results are compared with the reach of high luminosity LHC (HL-LHC). The final states of the exotic decays are categorized according to the number of resonances and possible topologies. 
The projected reach for those channels is shown in Fig.~\ref{fig:summary-Z}. In comparison with the HL-LHC, the future Z-factories can be more sensitive to many interesting decay modes. 

\subsection{Higgs exotic invisible decays}
\begin{figure}[!tbp]
   \centering	
   \includegraphics[width=0.47 \columnwidth]{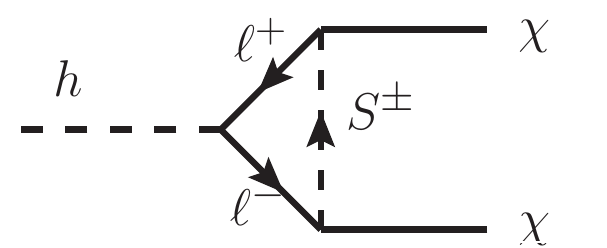}\qquad
   \includegraphics[width=0.47 \columnwidth]{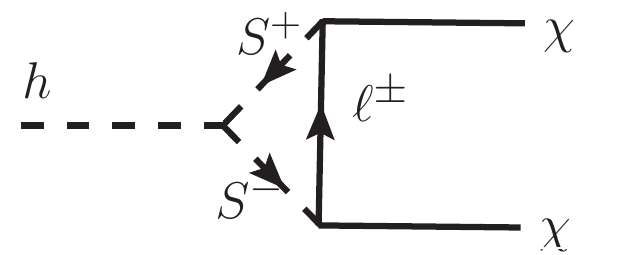}
   \caption{The one-loop induced Higgs invisible decay (from Ref.\cite{Liu:2021mhn}). The cross-diagrams for Majorana fermion $\chi$ are not shown here but are included in the calculation.}\label{fig:h_chichi}
\end{figure}

The Higgs invisible decay $h \to \chi \chi$ is induced by the two Feynman diagrams at one-loop level listed in Fig.~\ref{fig:h_chichi}, and is similar to the Higgs to neutralinos decay in the SUSY models \cite{Drees:1996pk, Djouadi:2001kba, Eberl:2001vb, Berlin:2015njh}. Due to the small lepton mass, the first diagram is usually negligible.

The most stringent constraint on the branching ratio for the invisible decay of the Higgs boson is $Br(h \to {\rm inv}) < 13\%$, ascertained by the ATLAS Run-II with an integrated luminosity of $139~{\rm fb}^{-1}$ \cite{ATLAS:2020cjb}. Projected advancements at the HL-LHC anticipate an enhanced sensitivity for the detection of invisible Higgs decay, with expectations set around $3.5\%$ \cite{Bernaciak:2014pna}. Furthermore, at the future $e^+e^-$ colliders such as the CEPC, the sensitivity could be refined to approximately $0.1\%$ with $20~{\rm ab}^{-1}$ integrated luminosity at its Higgs operation~\cite{CEPCStudyGroup:2018ghi, CEPCStudyGroup:2023quu}.

\begin{figure}[!tbp]
\centering	
\includegraphics[height=0.33\textheight, width=0.45 \columnwidth]{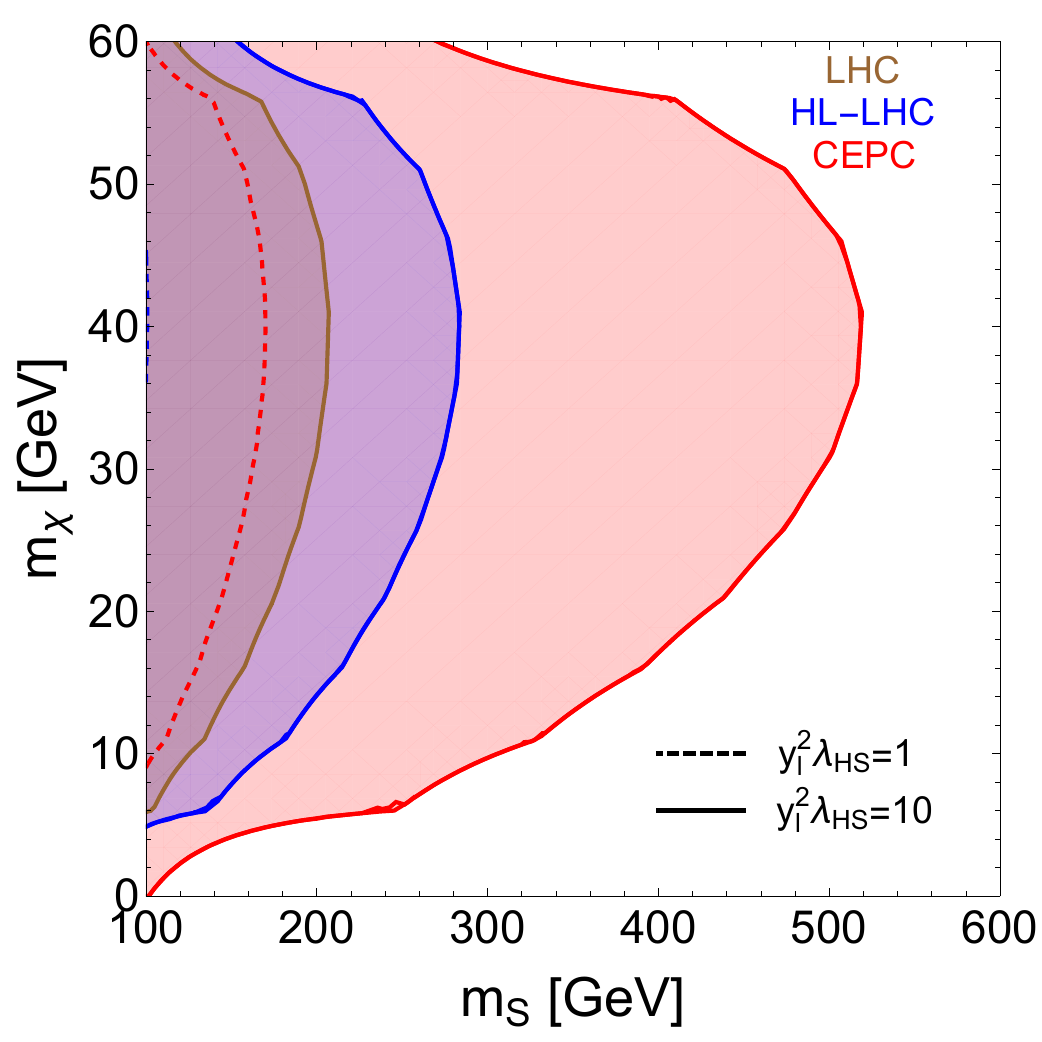}\qquad\quad
\includegraphics[height=0.33\textheight, width=0.47 \columnwidth]{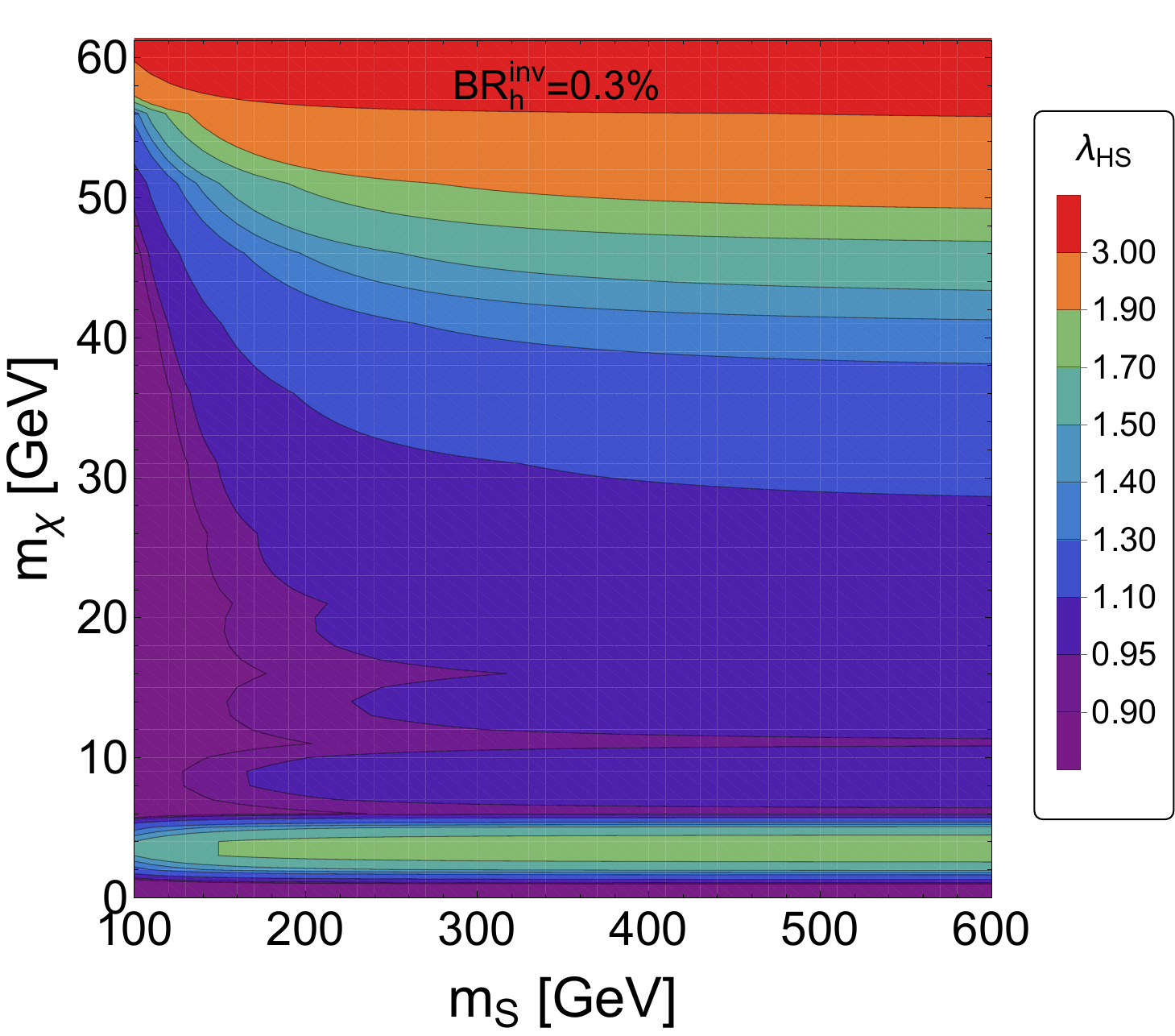}
\caption{\textit{Left}: The Higgs portal coupling $\lambda_{HS}$ (Eq.~\ref{eq:scalar-portal-coupling}) and the fermion portal coupling $y_\ell$ (Eq.~\ref{eq:fermion-portal-DM}) can induce the Higgs invisible decay $h \to \bar{\chi}\chi$ at one-loop level. Constraints on the coupling combination $y_\ell^2 \lambda_{HS}$ are derived from searches for invisible Higgs decays at the LHC, HL-LHC ($\text{BR}_{h \to \text{inv}} < 3.5\%$), and CEPC ($\text{BR}_{h \to \text{inv}} < 0.3\%$). The recent CEPC Technical Design Report (TDR)~\cite{CEPCStudyGroup:2023quu} projects an improved limit of $\text{BR}_{h \to \text{inv}} < 0.07\%$ with an upgraded integrated luminosity of $20~\text{ab}^{-1}$ during Higgs operation, which could further tighten the bound on $y_\ell^2 \lambda_{HS}$ by a factor of two.
\textit{Right}: For each mass point $(m_S, m_\chi)$, the fermion portal coupling $y_\ell$ is set to its thermal value $y_\ell^{\rm th}$, as required by the observed dark matter relic abundance. Constraints on the Higgs portal coupling $\lambda_{HS}$ are then derived using the CEPC limit on the invisible Higgs branching fraction $\Br(h \to \text{inv}) < 0.3\%$.
Figures from Ref.\cite{Liu:2021mhn}.
}
\label{fig:lambdaHS-from-BRinv-CEPC}
\end{figure}

Anticipated data from forthcoming collider experiments can set limits on $y_\ell^2\lambda_{HS}$, depending on the scalar mass $m_S$ and the DM candidate mass $m_\chi$. In the left panel of Fig.~\ref{fig:lambdaHS-from-BRinv-CEPC}, one can observe the sensitivity contours for $y_\ell^2\lambda_{HS}$ corresponding to the LHC (brown), the HL-LHC (blue), and the CEPC (red). The dashed and solid lines represent $y_\ell^2\lambda_{HS}= 1$ and $y_\ell^2\lambda_{HS}= 10$, respectively. It is shown that the prospective $e^+e^-$ collider exhibits superior sensitivity in comparison to the hadron collider alternatives.

Turning to the right panel of Fig.~\ref{fig:lambdaHS-from-BRinv-CEPC}, the Yukawa coupling $y_\ell$ is calibrated to its thermal value $y_\ell^{\rm th}$, which is requisite for satisfying the DM relic abundance criteria. By fixing the values of $m_S$ and $m_\chi$, the future sensitivity to $\lambda_{HS}$ can be extrapolated utilizing the projected CEPC sensitivity for the invisible Higgs decay branching ratio, $Br(h\to{\rm inv}) = 0.3\%$. The resulting sensitivity contours are illustrated accordingly. A notable characteristic is the reduction in sensitivity to $\lambda_{HS}$ for $m_\chi$ values below 6 GeV, attributable to the decay width's leading order term in the small $m_\chi$ expansion being linearly dependent on $m_\chi$. As $m_\chi$ diminishes further, the thermal value $y_\ell^{\rm th}$ increases to compensate the annihilation cross-section, thereby restoring and even enhancing the sensitivity to $\lambda_{HS}$. Consequently, the optimal sensitivity for $\lambda_{HS}$ is achieved in regions of small $m_S$ and moderate $m_\chi$.

\subsection{Decays into Long-Lived Particles}
\subsubsection{Higgs exotic decays into Long-Lived Particles}
\label{subsubsec:h2LLPs}
The CEPC is proposed to be a key facility in the search for new physics, particularly through the examination of Higgs boson decays into long-lived particles (LLPs). Recent comprehensive studies have provided valuable insights into key areas that could enhance our understanding of the potential of the CEPC to observe long-lived particles through Higgs boson decay.

Ref.~\cite{Zhang:2024bld} investigated the possibility of neutrally charged LLPs being produced through the exotic decay of the Higgs boson. By analyzing the process $e^+e^- \to ZH$, where the Z boson decays inclusively and the Higgs boson further decays into LLPs ($X_1$ and $X_2$), the study has employed advanced machine learning techniques to analyze an integrated luminosity of 20 ab$^{-1}$. These LLPs can decay into either a neutrino pair or a quark-antiquark pair, 
leading to distinct final states that were identified using Convolutional Neural Networks (CNN) and Graph Neural Networks (GNN). The findings have provided constraints on the branching ratio of Higgs boson decay to LLPs, offering a new observation on the Higgs boson's decay characteristics.

Furthermore, Ref.~\cite{Alipour-Fard:2018lsf} explored the production of long-lived scalar particles 
from Higgs exotic decays at the CEPC. The signal process involves a Higgsstrahlung event, followed by the decay of the Higgs boson into a new long-lived scalar boson $X$, which subsequently decays into a pair of quarks. This research has considered the Higgs bosons produced at CEPC, providing sensitivities to the branching ratio of $h \to XX$ and interpreting the results within the framework of the Higgs-portal Hidden Valley model and neutral-naturalness models.

Ref.~\cite{Cheung:2019qdr} looked into displaced-vertex signatures of scalar LLPs pair-produced from exotic Higgs decays has also been a significant focus. The study has examined two theoretical models, including a Higgs-portal model that predicts a very light scalar boson $h_s$, which decays into a pair of muons or pions, and a neutral-naturalness model that predicts 
the lightest mirror glueball with a mass of $\mathcal O(10)$ GeV. These models have been analyzed for their distinctive signatures at colliders, providing a comprehensive understanding of the potential LLP signatures.

Additionally, Ref.~\cite{Jeanty:2022cwr} addressed the sensitivity reach to massive LLPs within the context of the Hidden Valley model, where the Higgs boson decays into two long-lived Hidden-Valley particles that subsequently decay into b-quarks. The research has also investigated the sensitivity to long-lived dark photons produced in Higgsstrahlung events via the Higgs portal, $h \to\gamma_D\gamma_D$. The high statistical significance data generation at the CEPC, combined with the advanced analysis techniques, is expected to provide a more precise measurement of the Higgs boson's decay width, 
surpassing current measurements at the LHC. More detailed discussions can be found in Section~\ref{sec:LLP}.

\subsubsection{Z exotic decays into Long-Lived Particles}
\label{subsubsec:Z2LLPs}
At the CEPC, the high-luminosity Z-boson factory provides a unique opportunity to investigate Long-Lived Particles (LLPs) from Z-boson exotic decays.

Ref.~\cite{Wang:2019orr} examined the sensitivity of the CEPC to the decay of Z-bosons into long-lived lightest neutralinos (denoted as $\tilde \chi_1^0$) within the context of R-parity-violating supersymmetry. The lightest neutralino is predominantly bino-like with minor Higgsino components. The research emphasizes the $\lambda_{i j k}^{\prime} L_i \cdot Q_j \bar{D}_k$  operators, particularly the $\lambda_{112}^{\prime} L_1 \cdot Q_1 \bar{D}_2$  operator, which leads to the decay of the lightest neutralino into SM particles via a scalar-fermion exchange. For specific conditions, the lightest neutralino becomes long-lived, allowing it to travel a macroscopic distance before decaying.

Besides, the sensitivity estimation is provided for the CEPC in terms of contour curves on a plane of model parameters $\lambda_{112}^{\prime} / m_{\tilde{f}}^2$ versus $m_{\tilde{\chi}_1^0}$. It shows that for a branching ratio $\Br(Z \rightarrow \tilde{\chi}_1^0 \tilde{\chi}_1^0)$ of $10^{-3}$ and a neutralino mass of approximately 40 GeV, the model parameter $\lambda_{112}^{\prime} / m_{\tilde{f}}^2$ can be probed down to about $1.5 \times 10^{-14}$ ($3.9 \times 10^{-14}$) GeV$^{-2}$ at the CEPC with a center-of-mass energy of $\sqrt s = 91.2$ GeV and integrated luminosities of 150 (16) ab$^{-1}$.

Additionally, Ref.~\cite{Calibbi:2022izs} discussed the investigation of ALPs coupled to charged leptons in Z-boson decays at CEPC. The ALPs are assumed to have very long lifetimes and behave as missing energy. 
The study analyzes the signal process $e^-e^+ \to \mu^-\mu^+a$ and considers various background sources. Sensitivity reaches are presented for different integrated luminosities.

Lastly, Ref.~\cite{Wang:2019xvx} studied the sensitivity of different experiments to $Z$-boson decays to a pair of long-lived neutralinos in R-parity-violating supersymmetry. Assuming a negligible background, it presents the sensitivity reaches of different far detector (FD) designs at the CEPC for a branching ratio of $Z\to\tilde \chi_1^0\tilde \chi_1^0$ equal to $10^{-3}$. It also compares these sensitivity reaches with those of the main detector (MD) and other experiments. For other possibilities of $Z$ exotic decays to long-lived particles, see $e.g.$~\cite{Cheng:2019yai,Cheng:2021kjg}.

Overall, at the CEPC, the potential to probe BSM physics by studying $Z$-boson exotic decays could offer insights into long-lived particles and their properties. More detailed discussions can be found in Section~\ref{sec:LLP}.













\newcommand{\gev}{~\mathrm{GeV}}
\newcommand{\tev}{~\mathrm{TeV}}

\newcommand\refeqq[1]{Eq.~(\ref{#1})}
\newcommand\refeqs[1]{Eqs.~(\ref{#1})}
\newcommand\refta[1]{Tab.~\ref{#1}}
\newcommand\refse[1]{Sect.~\ref{#1}}
\newcommand\refses[1]{Sects.~\ref{#1}}
\newcommand\citere[1]{Ref.~\cite{#1}}
\newcommand\citeres[1]{Refs.~\cite{#1}}
\newcommand\refap[1]{App.~\ref{#1}}
\def\reffi#1{\mbox{Fig.~\ref{#1}}}

\newcommand{\sigCMS}{2.9}
\newcommand{\muCMS}{0.35}
\newcommand{\dmuCMSpl}{0.12}
\newcommand{\dmuCMSmi}{0.12}

\newcommand\plane[2]{$(#1, #2)$ plane}

\newcommand\hotf{\ensuremath{h_{125}}}
\newcommand\hnf{\ensuremath{h_{95}}}

\newcommand{\SH}[1]{{\color{black}#1}}
\newcommand{\SHc}[1]{{\color{red}[SH: #1]}}
\newcommand{\TB}[1]{{\color{magenta}#1}}
\newcommand{\TBc}[1]{{\color{red}[TB: #1]}}
\newcommand{\TBd}[1]{\TB{\st{#1}}}
\newcommand{\TBalt}[1]{{\color{green}#1}}
\renewcommand{\GW}[1]{{\color{black}#1}}
\newcommand{\GWc}[1]{{\color{red}[GW: #1]}}







\subsection{The 95 GeV Higgs boson at the CEPC}
\label{sec:h95}

CMS and ATLAS have performed searches for scalar
di-photon resonances using LHC Run~1 and~2 data.
CMS observed a local excess with $2.8\,\sigma$ significance at
95.3~GeV in the Run~1 result.
Recently, results based on the full Run~2 data set at $13\tev$
showed a local excess of $2.9\,(1.7)\,\sigma$ at $95.4 \gev$
at CMS~\cite{CMS:2024yhz} (ATLAS)~\cite{ATLAS:2023jzc}.
Using both ATLAS and CSM Run~2 results, and neglecting possible
correlations, in \citere{Biekotter:2023oen} 
a combined signal strength of
$\mu^\mathrm{exp}_{\gamma\gamma} = 0.24^{+0.09}_{-0.08}$
was obtained, corresponding to an excess of $3.1\,\sigma$.
LEP reported a local $2.3\,\sigma$ excess
in the~$e^+e^-\to Z(\phi\to b\bar{b})$
searches\,\cite{Barate:2003sz}, consistent with a
scalar resonance with a mass of about $95.4 \gev$
and a signal strength of
$\mu_{bb}^{\rm exp} = 0.117 \pm 0.057$~\cite{Cao:2016uwt,Azatov:2012bz}.
Here we consider the possiblity that these excesses
arise from the production of a single new particle --
a possible first sign of BSM physics in the Higgs sector.

\begin{figure}[htb]
\centering
\includegraphics[width=0.45\textwidth]{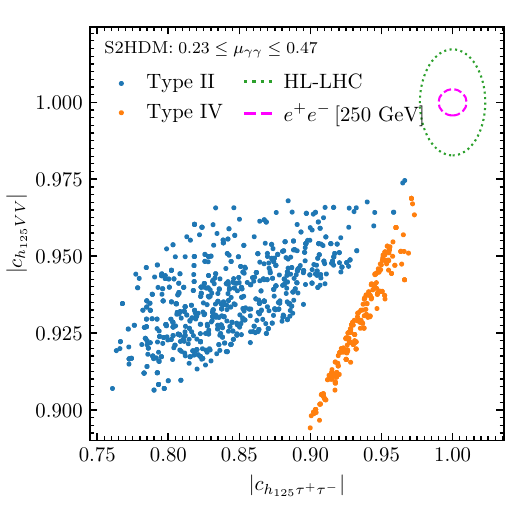}~~
\includegraphics[width=0.45\textwidth]{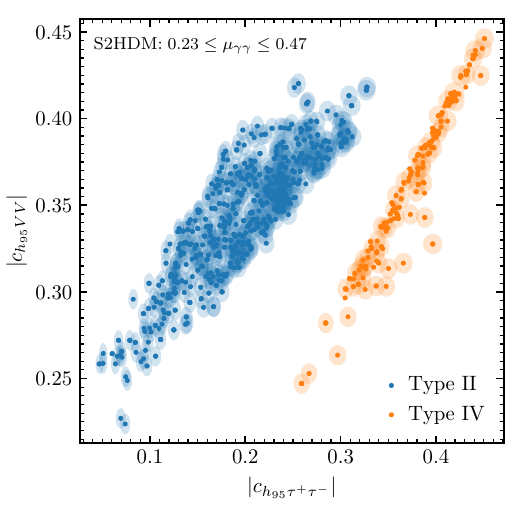}
\caption{\small
S2HDM parameter points passing the applied constraints for the di-photon
and $b\bar b$ signal strengths. (For the di-photon signal strength a
slightly higher value as $\mu^\mathrm{exp}_{\gamma\gamma}$ were used, as in
\citere{Biekotter:2023jld}, which is not expected to change
the results in a qualitative way.)
Blue (orange) points correspond to the S2HDM type~II (IV).
Left: \plane{|c_{h_{125} \tau^+ \tau^-}|}{|c_{h_{125} VV}|}.
The green dotted and the magenta dashed ellipses indicate the 
projected experimental precision of the
coupling measurements at the
HL-LHC~\cite{Cepeda:2019klc} and a future $e^+e^-$
collider operating at 250~GeV and assuming
$2~\mathrm{ab}^{-1}$ of integrated luminosity~\cite{Bambade:2019fyw},
respectively, with their centers located at the SM values.
Right: \plane{|c_{h_{95} \tau^+ \tau^-}|}{|c_{h_{95} VV}|} (where
$c_{\hnf xx}$ denotes the coupling strength relative to the SM
Higgs-boson coupling). The shaded ellipses around
the dots indicate the projected experimental
precision with which the couplings of
$h_{95}$ could be measured at a future $e^+e^-$ collider,
which we evaluated according to \citere{Heinemeyer:2021msz}.
Here $c_{\hnf xx}(c_{\hotf xx})$ denotes the coupling strength of
\hnf(\hotf) relative to the SM Higgs-boson coupling.}
\label{fig:h95}
\end{figure}

In \citeres{Biekotter:2023jld,Biekotter:2023oen} it was demonstrated that
the extension of the 2HDM by a complex singlet,
the S2HDM~\cite{Biekotter:2021ovi},
can give a perfect description of
these excesses, while being in agreement with LHC BSM Higgs searches and
LHC Higgs rate measurements.
The S2HDM represents a template for a broad class of models where
a mostly gauge-singlet scalar particle, \hnf,
obtains its couplings to fermions
and gauge bosons via the mixing with
the SM-like Higgs boson at $125\gev$.
In \citere{Biekotter:2023jld} it
was also demonstrated that a future $e^+ e^-$ collider
operating at 250~GeV could determine
the couplings of \hotf\ to a sufficiently high
precision to find deviations w.r.t.\ a SM Higgs boson
(see \reffi{fig:h95} left)
and thus test the proposed scenario.
Despite the suppressed couplings of the possible state at $95.4 \gev$
compared to \hotf, a future $e^+e^-$ Higgs factory could produce
\hnf\ in large numbers (see e.g.~\citere{Drechsel:2018mgd})
and determine its properties with high precision
(see \reffi{fig:h95} right).

\subsection{Top quark exotic decays}

In the context of the dark force model, as outlined in Ref.~\cite{Kong:2014jwa}, the decay of a top quark into a charged Higgs boson via the process $t \to b H^+$ is a significant opportunity for exploring new physics, particularly if the charged Higgs is relatively light. This charged Higgs, in turn, is proposed to predominantly decay into dark gauge bosons ($Z^\prime$s), 
which are key components of the dark sector.  
The CEPC, which has the upgrading program to increase its center of mass energy to 360 GeV after the Higgs operation, could potentially observe the decay chain $t\to bH^+\to bW^+ +Z^\prime s$.

\begin{figure}[!tbp]
\centering	
\includegraphics[width=0.7\textwidth,height=0.35\textheight]{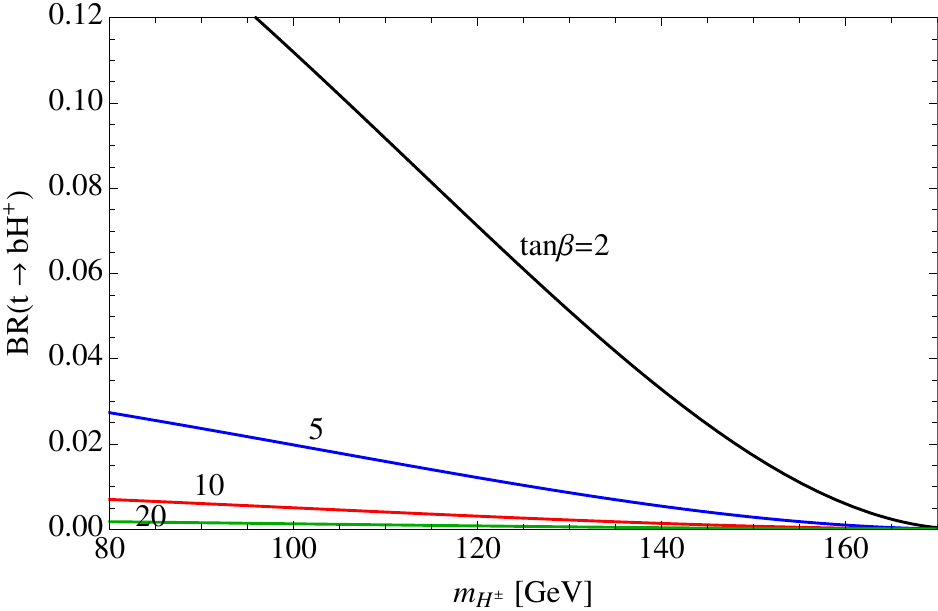}
\caption{$\Br(t \to b H^+)$ for $\tan\beta = 2$ (black), $5$ (blue), $10$ (red), $20$ (green). The $t$ decay into $H^+$ is larger for smaller $m_{H^\pm}$ and smaller $\tan\beta$, from Ref.\cite{Kong:2014jwa}.}\label{fig:chargedHiggs}
\end{figure}

Direct calculations demonstrate that for a charged Higgs mass of $m_{H^\pm}=140~\text{GeV}$, the branching ratio for the top quark decay into a charged Higgs, $\Br\left(t \rightarrow b H^{+}\right)$, is approximately between 0.03 and 0.0003, within the parameter range of $\tan\beta=2-20$, as shown in Fig.\ref{fig:chargedHiggs}. This range of branching ratios indicates that even a small proportion of top quark decays could produce a detectable number of $Z^\prime$ bosons at high-energy collider experiments.

The detection of such exotic decays depends on the precise measurement of the top quark's mass and decay width. It is highlighted that the present large uncertainty in the top quark decay width (around 25\%) leaves room for new decay modes that could originate from physics beyond the SM. Moreover, it underscores that the top quark's short lifetime and its decay before forming hadrons make it an ideal candidate for probing new physics.

A study of top quark mass measurements is presented at the $t\bar{t}$ threshold based on CEPC~\cite{Li:2022iav}. A center-of-mass energy scan near two times the top quark mass is performed, and the measurement precision of top quark mass, width and $\alpha_S$ are evaluated using the $t\bar{t}$ production rates. Realistic scan strategies at the threshold are discussed to maximize the sensitivity to the measurements individually and simultaneously in the CEPC scenarios, assuming a total luminosity limited to 100 fb$^{-1}$. With the optimal scan for individual property measurements, the top quark mass precision is expected to be 9 MeV, the top quark width precision is expected to be 25 MeV and $\alpha_S$ can be measured at a precision of 0.00034, considering only the statistical uncertainty. Taking into account the systematic uncertainties from theory, width, $\alpha_S$, experimental efficiency, background subtraction, beam energy and luminosity spectrum, the top quark mass can be measured at a precision of 24 MeV optimistically and 57 MeV conservatively at the CEPC, depending on systematic assumptions.

\subsection{Summary}

Exotic decays of Higgs, $Z$ bosons, and top quarks provide a crucial window into novel physics, often associated with Higgs portal models or exotic effective operators. These decays are vital for precision measurements at future Higgs factories and could benefit from machine learning technologies that enhance data analysis and exotic signal detection.

Lepton colliders like the CEPC are particularly well-suited for detecting Higgs boson decays into BSM particles, which may undergo further decays into multiple particle final states. Four distinct decay categories are considered. The recoil mass method used at lepton colliders allows for better separation of signal from background, compared to hadron colliders like the LHC. CEPC’s projected sensitivity of the decay branching ratio enhancements is expected to outperform the LHC by orders of magnitude, especially for channels that do not involve missing energy. The potential of the SM-$s$ model, where a new scalar particle ($s$) mixes with the Higgs boson, is discussed. This theoretical framework addresses issues like naturalness and dark matter interactions, which may address the hierarchy problem. A notable hadronic decay,  $h \rightarrow ss$ with $s$ further into quarks, which is difficult to detect at the LHC, can be studied more effectively at the CEPC due to its cleaner experimental environment and greater sensitivity. 
The exotic decays of the $ Z $ gauge boson at future Z-factories have been briefly discussed in this section, highlighting that many decay channels share similar features with exotic Higgs decays. Like exotic Higgs decays, the CEPC offers an improvement of several orders of magnitude in sensitivity to exotic $Z$ decays, making it a major physics objective at the CEPC.

Moreover, Supersymmetric extensions, such as the NMSSM and scNMSSM, are examined for their ability to generate exotic Higgs decays, including invisible decays into dark matter candidates like neutralinos. The CEPC’s capabilities in probing such decays, particularly final states like $h_1 \rightarrow 4\tau$, exceed those of the HL-LHC with only 0.26 fb$^{-1}$.
Additionally, the potential existence of a 95 GeV scalar boson is discussed, as suggested by excesses observed in CMS and ATLAS data from Run 1 and Run 2 of the LHC. This possible BSM particle, described within the S2HDM model, represents a gauge-singlet scalar that mixes with the SM Higgs boson. The CEPC could provide precise measurements to confirm or refute the presence of such a particle. Finally, exotic decays of the top quark, such as  $t \rightarrow bH^+$, where the charged Higgs decays into dark-sector particles like dark gauge bosons, are discussed. The CEPC, particularly at higher energies, presents an opportunity to detect these decays and investigate new physics related to the top quark’s interactions with the dark sector.
This section also discusses Higgs and $Z$ boson decays into long-lived particles (LLPs), which travel before decaying into visible SM particles. Lepton colliders like the CEPC, with high luminosity, are well-suited for such searches. Future  Z-factories (e.g., Giga $Z$, Tera $Z$) offer unique capabilities in detecting LLPs, which will be discussed in detail in Section~\ref{sec:LLP}. Invisible Higgs decays into dark matter candidates are also explored, with the CEPC expected to improve LHC constraints by an order of magnitude. Advanced machine learning techniques, such as convolutional and graph neural networks, enhance sensitivity to these exotic decays.

Overall, the CEPC offers extensive opportunities to explore exotic decays and probe BSM physics through precision measurements and improved detection capabilities, positioning it as a critical tool for advancing particle physics.

\clearpage
\section{Electroweak phase transition and gravitational waves}
\label{sec:EwptGw}

\subsection{Introduction}

The nonzero vacuum expectation value (vev) of the Higgs field spontaneously breaks the electroweak gauge group $SU(2)_L\times U(1)_Y$, thereby giving mass to the $W^\pm$ and $Z$ gauge bosons as well as the fermions in the Standard Model (SM). {At sufficiently high temperatures, however, thermal quantum corrections involving the particles in the early universe plasma change the shape of the Higgs potential such that -- in a purely Standard Model universe -- the minimum of energy lies at the origin with $\ave{h}=0$ and EW symmetry restored\cite{Dolan:1973qd}. }
The  history of the Higgs vev evolution from its value in the early Universe to today's $\ave{h}=v_{\rm EW}=246$ GeV, 
{and the associated EW symmetry-breaking transition\footnote{Following the Ehrenfest classification, we distinguish {\it bona fide} phase transitions, characterized by discontinuities in derivatives of thermodynamic quantities, from smooth crossover transitions devoid of such discontinuities.},}
 is not only of considerable interest but also of utmost importance. {In particular, the occurrence of a first order electroweak phase transition (FOEWPT), if sufficiently strong, can provide the necessary preconditions for generating the cosmic matter-antimatter asymmetry via electroweak baryogenesis and as a source of potentially observable gravitational waves. A FOEWPT may also impact the nature and dynamics of the dark matter.}

In the SM, the Higgs vev $\ave{h}$ smoothly transits from zero to $v_{\rm EW}$ as the Universe cools down. {Lattice simulations indicate that for $m_h\gtrsim 70$ GeV, this transition is a smooth crossover}~\cite{Kajantie:1996qd,Rummukainen:1998as,Laine:1998jb}. However, in the presence of new physics beyond the SM (BSM), $\ave{h}$ can undergo a discontinuous jump known as a first-order electroweak phase transition (FOEWPT)~\cite{Dine:1992wr}. During this process, bubbles containing the $\ave{h}\neq0$ vacuum form and expand within the background of the old vacuum where $\ave{h}= 0$. These bubbles eventually fill the entire space, converting the Universe from the old false vacuum to the new true vacuum. 

\begin{figure}[t]
\centering
\includegraphics[width=1.0\textwidth]{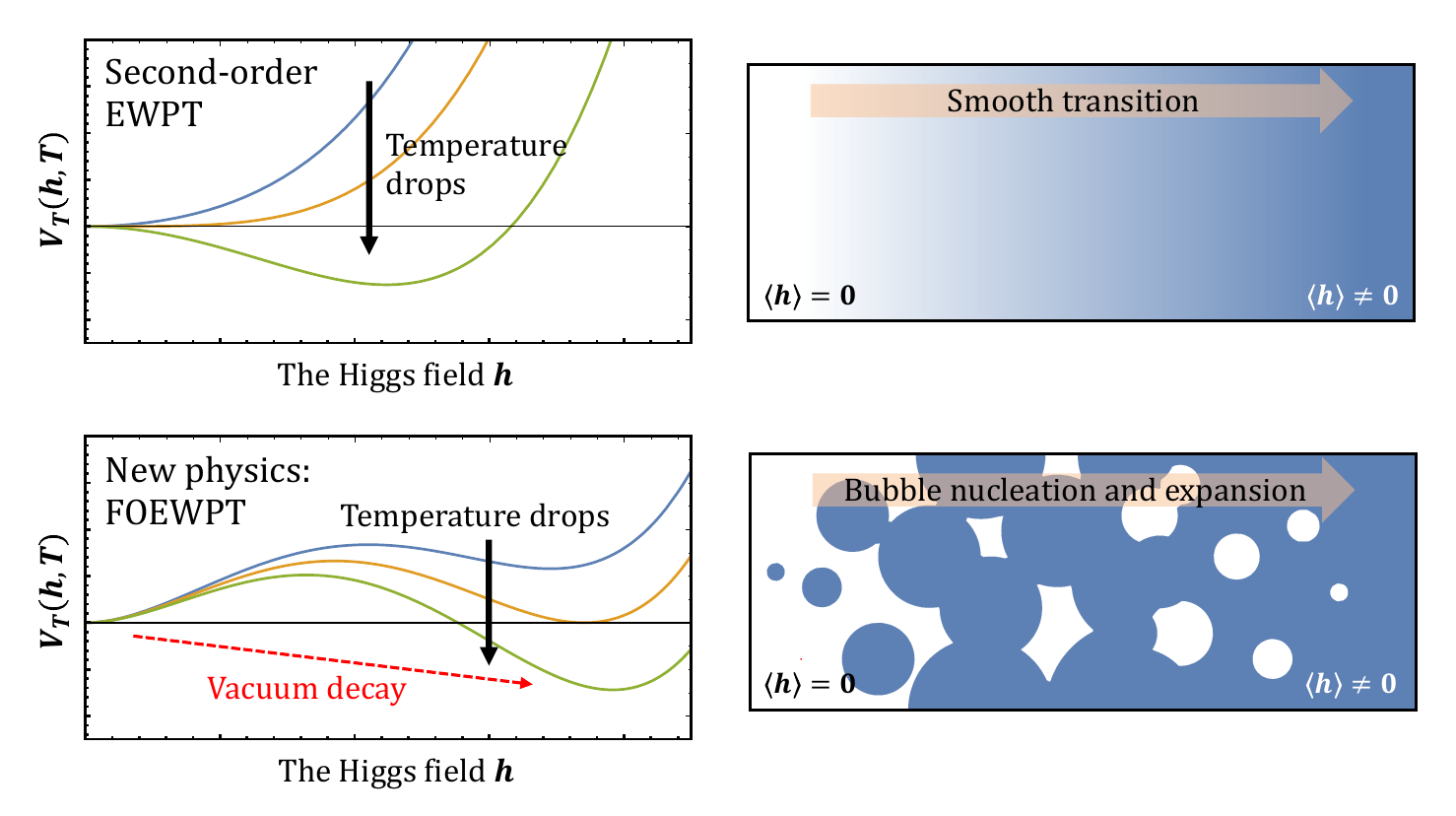}
\caption{Illustration of electroweak phase transition patterns. Top: in the SM, the transition is a smooth crossover. Bottom: in many new physics models, the scalar potential exhibits a barrier, allowing for a FOEWPT with bubble nucleation and expansion.}
\label{fig:thermalhist}
\end{figure}

While the existence of a crossover transition cannot be ascertained within a purely perturbative framework using the Higgs potential, for purposes of intuition it is still useful to compare the first order transition with the relatively \lq\lq smooth\rq\rq\, second order phase transition. These two different patterns of electroweak phase transition are illustrated in Fig.~\ref{fig:thermalhist}. The top left panel shows the thermal evolution of the Higgs potential in the presence of a second order phase transition, while the bottom left panel sketches the FOEWPT in the presence of new physics beyond the SM. The right panels illustrate the corresponding spacetime evolution from the symmetric to broken symmetry phases. The picture for a smooth crossover would be analogous to that of the top right panel. The underlying reason for a FOEWPT is the existence of a potential barrier that forbids a smooth transition. One should keep in mind, however, that the presence of a barrier in the potential is a necessary but not sufficient condition for a FOEWPT. For $m_h$ in the vicinity of the \lq\lq critical Higgs mass\rq\rq\, $\sim 70$ GeV, infrared contributions to thermal loops render the aforementioned perturbative description of EW symmetry-breaking invalid. The latter may occur via a smooth crossover even in the presence of a perturbative thermal barrier in the potential. Below, we highlight recent developments in non-perturbative studies of the EWPT in BSM scenarios and their relevance to future CEPC measurements.

The FOEWPT holds greater scientific interest compared to the SM crossover due to its profound cosmological implications. It can drive the Universe out of equilibrium facilitating electroweak baryogenesis to generate the matter-antimatter asymmetry (see Ref.~\cite{Morrissey:2012db} for a review). During this process, elementary particles engage in CP-violating scatterings with the bubble wall, ultimately resulting in the creation of a chiral asymmetry, which is then converted into a net baryon number by electroweak sphalerons. The baryon asymmetry is subsequently swept into the true vacuum through bubble expansion, and stored until today~\cite{Joyce:1994fu,Fromme:2006wx,Morrissey:2012db,Jiang:2015cwa,Chiang:2017nmu,Xie:2020wzn,Bian:2019kmg,Xie:2020bkl,Bell:2020gug,Bian:2017wfv,Huang:2018aja,Kanemura:2023juv,Enomoto:2024jyc,Huang:2025cqi}. This mechanism has been the primary motivation for studying FOEWPT over the past several decades. 

Furthermore, first-order phase transitions generally generate stochastic gravitational wave (GW) backgrounds through bubble collisions, sound wave motion, and turbulence in the plasma~\cite{Weir:2017wfa,Mazumdar:2018dfl,Athron:2023xlk,Caldwell:2022qsj,Guo:2021qcq}. A FOEWPT, which typically occurs at $T \sim 100$ GeV, sources GWs that peak at frequencies of $\mathcal{O}({\rm mHz})$ today~\cite{Grojean:2006bp,Caprini:2015zlo,Caprini:2019egz}. These frequencies fall within the sensitivity range of several near-future space-based detectors, such as LISA~\cite{LISA:2017pwj}, TianQin~\cite{TianQin:2015yph,TianQin:2020hid}, and Taiji~\cite{Hu:2017mde,Ruan:2018tsw}, and hence will be efficiently probed in the next decade. {As we discuss below, the EWPT constitutes an ideal \lq\lq laboratory\rq\rq\, for studying FOPT-catalyzed GWs as collider studies provide a complementary probe of the underlying particle physics. Such complementarity may not be available for GWs associated with other scales, such as those detectable in pulsar timing arrays.}

Additionally, recent studies suggest that first-order cosmic phase transitions, including FOEWPT, could be crucial in generating dark matter and matter-antimatter asymmetry through various processes as follows.
\begin{enumerate}
\item It could impact the dark matter production, decay and annihilation via the sudden change of particle mass before and after the phase transition~\cite{Baker:2016xzo,Baker:2017zwx,Wong:2023qon,Hambye:2018qjv}.
\item The interaction between the bubble wall and particles in the plasma can produce superheavy particles to be dark matter or generate the matter-antimatter asymmetry~\cite{Azatov:2021ifm,Azatov:2021irb,Ai:2024ikj,Baldes:2021vyz,Huang:2022vkf,Chun:2023ezg}.
\item Due to the mass gap between the two sides of the bubble walls, phase transitions can filter particles to form superheavy dark matter~\cite{Baker:2019ndr,Chway:2019kft,Chao:2020adk,Jiang:2023nkj}, or trap particles to form solitons or even primordial black holes (PBHs)~\cite{Huang:2017kzu,Bai:2018vik,Hong:2020est,Baker:2021nyl,Kawana:2021tde,Huang:2022him,Kawana:2022lba,Lu:2022paj,Lu:2022jnp,Kim:2023ixo,Xie:2024mxr,Cline:2022xhx,Gehrman:2023qjn}.
\item The bubble collision or over-densities arise from the randomness of bubble nucleation can collapse to PBHs~\cite{Liu:2021svg,Hashino:2021qoq,Kawana:2022olo,Hashino:2022tcs,Lewicki:2023ioy,Gouttenoire:2023naa,Gehrman:2023qjn,Kanemura:2024pae,Cai:2024nln,Arteaga:2024vde,Ai:2024cka,Goncalves:2024vkj}.
\end{enumerate}

As mentioned above and illustrated in Fig.~\ref{fig:thermalhist}, from a particle physics perspective, a FOEWPT may occur when there is a barrier separating the two local minima (vacua) of the finite temperature Higgs potential, denoted as $V_T(h,T)$, which originates from physics beyond the SM. This barrier {may} prevent a smooth transition from the false vacuum to the true vacuum, necessitating a thermal tunneling process known as FOEWPT. The probability of this tunneling is determined by the vacuum decay rate~\cite{Linde:1981zj}. Once the decay rate exceeds a certain threshold, bubbles begin to form, leading to FOEWPT.

\begin{figure}[t]
\centering
\includegraphics[width=1.0\textwidth]{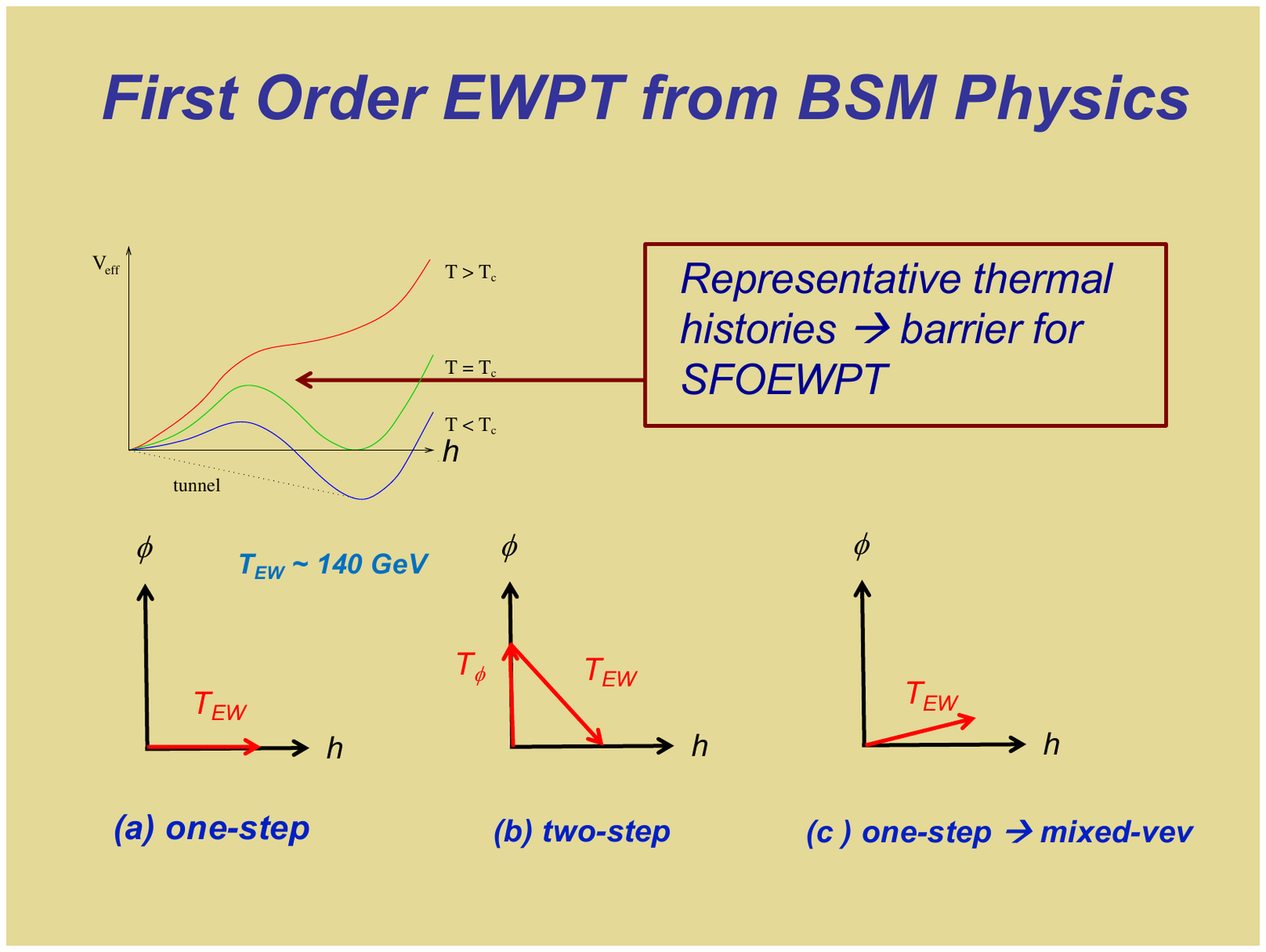}
\caption{Representative thermal histories allowing for a first order electroweak phase transition to the present Higgs phase at the electroweak temperature $T_{EW}$ in the present of an additional scalar $\phi$: (a) a one-step transition arising from thermal loops containing $\phi$ or zero-temperature, higher dimensional operators induced by $\phi$ and/or other new particles; (b) a two-step transition first at temperature $T_\phi$ to a phase in which $\phi$ obtains a non-zero vev, followed by a first order transition the Higgs phase; (c) a one step transition to the Higgs phase in which $\phi$ also obtains a non-zero vev (adapated from Ref.~\cite{Ramsey-Musolf:2019lsf}).}
\label{fig:hist}
\end{figure}

{It is useful to characterize the possibility of a FOEWPT in terms of both the thermal history of EW symmetry breaking and the underlying particle physics. In terms of the former, Fig.~\ref{fig:hist} displays three generic thermal histories in the presence of a new scalar field $\phi$. The latter may either by charged under SM gauge symmetries or a gauge singlet. Fig.~\ref{fig:hist}(a) indicates a one-step transition from the symmetric phase to the present Higgs phase. Fig \ref{fig:hist}(b) gives a two-step transition history, wherein the universe first goes to a phase associated with a non-zero vev for $\phi$ followed by a subsequent transition to the Higgs phase. Fig.~\ref{fig:hist}(c) indicates a one-step transition to a mixed-vev phase.}

{The particle physics driving the FOEWPT in these scenarios} can be categorized into four types based on the source of the potential barrier~\cite{Chung:2012vg}: I) thermally driven, IIA) tree-level with renormalizable operators, IIB) tree-level with high-dimensional operators, and III) {zero-temperature} loop-driven. Each type encompasses numerous new physics models with diverse cosmological implications and phenomenological signals. {Types I, IIB, and III are particularly relevant to the one-step transitions of Fig.~\ref{fig:hist}(a), while type IIA is more germane to the thermal histories of Figs. \ref{fig:hist}(b,c).} As we will see in the following subsections, the CEPC can efficiently probe type-IIB via Higgs precision measurements, and type-I, type-IIA via {both precision Higgs measurements and} Higgs exotic decay.

 {Collider} experiments provide an efficient approach to probe the underlying physics of the barrier and test the nature of the electroweak phase transition and also the associated new physics mechanisms with baryogenesis and/or dark matter. The typical phenomenology of FOEWPT includes the on-shell production of new particles or deviations in the properties of the Higgs boson~\cite{Ramsey-Musolf:2019lsf}. {Importantly, the globally envisioned future collider program, including the high luminosity LHC as well as future lepton and hadron colliders, is capable of providing a (nearly) definitive test of BSM-induced FOEWPT scenarios. The fundamental reason is that EW symmetry breaking in the Standard Model occurs at the \lq\lq electroweak temperature\rq\rq\, $T_\mathrm{EW}\sim 140$ GeV. Any BSM physics that changes the transition from a smooth crossover into a FOEWPT cannot be too heavy with respect to  $T_\mathrm{EW}$ and cannot couple too feebly to the Higgs boson. Generic arguments imply that the mass of the new particles responsible for a FOEWPT should be $\lesssim 700$ GeV and that the magnitude of associated changes in Higgs boson properties larger than $\mathcal{O}(1\%)$\cite{Ramsey-Musolf:2019lsf}. While exceptions occur, the vast majority of model-specific studies bear out these generic expectations. }

One example is the gauge singlet {scalar} extension of the Standard Model (referred to as the xSM, falling into type-IIA barrier). {For seminal studies, see Refs.~\cite{Pietroni:1992in,Choi:1993cv,Espinosa1993,Benson:1993qx,Ham:2004cf,Profumo:2007wc,Espinosa:2011ax} and Ref.~\cite{Ramsey-Musolf:2019lsf} for a comprehensive set of references. The xSM exhibits a rich collider phenomenology as well as}  correlations, {including} complementarity between high-energy collider searches for di-Higgs/di-boson processes and GW measurements~\cite{Alves:2018jsw,Alves:2018oct,Alves:2019igs,Alves:2020bpi,Alves:2020bpi,Liu:2021jyc,Ramsey-Musolf:2024ykk}. Similar analyses have been conducted for other models, such as the Georgi-Machacek model~\cite{Bian:2019bsn,Zhou:2018zli}, the inert doublet model (IDM)~\cite{Huang:2017rzf,Wang:2020wrk}, the dark gauged sector~\cite{Ghosh:2020ipy}, etc. See Ref.~\cite{Ramsey-Musolf:2019lsf} for a more extensive references to the extensive literature and studies in other models. {Below we highlight recent xSM studies particularly relevant to the CEPC.}

The LHC can investigate new particles associated with the physics of FOEWPT up to the TeV scale due to its exceptionally high collision energy. While the 240 GeV CEPC cannot directly probe heavy degrees of freedom, it presents an opportunity for precise measurements. CEPC can accurately determine the properties of the Higgs boson, indirectly providing insights into FOEWPT. Notably, deviation in the couplings involving the Higgs boson ($hWW$, $hZZ$, and $h^3$) may indicate potential dynamics associated with FOEWPT~\cite{Noble:2007kk,Profumo:2014opa,Huang:2015izx,Huang:2016cjm,Cao:2017oez,Alves:2018jsw,Zhou:2018zli,Liu:2021jyc,Chen:2019ebq,Bian:2019bsn,Su:2020pjw,Song:2022xts} . Furthermore, under certain circumstances, the CEPC can effectively explore scenarios involving light new degrees of freedom, potentially leading to their discovery. {The CEPC presents a particular opportunity to search for exotic Higgs decays into new light scalars associated with catalysis of a FOEWPT.} Subsequent subsections will delve into different FOEWPT scenarios, with a particular emphasis on their interplay with the CEPC.

\subsection{Higgs precision measurements}

The Higgs couplings can be influenced by new physics associated with the FOEWPT barrier.  {Recent studies demonstrate} that Higgs factories can effectively investigate FOEWPTs in various scenarios~\cite{Huang:2016cjm,Ramsey-Musolf:2019lsf}, including the general and $\mathbb{Z}_2$-symmetric xSMs (type-IIA barrier), the real triplet scalar extension (type I and type-IIA), stop-like scalar and heavy fermion extensions of the SM (type-I barrier). The correlation and complementarity between precise measurements of Higgs properties like the $hZZ$ and $h^3$ couplings, as well as GW signals, enable this probing. Remarkably, CEPC/ILC/FCC-ee can detect a significant portion of the FOEWPT parameter space by observing deviations in the $hZZ$ coupling. 
Additionally, the deviation of the $h^3$ coupling, can also be investigated. 
At the CEPC TDR nominal operation scenario, the $\sigma(ZH)$ is expected to be measured with a relative accuracy of 0.26\%~\cite{CEPCStudyGroup:2023quu,CEPCPhysicsStudyGroup:2022uwl}. With recent progress in AI-enhanced reconstruction and analysis, this accuracy is expected to improve significantly. Therefore, in the following discussion, we present the phase space coverage not only assuming the relative uncertainty of 0.26\%, but also considering an aggressive scenario of 0.1\%.

\begin{figure}[t]
\centering
\includegraphics[width=0.8\textwidth]{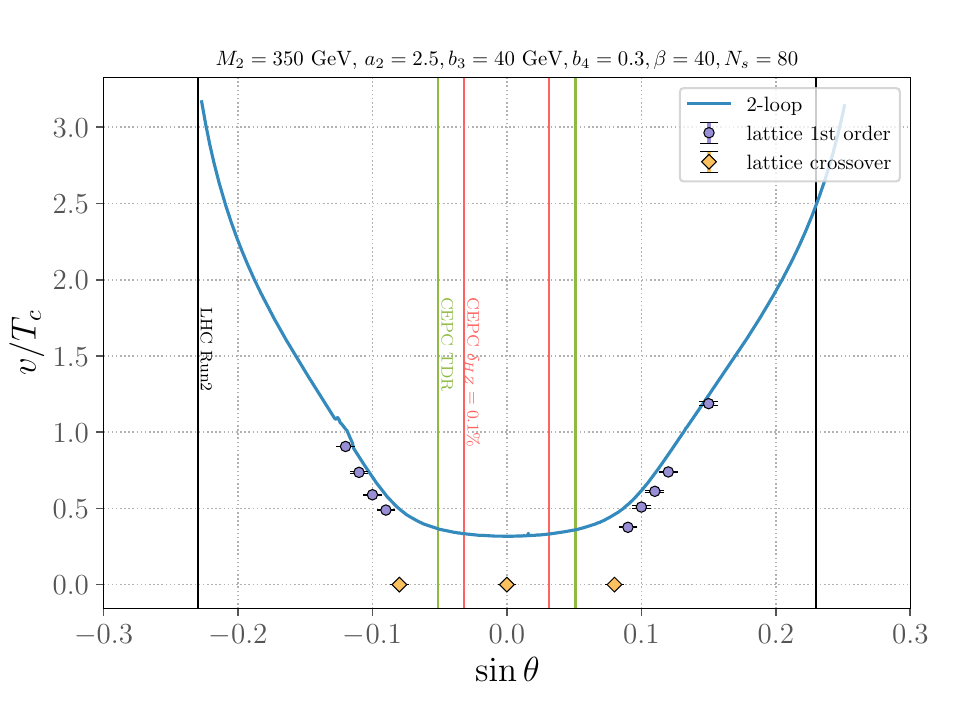}
\caption{Discontinuity in the Higgs vev ($v$) at the critical temperature ($T_c$) as  function of the doublet-singlet mixing angle $\theta$ in the real scalar singlet extension of the SM (xSM). Blue circles (yellow diamonds) give lattice results for a first order (crossover) transition, while blue curve is obtained from a two-loop perturbative computation using the $T>0$ EFT framework. Black and green vertical lines indicate $\sin\theta$ sensitivities of LHC Run 2 and the CEPC, respectively (adapted from Ref.~\cite{Niemi:2024axp} by G. Xia).}
\label{fig:lat}
\end{figure}

{Here, we illustrate the precision Higgs measurement probes of the FOEWPT using two explicit models as well as the SM effective field theory (EFT). Starting with the xSM, a combination of thermal effective field theory (EFT) and lattice studies have refined the confrontation between theory and experiment~\cite{Niemi:2024axp}. A key quantity that connects the phase transition and collider phenomenology is the singlet-doublet mixing angle, $\theta$. Fig.~\ref{fig:lat} shows the discontinuity in the order parameter $\langle \phi^\dag\phi\rangle$ as a function of the $\sin\theta$. The solid lines show the relationship as derived from perturbation theory, while the dots give the results of lattice simulations. The latter indicate for sufficiently small $|\sin\theta|$, the transition is a smooth crossover, whereas perturbation theory always implies the presence of a FOEWPT. The vertical lines show the $\sin\theta$ sensitivities of the HL LHC and CEPC, with the latter dominated by measurements of the associated production $e^+e^-\to ZH$ cross section. Importantly, the perturbative computations imply that there would always be a significant portion of FOEWPT-viable parameter space inaccessible to the CEPC, whereas the lattice results indicate otherwise. For larger (smaller) values of the Higgs portal coupling, the value of $|\sin\theta|$ at the crossover-FOEWPT boundary decreases (increases). In short, the state-of-the-art theory suggests that for the singlet-like scalar being relatively heavy, precision Higgs studies at the CEPC will provide a powerful test of the xSM FOEWPT when the latter is connected to non-vanishing singlet-doublet mixing. }

\begin{figure}[t]
\centering
\includegraphics[width=0.75\textwidth]{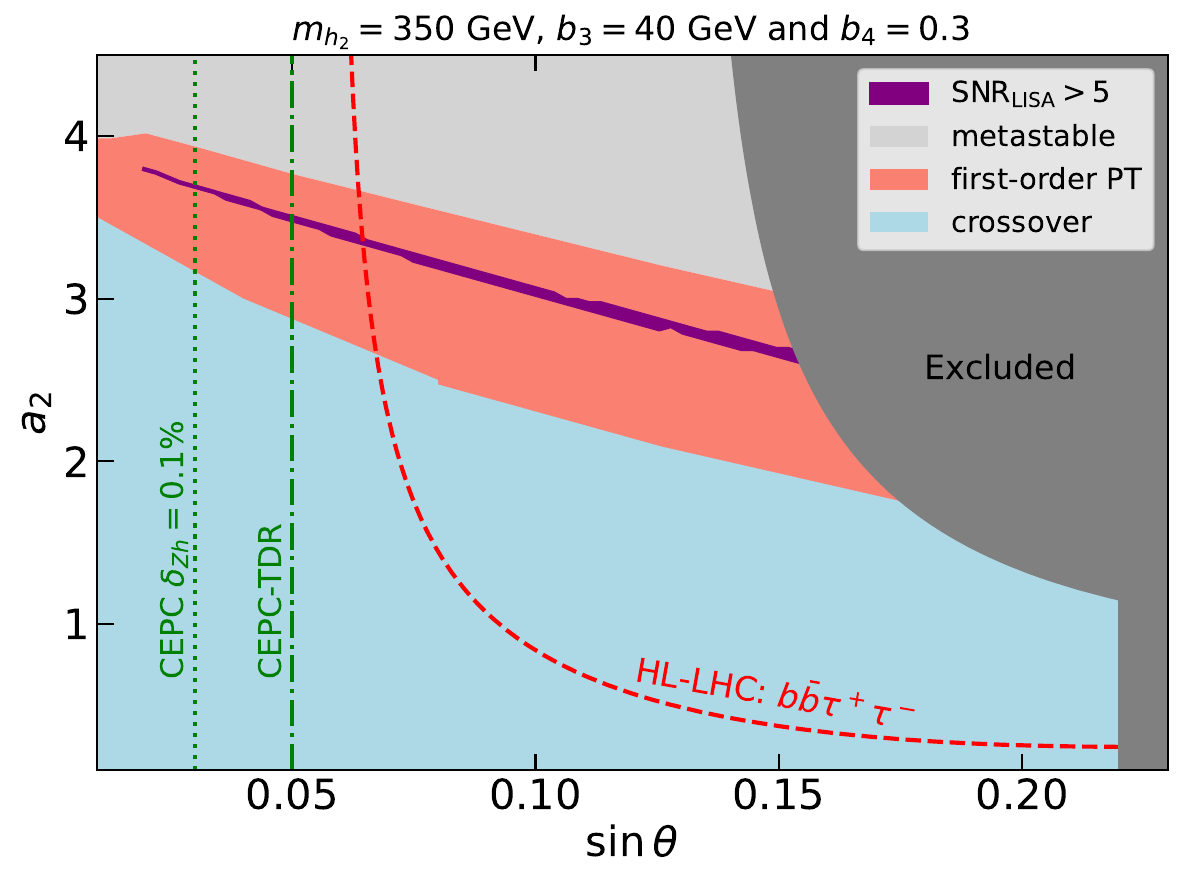}
\caption{Phase diagram for the real scalar singlet extension of the SM in the plane of the doublet-singlet mixing angle $\theta$ and double-singlet cross-quartic portal coupling $a_2$. Light blue and red regions indicate cross over and two-step EWPT regions, respectively, while the light grey region corresponds to a metastable electroweak vacuum. The dark grey region is experimentally excluded. Dashed red curve and dashed green lines indicate sensitivities of  high luminosity LHC resonant di-Higgs searches in the $b{\bar b}\tau^+\tau^-$ channel and different scenarios of the CEPC precision $\sigma(e^+e^-\to Zh)$ exclusion reach, respectively. Purple band shows parameter region consistent with a LISA GW observation with SNR $>5$. Dark grey region is experimentally excluded (adapted from Ref.~\cite{Ramsey-Musolf:2024ykk} by V.Q. Tran)}
\label{fig:xSM_GW_Collider}
\end{figure}

The interplay between the High Energy Frontier experiments---i.e., the LHC, HL-LHC, and future electron-positron Higgs factories---and future GW probes is illustrated in Fig.~\ref{fig:xSM_GW_Collider} (adapted from Ref.~\cite{Ramsey-Musolf:2024ykk}), which shows the xSM phase diagram in the $\sin\theta$-portal coupling ($a_2$) plane. The pink region gives the two-step EWPT viable region, while the purple band indicates the LISA sensitivity for a signal-to-noise ratio $\geq 5$ level. The  dashed curve shows the exclusion reach for resonant di-Higgs production at the HL-LHC in the  $b{\bar b} \tau^+\tau^-$ channel. The vertical line indicates the CEPC sensitivity to $\sin\theta$, which significantly extends the reach as compared to the HL-LHC. In the event of a  GW observation, and assuming the xSM is realized in nature, one could either anticipate a CEPC discovery of a significant singlet-doublet mixing or identify the narrow region of xSM parameter space consistent with both sets of experiments.

\begin{figure}[t]
\centering
\includegraphics[width=0.55\textwidth]{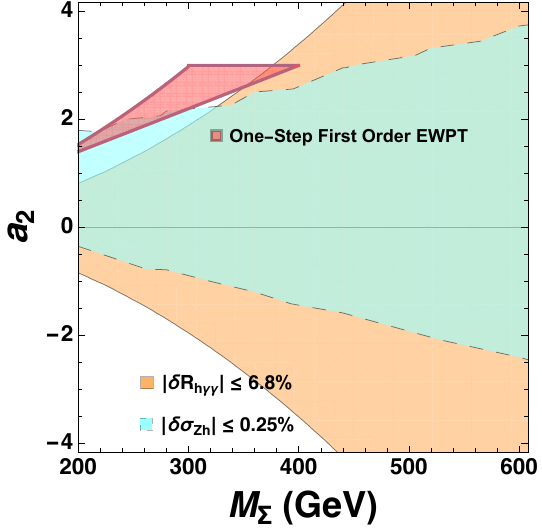}
\caption{Sensitivities of (a) the CEPC measurement of $\sigma(Zh)$ with 0.25\% precision (light blue) and (b) HL-LHC determination of the Higgs diphoton decay rate at 6.8\% precision (light brown) to the  FOEWPT in the real triplet extension of the SM. Vertical and horizontal axes give the doublet-triplet cross-quartic coupling and triplet mass, respectively. Red region indicates a single step FOEWPT (adapted from Ref.~\cite{Ramsey-Musolf:2021ldh} by G. Xia and J. Zhou).}
\label{fig:SigmaSM}
\end{figure}

{Combinations of lattice and thermal EFT studies have also been recently carried out for the real triplet scalar extension, or $\Sigma$SM \cite{Niemi:2018asa,Niemi:2020hto}. A SFOEWPT may arise via either thermal effects in the one-step case or tree-level barrier in the two-step case [Fig.~\ref{fig:hist}(b)]. The $\sigma$SM can also modify Higgs boson properties via loop effects in both the $h\to\gamma\gamma$ and $e^+e^-\to Zh$ processes. The relevant parameters are the triplet scalar mass, $M_\Sigma$, and portal coupling $\lambda_3$. As indicated in Fig.~\ref{fig:SigmaSM} (adapted from Ref.~\cite{Ramsey-Musolf:2021ldh}), the associated production process $e^+e^-\to Zh$ provides a particularly powerful probe of the one-step FOEWPT region, indicated in red. Assuming a $\sim 0.25\%$ determination of the cross section, the CEPC could probe all of the one-step FOEWPT-viable parameter space in this scenario for new physics.}

Below, we {also} use the SM effective field theory (EFT) as a representative example of the type-IIB barrier to show the correlation between the Higgs trilinear coupling and the FOEWPT.

\begin{figure}
    \centering
    \includegraphics[width=0.8\linewidth]{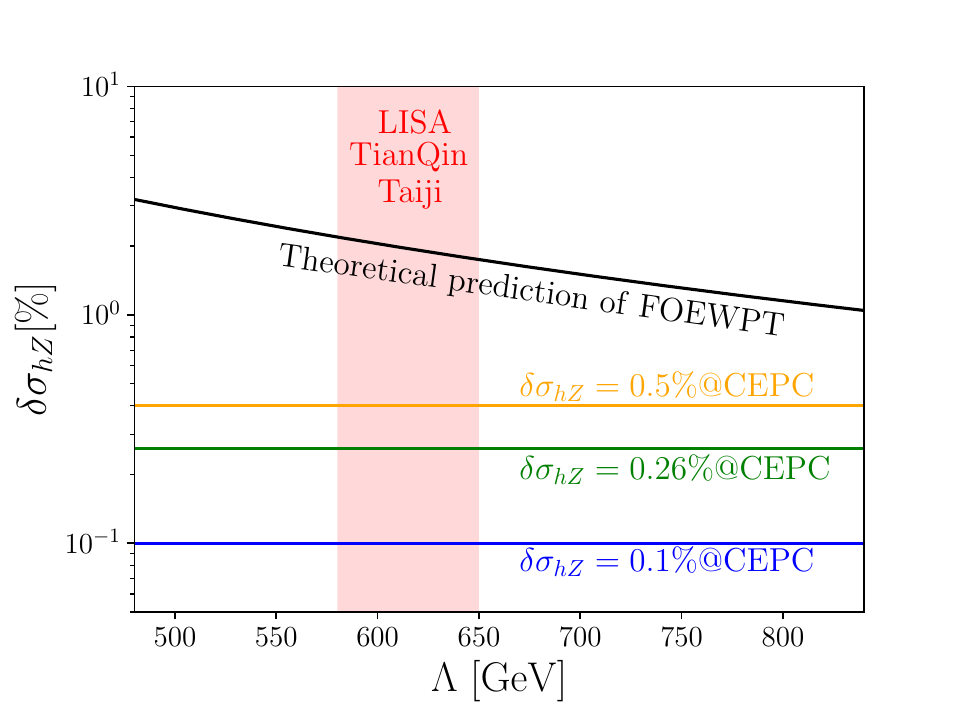}
    \caption{ The observational abilities of different experiments for the FOEWPT in the SMEFT.}
    \label{fig:ewpt_cepcrange}
\end{figure}

If new physics degrees of freedom are significantly heavier than the electroweak scale, they can be integrated out, allowing for an SM EFT description. In the scalar sector, a generic Higgs potential can be derived from the dimension-6 SM EFT framework, as indicated in~\cite{Huang:2015izx, Huang:2016odd, Cao:2017oez}.
\begin{equation}\label{vdim6}
V(h)=\frac{1}{2} \mu^2 h^2-\frac{\lambda}{4} h^4+\frac{1}{8\Lambda^2} h^6.
\end{equation}
This potential has been associated with various new physics models such as inert singlet, doublet, triplet, or composite Higgs models~\cite{Huang:2015izx, Huang:2016odd, Cao:2017oez}. 
Notably, the FOEWPT predicts a discernible deviation of the tri-linear Higgs coupling compared to the prediction of the SM.
At the one-loop level, the deviation of the tri-linear Higgs coupling $\delta_h $ could be obtained as $\delta_h \in(0.6 , 1.5)$, which could be tested by the precise measurements of the cross section of $e^+e^-\to h Z$ at CEPC~\cite{Huang:2015izx, Huang:2016odd, Cao:2017oez} as shown in Fig.~\ref{fig:ewpt_cepcrange}.
As the golden channel of Higgs factory, the $h Z $ production rate has been calculated very precisely, including one-loop and two-loop quantum corrections. Here, we refer $\sigma_{h Z}^{\mathrm{SM}}$ to the most state-of-the-art calculation of $h Z$ production in the SM~\cite{Gong:2016jys, Sun:2016bel}.The cross section of the $h Z$ channel $\sigma_{h Z}$ could be measured with an accuracy of $0.25 \%$ at CEPC. We define the deviation of cross section of the $h Z$ production, normalized to the SM cross section, as follows:
$$
\delta{\sigma_{h Z }} \equiv \frac{\sigma_{h Z}}{\sigma_{h Z}^{\mathrm{SM}}}-1 .
$$

In general, the operator $\mathcal{O}_6=|H|^6$ contributes to the $hZ$ cross section through loop corrections. Other dim-6 operators, e.g. $\mathcal{O}_H=\frac{1}{2}\left(\partial_\mu|H|^2\right)^2$ and $\mathcal{O}_T=\frac{1}{2}\left(H^{\dagger} \overleftrightarrow{D}_\mu H\right)^2$, contribute to the $h Z $ production at tree-level. 
At a lepton collider with a center of mass energy $\sqrt{s}=250~\mathrm{GeV}$, the high dimension operators' contribution to the $h Z$ production is approximately given by
\begin{multline}\label{zh_approx}
\delta \sigma_{h Z} \simeq  \left(0.26 c_{W W}+0.01 c_{B B}+0.04 c_{W B} -0.06 c_H-0.04 c_T\right.  \\
\left.+0.74 c_L^{(3) \ell}+0.28 c_{L L}^{(3) \ell}+1.03 c_L^{\ell}-0.76 c_R^e\right) \times \mathrm{TeV}^2+0.016 \delta_h.
\end{multline}
The contribution $\delta_h$ in the $h Z$ channel is often neglected due to its loop suppression in operator analyses. However, we argue against ignoring it in our FOEWPT study based on the following reasons. Firstly, the FOEWPT condition necessitates a significant coefficient $c_6$, leading to a substantial contribution of $0.96\% - 2.4\%$ to $\delta{\sigma_{hZ}}$. Importantly, the current experimental constraints on $c_6$ are practically non-existent. Secondly, compared to other dimension-6 operators, the coefficients of which face stricter constraints, $c_6$ stands out as being less constrained. As such, the contribution $\delta_h$ cannot be disregarded.
The above study demonstrates that the possibility of FOEWPT induced by the $|H|^6$ operator remains viable and consistent with current experimental data. The investigation into the dim-6 operators generated in three scalar extension models can be applied to a wide range of new physics models~\cite{Cao:2017oez}. Typically, a FOEWPT requires an Higgs portal coupling on the order of unity, and a large Higgs portal coupling may suggest a composite nature for the Higgs boson. If the Higgs boson is a pseudo-Goldstone boson originating from strong dynamics, the coefficients of dim-6 operators can be estimated using naive dimensional analysis. The estimated coefficients of dominant CP-conserving operators are presented below: 
$$
c_{W W} \sim c_{B B} \sim c_{W B} \sim \frac{1}{\Lambda^2} \sim \frac{1}{(4 \pi f)^2},\quad
c_H \sim c_T \sim\frac{1}{f^2},\quad
c_6 \sim-\frac{\Lambda^2 }{f^4}=-\frac{1}{(f / 4 \pi)^2}.
$$
If the EW phase transition is a FOEWPT, then one needs
$$
\frac{1}{(0.89~ \mathrm{TeV})^2}<-c_6<\frac{1}{(0.55 \mathrm{TeV})^2},~\text{or equivalently}~6.91~ \mathrm{TeV}<f<11.18~ \mathrm{TeV}.
$$
The coefficients $c_{W W, B B, W B, H, T}$ are consistent with current experiments if the scale $f$ is within the above range. Using Eq.~(\ref{zh_approx}), we find $\delta \sigma_{h Z } \sim 0.1 \%$ without the $\delta_h$ term. The FOEWPT condition requires $0.6<\delta_h<1.5$. Therefore, including the $\delta_h$ contribution yields $\delta \sigma_{hZ}$ in the range of $(0.96-2.4) \%$, which could be probed at future lepton colliders, such as the CEPC.

\begin{figure}[t]
	\centering
	\includegraphics[width=0.75\textwidth]{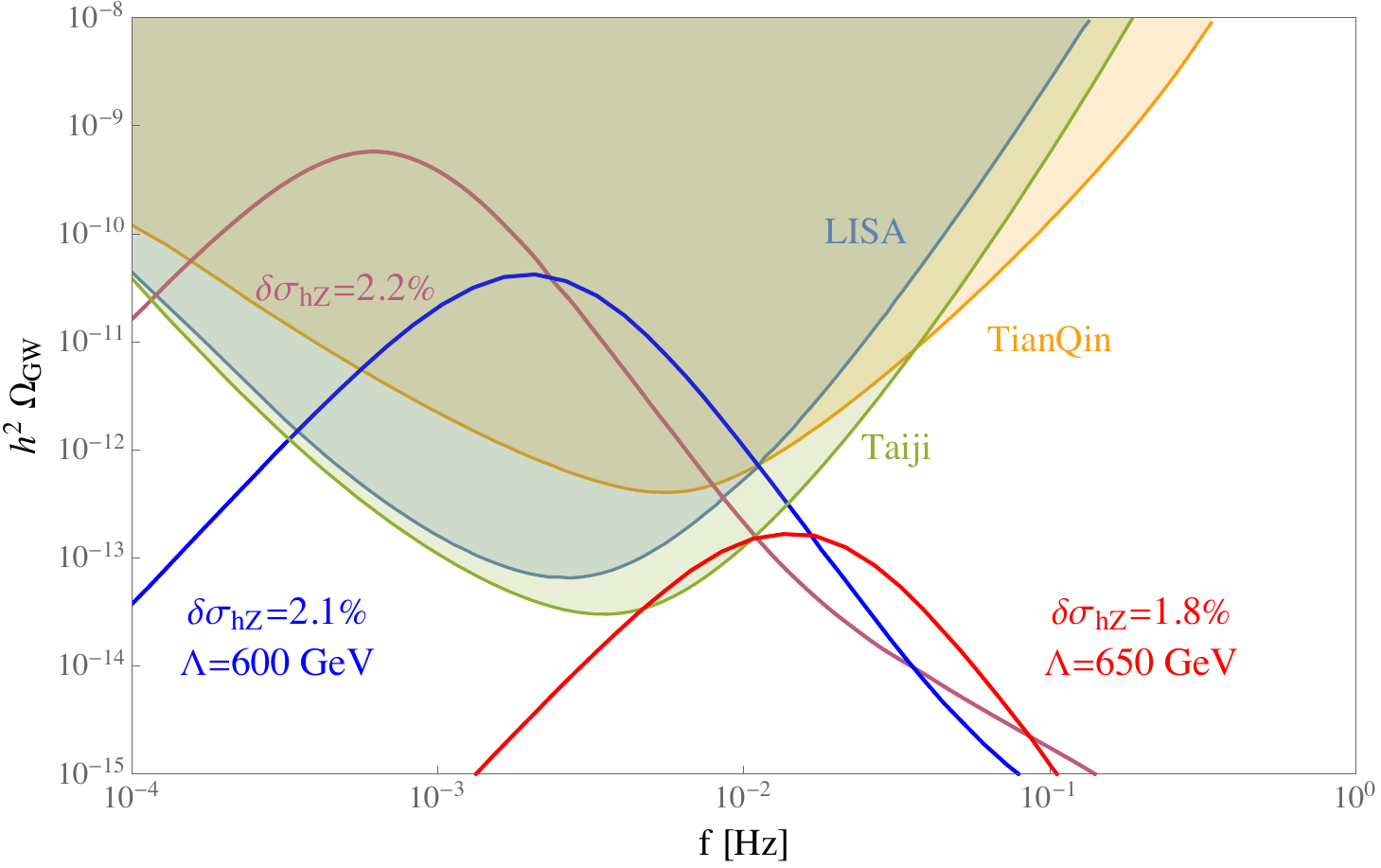}
	\renewcommand{\figurename}{Fig.}
	\caption{General prediction of $hZ$ cross section deviation to SM and its corresponding GW signals in the SMEFT under the condition of FOEWPT~\cite{Huang:2015izx, Huang:2016odd}.}
	\label{fig:gwdim6}
\end{figure}
Another important prediction of this type of new Higgs potential is the detectable phase transition GWs in near-future
space-based interferometers, such as LISA, TianQin, and Taiji. As shown in Fig.~\ref{fig:gwdim6}, the collider in synergy with the GW experiments could make complementary tests on the generic Higgs potential from generic new physics models, which is also directly connected with the electroweak baryogenesis~\cite{Huang:2015izx, Huang:2016odd, Cao:2017oez}.
\subsection{Higgs exotic decay}

It has been proposed that studying the exotic decay of the Higgs boson can effectively probe the FOEWPT, for two main reasons. Firstly, the Higgs boson has a very narrow width ($\Gamma_h\approx4.07$ MeV), which makes it highly sensitive to the BSM physics. Secondly, the FOEWPT requires significant interaction between the Higgs boson and the new physics sector. Consequently, if kinematically allowed, the Higgs boson can have a large decay branching ratio to the new physics particles. Therefore, accurately determining the decay properties of the Higgs boson at the CEPC would greatly assist in testing relevant models and studying the characteristics of the electroweak phase transition. See Section~\ref{sec:Higgsexotic} for more new physics implications from the Higgs exotic decay, while in this subsection we focus on the relation to the FOEWPT.

This concept has been explored in research on the xSM (i.e. real singlet expansion of the SM), where the potential barrier can be thermally driven (type-I) or tree-level driven with renormalizable operators (type-IIA)~\cite{Carena:2019une,Kozaczuk:2019pet,Carena:2022yvx}. In this scenario, when contribution of the singlet $s$ to the potential has a $\mathbb{Z}_2$ symmetry, or when $s$ has a small mixing angle $\theta$ with the Higgs boson $h$, the exotic decay $h\to ss$ partial width can be expressed as
\begin{equation}
\Gamma(h\to ss)\approx\frac{a_2^2v_{\rm EW}^2}{32\pi m_h}\sqrt{1-\frac{4m_s^2}{m_h^2}},
\end{equation}
where $m_h=125$ GeV is the Higgs mass, $m_s$ is the singlet mass, and $a_2$ is the Higgs-portal coupling of $\mathcal{L}\supset-a_2h^2s^2/4$. On the other hand, the necessary condition for FOPEWPT requires~\cite{Kozaczuk:2019pet}
\begin{equation}
a_2>\begin{dcases}~
\frac{2m_s^2}{v_{\rm EW}^2},&\text{$\mathbb{Z}_2$-odd $s$};\\
~\frac{m_s^2}{4v_{\rm EW}^2}\frac{\Delta}{1-\Delta},&\text{small mixing $s$, with $\Delta\gtrsim0.6-0.8$}.
\end{dcases}
\end{equation}
This clearly shows the relation between FOEWPT and Higgs exotic decay.

\begin{figure}
\centering
\includegraphics[width=0.49\textwidth]{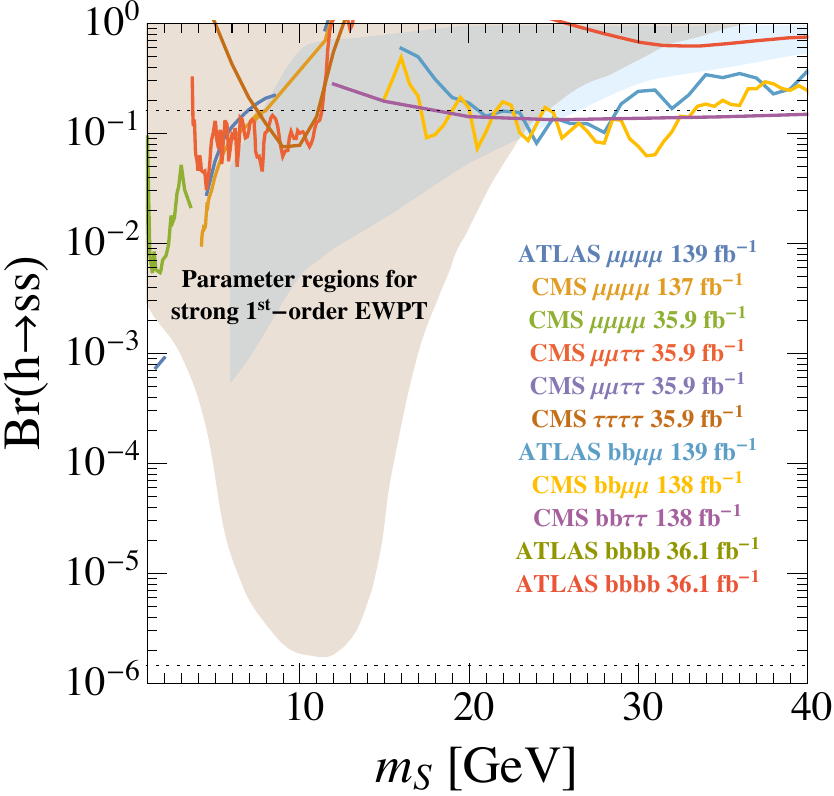}
\includegraphics[width=0.49\textwidth]{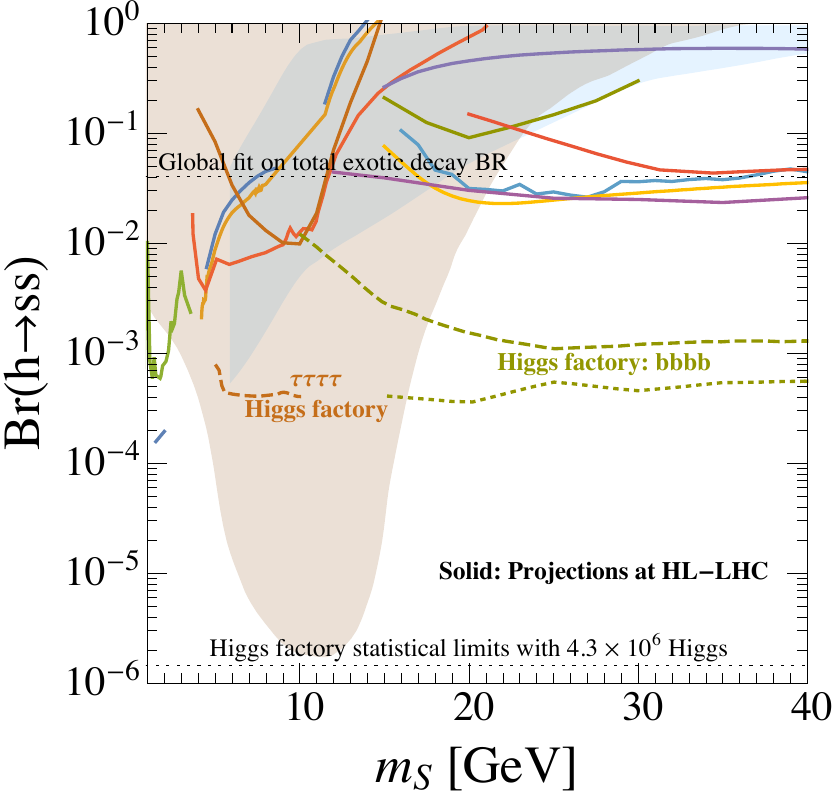}
\caption{Updated version of figure from Ref.~\cite{Carena:2022yvx}, the Higgs exotic decay $h\to ss\to XXYY$ as a probe for the FOEWPT, where $X$ and $Y$ denote the SM particles. Left: the current bounds; Right: the future projections. The $s\to$ SM decays are assumed to be mediated by $h$-$s$ mixing~\cite{Gershtein:2020mwi}. The expected HL-LHC reach for exotic decay branching ratio (4\%~\cite{deBlas:2019rxi}) and the statistical limit of $4.3\times 10^6$ Higgs at future lepton colliders are shown as upper and lower horizontal dotted lines, respectively. The FOEWPT parameter space is shown in brown and light blue shadowed regions for various benchmarks~\cite{Kozaczuk:2019pet,Carena:2019une}. Dashed lines are the projected reach of future lepton colliders. See the text for the details.}
\label{fig:bounds_2}
\end{figure}

For a $\mathbb{Z}_2$-symmetric potential, the $h\to ss$ process leads to the Higgs invisible decay. If $s$ mixes with $h$, then the decay chain would be $h\to ss\to XXYY$, where $X$ and $Y$ denote SM particles. Recent research, summarized in the left panel of Fig.~\ref{fig:bounds_2} from Ref.~\cite{Carena:2022yvx}, has constrained the FOEWPT parameter space using data from LHC searches targeting specific final states, namely $XXYY=\mu^+\mu^-\mu^+\mu^-$, $\mu^+\mu^-\tau^+\tau^-$, $\tau^+\tau^-\tau^+\tau^-$, $b\bar b\mu^+\mu^-$, $b\bar b\tau^+\tau^-$, and $b\bar bb\bar b$. The parameter space corresponds to both the spontaneous $\mathbb{Z}_2$-breaking model (brown region) and the general xSM with a mixing angle $\sin\theta=0.01$ between the $s$ and $h$ scalar bosons (blue region)\footnote{These regions have been obtained from perturbative treatments of the phase transition. The recent lattice study of Ref.\cite{Niemi:2024axp} suggests that the portions of these regions associated with very small $\mathrm{Br}(h\to ss)$ likely correspond to a crossover transition rather than a FOEWPT.}. These bounds assume that the decays $s\to XX$ and $s\to YY$ are mediated by the mixing between the Higgs boson $h$ and the new scalar boson $s$, with branching ratios obtained from Ref.~\cite{Gershtein:2020mwi}.

Projections for the HL-LHC are presented in the right panel of Fig.~\ref{fig:bounds_2}. Although the dominant decay channel for $s$ is $s\to b\bar b$ when $m_s\gtrsim10$ GeV, the reach of the $b$-relevant channels is constrained due to the large multi-jet background at the LHC. However, lepton colliders like CEPC can effectively measure this channel because of the clean collision environment. For example, CEPC operating at $\sqrt{s}=240$ GeV with an integrated luminosity of $5~{\rm ab}^{-1}$ enables probing of the $b\bar bb\bar b$ channel through associated production $e^+e^-\to hZ$, achieving a branching ratio sensitivity down to $10^{-3}$. This coverage extends to a substantial portion of the FOEWPT parameter space, as depicted by the dashed and dotted lines in the right panel of Fig.~\ref{fig:bounds_2} (from~\cite{Liu:2016zki} and~\cite{Wang:2023zys}, respectively). A similar sensitivity is found in an ILC-related research~\cite{Kato:2022}. Additionally, CEPC exhibits improved sensitivity for the $\tau^+\tau^-\tau^+\tau^-$ channel compared to the HL-LHC, particularly for $4~{\rm GeV}\lesssim m_s\lesssim10$ GeV~\cite{Shelton:2021xwo}. This is demonstrated by the orange dashed line in Fig.~\ref{fig:bounds_2}. By combining the $b\bar bb\bar b$ and $\tau^+\tau^-\tau^+\tau^-$ channels, CEPC has the potential to probe almost the entire FOEWPT parameter space for the general xSM with a mixing angle $\sin\theta=0.01$.

The above discussions are based on the prompt decays of $s$. A small mixing angle $\theta$ for the singlet $s$ can lead to its detection in long-lived particle (LLP) searches, as suggested in Ref.~\cite{Liu:2022nvk}. The use of LLP detectors at the LHC enables an extension of the sensitivity to FOEWPT, surpassing the coverage achievable through prompt searches for exotic Higgs decays. We expect the LLP search at future Higgs factories can also have significant capability in probing the FOEWPT scenario.

Although the discussions presented here are taking the xSM and real triplet extension as examples, they also generally apply to other BSM models. For example, a complex scalar $S^+$ that is an $SU(2)_L$ singlet but carries unit charge under $U(1)_Y$, which generally exists in lepton-portal dark matter models~\cite{Liu:2021mhn,Cermeno:2022rni,Jueid:2020yfj,Jueid:2021wla}, can also induce FOEWPT, and the corresponding parameter space can be probed by the $h\to S^+S^-$ decay~\cite{Liu:2021mhn}. Furthermore, the CEPC precise measurement on the Higgs CP property also helps to identify the BSM CP-violating phase~\cite{Chen:2017bff,Chen:2017nxp,Ge:2020mcl,Alonso-Gonzalez:2021jsa}, which is another necessary condition for electroweak baryogenesis. Recently, it has been shown that the Higgs exotic decay can even probe the MeV-scale phase transition accounting for the nano-Hertz GW excess~\cite{Li:2023bxy}.

\clearpage

\section{Dark Matter and Dark Sector}
\label{sec:DMDS}

There is substantial evidence for the existence of Dark Matter (DM) from astrophysical and cosmological observations. However, its particle nature remains unknown and awaits exploration. Conventional dark matter candidates, such as Weakly Interacting Massive Particles (WIMPs), have significant couplings to Standard Model particles, which enable the thermal freeze-out mechanism and result in the correct relic abundance \cite{Bertone:2016nfn, Roszkowski:2017nbc}. This has motivated searches for DM at high-energy hadron and lepton colliders, aiming to produce DM particles directly. These searches target both heavy DM particles up to the TeV scale and lighter ones down to the GeV scale. Collider searches offer a complementary cross-check to direct and indirect search experiments. A recent comprehensive review of DM phenomenology can be found in Ref.~\cite{Cirelli:2024ssz} and the review of DM collider phenomenology can be found in Ref.~\cite{Boveia:2018yeb}. 

Recently, a new class of models has garnered significant attention within the scientific community, proposing that DM does not directly couple to SM particles but resides in a ``dark sector" (or ``hidden sector") \cite{Pospelov:2007mp}. The dark sector interacts with the SM through portal particles, which could have tiny couplings to the SM sector to evade the null results from the direct detection, but still could provide the DM right relic abundance. Thus, for these interactions to enable secluded annihilation for DM relic abundance, the mass of the portal particles must be smaller than that of the DM particles. Due to their feeble couplings and low masses, these portal particles are ideal targets for searches at the luminosity frontier. Electron-positron colliders such as CEPC and FCC-ee, which function as high-luminosity $W$, $Z$, and Higgs boson factories and offer cleaner experimental environments compared to hadron colliders, are well-suited to search for these feebly interacting portal particles \cite{CEPCStudyGroup:2018ghi, CEPCPhysicsStudyGroup:2022uwl}.
For example, the CEPC can search for exotic decays related to the dark sector in both \(Z\) and Higgs decay events, as discussed in Section~\ref{sec:exotic-decay-in-DS}. Additionally, the CEPC is capable of searching for various DM particles, including those predicted by supersymmetric models, scalar portals, lepton portals, and gauge mixing portals. It can also investigate DM particles with millicharge, electromagnetic form factors, and those described by effective field theory (EFT) frameworks. The simplified models typically used in the LHC new physics searches can be found in Ref.~\cite{LHCNewPhysicsWorkingGroup:2011mji}.

\begin{table}[]
\scriptsize
\begin{tabular}{|c |c |c|c| c| c | c|}
  \hline
 Portal & Effective operator &  $\sqrt{s}$ [GeV] & $\mathcal{L}[ab^{-1}]$ & Sensitivity of CEPC (HL-LHC)& Figs. & Ref.   \\ \hline \hline
  Scalar & $\lambda_{HP} |H|^2 S^2 \rightarrow$ scalar mixing $\sin\theta$ & 250 & 5 & invisible S, $\sin\theta \approx 0.03 $ (0.20 global-fits) & \ref{fig:K-and-S:1} & \cite{Liu:2017lpo} \\ \hline 
 \multirow{3}*{Fermion} & $y_{\ell}\bar{\chi}_L S^{\dagger} \ell_R$  + H.c. &250 &5 & covering $100 \, {\rm GeV} < m_S < 170 \, {\rm GeV}$ & \ref{fig:leptonportal-DrellYan} & \cite{Liu:2021mhn} \\ \cline{2-7}
 & $\kappa \Phi {\overline{q^{\prime}_L}} \ell_R $ + H.c. (dark QCD) & 250 &5  & $m_\Phi \sim $ 10 TeV for $c\tau_{\rm dark pion}\in [1,10^3]$ cm (Null) & \ref{CEPC2} & \cite{Zhang:2021orr} \\ \cline{2-7}
 & $y \Phi \bar{F}_L \ell_R $ + H.c. & 240 & 5.6 & $y \theta_L \in [10^{-11},\; 10^{-7}]$ ($\lesssim 10^{-8}-10^{-9}$) & - & \cite{Cao:2023smj} \\ \hline
 \multirow{6}*{Vector} & 
 $ A'_{\mu} \left(e \epsilon J_{\mathrm{em}}^\mu + g_D \bar{\chi} \gamma^\mu \chi \right)$  & 250 & 5 & $\epsilon \sim 10^{-3}$ for $g_D =e$ and $m_{A'}<125$ GeV ($\epsilon \sim 0.02 $ ) & \ref{fig:K-and-S:2}, \ref{fig:summary1} & \cite{Liu:2017lpo} \\ \cline{2-7} 
 & \multirow{3}*{$ \varepsilon A_\mu \bar{\chi} \gamma^\mu \chi$, (millicharge DM) }&  250 &5   & $\epsilon \sim 0.1$ for $m_\chi \sim 50$ GeV & \multirow{3}*{ - } & \multirow{3}*{\cite{Liu:2019ogn}}\\ 
\cline{3-5}
& & 91.2 & 2.6  &  $\epsilon \sim 0.02$ for $m_\chi \sim 5$ GeV & & \\
\cline{3-5}
& & 160  & 16  & $\epsilon \sim 0.5$ for $m_\chi \sim 10$ GeV & & \\ 
\cline{2-7}
& \multirow{2}*{$\begin{array}{c}\frac{1}{2} \mu_\chi \bar{\chi} \sigma^{\mu \nu} \chi F_{\mu \nu}+\frac{i}{2} d_\chi \bar{\chi} \sigma^{\mu \nu} \gamma^5 \chi F_{\mu \nu}\\-a_\chi \bar{\chi} \gamma^\mu \gamma^5 \chi \partial^\nu F_{\mu \nu}+b_\chi \bar{\chi} \gamma^\mu \chi \partial^\nu F_{\mu \nu}\end{array}$}&  91.2 &  100   &
$\mu_\chi, d_\chi\sim 4\times 10^{-7} ~(4\times 10^{-6}) \mu_B {\rm ~for~} m_\chi <25 \, {\rm GeV}$ & \multirow{2}*{\ref{fig:DM-electromagnetic}} & \multirow{2}*{\cite{Zhang:2022ijx}} \\ 
\cline{3-5}
& &  240 &  20  & $a_\chi, b_\chi\sim 10^{-6} ~(2\times 10^{-6})\, \rm {GeV^{-2}}{\rm ~for~} m_\chi <80$ GeV  & &  \\ \hline
\multirow{3}*{EFT} & $\frac{1}{\Lambda^2}  \sum_i \left(\bar{\chi} \gamma_\mu (1-\gamma_5) \chi \right) \left(\bar{\ell} \gamma^\mu (1-\gamma_5) \ell\right)$ & 250 &5   & $\Lambda_i\sim 2$ TeV ($m_\chi= 0$) (Null) & 
\ref{fig:LDM} & \cite{Kundu:2021cmo}  \\ \cline{2-7}
& $\frac{1}{\Lambda_A^2} \bar{\chi} \gamma_\mu \gamma_5 \chi \bar{\ell} \gamma^\mu \gamma_5 \ell$ & 250 &5  & $\Lambda_A\sim 1.5$ TeV (Null) & \ref{fig:mq_rej:2}  & \cite{Liu:2019ogn} \\ 
\cline{2-7}
& $\begin{array} {c}\sum_i \frac{1}{\Lambda_i^2} \left(\bar{e} \Gamma_\mu e\right)\left(\bar{\nu}_L \Gamma^\mu \chi_L\right) + {\rm H.c.} \\ \Gamma_\mu= 1,\gamma_5,\gamma_\mu, \gamma_\mu\gamma_5, \sigma_{\mu\nu} \end{array}$ & 240 & 20 & $\Lambda_i\sim 1$ TeV ($m_\chi= 0$) (Null) & \ref{fig:sens:cosmo:collider} & \cite{Ge:2023wye}  \\ 
\hline
\end{tabular}
\caption{Recent results from the CEPC study on dark sector signals. The first column lists the signal signatures, the second column presents the corresponding relevant operators, the third and fourth columns provide the center-of-mass energy and the integrated luminosity. The fifth column presents the sensitivity to the coupling, suppression scale, or branching ratios at the CEPC. Where relevant HL-LHC sensitivities are available, they are shown in parentheses or with Null meaning HL-LHC constraints do not apply. The final two columns provide the corresponding figures in this draft and references for the readers' convenience.
}
\label{tab:DMDS-summary}
\end{table}

The phenomenology studies of DM and dark sector particles at the lepton colliders have been performed intensively recently. In this section, we will refer to previous study results in Refs.~\cite{CEPCStudyGroup:2018ghi, CEPCPhysicsStudyGroup:2022uwl} and focus exclusively on recent progress in the following benchmark scenarios to highlight the capabilities of CEPC. We categorize DM and dark sector models based on their portal connections to SM, specifically: (A) Scalar portal, (B) Fermion portal, (C) Vector portal, and (D) DM within the EFT framework.
In the last subsection (E), we discuss a scenario where dark matter does not require a portal but is instead directly charged under Standard Model gauge interactions. We also examine how the loop effects of this DM influence electroweak and Higgs precision observables, providing constraints on this specific model.
Note that the Axion portal is specifically discussed in the later section More Exotics~\ref{sec:exotica}. For each of these models, we discuss the sensitivity of the CEPC. A summary of the sensitivity for each model, along with corresponding figures and references, is summarized in Table~\ref{tab:DMDS-summary} for readers convenience.

\subsection{Scalar portal}
\label{sec:DM:Zdecay}

\begin{figure}[]
\includegraphics[width=0.55\textwidth]{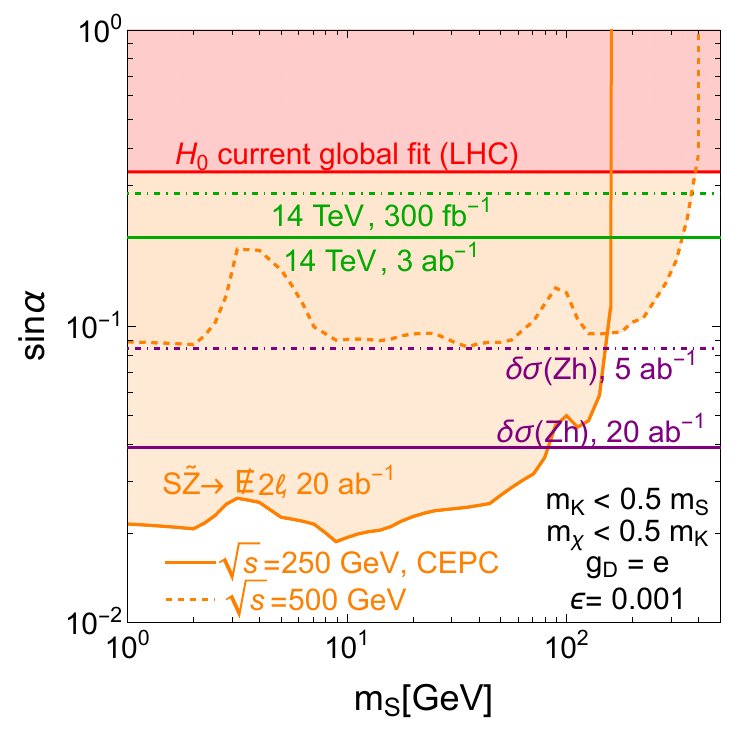}
\caption{
The SM Higgs mixes with a dark singlet scalar through the Higgs portal coupling, parameterized by the mixing angle $\alpha$. This mixing can be probed through two complementary approaches: (1) Precision measurements of Higgs couplings at the LHC (via global fits, shown as horizontal red lines, green dot-dashed lines, and green solid lines) and at CEPC (via $e^+e^- \to  Z h$ cross sections, represented by horizontal purple dot-dashed and solid lines) constrain $\sin \alpha$ through modifications of SM Higgs properties (independent of $m_S$); (2) Direct searches for $e^+e^- \to  S Z$ with $S \to {\rm inv}$ at CEPC, employing the recoil-mass technique similar to invisible Higgs searches, are shown as orange dashed and solid lines. Figure taken from Ref.~\cite{Liu:2017lpo}.
}
\label{fig:K-and-S:1}
\end{figure}

The scalar portal means the DM interact with SM particles via SM scalars, such as Higgs. Thus, hidden sector particles can be associated with the production of $Z$ and $H$ bosons at the Higgs factory. Ref.~\cite{Liu:2017lpo} studied the reach of an ultraviolet (UV) model, the Double Dark Portal model, at an electron-positron Higgs factory with $\sqrt{s} = 250/500$ GeV and $\mathcal{L}=5\,{\rm ab}^{-1}$. This model includes both gauge boson and scalar portals; in the scalar portal, there is a new dark sector scalar $S$ with mixing angle $\sin \alpha$ to the SM Higgs. 
The scalar mixing can lead to the associated production with the SM Higgs $H_0$ boson:  
$\tilde{Z} S$, where the tilde denotes the particles in their mass eigenstates. 
In this channel, we considered the leptonic decay of the $Z$ boson and the dark scalar $S$ decay into dark matter $\chi$, resulting in a final state of dilepton plus missing energy $\bar{\ell} \ell + \slashed{E}$. 
We show the exclusion sensitivities of this channel in Fig.~\ref{fig:K-and-S:1}, which provides the limits in the 2D parameter space of $\sin \alpha$-$m_S$ for the dark scalar. The phenomenology for the vector portal in this model is presented in Section~\ref{sec:DM:Zdecay:2}.

\subsection{Fermion portal}

Traditional WIMP DM generally come across a conflict between direct detection limit and relic abundance because they come from the same interaction and processes. In order to solve this conflict, many WIMP models carefully arrange the interactions to suppress the signals for direct and indirect detection \cite{Kumar:2013iva}. For instance, direct detection signals can be tuned to couple to nuclear spin, or be suppressed by small momentum transfer or low velocity of dark matter. Indirect detection signals can be $p$-wave suppressed by configuring the initial dark matter pair state to have an angular momentum quantum number \(L = 1\) or by being chirally suppressed. Therefore, collider searches become crucial and complementary, as the DM produced at colliders typically moves at velocities close to the speed of light, rendering the low dark matter velocity suppression ineffective. In this section, we will explore three critical models that have been studied in recent researches.

\subsubsection{Lepton portal DM} 
\label{sec:DM:lepton-portal}

\begin{figure}
\centering	
\includegraphics[width=0.45 \columnwidth]{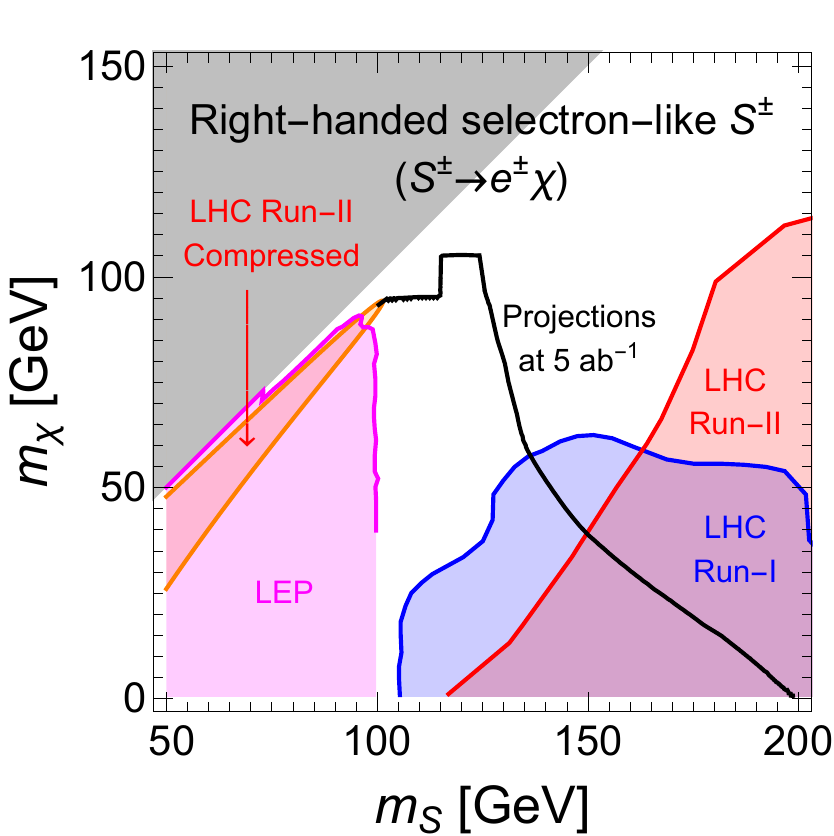}
\includegraphics[width=0.45 \columnwidth]{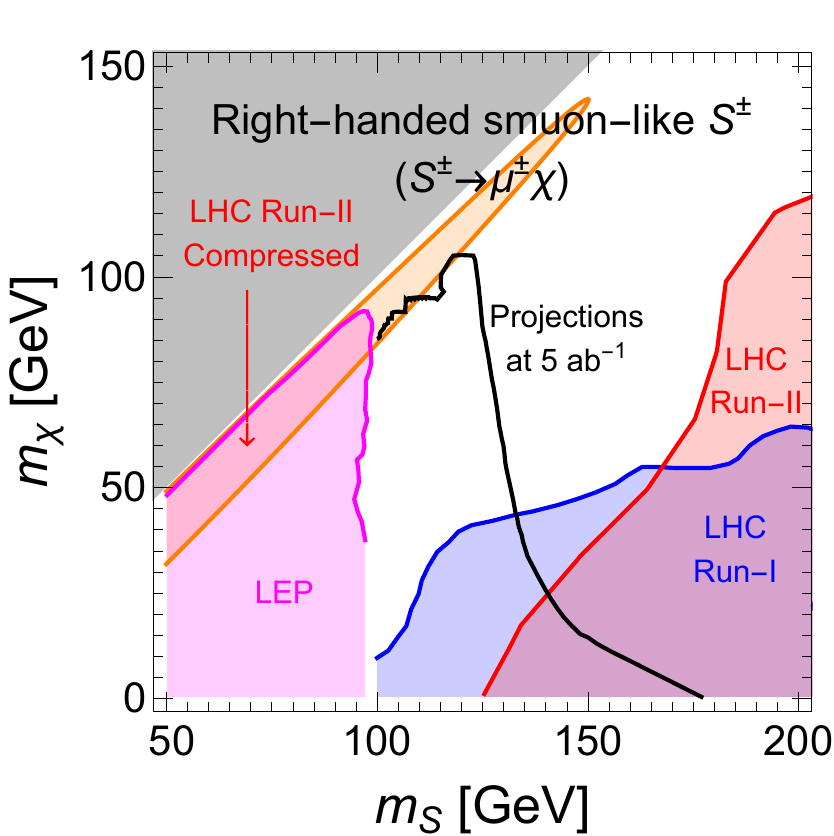}
\caption{
Current constraints on right-handed selectron-like (\textit{Left Panel}) and smuon-like (\textit{Right Panel}) scalar $S$ in the $(m_{\rm DM}, m_S)$ plane~\cite{Liu:2021mhn}. Black lines show CEPC projections at $5~\mathrm{ab}^{-1}$ from the Drell-Yan process $e^+e^- \to S^{\pm(*)} S^{\mp}$ with $S^\mp \to \ell^\mp \chi$ (allowing one off-shell $S^{\pm(*)}$). Sensitivity extends to $m_S > \sqrt{s}/2 = 120~\mathrm{GeV}$ via off-shell $S^{\pm (*)}$ contributions. The CEPC 360 GeV operation could further extend the constraints, complementing the coverage to LEP and LHC exclusions.
}
\label{fig:leptonportal-DrellYan}
\end{figure}

One relevant fermion portal example for the CEPC is the lepton portal DM model~\cite{Bai:2014osa}, in which a Majorana DM candidate, denoted as $\chi$, couples to the SM right-handed leptons $\ell_R$ via a complex charged scalar mediator $S$ through Yukawa-type coupling $y_\ell \bar{\chi}_L \ell_R S^\dagger$. 
Recently, Ref.~\cite{Liu:2021mhn} have studied the collider phenomenology and the interplay with the gravitational wave (GW) astronomy in an extension of the lepton portal DM model. The masses of DM and mediator $S$, as well as the lepton coupling $y_\ell$ and the Higgs portal coupling $ \lambda_{HS}$, which is related to the operator $ \lambda_{HS} (S^* S) (H^\dagger H)$, can be probed at the CEPC via the pair production of mediators $e^+e^-\to S^{\pm(*)} S^\mp\to\ell^+\chi \ell'^-\chi$, exotic decays of the Higgs or $Z$ boson, $h/Z\to S^{\pm (*)} S^{\mp (*)}\to\ell^+\chi\ell'^-\chi$ and $h\to \chi\chi$, and the Higgs couplings, including $h\ell^+\ell^-$, $h\gamma\gamma$ and $hZZ$. 
In the left and middle panels of Fig.~\ref{fig:leptonportal-DrellYan}, we show the constraints of CEPC at 240 GeV with $5~{\rm ab}^{-1}$ from the pair production of mediators $e^+e^-\to S^{\pm(*)} S^\mp$ with the dark scalar subsequently decay into $S^\mp \to \ell^\mp \chi $.

In addition to the collider signals, the model might trigger a first-order phase transition in the early Universe, provided that the mediator mass parameter $\mu_S^2$ is negative, and the Higgs portal coupling $\lambda_{HS}$ is large enough. Therefore, the first-order phase transition can induce significant gravitational wave signal, which can probe the parameter space of the scalar potential couplings.

On the other hand, the scalar $S$ can mediate 1-loop diagram with the same Higgs portal coupling, which modifies the Higgs decay process, such as $h \to \chi \chi$ invisible channel and $h \to \ell^+ \ell^-$ leptonic channels. For invisible channel $h \to \chi \chi$, the partial decay width is given as
\begin{align}
    \Gamma(h\to \chi\bar\chi)=\frac{g_{h\chi\chi}^2m_h}{8\pi}\left( 1-\frac{4m_\chi^2}{m_h^2}\right)^{\frac{3}{2}} \,, \quad g_{h\chi \chi} \approx  - \frac{ y_\ell^2 \lambda_{HS}m_\chi v}{16 \pi^2  m_S^2 }  
	+\mathcal{O}\left(m_S^{-3} \right).
\end{align}
Given the Higgs invisible BR sensitivity for HL-LHC around $3.5\%$ \cite{Bernaciak:2014pna} and CEPC around $0.3\%$ \cite{CEPCStudyGroup:2018ghi}, 
we show the sensitivity reach for HL-LHC and CEPC on the coupling $\lambda_{\rm HS} y_\ell^2$ as 2D function of $m_S$ and $m_\chi$ in the right panel of Fig.~\ref{fig:leptonportal-DrellYan}. For example, with $m_\chi = 40$ GeV and $m_S =200$ GeV, the corresponding constraints for HL-LHC and CEPC are $\lambda_{\rm HS} y_\ell^2 < 1.84$ and $< 0.54$, respectively.

\begin{figure}
\centering	
\includegraphics[width=0.32 \columnwidth]{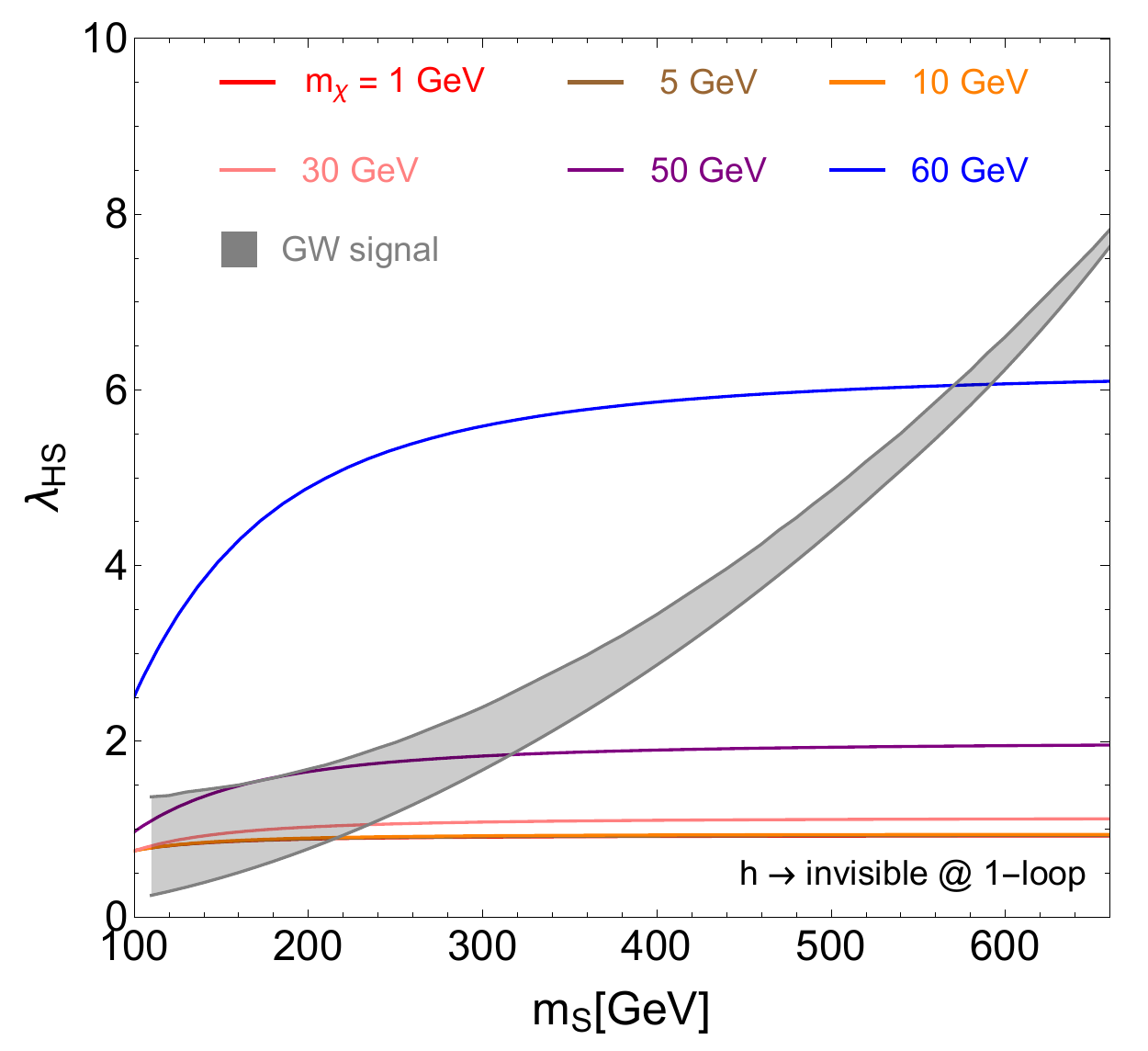}
\includegraphics[width=0.32 \columnwidth]{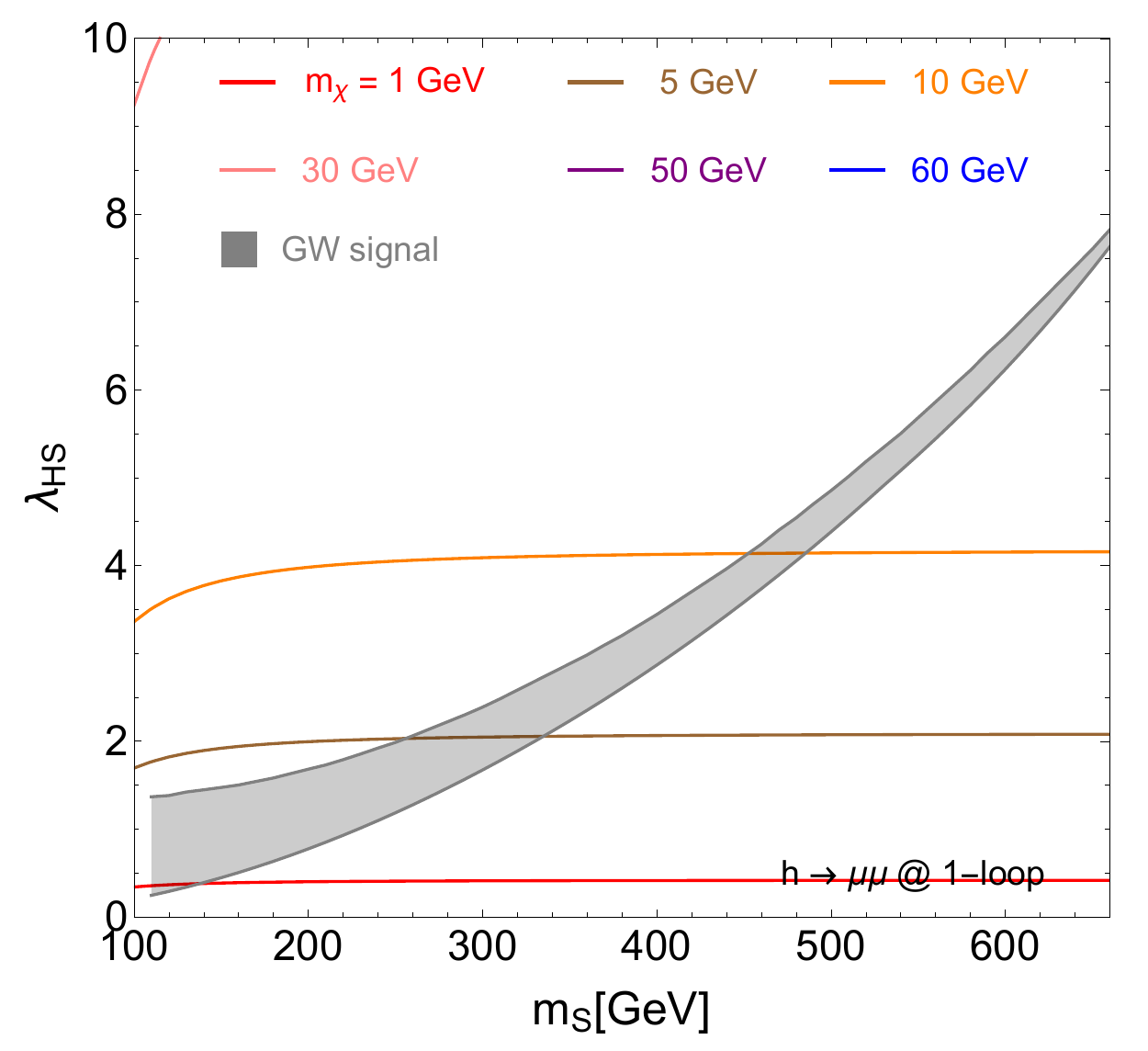}
\includegraphics[width=0.32 \columnwidth]{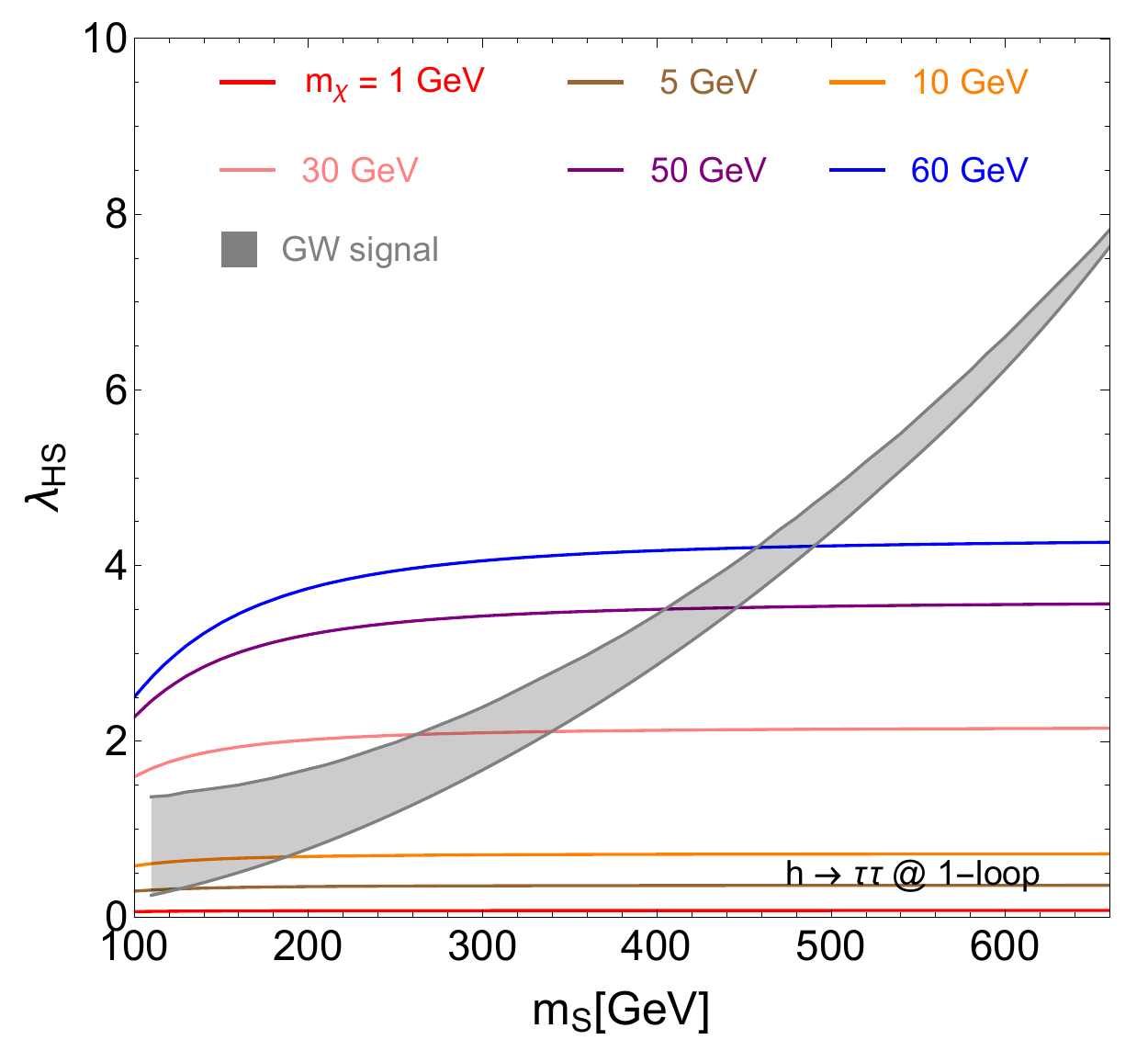}
\caption{
The Higgs portal coupling $\lambda_{HS}$ and lepton portal coupling $y_\ell$ can induce SM Higgs decays to dark matter pairs $\chi\bar{\chi}$ (invisible channel), $\mu^+\mu^-$, and $\tau^+\tau^-$ through 1-loop diagrams (see Fig.~\ref{fig:h_chichi}). The lepton portal coupling $y_\ell$ is fixed to its thermal value $y_\ell^{\rm th}$ to satisfy the DM relic abundance via $\bar{\chi}\chi \to \ell^+\ell^-$ annihilation. Therefore, precision Higgs measurements at CEPC constrain $\lambda_{HS}$ (shown as solid colored lines for different $m_\chi$ values), with sensitivity to an invisible branching ratio $\text{BR}(h \to \text{inv}) = 0.3\%$ and leptonic coupling precision reach $\delta\kappa_\mu < 8.7\%$ and $\delta\kappa_\tau < 1.5\%$.  Large $\lambda_{HS}$ values could generate gravitational wave signals from first-order electroweak phase transitions, with the gray region showing parameter space accessible to future LISA observations. Figures taken from Ref.~\cite{Liu:2021mhn}.
}
\label{fig:h-1loop-GW}
\end{figure}

In Fig.~\ref{fig:h-1loop-GW}, we show the LISA sensitivity projections and the CEPC projections for comparison, where we have assumed the leptonic coupling $y_{\ell}$ has been fixed by the relic abundance requirement. Therefore, we can directly compare the LISA and CEPC projections on the Higgs portal coupling $\lambda_{\rm HS}$, where the overlap of the parameter space reachable by the two probes can be used for crosschecking. The procedure of Ref.~\cite{Liu:2021mhn} can be generalized to other leptophilic WIMP models that are difficult to be probed in the direct and indirect detections, especially the models with scalar DM and/or mediators in which a first-order phase transition may happen.

\subsubsection{Asymmetric DM} 
\label{sec:DM:asymmetric}
In addition to DM, the observed baryon asymmetry of the Universe (BAU) is also a main puzzle in cosmology and particle physics.
Current measurements show that the abundance of baryon and DM are roughly at the same order of magnitude ($\Omega_{\text{DM}} \simeq 5\Omega_\text{B}$)~\cite{Akrami:2018vks,Ade:2015xua}. 
This coincidence motivates consider the so-called ``asymmetric DM" (ADM) model~\cite{Kaplan:2009ag,Petraki:2013wwa,Zurek:2013wia,Bai:2013xga}. 

A new ADM model has been proposed and studied in~\cite{Zhang:2021orr}. In this model, the dark sector is charged under a dark QCD,  $\rm SU(3)'$, 
and the mass of DM is generated via the dark confinement (so the DM is actually a ``dark baryon'').
Furthermore, to generate dark and visible asymmetry simultaneously, we introduce a scalar mediator (labeled as $\Phi$) that is charged under dark $\rm SU(3)'$ and SM $\rm U(1)_Y$. 
Mediator $\Phi$ couples to dark quark (labeled as $q'$) and SM right-handed leptons, and thus provides a portal for us to search for. 

\begin{figure}[!t]
\centering
{\includegraphics[width=3.0in]{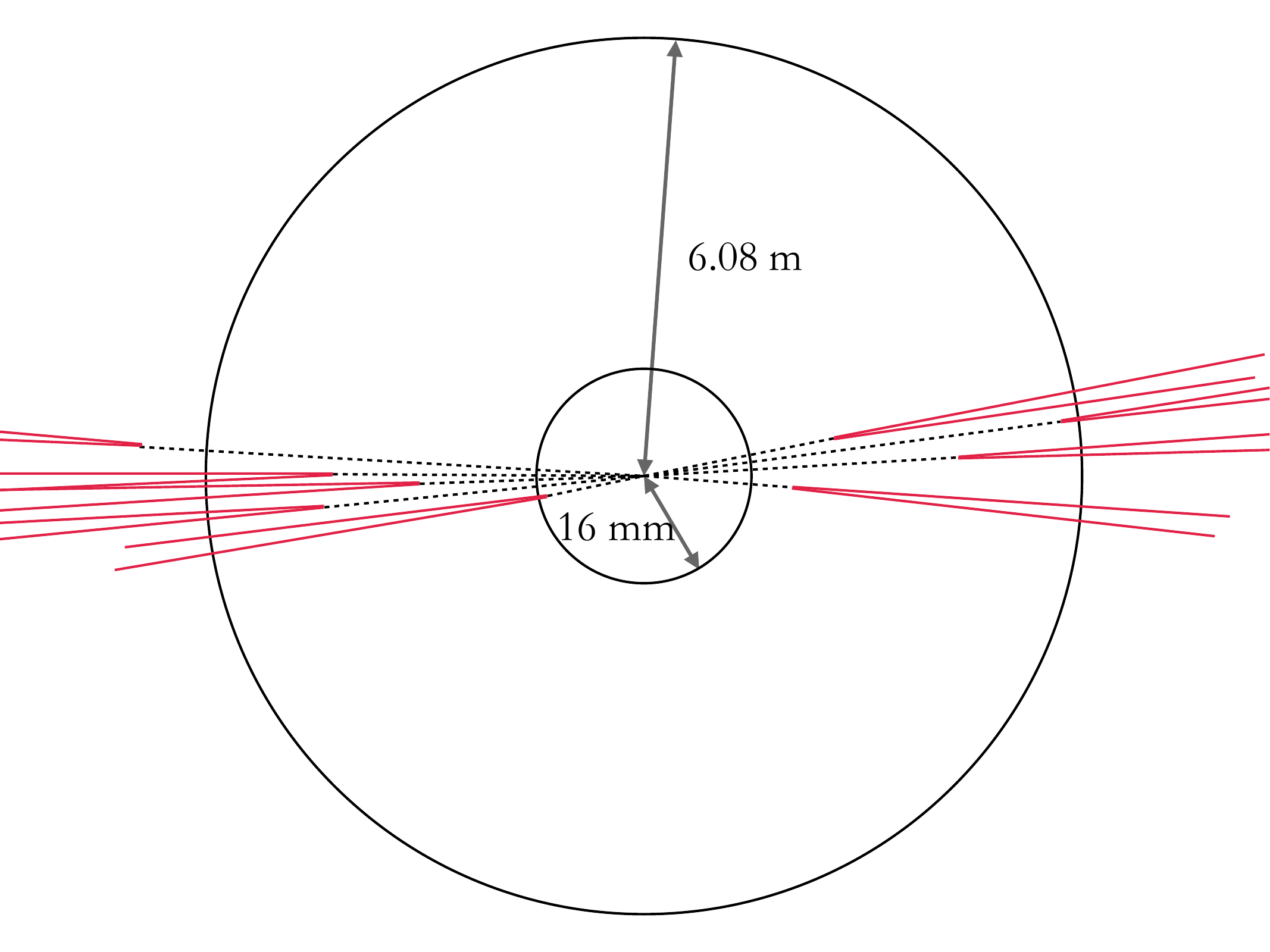}}
\quad {
\includegraphics[width=3.0in]{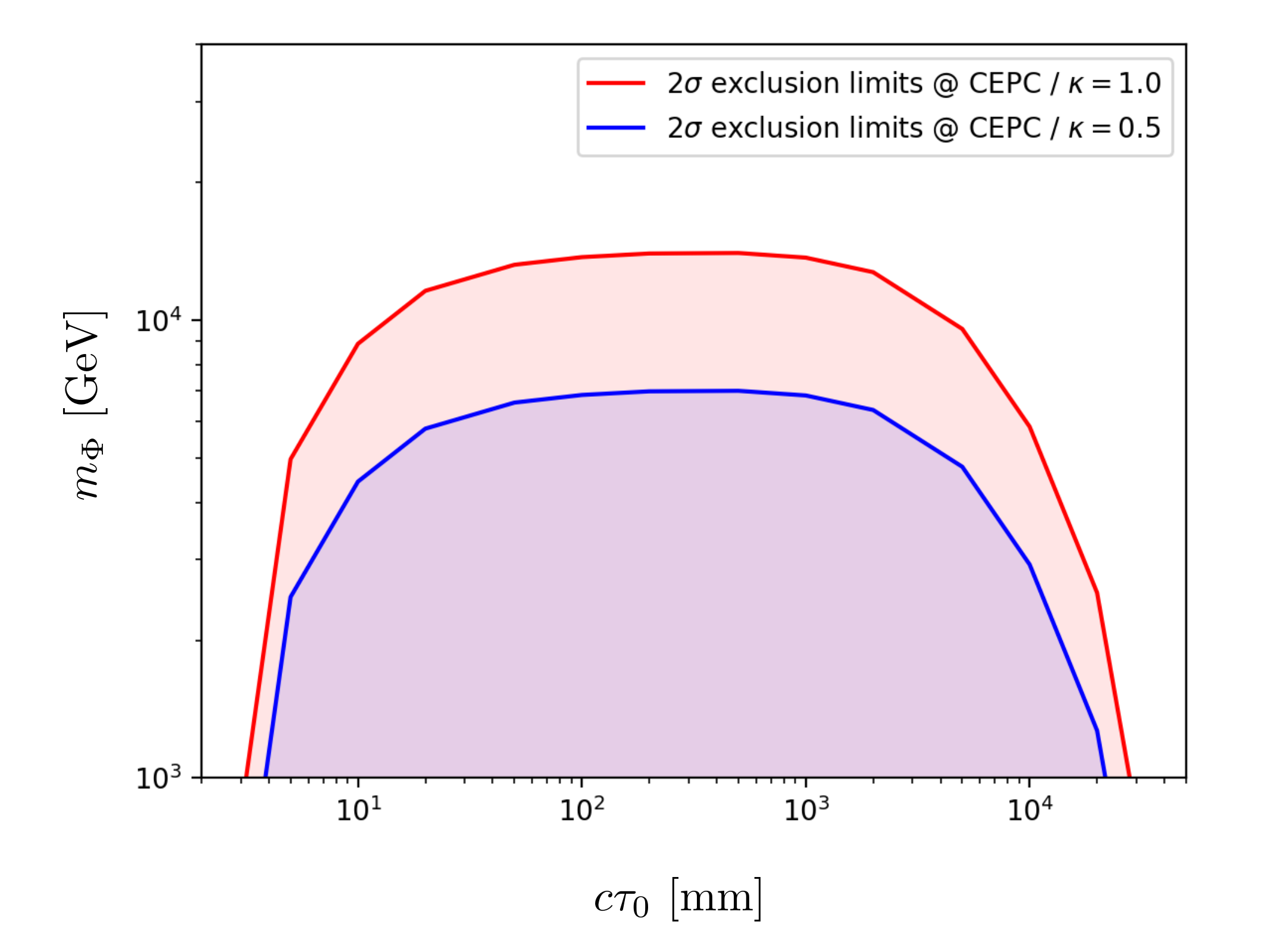}}
\caption{
Left: An illustration of the signal process at CEPC. Detector is represented by two circles. Black dotted lines and red solid lines are dark pions and muons, respectively.
Right: 2 $\sigma$ exclusion limits on the mediator mass $m_{\Phi}$ as a function of the dark pion proper decay length, with coupling $\kappa$ fixed to 0.5 and 1.0 respectively. The HL-LHC can only provide very weak constraints on heavy $\Phi$, thus are not shown here~\cite{Zhang:2021orr}.}
\label{CEPC2}
\end{figure}

The Lagrangian related to collider search is
\begin{eqnarray}
\mathcal{L} \supset \bar{q'} ( D \!\!\!\!/  - m_{q'} ) q'  + (D_{\mu} \Phi )^{\dagger}(D^{\mu} \Phi ) - m^2_{\Phi} \Phi^{\dagger} \Phi -\frac{1}{4} {G'}^{\mu\nu} {G'}_{\mu\nu} - ( \kappa \Phi \bar{q'}_L  {l}_R  +h.c.)~,~
\label{Lag_collider}
\end{eqnarray}
where $G'^{\mu\nu}$ is the field strength of dark gluon. 
Mediator $\Phi$ can be produced in pairs at LHC via the Hyper charge it carries.  
Then $\Phi$ decays to a SM lepton and a dark quark $q'$. 
While on CEPC, $q'\bar{q}'$ can be produced directly via a t-channel $\Phi$. 
Due to the dark confinement, $q'$ will hadronize to a cluster of dark mesons (labeled as $\pi'$). 
Dark meson $\pi'$ (long-lived) will decay to lepton pair via the $\Phi$ portal, and leave displaced vertex inside detector. 
Fig.~\ref{CEPC2} (left) shows the predicted signal process on CEPC for illustration. 
The study shows that CEPC has the ability to cover a large parameter space of this model, see Fig.~\ref{CEPC2} (right). 
The mass of mediator can be excluded up to $\mathcal{O}(10)$ TeV, if the proper lifetime of dark pion $\pi'$ is between 10 mm and 10 m. 
This bound is stronger than the limit from current ATLAS displaced lepton jet search result~\cite{ATLAS:2016jza}.

\subsubsection{Long-lived dark scalar} 
\label{sec:DM:long-lived VLL}

Recently, vector-like leptons (VLL) as a simple extension to the standard model have attracted widespread attention in both theory and experiments. The VLL model can include an additional dark sector scalar $\phi$, mediating the heavy vector-like lepton $F^\pm$ mixing with the first SM lepton generation~\cite{Cao:2023smj}. The relevant Lagrangian in the mass basis at leading order reads:
\begin{equation}
\label{eq:LLVLL-Lag}
\begin{aligned}
\mathcal{L}_\text{int} &\supset  \bar{F} (i\partial_\mu-e A_\mu+e\tan\theta_W Z_\mu)\gamma^\mu F 
    - m_F\bar{F}F  - m_\ell \bar{\ell} \ell \\
    & +\frac{1}{2}\frac{e}{\sin{\theta_W}\cos{\theta_W}}\theta_L Z_\mu (\bar{F_L}\gamma^\mu \ell_L +{\rm h.c.}) -\frac{e}{\sqrt{2}\sin\theta_W} \theta_L (W_\mu^+ \bar{\nu}_L\gamma^\mu F_L +{\rm h.c.})\\
& - y\phi\left(\bar{F}_L\ell_R + \bar{\ell}_R F_L+ \theta_R\bar{F} F - \theta_L \bar{\ell}\ell\right),
\end{aligned}
\end{equation}
where the mixing parameters $\theta_{L/R}$ hold the relation $\theta_L \simeq \frac{m_{\ell}}{m_F} \tan\theta_R \ll \theta_R$. Ref.~\cite{Cao:2023smj} focuses on the parameter space $m_F > m_\phi \gg m_\ell$ and $m_F > 200\,\text{GeV}$ to avoid constraints from multilepton and $Z$ boson searches. In this case, the scalar $\phi$ can only decay to a lepton pair $\bar{\ell} \ell$, but this decay is suppressed by the mass ratio $(m_{\ell}/m_F)^2$, thus $\phi$ can naturally be long-lived.

Regarding CEPC, one requires the lepton in Eq.~\eqref{eq:LLVLL-Lag} to be an electron. Therefore, the electron-positron collider can produce a pair of dark scalars $\phi \phi$ by exchanging $F^\pm$ via the $t$ channel, followed by subsequent decays into $e^+e^-$ pairs. Since the CEPC has a center of mass energy much smaller than the LHC, the direct production of the heavy VLL $F$ is not possible.

If $\phi$ is long-lived, its decay to an electron pair can lead to a displaced vertex (DV) signature. Since there are two $\phi$ scalars, there could be one displaced vertex without specifying where the other $\phi$ decays, leading to the inclusive displaced vertex (iDV) signature. It is also possible that both $\phi$ scalars decay in the inner tracker, allowing the reconstruction of two DVs from di-electrons.

The corresponding sensitivities for the iDV and 2DV at future CEPC are studied in Ref.~\cite{Cao:2023smj}, where the number of signal events $N=3$ is plotted. Compared to HL-LHC searches, the DV searches at CEPC can effectively probe low-mass $m_\phi$ but are less capable of detecting larger masses due to the lower center-of-mass energy of CEPC. In comparison with the LHC, which exclude very small coupling combinations $y \theta_L$ through the dilepton plus missing energy searches, CEPC shows complementary sensitivity for intermediate $y \theta_L \in [10^{-11}, 10^{-7}]$.

\subsection{Vector portal}
In general, the SM vector portal for DM involve the neutral mediators, the Z boson and the photon ($\gamma$), consistent with the fact that DM is neutral. In this section, we will consider these cases separately. 
\subsubsection{Dark sector particles from gauge boson associated production}
\label{sec:DM:Zdecay:2}

\begin{figure}[]
\includegraphics[width=0.55\textwidth]{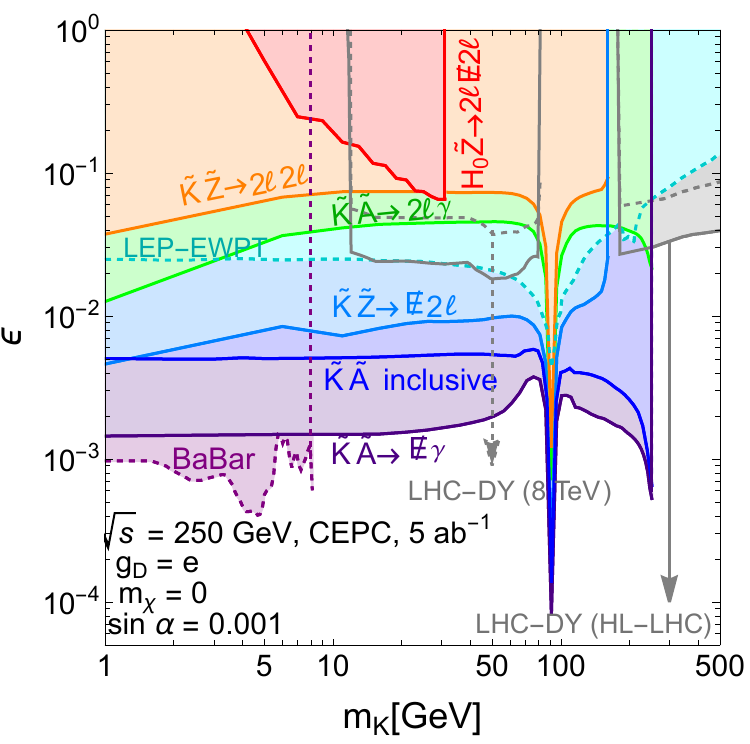}
\caption{
Projected exclusion regions at CEPC for the kinetic mixing dark photon model in the $\epsilon$ vs.~$m_K$ plane, where $K$, $Z$, and $A$ represent the dark photon, $Z$ boson, and photon (mass eigenstates denoted by tildes in the figure). Shown production channels include: $KA \to \slashed{E}\gamma$, $KA \to \ell^+\ell^-\gamma$, $KZ \to \slashed{E}\ell^+\ell^-$, and $KZ \to \ell^+\ell^-\ell^{'+}\ell^{'-}$, along with an inclusive mono-photon search $KA$ (inclusive) without specified $K$ decays. The sensitivity peak near $m_K \approx m_Z$ originates from kinetic mixing between the dark photon and hypercharge field, which induces significant $K$-$Z$ mixing effects. Constraints are compared with: (i) LHC and HL-LHC ($3\,\mathrm{ab}^{-1}$) Drell-Yan projections (gray dashed and solid lines, respectively), (ii) BaBar mono-photon searches (purple dashed), and (iii) LEP electroweak precision tests (EWPT, cyan dashed). Figure taken from Ref.~\cite{Liu:2017lpo}.
}
\label{fig:K-and-S:2}
\end{figure}

\begin{figure}[]
\includegraphics[width=0.48\textwidth]{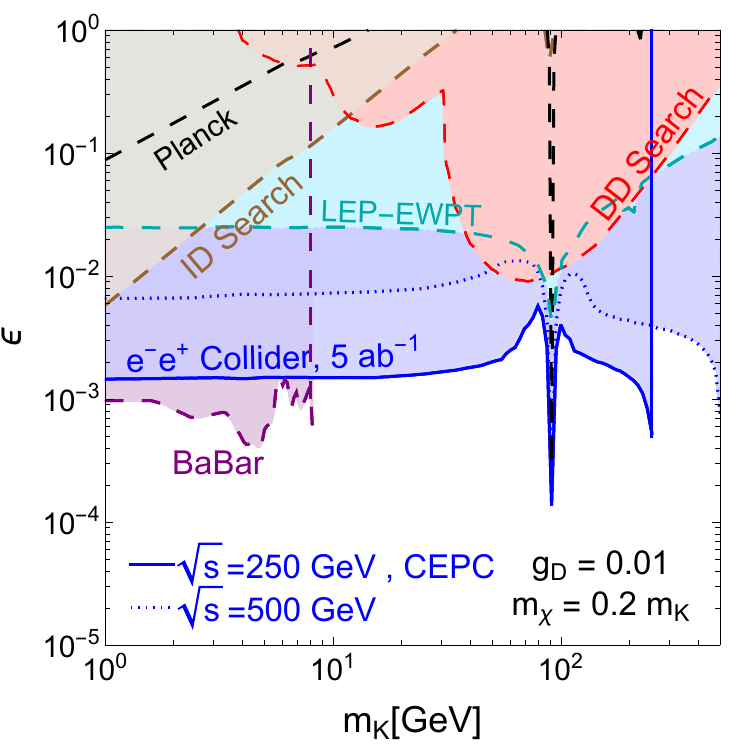}
\includegraphics[width=0.48\textwidth]{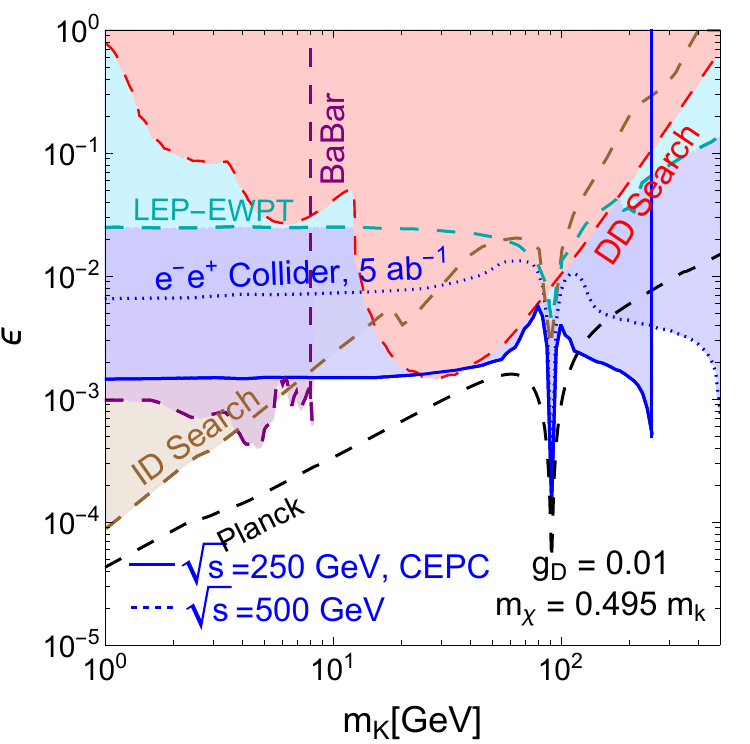}
\caption{The CEPC sensitivity to the kinetic mixing portal parameter $\epsilon$ in the dark photon mediated model is shown for the mono-photon channel $e^+e^- \to \gamma K \to \gamma \slashed{E}$, where the dark photon $K$ predominantly decays to dark matter pairs $\bar{\chi}\chi$ (invisible channel, represented by blue solid/dotted lines and shaded regions). Also shown are constraints from: DM direct detection (red dashed), DM indirect detection (brown dashed), BaBar mono-photon searches (purple dashed), and LEP electroweak precision tests (cyan dashed). The black dashed line indicates the parameter space where the thermal annihilation processes $\bar{\chi}\chi \to K \to \bar{f} f, W^+W^-$ produce the observed relic abundance, assuming a dark matter-dark photon coupling $g_D = 0.01$. In the right panel, the mass configuration $m_\chi \approx m_K/2$ enables resonant dark matter annihilation, substantially reducing the required $\epsilon$ (as evident from the shift in the dashed black lines). Figure adapted from Ref.~\cite{Liu:2017lpo}.
}
\label{fig:summary1}
\end{figure}

In the Double Dark Portal model, the vector portal features a new $U(1)'$ vector gauge boson $K$ with kinetic mixing parameter $\epsilon$, with the fermionic dark matter $\chi$ coupling to the $U(1)'$ vector~\cite{Liu:2017lpo}.
These new gauge portal interactions can lead to the following associated production with the SM $Z$ boson: $\tilde{Z} \tilde{K}$ and $\gamma \tilde{K}$.  
For the $\tilde{Z} \tilde{K}$ channel, we considered the following decays: one where $\tilde{Z} \to \bar{\ell} \ell$ and $\tilde{K} \to \bar{\chi} \chi$, resulting in dileptons plus missing energy; and another where $\tilde{Z} \to \bar{\ell} \ell$ and $\tilde{K} \to \bar{\ell} \ell$, resulting in a four-lepton final state. 
For the $\gamma \tilde{K}$ channel, we considered several decays: first, using the recoil mass method to look for $\tilde{K}$ inclusive decay; second, $\tilde{K} \to \bar{\ell} \ell$, resulting in dileptons plus one photon final state; and third, $\tilde{K} \to \bar{\chi} \chi$, resulting in a mono-photon final state.

The corresponding exclusion sensitivities of these channels are shown in Fig.~\ref{fig:K-and-S:2}, which provide the limits in the 2D parameter space $\epsilon$-$m_{K}$ for the dark vector. In Fig.~\ref{fig:summary1}, we compare the collider sensitivity with dark matter detection and indirect detection experiments, as well as the relic abundance requirement. We emphasize that the collider constraint is not sensitive to the coupling between the DM and the mediator, as long as the invisible decay of the dark photon dominates. Therefore, the reach of a future $e^+e^-$ collider will complement and potentially supersede that of dark matter searches.

There are also models that consider the Higgs boson as a portal to a dark sector with couplings to a gauge boson $Z_d$ of a broken U(1) symmetry. If $Z_d$ is light enough, precision Higgs decay can test $h\rightarrow Z_d Z$~\cite{Davoudiasl:2013aya,Davoudiasl:2012ag} and $h\rightarrow Z_d Z_d$~\cite{Curtin:2013fra,Curtin:2014cca} channels. Relatively light $Z_d$ can also decay into the SM particles via Higgs portal. An electron-positron collider such as the CEPC has the potential to open up new regions of parameter space. Ref.~\cite{Giffin:2020jtl} studied a Higgs ported $Z-Z_d$ mass mixing scenario, parametrized as $\frac{g}{c_W}m_{Z_d}\delta\cdot h X_{\mu}Z_d^{\mu}, (X\equiv Z,Z_d)$. The $e^+e^-\rightarrow h Z_d$ channel with subsequent $Z_d\rightarrow \ell\ell$ decay can be sensitive to the mixing parameter and cross section of $\delta\sim 8\times 10^{-3}$ and $\sigma(e^+e^-\rightarrow h Z_d$) $\sim$ 3-4 ab at $\sqrt{s}=240$ GeV with 5 ab$^{-1}$.

\subsubsection{Millicharge DM}
\label{sec:DM:millicharge}

The CEPC can shed light on the particle properties of dark matter (DM), which currently remain elusive despite the overwhelming evidence from cosmology and astrophysical measurements. Recently, Ref.~\cite{Liu:2019ogn} investigated the capability of CEPC in probing the parameter space of  millicharged DM. The interaction Lagrangian of this model is given by:
\begin{equation}
{\cal L} = e \varepsilon A_\mu \bar \chi \gamma^\mu \chi,
\label{eq:millicharge}
\end{equation}
where $\chi$ is the Dirac DM, $A_\mu$ is the photon, $e$ is the electromagnetic coupling strength, and $\varepsilon$ is the millicharge.

The monophoton signature at CEPC, when a dark matter pair is produced in association with a photon, demonstrates significant potential for improving DM constraints, providing a promising approach to investigate dark matter properties~\cite{Liu:2019ogn}. Sensitivity projections are calculated for three CEPC operational configurations: (i) $5.6~\mathrm{ab}^{-1}$ in $H$-mode, (ii) $16~\mathrm{ab}^{-1}$ in $WW$-mode, and (iii) $2.6~\mathrm{ab}^{-1}$ in $Z$-mode. The $Z$ and $H$ modes exhibit optimal sensitivity for millicharged DM, constraining the mixing parameter $\epsilon$ to within a few percent for masses below and above $40~\mathrm{GeV}$, respectively. These results represent an improvement of nearly one order of magnitude over existing collider limits for dark matter mass larger than 1 GeV~\cite{Davidson:2000hf, Soper:2014ska, Magill:2018tbb, Liu:2018jdi}.

\subsubsection{Dark sector particles with EM form factors}
\label{sec:DM:form-factor}
	
Primary importance for our understanding of elementary interactions is shedding light on the dark sector states. Recently, Ref.~\cite{Zhang:2022ijx} investigated the sensitivity of CEPC on the dark states with electromagnetic form factors for  magnetic dipole moments (MDM) and electric dipole moments (EDM) at mass-dimension 5, and anapole moment (AM) and charge radius interaction (CR) at mass-dimension 6.
	
In  Ref.~\cite{Zhang:2022ijx}, the fermionic dark state $\chi$  may have the effective interactions with the hypercharge gauge boson field $B_\mu$  \cite{Chu:2018qrm,Arina:2020mxo}, and the interactions can be  written with electromagnetic field strength tensor $F_{\mu\nu}\equiv\partial_\mu A_\nu-\partial_\nu A_\mu$ and $Z$ gauge field strength tensor $Z_{\mu\nu}\equiv\partial_\mu Z_\nu-\partial_\nu Z_\mu$ as
\begin{eqnarray}
		{\cal L}_\chi&=&\frac{1}{2} \mu_{\chi} \bar{\chi} \sigma^{\mu \nu} \chi F_{\mu \nu}
		+\frac{i}{2} d_{\chi} \bar{\chi} \sigma^{\mu \nu} \gamma^{5} \chi F_{\mu \nu}
		-a_{\chi} \bar{\chi} \gamma^{\mu} \gamma^{5} \chi \partial^{\nu} F_{\mu \nu}
		+b_{\chi} \bar{\chi} \gamma^{\mu} \chi \partial^{\nu} F_{\mu \nu} \\ \nonumber
		&+&\frac{1}{2} \mu_{\chi}^Z \bar{\chi} \sigma^{\mu \nu} \chi Z_{\mu \nu}
		+\frac{i}{2} d_{\chi}^Z \bar{\chi} \sigma^{\mu \nu} \gamma^{5} \chi Z_{\mu \nu}
		-a_{\chi}^Z \bar{\chi} \gamma^{\mu} \gamma^{5} \chi \partial^{\nu} Z_{\mu \nu}
		+b_{\chi}^Z \bar{\chi} \gamma^{\mu} \chi \partial^{\nu} Z_{\mu \nu}.
\end{eqnarray}
Here ${\cal C}_\chi^Z=-{\cal C}_\chi \sin\theta_W$ with ${\cal C}_\chi=\mu_\chi,\, d_\chi,\, a_\chi,\, b_\chi$, and $\theta_W$ denotes Weinberg angle. $\mu_\chi$ and $d_\chi$ are the dimensional coefficients of the mass-dimension 5 MDM and EDM interactions, expressed in units of the Bohr magneton $\mu_B$, and $\sigma_{\mu\nu}\equiv i[\gamma^\mu,\gamma^\nu]/2$; $a_\chi$ and $b_\chi$ are the dimensional coefficients of the mass-dimension 6 AM and CR interactions.

\begin{figure*}[!t]
		\centering
		\includegraphics[width=0.75 \columnwidth]{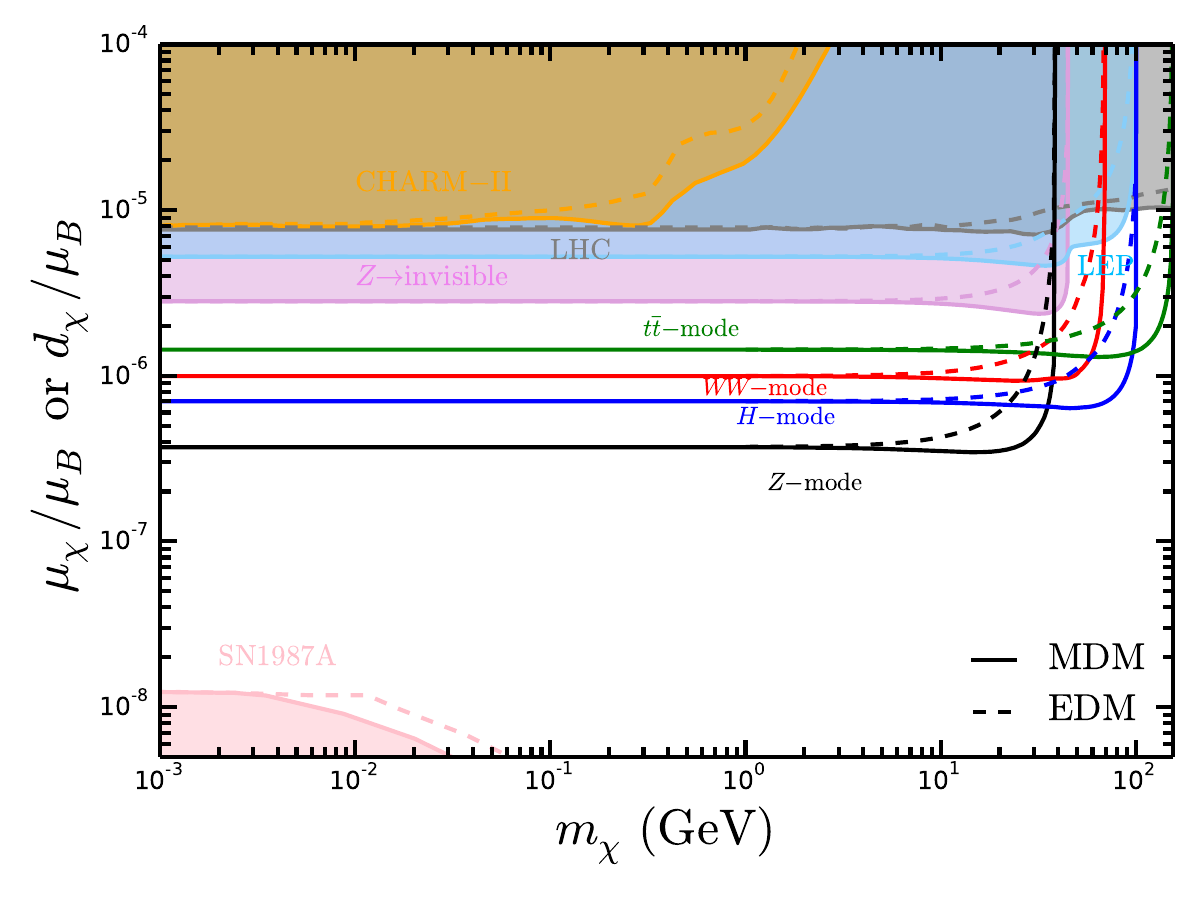}
		\caption{
			The expected 95\% C.L. exclusion limits on the electromagnetic form factors
			for mass-dimension 5 operators through MDM (solid) and EDM (dashed) in the four CEPC running modes~\cite{Zhang:2022ijx}.
			The landscape of current leading constraints are also shown  with shaded regions, exploiting from terrestrial experiments, such as proton-beam experiments CHARM-II \cite{Chu:2020ysb}, monophoton searches and $Z$-boson invisible decay at LEP, monojet searches at LHC \cite{Arina:2020mxo}, and astrophysics supernovae SN 1987A \cite{Chu:2019rok}.
		}
		\label{fig:DM-electromagnetic}
\end{figure*}
 
The dark states $\chi$ can be produced from the process $e^+ e^- \to \gamma/Z\to \chi \bar\chi$ at CEPC. In order to probe the dark state at CEPC, one should consider  $\chi\bar\chi$ pair production associated with a hard photon radiated from the initial state electron or positron ( $e^+ e^- \to \chi \bar\chi \gamma$), i.e., the typical monophoton signature. 

Fig.~\ref{fig:DM-electromagnetic} presents the 95\% C.L. upper bounds from CEPC's monophoton channel on the electromagnetic form factors for mass-dimension~5 operators, showing results for magnetic dipole moments (MDM, solid lines) and electric dipole moments (EDM, dashed lines). These limits are derived using: (i) $20~\mathrm{ab}^{-1}$ in $H$-mode, (ii) $6~\mathrm{ab}^{-1}$ in $WW$-mode, (iii) $100~\mathrm{ab}^{-1}$ in $Z$-mode, and (iv) $1~\mathrm{ab}^{-1}$ in $t\bar{t}$-mode. 

The $Z$-mode demonstrates optimal sensitivity for light dark states with mass-dimension~5 operators, probing MDM for $m_\chi \lesssim 35~\mathrm{GeV}$ and EDM for $m_\chi \lesssim 25~\mathrm{GeV}$, reaching couplings as low as $3.7 \times 10^{-7}\,\mu_B$. This represents an improvement of approximately one order of magnitude over HL-LHC projections (with 25\% systematic uncertainty), which can only probe these operators down to $4 \times 10^{-6}\,\mu_B$.
While the $H$-mode shows reduced sensitivity (by about a factor of two) for light dark states compared to the $Z$-mode, it provides superior coverage for DM masses between 45--62.5 GeV. Combined with the 350 GeV $t\bar{t}$ mode, CEPC will explore previously inaccessible parameter space for both mass-dimension~5 and mass-dimension~6 operators of dark states.

\subsection{DM in EFT framework}

\subsubsection{Leptophilic DM}
\label{sec:leptophilic_DM}

Most of the existing experimental constraints on DM crucially rely on its interactions with nucleons, and can therefore be largely evaded if the DM predominantly interacts with the SM leptons, but not quarks at tree-level. Such {\it leptophilic} DM (LDM) arises naturally in many BSM scenarios~\cite{Krauss:2002px, Baltz:2002we, Ma:2006km, Hambye:2006zn, Bernabei:2007gr, Cirelli:2008pk, Chen:2008dh, Bi:2009md, Ibarra:2009bm, Dev:2013hka, Chang:2014tea, Agrawal:2014ufa, Bell:2014tta, Freitas:2014jla,Cao:2014cda, Lu:2016ups, Duan:2017pkq,Madge:2018gfl,Junius:2019dci, Ghosh:2020fdc,  Chakraborti:2020zxt, Horigome:2021qof}, some of which could even explain various existing experimental anomalies, such as the muon anomalous magnetic moment
~\cite{Abi:2021ojo}, DAMA/LIBRA annual modulation~\cite{Bernabei:2020mon}, 
anomalous cosmic ray positron excess~\cite{Abdollahi:2017nat, DAMPE:2017fbg, Adriani:2018ktz, AMS:2021nhj}, 
the galactic center gamma-ray excess~\cite{TheFermi-LAT:2015kwa}, and XENON1T electron excess~\cite{XENON:2020rca}. Dedicated searches for LDM in direct detection~\cite{XENON100:2015tol, XENON:2019gfn, LZ:2021xov} and beam dump~\cite{Chen:2018vkr, Marsicano:2018vin} experiments have also been discussed. 
 
Lepton colliders provide an ideal testing ground for the direct production of LDM and its subsequent detection via either mono-photon~\cite{DELPHI:2003dlq, Birkedal:2004xn, Fox:2008kb,Konar:2009ae, Fox:2011fx,Bartels:2012ex, Dreiner:2012xm,Chae:2012bq, Liu:2019ogn,Habermehl:2020njb, Kalinowski:2021tyr, Barman:2021hhg, Kundu:2021cmo} or mono-$Z$~\cite{Wan:2014rhl, Yu:2014ula, Dutta:2017ljq, Grzadkowski:2020frj, Kundu:2021cmo} signatures. Here we will discuss the CEPC sensitivities to LDM in the monophoton channel following a model-independent EFT approach~\cite{Kundu:2021cmo}. See also Refs.~\cite{Kopp:2009et, Beltran:2010ww,Goodman:2010yf,Bai:2010hh, Goodman:2010ku, Fox:2011fx,Fox:2011pm, Rajaraman:2011wf,Chae:2012bq, Kahlhoefer:2017dnp, Penning:2017tmb, CMS:2017zts, ATLAS:2021kxv} for earlier works on collider searches for DM in the EFT framework. Here we assume the LDM to be fermionic and only show the results for the dimension-6 operators of vector-axialvector (V-A) type for illustration. Within the minimal EFT approach, the only relevant degrees of freedom are the DM mass $m_\chi$ and an effective cut-off scale $\Lambda$ which determines the strength of the four-Fermi operator given by 
 \begin{align}
    \mathcal{L}_{\rm eff}  =  \frac{1}{\Lambda^2}\sum_j\left(\overline{\chi}\Gamma^{j}_{\chi}\chi\right)\left(\overline{\ell}\Gamma^{j}_{\ell}\ell\right) \, , 
    \label{eq:EFT}
\end{align}
where  $\Gamma_{\chi}^{\mu} =  \left( c^{\chi}_{V}+ c^{\chi}_{A} \gamma_5 \right) \gamma^{\mu}$ and $
   \Gamma_{e\mu} = \left( c^{e}_{V}+ c^{e}_{A} \gamma_5 \right)\gamma_{\mu}$ for V-A type. For simplicity, we will also set $c_V^\chi=c_A^\chi=c_V^e=c_A^e=1$. For other choices of the couplings, our results for the sensitivity on $\Lambda$ can be easily scaled accordingly.  Note that the EFT results are valid as long as $\Lambda>{\rm max}\{\sqrt s, 3m_\chi\}$~\cite{Matsumoto:2016hbs}.

\begin{figure}[t!]
\centering 
\includegraphics[width=0.8\linewidth]{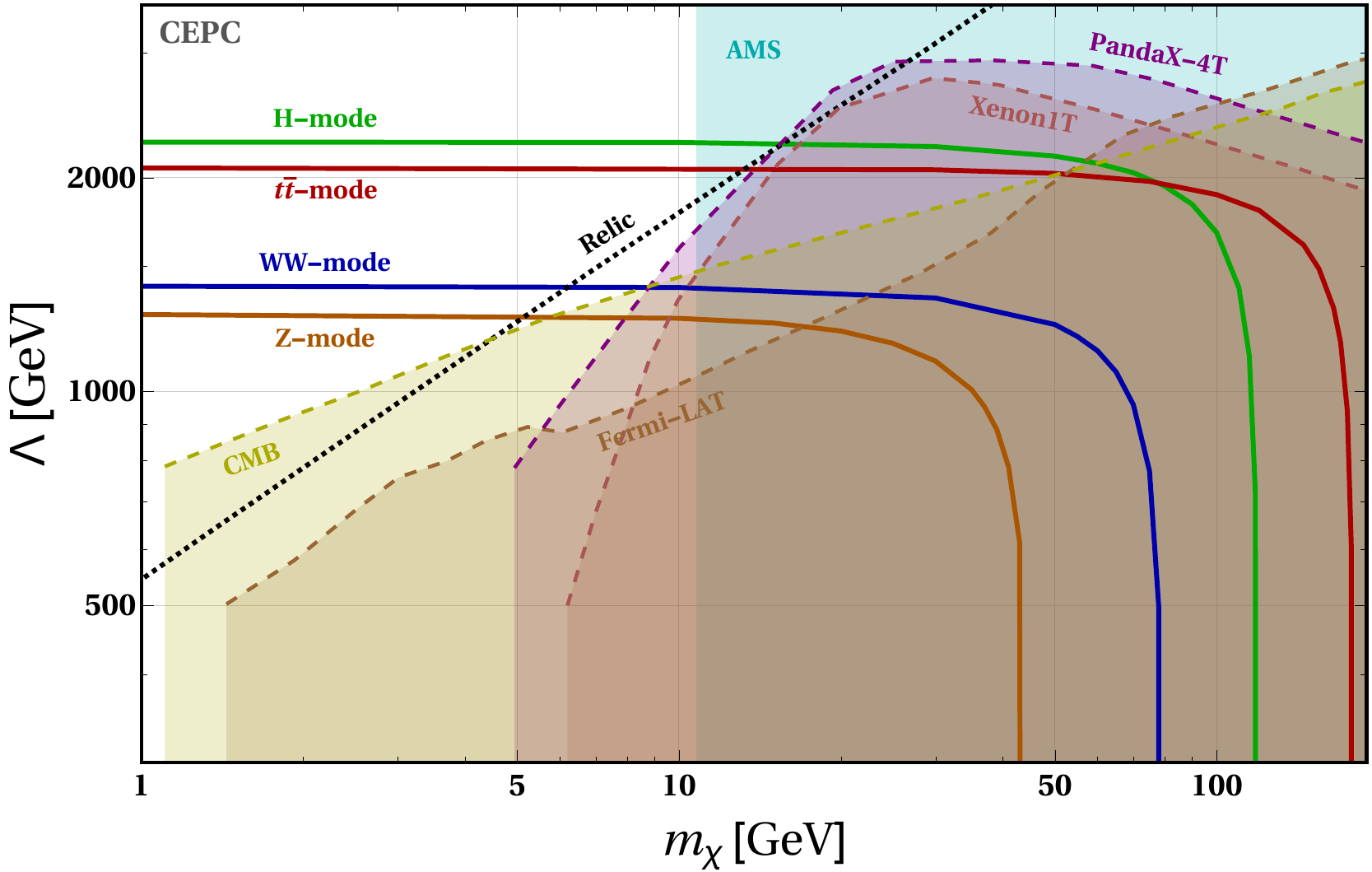}
\caption{The $3\sigma$ sensitivity contours in the mono-photon channel for the LDM with V-A operator structure with unpolarized $e^+e^-$ beams and different CEPC operation modes as given in Table~\ref{tab:Yield_T1}.  The various shaded regions are excluded by direct detection (XENON1T~\cite{XENON:2018voc}, PANDAX-4T~\cite{PandaX-4T:2021bab}), indirect detection (Fermi-LAT~\cite{Leane:2018kjk}, AMS~\cite{John:2021ugy}), astrophysics (SN1987A~\cite{Guha:2018mli}) and cosmology (CMB~\cite{Leane:2018kjk}) constraints. Along the dot-dashed line, the observed DM relic density is reproduced for a thermal DM assuming only DM-electron effective coupling. Figure updated from Ref.~\cite{Kundu:2021cmo} for CEPC.}
\label{fig:LDM}
\end{figure}

Our results for the mono-photon case  $e^+e^-\to\chi\overline{\chi}\gamma$ are shown in Fig.~\ref{fig:LDM}. The different solid contours correspond to different CEPC operation modes as given in Table~\ref{tab:Yield_T1}. The details of background estimations, signal selection and cut-based analysis can be found in Ref.~\cite{Kundu:2021cmo}. 
The various shaded regions are excluded by direct detection (XENON1T~\cite{XENON:2018voc}, PANDAX-4T~\cite{PandaX-4T:2021bab}), indirect detection (Fermi-LAT~\cite{Leane:2018kjk}, AMS~\cite{John:2021ugy}), astrophysics (SN1987A~\cite{Guha:2018mli}) and cosmology (CMB~\cite{Leane:2018kjk}) constraints. Along the dot-dashed line, the observed DM relic density is reproduced for a thermal DM assuming only DM-electron effective coupling. From Fig.~\ref{fig:LDM}, it is clear that CEPC can probe new LDM parameter space, especially in the low DM mass range (for $m_\chi \lesssim 10$ GeV).

\begin{figure}[!t]
    \centering
    \includegraphics[width=0.75\textwidth]{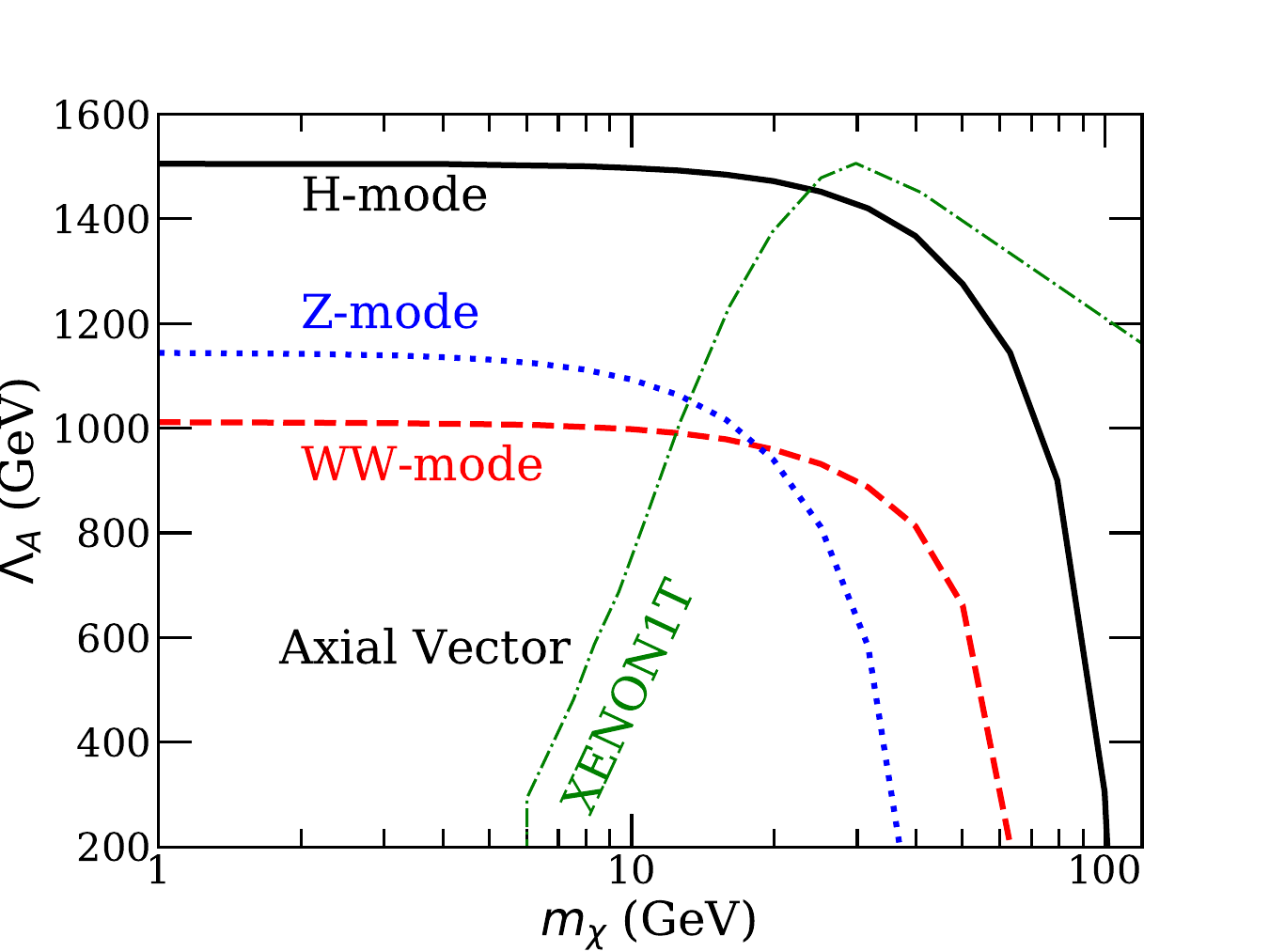}
    \caption{Expected 95\% CL constraints on $\Lambda_A$ in the three CEPC running modes. Also shown are the Xenon1T constraints on $\Lambda_A$~\cite{XENON:2019rxp}.}
    \label{fig:mq_rej:2}
\end{figure}

Fig.~\ref{fig:mq_rej:2} shows the 95\% CL lower bound on the new physics scale for one of the effective field theory interactions of DM in Ref.~\cite{Liu:2019ogn}. The interaction Lagrangian in this case is given by:
\begin{equation}
{\cal L}= \frac{1}{\Lambda_A^2} \bar{\chi}\gamma_\mu\gamma_5\chi\bar{\ell}\gamma^\mu\gamma_5 \ell,
\end{equation}
where $\chi$ is the Dirac DM, $\ell$ are charged leptons, and $\Lambda_A$ is the new physics scale. The $H$-mode yields the most stringent constraints on $\Lambda_A$. The three modes of CEPC are expected to lead to better limits in the DM mass range of $m_\chi\lesssim (10-25)$ GeV than Xenon1T \cite{XENON:2019rxp}, under the assumption that $\Lambda_A$ takes the same value for charged leptons and quarks.


\subsubsection{Interplay of dark particles with neutrinos}
\label{sec:DM:nu}

Interactions between dark matter (DM) and standard model (SM) particles have been extensively studied through direct detection methods \cite{Liu:2017drf, Schumann:2019eaa, Billard:2021uyg}, indirect detection techniques \cite{Gaskins:2016cha, Leane:2020liq, Slatyer:2021qgc}, and collider searches \cite{Boveia:2018yeb, Gori:2022vri, Lagouri:2022ier, Penning:2017tmb}. The null results and stringent constraints on the couplings have led to the postulation of a dark sector comprising particles with varying mass scales and feeble couplings to the SM.

One advantage of collider detection is the ability to search for heavy particles as long as their mass, $m_\chi$, is less than or equal to $\sqrt{s}$. Thus, colliders can probe not only the true DM particles that persist today but also any DS particles that can be directly generated. Additionally, most constraints focus on the DM coupling with nucleons and, consequently, quarks. Collider searches offer a tunable environment to distinguish between the leptophilic and hadrophilic natures of DM.

\begin{figure}[t]
\centering
\includegraphics[width=0.7\textwidth]{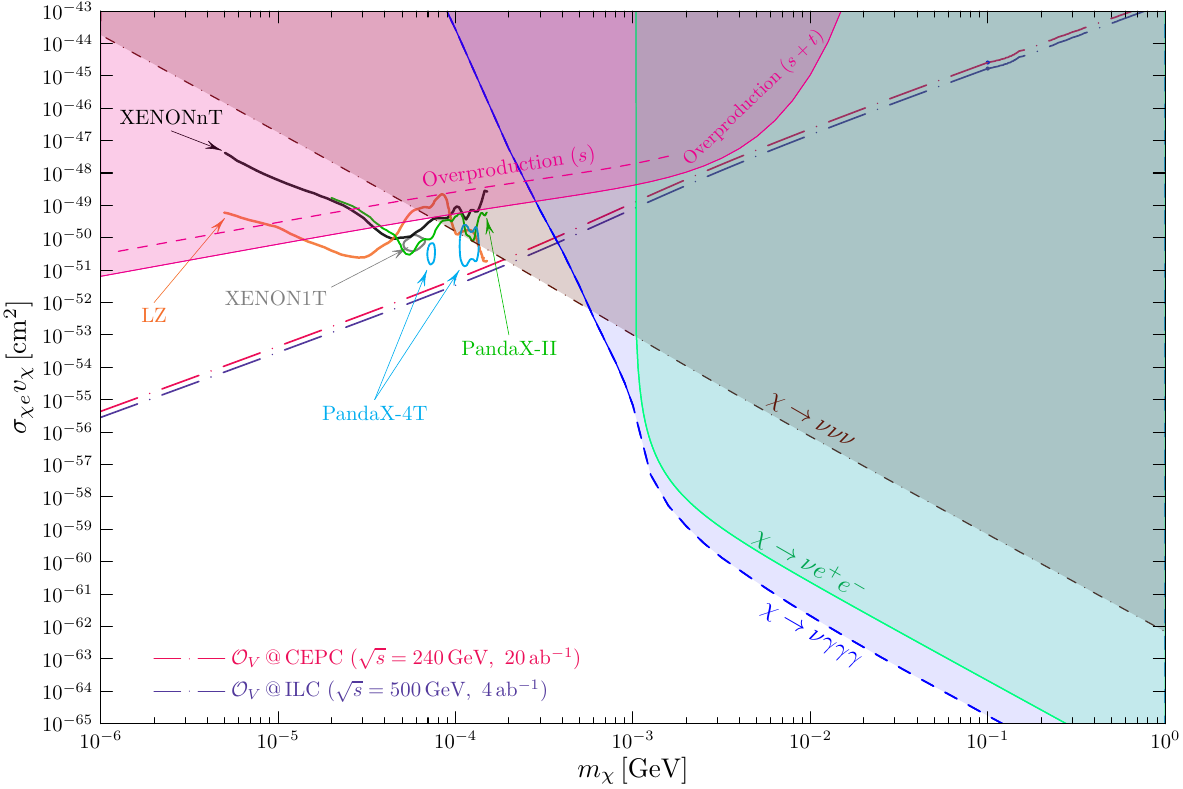}
\caption{Constraints on the fermionic
absorption DM from direct detection experiments
(PandaX-II, PandaX-4T, XENONnT, and LZ), cosmology,
astrophysics, and the projected
sensitivities at the future lepton colliders (CEPC, ILC,
and CLIC)~\cite{Ge:2023wye}. 
}
\label{fig:sens:cosmo:collider}
\end{figure}

Ref.~\cite{Ge:2023wye} focuses on absorption operators that couple a dark fermion with neutrinos and charged electrons/positrons. The study examines mono-photon and electron-positron pair productions associated with missing energy (a neutrino and a dark sector fermion) at future $e^+e^-$ colliders such as CEPC, FCC-ee, ILC, and CLIC~\cite{Ge:2023wye}. The findings indicate that mono-photon searches prevail at CEPC and ILC, while $e^+e^- + \slashed{E}$ dominates at CLIC. The combined sensitivity can reach well above 1\,TeV at CEPC/FCC-ee and ILC, and can further extend to 30\,TeV at CLIC.

Fig.~\ref{fig:sens:cosmo:collider} shows astrophysical $X/\gamma$-ray observations and cosmological constraints for sub-MeV absorption dark matter~\cite{Ge:2022ius}, demonstrating that collider searches are actually more sensitive. For a heavy dark fermion ($m_\chi > 2m_e$), the collider probe is generally weaker than astrophysical and cosmological constraints due to the increased decay width. However, this is only true when the dark fermion is assumed to be the genuine DM. The astrophysical and cosmological constraints can be relaxed by the presence of a large number of particles in the DS. In this sense, collider searches provide a complementary approach to addressing either light or heavy dark fermions.
	
\subsection{Dark matter and its loop effects at CEPC}
\label{sec:directchargeDM-loop}

Because of the limited center-of-mass energy, the massive particles present in dark matter (DM) models could hardly be directly produced at the CEPC. Nevertheless, the precise measurements of electroweak phenomena at the CEPC offer a propitious avenue for indirect detection of these particles via loop effects. Specifically, the CEPC is well-suited for the examination of DM models featuring supplementary electroweak multiplets.

A number of studies have been performed on this topic \cite{Cao:2014ita,Fedderke:2015txa,Cao:2016qgc,Cai:2016sjz,Liu:2017msv,Cai:2017wdu,Xiang:2017yfs,Wang:2017sxx,Gao:2021jip}, with a particular emphasis on two models \cite{Mahbubani:2005pt,Cohen:2011ec,Dedes:2014hga,Cai:2016sjz}: the singlet-doublet dark matter (SDDM) and doublet-triplet dark matter (DTDM) models. These two models can be regarded as the generalizations of electroweak sectors with proper UV completion. For instance, the SDDM and DTDM models are similar to the bino-Higgsino and Higgsino-wino sectors in SUSY models. The SDDM model features one singlet Weyl spinor with a Majorana mass term of $-m_s SS/2$, alongside two doublet Weyl spinors possessing opposite $U(1)_Y$ charges and a mass term of $-m_D \epsilon_{ij} D_1^j D_2^j$. Within this model, there exist two new Yukawa couplings $y_1 SD_1^i H_i - y_2 SD_2^i H_i^\dagger$, where $H_i$ denotes the SM Higgs doublet. The DTDM model involves one triplet and two doublet Weyl spinors. Both models are characterized by four independent parameters: two mass parameters and two Yukawa couplings. After electroweak symmetry breaking, the Yukawa terms induces the mixing of states. The lightest neutral particle potentially serves as a candidate for DM.

\begin{figure*}
  \includegraphics[width=0.49\textwidth]{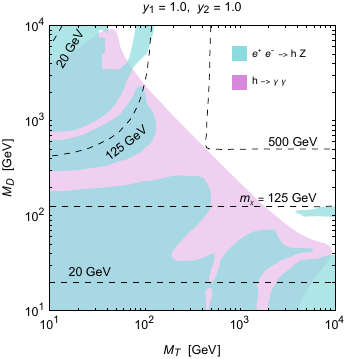}
  \includegraphics[width=0.49\textwidth]{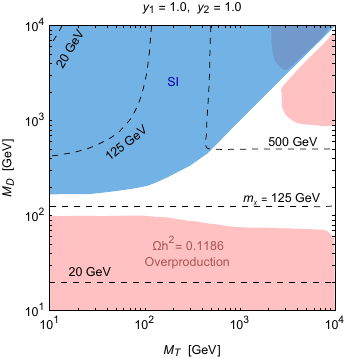}
  \caption{\textit{Left panel}: CEPC sensitivity in the doublet-triplet dark matter (DTDM) model with $y_1 = y_2 = 1$. The cyan and purple shaded regions show exclusions from CEPC through: (i) deviations in the $Zh$ production cross-section ($\Delta\sigma(e^+e^- \to hZ)/\sigma_0 > 0.5\%$), and (ii) modifications of the Higgs diphoton decay branching ratio ($\Delta\Gamma(h \to \gamma\gamma)/\Gamma_0 > 9.4\%$), respectively.
  \textit{Right panel}: Constraints from spin-independent DM-nucleon scattering (blue shaded) and DM relic abundance (red shaded). These DM constraints cannot probe the ``blind spot" region when $m_D < m_T$ under the $y_1 = y_2$ condition, where both the DM-$Z$ and DM-$h$ couplings vanish. CEPC effectively covers this parameter space, demonstrating strong complementarity with dark matter experiments.
  }
  \label{fig:DTDM}
\end{figure*}

The new particles introduced in these models have the potential to influence various electroweak phenomena at the CEPC, including Higgs decays \cite{Xiang:2017yfs,Wang:2017sxx}, oblique parameters \cite{Fedderke:2015txa,Cai:2016sjz,Cai:2017wdu}, and production of electroweak particles, such as $e^+e^- \rightarrow \mu^+\mu^-$, $ZZ$, $W^+W^-$, and $Z\gamma$, through loop effects \cite{Cao:2016qgc,Liu:2017msv,Gao:2021jip}. Notably, the CEPC, renowned as a Higgs factory, possesses a robust capacity to scrutinize these effects by means of measuring the $Zh$ associated production with a remarkable precision of $0.5\%$ \cite{Xiang:2017yfs,Wang:2017sxx,Gao:2021jip}. To quantify the impact of new physics models on the $Zh$ production cross section, we define a deviation parameter $\Delta \sigma /\sigma_0 \equiv |\sigma_\mathrm{NP} - \sigma_\mathrm{SM}|/\sigma_\mathrm{SM}$. Another promising avenue for the detection of new particles in the dark sector is the diphoton decay of the Higgs boson. For this process, we define a deviation $\Delta \Gamma /\Gamma_0 \equiv |\Gamma_\mathrm{NP} - \Gamma_\mathrm{SM}|/\Gamma_\mathrm{SM}$ for this process. In the left panel of Fig.~\ref{fig:DTDM}, we display the parameter regions in the DTDM model for $y_1=y_2=1$ with notable deviations $\Delta\sigma/\sigma_0 > 0.5\%$ for the $Zh$ production and $\Delta\Gamma/\Gamma_0 > 9.4\%$ for the Higgs diphoton decay at the CEPC with an integrated luminosity of $5~\mathrm{ab}^{-1}$.

In the right panel of Fig.~\ref{fig:DTDM}, we display the parameter regions excluded by the observed DM relic density \cite{Planck:2018vyg} and the LZ searches for DM-nucleon spin-independent scattering \cite{LZtalk}.
It is important to highlight that the exploration of the parameter space at the CEPC can serve as a valuable complement to other DM detection experiments, as illustrated in Fig.~\ref{fig:DTDM}.   
Despite the inclusion of DM within the electroweak multiplets, the interactions between DM particles and nucleons may be suppressed in certain parameter regions, leading to what is commonly referred as ``blind spots'' for direct detection \cite{Cheung:2012qy,Huang:2014xua}. Within the framework of the DTDM model, a global custodial symmetry emerges in the scenario where $y_1 = y_2$ \cite{Dedes:2014hga, Cai:2016sjz}, resulting in a null coupling between DM and the $Z$ boson, thereby yielding vanishing spin-dependent scattering. Furthermore, when $m_D < m_T$, the condition of $y_1 = y_2$ gives rise to a null coupling between DM and the Higgs boson, resulting in vanishing spin-independent scattering. The SDDM model exhibits similar characteristics to the DTDM model.

\subsection{Summary}

\begin{figure}[t]
\centering
\includegraphics[width=1.0\textwidth]{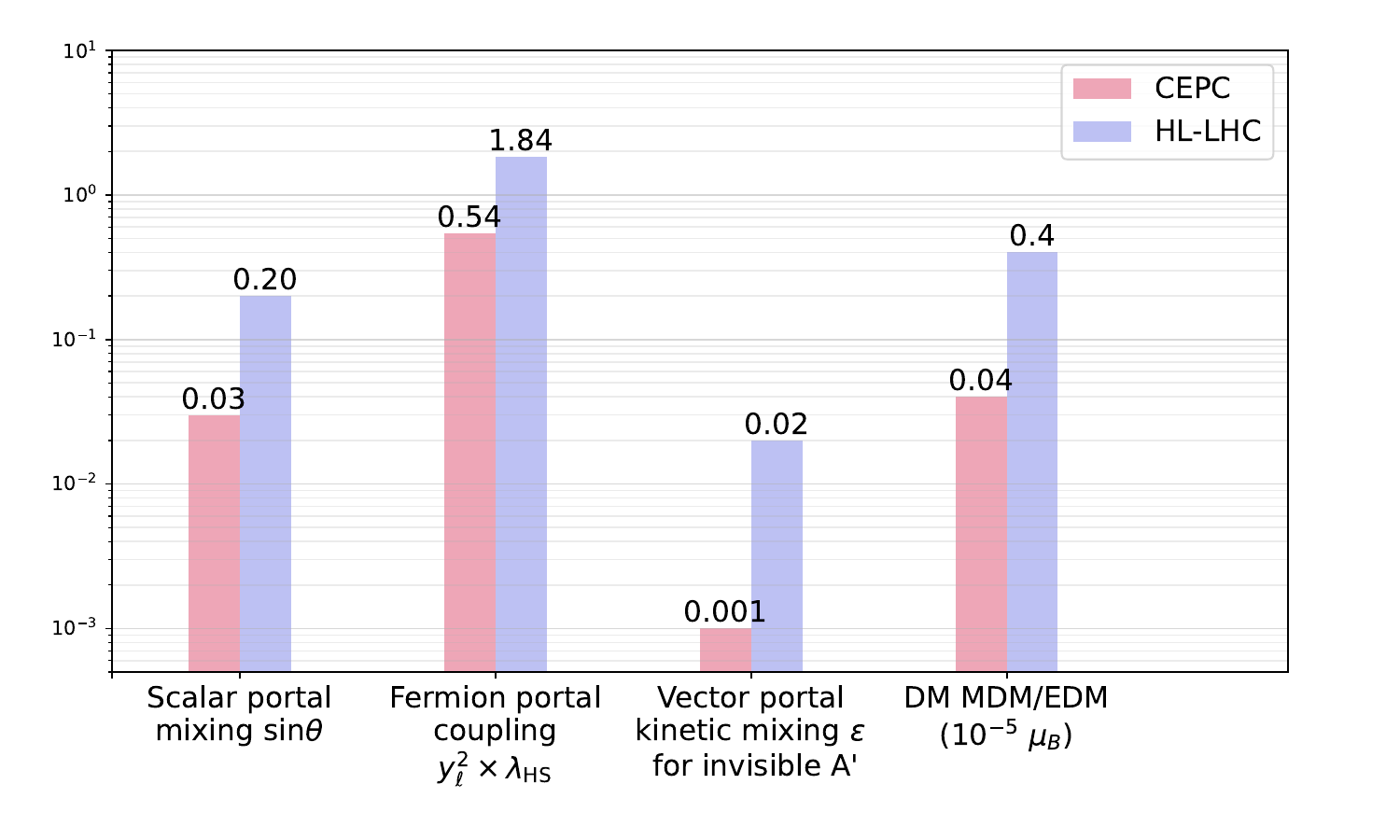}
\caption{The sensitivities for scalar, fermion, and vector portals, as well as dark matter magnetic dipole moment and electric dipole moment operators for CEPC and HL-LHC.}
\label{fig:Dark-Sector-LHC-CEPC}
\end{figure}

In this section, we explore the capabilities of the CEPC in searching for DM through various interaction portals, including the scalar portal, the fermion portal, the vector portal, and EFT operators. 
The clean experimental environment and clear center-of-mass energy of CEPC provide distinct advantages over the HL-LHC for DM searches. 

Specifically, as summarized in Fig.~\ref{fig:Dark-Sector-LHC-CEPC}, the CEPC exhibits improved sensitivity by approximately one order of magnitude compared to HL-LHC, as well as in the investigation of the magnetic dipole moment (MDM) and electric dipole moment (EDM). These enhancements originate from the reduction of background noise and targeted searches such as Higgs and $Z$ bosons. In contrast, the HL-LHC demonstrates superior sensitivity for direct search of heavy particles due to its larger cross sections from hadronic collisions, allowing significant production of dark matter signals from their decay. 

\begin{figure}[htb]
\centering
\includegraphics[width=0.95 \textwidth]{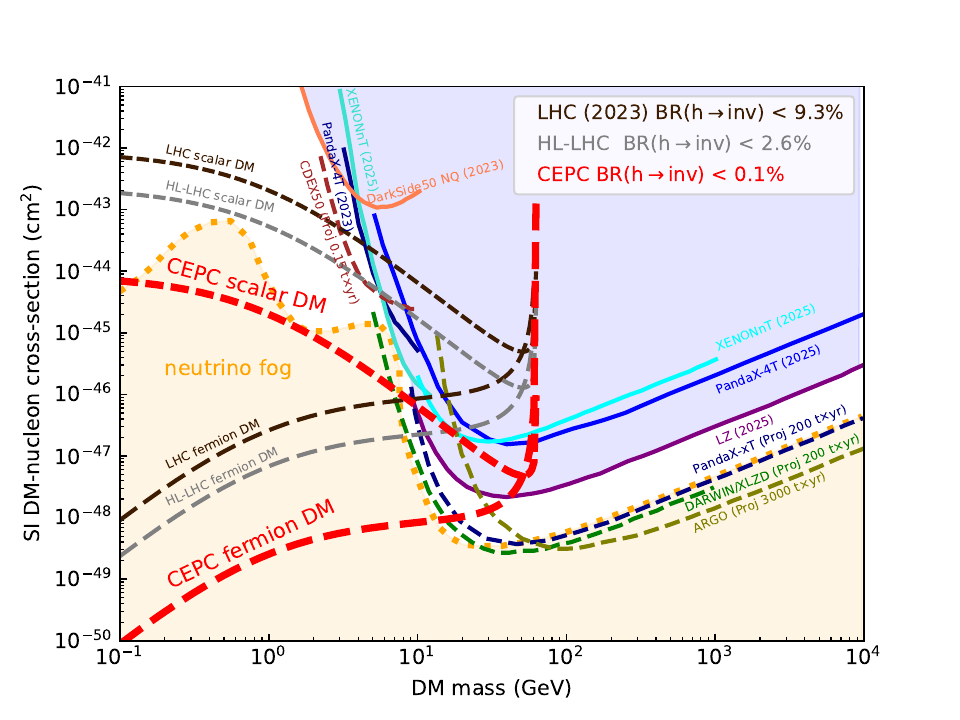}
\caption{
CEPC sensitivity to scalar and fermion DM via a Higgs-portal mediator, shown in the DM mass and spin-independent DM-nucleon cross-sections plane and compared with hadron-collider and direct-detection searches. \emph{Higgs invisible decays:} black and gray dashed lines indicate current LHC limits \cite{ATLAS:2023tkt} and HL-LHC projections \cite{deBlas:2019rxi}. \emph{Direct detection:} solid colored lines show current constraints from DarkSide-50 (2023) \cite{DarkSide-50:2022qzh}, LZ (2025) \cite{LZ:2024zvo}, PandaX-4T (2023, 2025) \cite{PandaX:2022aac,PandaX:2024qfu}, and XENONnT (2025) \cite{XENON:2024hup,XENON:2025vwd}; dashed colored lines give projected sensitivities for ARGO \cite{McDonald:2024osu}, CDEX-50 \cite{CDEX:2023vvc}, DARWIN/XLZD \cite{Baudis:2024jnk}, and PandaX-xT \cite{PANDA-X:2024dlo}. The neutrino fog follows Ref.~\cite{OHare:2021utq}.
}
\label{fig:Dark-Sector-Collider-DMDD}
\end{figure}

Beyond the comparison with the HL-LHC, Fig.~\ref{fig:Dark-Sector-Collider-DMDD} highlights the complementarity between the CEPC and direct-detection DM searches. The benchmark DM model is scalar/fermion DM with canonical Higgs-portal mediation to the SM sector. CEPC and (HL-)LHC searches provide strong limits at low DM mass ($\lesssim 10~\mathrm{GeV}$), complementing various direct-detection experiments. The CEPC achieves stronger projections than the HL-LHC, extending sensitivity into the neutrino fog for both scalar and fermion DM scenarios.


Overall, this comparison shows that the CEPC and HL-LHC are highly complementary in DM searches, where the CEPC’s order-of-magnitude improvement in sensitivity across a broad class of models (phase space) could significantly enhance our understanding of the nature of DM.

\clearpage
\section{Long-lived Particle Searches}
\label{sec:LLP}

\subsection{Introduction}

Recent years have witnessed a surge of interest in new fundamental particles with a relatively long lifetime which are predicted in many theoretical incarnations of physics beyond the Standard Model (BSM) and are often dubbed ``long-lived particles'' (LLPs).
Such particles are supposed to become long-lived, for various reasons including feeble couplings to other particles, phase space suppression, approximate symmetry, and heavy mediators.
Moreover, LLPs can resolve multiple fundamental problems of the Standard Model (SM) such as the non-vanishing neutrino mass, dark matter, baryogenesis, and naturalness.
Examples include heavy neutral leptons, dark photons, ALPs, and dark Higgs bosons (see, for instance, Refs.~\cite{Essig:2013lka,Alekhin:2015byh,Beacham:2019nyx,Curtin:2018mvb,Antel:2023hkf} for reviews of different models predicting LLPs).
On the other hand, at colliders such as the LHC, searches for heavy new particles have been going on, with no concrete discovery except stringent lower bounds on the mass of these particles such as squarks at $2 - 3$ TeV~\cite{ATLAS:2018nud,CMS:2017brl,CMS:2019vzo,CMS:2019zmd,ATLAS:2020xgt}; this situation has also shifted much attention in the community towards other BSM signatures including those associated with the LLPs~\cite{Lee:2018pag,Alimena:2019zri,DeRoeck:2019yvg,Alimena:2021mdu,Knapen:2022afb}.
For these reasons, collider searches for LLPs are becoming an increasingly important approach to probing BSM physics.

While the LHC has observed the SM Higgs boson in 2012~\cite{ATLAS:2012yve,CMS:2012qbp}, entered the Run 3 phase recently, and is expected to complete its high-luminosity stage by mid-2030s, intensive discussion concerning the next-generation high-energy colliders has never ceased since decades ago.
Such discussion mainly centers around these colliders' potential in discovering new fundamental particles and interactions.
These future collider-experiments include the Circular Electron Positron Collider (CEPC) in China~\cite{CEPCStudyGroup:2018rmc,CEPCStudyGroup:2018ghi,An:2018dwb, CEPCAcceleratorStudyGroup:2019myu, CEPCStudyGroup:2023quu}, 
the International Linear Collider (ILC) in Japan~\cite{Phinney:2007gp, Behnke:2013xla, Baer:2013cma, Behnke:2013lya, Fujii:2017vwa},
and the Future Circular Collider (FCC)~\cite{FCC:2018byv, FCC:2018evy, FCC:2018vvp, FCC:2018bvk} and 
the Compact Linear Collider (CLIC)~\cite{Linssen:2012hp, Klamka:2021cjt} at CERN.
The ILC and CLIC would be linear electron-positron colliders, while both the CEPC and the FCC in its initial stage, called the FCC-ee, would be operated as circular colliders of electron and positron beams.
These future colliders would be running at the center-of-mass (CM) energies ranging from about 91.2 GeV at the $Z$-pole, 240-250 GeV as a Higgs factory, 160 GeV as a $W$-boson factory, 350 GeV as a top-quark factory, to even higher energies up to the TeV scale.
The corresponding gauge and Higgs bosons are thus produced in a clean environment (i.e.~with only little background contamination).
We note that for the Higgs-factory mode at the CM energy of 240 GeV, sensitivities shown in this section usually correspond to the integrated luminosity of 5.6 ab$^{-1}$ for the CEPC,  
and the upgraded luminosity of 20 ab$^{-1}$ can lead to stronger discovery sensitivities. 

Compared to hadron colliders such as the LHC, the high-energy electron-positron colliders usually have higher integrated luminosities and looser trigger conditions.
This allows for not only precision measurements of the SM particles and parameters, but also searches for new particles including LLPs e.g.~via rare (electroweak) decays of the SM particles such as the $Z$- and Higgs bosons.
Since the $e^- e^+$ colliders have definite colliding-parton energies, recoil strategy can be adopted in collider searches.
Phenomenological studies on their sensitivities to LLPs have been performed extensively; see e.g.~Ref.~\cite{Blondel:2022qqo} for recent reviews of LLP experiments at the proposed FCC-ee collider.
These works have considered not only the default detector located at the interaction point (IP), dubbed ``\textit{main detector}'' (MD) or ``\textit{near detector}'' (ND) in this work, but also the proposed external detectors (or ``\textit{far detectors}'' (FDs) away from the IP), for searching for LLPs.
Moreover, beam-dump experiments have also been suggested for construction at future $e^- e^+$ colliders, aiming primarily at finding new exotic states with a long lifetime.
We note that the LLPs produced at the IP of beam-energy-symmetric $e^- e^+$ colliders tend to travel along the transverse direction in the laboratory frame, while at $pp$ colliders the LLPs often employ a large boost in the forward/longitudinal direction as a result of the proton's parton distribution.

In Table~\ref{tab:summaryLLPs}, we summarize results from recent CEPC's studies on LLPs. 
In the table, the first column lists the types of the LLPs; 
the second column presents the corresponding signal signature;
the third and fourth columns provide the center-of-mass energy and the integrated luminosity;
the fifth column indicates the considered main detector (MD) or far detectors (FD3 or LAYCAST);
the sixth column shows sensitivities on the couplings, suppression scales, branching ratios, or production cross sections with assumptions of the LLP's mass ($m$), lefetime ($\tau$) and others; 
the last two column provide the references.
Check the main text for the meanings of symbols and abbreviations.

\begin{table}[h]
\scalebox{0.65}{
\begin{tabular}{|c|c|c|c|c|c|c|c|}
\hline\hline
LLP & 
\multirow{2}{*}{Signal Signature} &  
$\sqrt{s}$ &
$\mathcal{L}$ & 
\multirow{2}{*}{Detector} &
Sensitivities on parameters & 
\multirow{2}{*}{Figs.} & 
\multirow{2}{*}{Refs.}   \\
Type & & [GeV]  & [ab$^{-1}$] &  & [Assumptions] & & \\ 

\hline
\multirow{8}{0.15\textwidth}{New scalar particles ($X$)} 
& $Z(\to {\rm incl.})\, h(\to X X)$, & \multirow{2}{*}{240} & \multirow{2}{*}{20} & \multirow{2}{*}{MD} & Br$(h \to X X) \sim 10^{-6} $ & \multirow{2}{*}{\ref{fig:ML-LLPs-Limits}} & \multirow{2}{*}{\cite{Zhang:2024bld}} \\ 
& $X \to q\bar{q}/\nu \bar{\nu}$ & & & & [$m \in (1, 50)$ GeV, $\tau \in (10^{-3}, 10^{-1})$ ns] & & \\

\cline{2-8}
&  & \multirow{6}{*}{240} & \multirow{6}{*}{5.6} 
& \multirow{2}{*}{MD} &  Br$(h \to XX) \sim 3 \times 10^{-6}$ & \multirow{2}{*}{\ref{fig:FDs-H2XX-0p5}} & \multirow{2}{*}{\cite{Wang:2019xvx}} \\ 
&  & & & & [$m = 0.5$ GeV, $c \tau \sim 5 \times 10^{-3}$ m] & & \\

\cline{5-8}
& $Z(\to {\rm incl.})\, h(\to X X)$, & & 
& \multirow{2}{*}{FD3} &  Br$(h \to XX) \sim 7 \times 10^{-5}$ & \multirow{2}{*}{\ref{fig:FDs-H2XX-0p5}} & \multirow{2}{*}{\cite{Wang:2019xvx}} \\ 
& $X \to {\rm incl.}$ & & & & [$m = 0.5$ GeV, $c \tau \sim$ 1 m] & & \\

\cline{5-8}
& & & 
& \multirow{2}{*}{LAYCAST} &  Br$(h \to XX) \sim 5 \times 10^{-6}$ & \multirow{2}{*}{\ref{fig:FDs-H2XX-0p5}} & \multirow{2}{*}{\cite{Lu:2024fxs}} \\ 
& & & & & [$m = 0.5$ GeV, $c \tau \sim 10^{-1}$ m] & & \\

\hline
\multirow{6}{0.15\textwidth}{RPV-SUSY neutralinos ($\tilde{\chi}_1^0$)} 
&  & \multirow{6}{*}{91.2} & \multirow{6}{*}{150} & \multirow{2}{*}{MD}   
& $\lambda'_{112}/m_{\tilde f}^2 \in (2 \times 10^{-14}, 10^{-8})$ GeV$^{-2}$ & \multirow{2}{*}{\ref{fig:NDs-Z2chi1chi1-RPVSUSY}} & \multirow{2}{*}{\cite{Wang:2019xvx}} \\
&   &  &  & 
& [$m \sim 40$ GeV, Br($Z\to \tilde{\chi}_1^0 \tilde{\chi}_1^0$) = $10^{-3}$] & &  \\ 

\cline{5-8}
& $Z\to \tilde{\chi}_1^0 \tilde{\chi}_1^0$, & & & \multirow{2}{*}{FD3} 
&  $\lambda'_{112}/m_{\tilde f}^2 \in (10^{-14}, ~10^{-9})$ GeV$^{-2}$ & \multirow{2}{*}{\ref{fig:FDs-Z2n1n1}} & \multirow{2}{*}{\cite{Wang:2019xvx}} \\ 
& $\tilde{\chi}_1^0 \to$ incl. & & & 
& [$m \sim 40$ GeV, Br($Z\to \tilde{\chi}_1^0 \tilde{\chi}_1^0$) = $10^{-3}$] & &  \\ 

\cline{5-8}
& & & & \multirow{2}{*}{LAYCAST} 
&  $\lambda'_{112}/m_{\tilde f}^2 \in (7\times10^{-15}, ~10^{-9})$ GeV$^{-2}$ & \multirow{2}{*}{\ref{fig:FDs-Z2n1n1}} & \multirow{2}{*}{\cite{Lu:2024fxs}} \\ 
& & & & 
& [$m \sim 40$ GeV, Br($Z\to \tilde{\chi}_1^0 \tilde{\chi}_1^0$) = $10^{-3}$] & &  \\ 

\hline
\multirow{7}{0.15\textwidth}{ALPs ($a$)} 
& $Z^{(*)} \to \mu^- \mu^+ a$ & \multirow{1}{*}{91} & \multirow{1}{*}{150} & \multirow{1}{*}{MD} & $f_a/C^A_{\mu\mu} \lesssim 950$ GeV  & \ref{fig:NDs-Z2LLa} & \cite{Calibbi:2022izs} \\ 

\cline{2-8}
&  & \multirow{6}{*}{91.2} & \multirow{6}{*}{150} & \multirow{2}{*}{MD} 
&  $C_{\gamma\gamma}/\Lambda \sim 10^{-3}$ TeV$^{-1}$ & \multirow{2}{*}{\ref{fig:FDs-ALP-FADEPC}} & \multirow{2}{*}{\cite{Lu:2024fxs}} \\ 
& & & & 
& [$C_{\gamma Z} = 0$, $m \sim 2$ GeV] & &  \\

\cline{5-8}
& $e^+ e^- \to \gamma\, a$, &  &  & \multirow{2}{*}{FD3} 
& $C_{\gamma\gamma}/\Lambda \sim 6 \times 10^{-3} $ TeV$^{-1}$ & \multirow{2}{*}{\ref{fig:FDs-ALP-FADEPC}} & \multirow{2}{*}{\cite{Tian:2022rsi}} \\ 
& $a \to \gamma \gamma$ & & & 
& [$C_{\gamma Z} = 0$, $m \sim 0.3$ GeV] & &  \\

\cline{5-8}
&  &  &  & \multirow{2}{*}{LAYCAST} 
&  $C_{\gamma\gamma}/\Lambda \sim 2 \times 10^{-3}$ TeV$^{-1}$ & \multirow{2}{*}{\ref{fig:FDs-ALP-FADEPC}} & \multirow{2}{*}{\cite{Lu:2024fxs}} \\ 
& & & & 
& [$C_{\gamma Z} = 0$, $m \sim 0.7$ GeV] & &  \\

\hline
\multirow{2}{0.16\textwidth}{Hidden valley particles ($\pi_V^0$)} & $Z\, h(\to \pi_V^0 \pi_V^0)$, & \multirow{2}{*}{350} & \multirow{2}{*}{1.0} & \multirow{2}{*}{MD}  &  $\sigma(h) \times {\rm BR} (h \to \pi_v^0 \pi_v^0) \sim 10^{-4}$ pb & \multirow{2}{*}{\ref{fig:upl}} & \multirow{2}{*}{\cite{Kucharczyk:2022pie}} \\ 
& $\pi_V^0 \to b \bar{b}$ & & & & [$m \in (25, 50)$ GeV, $\tau \sim 10^2$ ps] & & \\
\hline

\multirow{2}{0.15\textwidth}{Dark photons ($\gamma_D$)} 
& $Z(\to q \bar{q} )\, h(\to \gamma_D \gamma_D)$, & \multirow{2}{*}{250} & \multirow{2}{*}{2.0} & \multirow{2}{*}{MD} &  Br$(h \to \gamma_D \gamma_D) \sim 10^{-5}$, & \multirow{2}{*}{\ref{fig:NDs-h2dpdp}} & \multirow{2}{*}{\cite{Jeanty:2022cwr}}\\ 
& $\gamma_D \to \ell^- \ell^+ / q \bar{q}$ & & & & [$m \in (5, 10)$ GeV, $\tau \sim 10^2$ ps, $\epsilon \in (10^{-6}, 10^{-7})$] & & \\
\hline
\hline 
\end{tabular}
}
\caption{Summary  of results from recent CEPC's studies on LLPs. The first column lists the types of the LLPs; the second column presents the corresponding signal signature; the third and fourth columns provide the center-of-mass energy and the integrated luminosity; the fifth column indicates the considered main detector (MD) or far detectors (FD3 or LAYCAST); the sixth column shows sensitivities on the couplings, suppression scales, branching ratios, or production cross sections with assumptions of the LLP's mass ($m$), lefetime ($\tau$) and others; the last two column provide the references. Check the main text for the meanings of symbols and abbreviations.}
Summary  of results from recent CEPC's studies on LLPs. 
The first column lists the types of the LLPs; 
the second column presents the corresponding signal signature;
the third and fourth columns provide the center-of-mass energy and the integrated luminosity;
the fifth column indicates the considered main detector (MD) or far detectors (FD3 or LAYCAST);
the sixth column shows sensitivities on the couplings, suppression scales, branching ratios, or production cross sections with assumptions of the LLP's mass ($m$), lefetime ($\tau$) and others; 
the last two column provide the references.
Check the main text for the meanings of symbols and abbreviations.

\label{tab:summaryLLPs}
\end{table}


This section is organized as follows.
In Section~\ref{subsec:computationLLPsColliders} we present a computation procedure of LLP signal-event rates at colliders.
Then in Sections~\ref{subsec:NDs}, \ref{subsec:studiesFDs}, and \ref{subsec:beam_dump}, we review LLP studies and summarize their results at main detectors, proposed far detectors, and possible beam-dump experiments, respectively, for future high-energy $e^- e^+$ colliders.
Finally, we conclude in Section~\ref{subsec:summary} with a summary and an outlook.

\subsection{Computation of LLP signal-event rates}
\label{subsec:computationLLPsColliders}

In this section, we present and discuss a simplified and widely used computation procedure of the theory-predicted signal-event rates of the LLP.

\begin{figure}[t]
\centering
 \includegraphics[width=0.6\linewidth]{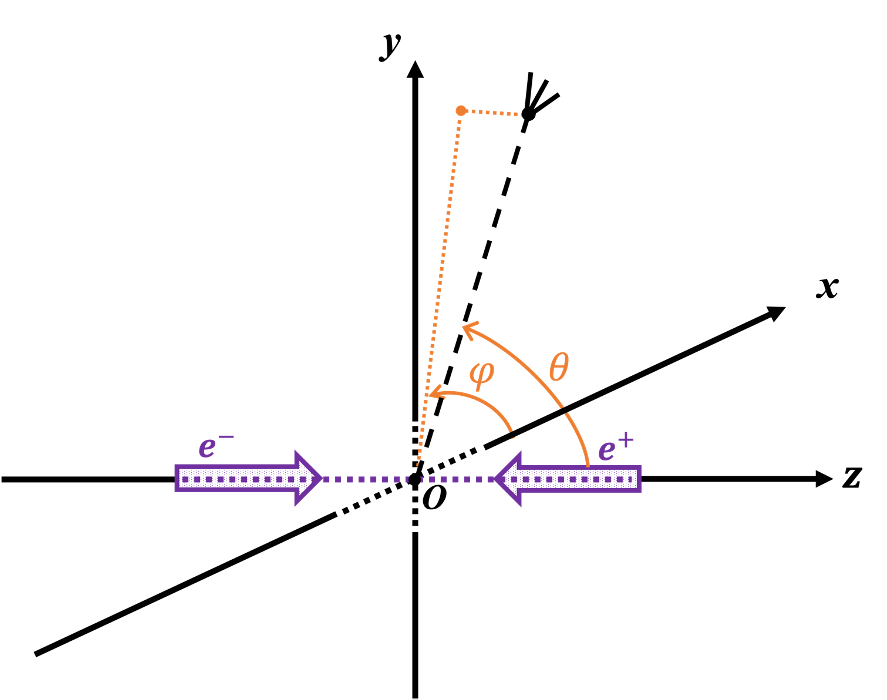}
\caption{Sketch of the production and decay of a LLP at electron-positron colliders.}
\label{fig:sketchLLPee}
\end{figure}

As shown in Fig.~\ref{fig:sketchLLPee}, we assume that the LLP is produced (essentially) at the IP of a collider, and travels a macroscopic distance with a constant velocity\footnote{This is true for electrically neutral LLPs; for electrically charged LLPs we assume that they have large enough transverse momentum $p_T$ so that the bending effect of the magnetic field at the MD is unimportant and that the potential ionization effect in material is negligible.} before decaying into SM or other new particles.
The survival probability of the LLP after traveling a distance $D$ can be estimated with the exponential decay law $P(D)=e^{-D/\lambda }$, where $\lambda$ is the LLP's boosted decay length in the laboratory frame.
Considering the effect of special relativity, $\lambda$ can be calculated as 
\begin{equation}
\lambda = \beta \gamma\, c \tau = \frac{p}{E}\, \frac{E}{m}\, c \tau = \frac{p}{m}\, c \tau \,\, ,
\label{eqn:decayL}
\end{equation}
where $c$ is the speed of light, while $p$, $E$, and $m$ are the magnitude of the three-momentum $\vec{p}$, energy, and mass of the LLP, respectively.
$\beta = p/E$ and $\gamma =  E/m$ are the speed and boost factor of the LLP, and $\tau$ is the LLP's lifetime in its rest frame and can be predicted from theoretical model parameters.
With the knowledge of the boosted decay length of an LLP denoted with the index $i$, it is then possible to compute its decay probability inside the fiducial volume (f.v.) of a detector, if the moving direction of the LLP points towards it,
\begin{equation}
P[(\text{LLP})_i\text{ in f.v.}] = e^{\left( -{ D_i^e }/{ \lambda_i } \right)} - e^{\left( -{D_i^l}/{\lambda_i} \right)}\,\, ,
\label{eqn:decayProb}
\end{equation}
where $D_i^e$ ($D_i^l$) is the distance from the IP to the point on the detector surface where the LLP would enter (leave) the detector if not having decayed beforehand, and $\lambda_i$ is the LLP's boosted decay length.
Note that $D_i^e < D_i^l$ by definition.
Obviously, $P[(\text{LLP})_i\text{ in f.v.}]=0$, if the LLP travels outside the fiducial volume's window.
In practice, the cylindrical/azimuthal symmetry, if (approximately) present for the relative position and orientation between the detector and the IP, can be made use of since the LLP kinematic distribution in the azimuthal angle is almost always homogeneous, as an overall factor applied to the right-hand side of Eq.~\eqref{eqn:decayProb}.
Taking advantage of the azimuthal symmetry allows for obtaining robust results with fewer MC simulation events required.

In a phenomenological analysis, the average decay probability of an LLP in a detector's fiducial volume, $\langle P[\rm{LLP\,\, in\,\, f.v.}]\rangle$, is often required in order to predict the signal-event rates.
To achieve sufficient precision in the prediction, it is usually required to perform Monte Carlo (MC) simulation, unless the signal-event kinematics can be analytically derived.
We thus compute $\langle P[\rm{LLP\,\, in\,\, f.v.}]\rangle$ with
\begin{eqnarray}
\langle P[\text{LLP}\text{ in f.v.}]\rangle=\frac{1}{N^{\text{MC}}_{\text{LLP}}}\sum_{i=1}^{N^{\text{MC}}_{\text{LLP}}}
                P[(\text{LLP})_i\text{ in f.v.}]\,\, .
                \label{eqn:AveDecayProb}
\end{eqnarray}
Here, $N^{\text{MC}}_{\text{LLP}}$ labels the total number of the LLPs generated with the MC simulation program, and $P[(\text{LLP})_i\text{ in f.v.}]$ is obtained with Eq.~\eqref{eqn:decayProb}.

We can therefore express the expected signal-event number with,
\begin{equation}
N^{\rm exp}_{\rm LLP} = N^{\rm prod}_{\rm LLP} \cdot \langle P[\rm{LLP\,\, in\,\, f.v.}]\rangle \cdot {\rm Br(LLP \to vis)} \cdot \epsilon_{\rm det} \,\, ,
\label{eqn:NexpLLP}
\end{equation}
where $N^{\rm prod}_{\rm LLP} = \sigma^{\rm prod}_{\rm LLP} \cdot \mathcal{L}$ is the total number of the LLPs produced at the collider and is determined by the LLP's production cross section $\sigma^{\rm prod}_{\rm LLP}$ and the collider's integrated luminosity $\mathcal{L}$.
It is important that the kinematic cuts (if there is any) to which $N^{\rm prod}_{\rm LLP}$ corresponds should be aligned with those imposed in the MC simulations for computing $\langle P[\rm{LLP\,\, in\,\, f.v.}]\rangle$.
$\rm Br(LLP \to vis)$ is the branching ratio of LLP decaying into visible products.
$\epsilon_{\rm det}$ denotes the detector efficiency for the visible final state.
For simplicity, we do not include the possible dependence of $\epsilon_{\rm det}$ on the momentum, energy, or production position of the final-state particles from LLP decays.
Further, LLP searches often impose cut selections on observables typical for LLP decay products, such as requiring a large transverse impact parameter.
If such selections and detector efficiencies specific for LLPs are to be imposed, it is required to simulate the LLP decays with correct decay BRs including the decay positions handled by the MC simulation tool used.

Eq.~\eqref{eqn:NexpLLP} depends in a complicated way on not only the collider setups including the detector's design but also the theoretical model parameters such as the LLP's mass and its couplings to other particles.
The LLP production cross section, $\sigma^{\rm prod}_{\rm LLP}$, is affected by both the collider beam energies and the coupling(s) inducing the LLP's production, and $\mathcal{L}$ reflects linearly the volume of the collected data.
To compute the average decay probability in the detector's fiducial volume, $\langle P[\rm{LLP\,\, in\,\, f.v.}]\rangle$, we should take into account the detector's geometry such as its position, shape, and volume, as well as the LLP's kinematics determined by beam energies, LLP mass, as well as its proper lifetime which further depends on its mass and decay couplings.
The visible decay branching ratio ${\rm Br(LLP \to vis.)}$ can be predicted by the LLP's mass and sometimes also its decay couplings' strengths.
Finally, the detector efficiency $\epsilon_{\rm det}$ can often be modeled as a function of the LLP's energy and travelling direction, and is essentially determined by the detector's design.

In the large decay-length limit (which is usually of the most interest in LLP studies) such that the boosted decay length $\lambda$ of the LLP is mostly dominant over the distance between the IP and the detector, the right-hand side of Eq.~\eqref{eqn:decayProb} can be expanded and as a result, the LLP decay probabilities in the fiducial volume and hence the signal-event number $N_{\text{LLP}}^{\text{exp}}$ become essentially, to the first order, proportional to $\Delta D=D_i^l - D_i^e$ times an overall factor accounting for the proportion of the generated LLPs travelling inside the window/solid-angle coverage of the detector.
Assuming a fixed $\Delta D$, the MD has the advantage of a huge solid-angle coverage compared to a FD, which can be offset, in principle, by building a FD with a large volume depending on its distance to the IP and the available space.
However, for a FD, it is possible to implement shielding measures such as rock and lead in the space between the IP and the FD, thus removing potential background sources that would weaken the sensitivity reach especially for the MD.
For beam-dump experiments, a boost of the LLP in the forward direction should be exploited and thus increasing the length of the detector can linearly strengthen the signal-event rates.

We note that 
advanced neural networks, trained using deep learning techniques to exploit the distinct LLP signatures and topologies, can further differentiate signals from SM backgrounds, especially in a clean environment provided by lepton colliders. Ref.~\cite{Zhang:2024bld} demonstrates that LLP searches can reach their full physics potential by harnessing the power of machine learning (ML). The LLP signal efficiency with an ML-based approach can increase significantly compared to traditional selection-based methods, while maintaining a background-free environment.

We conclude the section with a brief discussion on typical potential background sources for LLP searches at colliders.
Although in general collider searches for LLPs suffer from relatively few background sources compared to prompt searches, there remain several typical origins of background events~\cite{Alimena:2019zri,Knapen:2022afb}.
Particle collisions induce scattering processes that produce certain SM particles that have a relatively long lifetime such as charm and bottom mesons, leading to collider signatures similar to those of BSM LLPs.
Non-collision background sources include material interactions with SM particles, fake tracks and DVs, detector noise, and cosmic-ray muons.
While the SM irreducible background events can often be simulated in a reliable way, the non-collision type of background sources can usually be estimated only with non-traditional methods such as data-driven approaches, especially for the MD experiments.
For FD and beam-dump experiments, the large space available between the IP/beam dump and the detector can often allow for installation of shielding, removing some of these backgrounds to a large extent such as long-lived SM particles.
In practice, in many phenomenological analysis on collider searches for LLPs, in particular in the context of FD or beam-dump experiments, the assumption of zero background event is often made, expecting novel experimental strategies and methods, as well as instrumented apparatus such as shielding with lead or rock, can remove essentially all the background events.

\subsection{Studies with the main detector}
\label{subsec:NDs}

\begin{figure}[t]
\centering
\includegraphics[width=0.8\textwidth]{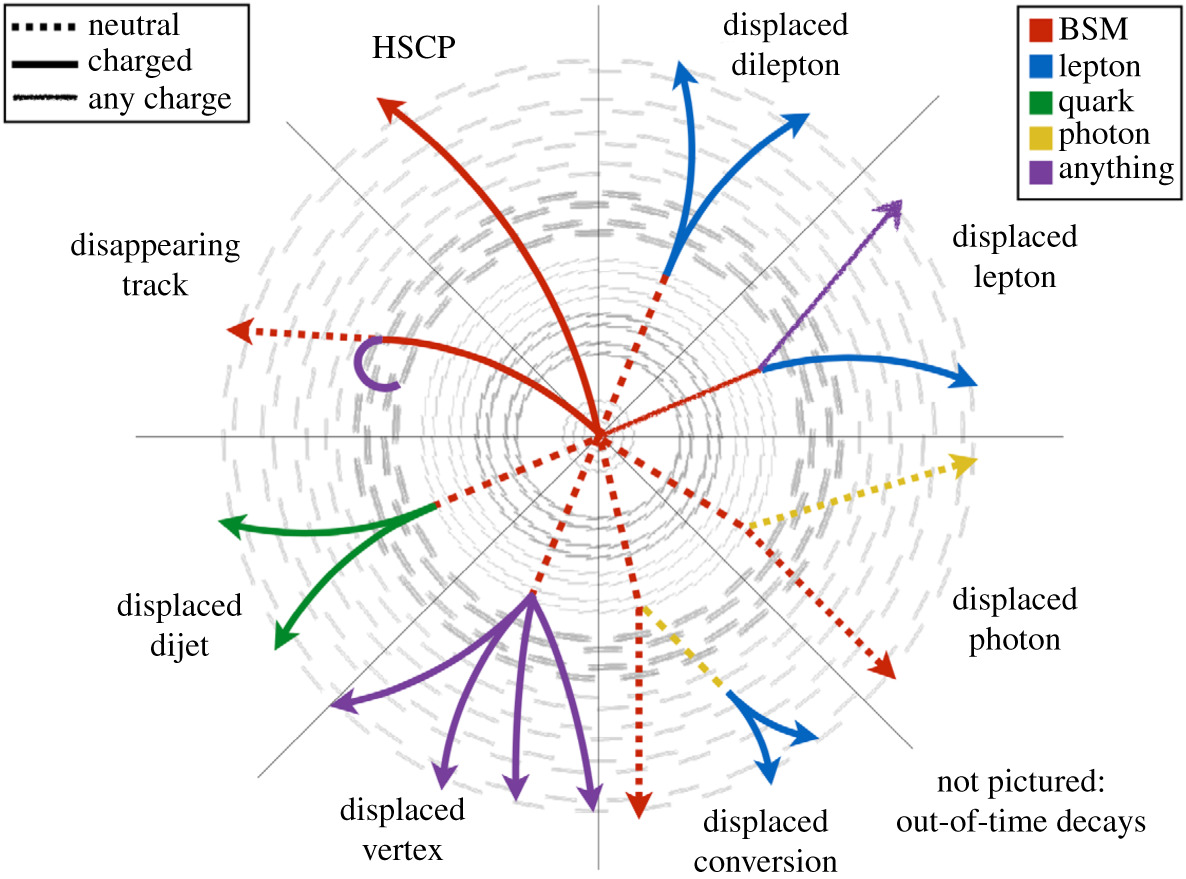}
\caption{Typical LLP signatures at the MD of a collider.
Taken from Ref.~\cite{DeRoeck:2019yvg}.
}
\label{fig:signature-LLP-ND}
\end{figure}

LLPs can manifest themselves via different signatures at the MD of a collider.
This is illustrated in Fig.~\ref{fig:signature-LLP-ND} extracted from Ref.~\cite{DeRoeck:2019yvg}.
Charged LLPs, if not too soft, can leave a visible track inside the MD.
For example, heavy stable charged particles (HSCPs) can travel through the whole detector without decaying, leaving a complete track.
Similarly, a charged LLP can decay inside the MD into charge-neutral or soft charged final states, resulting in a disappearing track.
Neutral LLPs often couple only very feebly with SM particles, so that they do not interact with the detector material.
If they leave the MD without decaying, they appear simply as missing energy or transverse momentum, and can only be searched for if they are produced in association with visible objects.
However, if they decay inside the MD but with a macroscopic distance from the IP, exotic signatures can arise including displaced vertex, displaced leptons, displaced jets, and non-pointing photons.

In the rest of this section, we review past LLP studies with the MD at future high-energy electron-positron colliders, which cover various theoretical models and collider signatures.
LLP studies related to the flavor physics are discussed in Section~\ref{sec:FlavorPortal}.
Sensitivity results for long-lived heavy neutral leptons, doubly-charged scalars in seesaw models, electrophilic ALPs ($e$ALPs) are presented in Sections~\ref{subsubsec:NwithND}, \ref{subsec:LL-doublyChargedHiggs}, and \ref
{subsec:moreExotics-ALP}, respectively.

\subsubsection{Higgs boson decays}
\label{subsubsec:ND-HiggsDecay}

LLPs can be produced from exotic decays of the Higgs boson, c.f. Section~\ref{subsubsec:h2LLPs}. Recent studies are summarized as follows.

\noindent\underline{New scalar particles:}

Ref.~\cite{Zhang:2024bld} performs a search for neutrally charged LLPs ($X_{1}$, $X_{2}$) produced via the rare decay of the Higgs boson. The signal process is $e^+e^-\to ZH (Z\to\text{inclusive}, H\to X_{1}+X_{2}$) at $\sqrt{s} = 240$ GeV. \( X_1 \) and \( X_2 \) can each decay into a $\nu\bar{\nu}$ pair or a $q\bar{q}$ pair, resulting in final states with either two jets (type-I signal) or four jets (type-II signal)~\footnote{The total invisible decay mode is not considered here.}. The study is conducted using full simulation MC samples corresponding to an integrated luminosity of $20.0\text{~ab}^{-1}$ and about \(4 \times 10^6\) Higgs bosons. Both Convolutional Neural Networks (CNN) and Graph Neural Networks (GNN) have been trained using full MC signal and background samples. The results of two neural networks agree with each other.

\begin{figure}[t]
\centering
\includegraphics[width=0.49\textwidth]{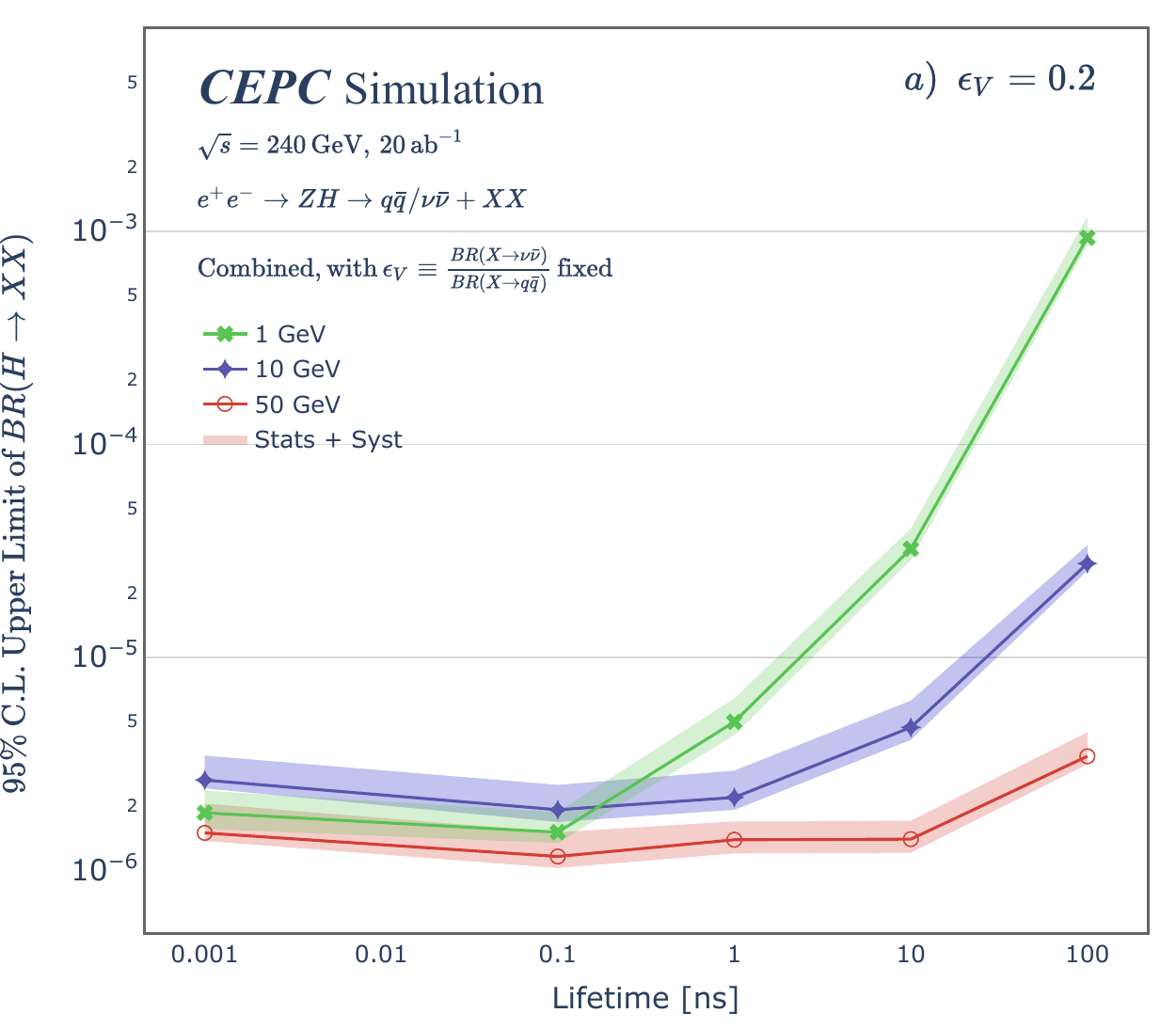}
\includegraphics[width=0.49\textwidth]{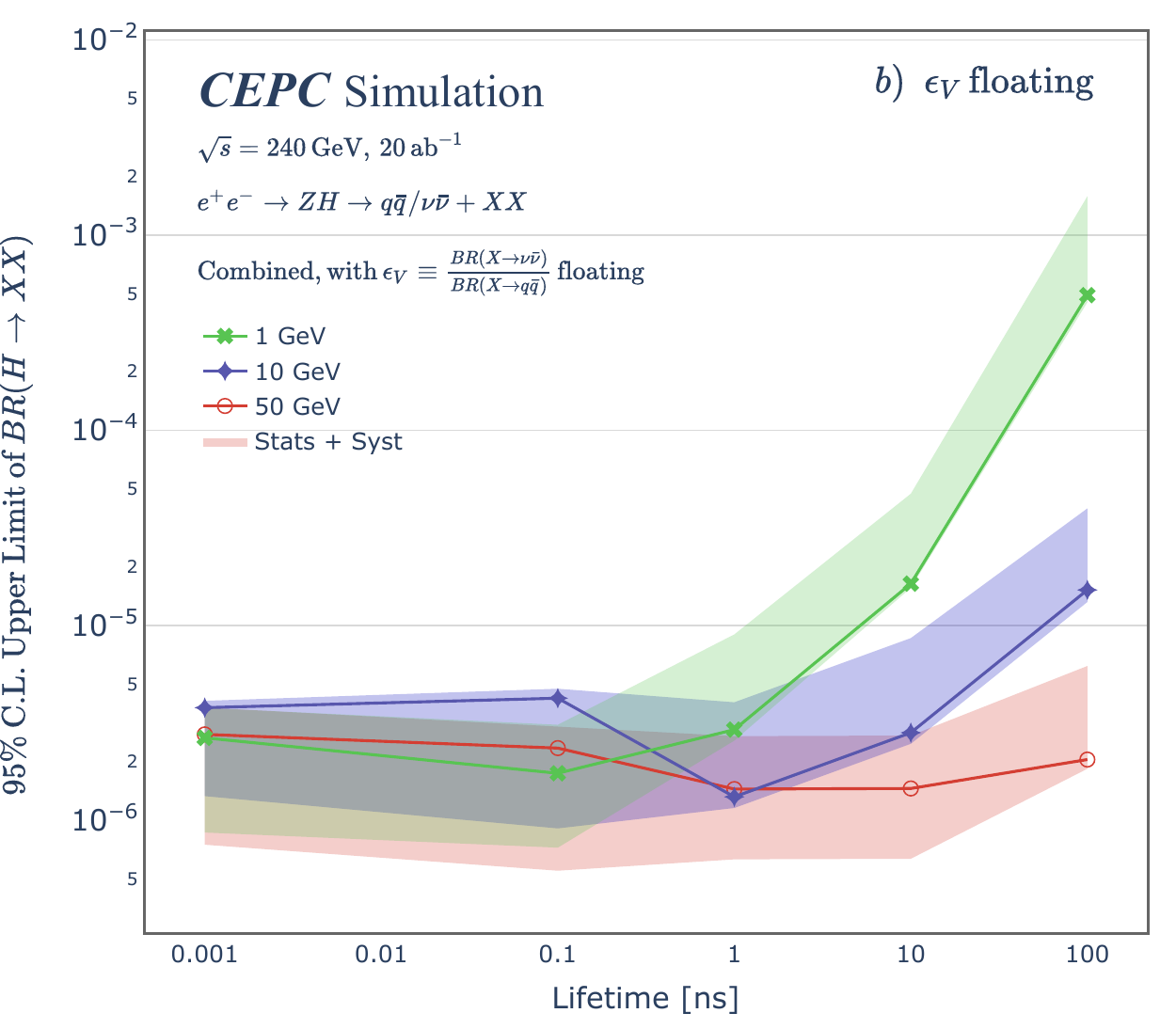}
\caption{
       (Color online) 95\% C.L. upper limit on the branching ratio (BR) for 
        the Higgs boson (\( H \)) decay into pairs of LLPs (\( X_1 X_2 \)) via \( e^+e^- \to Z H \), 
        where \( \epsilon_{V} \) is the ratio \( \frac{BR(X \to \nu\bar{\nu})}{BR(X \to q\bar{q})} \). 
        \textbf{a)}: a fixed ratio \( \epsilon_{V} = 0.2 \), 
        \textbf{b)}: a floating \( \epsilon_{V} \). 
        The shaded areas indicate statistical and systematic uncertainties combined.
        Taken from Ref.~\cite{Zhang:2024bld}. 
}
\label{fig:ML-LLPs-Limits}
\end{figure}

Fig.~\ref{fig:ML-LLPs-Limits} shows constraints on the branching ratio of Higgs boson decay to LLPs. 
For the two LLPs signal types, a parameter $\epsilon_{V}:= \frac{BR(X \to \nu\bar{\nu})}{BR(X \to q\bar{q})}$ is defined. In the case of Type I and Type II signal yields having a fixed ratio, a value of 0.2 is set and a one-dimensional 95\% Confidence Level upper limit on \( \mathcal{B}(H \rightarrow \text{LLPs}) \) is derived and shown in Fig.~\ref{fig:ML-LLPs-Limits}a).
In the case of Type I and Type II signal yields having a floating ratio  \( \epsilon_V \) with an allowed range of $10^{-6}$ and $100$, a one-dimensional 95\% Confidence Level upper limit on \( \mathcal{B}(H \rightarrow \text{LLPs}) \) is derived and shown in Fig.~\ref{fig:ML-LLPs-Limits}b). In the fixed ratio case, the upper limit results have significantly smaller uncertainties than the floating ratio case. The best upper limit result in the fix ratio case is \(1.2 \times 10^{-6}\) on \( \mathcal{B}(H \rightarrow \text{LLPs}) \) with a statistics of \(4 \times 10^6\) Higgs bosons.

\begin{figure}[t]
\centering
\includegraphics[width=0.9\textwidth]{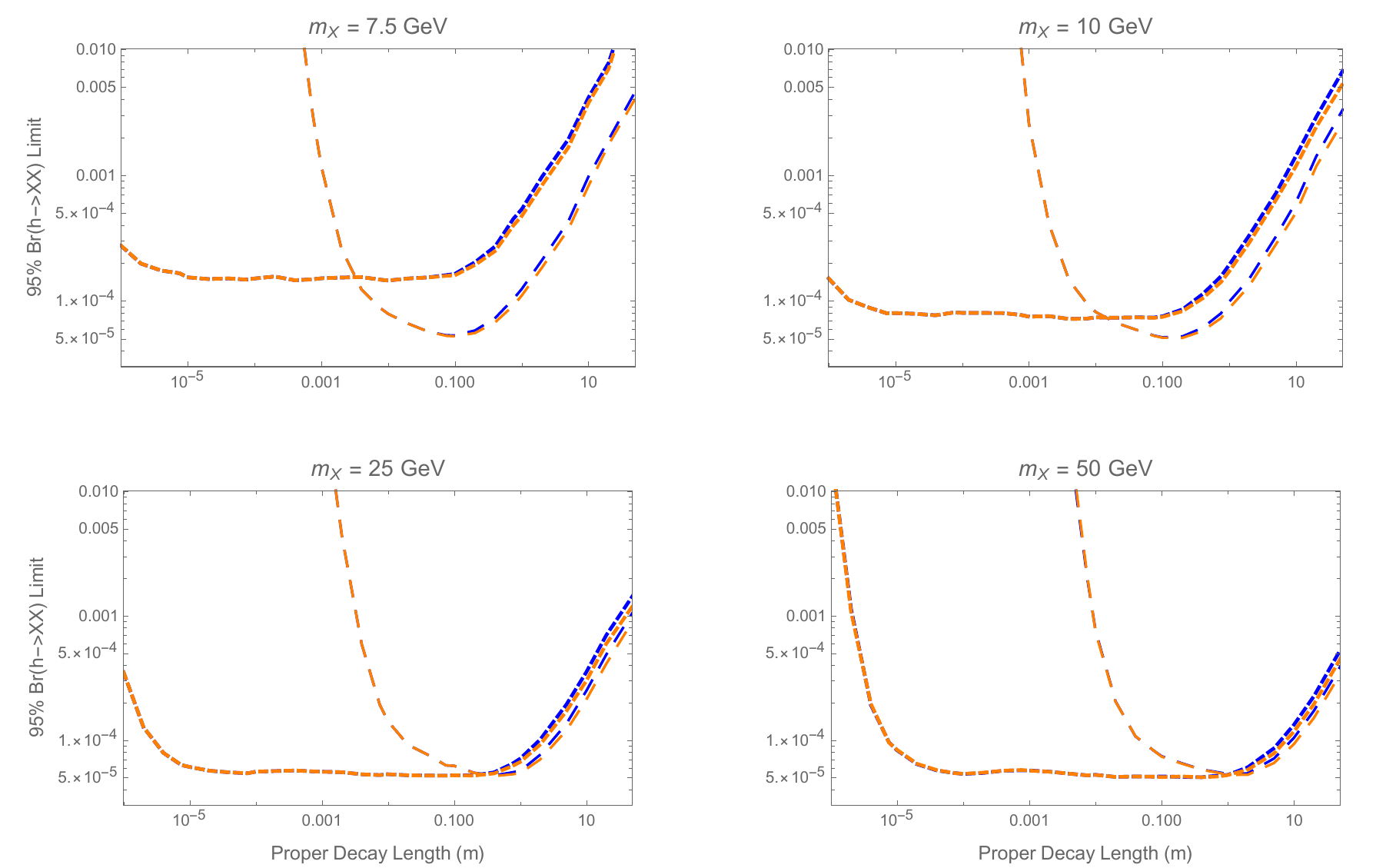}
\caption{Projected $95\%$ $h \to X X$ branching ratio limits as a function of proper decay length for a variety of $X$ masses.
Blue lines are for CEPC and orange lines are for FCC-ee, and where only one is visible they overlap.
The larger dashes are the `long lifetime' analysis and the smaller dashes are the `large mass' analysis.
Taken from Ref.~\cite{Alipour-Fard:2018lsf}.
}
\label{fig:ND-HiggsD-newS}
\end{figure}

Ref.~\cite{Alipour-Fard:2018lsf} investigates long-lived scalar particles produced from exotic Higgs-boson decays at the CEPC and FCC-ee.
The signal process is Higgsstrahlung $e^- e^+ \to  h Z$ at $\sqrt{s} = 250$ GeV, followed with $h \to X X$ and $Z\to (\ell^- \ell^+)$, where $\ell=e, \mu$ and $X$ is the long-lived new scalar boson.
$X$ is assumed to be produced at the IP and is required to decay inside the inner tracker into a pair of quarks, leading to displaced hadronic final states.
The $X$ particle is required to decay at a position with a distance larger than 3 cm (\SI{5}{\micro\metre})
to the IP, for the ``long lifetime'' (``large mass'') analysis.
Further, this displacement distance is required to be within the outer radius of the tracker which is 1.81 (2.14) m for the CEPC (FCC-ee) detector.
The dominant SM background processes, $e^-e^+ \to h Z \to (b\bar{b}) (\ell^+\ell^-)$ and $e^- e^+ \to Z Z \to (b\bar{b}) (\ell^+\ell^-)$, are investigated and selection cuts are imposed in order to eliminate such background.
The numbers of Higgs bosons produced at both CEPC and FCC-ee are considered to be $1.1 \times 10^6$.
In addition to forecasting sensitivities to the branching ratio of $h \to XX$ as shown in Fig.~\ref{fig:ND-HiggsD-newS}, results are also interpreted in the parameter space of theory models including the Higgs-portal Hidden Valley model and various incarnations of neutral-naturalness models.

Ref.~\cite{Cheung:2019qdr} studies displaced-vertex signatures of scalar LLPs pair-produced from exotic Higgs decays at the CEPC and FCC-ee with CM energy $\sqrt{s} = 240$ GeV and an integrated luminosity $\mathcal{L}_h =$ 5.6 ab$^{-1}$.
These charge-neutral LLPs decay into a pair of leptons or quarks at the partonic level.
Two theoretical models are investigated: a Higgs-portal model and a neutral-naturalness model.
These two models feature two representative mass ranges for scalar LLPs, corresponding to different characteristic signatures at colliders.

The Higgs-portal model includes a very light scalar boson, $h_s$, in the sub-GeV mass regime, stemming from a singlet scalar field appended to the SM.
Such a light scalar LLP decays into a pair of muons or pions, giving rise to a distinctive signature of collimated muon-jet or pion-jet, thanks to the sub-GeV mass. 
Thus, for this model, the signal process is $h \to h_s h_s, h_s \to \mu^- \mu^+, \pi^- \pi^+$ where the SM Higgs boson $h$ is produced in Higgsstrahlung process.
For the dimuon decay of $h_s$, displaced muons are detected in either the inner tracker (IT) detector or muon spectrometer (MS).
Further, for $h_s$ decays into a pair of charged pions, displaced vertices are considered to be reconstructed in the IT, HCAL, or MS.
The background is assumed to be negligible after event selections on the opening angle of the LLP decay products are imposed, and $95\%$ C.L.~sensitivity reaches are presented in terms of three-signal-event contour curves in the plane spanned by the $h_s$ mass and the Higgs bosons mixing angle $\theta$, for two benchmark choices of the new scalar-single field's vacuum expectation value (vev) $\langle \chi \rangle=10, 100$ GeV.

\begin{figure}[t]
\centering
    \includegraphics[width=0.49\textwidth]{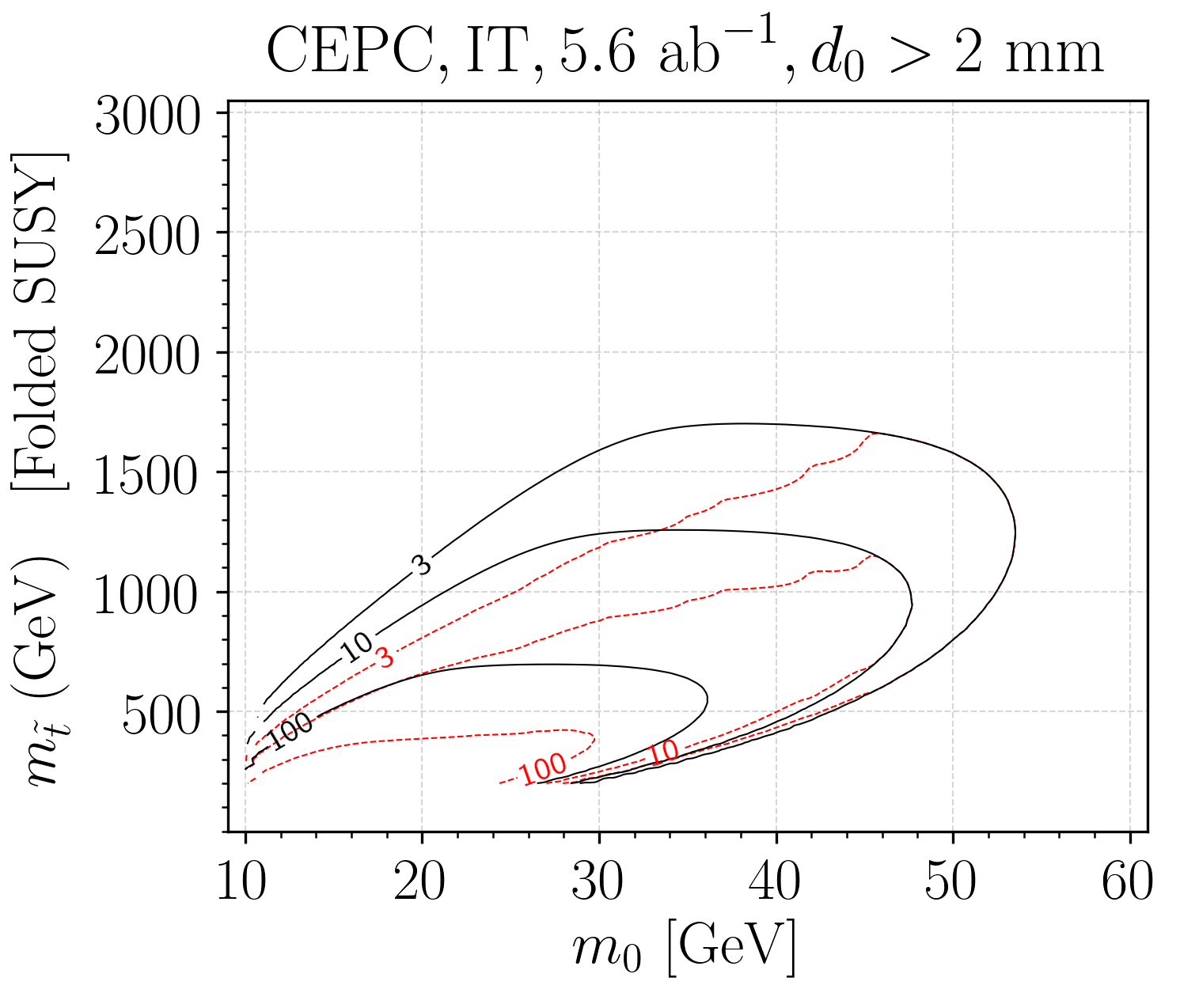}
\caption{Sensitivity reaches of log10$(N_{\text{signal}})$ at the CEPC
for the Folded SUSY model.
The black(red) curves correspond to $\kappa = \kappa_{\text{max}} (\kappa = \kappa_{\text{min}})$.
Taken from Ref.~\cite{Cheung:2019qdr}.
}
\label{fig:sensitivity_G_bbbar}
\end{figure}

On the other hand, the neutral-naturalness model, e.g.~folded supersymmetry, predicts the lightest mirror glueball $0^{++}$ of mass $\mathcal{O}(10)$ GeV, leading to displaced decays with a large transverse impact parameter because of the relatively large mass.
The mirror glueball in the mass range of $\mathcal{O}(10)$ GeV dominantly decays into a pair of $b$-jets, which is taken to be the signal channel.
Thus, for this model, the signal process is $h \to 0^{++} 0^{++}, 0^{++} \to b\bar{b}$.
Two major background processes are taken into account: $e^- e^+ \to Z Z \to (\ell^+ \ell^-, j j) (b \bar{b})$ and $e^- e^+ \to Z h \to (\ell^+ \ell^-, j j) (b \bar{b})$, in which the $b \bar{b}$ pair comes from prompt $Z$-boson's or SM Higgs boson's decay. 
Sensitivity reaches are shown in Fig.~\ref{fig:sensitivity_G_bbbar} in terms of the contour curves with different numbers of signal events, in the $(m_0,m_{\tilde{t}})$ plane, where $m_{\tilde{t}}$ is the stop mass, for two possible parameterizations of $\kappa$, which is a parameter taking into account the effect of
the glueball hadronization and nonperturbative mixing
effects between the excited glueball states and the SM Higgs boson.

\begin{figure}[t]
\centering
\adjustbox{valign=c}{\includegraphics[scale=0.36]{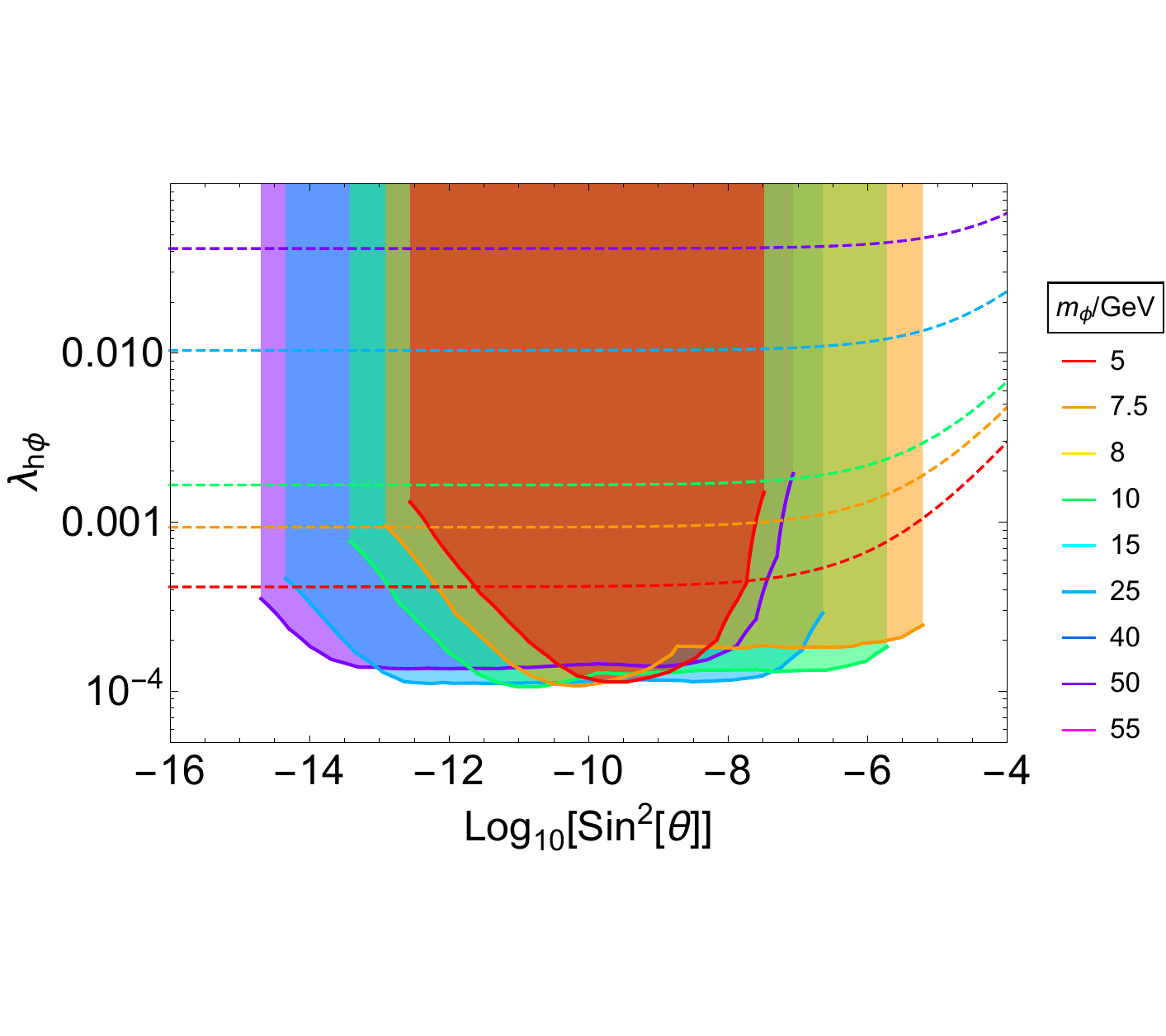}}
\adjustbox{valign=c}{\includegraphics[scale=0.42]{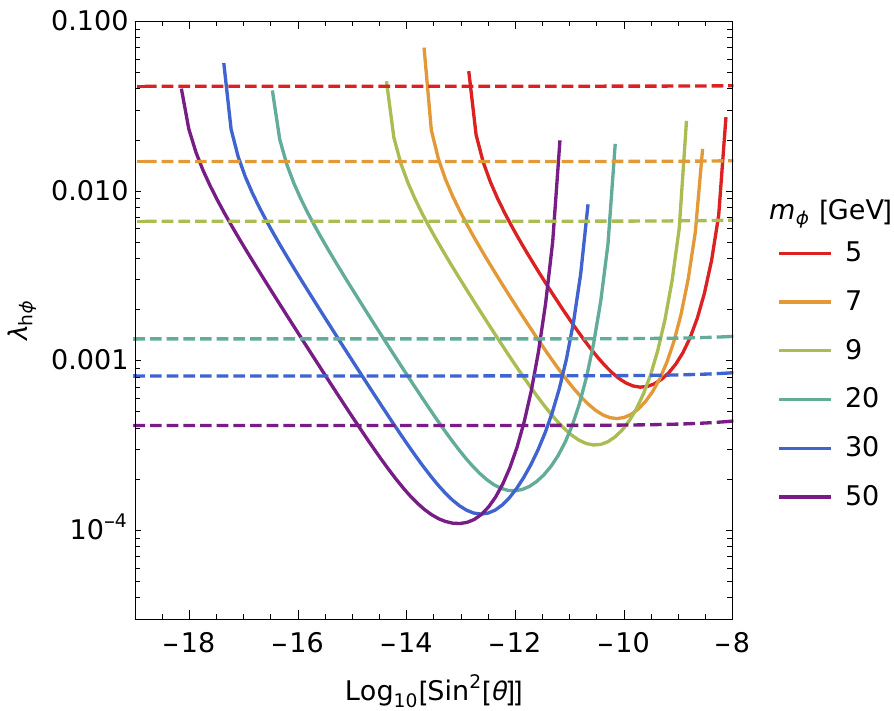}}
\caption{Left: bounds on $\lambda_{h\phi}$ and $\sin^2 \theta$ for various singlet masses arising from searches for displaced jets in Higgs decays at the FCC-ee with $\sqrt{s} =$ 240 GeV and integrated luminosity $\mathcal{L}_h =$ 5 ab$^{-1}$;
the dashed lines show the upper naturalness limit $\lambda_{h\phi}^{\rm{max}}=m_\phi^2/v^2 + 4\pi m_{\phi} \sin{\theta}/v$. 
Right: bounds on $\lambda_{h\phi}$ and $\sin^2 \theta$ for various singlet masses arising from searches for delayed jets in Higgs decays; the dashed lines show the upper naturalness limit $\lambda_{h\phi}^{\rm max}$ of for each mass.
Since physics is the same, sensitivities can be interpreted as the results at the CEPC.Taken from Ref.~\cite{Fuchs:2020cmm}.}
\label{fig:FDs-h2phiphi-dispDelayJets}
\end{figure}

Ref.~\cite{Fuchs:2020cmm} computes collider sensitivities to long-lived singlet scalar particles produced from SM Higgs-boson decays, $h \to \phi \phi$, considering signatures of invisible decays, displaced and delayed jets, and coupling fits of untagged decays. 
Results from the searches of displaced and delayed jets are shown in Fig.~\ref{fig:FDs-h2phiphi-dispDelayJets} for FCC-ee with $\sqrt{s} =$ 240 GeV and integrated luminosity $\mathcal{L}_h =$ 5 ab$^{-1}$.
Results for invisible decays are also given in this study.
Since physics is the same, sensitivities can be interpreted as the results at the CEPC.

\begin{figure}[t]
\begin{center}
\begin{overpic}[width=0.45\linewidth]{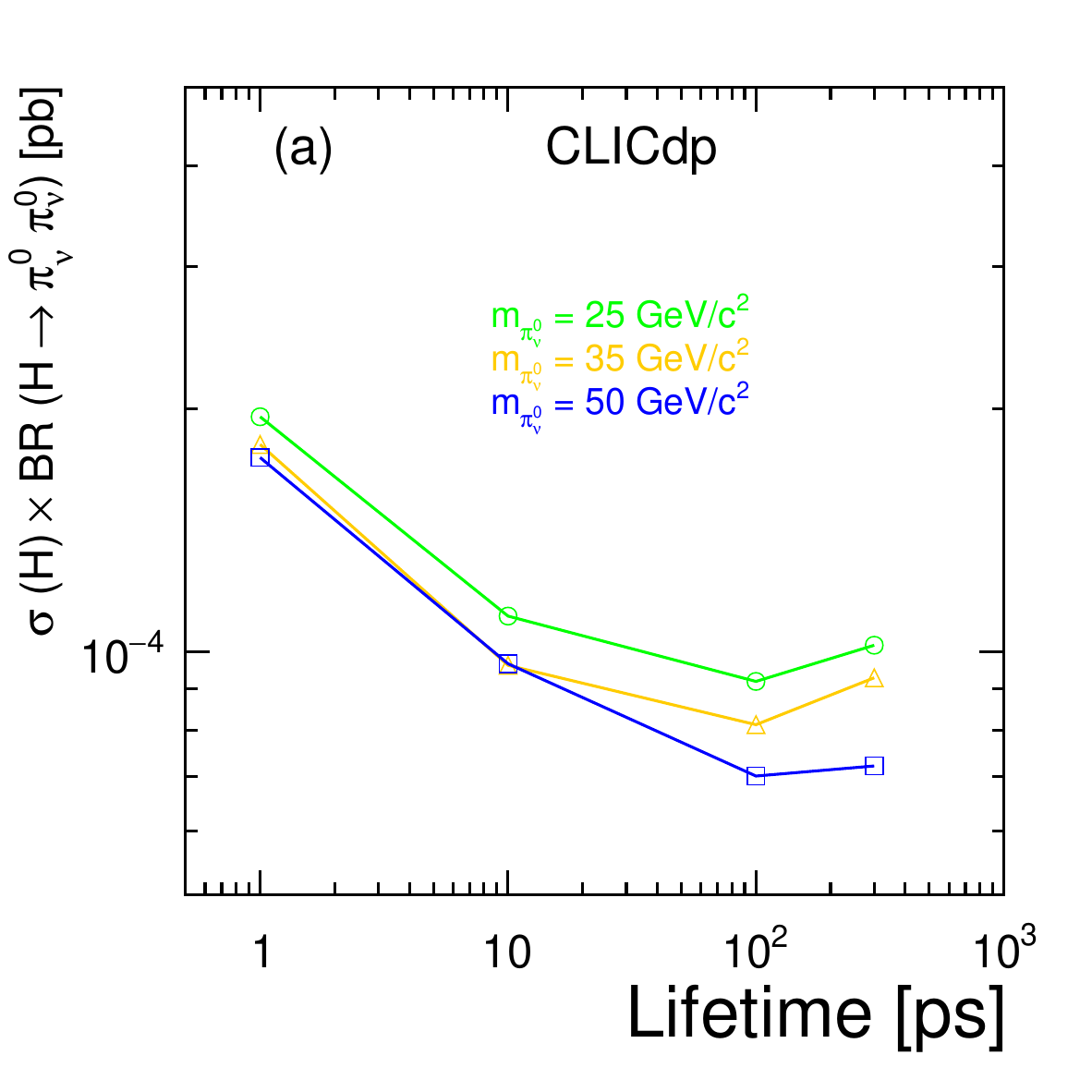}
\put(45,80){\scriptsize$\sqrt{s} = 350$ GeV}%
\end{overpic}
\begin{overpic}[width=0.45\linewidth]{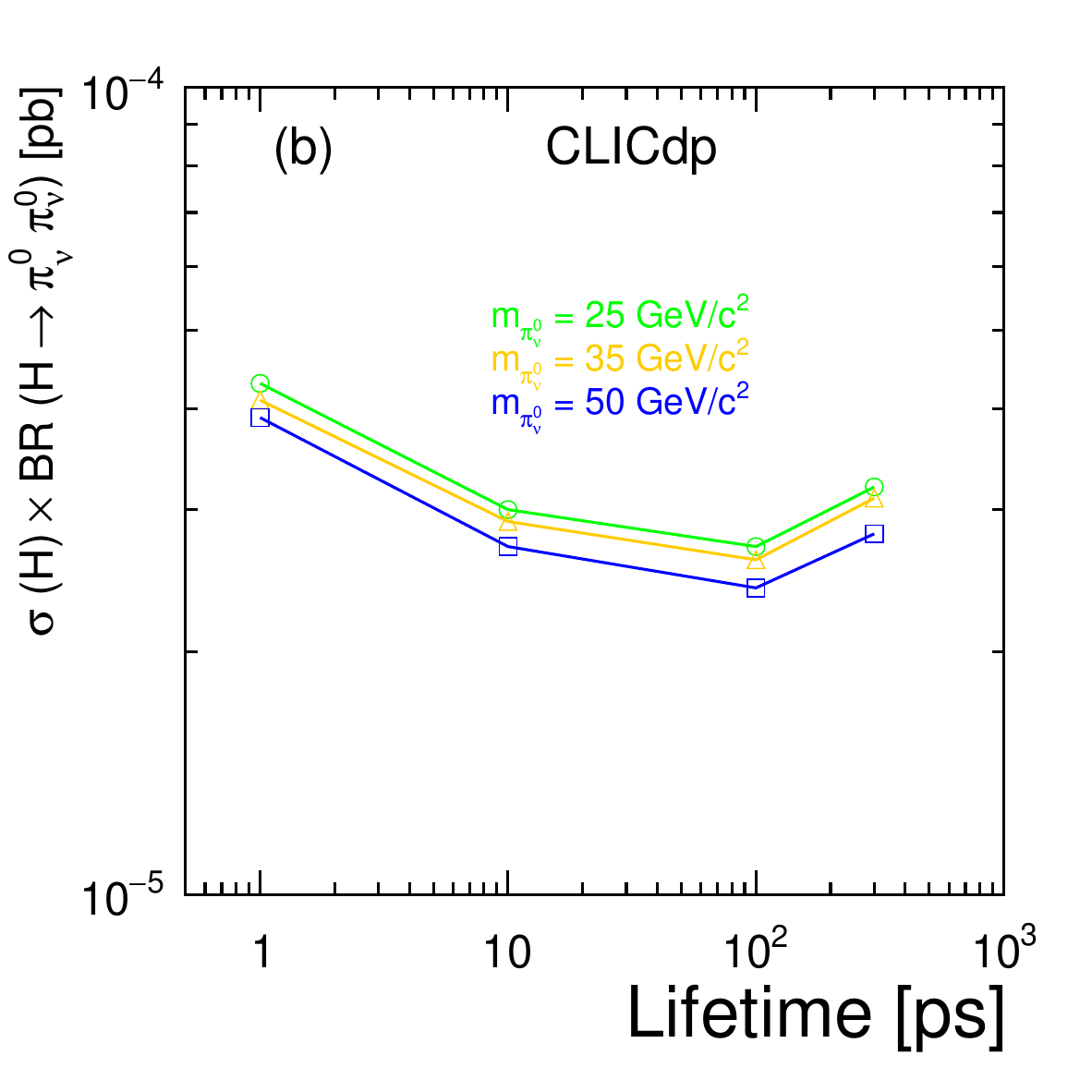}
\put(45,80){\scriptsize$\sqrt{s} = 3$ TeV}%
\end{overpic}
\end{center}
\caption{Expected 95\% CL cross-section upper limits on the $\sigma(H) \times BR(H \rightarrow \pi^0_v \pi^0_v)$, within the model~\cite{Strassler:2006ri}, for three different $\pi^0_v$ masses: 25~GeV (green), 35~GeV (yellow), 50~GeV (blue), as a function of $\pi^0_v$ lifetime for $\sqrt{s}=350$~GeV (a) and $\sqrt{s}=3$~TeV (b). 
Since physics is the same, sensitivities can be interpreted as the results at the CEPC with upgraded CM energies.Taken from Ref.~\cite{Kucharczyk:2022pie}.}
\label{fig:upl}
\end{figure}

\bigskip
\noindent\underline{Hidden valley particles:}

Ref.~\cite{Kucharczyk:2022pie} works on the sensitivity reach to massive LLPs using the ILD detector at CLIC with $\sqrt{s} = 350$ GeV and 3 TeV, and an integrated luminosity of 1 ab$^{-1}$ and 3 ab$^{-1}$, respectively.
The study is in the context of the Hidden Valley model.
In this work, two long-lived Hidden-Valley particles are pair produced from the SM Higgs boson decays, and subsequently decay into $b$-quarks, i.e.~$h \to \pi_V^0 \pi_V^0 \to (b \bar{b}) (b \bar{b})$, providing four $b$-jets in the final state.
At $\sqrt{s} =$ 350 GeV, Higgs bosons are dominantly produced from the Higgsstrahlung process ($e^- e^+ \to Z h$), while at $\sqrt{s} =$ 3 TeV, the dominant production channel is the $WW$-fusion ($e^- e^+ \to \nu \bar{\nu} h$).
Signal samples with $\pi_v^0$ lifetimes from 1 to 300 ps, masses between 25 and 50 GeV, and the parent Higgs mass of 126 GeV, are generated, while background samples of $q \bar{q}$, $q \bar{q} \nu \bar{\nu}$, $q \bar{q} q \bar{q}$, $q \bar{q} q \bar{q} \nu \bar{\nu}$ are generated, with additional samples of $t \bar{t}$ and $WWZ$ for $\sqrt{s} =$ 350 GeV.
The observables based on reconstructed displaced vertices are input for performing multivariate analysis and reducing the SM background.
Sensitivity results are presented for the production cross-section ($\sigma(h) \times {\rm BR} (h \to \pi_v^0 \pi_v^0)$) as a function of the LLP's lifetime for three different $\pi_v^0$ masses: 25, 35, 50 GeV.
We reproduce them in Fig.~\ref{fig:upl}.

\noindent\underline{Dark photons:}
\begin{figure}[t]
\centering
\includegraphics[width=0.7\textwidth]{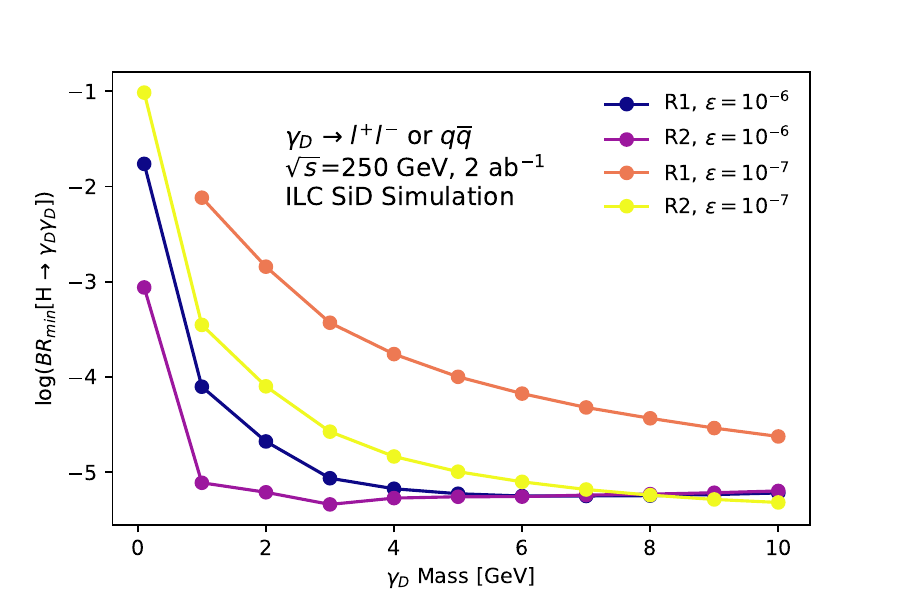}
\caption{The minimum branching ratio $H \rightarrow \gamma_{D} \gamma_{D}$ to which SiD will be sensitive for $\sqrt{s} = 250$ GeV and 2 ab$^{-1}$, when both leptonic and hadronic decays are reconstructed within the regions $R1$ and $R2$, for $\epsilon=10^{-6},10^{-7}$.
Since physics is similar, sensitivities can be interpreted as the results at the CEPC with the same CM energy and integrated luminosity.
Taken from Ref.~\cite{Jeanty:2022cwr}.
}
\label{fig:NDs-h2dpdp}
\end{figure}

Ref.~\cite{Jeanty:2022cwr} studies sensitivity to long-lived dark photons produced in Higgsstrahlung events via the Higgs portal, $h \to \gamma_D \gamma_D$, with the Silicon Detector (SiD) at ILC. 
The considered signal process is $e^- e^+ \to Z h \to (q \bar{q})\, (\gamma_D \gamma_D), \gamma_D \to \ell^- \ell^+ / q \bar{q}$ at $\sqrt{s} = 250$ GeV with an integrated luminosity of 2 ab$^{-1}$ for each of two polarization cases $e_L^- e_R^+$ and $e_R^- e_L^+$ at nominal ILC TDR polarization fractions, and 80\% electron polarization and 30\% positron polarization, respectively.
The following SM background processes are considered and generated: 2-fermion states $e^- e^+ \to f \bar{f}$, 3-fermion states $e \gamma \to e Z, \nu W \to 3 f$, and 4-fermion states $e^- e^+ \to WW, e \nu W, Z Z, e e Z, \nu \bar{\nu} Z \to 4 f$.
Two requirements for fiducial regions ``$R1$'' and ``$R2$'' are taken into account.
It is found that the requirement of a displaced vertex formed from tracks with measurably large impact parameter in the clean ILC event environment likely suppresses background events to a negligible level.
Assuming no background with the selections outlined in the study, baseline sensitivities on the minimal branching ratio $h \to \gamma_D \gamma_D$ as a function of $\gamma_D$ mass are presented in Fig.~\ref{fig:NDs-h2dpdp} for several choices of the kinetic mixing parater $\epsilon$.
Additionally, this study also presents the first full simulation of LLPs for SiD.
Since physics is similar, sensitivities can be interpreted as the results at the CEPC with the same CM energy and integrated luminosity.

\begin{figure}[t]
\centering
\includegraphics[width=0.49\textwidth]{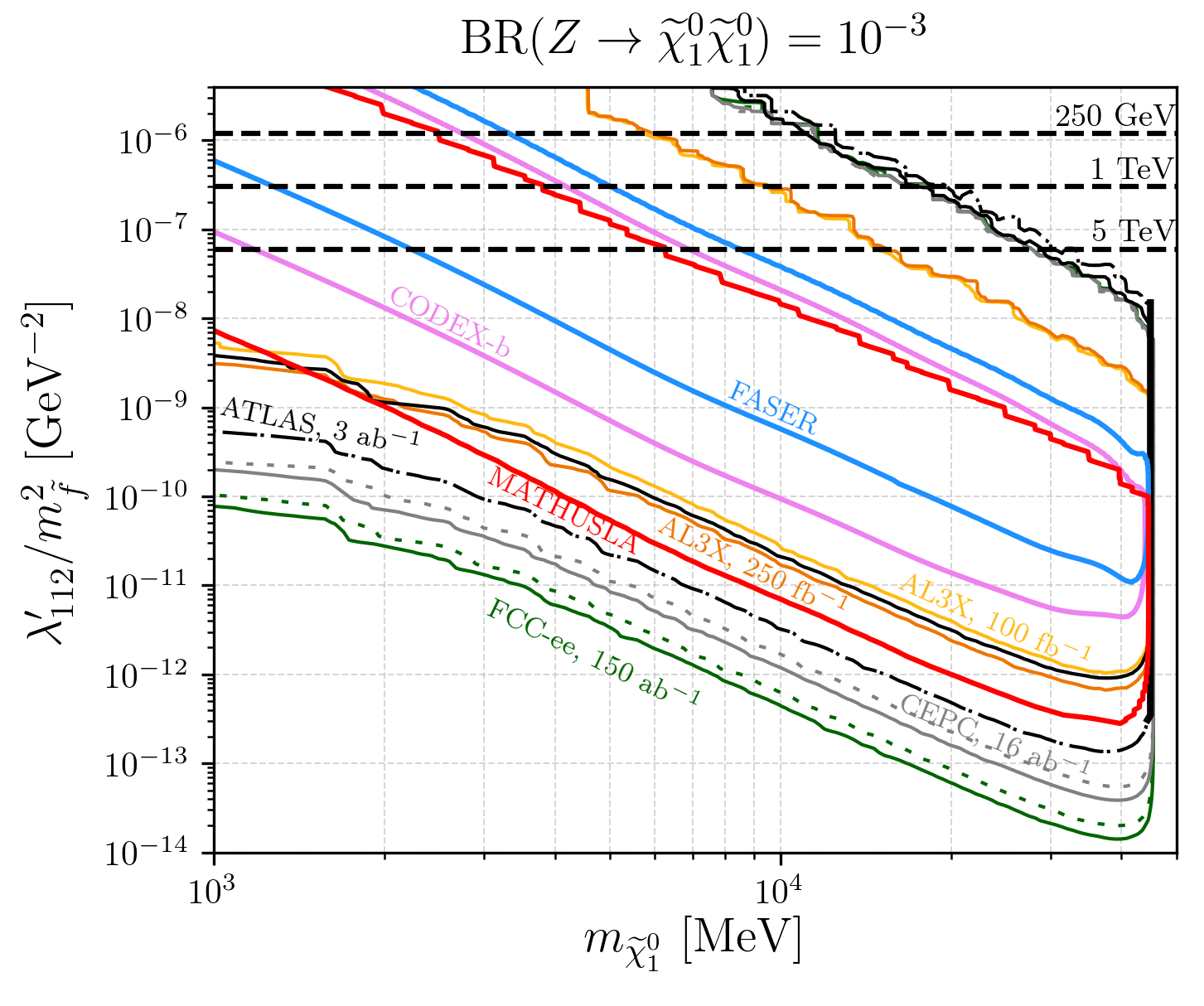} 
\vspace{0.8cm} 
\includegraphics[width=0.49\textwidth]{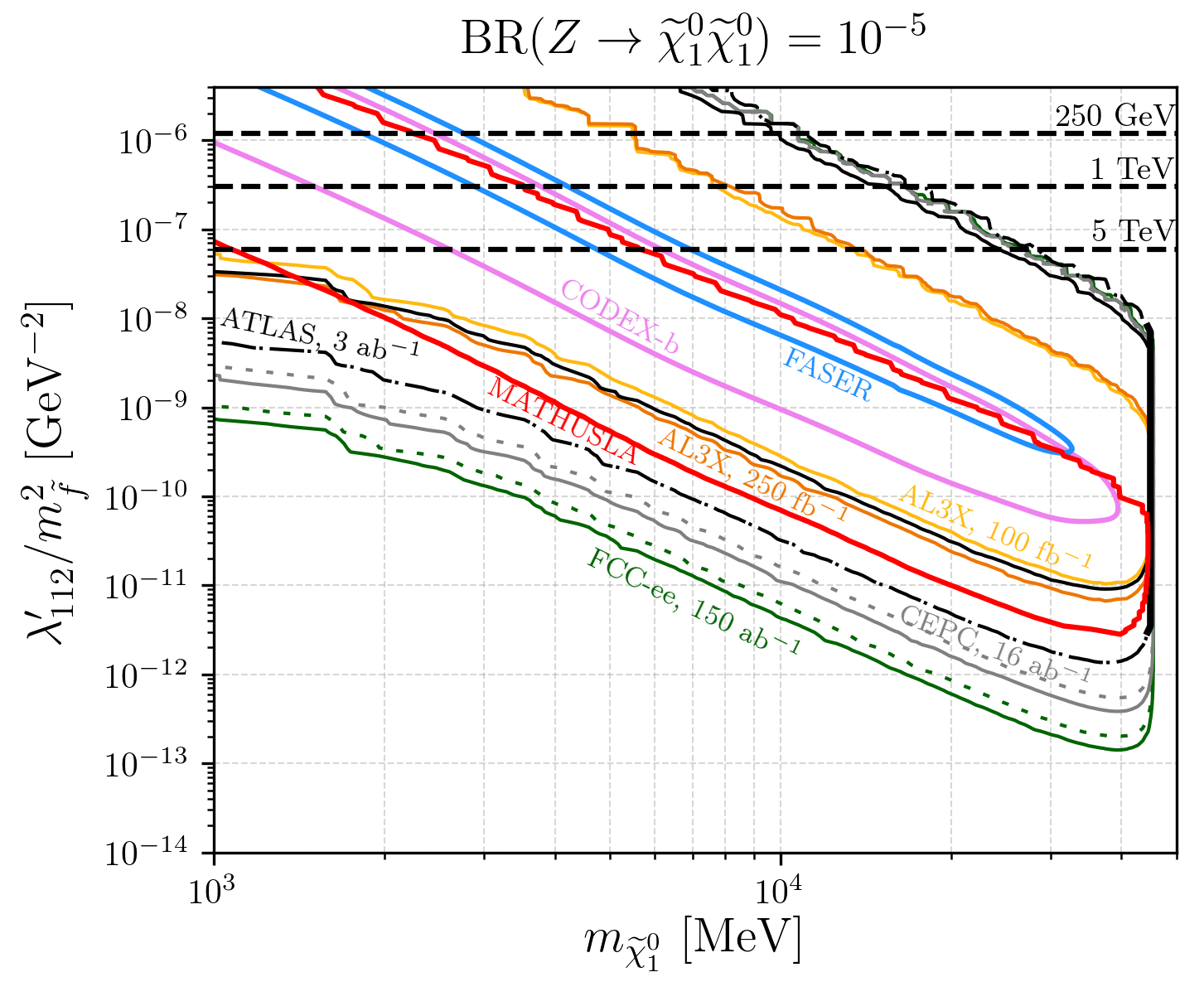}
\caption{The sensitivity estimate of the CEPC (grey) and the FCC-ee (green) presented in the 2D plane of $\lambda'_{112}/m_{\tilde f}^2$ vs. $m_{\tilde{\chi}_1^0}$ assuming BR($Z\rightarrow \tilde{\chi}_1^0 \tilde{\chi}_1^0$) = $10^{-3}$ (left) and $10^{-5}$ (right), respectively.
The solid contour curves correspond to three decay events in the fiducial volume when considering all decay modes of $\tilde{\chi}_1^0$, while the dashed lines include only visible/charged decay modes ($K^{(*)\pm} e^\mp$, $e^-us$ or $e^+\bar{u}\bar{s}$).
The estimates for experiments at the LHC: AL3X, CODEX-b, FASER and MATHUSLA, are reproduced from Refs.~\cite{Helo:2018qej,Dercks:2018wum}.
The ATLAS results correspond to HL-LHC for $\sqrt{s}=14$ TeV and 3 $\iab$ integrated luminosity.
The black horizontal dashed lines correspond to the current RPV bounds on the single coupling $\lambda^\prime_{112}$~\cite{Kao:2009fg} for three different degenerate sfermion masses $m_{\tilde{f}}=250$ GeV, 1 TeV, and 5 TeV as labelled.
Taken from Ref.~\cite{Wang:2019orr}.
}
\label{fig:NDs-Z2chi1chi1-RPVSUSY}
\end{figure}

\subsubsection{$Z$-boson decays}

Future lepton colliders such as the CEPC and FCC-ee would run as high-luminosity $Z$-boson factories, which offer a unique opportunity to study LLPs from rare $Z$-boson decays, c.f. Section~\ref{subsubsec:Z2LLPs}.
Recent studies are summarized as follows.

Ref.~\cite{Wang:2019orr} considers the physics scenario where the long-lived lightest neutralino pair $\neutralino1 \neutralino1$ is produced from $Z$-decays in the context of the R-parity-violating supersymmetry model.
The lightest neutralino $\neutralino1$ is dominantly bino-like with small Higgsino components.
The study focuses on the $\lambda'_{ijk} L_i \cdot Q_j \bar{D}_k$ operators, and $\lambda'_{112}$ $L_1 \cdot Q_1 \bar{D}_2$ is chosen to be the only nonvanishing RPV operator, which leads to the lightest neutralino decays to SM particles via a sfermion exchange.
For $m_{\tilde{\chi}_1^0}<m_Z/2$ and small $\lambda'$ couplings, the lightest neutralino becomes long-lived and decays after having travelled a macroscopic distance.
The signal process is $Z \to \neutralino1 \neutralino1,\,\,\, \neutralino1 \to e^{\mp} K^{(*)\pm} / e^{\mp} j j $ at $\sqrt{s} = 91.2$ GeV.
The fiducial volume of the detectors is considered as the inner detector consisting of the vertex detector and the tracker. This choice is conservative and ensures that the electrons produced from the neutralino decays could be reconstructed. 
The signal events require at least one neutralino decaying inside the inner detector, while the other could decay either inside or outside the inner detector.\par 
The background is assumed to be negligible, and sensitivity reaches are presented in terms of three-signal-event contour curves on the model parameters.
Fig.~\ref{fig:NDs-Z2chi1chi1-RPVSUSY} shows the sensitivity estimate of the CEPC (grey) and the FCC-ee (green) in the 2D plane of $\lambda'_{112}/m_{\tilde f}^2$ vs.~$m_{\tilde{\chi}_1^0}$  for two different benchmark values of BR($Z\rightarrow \tilde{\chi}_1^0 \tilde{\chi}_1^0$), respectively.
The analyses indicates that when assuming Br$(Z\rightarrow \tilde{\chi}_1^0\tilde{\chi}_1^0) = 10^{-3}$ and $m_{\tilde{\chi}_1^0} \sim 40$ GeV, the model parameter $\lambda'_{112} / m^2_{\tilde{f}}$ can be probed down to as low as $\sim 1.5 \times 10^{-14}$ ($3.9 \times 10^{-14}$) GeV$^{-2}$ at the FCC-ee (CEPC) with CM energy $\sqrt{s} = 91.2$ GeV and 150 (16) $\iab$ integrated luminosity.
Sensitivity results are compared with the projected sensitivity reaches of the ATLAS experiment at the HL-LHC~\cite{ATLAS:2008xda} and the proposed LHC experiments with far detectors (AL3X~\cite{Gligorov:2018vkc}, CODEX-b~\cite{Gligorov:2017nwh}, FASER~\cite{Feng:2017uoz}, and MATHUSLA~\cite{Curtin:2018mvb}).

\begin{figure}[t]
\centering
\includegraphics[width=0.75\textwidth]{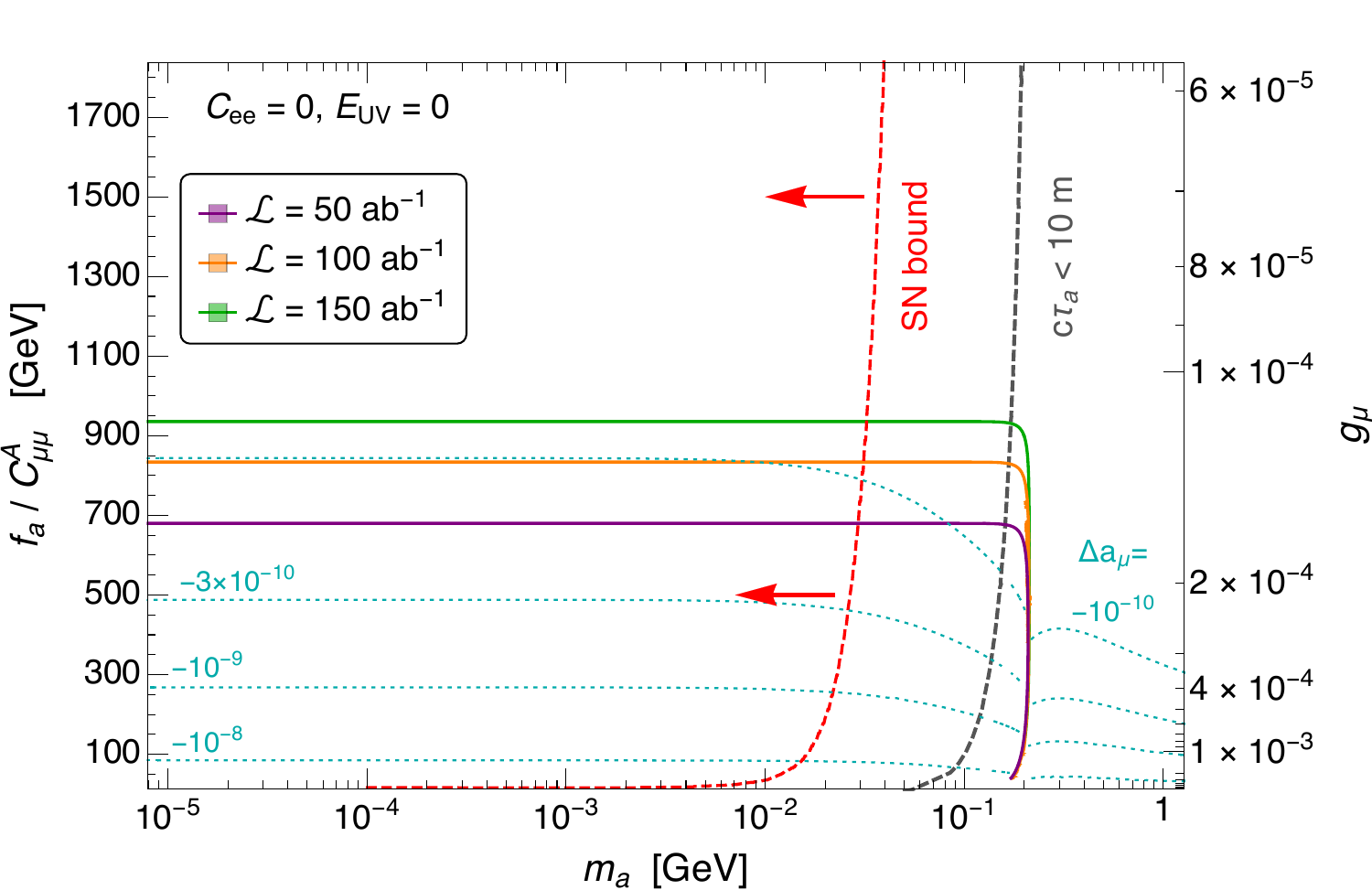}
\caption{Prospected CEPC/FCC-ee 95\% CL exclusion on the $(m_a,\,f_a/C^A_{\mu\mu})$ plane for a muonic ALP ($C^A_{ee}=0$, $E_\text{UV}=0$) with different assumptions for the integrated luminosity $\mathcal{L}$, as indicated. On the right side of the dashed grey line the proper decay length of the ALP
is $c\tau_a <10$~m. The region to the left of the red dashed line is excluded by SN1987A data, according to the analysis in \cite{Croon:2020lrf}. The dotted cyan contours show the ALP contribution to the anomalous magnetic moment of the muon, $\Delta a_\mu \equiv (g-2)_\mu/2$.
Taken from Ref.~\cite{Calibbi:2022izs}.
}
\label{fig:NDs-Z2LLa}
\end{figure}

Ref.~\cite{Calibbi:2022izs} probes ALPs coupled to charged leptons in leptonic decays of the $Z$-boson at CEPC and FCC-ee.
The ALPs are assumed to have very long lifetime, travel through the main detector, and behave as missing energy.
The signal process is $e^- e^+ \to Z^{(*)} \to \ell^- \ell^+ a$ at $\sqrt{s} = 91$ GeV.
This study analyzes the signal process of $e^- e^+ \to \mu^- \mu^+ a$, and considers background sources including $\tau^+ \tau^-$, $\mu^+ \mu^- \nu \bar{\nu}$, and $\mu^+ \mu^- \gamma$ (where $\gamma$ gets undetected).
Fig.~\ref{fig:NDs-Z2LLa}, extracted from Ref.~\cite{Calibbi:2022izs}, presents sensitivity reaches in terms of the contour curves in the $f_a/C^A_{\mu\mu}$ vs.~$m_a$ plane for various integrated luminosities of 50, 100 and 150 $\iab$.
Limits based on the ALP contribution to the anomalous magnetic moment of the muon are also shown together.
The sensitivity reaches for the ALP with a large coupling to photons and negative $C^A_{\mu\mu}$ are also derived in this study.

\subsubsection{Supersymmetry (SUSY)}

Ref.~\cite{Heisig:2012px} reviews studies on LLPs at the ILC and CLIC in the context of SUSY models.
The long-lived next-to-lightest supersymmetric particle (NLSP) is the stau $\tilde{\tau}$, and the lightest supersymmetric particle (LSP) is the gravitino $\tilde{G}$ or axino $\tilde{a}$.
The signal includes both the 2-body decay process $\tilde{\tau} \to \tau \tilde{G} / \tau \tilde{a}$ and 3-body decay process $\tilde{\tau} \to \gamma \tau  \tilde{G} / \gamma \tau \tilde{a}$.

Ref.~\cite{Martyn:2007mj} studies the SUSY scenario at the ILC where the gravitino $\tilde{G}$ is the LSP and a charged stau $\tilde{\tau}$ is the long-lived, metastable NLSP.
The signal process is $\tilde{\tau} \to \tau \tilde{G}$.
In the analyses, stau detection and measurement principle consists of several steps: identify a $\tilde{\tau}$ and determine its mass from kinematics; follow the track until it is trapped inside the detector; observe the stopping point until a decay $\tilde{\tau} \to \tau \tilde{G}$ is triggered by a large energy release uncorrelated to beam collisions; record the decay time to determine the $\tilde{\tau}$ lifetime; finally, measure the $\tau$ recoil energy to get the gravitino mass.
The case study assumes the ILC running at $\sqrt{s} =$ 500 GeV and an integrated luminosity $\mathcal{L} =$ 100 fb$^{-1}$.

Ref.~\cite{Ibarra:2006sz} investigates the prospects of observing lepton flavour violation at future $e^- e^+$ and $e^- e^-$ linear colliders in scenarios where the gravitino $\tilde{G}$ is the LSP, and the long-lived stau $\tilde{\tau}$ is the NLSP.
Since the NLSP can only decay gravitationally into gravitinos and charged leptons, the decay rate is very suppressed and the NLSP could traverse several layers of the vertex detector before decaying or even being stopped and trapped in it.
The signals of lepton flavor violation would consist of two heavily ionizing tracks owing to the long-lived staus accompanied by two or four charged leptons.
The signals consist of multilepton final states with two heavily ionizing charged tracks produced by the long-lived staus. 
The numerical analyses are performed at the ILC with $\sqrt{s} =$ 500 GeV and an integrated luminosity of 500 fb$^{-1}$ assuming the beams are unpolarized.
The sensitivity reaches to lepton flavor violation are presented and compared with the present and future constraints on lepton flavor violation stemming from the non-observation of rare leptonic decays.

\subsubsection{Vector-like leptons with scalar}

Ref.~\cite{Cao:2023smj} considers vector-like leptons (VLLs) $F^\pm$ as a simple extension to the SM, with an accompanying scalar particle $\phi$ at future electron-positron colliders.
Assuming $F^\pm$ and $\phi$ mix only with the first-generation SM leptons, the authors focus on CEPC with CM energy 240 GeV.
The scalar particle $\phi$ is long-lived; it is pair produced and subsequently decays into $e^- e^+$.
Employing the inner tracker for reconstructing the displaced vertex and applying appropriate event-selection cuts to suppress background from the SM $Z$- and Higgs bosons' prompt decays, the analysis finds good performance of CEPC for $m_\phi < 70$ GeV and $m_F< 1$ TeV.
Details of this study can be found in Section~\ref{sec:DM:long-lived VLL}.

\subsection{Studies with far detectors}
\label{subsec:studiesFDs}

It has been well accepted among the high-energy-physics community that at colliders such as LHC, besides the traditional MD installed at the IP, FDs, can also be constructed for operation away from the IP by up to $\mathcal{O}(100)$ m.
When the decay lengths of LLPs are very long (e.g.~$\lambda \gtrsim \mathcal{O}(100)$ m), they can have a sizable probability to travel through the MD, acting as missing energy.
In this case, a far detector could have a better chance to observe their decay processes if $\lambda$ of the LLPs falls roughly around its distance to the IP, and could reconstruct the information of time, position, energy, momentum, mass, etc.
Moreover, the large space between the FD and the IP allows for sufficient shielding which can effectively remove background sources.
Therefore, in principle, such far detectors can enhance the discovery potential for LLPs with very long decay lengths.

\subsubsection{Far detectors at hadron colliders}
\begin{table}[htbp]
\centering
\begin{tabular}{cccc}
\hline
\hline
                 &    ANUBIS          &        FASER   &     FASER2        \\
IP               &     ATLAS                 &        ATLAS              &         ATLAS     \\
int.~lumi.~[fb$^{-1}$]      &   3000                    &       150               &         3000       \\
\hline
                 &           FACET     &     MATHUSLA    &       AL3X       \\
IP               &         CMS             &           CMS           &        ALICE    \\
int.~lumi.~[fb$^{-1}$]       &      3000                &   3000                  &      100, 250               \\
\hline
                &      CODEX-b        &          MoEDAL-MAPP1    &       MoEDAL-MAPP2      \\
IP               &          LHCb            &             LHCb         &         LHCb            \\
int.~lumi.~[fb$^{-1}$]       &      300                &       30               &           300   \\
\hline
\hline
\end{tabular}
\caption{Summary table of the LLP FDs at the LHC, listing the associated interaction point and the projected integrated luminosity.
}
\label{table:LLPFD}
\end{table}

\begin{figure}[t]
\centering
\includegraphics[width=0.75\textwidth]{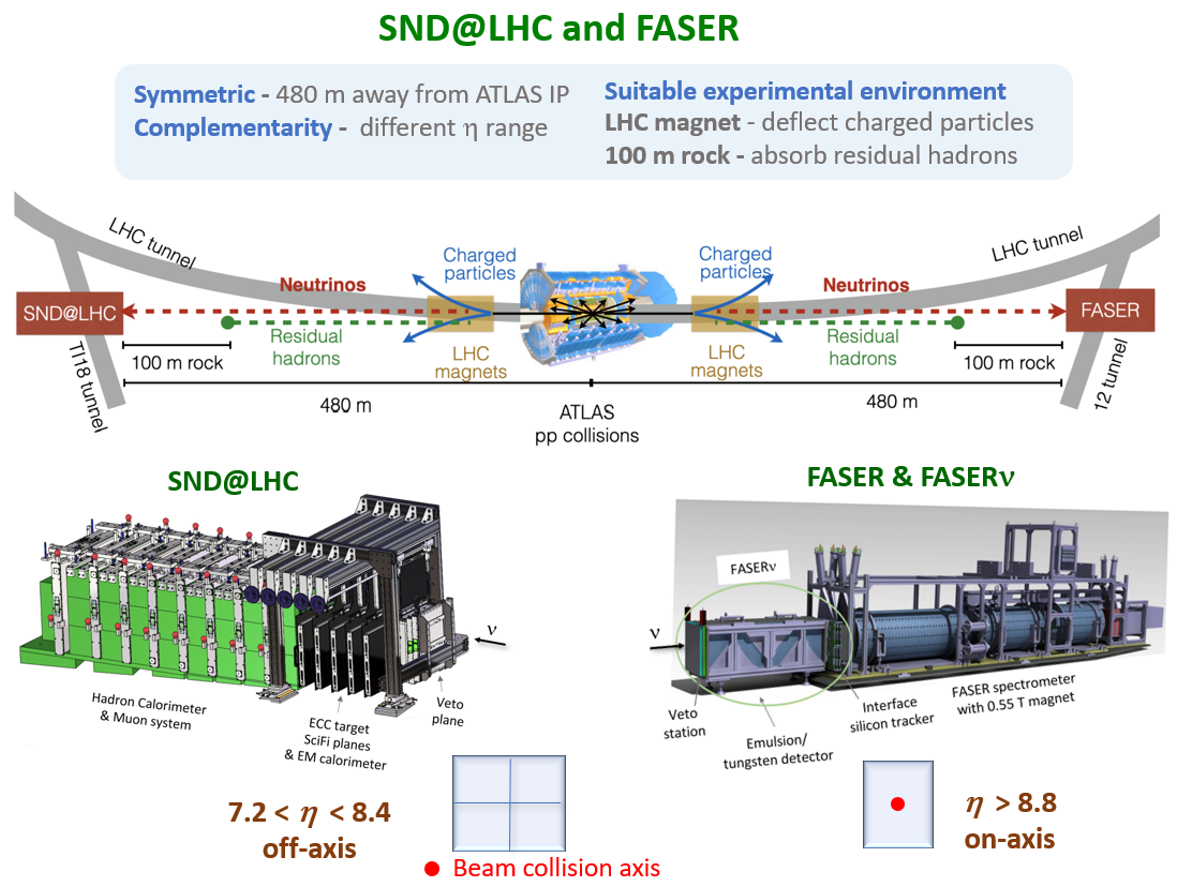}
\caption{Schematic pictures of the SND@LHC, FASER$\nu$, and FASER experiments.
Taken from Ref.~\cite{SNDatLHC:2022}.
}
\label{fig:SND-FASER}
\end{figure}

Two classes of FDs are currently operating or have been proposed at the LHC.
The first class is a series of detectors that are composed of dense materials such as tungsten or liquid argon as targets and mainly purposed for detecting high-energy active neutrinos originating from the LHC IPs through neutrino-nucleus deep inelastic scattering.
Among them, SND@LHC~\cite{SHiP:2020sos,SNDLHC:2022ihg} and FASER$\nu$~\cite{FASER:2019dxq,FASER:2020gpr} are collecting data during the LHC Run 3.
AdvSND~\cite{Feng:2022inv,MammenAbraham:2022xoc}, FASER$\nu$2~\cite{MammenAbraham:2020hex,Anchordoqui:2021ghd,Feng:2022inv}, and FLArE-10/100~\cite{Feng:2022inv,Batell:2021blf} would be further upgraded experiments and have been proposed to be running during the high-luminosity LHC (HL-LHC) era at the proposed Forward Physics Facility (FPF)~\cite{Feng:2022inv}.
These FPF experiments would be installed in the very forward direction on/off-axis, about 600 meters away from the ATLAS IP.

\begin{figure}[t]
\centering
\includegraphics[width=0.75\textwidth]{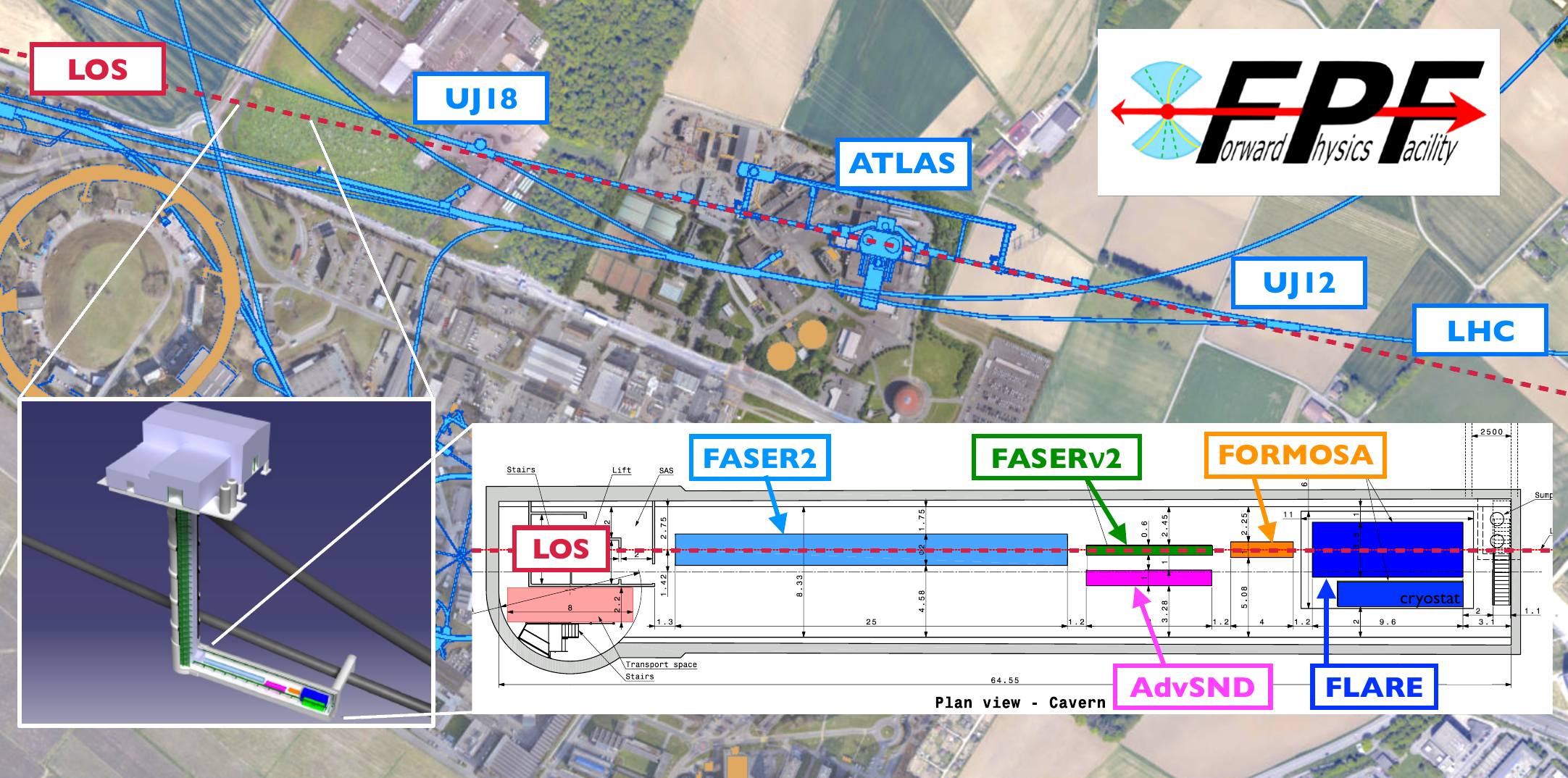}
\caption{The preferred location for the FPF, a proposed new cavern for the HL-LHC era.
The FPF will be 65~m-long and 8.5~m-wide and will house a diverse set of experiments to explore the many physics opportunities in the far-forward region.
Taken from Ref.~\cite{Feng:2022inv}.
}
\label{fig:FPF}
\end{figure}

Another type of FDs at the LHC aim primarily at searching for displaced decays of LLPs into charged final-state particles.
These experiments include and FASER and FASER2~\cite{Feng:2017uoz,FASER:2019aik,FASER:2022hcn}, MATHUSLA~\cite{Chou:2016lxi,Curtin:2018mvb,MATHUSLA:2020uve},  ANUBIS~\cite{Bauer:2019vqk}, AL3X~\cite{Gligorov:2018vkc}, FACET~\cite{Cerci:2021nlb}, CODEX-b~\cite{Gligorov:2017nwh,Aielli:2019ivi},  MoEDAL-MAPP1 and MAPP2~\cite{Pinfold:2019nqj,Pinfold:2019zwp}, which suggest to install auxilliary detectors at positions $\mathcal{O}(5-500)$ m away from the IPs of the ATLAS, CMS, or LHCb experiments.
For example, FASER is a small cylindrical detector installed right behind FASER$\nu$ and is currently operating during Run 3.
FASER2 would be the upgraded program of FASER, installed at FPF with a distance of 620 m from the ATLAS IP.
Alternatively, in one of the service shafts above the ATLAS IP, another detector called ANUBIS has been suggested to be constructed; it also has a cylindrical shape but faces vertically.
MATHUSLA has been proposed to be constructed about $\sim$ 100 m above the CMS IP, with a mostly empty decay volume monitored by trackers for reconstruction of LLP decays into charged particles.
In the forward direction of the CMS IP, FACET has been brought up to be placed surrounding the beam pipe.
In the vicinity of the ALICE IP, AL3X has been suggested.
Finally, for the LHCb IP, some far-detector proposals currently exist: CODEX-b, MoEDAL-MAPP1 and MAPP2.
For a summary of these detectors including their geometries and corresponding integrated luminosities, see e.g.~Refs.~\cite{DeVries:2020jbs,Cerci:2021nlb,Mao:2023zzk}.
We list these LLP FDs in Table~\ref{table:LLPFD} including their associated interaction point and the integrated luminosity.
In Fig.~\ref{fig:SND-FASER}, schematic pictures of SND@LHC, FASER$\nu$, and FASER, extracted from Ref.~\cite{SNDatLHC:2022}, are also shown, and in Fig.~\ref{fig:FPF} the location of the FPF is illustrated together with the setup of the experiments proposed to be hosted there, reproduced from Ref.~\cite{Feng:2022inv}.

\subsubsection{Proposed far detectors at lepton colliders}

\begin{figure}[t]
\centering
\includegraphics[width=0.5\textwidth]{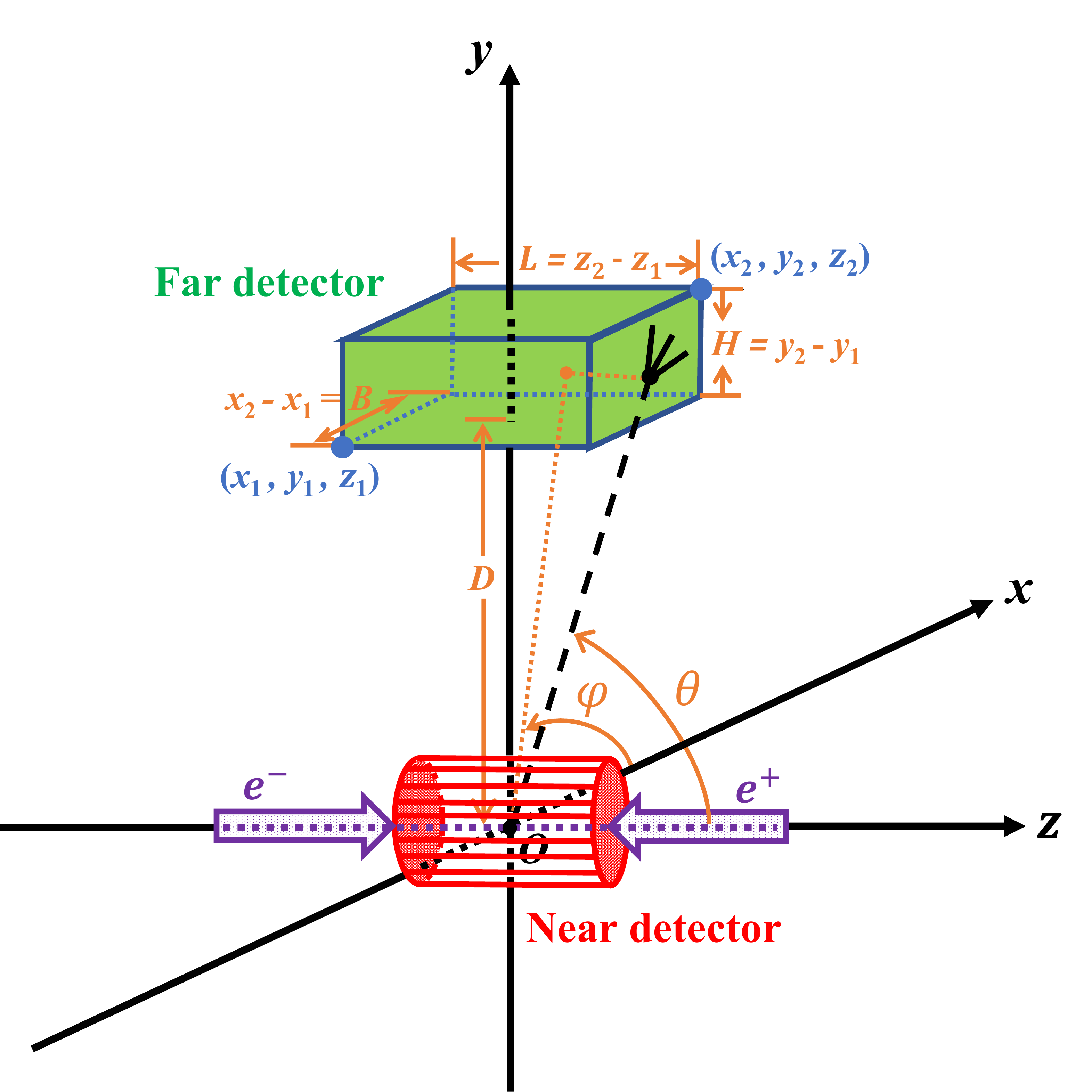}
\caption{The sketch displays the main detector (near detector, abbreviated as ND) and an example far detector.
The dashed line indicates that one LLP is produced at the IP, travels through the main detector and decay insider the far detector.
Taken from Ref.~\cite{Wang:2019xvx}.
}
\label{fig:FDsketchEEc}
\end{figure}

Besides at the LHC, FDs for detecting LLPs have also been proposed for operation at future $e^- e^+$ colliders, with the potential of enhancing the sensitivity reach to LLPs~\cite{Wang:2019xvx,Chrzaszcz:2020emg,Schafer:2022shi}.
In particular, Ref.~\cite{Wang:2019xvx} firstly proposes the installation of FAr Detectors at the Electron Positron Collider (FADEPC) and investigates their basic designs and the corresponding discovery potentials for LLPs in several theoretical scenarios.

For planning the construction of FDs, since the tunnel of the LHC has already been constructed, there is little free space that could be utilized.
However, for future $e^- e^+$ colliders the situation is more optimistic and open; as their construction plan is still under development, there is large freedom in the geometry and location of such FDs to be deployed.
Therefore, it is both important and practical to start to design such FDs already now, so that it would be possible to have them built during the construction of the main experiment.
Further, at the LHC with proton-proton collisions, owing to the parton distributions in the protons, there is typically a large boost in the very forward direction for the LLPs, and as a result, all the proposed FD experiments there should have some longitudinal distance from the IP.
However, it is a different case for symmetrical electron-positron colliders, where the LLPs produced would tend to travel in the transverse direction.
%
Given the difference in the LLP kinematics between the LHC and future high-energy electron-positron colliders, as well as the currently large freedom for locating both the experimental hall and the FDs, we argue that auxiliary FDs at future $e^- e^+$ colliders could play a unique role in searching for LLPs.
Fig.~\ref{fig:FDsketchEEc} from Ref.~\cite{Wang:2019xvx} is the sketch of an example FD at future $e^-e^+$ colliders.
The coordinate system is set up as follows: the origin $O$ is the IP; the injected electron and positron beams travel along the $z$ axis, while the $+z$ direction is defined as the electron beam outgoing direction; the vertical and horizontal axes orthogonal to the $z$-axis are set to be $y$- and $x$-axes, respectively;  the $+y$ direction are chosen to be upward.
The red cylinder enclosing the IP depicts the main detector (near detector, abbreviated as ND), while the green cuboid illustrates a far detector located with a distance from the IP. The distance of the FD to the IP is labeled with $D$. \par
Refs.~\cite{Wang:2019xvx,Tian:2022rsi} consider various locations and geometrical setups of far detectors (FD1–FD8) at future $e^-e^+$ colliders and investigate their potentials for discovering LLPs in the physics scenarios including exotic Higgs decays, the lightest neutralinos in the R-parity-violating supersymmetry (RPV-SUSY), heavy neutral leptons, and ALPs. Besides, to compare discovery sensitivities between the FDs and NDs, Ref.~\cite{Wang:2019xvx} also derives sensitivity reach to the LLPs at the NDs of future $e^-e^+$ colliders and LHC FD experiments such as AL3X, CODEX-b, and MATHUSLA100.
For the NDs at future $e^-e^+$ colliders, the CEPC's baseline detector setup are chosen.
For the FDs at future $e^-e^+$ colliders, their shapes are assumed as cuboid and the locations of the FDs are $\sim 5-100$ m away in the transverse direction from the IP.
%
For example, the FD1 design in this study is about $5-10$ m from the IP and employs a volume of $5 \times 10^3$ m$^3$.
It can be placed inside the experiment hall if the hall is big enough, or in a cavern or shaft near the experiment hall.
The volume of the other FD designs is large, and they are $50-100$~m from the IP.
%
Ref.~\cite{Wang:2019xvx} finds that for searching for LLPs, FDs at future lepton colliders can extend and complement the sensitivity reaches of the default MD and the present and future LHC experiments.
In particular, for the theoretical models considered, a MATHUSLA-sized far detector would give a modest improvement compared to the case with a main detector only at future lepton colliders.
\begin{figure}[t]
    \centering
    \includegraphics[width=0.75\linewidth]{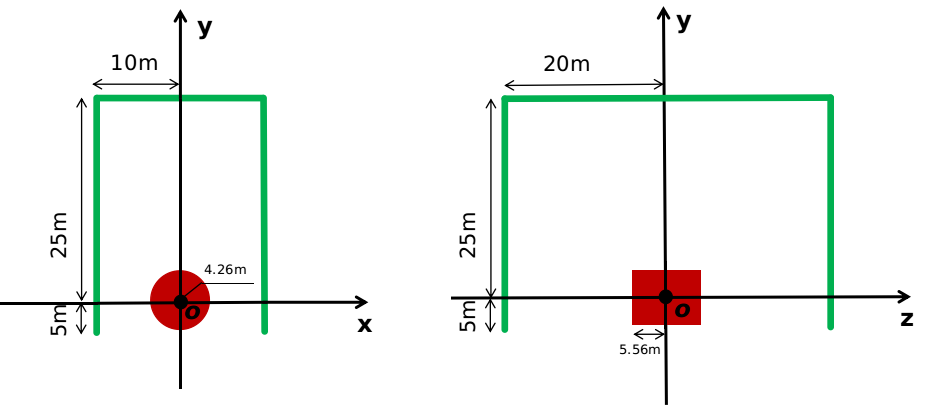}
    \caption{
     The left and right plots show section views of the experimental setup in the $xOy$ and $yOz$ planes, respectively.
    In the plots, the red cylinder enclosing the coordinate origin represents the main detector of CEPC/FCC-ee.
    The green area depicts the proposed new far detector, LAYCAST, in the shape of a thin layer to be instrumented on the cavern surface.
    Take from Ref.~\cite{Lu:2024fxs}.}
    \label{fig:FD902}
\end{figure}
\par
Inspired by the previous proof-of-principle study~\cite{Chrzaszcz:2020emg}, recently Ref.~\cite{Lu:2024fxs} propose a similar tracker detector, named as the LAYered CAvern Surface Tracker (LAYCAST), to be installed on the wall and ceiling of the main cavern 
at future electron-positron colliders such as CEPC and FCC-ee.
The fiducial volume is taken to be the space between the main detector and the cavern's surface.
The setup of LAYCAST is shown in Fig.~\ref{fig:FD902}, where the coordinate system is the same as in Fig.~\ref{fig:FDsketchEEc}.
The shape of the experimental hall is simplified into a cuboid.
Considering that the floor of the experiment hall cannot be installed owing to load bearing and other reasons, the LAYCAST would be mounted on the roof surface and four vertical walls of the experimental hall. 
The LLPs produced at the IP, if decaying inside the main detector, can potentially be observed therein via the decay products.
If they traverse the main detector and decay before reaching the cavern's inner surface, they may be detected by the LAYCAST experiment.

\begin{figure}[hbtp]
     \centering
     \includegraphics[width=0.8\linewidth]{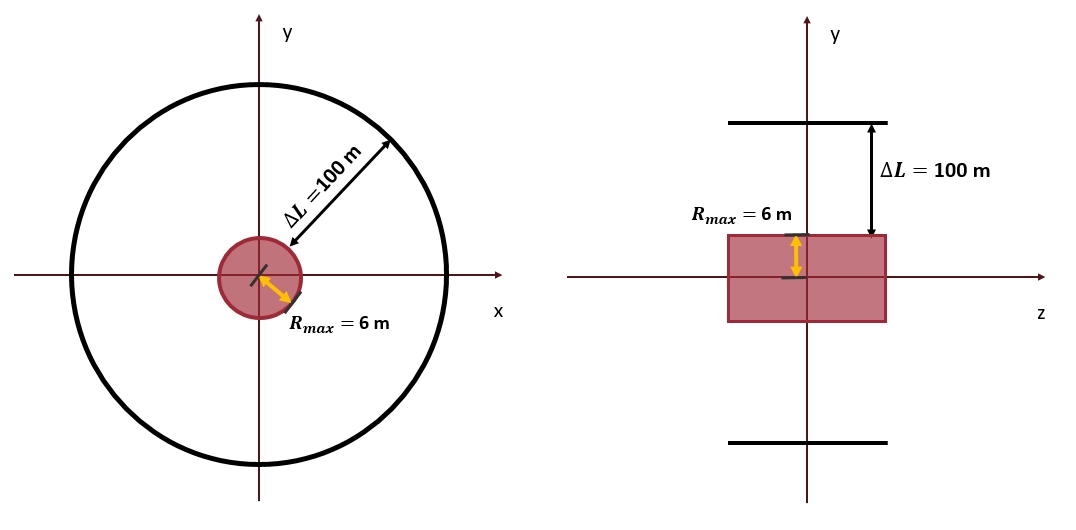}
     \caption{The left and right plots show the layout of the Far Barrel Detector (solid lines) relative to the main detector (red) in the $x$-$y$ (left) and $y$-$z$ (right) planes. Taken from Ref.~\cite{Zhang:2024bld}.}
     \label{fig:llp_fbd}
 \end{figure}
 
Ref.~\cite{Zhang:2024bld} explores the potential to improve sensitivity for long-lived particle detection by deploying a far detector positioned substantially farther from the IP than the main detector. The proposed far detector may consist of stacked, multi-layer scintillator arrays in the barrel region outside the main detector, also known as the Far Barrel Detector (FBD). Fig.~\ref{fig:llp_fbd} shows the configuration of the far and main detectors, where $R_{max}$ represents the outer radius of the main detector and the $\Delta L$ is the gap between the far detector and the main detector.
 
Recent LLP studies with various FDs are summarized as follows.
Sensitivity results for long-lived heavy neutral leptons are presented in Section~\ref{subsubsec:FDs-HNL}.

\subsubsection{Higgs boson decays}

\begin{figure}[htbp]
\centering
\includegraphics[width=0.49\textwidth, height=6.8cm]{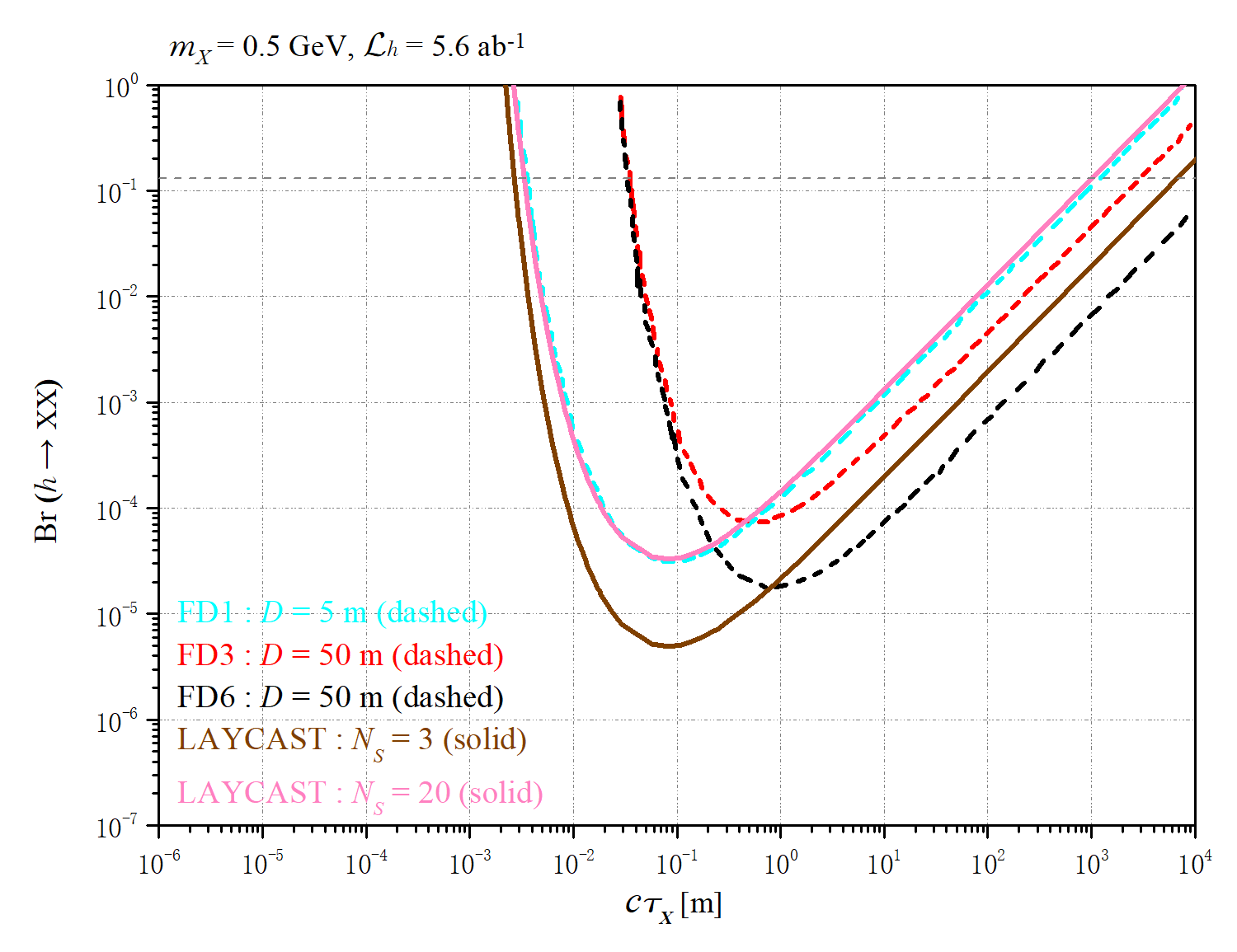}
\includegraphics[width=0.49\textwidth]{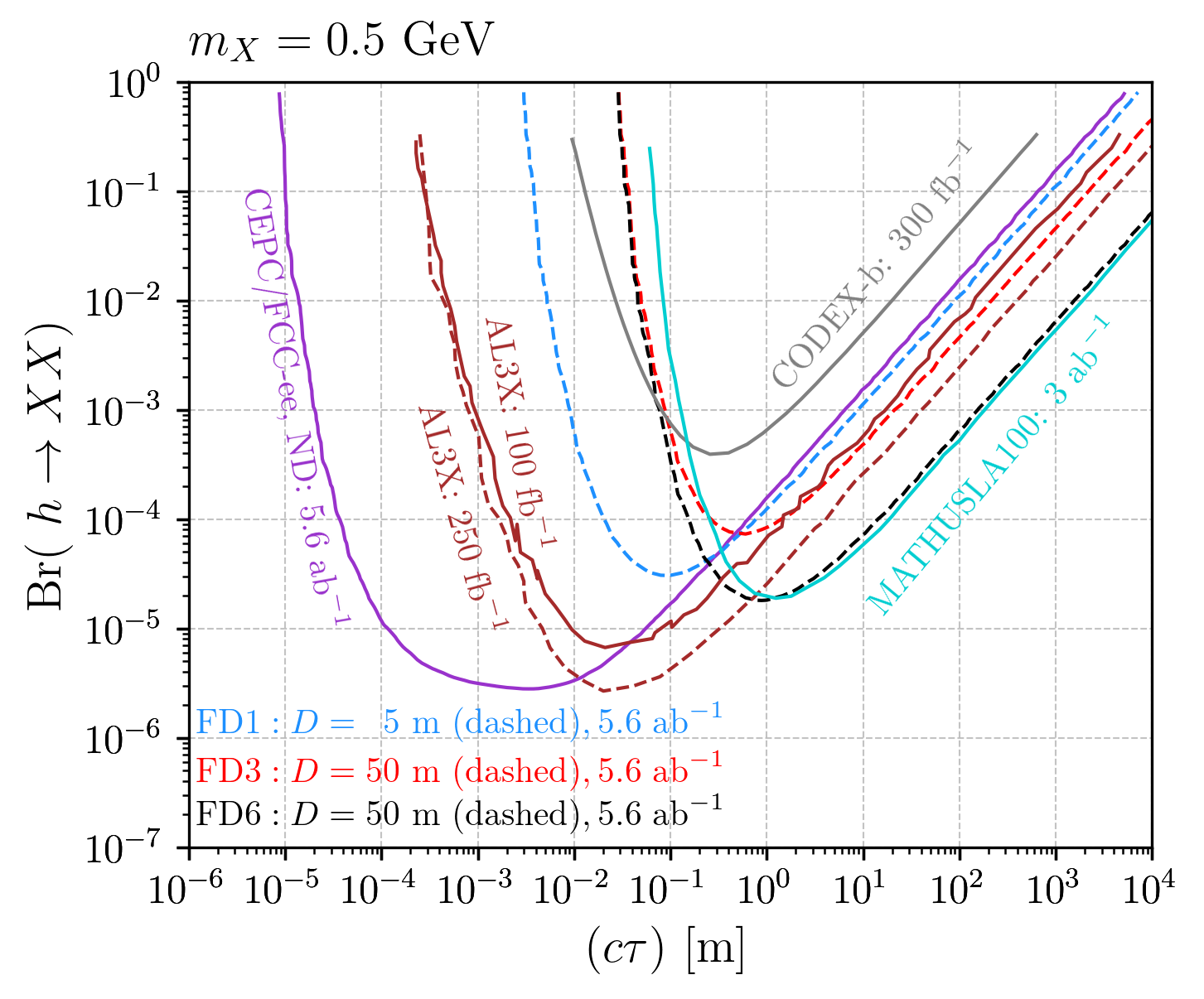}
\caption{Left: Sensitivity reaches of the CEPC/FCC-ee's far detectors FD1, FD3, FD6 and LAYCAST in the Br($h \rightarrow XX$) vs. $c\tau$ plane for $m_X = 0.5$ GeV. 
Right: Sensitivity reaches of the CEPC/FCC-ee's far detectors FD1, FD3, FD6, compared with predictions for the CEPC/FCC-ee’s main detector (near detector, abbreviated as ND) and for AL3X, CODEX-b and MATHUSLA100.  
Taken from Ref.~\cite{Wang:2019xvx,Lu:2024fxs}.
}
\label{fig:FDs-H2XX-0p5}
\end{figure}

Refs.~\cite{Wang:2019xvx,Lu:2024fxs} study a pair of long-lived light scalars $X$ produced from the SM Higgs boson decays, $h \to X X$, at $\sqrt{s} =$ 240 GeV.
The total number of the SM Higgs bosons produced at either the CEPC or FCC-ee is specified as $N_h = 1.14 \times 10^6$ with an integrated luminosity $\mathcal{L} =$ 5.6 ab$^{-1}$. 
Sensitivity results in terms of 3-signal-event contour curves are presented, corresponding to 95\% C.L.~limits with zero background events. 
Besides, sensitivity results in terms of 20-signal-event contour curves for LAYCAST are also presented, corresponding to 95\% C.L.~limits with about 100 background events, to estimate the effect of non-zero background.
Fig.~\ref{fig:FDs-H2XX-0p5}, extracted from Refs.~\cite{Wang:2019xvx,Lu:2024fxs}, shows sensitivity reaches in the branching ratio Br($h \rightarrow XX$) vs.~proper decay length $c\tau$ plane for the the light scalar mass $m_X = 0.5$ GeV. 
Sensitivity reaches for $m_X = 10$ GeV are also given in Ref.~\cite{Wang:2019xvx,Lu:2024fxs}.

Ref.~\cite{Zhang:2024bld} quantifies the enhancement in sensitivity by a gain factor \(F_\textrm{gain}\) as follows:
\begin{equation}
    F_\textrm{gain} = \frac{N_\textrm{obs}}{N_\textrm{gen}} 
    = \frac{\Delta\Omega}{4\pi}\left(\frac{e^{-R_\textrm{max}/d}-e^{-(R_\textrm{max}+\Delta L)/d}}{1 - e^{-R_\textrm{max}/d}}\right) + 1,
\end{equation}
 where \(N_\textrm{obs}\) is the number of observed LLP events, \(N_\textrm{gen}\) is the number of generated LLP event, \(\Delta \Omega / 4\pi \) is the far detector's geometric acceptance factor and \(d\) is the decay length of the LLPs in the laboratory frame (\(R_\textrm{max}\) and \(\Delta L\) are as shown in Fig.~\ref{fig:llp_fbd}). The estimated sensitivity gain factor, \(F_\textrm{gain}\), is obtained by considering the far detector in the barrel region, which has a geometric acceptance factor of approximately 0.7, with \(R_\textrm{max}\) set to 6 meters and \(\Delta L\) set to 100 meters. The calculated gain factors are summarized in Table~\ref{tab:f_gain}, 
 highlighting significant sensitivity improvements for LLPs with lower masses and longer lifetimes. 

\begin{table}[!hbtp]
  \centering
  \caption{Sensitivity gain factor \(F_\textrm{gain}\) estimated for different LLP masses and lifetimes with the far detector. Taken from Ref.~\cite{Zhang:2024bld}.}
  \begin{tabular}{ccccccc}
  \toprule
    \toprule
   \multirow{2}{*} & $F_\textrm{gain}$  & \multicolumn{5}{c}{Lifetime [ns]} \\
  \cmidrule{2-7}
   & Mass [GeV] & 0.001 & 0.1 & 1 & 10 & 100 \\ 
   \midrule
   \multirow{2}{*} & 1 & 1     & 1   & 2.8 & 9.9 & 13.7 \\
     & 10 & 1     & 1   & 1   & 2.9  & 10.1 \\ 
     & 50 & 1     & 1   & 1   & 1.1  & 3.3 \\ 
    \bottomrule
  \end{tabular}
  \label{tab:f_gain}
\end{table}

\subsubsection{$Z-$boson decays}

\begin{figure}[t]
\centering
\includegraphics[width=0.49\textwidth, height=6.8cm]{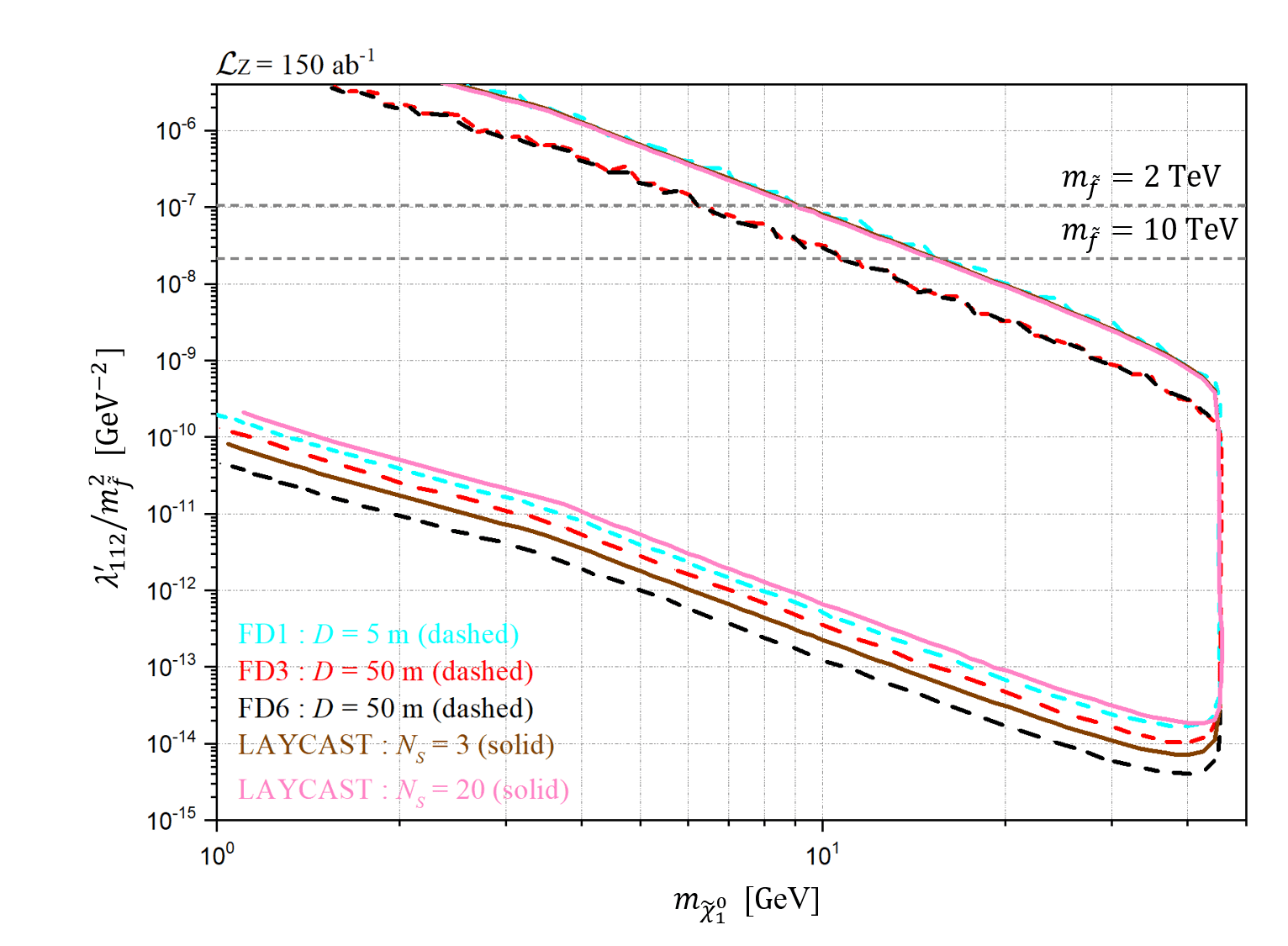}
\includegraphics[width=0.49\textwidth]{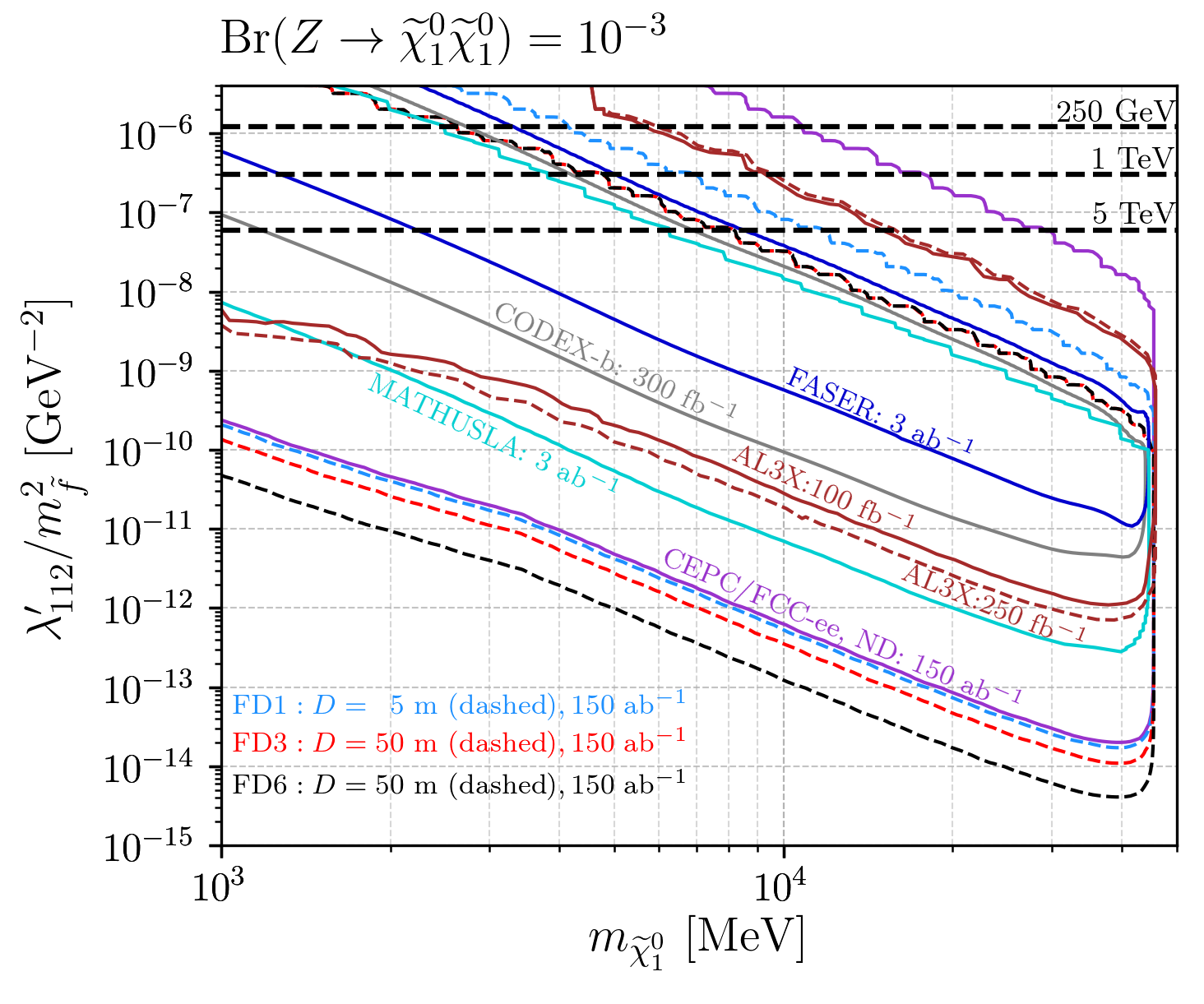}
\caption{Sensitivity reaches of different experiments assuming Br($Z\rightarrow\tilde{\chi}_1^0\tilde{\chi}_1^0$) = $10^{-3}$.  
Left: results of the CEPC/FCC-ee's far detectors FD1, FD3, FD6 and LAYCAST in the $\lambda'_{112}/m^2_{\tilde{f}}$ vs. $m_{\tilde{\chi}_1^0}$ plane.
Right: results of the CEPC/FCC-ee's far detectors FD1, FD3, FD6, compared with predictions for the CEPC/FCC-ee's main detector (near detector, abbreviated as ND) and other experiments.
Taken from Ref.~\cite{Wang:2019xvx,Lu:2024fxs}.
}
\label{fig:FDs-Z2n1n1}
\end{figure}

Ref.~\cite{Wang:2019xvx,Lu:2024fxs} consider $Z$-boson decays to  a pair of long-lived neutralinos in the RPV-SUSY, $Z \to \neutralino1 \neutralino1$, at $\sqrt{s} =$ 91.2 GeV.
The lightest neutralino is mostly bino with tiny components of Higgsinos.
In the analyses, the total number of the $Z$-bosons produced at the CEPC is specified as $N_Z^{\mathrm{CEPC}}=7.0 \times 10^{11}$ corresponding to a total integrated luminosity of $\mathcal{L}_{Z}^{\mathrm{CEPC}}=16$ ab$^{-1}$, while $N_Z^{\mathrm{FCC-ee}}=5.0 \times 10^{12}$ corresponding to $\mathcal{L}_Z^{\text{FCC-ee}}=150$ ab$^{-1}$. 
Sensitivity results in terms of 3-signal-event contour curves are presented, corresponding to 95\% C.L.~limits with vanishing background. 
Besides, sensitivity results in terms of 20-signal-event contour curves for LAYCAST are also presented, corresponding to 95\% C.L.~limits with about 100 background events, to estimate the effect of non-zero background.
Fig.~\ref{fig:FDs-Z2n1n1} is reproduced from Ref.~\cite{Wang:2019xvx,Lu:2024fxs}, and it show sensitivity reaches of different FD designs assuming Br($Z\rightarrow\tilde{\chi}_1^0\tilde{\chi}_1^0$) = $10^{-3}$ for $m_{\neutralino1}\ll m_Z/2$.
Sensitivity reaches of both the FDs and NDs at the CEPC/FCC-ee with different integrated luminosities of $\mathcal{L} =$ 16, 150, and 750 ab$^{-1}$, are also presented and compared in Ref.~\cite{Wang:2019xvx,Lu:2024fxs}.

\subsubsection{Axion-like particles}

\begin{figure}[t]
\centering
\includegraphics[width=0.49\textwidth]{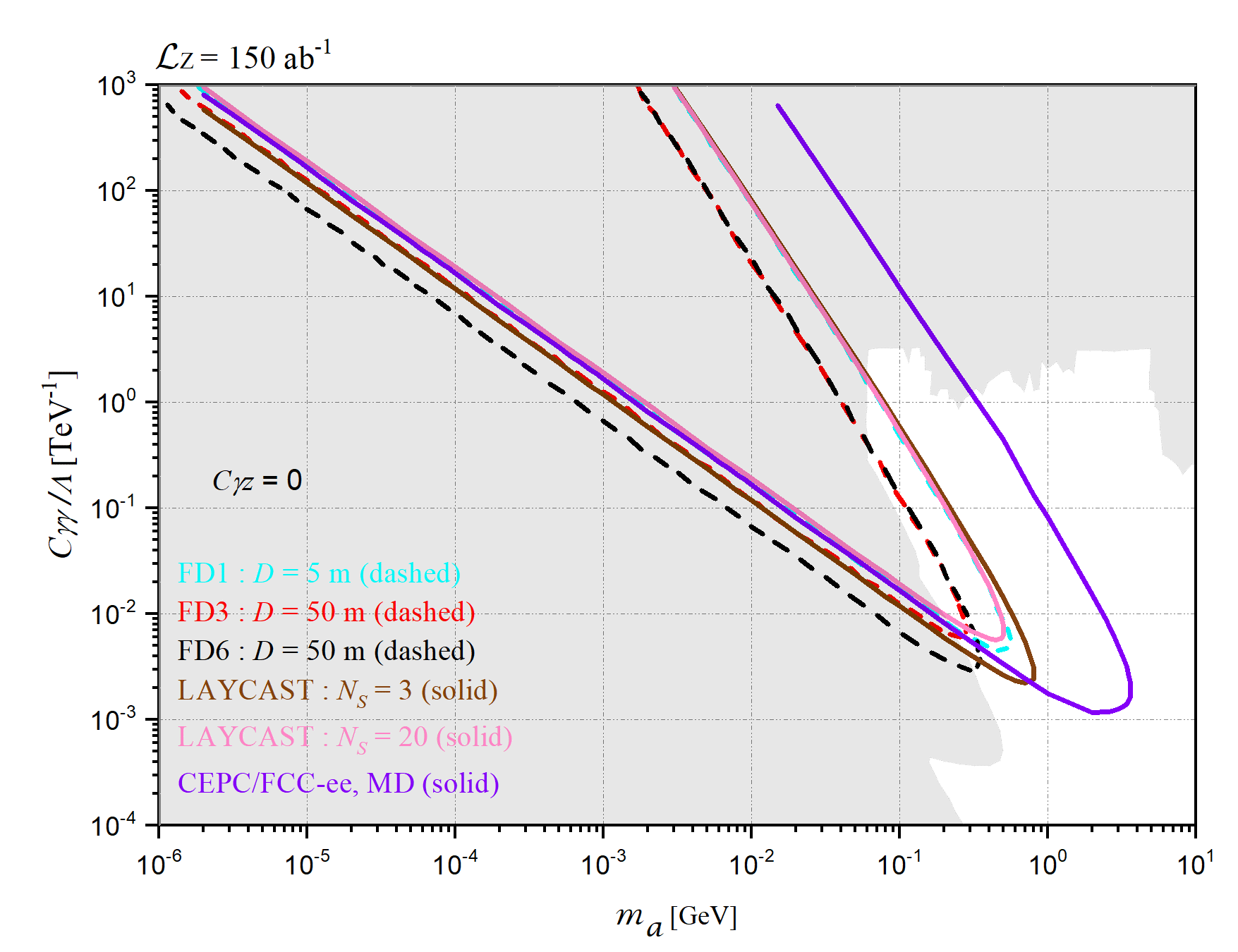}
\includegraphics[width=0.49\textwidth]{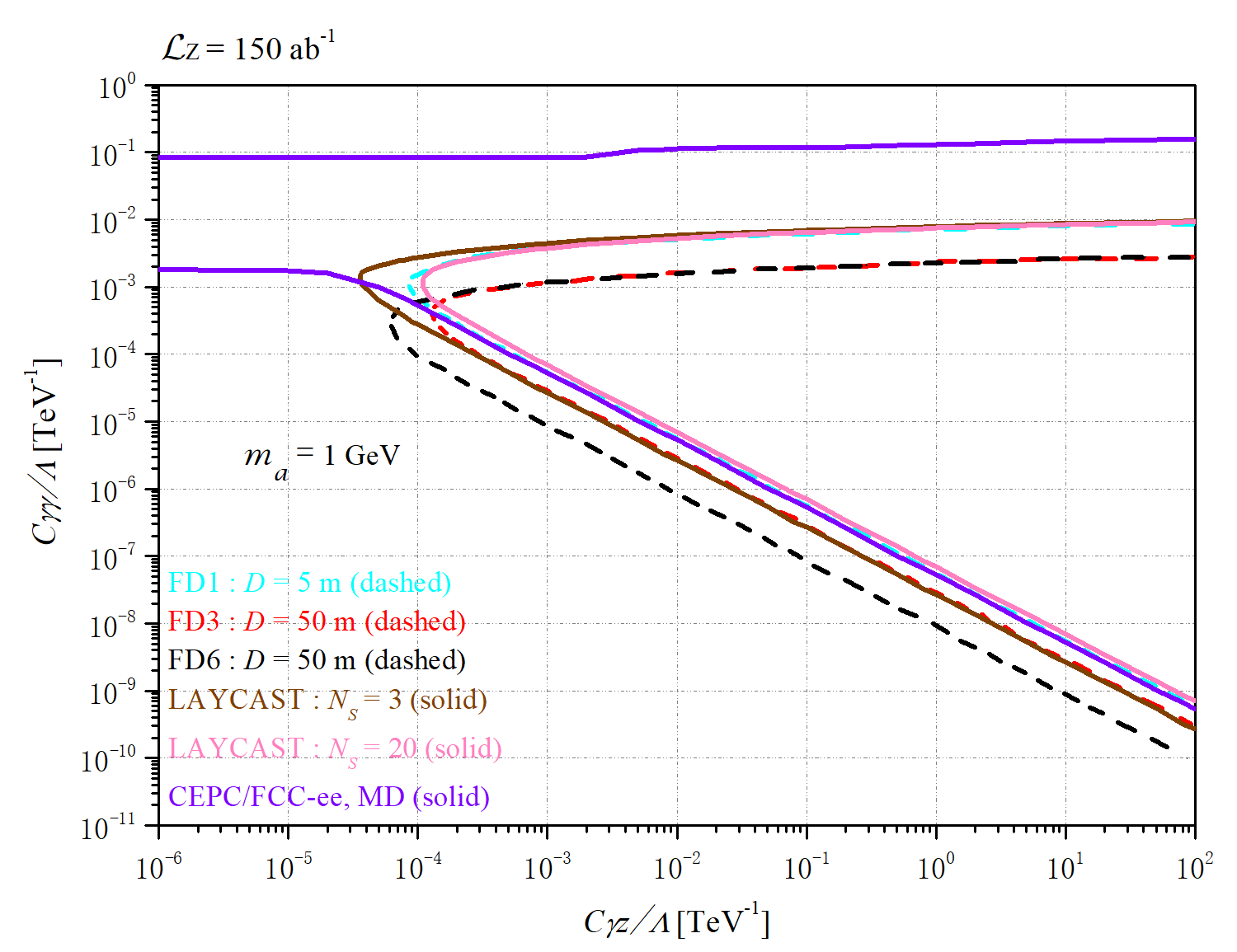}
\caption{Left: when $C_{\gamma Z} = 0$, sensitivity reaches of representative far detectors FD1, FD3, FD6, LAYCAST and the CEPC/FCC-ee's main detector (MD) with integrated luminosity $\mathcal{L}_Z$ = 150 ab$^{-1}$ in the $C_{\gamma\gamma}/\Lambda$ vs. $m_a$ plane.
Right: when both $C_{\gamma Z}$ and $C_{\gamma \gamma}$ are free parameters, sensitivity reaches  
with $m_a =$ 1 GeV in the $C_{\gamma\gamma}/\Lambda$ vs.~$C_{\gamma Z}/\Lambda$ plane with $\mathcal{L}_Z$ = 150 ab$^{-1}$.
Taken from Ref.~\cite{Tian:2022rsi,Lu:2024fxs}.
}
\label{fig:FDs-ALP-FADEPC}
\end{figure}

Refs.~\cite{Tian:2022rsi,Lu:2024fxs} are follow-up works of Ref.~\cite{Wang:2019xvx}.
Ref.~\cite{Tian:2022rsi} considers the eight designs of FDs with different locations, volume, and geometries proposed in Ref.~\cite{Wang:2019xvx}, while Ref.~\cite{Lu:2024fxs} considers another far detector design, the LAYCAST.
They investigate the potential of different far detector designs for discovering long-lived ALPs via the process $e^- e^+ \to \gamma\, a, a \to \gamma \gamma$ at future $e^- e^+$ colliders running at the CM energy of 91.2 GeV and integrated luminosities of 16, 150, and 750 ab$^{-1}$. 
Sensitivities to the model parameters in terms of the effective ALP-photon-photon coupling $C_{\gamma\gamma} / \Lambda$ ($\Lambda$ is the effective cutoff scale), the effective ALP-photon-$Z$ coupling $C_{\gamma Z}/\Lambda$, and the ALP mass $m_a$, are presented for three physics scenarios: $C_{\gamma Z}  = 0$, $C_{\gamma Z} = C_{\gamma\gamma}$, and both $C_{\gamma Z}$ and $C_{\gamma \gamma}$ are independent parameters.
Sensitivity results in terms of 3-signal-event contour curves are presented, corresponding to 95\% C.L.~limits with vanishing background. 
Besides, sensitivity results in terms of 20-signal-event contour curves for LAYCAST are also presented, corresponding to 95\% C.L.~limits with about 100 background events, to estimate the effect of non-zero background.
Fig.~\ref{fig:FDs-ALP-FADEPC} is from Refs.~\cite{Tian:2022rsi,Lu:2024fxs} and two plots compare the performances of the representative far detectors FD1, FD3, FD6, LAYCAST and the CEPC/FCC-ee's main detector (MD).
Sensitivity results for the case that $C_{\gamma Z} = C_{\gamma\gamma}$ are also given in these studies.

\begin{figure}[t]
\centering
\includegraphics[width=0.49\textwidth]{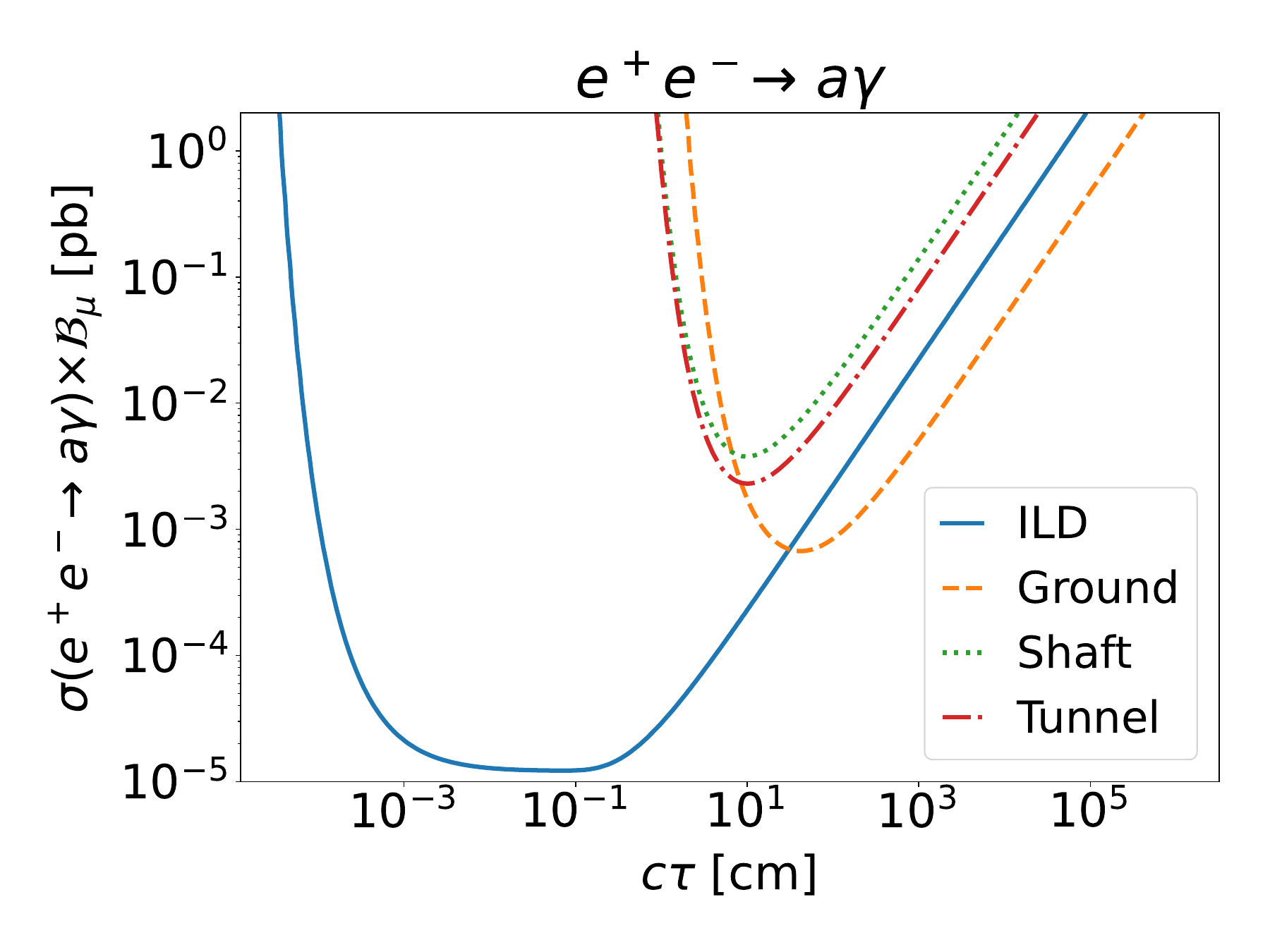}\hspace*{0.1cm}
\includegraphics[width=0.49\textwidth]{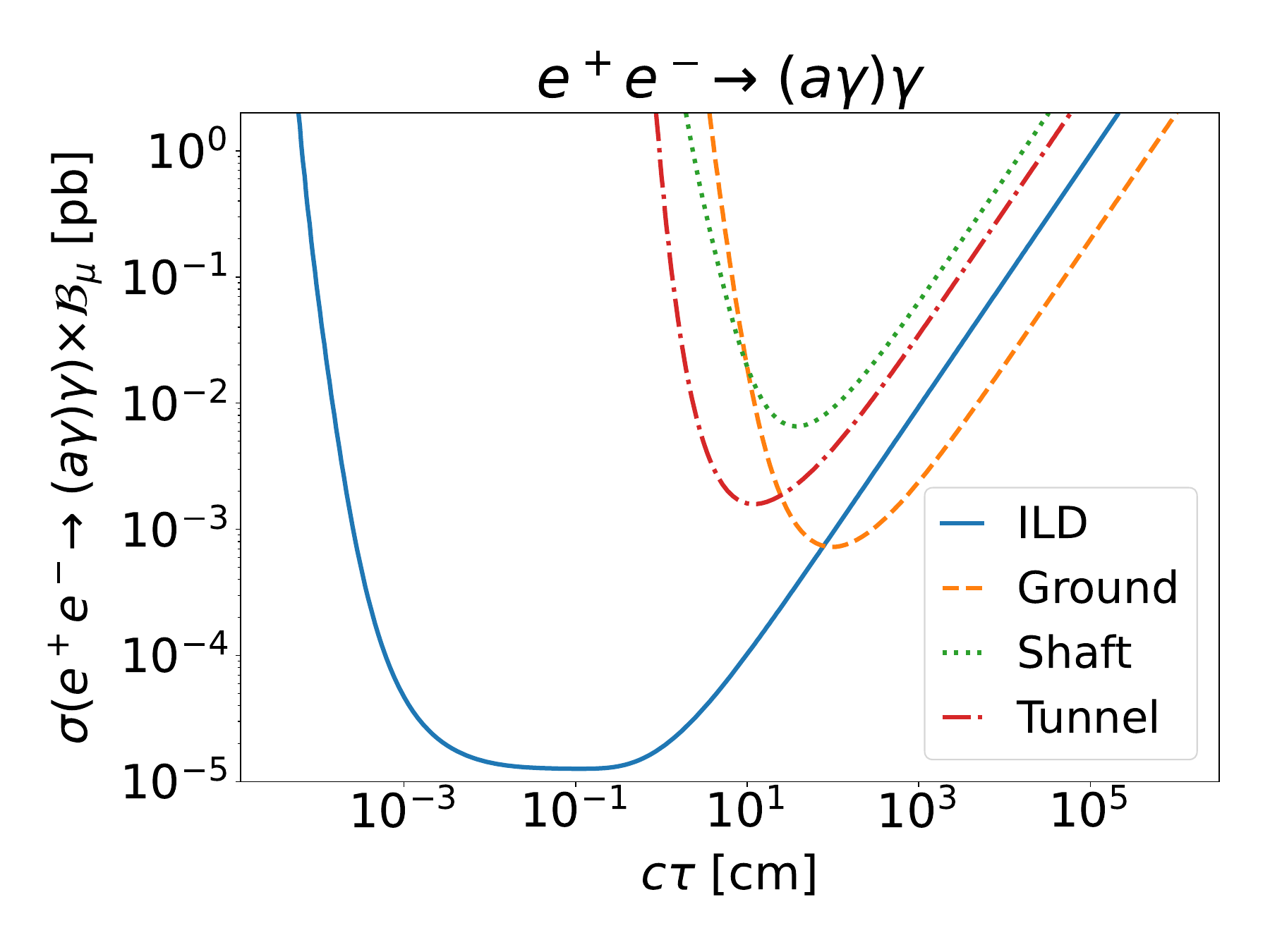}
\caption{Contours of $N_a =$ 3 ALPs with $m_a =$ 300 MeV decaying within various ILC detectors, as a function of the production cross section, $\sigma$, and the proper lifetime, $c\tau_a$. 
Shown are the production channels $e^+e^- \to a\gamma$ (left) and $e^+e^- \to Z\gamma \to (a\gamma)\gamma$ (right) at $\sqrt{s} =$ 250 GeV and with $\mathcal{L} =$ 250 fb$^{-1}$. 
Predictions are made for the ILD (blue, plain) and far detectors placed in the Shaft (green, dotted), in the Tunnel (red, dot-dashed) and on the Ground (orange, dotted). 
The branching ratio of the ALP into muons is indicated by $\mathcal{B}_{\mu}$.
Taken from Ref.~\cite{Schafer:2022shi}.  
}
\label{fig:FDs-ALPs-ILC}
\end{figure}

Ref.~\cite{Schafer:2022shi} explores the discovery potential of FDs for long-lived ALPs at future $e^- e^+$ colliders, such as the ILC. 
Three possible setups of FDs are proposed, and could be installed in planned underground cavities around the ILC detector hall or on the ground.
The authors consider cuboid for the shape of the FDs.
The first type ``Shaft (S)'' is located in the vertical shaft above the collision point, which will be used to lower the main ILD and SiD detectors into the detector hall. 
Its position is centered around the coordinate $(x, y, z) = (0, 45, 0)$ m, and the geometry is 18 m $\times$ 30 m $\times$ 18 m.
The second type ``Tunnel (T)'' is located inside the access tunnel that surrounds the detector hall. 
Its position is centered around the coordinate (0, -5, -35) m, and the geometry is 140 m $\times$ 10 m $\times$ 18 m.
The third type ``Ground (G)'' is a large detector placed on the ground above the detector hall. 
Its position is centered around the coordinate (0, 75, 0) m, and the geometry is 1000 m $\times$ 10 m $\times$ 1000 m.

This study considers sub-GeV long-lived ALPs produced via $e^- e^+ \to \gamma\, a$ or $e^- e^+ \to \gamma\, Z \to \gamma\, (\gamma a)$ process and decaying into pairs of charged leptons at the ILC with $\sqrt{s} =$ 250 GeV.
The background is assumed to be negligible, and sensitivity reaches are shown in terms of three-signal-event contour curves.
We extract Fig.~\ref{fig:FDs-ALPs-ILC} from Ref.~\cite{Schafer:2022shi}, which shows the sensitivity reach of the ILD main detector and the three FD designs to the production cross sections of ALPs with mass $m_a =$ 300 MeV.
In this study, the results are also compared with searches for long-lived ALPs produced from meson decays at Belle II.

\subsection{Studies with beam dumps}
\label{subsec:beam_dump}

Future $e^-e^+$ colliders employ high-energy electron and positron beams.
The beam dump can result in copious production of LLPs.
There exist multiple studies considering a beam-dump experiment to search for LLPs at, e.g.~the ILC.
It is easily conceivable that similar experimental setups can also be instrumented at other $e^- e^+$ colliders such as the CEPC, FCC-ee, and CLIC, despite the different beam energies.
\begin{figure}[t]
\begin{center}
\includegraphics[width=0.75\textwidth]{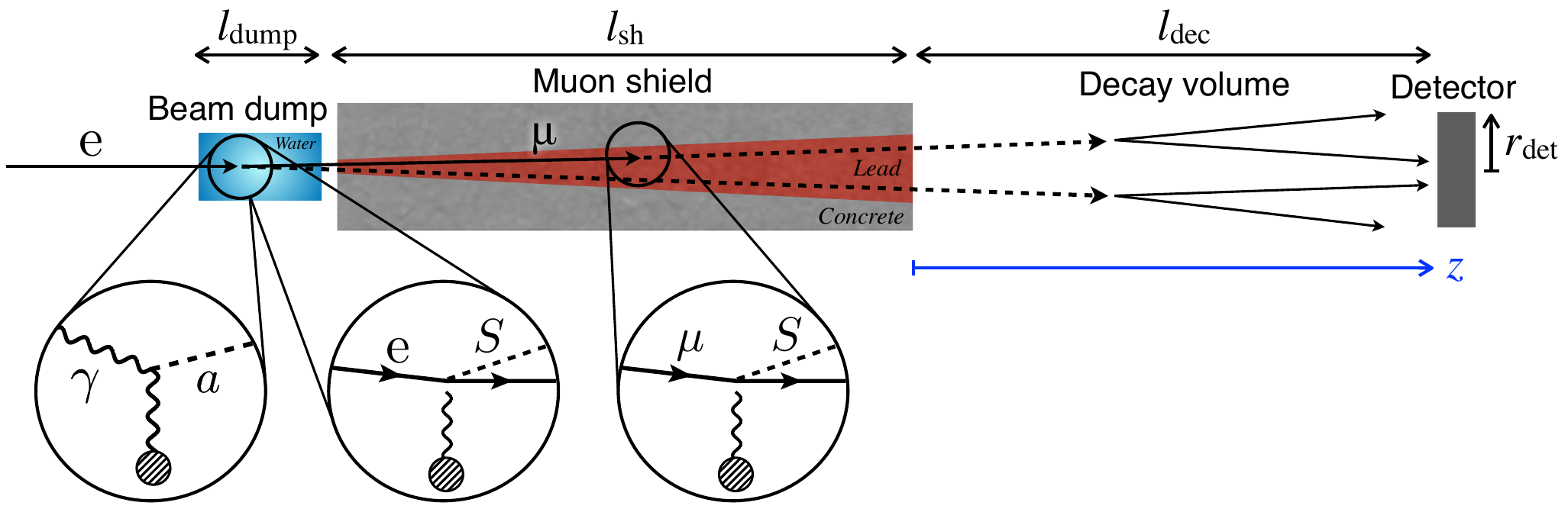}
\caption{The setup of the beam-dump experiment at the ILC
consists of four parts: the main beam dump, a muon shield, a decay volume, and a detector.
The figure depicts LLP signals including ALP ($a$) emissions via the photon interaction and light scalar particle ($S$) emissions via the electron and muon interactions.
Taken from Ref.~\cite{Sakaki:2020mqb}.
}
\label{fig:expSetupBD}
\end{center}
\end{figure}
In particular, Ref.~\cite{Sakaki:2020mqb} shows Fig.~\ref{fig:expSetupBD} illustrating a sample setup of a beam-dump experiment at the ILC, which consists of four parts: the main beam dump, a muon shield, a decay volume, and a detector. 
Water is planned as the absorber in the main beam dump of the ILC~\cite{Satyamurthy:2012zz}. 
The length of water cylinder along the beam axis is $l_{\rm dump} =$ 11 m.
Inside the main beam dump, electromagnetic shower produces electrons, positrons, and photons.
The muon shield length and the decay volume length are $l_{\rm sh} =$ 70 m and $l_{\rm dec} =$ 50 m, respectively.
The muon shield could consist of the lead shield and the active shield.
The shape of the detector is assumed as a cylinder, and its axis should be aligned with the beam axis. 
The radius of the detector $r_{\rm det}$ is set to be 2 to 3 meters.
Recent LLP studies with beam dump experiments are summarized as follows.

\subsubsection{ALPs and new scalar particles}

The authors of Ref.~\cite{Sakaki:2020mqb} investigate the sensitivities of a beam-dump experiment at the ILC to a long-lived ALP and a light scalar particle coupled to charged leptons. 
In their analysis, the lengths of the beam dump region $l_{\rm dump}$, the muon shield $l_{\rm sh}$, and the decay volume  $l_{\rm dec}$ are set to be 11 m, 70 m and 50 m, respectively.
The radius of the detector $r_{\rm det}$ is set to 2 m and the detection efficiency is assumed to be 100\%.
The authors consider the case of ILC-250 GeV with the beam energy $E_{\rm beam} =$ 125 GeV.
The number of incident electrons into the beam dump is assumed to be $N_{\rm EOT} = 4 \times 10^{21}$ per year.

The signal production process is illustrated in Fig.~\ref{fig:expSetupBD}.
The ALPs are emitted by the photons in the beam dump, pass through the muon shield, and decay in the decay volume into photon pairs, which reach the detector at the end recording a signal event.
New scalar particles are emitted via electron interactions with the oxygen nuclei in the beam dump and via muon interaction with the lead nuclei in the muon shield.
The generated scalar particle decays into photons, electron-positron, and muon pair in the decay volume, which reach the detector and are observed as signal events.
\begin{figure}[t]
\begin{center}
\includegraphics[width=0.47\textwidth]{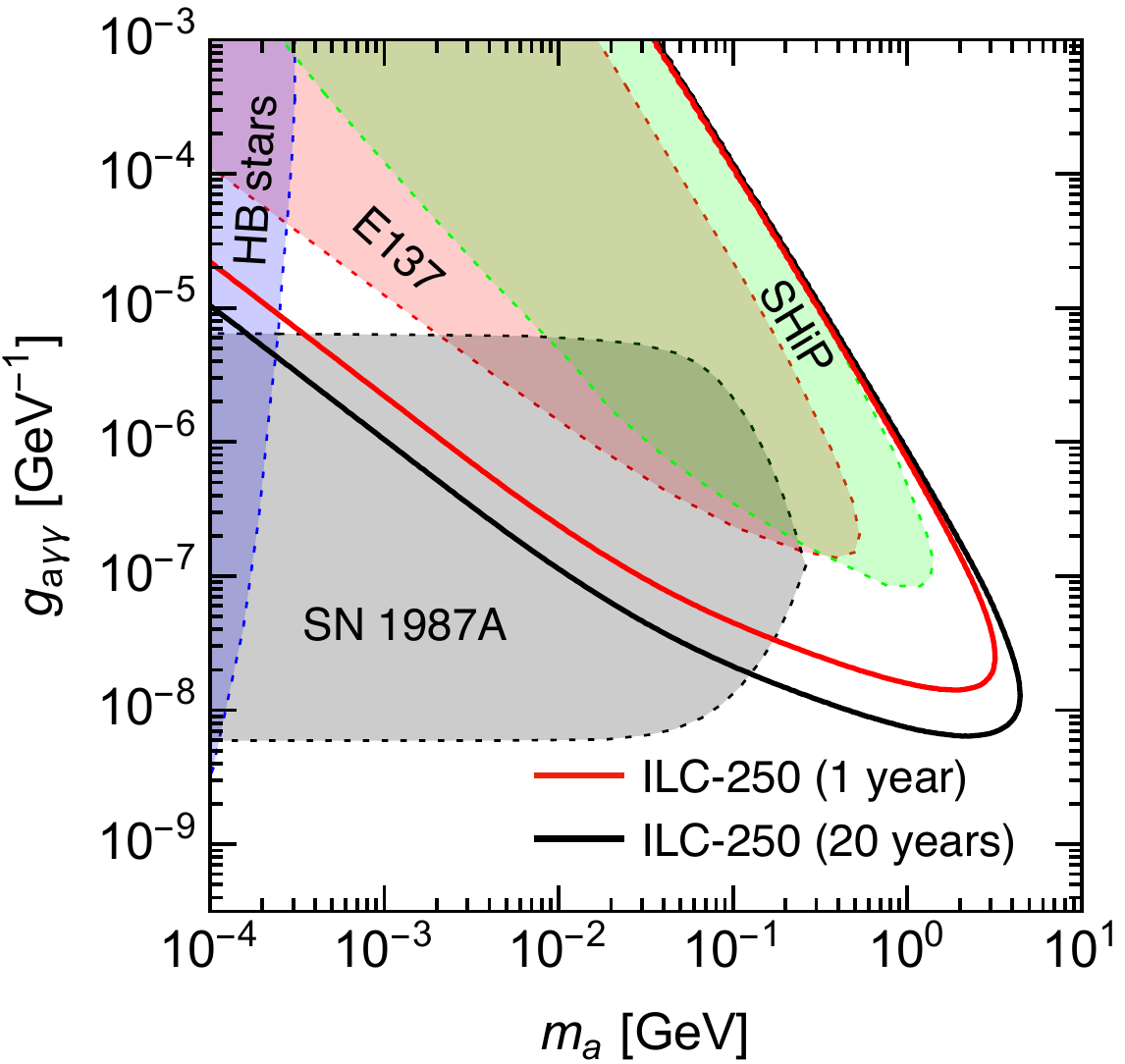}
\includegraphics[width=0.45\textwidth]{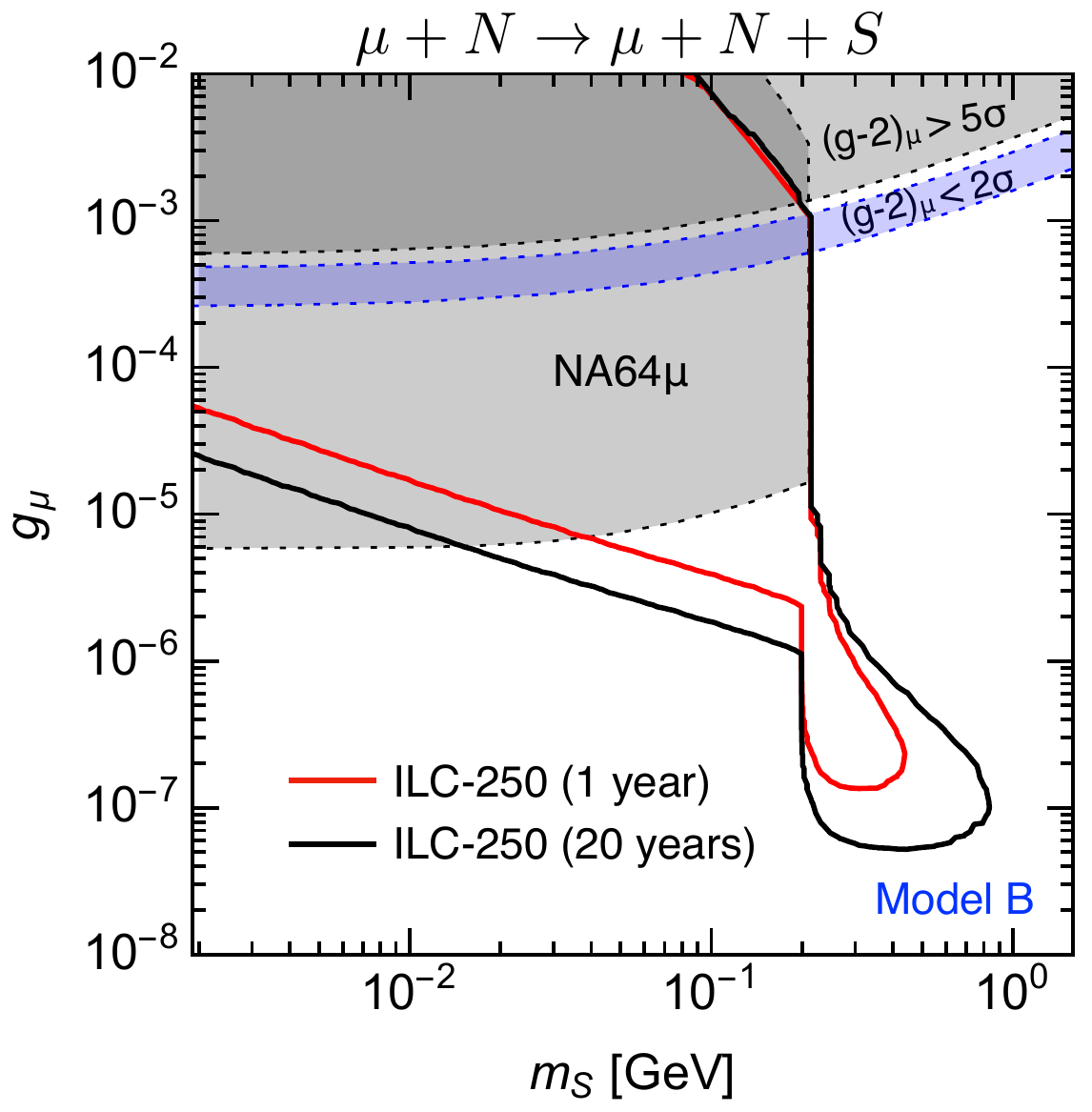}
\caption{Both plots are extracted from Ref.~\cite{Sakaki:2020mqb} and show discovery sensitivities of the beam-dump experiment at the ILC to the ALPs and new scalar particles.
Left: sensitivities in the ALP-photon-photon coupling $g_{a\gamma\gamma}$ vs.~ALP mass $m_a$ plane, where red and black curves correspond to the bounds of sensitivity for ILC-250 GeV at 95\% C.L.~with 1- and 20-year statistics;
the shaded regions are constraints for E137 from Ref.~\cite{Dolan:2017osp}, SN 1987A from Ref.~\cite{Dolan:2017osp,Jaeckel:2017tud}, HB stars from Ref.~\cite{Cadamuro:2011fd}, and SHiP from Ref.~\cite{Dolan:2017osp,Dobrich:2015jyk,Dobrich:2019dxc}.  
Right: sensitivities in the $g_\mu$ vs.~$m_S$ plane for Model B ($S$ couples to muons only, i.e.~$g_\mu \neq 0,\,\, g_e = g_\tau = 0$), where the signal process contains a muon in the initial state (i.e.~$\mu + N \to \mu + N + S$);
the gray shaded regions are constraints from NA64$\mu$ and muon $g - 2$ from Ref.~\cite{Chen:2017awl}; note that, although the results for $m_S > 2 m_\mu$ are absent for NA64$\mu$, it would also have a sensitivity in that region generally.
}
\label{fig:ALPs-BD-ILC}
\end{center}
\end{figure}

Background events are assumed to be removed with veto counters located behind the shield and in front of and around the detector.
Fig.~\ref{fig:ALPs-BD-ILC} is extracted from Ref.~\cite{Sakaki:2020mqb}.
The left plot shows discovery sensitivities of the beam-dump experiment at the ILC to the ALPs coupling to photons, where the red and black curves correspond to  ILC-250 GeV at 95\% C.L. with 1- and 20- year statistics, respectively.
The right panel is for the sensitivity reach muon-coupled light scalar particle.

\subsubsection{New neutral gauge bosons}

\begin{figure}[t]
\centering
\includegraphics[width=0.49\textwidth]{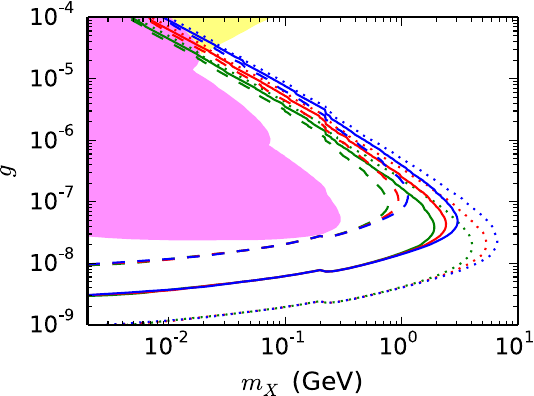}
\includegraphics[width=0.49\textwidth]{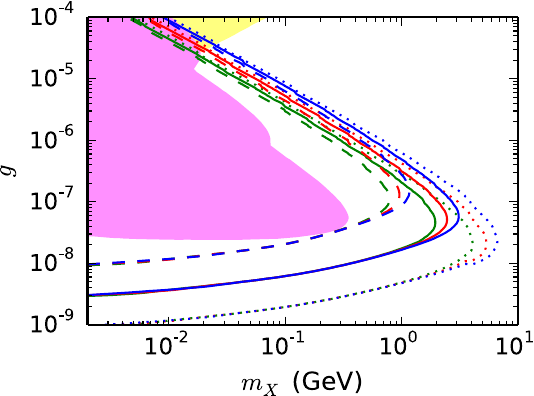}
\caption{Left: contours of expected number of signal events for the $U(1)_{e-\mu}$ model;
the beam energy is taken to be $E_{\rm beam}=125$ (green), $250$ (red), and $500\, {\rm GeV}$ (blue);
the dotted, solid, and dashed lines are for $N_{\rm sig}=10^{-2}$, $1$, and $10^2$, respectively, taking $N_e=4\times 10^{21}$;
the mixing parameter is taken to be $\kappa_\epsilon=1$;
the pink and yellow shaded regions are excluded by beam-dump and neutrino-electron scattering experiments, respectively.
Right:  same as the left plot, but for the $U(1)_{e-\tau}$ model. 
Taken from Ref.~\cite{Asai:2021xtg}.
}
\label{fig:lepNGB-BD-ILC}
\end{figure}

Ref.~\cite{Asai:2021xtg} studies the prospects of searching for light and long-lived leptophilic gauge bosons (LGBs) in the beam-dump experiment using  $e^\mp$ beams at the ILC.
The experimental setup is similar to that shown in Fig.~\ref{fig:expSetupBD}.
The authors consider LGBs coupled to leptons $e, \mu, \tau$ with charges $Q_e, Q_\mu, Q_\tau$, respectively.
Three cases of $(Q_e, Q_\mu, Q_\tau) =$ (1, -1, 0), (1, 0, -1), or (0, 1, -1) are taken into account, corresponding to U(1)$_{e-\mu}$, U(1)$_{e-\tau}$, and U(1)$_{\mu-\tau}$ models, respectively.
With one-year operation, about $4 \times 10^{21}$ electrons and positrons are injected into the dump.
With the injection of the electron (or position) beam into the dump, the LGB (denoted as $X$) can be produced by the scattering process $e^\pm N \to e^\pm N^\prime X$ (with $N$ and $N^\prime$ being nuclei).
SM background is assumed to be negligible.

We extract Fig.~\ref{fig:lepNGB-BD-ILC} from Ref.~\cite{Asai:2021xtg} which shows the expected number of signal events $N_{\rm sig}$ for the cases of U(1)$_{e-\mu}$ and U(1)$_{e-\tau}$ models in the $g$ vs.~$m_X$ plane, where $g$ and $m_X$ denote the $X$ coupling to the charged leptons and the mass of $X$, respectively.
The dotted, solid, and dashed lines correspond to $N_{\rm sig} = 10^{-2}$, 1 and $10^2$, respectively, for the beam energy taken to be $E_{\rm beam} =$ 125 (green), 250 (red), and 500 GeV (blue).
Results for the $U(1)_{\mu-\tau}$ case are also given in Ref.~\cite{Asai:2021xtg}.

\begin{figure}[t]
\begin{center}
\includegraphics[height=0.3\textheight, width=0.75\textwidth]{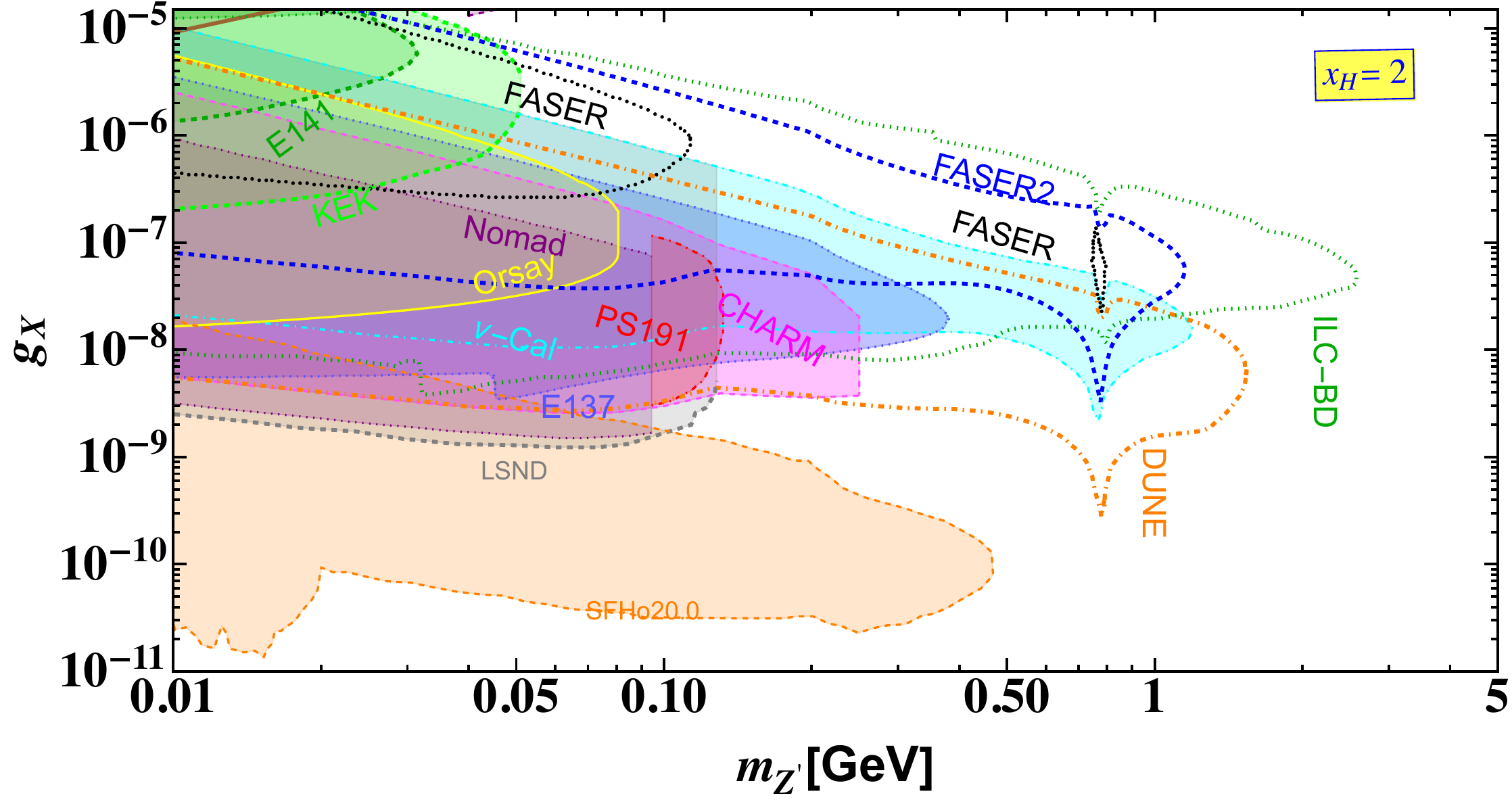}
\caption{Limits in the $m_{Z^\prime}^{}-g_X^{}$ plane for $x_H^{}> 0$ and $x_\Phi^{}=1$ considering $10$ MeV $\leq m_{Z^\prime}^{} \leq 5$ GeV, showing the regions that could be probed by FASER, FASER2, ILC-Beam dump, and DUNE.
The parameter space is compared with existing bounds from different beam-dump experiments and a cosmological observation of supernova SN1987A (SFH020.0), respectively.
Taken from Ref.~\cite{Asai:2022zxw}.
}
\label{fig:Zprime-BD-ILC-DUNE}
\end{center}
\end{figure}
Ref.~\cite{Asai:2022zxw} obtains bounds and sensitivities on the chiral $Z^\prime$ gauge boson in the electron-positron beam-dump experiment at the ILC, in the proton beam-dump experiment of DUNE, and the experiments of FASER(2).
It assumes 10-year running for the calculations of the future beam-dump experiments, and 150 fb$^{-1}$ (3 ab$^{-1}$) for LHC Run 3 (high-luminosity LHC).
The $Z^\prime$ particle, being lighter than 5 GeV, is considered to be produced in rare meson decay and bremsstrahlung processes for all kinds of beam-dump experiments and additionally pair annihilation process for electron and positron beam-dump experiments.
Sensitivities in the parameter space projected for FASER, FASER2, DUNE, and ILC beam dumps are obtained and compared with the existing bounds from Orsay, Nomad, PS191, KEK, LSND, CHARM, and SN1987A.
Fig.~\ref{fig:Zprime-BD-ILC-DUNE} is taken from Ref.~\cite{Asai:2022zxw}, showing the limits in the $m_{Z^\prime}^{}-g_X^{}$ plane for the case with $x_H= 2$ and $x_\Phi=1$, where $x_H$ and $x_\Phi$ are $U(1)_X$ charge parameters and $g_X$ is the $U(1)_X$ coupling.
Further results with different $x_H$ values of 1 and 0.5 are also provided in Ref.~\cite{Asai:2022zxw}.

\subsection{Summary and Discussion}
\label{subsec:summary}

Searching for new particles is one of the key focuses of current collider-based particle physics.
Beyond precision measurements, these future experiments can also probe BSM physics in terms of light new physics, by searching for light, exotic states.
In particular, light new physics is predicted in many BSM theories to manifest itself as LLPs leading to striking collider signatures different from the conventional ones.
Such LLPs have a relatively long lifetime for various reasons including feeble couplings to other particles and suppressed phase space, and have been proposed for solving multiple issues present in the SM such as the non-vanishing neutrino masses and the dark matter.

\begin{figure}[t]
\begin{center}
\includegraphics[width=0.75\textwidth]{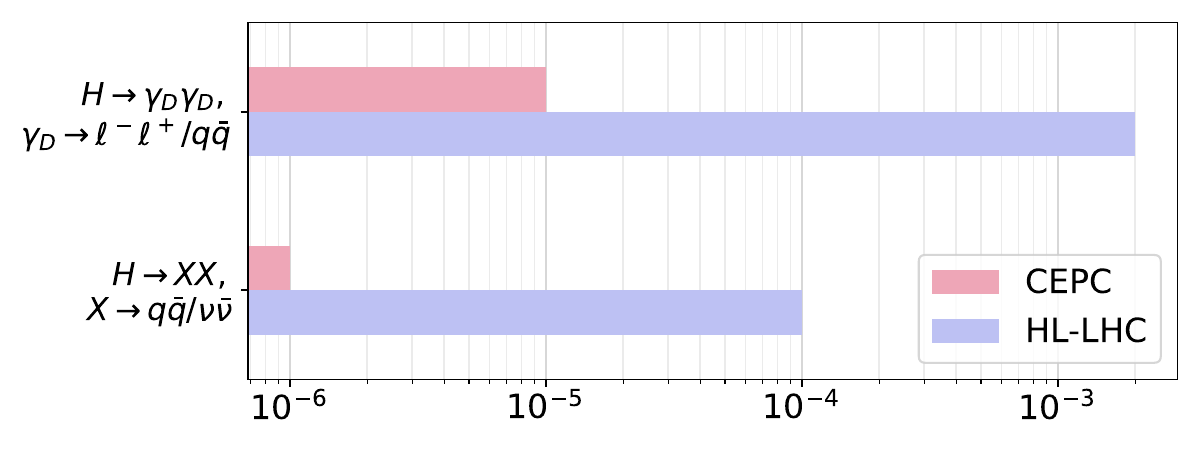}
\caption{The sensitivity reach for the Higgs decay to a pair of long-lived particle $X$ with subsequent decay $X \to \bar{b}b, \bar{\nu} \nu$ and a pair of long-lived dark photon $\gamma_D$ with subsequent decay $\gamma_D \to \bar{q}q, \bar{\ell} \ell$ for CEPC and HL-LHC projections.
}
\label{fig:sec-LLP-summary}
\end{center}
\end{figure}

Compared to hadron colliders such as the LHC, $e^- e^+$ colliders have unique characteristics; higher luminosity is expected, the collider environment is cleaner, the trigger requirement is typically looser, the absence of parton distribution in the electrons fixes the parton-level collision energy, etc.
These features are appealing for collider searches for BSM physics especially LLPs.
Indeed, various phenomenological analyses have been performed for LLP searches at future $e^- e^+$ colliders, and thus, as these experiments are currently in the stage of designing and planning, we find it timely to summarize the current status of these phenomenological studies and provide an outlook. In Fig.~\ref{fig:sec-LLP-summary}, we show the sensitivity reach for the Higgs decay to a pair of long-lived particle $X$ with subsequent decay $X \to \bar{b}b, \bar{\nu} \nu$ and a pair of long-lived dark photon $\gamma_D$ with subsequent decay $\gamma_D \to \bar{q}q, \bar{\ell} \ell$ for CEPC and HL-LHC projections~\cite{Liu:2018wte, ATLAS:2022hsp}, where CEPC is about two orders of magnitude better than HL-LHC.

We start with a detailed explanation of the typical computation procedure for the number of LLP signal events inside the fiducial volume of a detector.
This approach has been widely applied in the literature.
We then have reviewed the existing LLP phenomenological studies at future $e^- e^+$ colliders according to the associated detector type separately.
Mainly three types of experimental setups have been under discussion for LLP searches at future electron-positron colliders, namely main detector, far detector, and beam-dump experiment.
The main detector (near detector, abbreviated as ND) refers to the default local detector enclosing the IP, the far detector is an external, auxiliary detector with a macroscopic distance from the IP, and the beam-dump experiment puts a detector tens of meters behind the beam dump.
For the MD experiments, we have reviewed existing works considering mainly LLPs from rare Higgs and $Z$-bosons' decays, including new light scalar particles, dark photons, lightest neutralino in the RPV-SUSY, and ALPs.
Heavy Neutral Leptons (HNLs) have also been studied, produced either via direct collision or rare $Z$-boson decays.
Further studies include a long-lived stau in the SUSY, and a vector-like lepton with a long-lived accompanying scalar.
Then, FDs with various geometrical configurations have been investigated for their sensitivity reach to a new scalar particle mixed with the SM Higgs boson, HNLs or light neutralinos in the RPV-SUSY from $Z$-boson decays, as well as ALPs coupled to the SM photon or the $Z$-boson.
In general, we expect a modest improvement on the sensitivity reaches compared to the case with the default MD only.
Finally, a beam-dump experiment has been proposed mainly for electron-positron colliders with a CM energy between 250 GeV and 3 TeV.
Various theoretical scenarios including the HNLs, ALPs, new scalar particles, and new neutral gauge bosons have been extensively studied, showing excellent sensitivity reach compared to other existing or proposed experiments.

The MD usually has the best acceptance rates for the LLP signal events; however, its sensitivity reach often still suffers from certain background sources to various extent despite the relatively clean environment.
The proposed FDs, have a lower acceptance rate, because they usually cannot have an almost $4\pi$ solid-angle coverage; however, with a large distance between the FD and the IP, shielding such as lead and concrete can be instrumented in the space in between, effectively removing the background events.
For the FDs, in general we observe sensitivity reach to certain parts of the models' parameter space mildly beyond the region that could be probed by the MD.
The beam-dump experiments, set up at e.g.~the ILC, have been shown to be able to probe large parts of the parameter space where the other existing and proposed experiments are insensitive.

\begin{figure}[htbp]
    \centering
    \includegraphics[width=0.6\linewidth]{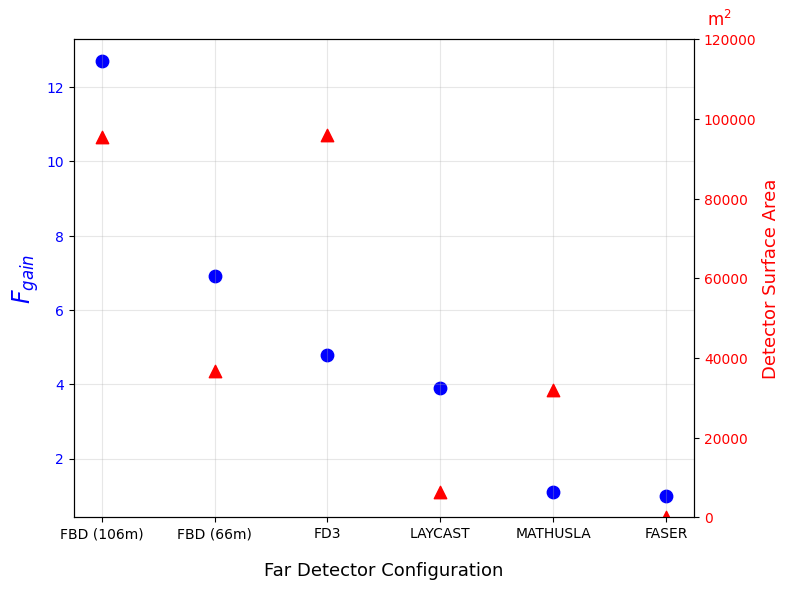}
    \caption{
     The gain factor \(F_\textrm{gain}\) (left axis) and the corresponding detector surface area (right axis) are shown for various far detector configurations. FBD (106m) has \(\Delta L\) set as 100m and FBD (66m) has \(\Delta L\) set as 60m.}
    \label{fig:gain_area_experiments}
\end{figure}

Fig.~\ref{fig:gain_area_experiments} compares the LLP detection sensitivity and surface area requirements for various far detector configurations, including the Far Barrel Detector (FBD)~\cite{Zhang:2024bld}, FD3 of the FADEPC~\cite{Wang:2019xvx}, 
LAYCAST~\cite{Lu:2024fxs},
MATHUSLA~\cite{MATHUSLA:2020uve}, and FASER~\cite{FASER:2019aik}. 
The gain factor \(F_\textrm{gain}\), as defined in Ref.~[69], and the corresponding detector surface area have been calculated and presented. The results demonstrate that achieving a higher gain factor typically requires a larger detector surface area. The findings also emphasize the important role that geometric coverage plays in enhancing sensitivity. Far detectors with limited geometric coverage, such as MATHUSLA and FASER, offer restricted sensitivity improvements compared to the main detector with full geometric coverage.

Before ending this section, we briefly discuss the requirements for program tools, civil engineering and detector technology to promote the LLP studies at high-energy electron-positron colliders.

\underline{Requirements for the program tools:}
performing such phenomenological studies often requires MC simulation.
Currently the tools publicly available and specifically for computing LLP signal-event rates, such as \texttt{FORESEE}~\cite{Kling:2021fwx}, \texttt{DDC}~\cite{Domingo:2023dew}, and \texttt{SensCalc}~\cite{Ovchynnikov:2023cry}, do not include detector-level effects such as smearing and detector efficiencies.
In order to perform realistic estimates for LLP sensitivity reach with fast-simulation tools, these effects should be included within a sophisticated framework.

\underline{Requirements for the civil engineering:}
the FD and beam dump experiments require large spaces for installing the experimental facilities.
Depending on the available space, the FD can be placed in a cavern inside or in the vicinity of the experimental hall, in the transverse direction with respect to the IP.
It is important that designing and planning of such experiments should already proceed before the construction of the main experiment starts.
The same conclusion applies for beam-dump experiment proposals, too.

\underline{Requirements for the detector technology:}
The detector technologies need also further development.
For reconstructing a displaced vertex, the trackers should be equipped with better tracking resolution, and the relevant analysis algorithms should be further developed as well.
In order for the FDs to have good sensitivities, a large solid-angle coverage is required; given the macroscopic distance to the IP, the FD should have a large volume.
If displaced signatures arise pointing to new physics, it would be important to determine the properties of the observed LLP such as its mass and to identify the LLP decay products.
For these purposes, installing magnetic fields and implementing PID (Particle IDentification) strategies and methods such as ionization measurement and Cherenkov imaging would be helpful.
Moreover, if timing detectors with a timing resolution in the picosecond regime can be installed, multiple purposes can be fulfilled, such as providing timing information for the LLPs, realizing event correlation between the MD and FD, and removing potential background events from the cosmic rays.
See, for example, Refs.~\cite{Anchordoqui:2021ghd,Aielli:2019ivi,Mao:2023zzk} for similar discussion on precision timing at the LHC FDs.
We further emphasize the importance of shielding that should be instrumented for FDs and the beam-dump experiments; it can consist of concrete, lead, metal, or magnetic field, and sufficiently remove background events.
Finally, the associated cost should be contained to an acceptable level.

In this section, we clearly observe the mutual complementarity between the various experimental setups at future high-energy $e^- e^+$ colliders, for LLP searches.
In the future, beyond the usual cut-based studies, machine-learning techniques can also be applied to further improve the discovery potential of these experiments; see, e.g.~Ref.~\cite{Zhang:2024bld} for a relevant discussion.
Moreover, additional collider experiments with other beam-type setups, can be considered for LLP searches as well, including $\mu^+ \mu^-$~\cite{Delahaye:2019omf,Long:2020wfp,AlAli:2021let,Accettura:2023ked,MuonCollider:2022xlm}, electron-muon~\cite{Lu:2020dkx,Bouzas:2021sif}, electron-proton~\cite{Dainton:2006wd,LHeCStudyGroup:2012zhm,Klein:2018rhq,FCC:2018byv,FCC:2018vvp}, and muon-proton~\cite{Cheung:2021iev,Caliskan:2017meb,Acar:2017eli} collisions.
We expect them to be sensitive to parameter space inaccessible by hadron or $e^- e^+$ colliders.
\clearpage

\section{Supersymmetry}
\label{sec:SUSY}

\subsection{Introduction}
Supersymmetry (SUSY) provides an intriguing candidate to solve the gauge hierarchy problem in the Standard Model (SM). 
The Supersymmetric Standard Models (SSMs) introduce many appealing features, including gauge coupling unification, and a natural mechanism for electroweak symmetry breaking. In addition, the SSM provides a comprehensive theory framework for novel phenomena. For example, the Lightest Supersymmetric Particle (LSP) can serve as a viable dark matter (DM) candidate with R-parity conservation. The SUSY searches at the LHC have already set strong constraints on SSMs~\cite{ATLAS:2020xgt, Aad:2020sgw,Aad:2019vvi, Aad:2019qnd}. While the CEPC is designed for lower energy, it covers important parameter spaces that are difficult for a high-energy $pp$ collider to reach. This is particularly important for the search for a number of new physics particles~\cite{Han:2013usa,Han:2014xoa,Kobakhidze:2016mdx,Han:2016gvr,Abdughani:2017dqs,Ren:2017ymm,Duan:2017ucw,Duan:2018rls,Abdughani:2019wss,Abdughani:2019wai,Gu:2020ozv,Abdughani:2021pdc,Wang:2021bcx,Gu:2021lni,Lv:2022pme,Flambaum:2022lkc}. 
In addition, the precision measurements at the CEPC can probe SUSY even without directly producing new particles~\cite{Hu:2014eia,Cao:2014rma,Liu:2017msv,Duan:2018cgb,Han:2020exx,Athron:2022uzz}. 

In this section, we will focus on recent studies of the CEPC on several scenarios with light electroweakino and sleptons, and SUSY dark matter and long-lived SUSY particle searches are discussed separately in Section~\ref{sec:DMDS} and  Section~\ref{sec:LLP}. 
These scenarios can have various physics motivations (for some examples, see Refs.~\cite{Han:2013usa,Leggett:2014mza, Leggett:2014hha, Du:2015una, Li:2015dil}). For both electroweakino and slepton searches, the discovery potential can reach up to the kinematic limit of the collider $\sqrt{s}/2$, and it is capable of covering interesting compressed mass parameter regions. Heavier selectron particle can be searched from light bino pair production via a t-channel selectron, which can break through the collision energy limits. 
Moreover, these studies also serve as references and benchmark searches at other proposed electron-positron colliders, such as the FCC-ee and the ILC, given their similar nature in the physical process, center-of-mass energies, and target luminosities. 

\subsection{Light electroweakino searches} 
\begin{figure}[t]
  \centering
      {\includegraphics[height=0.26\textheight, width=.4\textwidth]{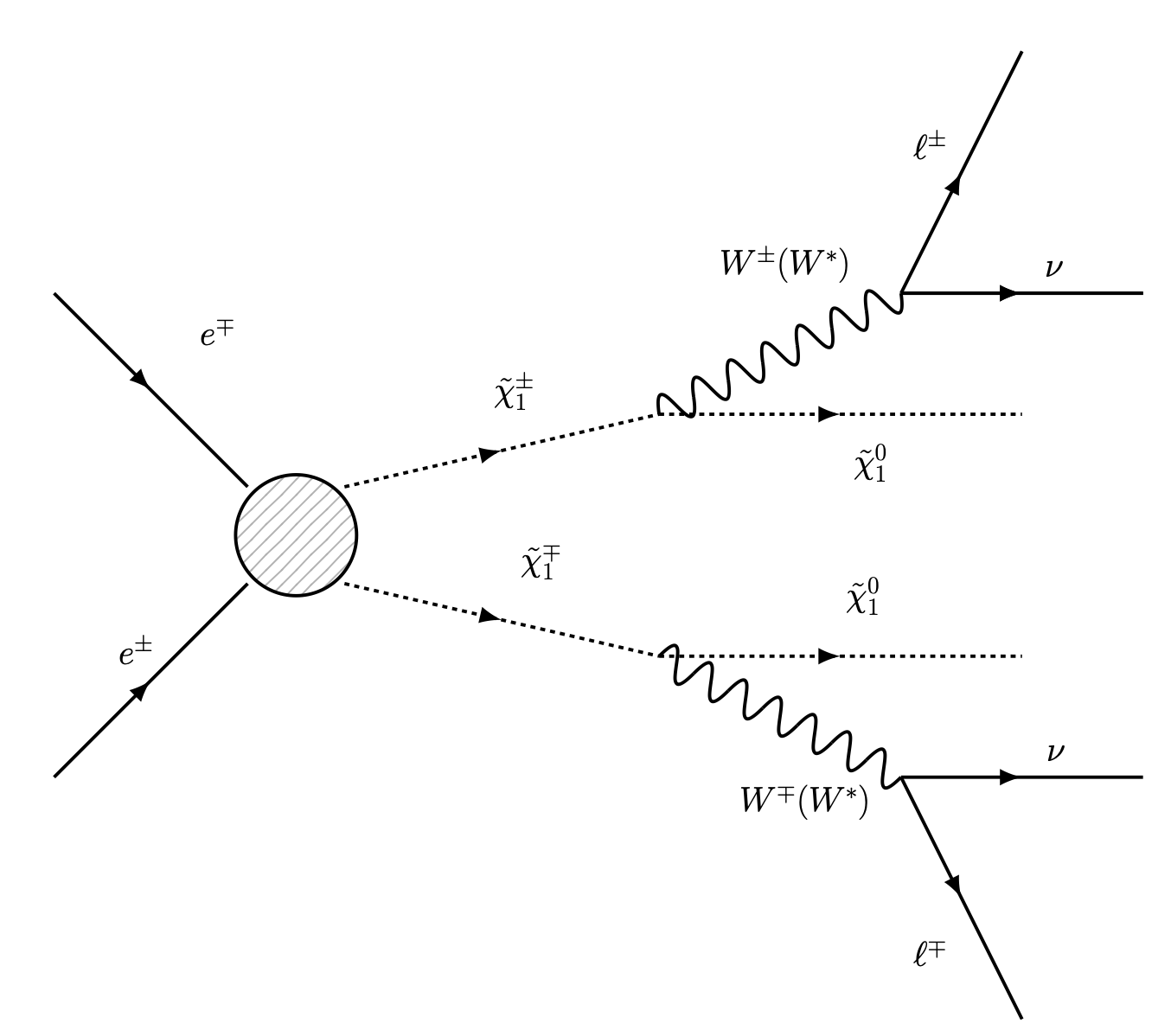}}
    {\includegraphics[width=0.7\textwidth]{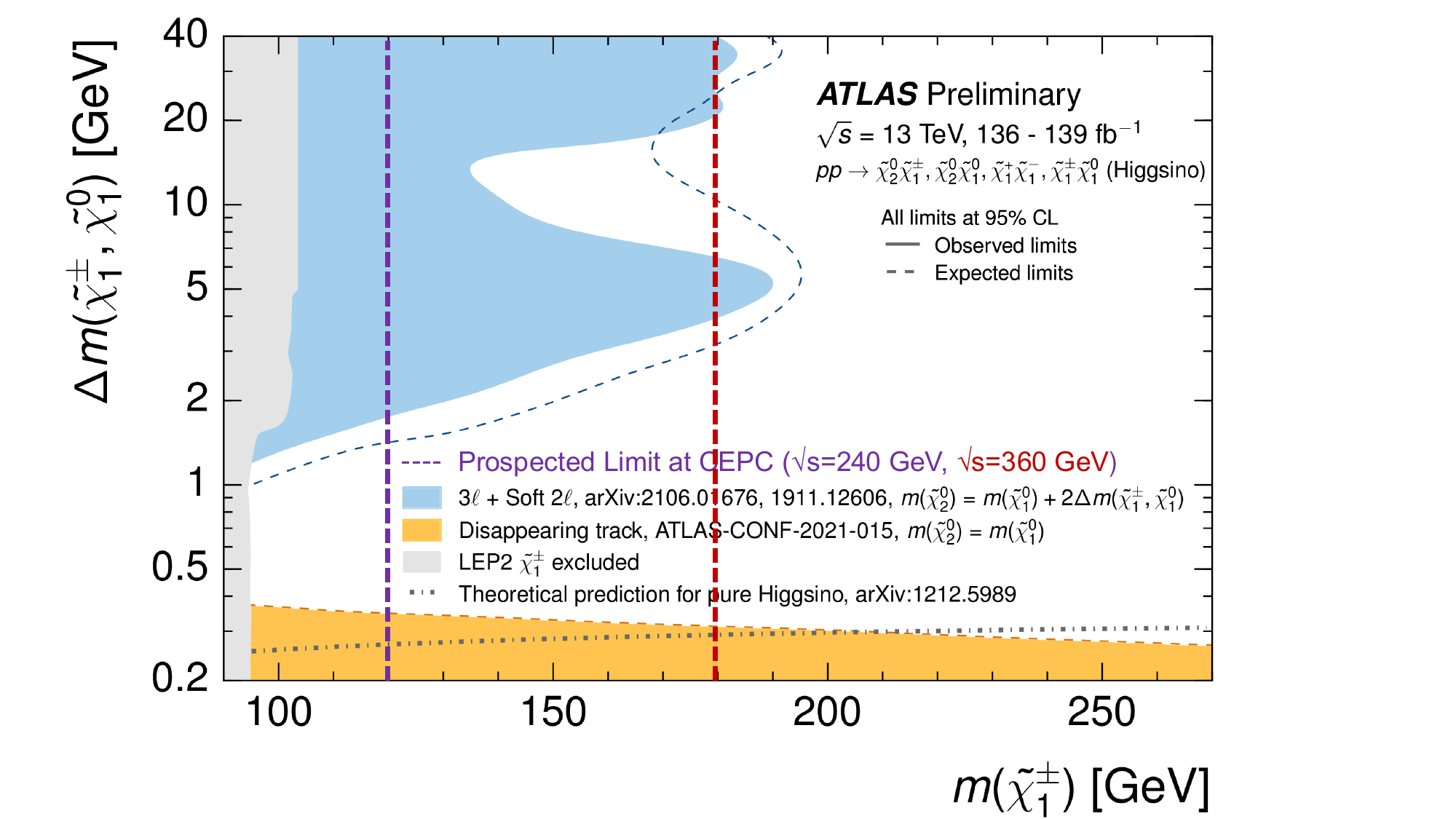}}
    \caption{\textit{Top}: Representative diagram for the pair production of charginos and subsequent decay into dilepton final states. \textit{Bottom}: Projected limits at the CEPC (purple dotted) in comparison ATLAS observed and expected exclusion limits on simplified SUSY models for chargino-pair production with Higgsino-like LSP. Limits from the LEP are shown in light grey.}
    \label{fig:ewksl}
  \end{figure}

The light Higgsino particles, well-motivated by naturalness conditions, tend to have small mass splitting among the chargino and neutralino~\cite{Han:2013usa}. Therefore, they are quite challenging to be probed in the LHC experiments due to their very soft decay products. The sensitivity studies for chargino pair production by considering scenarios for both a Bino-like and a Higgsino-like neutralino as the LSP have been performed and published in Ref.~\cite{Yuan:2022ewk}. With a cleaner collision environment and better low-energy particle reconstruction, the CEPC has shown the capability of probing the very compressed region, as shown in Fig.~\ref{fig:ewksl}. 

\begin{figure*}[h!]
\centering
    {\includegraphics[height=0.32\textheight, width=.45\textwidth]{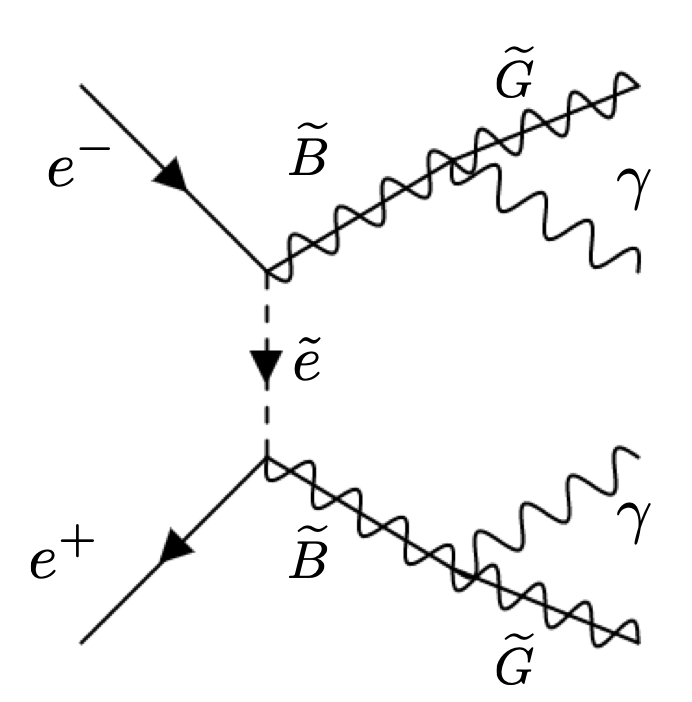}}
\includegraphics[height=0.38\textheight,width=1.0\textwidth]{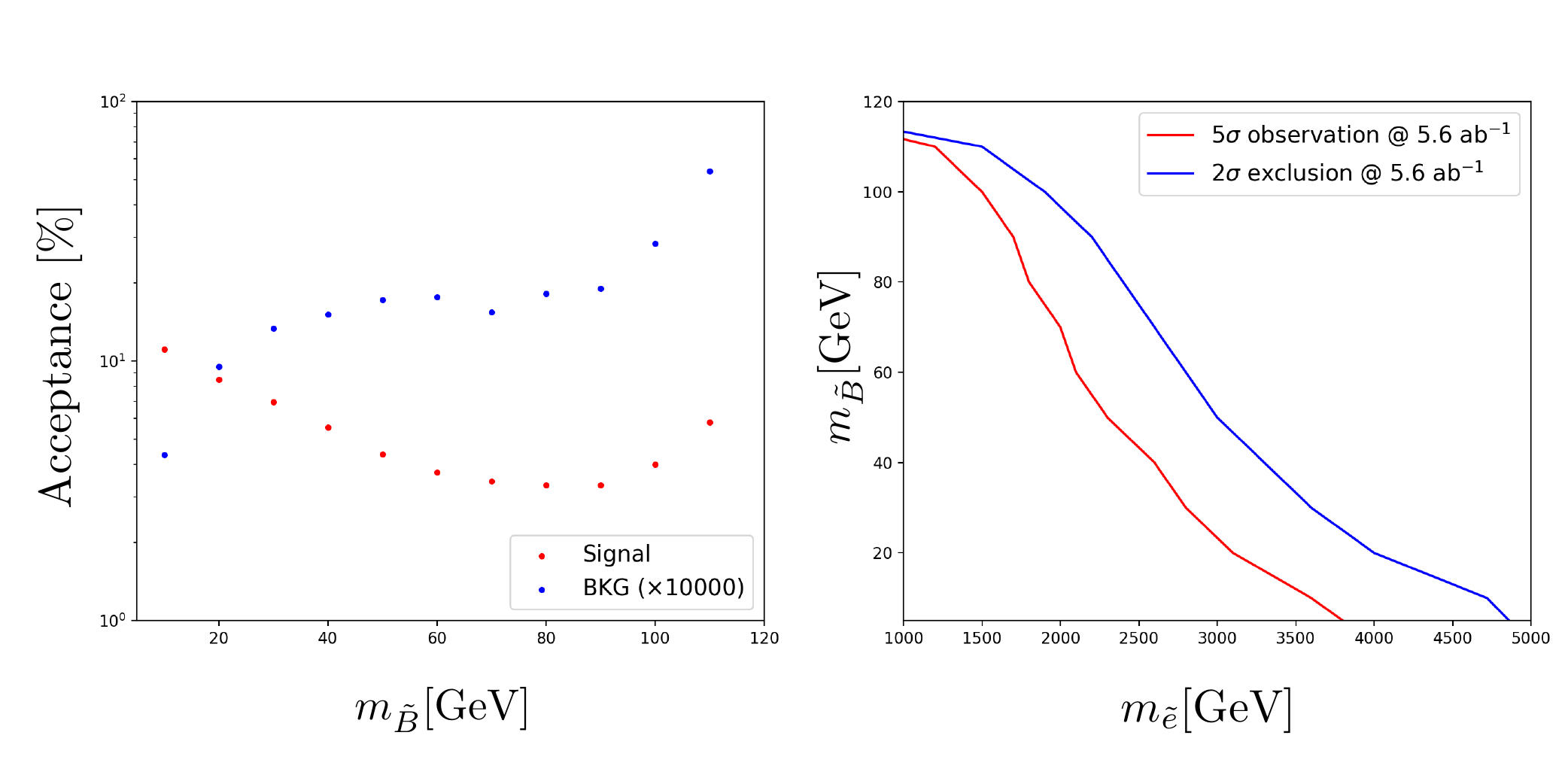}
\vspace*{-1.0cm}
\caption{
\textit{Top}: representative diagram illustrating light bino production.
\textit{Bottom left}: acceptance of signal and background processes as functions of $m_{\tilde{B}}$. 
Here the acceptance of the background process has been multiplied by 10,000. 
\textit{Bottom right}: 2$\sigma$ exclusion and 5$\sigma$ observation limits on the $m_{\tilde{e}} - m_{\tilde{B}}$ plane 
at a future lepton collider running with an integral luminosity 5.6 ab$^{-1}$ and center-of-mass energy 240 GeV. 
Regions below the red (blue) curves are observable (excluded).}
\label{limits} 
\end{figure*}

A light bino ($m_{\tilde{B}}\sim \mathcal{O}(10)$ GeV) scenario is motivated by Gauge-Mediated SUSY Breaking~\cite{Pagels:1981ke}. This scenario has been studied for the CEPC~\cite{Chen:2021omv}, where bino is the next-to-lightest supersymmetric particle (NLSP), while the lightest supersymmetric particle (LSP) and dark matter candidate is the sub-GeV gravitino ($\tilde{G}$). 
The process of bino pair production via a $t$-channel selectron ($\tilde{e}$), where bino subsequently decays to gravitino and a photon, has been considered, namely $e^+ e^- \to \tilde{B}\tilde{B} \to \gamma\gamma \tilde{G}\tilde{G}$.
The corresponding dominant background process is $e^+ e^- \to \gamma\gamma \nu\bar{\nu}$ (via $Z$ boson invisible decay), which has been suppressed by a dedicated cut-flow using a list of kinematic variables with good signal and background separation power.
The study shows that the CEPC can be sensitive to selectron lighter than 4.5 TeV (2 TeV) with bino mass around 10 GeV (100 GeV), as shown in Fig.~\ref{limits}. This is much larger than the current LHC bound which excludes selectron mass only up to several hundred GeV.

\subsection{Light slepton searches} 

\begin{figure}[t]
  \centering
      {\includegraphics[height=0.3\textheight, width=.8\textwidth]{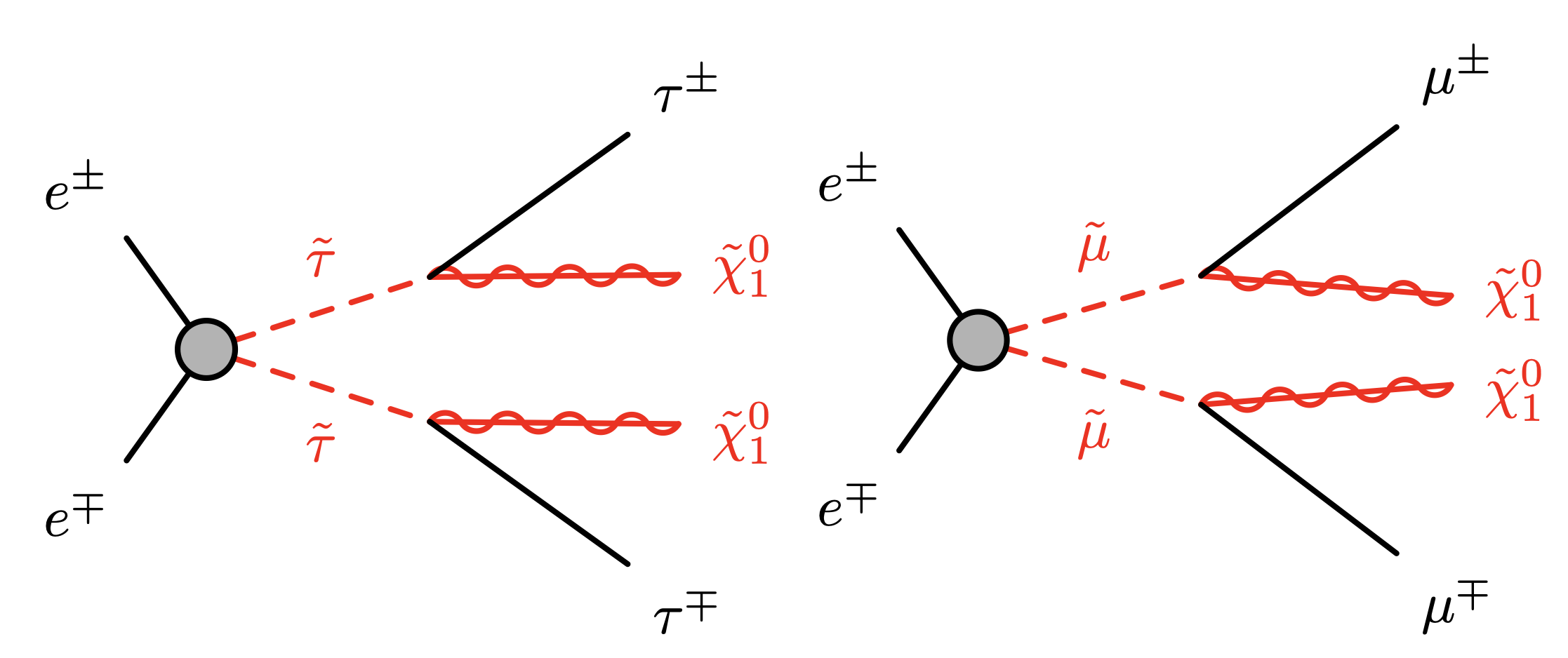}}
    {\includegraphics[height=0.3\textheight, width=.49\textwidth]{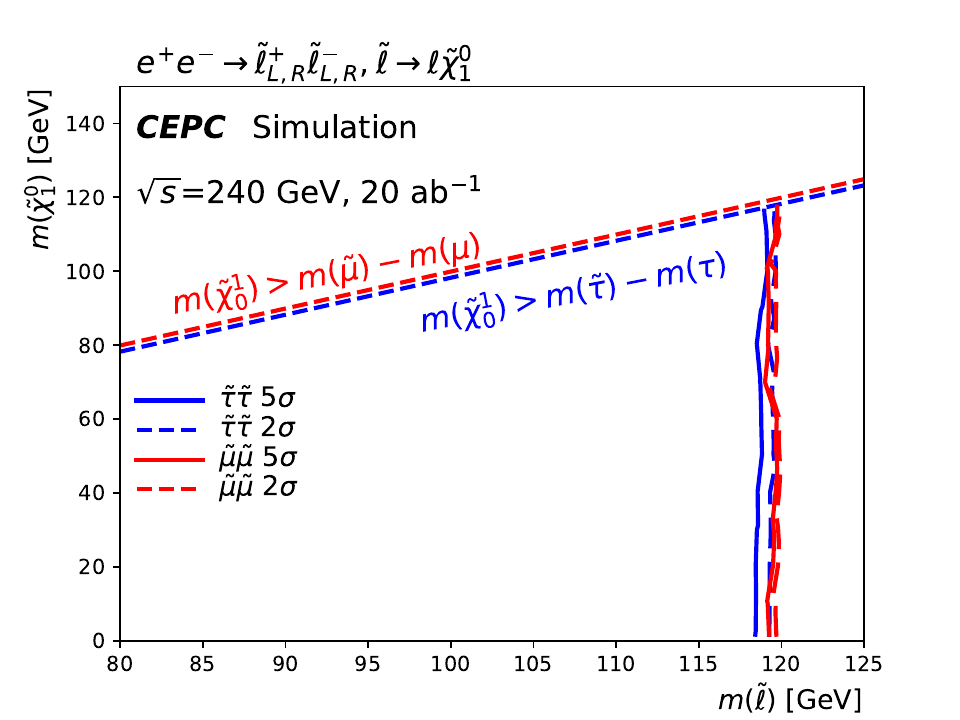}}
    {\includegraphics[height=0.3\textheight, width=.49\textwidth]{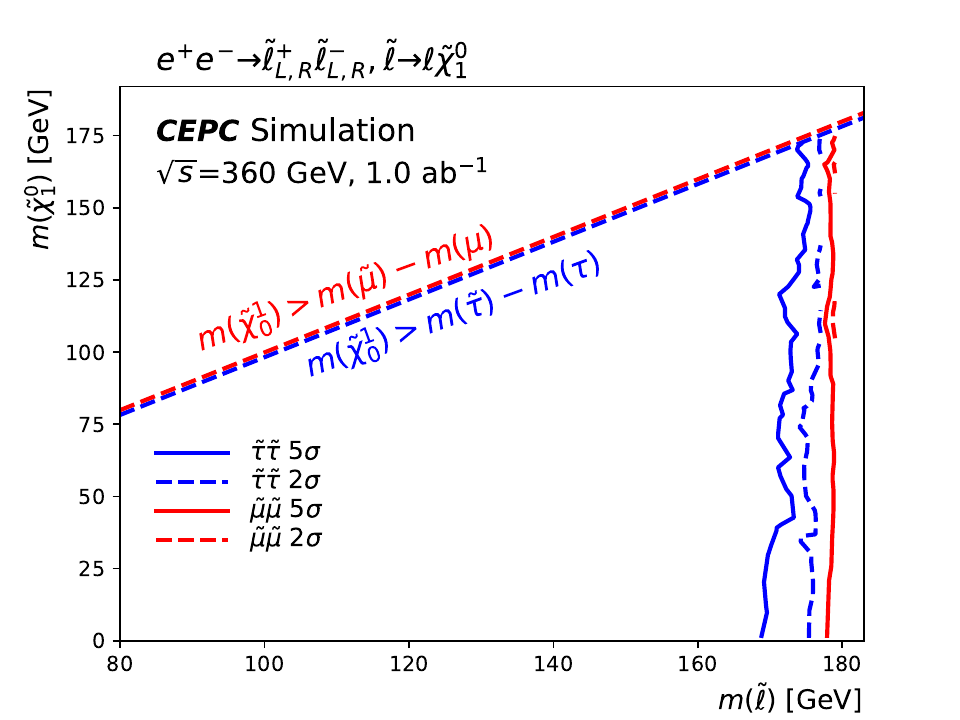}}
    \caption{\textit{Top}: representative diagram illustrating the pair production of charged staus (smuons) and subsequent decay into a two-tau (two-muon) final state.
    \textit{Bottom}: The 5$\sigma$ discovery contour (solid line) and 2$\sigma$ exclusion contour (dashed line) for the direct stau production and direct smuon production with 5\% flat systematic uncertainty. Left (Right) plot presents the center-of-mass energy of 240 (360) GeV. }
    \label{fig:stausmuon}
  \end{figure}

Light smuon and stau particles are interesting to search for in their own right, and they are also favored by SUSY explanations to the latest muon $g$-2 excess. Similar to case with light electroweakinos, it is challenging to search for light smuons and staus at the LHC, especially in the compressed region where their masses are close to that of the LSP. Note that such compressed regions are favored by dark matter relic density requirements, and they have been explored for the CEPC~\cite {Yuan:2022ykg}. This study assumed a flat 5\% systematic uncertainty, the discovery sensitivity was predicted to reach 119 (118) GeV for smuon (stau) mass via direct smuon (stau) production, and the results are  shown in Fig.~\ref{fig:stausmuon} (left). This demonstrate that the CEPC can fill a significant  gap to narrow down on the compressed region. 

The center-of-mass energy of the CEPC will be upgraded to 360 GeV after its ten-year running at 240 GeV as a Higgs factory. A dedicated sensitivity study on direct stau and smuon pair production at the CEPC with $\sqrt{s}$ = 360~GeV has been carried out in Ref.~\cite {Lyu:2025slp}. This study assumed 1.0 ab$^{-1}$ integrated luminosity and a flat 5\% systematic uncertainty. The CEPC at 360 GeV is predicted to be capable of discovering the production of combined left-handed and right-handed stau up to 170~GeV; the discovery potential of direct smuon reaches up to 178 GeV with the same assumptions, as shown in Fig.~\ref{fig:stausmuon} (right). This result also gives a strong motivation to raise the center-of-mass energy of CEPC from 240 GeV to 360 GeV.

\begin{figure}[t]
  \centering
      {\includegraphics[height=0.25\textheight, width=.8\textwidth]{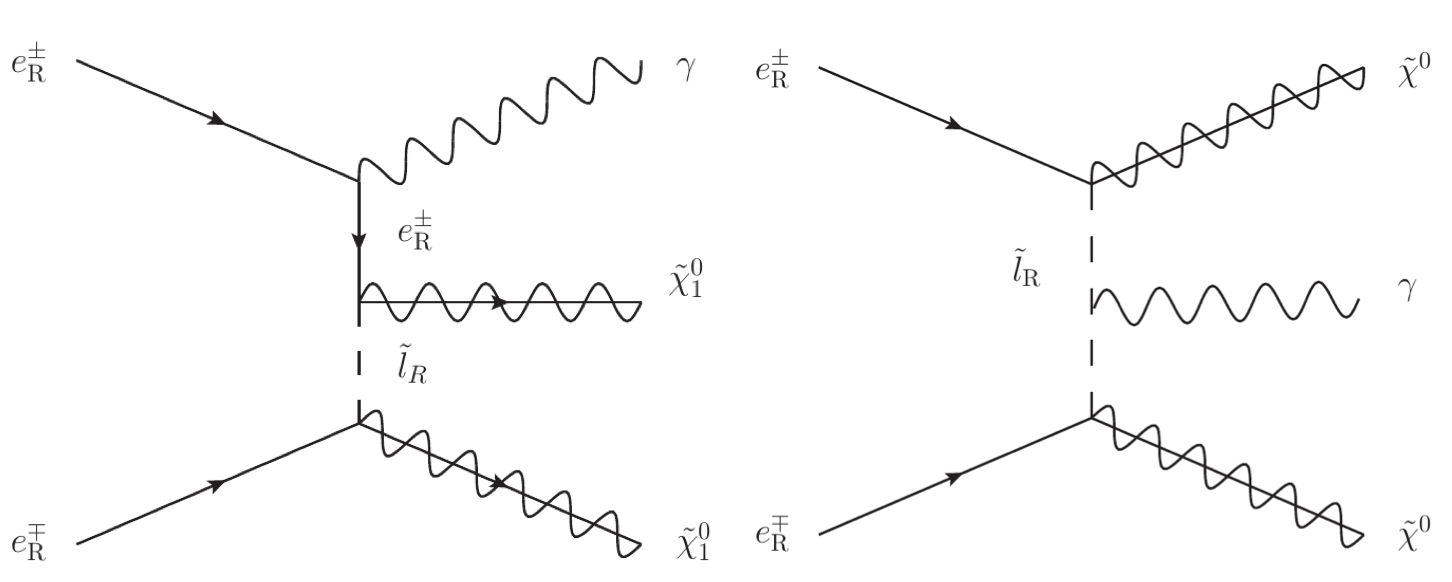}}
      {\includegraphics[height=0.35\textheight, width=.8\textwidth]{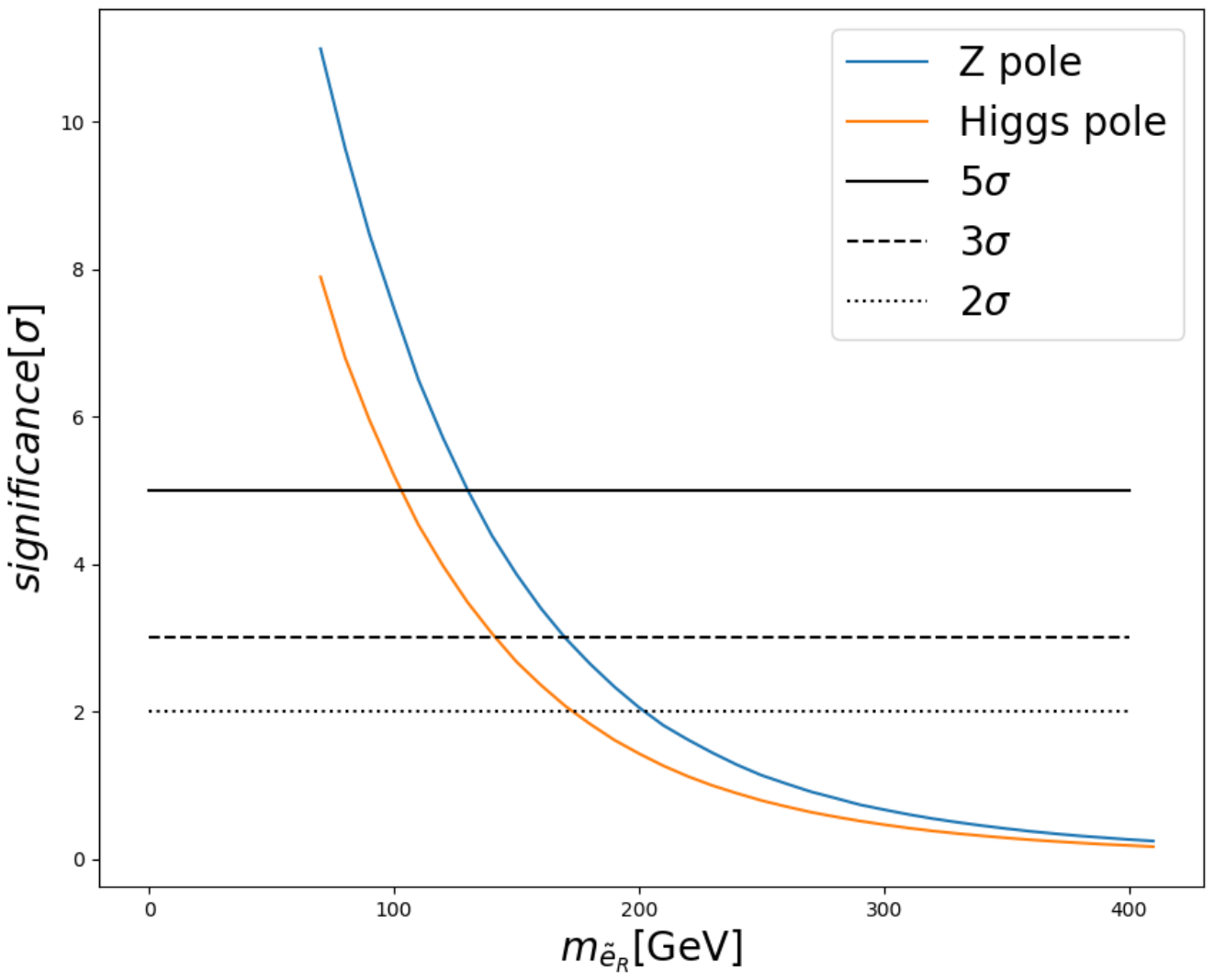}}
    \caption{Top: Representative diagram illustrating heavier selectron production. Bottom: Significance of exclusion ability.}
    \label{fig:heavierse}
  \end{figure}

Relatively heavier selectrons, especially above the kinematic limit for direct production, can also be searched for in the process~\cite{Ahmed:2022ude}: $e^{+}_{R} e^{-}_{R}\rightarrow {\tilde \chi_{1}^{0}({\rm bino})}+{\tilde \chi_{1}^{0}({\rm bino})}+{\gamma}$, as shown in Fig.~\ref{fig:heavierse}. The reach depends on the model assumptions. For example, if the relic abundance requirement is satisfied by the LSP annihilating through the Z-pole,  the right-handed selectron can be excluded up to 180 (210) GeV respectively at 3(2)$\sigma$. On the other hand, if the annihilation through the Higgs pole dominates, right-handed selectron will be excluded up to 140 (180) GeV at 3(2)$\sigma$.

\begin{figure*}[ht!]    	
 \begin{center} 
\includegraphics[width=0.6\linewidth]{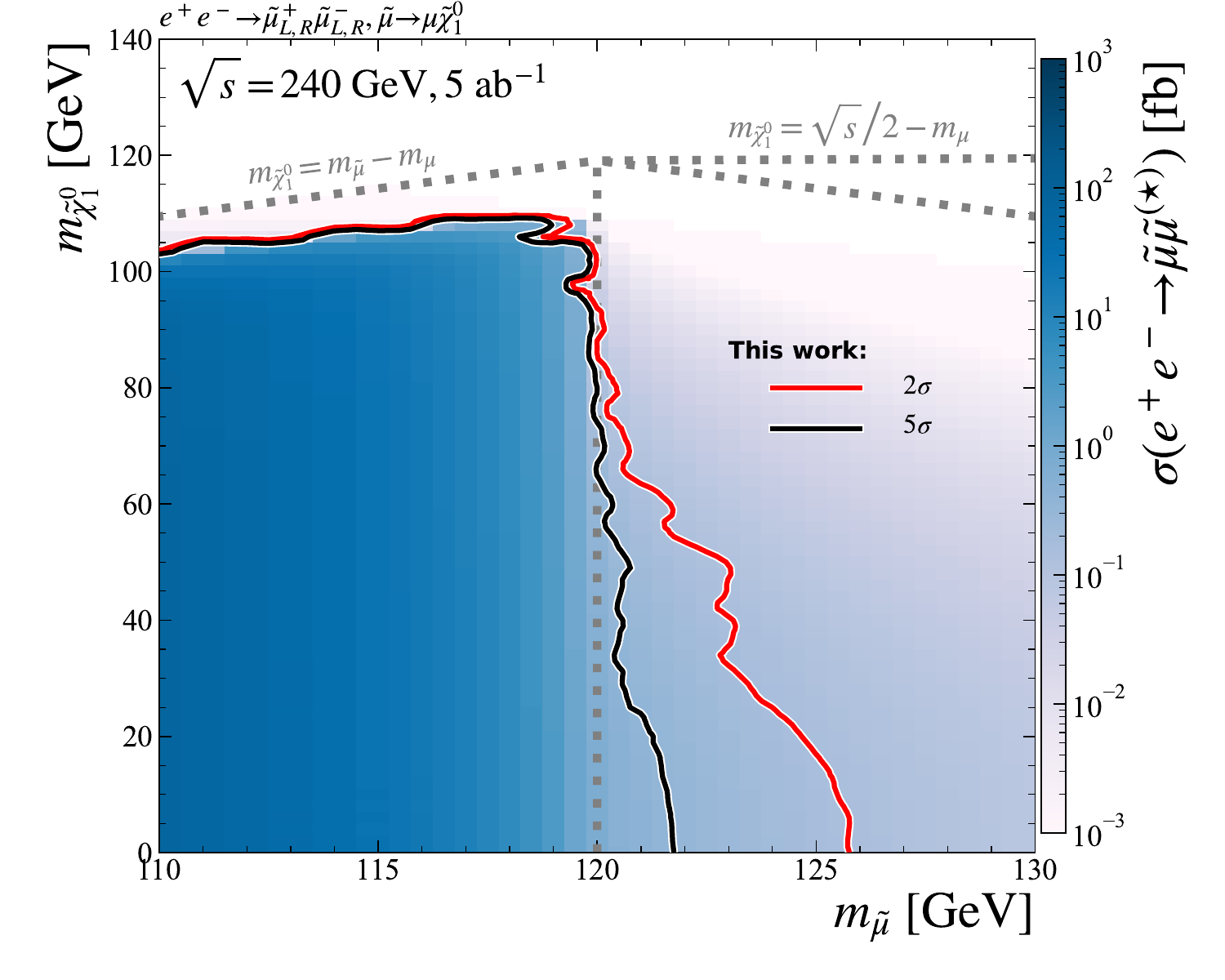}
 \end{center} 
\caption{The prospected exclusion contour and discovery contour at CEPC for the direct smuon production with $5\%$ flat systematic uncertainty.}
\label{fig:dect_cepc}
\end{figure*}

Another in-depth exploration is performed for the off-shell sparticle pair production at the CEPC~\cite{Yang:2022qga}. Assuming a flat $5\%$ systematic uncertainty, the discovery sensitivity reaches up to 126 GeV (122 GeV) for the smuon mass, as shown in Fig.~\ref{fig:dect_cepc}, which  
can break through the limits of the on-shell kinematic limit of $\sqrt{s}\big/2$ and enter the off-shell region in detecting new physical processes.

\begin{figure}[t]
    \centering
    \includegraphics[height=0.3\textheight, width=0.49\textwidth]{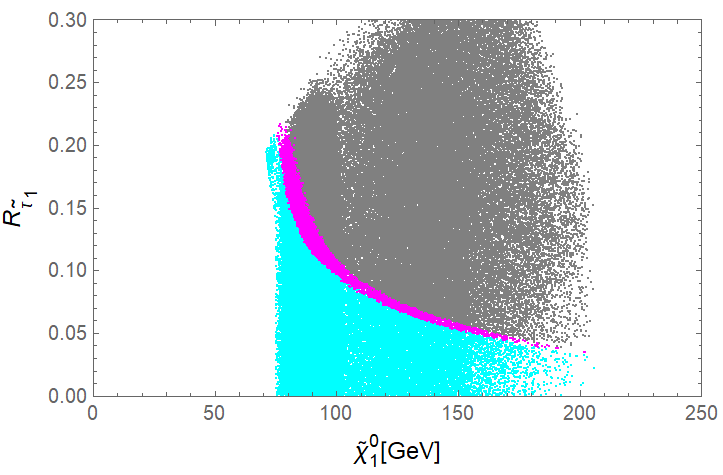}
    {\includegraphics[height=0.3\textheight, width=0.49\textwidth]{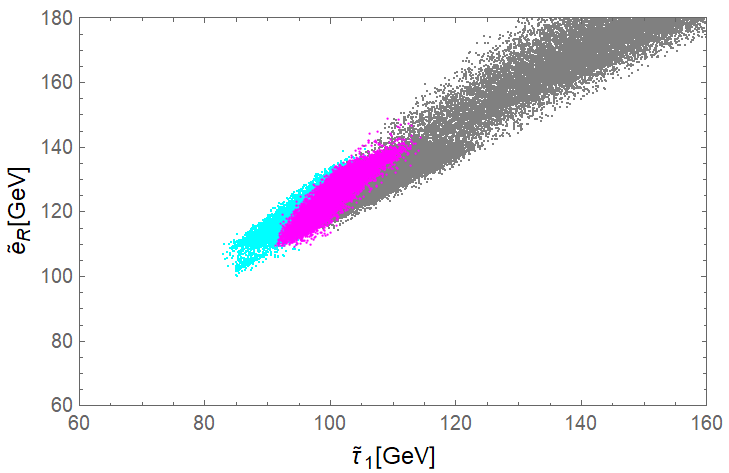}}
    \caption{Left: Bulk region in Generalized No-Scale $\mathcal{F}$-$SU(5)$.
    Right: Light right-handed slepton masses in this bulk region. Cyan, magenta, and gray points correspond to under-saturated, saturated, and over-saturated DM relic density.}
    \label{FSU5-stau1}
\end{figure}

A recent study~\cite{Li:2023pab} considered ${\cal F}$-$SU(5)$, {\it i.e.}, the flipped $SU(5)\times U(1)_X$ GUT model~\cite{Barr:1981qv} with extra TeV-scale vector-like particles~\cite{Jiang:2006hf} that have been constructed systematically in local F-theory model building~\cite{Jiang:2008xrg, Jiang:2009za}. Alternatively, these models can also be realized in free fermionic string constructions~\cite{Lopez:1992kg}. Super-Natural SUSY via No-Scale SUGRA is a natural resolution to the SUSY EW fine-tuning (EWFT)
problem, but it can not realize
the specific scenario with a light bino LSP due to a correlation of the bino mass with the Wino and gluino masses. Therefore, the Generalized No-Scale SUGRA is proposed, where effective super-natural SUSY can be realized. To uncover the `bulk region' for relic density, only light right-handed sleptons can be considered given that the LHC SUSY searches indicate that all other sfermions must be heavy. 
It is shown that the ratio of the mass difference $\mathcal{R}_{\phi}\equiv (m_{\phi}-m_{\Tilde{\chi}_{1}^{0}})/m_{\Tilde{\chi}_{1}^{0}}$ plays an important role, and $\mathcal{R}_{\phi} \gtrsim 10\%$ is a conservative criterion to formulate the bulk region via traditional annihilation-dominated thermal freeze-out.
The bulk region stau and selectron mass relations are illustrated in FIG.~\ref{FSU5-stau1}, where all points satisfy current experimental constraints, and $\mathcal{R}_{\Tilde{\tau}_{1}}$ plots as a function of the Bino-like neutralino $m_{\Tilde{\chi}_{1}^{0}}$. 
The mass hierarchy in $\mathcal{F}$-$SU(5)$ is $m_{\Tilde{\chi}_{1}^{0}}\textless m_{\Tilde{\tau}_{1}} \textless m_{\Tilde{e}_{R}} = m_{\Tilde{\mu}_{R}}$, hence, $\mathcal{R}_{\Tilde{e}_{R}}$ always exceeds $\mathcal{R}_{\Tilde{\tau}_{1}}$. If the Bino contributes all the DM abundance, the ratio $\mathcal{R}_{\Tilde{\tau}_{1}}\gtrsim 10\%$ implies $m_{\Tilde{\chi}_{1}^{0}}\leq 103.0$ GeV. 
The upper limits on the $\Tilde{\tau}_{1}$ and $\Tilde{e}_{R}$ masses are around 115 GeV and 150 GeV, respectively. 
Recognizing that these right-handed sleptons and Bino LSP are $naturally$ light, thus, the LSP has not been fine-tuned to fortuitously conform to the Planck satellite $5 \sigma$ relic density observations. These light sleptons could conceivably be observed at the Future Circular Collider (FCC-ee)~\cite{FCC:2018byv,FCC:2018evy} at CERN and the Circular Electron-Positron Collider (CEPC)~\cite{CEPCStudyGroup:2018ghi} with its sensitivity specified in Ref.~\cite{Yuan:2022ykg}.

\subsection{Summary}
\begin{figure}[t]
    \centering
    \includegraphics[width=0.8\textwidth]{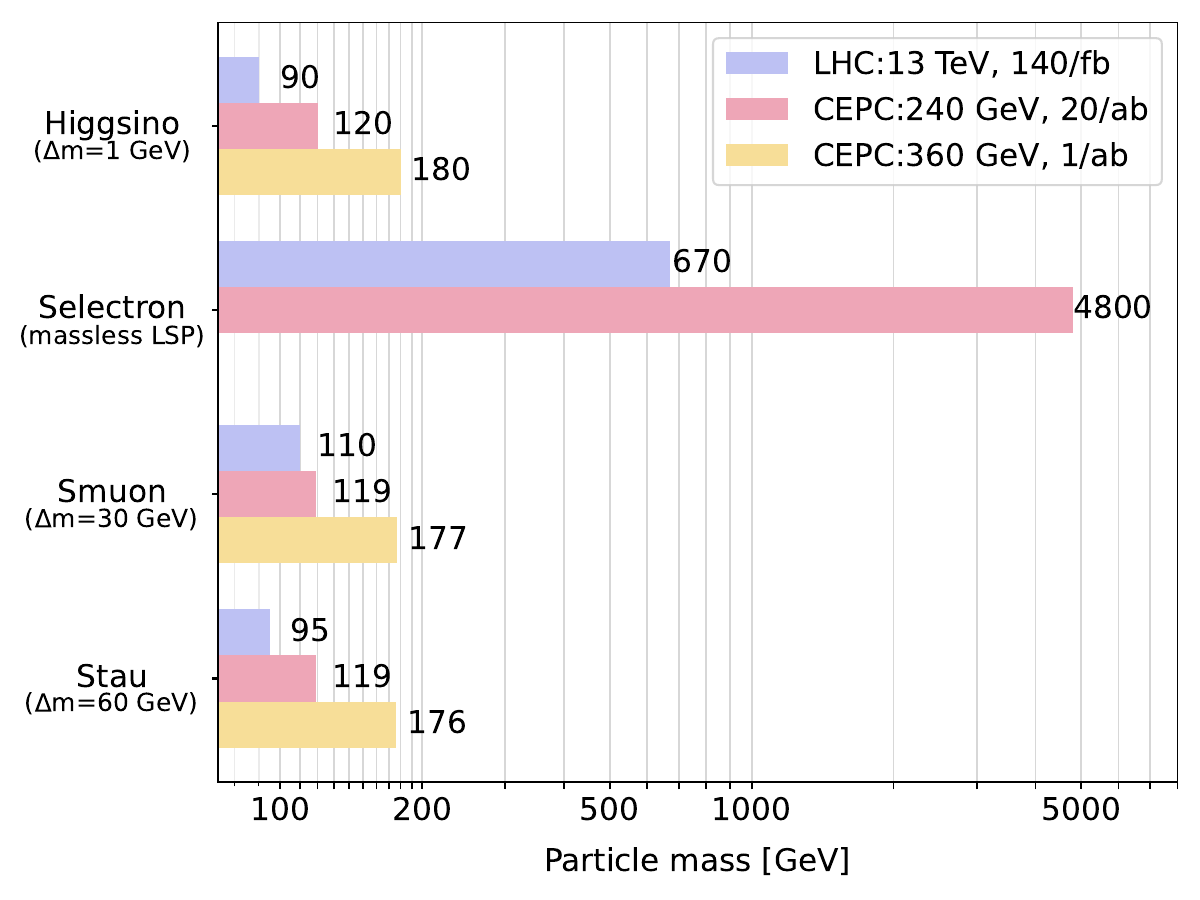}
    \caption{The exclusion reaches for higgsino and slepton at CEPC and LHC. The compressed SUSY particles are picked for the comparison except for selectron.}
    \label{SusySummary}
\end{figure}

In summary, among all the searches discussed in this chapter, which summarized in Table \ref{table:SUSYSummary}, it is observed that the discovery potential is primarily constrained by the detector kinematics for $s$-channel SUSY production. However, when a supersymmetric particle can be produced via the $t$-channel or when an off-shell supersymmetric particle is produced via the s-channel, it is possible to break through the collision energy limits and probe heavier supersymmetric particles. Furthermore, for most of the studies presented below, the results are based on a center-of-mass energy of 240 GeV. A dedicated search for light smuons and staus extended this study to 360 GeV, reflecting an increase in sensitivity. A similar conclusion of enhanced sensitivity is expected for other searches when the center-of-mass energy reaches 360 GeV. CEPC is an excellent discovery machine for light SUSY particles, especially has absolutely advantage at compressed scenarios with small mass splitting between NLSP and LSP compared with LHC, as shown in Fig.~\ref{SusySummary}.

\begin{table}[]
\scriptsize
\begin{tabular}{|c |c |c|c| c| c | c|}
 \hline\hline
 Search & Production &  $\sqrt{s}$ [GeV] & $\mathcal{L}[ab^{-1}]$ &Sensitivity & Figs. & Ref.   \\ \hline
 
  \multirow{2}*{Light electroweakino} & chargino pair  & 240 & 5.05 & chargino excluded up to 120~GeV & \ref{fig:ewksl} & \cite{Yuan:2022ewk} \\ \cline{2-7}
 &  $e^+ e^- \to \tilde{B}\tilde{B} \to \gamma\gamma \tilde{G}\tilde{G}$. & 240 & 5.6 & selectron excluded up to 4.5 TeV  & \ref{limits}  & \cite{Chen:2021omv} \\ \hline

 \multirow{3}*{Light slepton} & smuon pair &240 &20 & smuon excluded up  119~GeV & \ref{fig:stausmuon} & \cite{Yuan:2022ykg} \\ \cline{2-7}
 & stau pair &240 &20 & stau excluded up 119~GeV & \ref{fig:stausmuon} & \cite{Yuan:2022ykg} \\ \cline{2-7}
 & smuon pair &360 &1 & smuon excluded up  177~GeV & \ref{fig:stausmuon}& \cite{Lyu:2025slp} \\ \cline{2-7}
 & stau pair &360 &1 & stau excluded up  176~GeV & \ref{fig:stausmuon} & \cite{Lyu:2025slp} \\ \cline{2-7}
   
 &  $e^{+}_{R} e^{-}_{R}\rightarrow {\tilde \chi_{1}^{0}({\rm bino})}+{\tilde \chi_{1}^{0}({\rm bino})}+{\gamma}$ & 240 & 3 & right-handed selectron excluded up to 210~GeV & \ref{fig:heavierse} & \cite{Ahmed:2022ude}\\  \cline{2-7}

  & off-shell smuon pair & 240 &5 & smuon excluded up 126~GeV & \ref{fig:dect_cepc} & \cite{Yang:2022qga} \\ \cline{2-7}

& ${\cal F}$-$SU(5)$ & - & - & upper limits on  $\Tilde{\tau}_{1}$  up to 115 GeV  & \ref{FSU5-stau1} & \cite{Li:2023pab}  \\  \cline{2-7}
  & ${\cal F}$-$SU(5)$ & - & - & upper limits on   $\Tilde{e}_{R}$ up to  150 GeV & \ref{FSU5-stau1} & \cite{Li:2023pab}  \\ \hline
\hline
\end{tabular}
\caption{Recent results from CEPC studies on SUSY. The first column lists the signal signatures, the second column presents the corresponding production modes, the third and fourth columns provide the center-of-mass energy and the integrated luminosity, the fifth column shows the sensitivity to the coupling, suppression scale, or branching ratios, and the last column provides the references. }
\label{table:SUSYSummary}
\end{table}

\clearpage
\section{Flavor Portal New Physics}
\label{sec:FlavorPortal}

The CEPC, when running at the $Z$ pole, allows us to probe the flavor structures of the $Z$ couplings to matter fields with extremely high precision. This also allows us to get very large samples of all $b$-flavored hadrons, charmed hadrons, and $\tau$ leptons, with large boost in a very clean environment. These features make the CEPC also a flavor factory, which has a unique sensitivity to a large number of flavor processes that are generally not accessible at the current LHCb and Belle II experiments~\cite{CEPCStudyGroup:2018rmc,CEPCFlavorWP}. It is also stated in many dedicated studies~\cite{CEPCFlavorWP} that the CEPC's higher energy (operating at the $Z$ pole or even higher), clean lepton collision environment and advanced detector system provide many powerful, sometimes unique, flavor physics probes (see also~\cite{Zheng:2020ult,Li:2022tlo,Aleksan:2021gii,Aleksan:2021fbx,Amhis:2021cfy,Kamenik:2017ghi,Li:2020bvr,Monteil:2021ith,Chrzaszcz:2021nuk,Dam:2018rfz,Qin:2017aju,Li:2018cod,Calibbi:2021pyh,Altmannshofer:2023tsa} for examples at CEPC or parallel proposals). 

The CEPC holds significant potential in the realm of flavor physics that enables the imposition of various indirect constraints on BSM. A crucial aspect to be mentioned is the role of the Cabibbo–Kobayashi–Maskawa (CKM) matrix, which, in the SM, is defined and primarily measured via charged-current couplings at the electroweak scale or lower. This matrix is particularly sensitive to the general NP couplings, especially since many high-precision or highly sensitive flavor processes are not charged currents in nature. Governed by electroweak gauge coupling and unitarity, the CKM matrix imposes stringent restrictions on the BSM couplings to quarks. Therefore, unless the BSM physics introduces couplings to the SM quarks that adhere to the specific patterns of the CKM matrix, rather than being of arbitrary flavor structures, the scales of NP probed by flavor physics will be significantly elevated, extending well beyond the TeV range.

The sensitivity of the flavor sector in the SM to NP is underscored by several factors. Firstly, the suppression of flavor-changing neutral-current (FCNC) processes by the Glashow–Iliopoulos–Maiani (GIM) mechanism~\cite{Glashow:1970gm} in the SM is a key aspect. This makes the FCNC processes very rare in the SM and, at the same time, quite sensitive to various BSM physics. A typical example here is the $B_{d,s}^0-\bar{B}_{d,s}^0$ mixings. Secondly, the charged lepton-flavor violation (cLFV) represents a special case of FCNC, exhibiting even greater suppression due to the light neutrino masses. Any observation of a cLFV process is definitely a distinct sign of BSM physics. Thirdly, the flavor sector of the SM is characterized by several (approximate) symmetries. One of them is the approximate, accidental symmetry among the three generations, leading to lepton flavor universality (LFU) that is only slightly violated by the lepton Yukawa couplings. Other examples are the conservation of lepton and baryon numbers in collider environments. Violations of these symmetries are therefore indicative of BSM physics. Lastly, the decays of heavy flavored states in the SM are markedly suppressed, both by the small off-diagonal elements of the CKM matrix and the hierarchy between fermion masses and the electroweak scale. This results in narrow decay widths ($\lesssim 10^{-12}$ GeV) for many flavored particles, rendering them long-lived. Consequently, the small SM widths significantly amplify the impact of any NP amplitudes. To better demonstrate the sensitivity of the flavor sector to NP, one can write the reach of the BSM scale $\Lambda_{\rm NP}$ from dim-6 operators via charged-current $b\to c$ transitions as:
\begin{equation}
    \Lambda_{\rm NP} \sim (G_F|V_{cb}|\delta_{\rm exp})^{-\frac{1}{2}}  \sim (1.5~\text{TeV})\times \delta^{-\frac{1}{2}}_{\rm exp}\,,
    \label{eq:deltaFCCC}
\end{equation}
where $\delta_{\rm exp}$ is the relative precision reached in the experiment. Similarly, from the FCNC $b\to s $ transitions one may obtain
\begin{equation}
    \Lambda_{\rm NP} \sim \Big(\frac{\alpha}{4\pi} \frac{m_t^2}{m_W^2}G_F|V_{tb}V_{ts}^\ast|\delta_{\rm exp}\Big)^{-\frac{1}{2}}  \sim (30~\text{TeV})\times \delta^{-\frac{1}{2}}_{\rm exp}~.
    \label{eq:deltaFCNC}
\end{equation}
In both cases the interpretation is model-dependent, which are common in indirect searches.

The aforementioned arguments are actually pertinent to both direct and indirect search methodologies. Indirect probes, particularly abundant in the flavor sector, transform effectively all flavor physics studies into potential avenues for BSM investigation. This is due to the fact that certain processes are more germane to NP searches for a variety of reasons. Conversely, light BSM degrees of freedom may be produced through their interactions with the SM fermions. This possibility is imminent since the SM gauge interactions of these light states must be sufficiently suppressed to align with the existing experimental data. The following highlights are some of the most prominent examples of BSM searches at CEPC interfacing with flavor physics, with key numbers summarized in Table~\ref{tab:flavorBSM}. For a more thorough review on the potential and phenomenology of flavor physics at CEPC, we refer the readers to Ref.~\cite{CEPCFlavorWP}.

\begin{table}[h!]
\centering
\resizebox{0.67\textwidth}{!}{
\renewcommand{\arraystretch}{1.2}
\tabcolsep=0.35cm
\begin{tabular}{ccc}
\toprule[1pt]
Measurement & Current Limit & CEPC \\ 
\midrule
BR($Z \to \tau \mu $) & $<6.5\times 10^{-6}$ &   $\mathcal{O} (10^{-9})$ \\
BR($Z \to \tau e $) & $<5.0\times 10^{-6}$ &   $\mathcal{O} (10^{-9})$ \\
BR($Z \to \mu e $) & $<7.5\times 10^{-7}$ &   $10^{-8}-10^{-10}$ \\
\midrule
BR($\tau \to \mu\mu\mu$) &$< 2.1\times 10^{-8}$& $\mathcal{O}(10^{-10})$ \\
BR($\tau \to eee$)  & $<2.7\times 10^{-8}$ &  $\mathcal{O}(10^{-10})$ \\
BR($\tau \to e\mu\mu $) &  $<2.7\times 10^{-8}$ & $\mathcal{O}(10^{-10})$ \\
BR($\tau \to \mu ee $) &   $<1.8\times 10^{-8}$ & $\mathcal{O}(10^{-10})$ \\
BR($\tau \to \mu\gamma$) & $<4.4\times 10^{-8}$ & $\mathcal{O}(10^{-10})$ \\
BR($\tau \to e\gamma $) & $<3.3\times 10^{-8}$ &   $\mathcal{O}(10^{-10})$ \\
\midrule
BR($B_s \to \phi\nu\bar{\nu} $) & $<5.4\times 10^{-3}$ &   $\lesssim 1\%$ (relative)  \\
BR($B^0 \to K^{\ast 0 }\tau^+ \tau^- $) & - &   $\lesssim \mathcal{O}(10^{-6})$ \\
BR($B_s \to \phi\tau^+ \tau^- $) & - &   $\lesssim \mathcal{O}(10^{-6})$  \\
BR($B^+ \to K^+\tau^+ \tau^- $) & $<2.25\times 10^{-3}$ &   $\lesssim \mathcal{O}(10^{-6})$ \\
BR($B_s \to \tau^+ \tau^- $) & $<6.8\times 10^{-3}$ &   $\lesssim \mathcal{O}(10^{-5})$ \\
BR($B^0 \to 2\pi^0 $) & $\pm 16\%$  (relative) &   $\pm 0.25\%$ (relative)  \\
$C_{CP}(B^0 \to 2\pi^0 $) & $\pm 0.22$  (relative) &   $\pm 0.01$ (relative)  \\
\midrule 
BR($B_c\to \tau \nu$) & $\lesssim$ 30\% & $\pm$ 0.5\% (relative)\\
BR($B_c\to J/\psi\tau \nu$)/BR($B_c\to J/\psi\mu \nu$) & $\pm$ 0.17 $\pm$ 0.18 &  $\pm 2.5\%$ (relative) \\
BR($B_s\to D_s^{(\ast)}\tau \nu$)/BR($B_s\to D_s^{(\ast)}\mu \nu$) & - &  $\pm 0.2\%$ (relative) \\
BR($\Lambda_b\to \Lambda_c\tau \nu$)/BR($B_c\to \Lambda_c\mu \nu$) & $\pm$ 0.076 &  $\pm 0.05\%$ (relative) \\
 \midrule
   BR$(\tau \to \mu X_{\rm inv})$ & $7\times 10^{-4}$ & (3-5)$\times 10^{-6}$ \\ 
   BR$(B \to \mu X_{\rm LLP}(\to \mu\mu))$ & - & $\mathcal{O}(10^{-10})$ (optimal) \\ 
\bottomrule[1pt]
\end{tabular}}
\caption{Preliminary sensitivities of flavor physics probes at CEPC to BSM physics, adapted from Ref.~\cite{CEPCFlavorWP}. The notation $X_{\rm inv}$ stands for an invisible narrow resonance and $X_{\rm LLP}$ represents a long-lived BSM particle. The limit for the LLP particle is obtained when its lifetime is optimal, as shown in the right panel of Fig.~\ref{fig:cLFV}. For some channels with extremely high precision expected, the actual sensitivities will be mostly determined by systematic effects, which cannot be precisely evaluated at the current stage. Consequently, only the rough sensitivity levels are reported. See Ref.~\cite{CEPCFlavorWP} for more details.}
\label{tab:flavorBSM}
\end{table}

\subsection{cLFV processes}  
Reflecting on the cLFV searches at CEPC, the facility's significant integrated luminosity in the $Z$-factory mode notably enhances its capacity for exploring cLFV directly from $Z$ and $\tau$ decays. These two modes benefit immensely from the CEPC's design and energy range, making it an ideal platform for such investigations.

The cLFV in $Z$ decays is a primary focus, particularly in modes like $Z \to \mu e$, $Z \to \tau \mu$, and $Z \to \tau e$. Each of these decay processes provides a unique window into potential BSM physics. When searching for $Z\to \mu e$ decays, the bottleneck of detection comes from the small but non-negligible lepton misidentification rate. Conversely, in modes involving the $\tau$ lepton, the extra neutrino from $\tau$ decays, puts the detector's momentum/energy resolution in a more significant position. In general, the expected limits at CEPC will exceed the current best ones by more than two orders of magnitude. Note that testing cLFV NP at higher scales is also possible at CEPC. One of the most prominent examples would be the cLFV Higgs decays. The phenomenology of these modes, while being analogous, is slightly more intricate due to the additional production of $Z$ in the $e^+e^- \to HZ$ process. Although the absolute sensitivity in these Higgs decay modes might be limited by smaller statistics, their higher energy scales and couplings could provide valuable complementarity for NP searches.

In terms of $\tau$ decays, many CEPC flavor physics projects are proposed to investigate a wide array of cLFV processes. The decay modes like $\tau \to 3\mu$, $\tau \to \mu \gamma$, and $\tau \to e \gamma$ are of significant interest, since direct resonance peaks can be reconstructed in their final states. Due to the clarity of their phenomenology, they are chosen as benchmark channels. Besides, the heavy mass of the $\tau$ lepton compared to other leptons makes its collider behavior more akin to flavored hadrons, fitting well with the energy range and detector design of the CEPC. This aspect of $\tau$ decays offers an array of valuable channels for cLFV studies, further underscoring the CEPC's discovery potential.

\subsection {Decays of $b$-flavored and charmed hadrons}
Rare decays of $b$-flavored and charmed hadrons induced by FCNC transitions are inherently loop-suppressed in the SM, rendering them highly sensitive to potential contributions from BSM. These FCNC processes are intrinsically fascinating in the context of flavor physics. Within the SM, they provide insights into hadronic properties and offer a means to constrain critical parameters like the CKM matrix elements. Meanwhile, for some of the charged-current induced decays of these hadrons, especially the (semi)leptonic decays involving the $\tau$ lepton, evidences from several experiments during the past years suggest that LFU is violated in the tau sector. The magnitude of such a violation, if fully confirmed, is a clear indication of BSM physics. Investigating such possibilities with the highest precision is then a great opportunity to explore the nature of leptons. 

As a powerful facility for flavor physics, CEPC also presents unique opportunities for studying many rare FCNC decays of $b$-flavored and charmed hadrons, which may be challenging to probe at other facilities. Such modes often involve final-state particles that demand exceptional energy/momentum resolution for precise reconstruction. Additionally, the inherently rare nature of these decays necessitates a very low background level for effective study, a condition that is readily met by the capabilities of CEPC. Importantly, many FCNC processes in the SM exhibit non-trivial CP-violating properties, and play a pivotal role in flavor physics. Measuring these CP properties not only deepens our understanding of flavor physics, but also serves as a powerful tool for identifying new sources of CP violation beyond the SM CKM matrix. Prominent examples of such processes include $b \to s\tau\tau$ transitions, which involve multiple neutrinos in the final state, and $B_{(s)} \to \pi^0\pi^0 \to 4\gamma$ decays. The study of these processes at CEPC could provide crucial insights into the intricacies of flavor physics and CP violation, shedding light on phenomena that lie beyond our current understanding of the SM.

In parallel with the phenomenological studies of various FCNC transitions, the charged-current induced transitions of $b$-flavored hadrons at CEPC, especially the $b\to c\ell(\tau)\nu$ processes, are being studied. Some of the most prominent examples include the $B_c\to \tau \nu$ and $B_c\to \ell \nu$ decays, which are challenging to measure precisely at other facilities. This is due to the rarity of $B_c$ meson and the elusive nature of $\ell(\tau)+\nu$ final states. To better constrain the forms of BSM physics behind, other relevant modes with distinct hadron dynamics like $B_s\to D_s^{(\ast)}\ell(\tau)\nu$, $B_c\to J/\psi\ell(\tau)\nu$, and $\Lambda_b\to \Lambda_c \ell(\tau)\nu$ are also studied. The characteristically relative sensitivity to these decays achieved at CEPC is of $\mathcal{O}(10^{-2})$ or less, which can also be interpreted in terms of EFT operators with a scale of multi-TeV according to Eq.~\eqref{eq:deltaFCCC}. 

\subsection{Light BSM degrees of freedom from flavor transitions} 

\begin{figure}[t]

    \adjustbox{valign=c}{\includegraphics[scale=0.36]{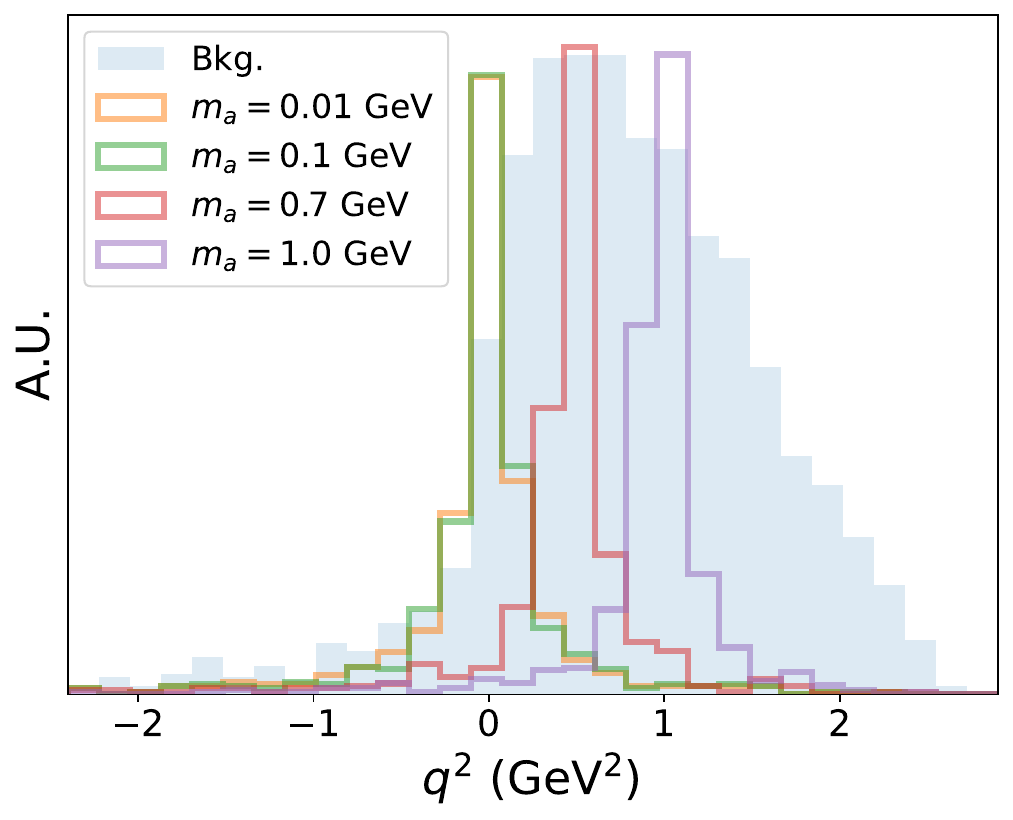}}
    \adjustbox{valign=c}{\includegraphics[scale=0.72]{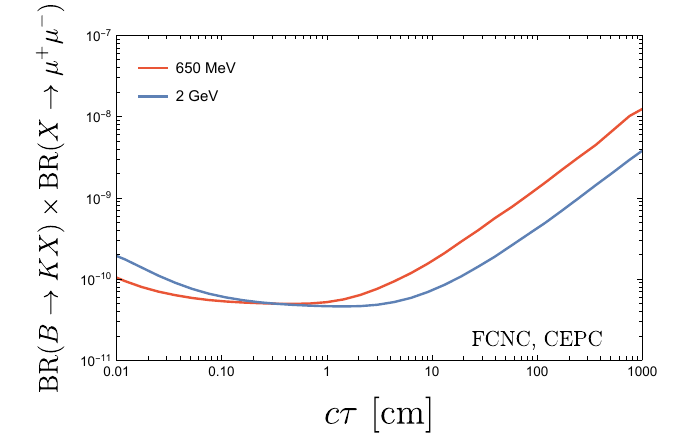}}
    
    \caption{Left: Reconstruction of invisible particle mass squared $q^2 \equiv (p_\tau-p_\mu)^2$ from $\tau$ decays to $\mu$ and an invisible BSM particle. Right: Projected sensitivity on FCNC $B\to K X_{\rm LLP}(\to \mu^+\mu^-)$ as a function of $X_{\rm LLP}$ lifetime. Two mass benchmarks are shown as blue and red curves, respectively. Both plots are taken from~\cite{CEPCFlavorWP}.}
    \label{fig:cLFV}
\end{figure}

Among the ongoing research at CEPC, two benchmark cases are currently in the preliminary stage, focusing on detecting BSM states emerging from either cLFV or quark FCNC processes. These benchmarks are distinguished by the nature of the BSM state involved: one involves an invisible BSM state, while the other involves an LLP in the final state.

For the scenarios involving invisible new particles, the reconstruction process presents a significant challenge, particularly in the preliminary studies that utilize fast detector simulation instead of a full simulation. The crux of reconstructing such final states lies in relying on the global energy-momentum conservation within the detector system. Supplementary information, such as track impact parameters, is also crucial as it provides additional constraints for the reconstruction process. Given the potential complexity of the reconstruction algorithms required, current studies at CEPC are focusing on inclusive $Z \to \tau \tau + X_{\rm inv}$ processes, where $X_{\rm inv}$ is the invisible new particle. In these processes, the extra particle $X$ is either generated via cLFV decays of the form $\tau \to \ell X$ or emerges from final-state radiations. This particle could be an ALP or a dark photon, characterized by suppressed or invisible decays.

While other neutrinoless processes are also relevant, they necessitate different reconstruction procedures. By integrating detailed information about the collision point, track momenta and impact parameters, along with global energy data, the current study demonstrates the feasibility of completely reconstructing a light invisible state from $\tau$ decays, provided all these data elements are properly integrated. This achievement is illustrated in the left panel of Fig.~\ref{fig:cLFV}, showcasing the capability of CEPC in detecting and analyzing such elusive BSM states. Depending on the new particle's mass, the expected sensitivity on BR($\tau\to \mu X_{\rm inv}$) may reach a level of $\mathcal{O}(10^{-6})$ or less.

Following the first benchmark study at CEPC focusing on invisible light BSM states, the second benchmark study shifts attention to light LLPs emerging from heavy-quark FCNC transitions. These light LLPs could be exemplified by ALPs, dark photons, or other NP models. A key aspect of this study is the model-independent nature of the search, with the LLPs being describable by a few effective parameters, such as their mass and decay length. In this context, the underlying UV physics does not significantly alter the search strategy.
The most distinctive signal feature in this study is a clearly reconstructed displaced decay vertex. This is vital for identifying the presence of the LLP. However, due to the extremely low target rate - with current constraints on exotic branching ratios producing LLPs from $B$-meson decays typically below $10^{-5}$ - it is unlikely that more than one displaced vertex will be observed in a given event.

The current analysis primarily focuses on the decay mode $X_{\rm LLP} \to \mu^+ \mu^-$, where the phenomenology at CEPC is expected to be straightforward, characterized by a low background level and a high signal efficiency. The production of the LLP is hypothesized to occur via the exclusive decay $B \to K^{(\ast)} + X_{\rm LLP}$. Notably, such processes can be generated even in the absence of flavor-violating interactions in some NP models. The preliminary results of this study are concisely summarized in the right panel of Fig.~\ref{fig:cLFV}. At CEPC the light LLPs could also be produced from Higgs decays or, more importantly, $Z$ exotic decays~\cite{Cheng:2021kjg,Cheng:2024hvq,Cheng:2024aco}. The flavor portal production then serves as a complement to LLP searches. 

Crucially, the sensitivity of this search to the LLP signal is closely tied to the particle's proper decay length. This sensitivity varies depending on the distance the LLP travels before decaying, and is found to peak around the centimeter scale. This dependency highlights the intricate interplay between the LLP's decay length and the detector's capability to effectively search for and analyze these rare events at CEPC.

\subsection{Summary}
The $Z$-pole operation of CEPC also serves as a flavor factory, generating and measuring flavored hadrons and leptons with high statistics and resolution. It will become the leading flavor physics experiment, refreshing our knowledge of flavor physics. We present a few benchmarks for investigating feeble BSM effects with flavor physics probes above. In these cases, the SM amplitudes are generically suppressed due to various reasons. Consequently, the relative importance of NP is enhanced, and a high signal-to-noise ratio may be reached at CEPC. Examples include cLFV transition processes, rare FCNC decay processes, and light BSM particle production from its interaction with the SM flavor sector.

\begin{figure}[t]
   \centering
   \includegraphics[width=0.9\linewidth]{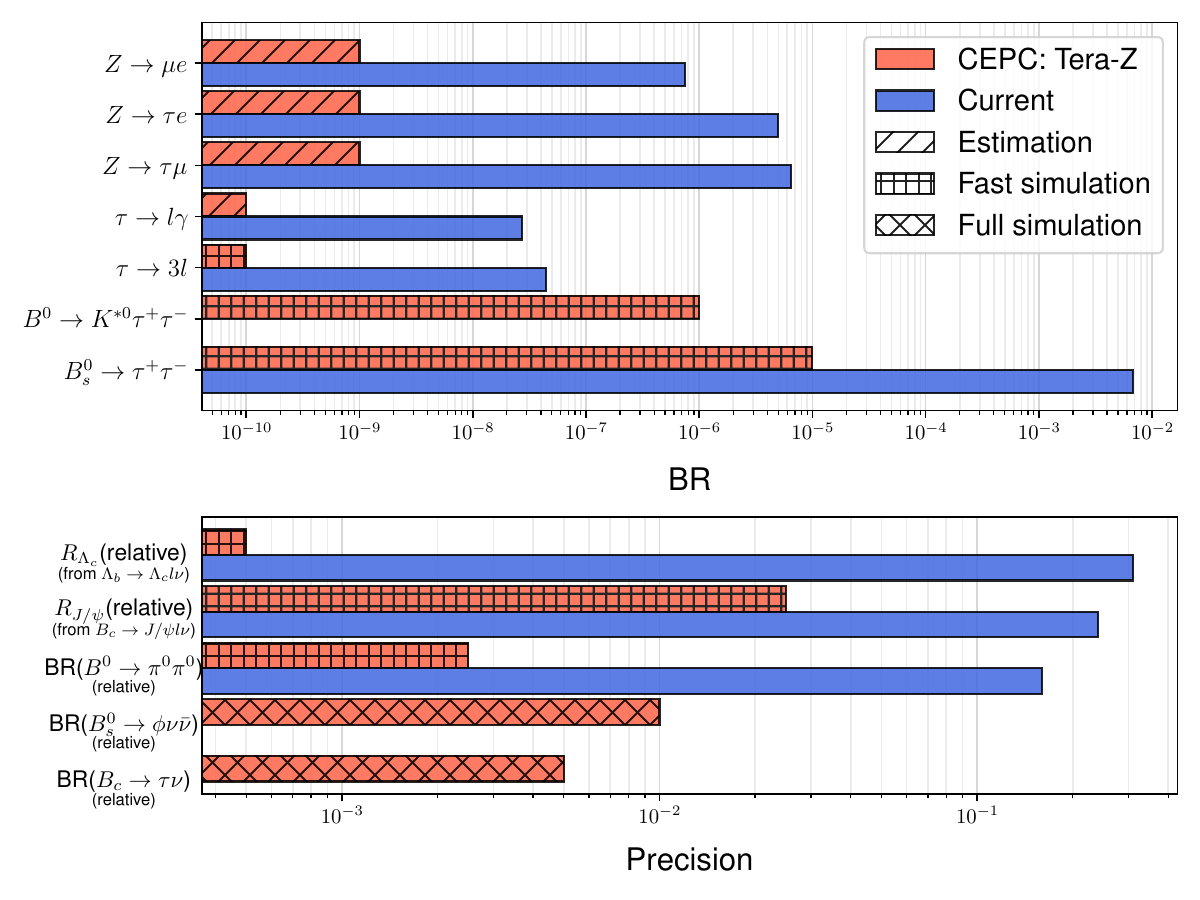}
   \caption{Summary plot of relevant flavor physics probes at CEPC. The upper and lower parts of the plot correspond to the BR upper limits reached and the sensitivities of SM processes. The current limits of $\tau \to 3\ell$ and $\ell \gamma$ channels are taken as the best one among all lepton flavor combinations.}
   \label{fig:flavorsummary}
\end{figure}

We list in Table~\ref{tab:flavorBSM} several channels where CEPC contributes potentially to BSM physics. However, the interpretation of BSM physics is highly model-dependent since the BSM impacts are indirect here. In Fig.~\ref{fig:flavorsummary}, we illustrate the target precision at CEPC in a model-independent way, together with the current best limits. Here we present only the conservative upper limits for many processes, because the systematic uncertainties cannot be precisely determined at the current stage.

\clearpage

\section{Neutrino Physics}
\label{sec:neutrino}

Neutrino oscillation~\cite{Fukuda:1998mi, Ahmad:2002jz} has firmly established the first and the only laboratory BSM evidence to date. Within the SM, the neutrinos are strictly massless to all orders in perturbation theory, and even non-perturbative effects cannot induce neutrino mass due to the SM's unbroken global $B-L$ symmetry. Therefore, studying neutrino mass generation mechanisms will pave the way for discovering underlying BSM physics. 

Under collider environments, neutrinos behave like missing energy and they are typically undetectable at calorimeters. The first direct observation of neutrino interactions at a particle collider experiment was only recently achieved by the FASER collaboration~\cite{FASER:2023zcr}. However, any nonstandard neutrino interaction effect is yet to be seen. In this section, we outline several possible BSM signals associated with the neutrino sector that potentially become observable at the CEPC. We also discuss their far-reaching implications for other outstanding puzzles of the SM, such as the origin of dark matter and visible matter. For reviews of neutrino physics at colliders, see Refs.~\cite{Deppisch:2015qwa, Cai:2017mow}, etc. 

Perhaps the most striking collider signals associated with neutrino mass generation mechanism are the lepton-number violating signals with same-sign charged lepton pairs, e.g.~the Keung-Senjanovi\'{c} process $pp\to \ell^\pm \ell^\pm jj$ for hadron colliders~\cite{Keung:1983uu}, which is the high-energy analog of the neutrinoless double beta decay ($0\nu\beta\beta$) process, but now with any lepton flavor. This is generically expected in theories of Majorana neutrinos, when we parametrize the $(B-L)$-breaking effects through an effective dimension-5 Weinberg operator $LLHH/\Lambda$~\cite{Weinberg:1979sa}. A well-known UV-complete example is the type-I seesaw mechanism~\cite{Minkowski:1977sc, Mohapatra:1979ia, Yanagida:1979as,GellMann:1980vs, Glashow:1979nm, Schechter:1980gr} with heavy right-handed neutrinos (RHNs). This can be realized by just adding the Majorana RHNs to the SM particle content, or in more natural ways by extending the SM gauge group to higher gauge groups like $U(1)_{B-L}$~\cite{Davidson:1979wr, Marshak:1979fm, Buchmuller:1991ce}, $SU(2)_L\times SU(2)_R\times U(1)_{B-L}$~\cite{Mohapatra:1974hk, Mohapatra:1974gc, Senjanovic:1975rk} or $SO(10)$~\cite{Fritzsch:1974nn, Witten:1979nr}, where the RHNs are necessary for anomaly cancelation. Thus ensues a very rich collider phenomenology of the RHNs~\cite{Deppisch:2015qwa}, if the seesaw scale happens to be around the electroweak scale. Here we will only cover some aspects of RHN phenomenology directly relevant to CEPC. 

Other simple well-motivated tree-level seesaw models include: the type-II seesaw where a left-handed triplet $\Delta_L$ is introduced to realize the seesaw framework~\cite{Konetschny:1977bn,Magg:1980ut,Schechter:1980gr,Cheng:1980qt,Mohapatra:1980yp,Lazarides:1980nt}; the type-III seesaw which is similar to the type-I seesaw, with the singlet fermions $N$ replaced by triplet fermions~\cite{Foot:1988aq,Ma:2002pf}. The tiny neutrino masses can also radiatively generated at loop levels, e.g. in the Zee model~\cite{Zee:1980ai} and Ma model~\cite{Ma:2006km} at 1-loop level and the Zee-Babu model~\cite{Zee:1985id,Babu:1988ki} at 2-loop level, where some beyond SM particles, including DM partcles in some models, are introduced to generate the neutrino masses. A comprehensive review of the tree and loop level of neutrino models can be found e.g. in Ref.~\cite{Cai:2017jrq}. The phenomenology of the neutrino models above are closely correlated with the high-precision low-energy physics, e.g. the lepton number violating neutrinoless double-beta decays ($0\nu\beta\beta$) and the parity-violating MOLLER experiment~\cite{Dev:2018sel,Cirigliano:2004tc,Abada:2007ux,Tello:2010am,Chakrabortty:2012mh,Barry:2013xxa,BhupalDev:2014qbx,Awasthi:2015ota,Bambhaniya:2015ipg,Borah:2016iqd,Li:2020flq,Li:2022cuq,deVries:2022nyh,Li:2024djp}.

\begin{figure}[!t]
\centering
\includegraphics[width=0.7\textwidth]{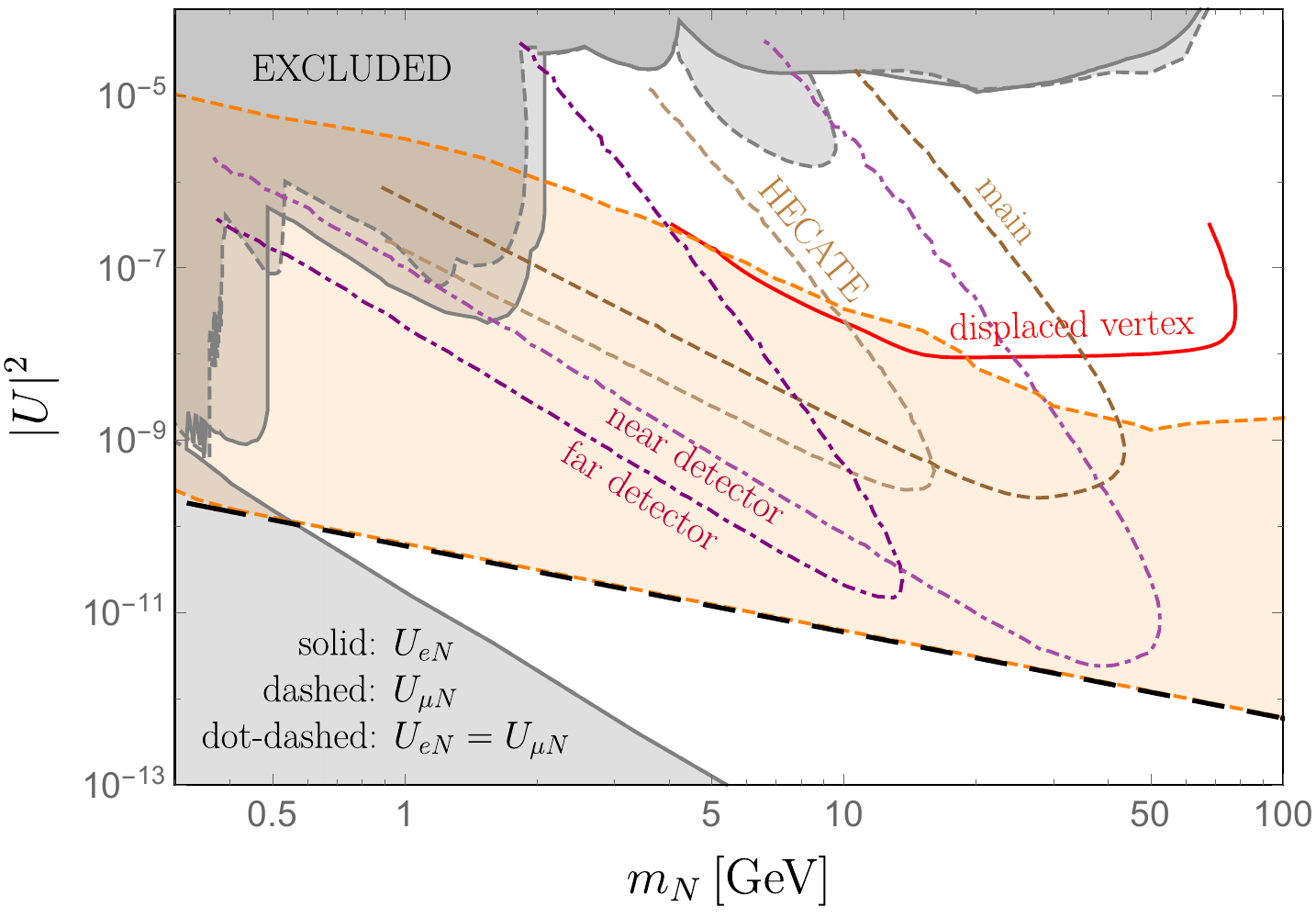}
\caption{Summary of prospects of heavy neutrino mass $m_N$ and the heavy-light neutrino mixing $|U|^2$ at the CEPC. The solid lines are for $U_{eN}$: displaced vertex (red)~\cite{Antusch:2016vyf}. 
The dashed lines are for $U_{\mu N}$: main detector and HECATE (darker and lighter brown)~\cite{Chrzaszcz:2020emg}. The dot-dashed lines are for $U_{eN} = U_{\mu N}$: main detector (near detector) and far detectors (lighter and darker purple)~\cite{Wang:2019xvx}. All the shaded regions are excluded by current limits, with solid for $U_{eN}$ and dashed for $U_{\mu N}$~\cite{neutrinolimits}.
The long dashed black line indicate the parameter space for the seesaw mechanism, while the shaded orange region is favored by the leptogenesis mechanism~\cite{Abdullahi:2022jlv}.
}
\label{fig:heavyneutrino:summary}
\end{figure}

In this section we focus mainly on the prospects of neutrino relevant physics at the CEPC. Subsection~\ref{subsec:heavyneutrino} is on the prospects of heavy neutrino $N$ at the CEPC. The heavy neutrino $N$ can be produced at high-energy lepton colliders via the process $e^+ e^- \to \nu N$ through the heavy-light neutrino mixing $U_{\alpha N}$ with the flavor index $\alpha = e,\, \mu,\, \tau$,  
and tends to be long-lived at the CEPC for its mass $m_N$ below tens of GeV. This makes the heavy neutrino a good candidate for the LLP signals in Section~\ref{sec:LLP}. It is also possible for the heavy neutrinos to induce prompt signals at the high-energy lepton colliders.
The prospects of $m_N$ and $|U|^2$ 
at CEPC are collected in Fig.~\ref{fig:heavyneutrino:summary}. As clearly seen in this figure, the mixing angle $|U|^2$ can be probed down to ${\cal O} (10^{-11})$ at the CEPC, covering a substantial region that is favored by the ``low-energy'' leptogenesis~\cite{Abdullahi:2022jlv}. 
The heavy neutrino can be pair produced from SM Higgs decay, i.e. $h \to NN$. It is found that the corresponding branching fraction can be probed up to ${\cal O} (10^{-4})$ at the CEPC at the $2\sigma$ C.L. (cf. Fig.~\ref{fig:hNNlimits})~\cite{Gao:2021one}, and the sensitivity of heavy-light neutrino mixing angle $|U|^2$ can be significantly improved to ${\cal O} (10^{-14})$ in this channel (cf. Fig.~\ref{fig:hNNsen})~\cite{Deppisch:2018eth}. In some scenarios, $N$ can also be produced at the lepton colliders via other BSM particles, e.g. the $Z'$ boson in $U(1)$ models or the $SU(2)_R$-breaking scalar in the left-right symmetric model (LRSM) (cf. Figs.~\ref{fig2} and \ref{figNev_epemD_NoB})~\cite{Das:2022rbl,Nemevsek:2016enw}. 
The prospects of non-standard neutrino interactions (NSIs) is presented in Subsection~\ref{subsec:NSI}. It is found that the precision of such interactions can be improved by roughly a factor of 50 at the CEPC with respect to current constraints (cf. Fig.~\ref{fig:cepceelr})~\cite{Liao:2021rjw}. The active-sterile neutrino transition magnetic moments can be probed up to ${\cal O} (10^{-7} \; {\rm GeV}^{-1})$, as shown in Fig.~\ref{fig:total} of Section~\ref{sebsec:neutrino:NMM}~\cite{Zhang:2023nxy}. The sensitivities of neutral and doubly-charged scalars in seesaw models is given in Section~\ref{subsec:LL-doublyChargedHiggs}. The Yukawa coupling of neutral scalar $H$ to charged leptons can be probed up to ${\cal O} (10^{-4})$ at the CEPC in the on-shell channels (cf. Fig.~\ref{fig:H3:prospect1}), while the neutral and doubly-charged scalar masses can be probed up to the few-TeV range in the off-shell channels (cf. Figs.~\ref{fig:H3:prospect2} and \ref{fig:prospect:Hpp:2})~\cite{Dev:2017ftk,BhupalDev:2018vpr}. The connection of neutrino physics to DM and baryon asymmetry in the Universe is investigated in Section~\ref{subsec:neutrino:DM}. It is found that the CEPC can probe wide region of $m_N$ and $|U|^2$ which is consistent with seesaw mechanism and leptogenesis (cf. Fig.~\ref{fig:lep})~\cite{Abdullahi:2022jlv}.

\subsection{Prospects of heavy neutrinos}
\label{subsec:heavyneutrino}

\subsubsection{Heavy neutrinos at the main detector}
\label{subsubsec:NwithND}

In Ref.~\cite{Antusch:2016vyf}, the authors investigate the sensitivity of future lepton colliders to long-lived heavy (almost sterile) neutrinos $N$ with electroweak scale masses and detectable time of flight, via displaced vertex signatures.
The theoretical framework is an explicit low-scale seesaw, the Symmetry Protected Seesaw Scenario (SPSS) model~\cite{Antusch:2015mia}. The signal process is $e^- e^+ \to \nu N$ running at the $Z$-pole and $\sqrt{s} =$ 240 GeV at the CEPC. The ILC’s Silicon Detector (SiD) is used as the benchmark detector at future lepton colliders~\cite{Behnke:2013lya}. 
The background processes are analyzed, and the heavy neutrino $N$ decays with a vertex displacement between 10 $\mu$m and 249 cm (the outer radius of the HCAL) are considered to be free of backgrounds and detectable by SiD.
Using 4 million Higgs bosons, the squared mixing angle $|\theta|^2$ of heavy neutrino could be constrained to $10^{-9}$ level, with $m_N$ as high as 70 GeV, 
and the $Z$ pole operation could even limit theta to $10^{-11}$ level for $m_N <$ 50 GeV.

Here we list some other studies of heavy neutrinos $N$ appearing as displaced vertices at future lepton colliders. 
Ref.~\cite{Barducci:2022hll} studies heavy neutrinos $N$ at future lepton colliders in the framework of the neutrino-extended Standard Model Effective Field Theory ($\nu$SMEFT).
The study focuses on four representative running modes: the FCC-ee with $\sqrt{s} =$ 90 and 240 GeV, the CLIC with $\sqrt{s} =$ 3 TeV, and a representative muon collider with $\sqrt{s} =$ 3 TeV.
With dimension-six EFT operators included, additional production and decay modes for the heavy neutrinos are present besides those arising from the mixing with the active neutrinos. 
The authors consider single- and pair-production of $N$ via four-fermion operators, and the most relevant additional decay modes are identified to be $N \to \nu \gamma$ (induced at 1-loop level) when $m_N \lesssim$ 15 GeV and $N \to 3f$ for larger masses, where $3f$ denotes various possible three-fermion combinations.
Depending on the heavy neutrino mass and the cutoff scale $\Lambda$ at which the EFT breaks down, the heavy neutrinos $N$ can decay either promptly or with a macroscopic distance, or appear stable at the detector level.
For the displaced vertex searches, the decay vertex is required to lie between 1 cm and 100 cm from the primary vertex.
The background is assumed to be negligible, and in both the two decay modes $N \to \nu \gamma,\; 3f$, the cutoff scaled can be probed up to roughly 20 TeV at the FCC-ee 240 GeV with an integrated luminosity of 5 ab$^{-1}$.

Similarly, Ref.~\cite{Rygaard:2022qms} focuses on long-lived heavy neutrinos via displaced vertex signature at the FCC-ee running at the $Z$-pole.
This study assumes that only one Majorana heavy neutrino $N$ mixes with the electron neutrino.
The signal process is $e^- e^+ \to \nu N \to \nu (e^+ e^- \nu)$, and the background processes include $Z \to e^+ e^-, \tau^+ \tau^-, b \bar{b}, c\bar{c}, u d s$.
For LLPs decaying at a displaced vertex, the transverse impact parameter $d_0$ of the displaced particles can be used as a complementary variable to the decay length. 
$d_0$ is given as the distance from the beam line to the projected back-trace of the displaced tracks in the transverse plane.
LLPs' decay products are expected to exhibit larger values of $d_0$.
This study selects long-lived heavy neutrinos by requiring both final-state electrons to have $d_0 > 0.5$ mm. Assuming $Z$-pole running with an integrated luminosity of 150 ab$^{-1}$, it is found that the squared mixing $|V_{eN}|^2$ can be probed up to ${\cal O}(10^{-8})$ for a heavy neutrino with mass of ${\cal O} (10\; {\rm GeV})$.

Ref.~\cite{Drewes:2022rsk} investigates methods to observe lepton number violation (LNV) and distinguish Dirac and Majorana heavy neutrinos $N$ at future lepton colliders (see also Refs.~\cite{Antusch:2022ceb,Antusch:2023nqd,Antusch:2023jsa,Antusch:2024otj,Ajmal:2024kwi,Bellagamba:2025xpd}).
These methods include the angular distribution and spectrum of the heavy neutrino's decay products as well as their lifetime.
The latter exploits the fact that the total decay width of $N$ differs by a factor of two between the Majorana and Dirac cases, leading to a decay length $\lambda$ differing in displaced vertex searches at colliders. 
Therefore, according to Eq.~\eqref{eqn:decayProb}, the decay probabilities inside the same detector should be different between the Majorana and Dirac cases as well.
This implies different numbers of observed signal events in displaced-vertex searches, and can be used to distinguish Dirac and Majorana heavy neutrinos.
The analytic estimates for the number of events and sensitivity regions during the $Z$-pole run for both Majorana and Dirac HNLs are also present in this study.
It should be noted that the analytic formulae that are used to generate Fig.~1b in Ref.~\cite{Drewes:2022rsk} can be generalized to other LLPs in a straightforward manner, and considerably simplify the computation of the corresponding event rates, e.g. those in Section~\ref{subsec:computationLLPsColliders}.

\begin{figure}[t]
\centering
\includegraphics[width=0.8\textwidth]{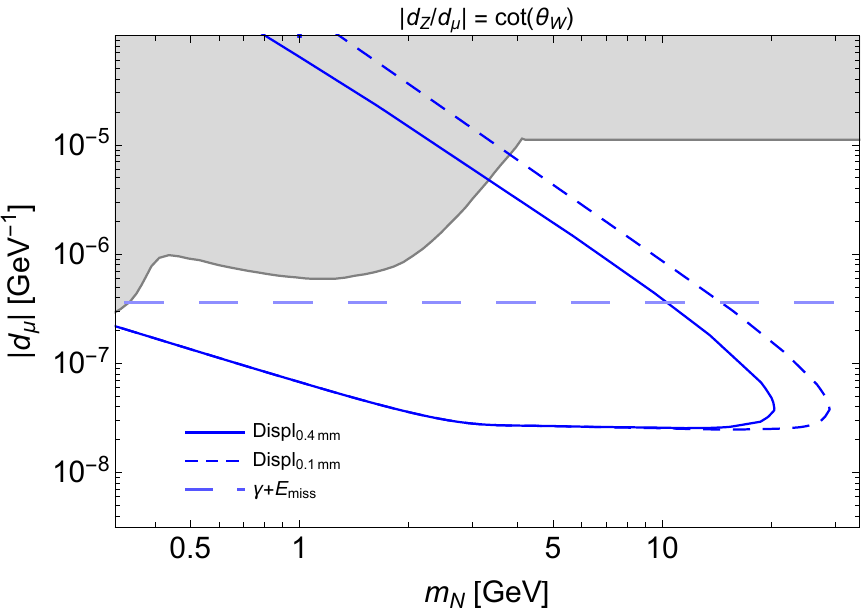}
\caption{The potential of FCC-ee to $m_N$ and the dipole coupling $|d_\mu|$ at the 90\% C.L. at the $Z$-pole with a luminosity of 150 ab$^{-1}$. 
The solid and short-dashed blue lines correspond respectively to the vertex transverse displacement of 0.4 mm and 0.1 mm, 
and the long-dashed light blue line denotes the sensitivity corresponding to the $\gamma$+missing energy signature. The sensitivities at CEPC can be obtained by a simple rescaling of the luminosity via $\sqrt{150/100}\simeq 1.22$. 
Taken from Ref.~\cite{Ovchynnikov:2023wgg}.
}
\label{fig:NDs-HNL-dipole-FCCee}
\end{figure}

Ref.~\cite{Ovchynnikov:2023wgg} studies the potential of future colliders to explore the parameter space of heavy neutrinos through the dipole portal. 
This work considers various signatures for the HNLs including missing energy and displaced decays, and discusses the complementarity between the hadron and lepton colliders.
At lepton colliders, the signal process is $e^- e^+ \to Z \to \nu N, N \to \nu e^- e^+, \nu \gamma$ at $\sqrt{s} =$ 91.2 GeV.
For the displaced vertex searches, the decay volume is considered to be the Innovative Detector for Electron–positron Accelerators (IDEA), which is a cylinder with radius $r$ = 4.5 m and longitudinal size $L$ = 11 m~\cite{Antonello:2020tzq}.
For the $e^- e^+$ final state, the background processes include $Z \to e^- e^+$ + inv., $e^- e^+ \to e^- e^+ \nu \bar{\nu}$. 
To reject the background by exploiting the long-lived signature of $N$, the following two choices of a displacement cut are applied: 
$r_{\rm displ} >$ 0.4 mm and 0.1 mm, depending on the spatial resolution of the tracker. Here $r_{\rm displ}$ is the vertex transverse displacement from the collision point.
Fig.~\ref{fig:NDs-HNL-dipole-FCCee} shows the FCC-ee potential to the parameter space of the heavy neutrino mass $m_N$ and the dipole coupling $|d_\mu|$, assuming $5 \times 10^{12}$ $Z$-bosons produced in total, corresponding to a luminosity of 150 ab$^{-1}$. Here the active neutrino flavor has been chosen to be muonic, which applies also to the electron and tauon flavors, as the FCC-ee sensitivity is flavor universal. For the case of CEPC, the nominal luminosity at the $Z$-pole is expected to be 100 ab$^{-1}$. The corresponding sensitivities of $d_\mu$ at CEPC will then be weaker than those at FCC-ee by a factor of $\sqrt{150/100} \simeq 1.22$.
Sensitivity reaches of the HL-LHC and FCC-hh are also given for comparison purpose in this work.

\subsubsection{Heavy neutrinos at far detectors}
\label{subsubsec:FDs-HNL}

\begin{figure}[t]
\centering
\includegraphics[width=0.6\textwidth]{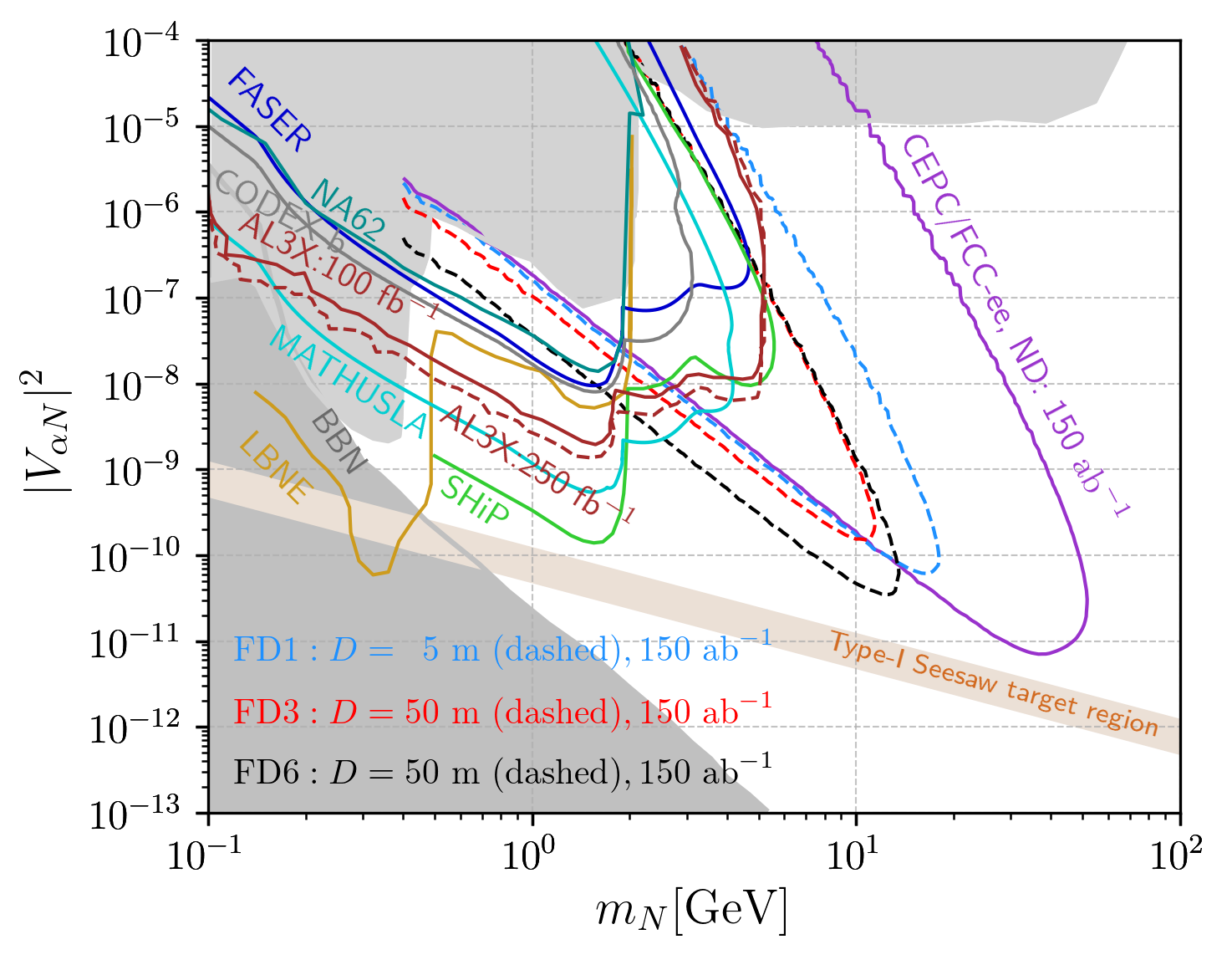}
\includegraphics[width=0.6\textwidth]     {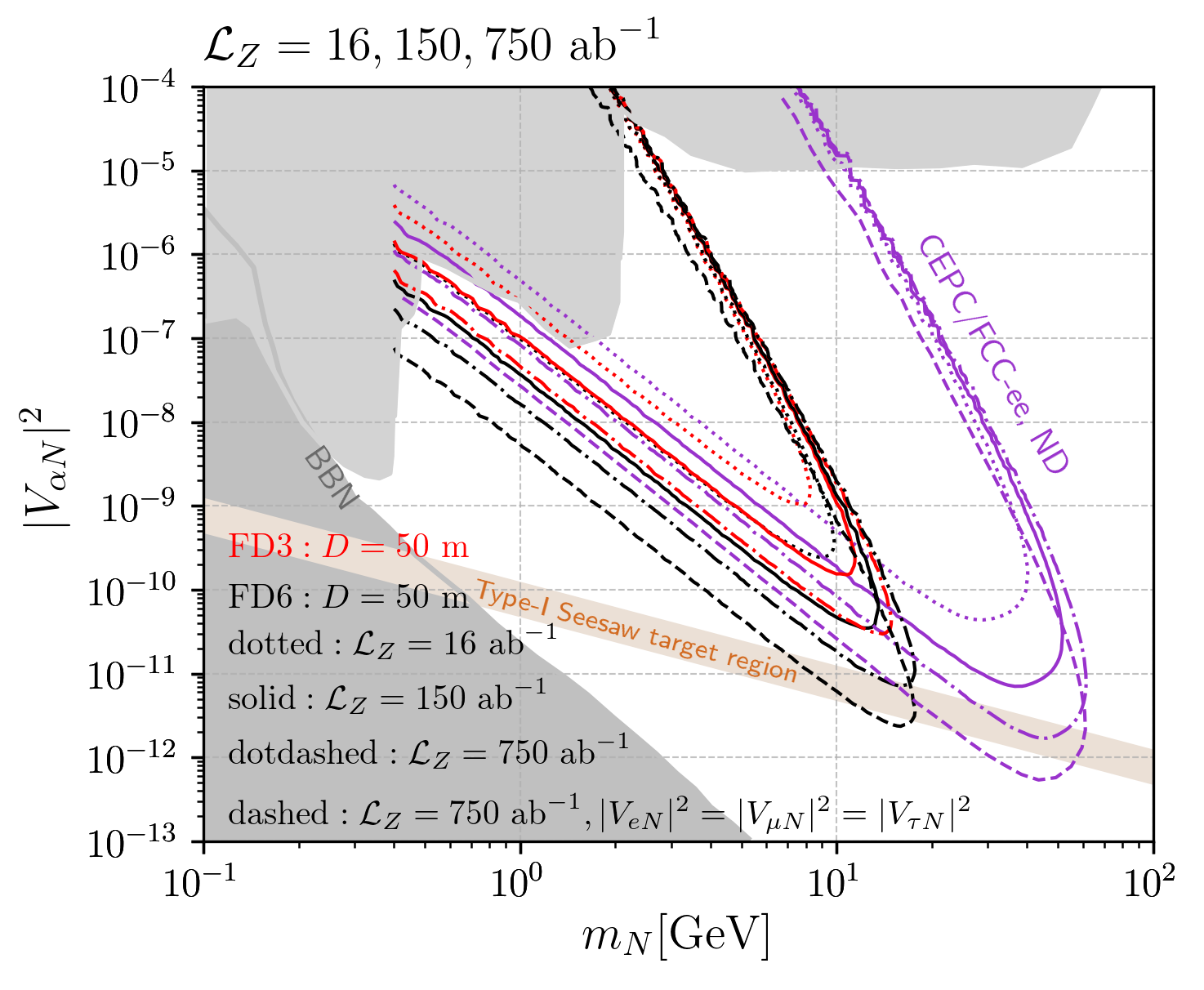}
\caption{ 
{\it Upper panel}: Sensitivity reaches of the CEPC/FCC-ee's far detectors FD1, FD3, FD6, in comparison with prospects at the main detector (near detector, abbreviated as ND) and other experiments. 
{\it Lower panel}: Sensitivity reaches of ND, FD3 and FD6 at the CEPC/FCC-ee with three different integrated luminosities $\mathcal{L}_Z = 16$, 150 and 750 fb$^{-1}$. The gray regions are excluded by current constraints. 
Taken from Ref.~\cite{Wang:2019xvx}.}
\label{fig:FDs-Z2nuN}
\end{figure}

Ref.~\cite{Wang:2019xvx} considers $Z$ boson decays to an active neutrino and a long-lived heavy neutrino $N$, $Z \to \nu N$, at $\sqrt{s} =$ 91.2 GeV.
In the analysis, the total number of the $Z$ bosons produced at the CEPC is specified as $N_Z^{\mathrm{CEPC}}=7.0 \times 10^{11}$ corresponding to a total integrated luminosity of $\mathcal{L}_{Z}^{\mathrm{CEPC}}=16$ ab$^{-1}$, while $N_Z^{\mathrm{FCC-ee}}=5.0 \times 10^{12}$ corresponding to $\mathcal{L}_Z^{\text{FCC-ee}}=150$ ab$^{-1}$.
The background is assumed to be negligible.
Sensitivity results at the 95\% C.L. in terms of 3-signal-event contour curves are presented in Fig.~\ref{fig:FDs-Z2nuN}, reproduced from Ref.~\cite{Wang:2019xvx}.
The sensitivity reaches of the CEPC/FCC-ee's far detectors FD1, FD3 and FD6 in the plane of $m_N$ and $|V_{\alpha N}|^2$ (with $\alpha = e$ or $\mu$) are shown in the upper panel, in comparison with the current constraints in gray and prospects at the main detector (near detector, abbreviated as ND) and other experiments including the LHC FDs. The luminosity has been taken to be 150 ab$^{-1}$. The sensitivities of the main detector (near detector, abbreviated as ND), FD3 and FD6 with three different luminosities of ${\cal L} = 16$, 150 and 750 ab$^{-1}$ are presented in the lower panel. The prospects for the case of heavy neutrino with equal mixings with all three active neutrino generations, i.e.~$|V_{e N}|^2=|V_{\mu N}|^2=|V_{\tau N}|^2$, are shown as the dashed lines in the lower panel, with $\mathcal{L}_Z=750$ ab$^{-1}$. It is clear that the main detector and FD6 at the CEPC or FCC-ee may probe the type-I seesaw limits on $|V_{\alpha N}|^2$, if $m_N$ lies between 10 GeV and 60 GeV. 

We note that if high-precision timing $\sim\mathcal{O}(\text{picosecond})$information can be obtained, it is possible to correlate the activities at the MD and the FD that stem from the same collision event at the IP.
Achieving this event correlation would allow for observing lepton-number-violating (LNV) processes that could arise, e.g.~from long-lived HNLs.
Ref.~\cite{Mao:2023zzk}, for instance, shows its feasibility with the proposed LHC far detectors; if observed, such LNV processes can pin down the Majorana nature of the neutrinos.
In principle, similar strategies can also be implemented at high-energy $e^+ e^-$ colliders.


\begin{figure}[t]
\centering
\includegraphics[width=0.7\textwidth]{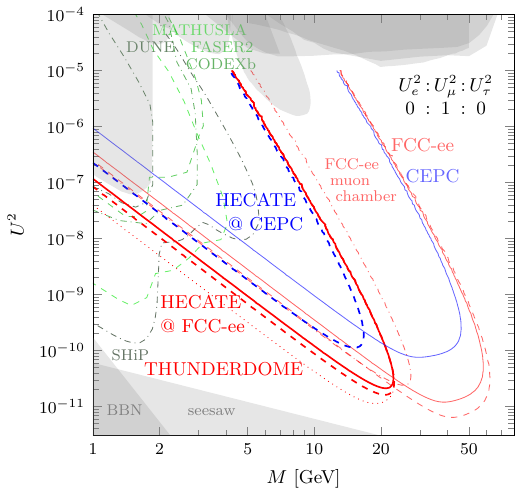}
\caption{Sensitivities for nine signal events that can be achieved at FCC-ee (red) and CEPC (blue). The difference of the FCC-ee and CEPC sensitivities is mainly due to the different luminosities at the two colliders. According to latest setup of CEPC-TDR, the coverage of FCC-ee and CEPC will be much closer. The faint and thick curves are the prospects for the main detectors and HECATE, respectively. The shaded regions are excluded by current experiments, and the expected sensitivities of selected other experiments are indicated by the green curves. See the text for more details. 
Taken from Ref.~\cite{Chrzaszcz:2020emg}.
} 
\label{fig:FDs-HNL-HECATE}
\end{figure}

Ref.~\cite{Chrzaszcz:2020emg} proposes to install a HErmetic CAvern TrackEr (HECATE) at the CEPC and FCC-ee, which is similar to the far detectors FD3 and LAYCAST in Section~\ref{sec:LLP}.
The HECATE detector would consist of resistive plate chambers (RPCs) or scintillator plates, constructed from extruded scintillating bars, located around the cavern walls and forming a 4$\pi$ detector. 
In order to obtain timing information and to distinguish particles from cosmic background, the HECATE detector should have at least two layers of detector material separated by a sizable distance.
For reliable tracking, at least four layers, along with a smaller size or optimized geometry of the detector plates, would be required.

This study estimates the HECATE sensitivity for long-lived heavy neutrino produced from $Z$ boson decays $Z \to \nu N$ at $\sqrt{s} =$ 91.2 GeV.
Similar to Eq.~\eqref{eqn:decayProb}, in their analyses, the decay probability of long-lived heavy neutrino inside the detector's fiducial volume is related to $\exp \{-l_0/\lambda_N \} - \exp \{ -l_1/\lambda_N \}$. 
The total number of $Z$ bosons are taken to be $3.5 \times 10^{11}$ and $2.5 \times 10^{12}$ at CEPC and FCC-ee, respectively. We extract Fig.~\ref{fig:FDs-HNL-HECATE} from Ref.~\cite{Chrzaszcz:2020emg} which shows the isocurve for nine signal events with the HECATE at CEPC and FCC-ee, which are shown respectively as the blue and red lines. 
Two setups of HECATE are investigated with $l_0 =$ 4 m and $l_1 =$ 15 m (solid) or 25 m (dashed).
Sensitivities are also compared with multiple other experiments in this study. The faint solid curves show the main detector sensitivity ($l_0 =$ 5 mm, $l_1 =$ 1.22 m). The faint dash-dotted curve indicates the additional gain if the muon chambers are used at the FCC-ee ($l_0 =$ 1.22 m, $l_1 =$ 4 m).
The thick curves show the sensitivity of  HECATE with $l_0 =$ 4 m, $l_1 =$ 15 m (solid) and $l_0 =$ 4 m, $l_1 =$ 25 m (dashed), respectively.
Finally, the faint dashed red line shows the FCC-ee main detector sensitivity with $5 \times 10^{12}$ $Z$ bosons, corresponding to the luminosity at two IPs. The difference of the FCC-ee and CEPC sensitivities is mainly due to the larger luminosity at the FCC-ee.
For comparison we indicate the expected sensitivity of selected other experiments with the different green curves as indicated in the plot~\cite{Chou:2016lxi,Feng:2017uoz,Gligorov:2017nwh,SHiP:2018xqw,Ballett:2019bgd}.
The gray areas in the upper part of the plot show the region excluded by past experiments~\cite{CHARM:1985nku, CMS:2018iaf, ATLAS:2019kpx, DELPHI:1996qcc, E949:2014gsn,LHCb:2016inz, Antusch:2017hhu, NuTeV:1999kej, Bernardi:1987ek}, while the grey areas at the bottom mark the regions that are disfavoured by BBN~\cite{Boyarsky:2020dzc} and neutrino oscillation data in the Neutrino
Minimal Standard Model ($\nu$MSM, labelled as ``seesaw'' in the figure)~\cite{Asaka:2005an,Asaka:2005pn}.


\subsubsection{SM Higgs decay $h \to NN$} 

CEPC can search for heavy $N$ within the reach of its center of mass energy. There have been studies on the weak single $N$ production at CEPC in the process $e^- e^+ \to \nu N$ for center-of-mass energy $\sqrt{s} = 240$ GeV~\cite{Liao:2017jiz}, and on high luminosity $Z$-pole running mode~\cite{Ding:2019tqq, Blondel:2021mss}. As $N$ has a large Majorana mass, lepton number violation occurs in $N$ decay. Same-sign, same flavor dileptons, and a reconstructable $N$ mass peak of final state lepton-jet system are the `smoking gun' signals for heavy $N$ search~\cite{Shoemaker:2010fg}. Meanwhile, CEPC is designed to yield $\sim$4M Higgs events. The high identification efficiency for soft leptons and low hadronic background at the CEPC offers a clean search opportunity for $h\rightarrow NN$. The dominant Higgs production channel at CEPC is $e^+e^-\rightarrow Zh$. The associated $Z$ complicates the signal and background analysis, as the $Z$ boson's decay products can be confused with those from heavy $N$ decay. On the other hand, with an extra $Z$ boson, the SM backgrounds can also be suppressed by requiring one more weak vertex. The leading SM backgrounds are from multi-tau production with one or two associated weak vector bosons ($V=W,Z$), e.g. $4\tau, 4\tau V, 2\tau 2\ell V$, etc., in which non-isolated and missing leptons can lead to same-sign same-flavor lepton pairs.

\begin{figure}[t!]
\centering
\includegraphics[height=0.3\textheight,width=0.7\textwidth]{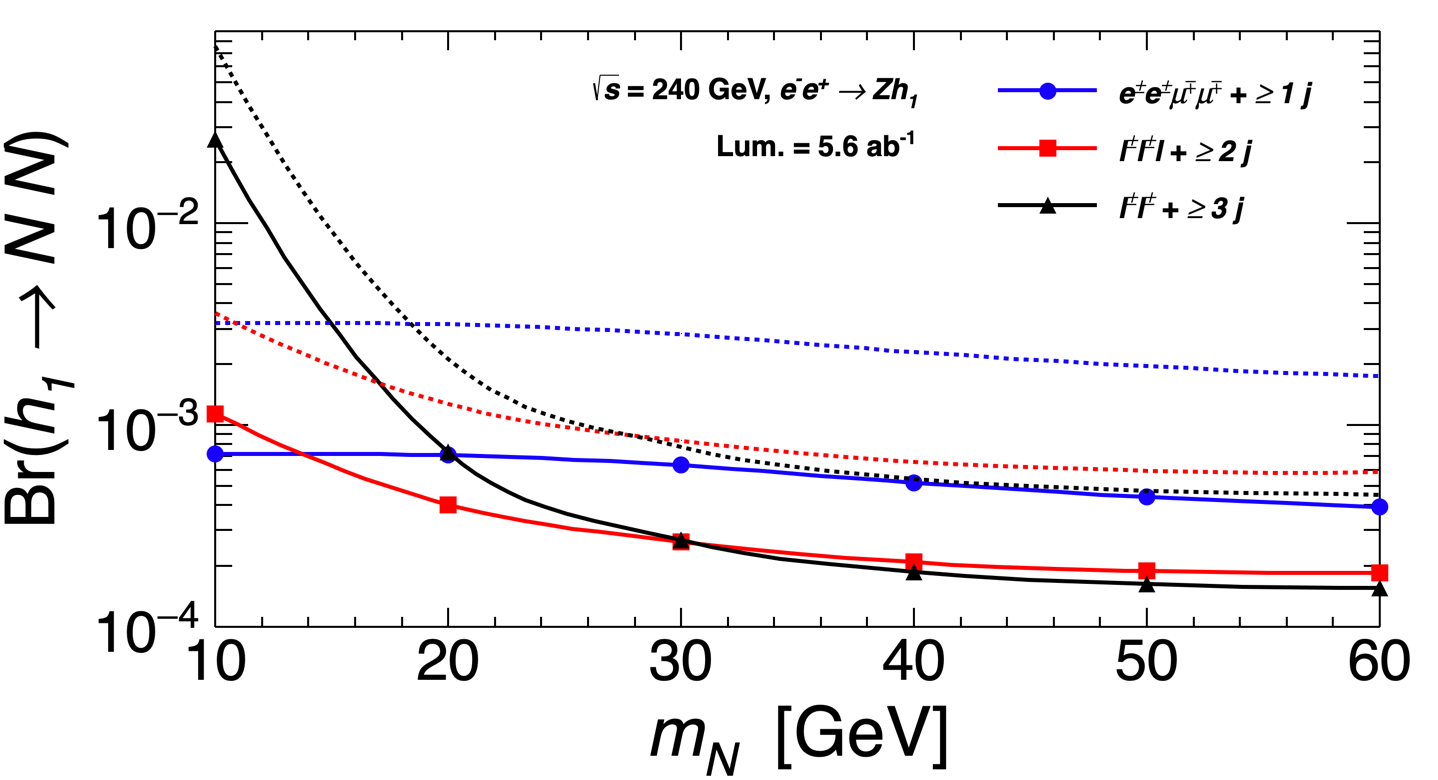}
\caption{
Projected CEPC sensitivities to the decay branching ratio of $h\rightarrow NN$ for 2-4$\ell$ channels at $\sqrt{s}=240$ GeV with a luminosity of 5.6 ab$^{-1}$. The $2\sigma$ and $5\sigma$ sensitivities are shown as the solid and dashed lines, respectively. Adapted from Ref.~\cite{Gao:2021one}.
}
\label{fig:hNNlimits}
\end{figure}

SM background analysis of the $n$-lepton ($n\ge 2$) channels with at least one set of same-sign dileptons~\cite{Gao:2021one} shows that the semileptonic heavy $N$ decay, requiring only one same-sign lepton pair, gives higher sensitivity than fully leptonic $N$ decay channels. Jet and lepton number counting plays an essential role 
in removing the SM background contamination. Leptonic decay of the associated $Z$ boson also leads to a same-sign same-flavor trilepton signal. For CEPC 240 GeV at 5.6 ab$^{-1}$ luminosity, multi-lepton rare decay search for $h\rightarrow NN$ will be  sensitive to Higgs-BSM scalar mixing angle up to around $|\sin\alpha|^2\le 10^{-4}$. The reaches on multi-lepton Higgs rare decay branching ratios are shown in Fig.~\ref{fig:hNNlimits}.

In the meantime, $h \rightarrow N N$ channel can also be used to test the origin of neutrino masses, i.e. seesaw mechanisms. In Ref.~\cite{Deppisch:2018eth}, pair-produced long-lived $N$ from Higgs decays is searched at colliders, including CEPC and ILC, for the $U(1)_{B-L}$ model. 
At the CEPC, with a center-of-mass energy $\sqrt{s} = 250$~GeV, the dominant Higgs production process is Higgs-Strahlung, $e^+e^- \to Z^* \to Z h$, which cross section is $\sigma\sim 240$~fb for $\sqrt{s} = 250$~GeV, and reduced by the Higgs-BSM mixing angle, $\propto \cos^2\alpha$. Comparing to the LHC, the Higgs production is about 200 times smaller, but CEPC has larger luminosity and clean background. 
The SM Higgs can decay into a pair of heavy $N$, with 
\begin{align}
\label{prod}
	\text{Br}(h \to NN) = 
		\frac{\Gamma(h\to NN)}{{\Gamma(h)}_\text{SM}\cos\alpha^2 
		+ \Gamma(h \to NN)},
\end{align}
where $\Gamma(h)_\text{SM}$ $\approx$ 4.2 $\times$ $10^{-3}$~GeV is the total decay width of the SM Higgs, and $\tilde{x}$ is the vev of the $B-L$ scalar, and
\begin{align}
	\Gamma(h\to NN) = \frac{2}{3}\sin^2\alpha\frac{M_N^2}{\tilde{x}^2}\frac{m_{h}}{8\pi}
	\left(1 - \frac{4M_N^2}{m_{h}^2}\right)^{3/2}.
\end{align}
Hence, the cross section of pair-produced heavy $N$ at the CEPC, from $(e^+e^- \to Z \to Z h \to Z + NN)$ dependent both on $M_N$ and $\sin \alpha$, when we fix $\tilde{x} =$ 3.75 TeV. 
It is found that the production cross section peaks where $M_N \approx $ 40 GeV, and can reach $\mathcal{O}(0.1)$ fb, when the Higgs-BSM mixing is at the current upper limits, $\sin \alpha \sim 0.3$~\cite{Robens:2022cun}.

The heavy $N$, can subsequently decay into SM states, via the active-sterile mixing. Giving our interested parameter space, $M_N \lesssim 60$ GeV, $|V_{\ell N}|^2 \sim m_\nu/M_N \approx 10^{-12}$, and $N$ mainly decays via three-body processes such as $N\to \ell^\pm q\bar{q}$ and $N\to \ell^+\ell^-\nu$. Hence, $N$ can be long-lived, and the resulting decay length for $M_N \lesssim m_Z$ can be expressed as %
\begin{align}
\label{lengthapproxi}
	L_N \approx 0.025~\text{m} 
	\cdot \left(\frac{10^{-12}}{|{V_{\ell N}|}^2}\right)
	\cdot \left(\frac{100~\text{GeV}}{M_N}\right)^5.
\end{align} 
Therefore, the $N$ can possesses decay length $\mathcal{O}(0.1)$ m, leading to displaced vertex signatures at the CEPC. 
To estimate the events of displaced vertex signals, we simplify the geometry and detector response of the CEPC detector. In Table~\ref{tab:SiD}, we show the size and resolution of the CEPC detector~\cite{CEPC-SPPCStudyGroup:2015csa}, comparing to the ILC \cite{Behnke:2013lya, Aihara:2009ad}. Here, $\sigma_d^t$ is the resolution of the detector in transverse direction, and $d_0$ is the transverse distance between the heavy $N$ and lepton in the final states, such as $|d_0| = |x p_y - y p_x|/p_T$, 
where $x$ and $y$ are the position where the heavy $N$ decayed, and $p_x$, $p_y$, $p_T$ are the components of momentum and transverse momentum of the final particles $\ell$, and $L_{xy}$ and $L_z$ are the transverse and longitudinal decay lengths of the HNLs, respectively. The Region 1 and 2, are approximated the tracker and muon systems of the corresponding detector. Given they can detect muons better, for the later results, we take $\ell = \mu$.
For heavy $N$, if it is decayed either inside the Region 1 or 2, and $|d_0|/\sigma_d^t$ is larger than required, we assume it can be detected by the detector with 100\% efficiency.

\begin{table}[t!]
	\centering	
	\begin{tabular}{c|ccccc}
		\hline
		Region & Inner Radius & Outer Radius & z-Extent & $|d_0|/\sigma_d^t$ & $\sigma_d^t$ 
		\\ \hline
		ILC Region 1  & \phantom{0}22 & 120 & 152 & 12 & 0.002 \\
		ILC Region 2  & 120 & 330 & 300 &  4 & 2     \\\hline
		CEPC Region 1 & \phantom{0}15 & 180 & 240 & 12 & 0.007 \\ 
		CEPC Region 2 & 180 & 440 & 400 &  4 & 2     \\\hline
	\end{tabular}
	\caption{Parameters of simplified detector geometries representing future detectors, namely ILC \cite{Behnke:2013lya, Aihara:2009ad}, CEPC \cite{CEPC-SPPCStudyGroup:2015csa}. All length units are in cm.}
	\label{tab:SiD}
\end{table}

\begin{figure}[!t]
\centering
\includegraphics[width=0.65\textwidth]{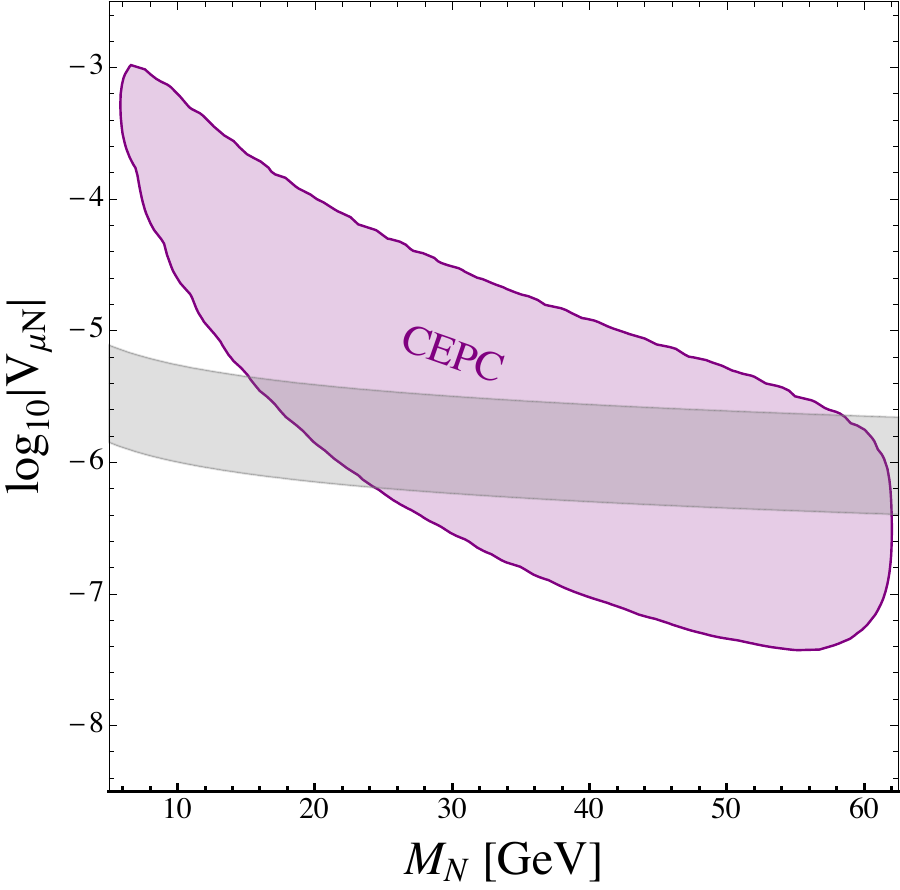}
\caption{Excluded regions in the $(M_N, |V_{\mu N}|)$ parameter space at 95\%~CL assuming no observation of a single displaced vertex for the 20~ab$^{-1}$ CEPC (purple). The grey band indicates the parameter region where a light neutrino mass in the interesting range is generated, $0.01~\text{eV} < m_\nu = |V_{\mu N}|^2 m_N < 0.3$~eV. 
 We fix the Higgs-BSM mixing $\sin \alpha = 0.3$. Adapted from Ref.~\cite{Deppisch:2018eth}. 
}
\label{fig:hNNsen}
\end{figure}

With such long-lived $N$, we assume the background can be negligible, so the sensitivity can be estimated by requiring the number of signal events, $N_S \gtrsim $ 3, at 95\% CL. And the results of CEPC with 20~ab$^{-1}$ integrated luminosity, is shown in Fig.~\ref{fig:hNNsen}. Assuming no observation of a single displaced vertex, the excluded regions in the $(M_N, V_{\mu N})$ parameter space at 95\%~CL is shown. The grey band indicates the parameter region where a light neutrino mass in the interesting range is generated, $0.01~\text{eV} < m_\nu = |V_{\mu N}|^2 M_N < 0.3$~eV. From the figure, we find the sensitivity region roughly tracks where decay length $L_N \sim \mathcal{O}(1)$ m. CEPC can reach the parameter space where type-I seesaw predicts, and even below it. Therefore, we have shown that CEPC have potential in revealing the nature of neutrino masses. However, such statement rely on the existence of significant Higgs-BSM mixing angle, which might be excluded by the precision measurement of Higgs signal rates at the CEPC~\cite{Gu:2017ckc}.

\subsubsection{Prospects of heavy neutrinos in $U(1)$ models}
\label{sec:N:U(1)}

\begin{figure*}
\includegraphics[width=0.8\textwidth]{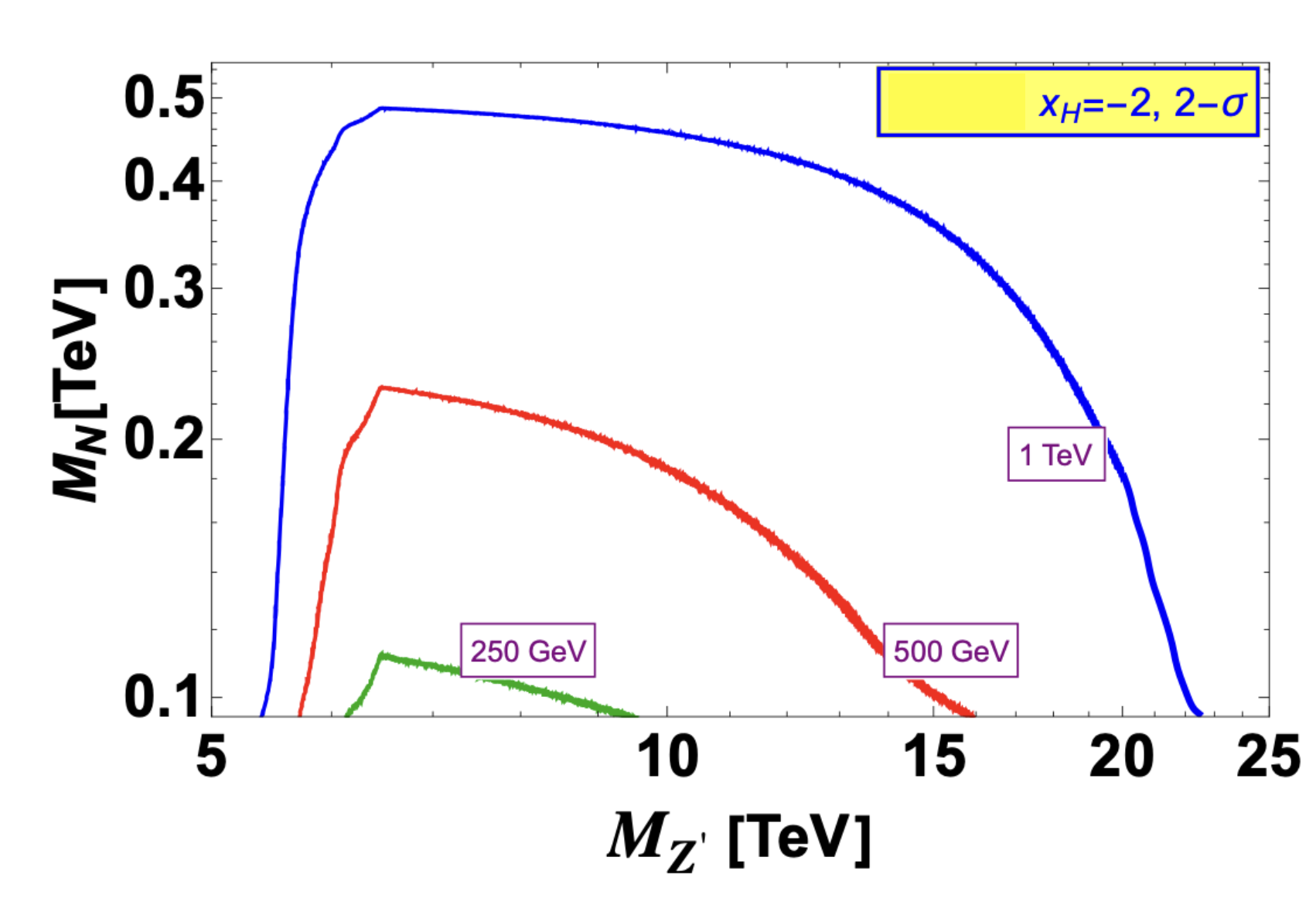}
\caption{2-$\sigma$ contour of the $M_N-M_{Z^\prime}$ plane at the electron-positron colliders at different center of mass energies studying same-sign dilepton plus four jet final states. Taken from Ref.~\cite{Das:2022rbl}}
\label{fig2}
\end{figure*}

Under the general $U(1)$ framework, a neutral BSM gauge boson $(Z^\prime)$ is evolved. Such a $Z^\prime$ gauge boson can be tested at the high energy experiments. We find that if $e^+ e^-$ colliders are built then we can study forward-backward (FB), left-right (LR) and left-right forward-backward (LR-FB) asymmetries at different center of mass energies~\cite{Das:2021esm}. The $Z^\prime$ in this scenario can directly interact with the right-handed neutrinos (RHNs). Hence we can study the pair production of RHNs from the $Z^\prime$ at the LHC and other proton-proton colliders at $\sqrt{s}=27$ TeV and $100$ TeV from prompt and displaced scenarios after the decay of RHNs. We find that the RHNs pair production from the $Z^\prime$ can be enhanced at $x_H=-1.2$ which is the general $U(1)$ charge of SM Higgs doublet~\cite{Das:2017flq,Das:2022rbl}. We find the branching ratio of $Z^\prime$ into a pair of RHNs $({\rm BR}(Z^\prime \to 2 N))$ is nearly one order of magnitude larger than the branching ratio of $Z^\prime$ into lepton doublets $({\rm BR}(Z^\prime \to 2 \ell))$. 
We produce the RHNs from $Z^\prime$ at the $e^-e^+$ collider following
\begin{equation}
\label{RHN-ee2}
\sigma(e^+e^- \to {Z^\prime}^\ast \to N_i N_i) \simeq \Big(\frac{{g^\prime}^4}{{M_{Z^\prime}}^4}\Big) \frac{ s(8+ 12 x_H + 5 x_H^2)}{192 \pi} \Big(1-\frac{4M_N^2}{s}\Big)^{3/2}  \,.
\end{equation}
Studying the signal of same sign dilepton plus four jets and corresponding SM backgrounds, we show the 2$-\sigma$ contours on the $M_{Z^\prime}-M_N$ plane in Fig.~\ref{fig2} for $x_H=-2$. The luminosities are respectively 2 ab$^{-1}$, 4 ab$^{-1}$ and 8 ab$^{-1}$ for 250 GeV, 500 GeV and 1 TeV. It is clear in this figure that a higher center-of-mass energy, such as the ILC and CLIC, can improve significantly the prospects of $m_N$. The contours for other values of $x_H$ can be found in Ref.~\cite{Das:2022rbl}.

\subsubsection{Prospects of heavy neutrinos in the LRSM}

\begin{figure}
\includegraphics[width=0.8\textwidth]{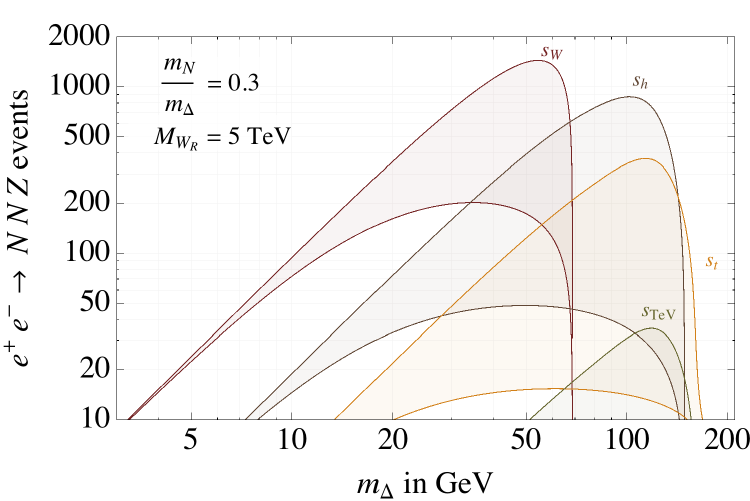}
\caption{Signal event rates for $e^+ e^- \to ZNN$ productions at lepton colliders for different $\sqrt{s}$, with a universal luminosity of ${\cal L} = 1$ ab$^{-1}$ and $m_N = m_\Delta/3$. The shaded regions cover the range of $\sin\theta \in (0.01,\, 0.1)$. Figure from Ref.~\cite{Nemevsek:2016enw}.}
\label{figNev_epemD_NoB}
\end{figure}

The production of $N$ in the minimal LRSM~\cite{Pati:1974yy,Mohapatra:1974gc,Senjanovic:1975rk} can be sizeable at lepton colliders, even for relatively high $W_R$ mass and small Higgs mixing. In particular, the neutral component $\Delta_0$ of the right-handed triplet $\Delta_R$ couples directly with the heavy neutrinos, and could mix with the SM Higgs. One of the resultant physical scalars, $\Delta$, is predominantly from $\Delta_0$. 
For $\sqrt s < \mathcal O(100) \text{ GeV}$, the dominant production of $\Delta$ at $e^+ e^-$ colliders occurs in the associated $\Delta Z$ production, and the corresponding leading-order cross section is 
\begin{equation}
  \sigma (e^+e^- \to \Delta Z) = s_\theta^2 \frac{G_F^2 M_Z^4}{96 \pi s} \left( \hat v_e^2 + \hat a_e^2 \right) \sqrt \lambda \,
  \frac{\lambda + 12 \left(M_Z^2/s \right)}{1 - \left(M_Z^2/s \right)^2} \,,
\end{equation}
where $\lambda = \left(1 - m_\Delta^2/s - M_Z^2/s \right)^2 - 4 m_\Delta^2 M_Z^2/s^2$, with $m_\Delta$ the scalar mass, $\hat v_e = -1$, $\hat a_e = -1 + 4 s_w^2$, and the $\Delta \nu \nu$ one via $WW$ fusion, see  Ref.~\cite{Djouadi:2005gi}. The decay $\Delta \to NN$ leads to the $N N Z$ final state with up to four leptons and no missing energy when $Z$ decays leptonically. 
The total integrated luminosity at LEP was too small to find more than $\sim 2$ $NNZ$ events from the collected data. On the other hand, the future $e^+ e^-$ machines may have sufficient sensitivity to look for heavy neutrinos from $\Delta$ decays. Various production c.m. energies are currently under consideration from the $Z$ pole at 90 GeV all the way to a $3 \text{ TeV}$ machine, with $\sqrt{s_{W,h,t,\text{TeV}}}=\{0.16, 0.24, 0.35, 1\} \text{ TeV}$~\cite{Gomez-Ceballos:2013zzn}.
The backgrounds depend on the c.m. energy and are particularly low below the $t \overline t$ threshold. Moreover, they can be reduced with cuts to a small level even above this energy. Conversely, for TeV machines, the $W$ VBF channel takes over and the $NN \nu \overline \nu$ final state dominates. 
The exact capabilities of the detectors are presently unknown, therefore we only show the signal event counts for different $\sqrt s$ cases in  Fig.~\ref{figNev_epemD_NoB}, as function of the scalar mass $m_\Delta$~\cite{Nemevsek:2016enw}.
With a luminosity of 1 ab$^{-1}$, over 1000 events can be produced at the $W$ pair threshold of 160 GeV, and the scalar mixing angle $\sin\theta$ can be probed down to 0.01. Given the total luminosity of 20 ab$^{-1}$ at the CEPC, the sensitivity of scalar mixing angle can be improved up to 0.002. Furthermore, at the running of 240 GeV, the sensitivity of heavy neutrino $N$ can be extended to over 100 GeV.

\subsection{Non-standard neutrino interactions} 
\label{subsec:NSI}

The presence of nonstandard neutrino interactions (NSI) has a large effect on the precision measurements at next-generation neutrino oscillation experiments, and other types of experiments can also constrain the NSI parameter space.
Ref.~\cite{Liao:2021rjw} considered the monophoton channel at the CEPC. The Lagrangian of neutral current (NC)  NSI with electrons can be written as, 
\begin{eqnarray}
	\label{eq:NSIee}
	\mathcal{L}_{\rm{NSI}}^{\rm NC,e}
	&=&-2\sqrt{2}G_F\epsilon_{\alpha\beta}^{eL}(\bar{\nu}_\alpha\gamma^\mu P_L \nu_\beta) (\bar{e}\gamma_\mu P_L e)
	-2\sqrt{2}G_F\epsilon_{\alpha\beta}^{eR}(\bar{\nu}_\alpha\gamma^\mu P_L \nu_\beta) (\bar{e}\gamma_\mu P_R e),
\end{eqnarray}
where $\alpha$, $\beta$ label the lepton flavors ($e, \mu, \tau$).

\begin{figure}[!t]
	\begin{centering}
	\includegraphics[width=0.46\textwidth]{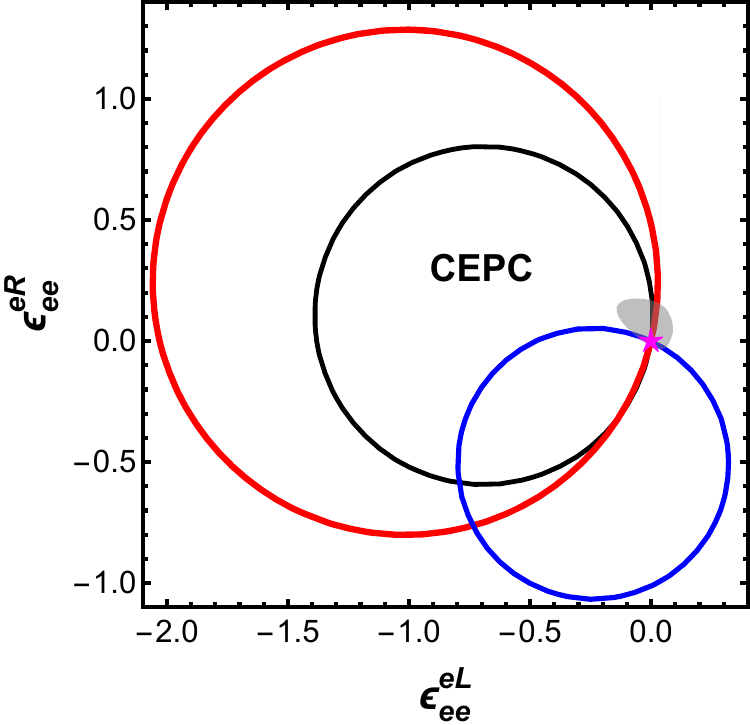}
	\includegraphics[width=0.49\textwidth]{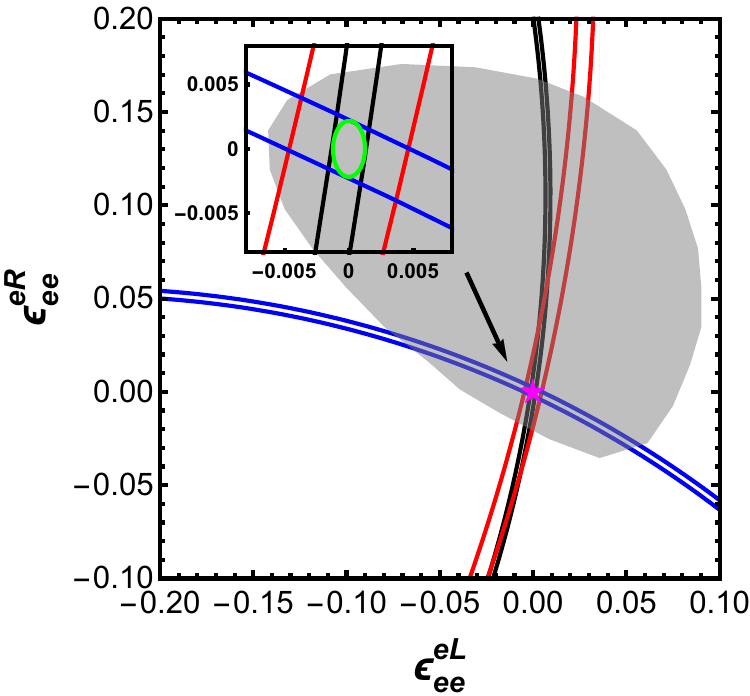}
	\caption{{\it Left panel}: The allowed  90\% C.L. region for electron-type neutrino NSI in the planes of $(\epsilon_{ee}^{eL},\ \epsilon_{ee}^{eR})$ at future CEPC with 5.6 ab$^{-1}$ data of $\sqrt{s}=240$ GeV (Black), with 2.6 ab$^{-1}$ data of $\sqrt{s}=160$ GeV (Red), and with 16 ab$^{-1}$ data of $\sqrt{s}=91.2$ GeV (Blue), respectively. The allowed 90\% C.L. regions arising from the global analysis of the LEP, CHARM, LSND, and reactor data \cite{Barranco:2007ej},  are shown in the shaded gray regions. {\it Right panel}: With all the data collected in all three running modes, the combined result is shown as the green region. Figure from Ref.~\cite{Liao:2021rjw}
	}
	\label{fig:cepceelr}
\end{centering}
\end{figure}

With the monophoton searches, Fig.~\ref{fig:cepceelr} shows the allowed 90\% C.L. region for NSI with electrons in the plane of $(\epsilon_{ee}^{eL},\ \epsilon_{ee}^{eR})$ at  CEPC with 5.6 ab$^{-1}$ data of $\sqrt{s}=240$ GeV (Black), with 2.6 ab$^{-1}$ data of $\sqrt{s}=160$ GeV (Red), and with 16 ab$^{-1}$ data of $\sqrt{s}=91.2$ GeV (Blue), respectively, from the production of single photon associated with neutrino pair $e^+ e^- \to \nu \bar\nu \gamma$.
From the left side of Fig.~\ref{fig:cepceelr}, one can see that the allowed region for each running mode lies between the two concentric circles, which can be a good complementary with current global analysis in constraining $(\epsilon_{ee}^{eL},\ \epsilon_{ee}^{eR})$. The coordinates of circle center for the contour of $(\epsilon_{ee}^{eL},\ \epsilon_{ee}^{eR})$ are dependent on $\sqrt{s}$. We can find that the direction from the SM point (0,0) to the circle center with $\sqrt{s}=91.2\ \mathrm{GeV}$ is approximately perpendicular to that with the other two running modes. Thus,  by combining the data from the three different running modes, the allowed regions for  NSI parameters with electrons can be severely constrained as compared to the global analysis, which is shown on the right side of Fig.~\ref{fig:cepceelr} with a green curve.
Even if both $\epsilon_{ee}^{eL}$ and $\epsilon_{ee}^{eR}$ are present, the allowed ranges for $|\epsilon_{ee}^{eL}|$ or $|\epsilon_{ee}^{eR}|$ can be constrained to be smaller than 0.002.

\subsection{Active-sterile neutrino transition magnetic moments} 
\label{sebsec:neutrino:NMM}

The discovery that neutrinos oscillate, and therefore neutrinos have distinct mass and flavor eigenstates, has
proven to be one of the most definitive pieces of evidence for physics beyond the Standard Model (BSM) in
the last two decades, which can be explained by including the additional heavy neutral leptons $N$ (often referred to sterile neutrinos). 
Ref.~\cite{Zhang:2023nxy} studied the active-sterile neutrino transition magnetic moments. The relevant operators respecting the ${\rm SU}(2)_L \otimes {\rm U}(1)_Y$ gauge symmetry can be written as~\cite{Magill:2018jla}
\begin{equation}
	\mathcal{L} \supset \bar{L}^k(d_{\cal W}^k {\cal W}_{\mu \nu}^a \tau^a + d_B^k B^{\mu \nu}) \tilde{H}\sigma_{\mu \nu} N+\mathrm{H.c.},
	\label{eq:LWB}
\end{equation}
where $\tilde{H}=i\sigma_2H^*$, $\tau^a=\sigma^a/2$ with $\sigma^a$ being Pauli matrices, ${\cal W}_{\mu \nu}^a$ and $B^{\mu \nu}$ denote the
${\rm SU}(2)_L$ and ${\rm U(1)}_Y$ field strength tensors with ${\cal W}_{\mu \nu}^a\equiv\partial_\mu {\cal W}_\nu^a-\partial_\nu {\cal W}_\mu^a+g\epsilon^{abc}W_\mu^b W_\nu^c$ and $B_{\mu\nu}\equiv\partial_\mu B_\nu-\partial_\nu B_\mu$, and $L$ are the SM lepton doublets.
After electroweak symmetry breaking  with the Higgs vacuum expectation value $v$, one obtains 
\begin{equation}
	\mathcal{L} \supset d_W^k(\bar{\ell}^k W^-_{\mu \nu} \sigma_{\mu \nu} N) + \bar{\nu}^k_L(d_\gamma^k F_{\mu \nu}-d_Z^k Z_{\mu \nu}) \sigma_{\mu \nu} N +\mathrm{H.c.},
	\label{eq:LWZk}
\end{equation}
which can induce  dipole operators to SM photon, the weak boson $Z$ and $W$, with
\begin{equation}
		d_\gamma^k=\frac{v}{\sqrt{2}}\left(d_{ B}^k\cos\theta_{w}+\frac{d_{{\cal W}}^k}{2}\sin\theta_w\right),\quad 
		d_Z^k=\frac{v}{\sqrt{2}}\left(\frac{d_{{\cal W}}^k}{2}\cos\theta_{w}-d_{ B}^k\sin\theta_w\right),\quad
		d_W^k=\frac{v}{2}d_{{\cal W}}^k,
	\label{eq:broken_unbroken_relation} 
\end{equation}
where $\theta_w$ is the weak mixing angle.

At CEPC, the sterile neutrino $N$ production will proceed from the process $e^+e^-\to N\bar\nu_k +{\rm H.c.}$  via either $Z$ or $\gamma$ mediator in $s$-channel depending on dipole portal couplings $d_Z^k$, $d_\gamma^k$ with $k=e,\mu,\tau$, 
or  via $W$ mediator in $t$-channel depending on electron neutrino dipole portal coupling  $d_W^e$ in Eq. (\ref{eq:LWZk}), respectively. 
With the subsequent decay channel $N\to\nu\gamma$ in the detector, the signature of a single photon final state with missing energy can be searched for at  CEPC.

\begin{figure}[!t]
	\begin{centering}
		\includegraphics[width=0.5\textwidth]{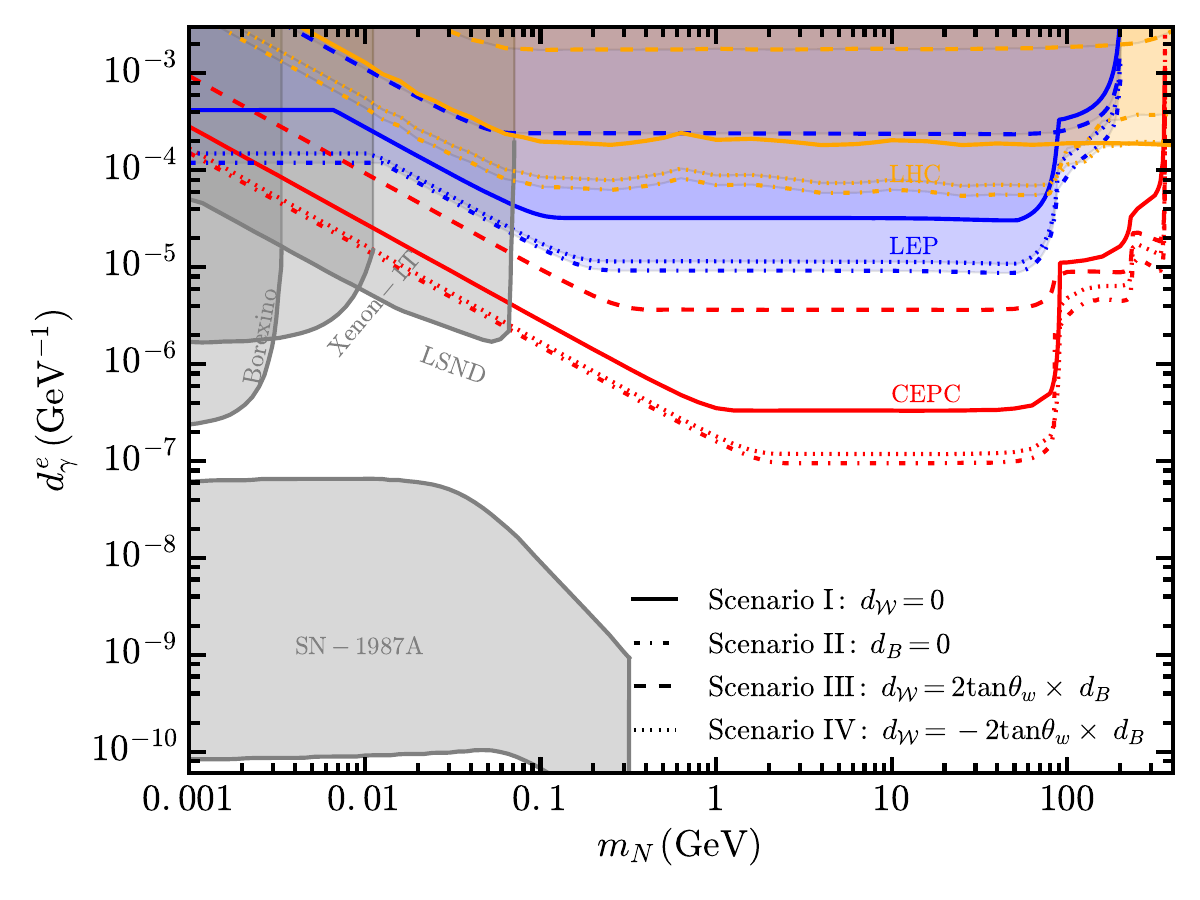} 
		\includegraphics[width=0.5\textwidth]{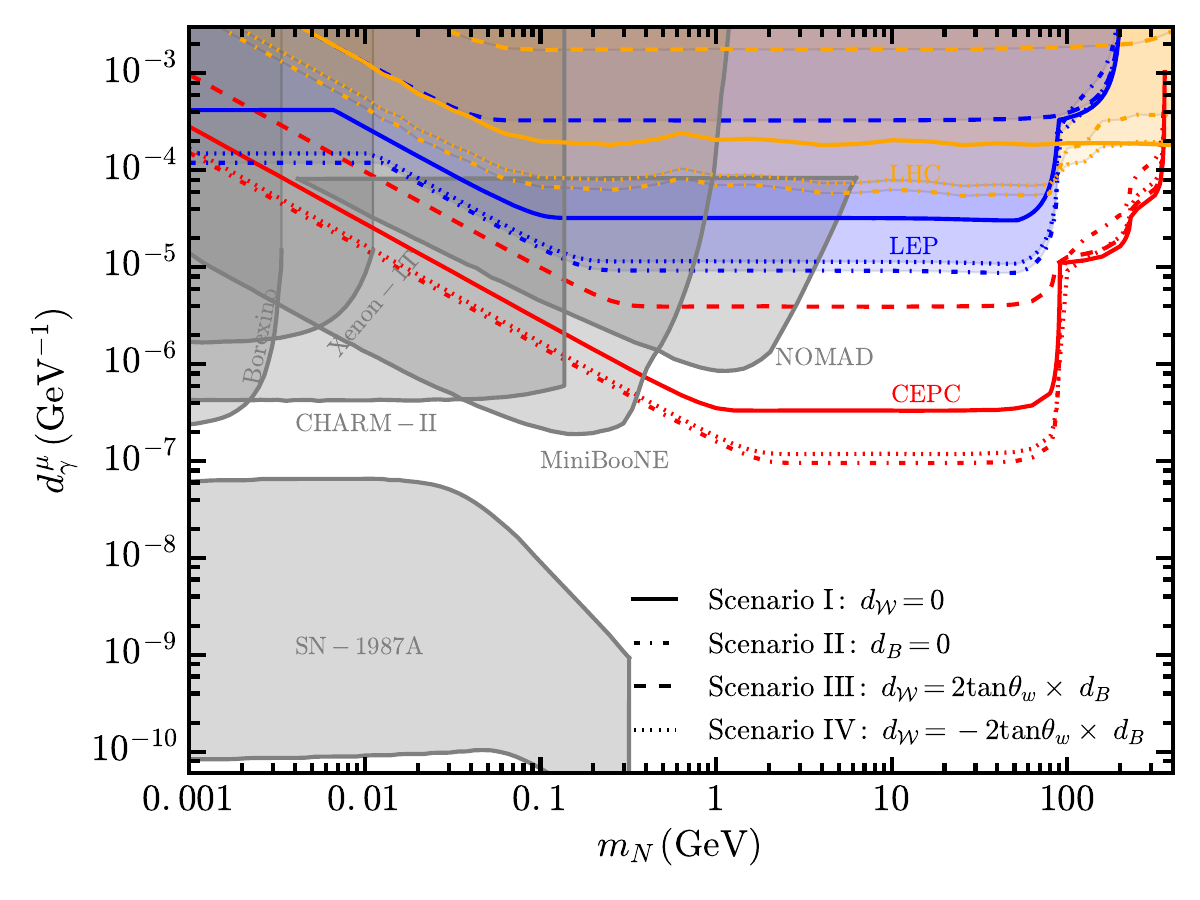} 
		\includegraphics[width=0.5\textwidth]{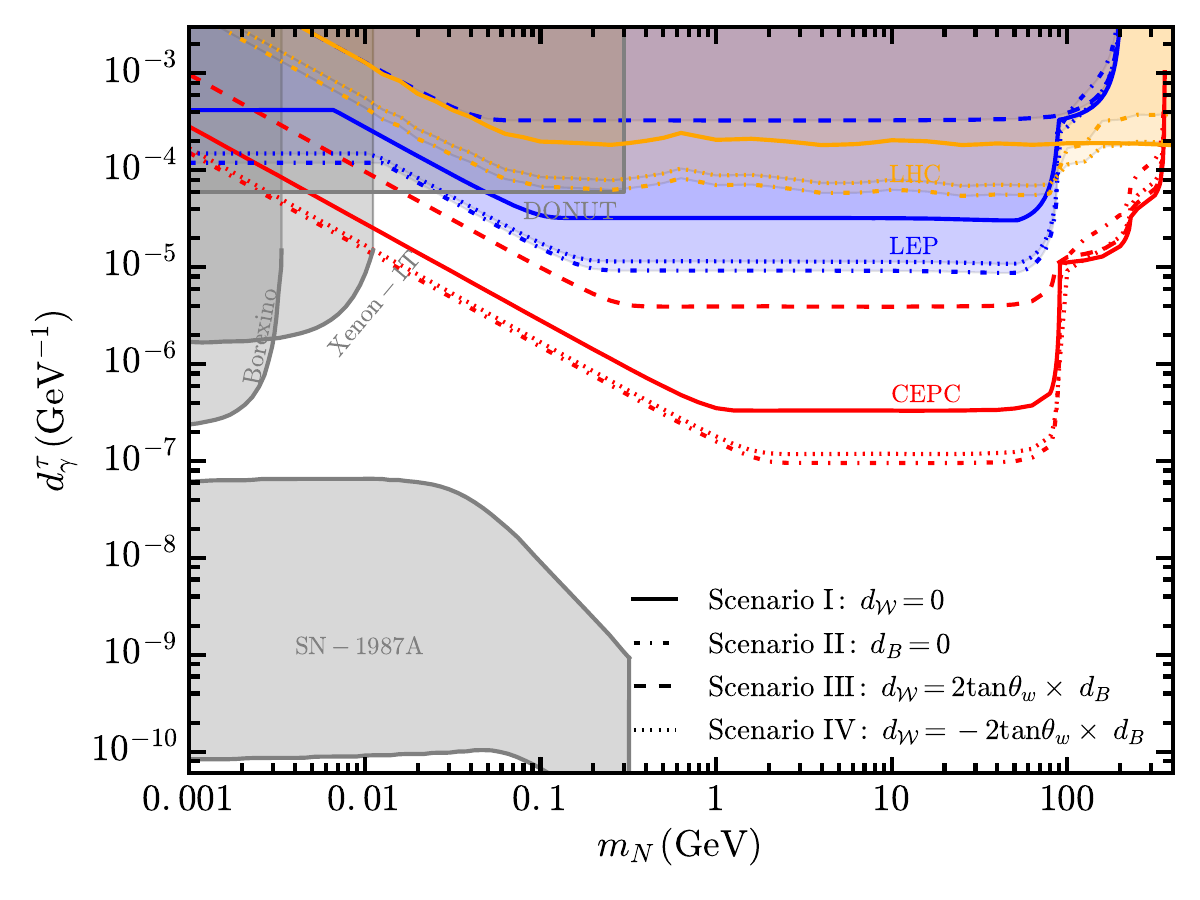}			
		\caption{
			The expected 95\% C.L. exclusion limits on active-sterile neutrino transition magnetic moment $d_\gamma^k$  in four scenarios at CEPC (red lines), which are the combination of the best constraints from four running modes at CEPC, for the three lepton flavors $e$ (upper), $\mu$ (middle) and $\tau$ (lower). The landscape of current leading constraints are also shown with shaded regions. Figure from Ref.~\cite{Zhang:2023nxy}.
		}
		\label{fig:total}
	\end{centering}
\end{figure}

The 95\% C.L. upper bounds on the neutrino dipole portal 
couplings $d_\gamma^k$ for the three lepton flavors $k = e,\; \mu,\; \tau$ at CEPC are shown in the upper, middle and lower panels of Fig.~\ref{fig:total}, respectively. And the limits from LEP~\cite{DELPHI:2003dlq, OPAL:1994kgw} and LHC~\cite{ATLAS:2020uiq, CMS:2018fon} are shown for comparison~\footnote{At the LHC, the limits for $d_{\cal W}= 2\tan\theta_w d_B$~($d_Z=0$) for $\tau$ flavor are not shown, as $\tau$ final states are not considered by the CMS search~\cite{ CMS:2018fon}, and we mainly focus on $pp \rightarrow Z$ channels for the ATLAS search~\cite{ATLAS:2020uiq}.}. The four scenarios with assumptions of $d_{\cal W}=0$, $d_B=0$ and $d_{\cal W}=\pm 2\tan\theta_w d_B$ are considered.
The combination of the best constraints from four running modes at CEPC with the total luminosity of $20\ \mathrm{ab}^{-1}$ data in the Higgs-mode, $6\ \mathrm{ab}^{-1}$  in	the $WW$-mode, $100\ \mathrm{ab}^{-1}$  
in the $Z$-mode, and $1\ \mathrm{ab}^{-1}$  in the $t\bar t$-mode is presented.
For  light sterile neutrino $N$, the $Z$-mode has the best sensitivity in all four scenarios. It is expected similar sensitivities could be reached at the FCC-ee.
One can see that depending on the the ratio $d_{\cal W}/d_B$, the constraints on  $d_\gamma^k$ can be fairly different.
While the current constraints on  $d_\gamma^k$ from terrestrial experiments such as  Borexino, Xenon-1T, CHARM-II, MiniBooNE, LSND, NOMAD,  and DONUT, and  astrophysics supernovae SN 1987A \cite{Magill:2018jla},  basically do not dependent on the ratio $d_{\cal W}/d_B$, since the typical scattering energies are far less than the electroweak scale.
The constraints from the monophoton searches at CEPC  are  in principle different on  $d_\gamma^e$ and on  $d_\gamma^{\mu,\tau}$ when $d_{\cal W}\neq0$, because there will be additional contributions from $W$-mediator.
In summary, CEPC can explore the previously unconstrained parameter region and will greatly improve the  limits on active-sterile neutrino transition magnetic moment  $d_\gamma^k$ compared to current experiments.

\subsection{Neutral and doubly-charged scalars in seesaw models}
\label{subsec:LL-doublyChargedHiggs}


Future lepton colliders can provide unique insight into the scalar sector of TeV scale models for neutrino masses with local $B-L$ symmetry. Our specific focus is on the TeV scale LRSM~\cite{Pati:1974yy,Mohapatra:1974gc,Senjanovic:1975rk}, which naturally embeds this $B-L$ symmetry. 
Due to mixing with other scalars, the neutral scalar $H_3$ from the right-handed triplet scalar $\Delta_R$ could acquire sizable flavor violating couplings to the charged leptons. Produced on-shell or off-shell at the planned $e^+e^-$ colliders, it would induce distinct lepton flavor violating (LFV) signals like $e^+e^- \to \ell_\alpha^\pm \ell_\beta^\mp ~ (+H_3)$ ($\alpha,\; \beta = e,\; \mu,\; \tau$), with the couplings $h_{\alpha\beta}$ probed up to $\sim 10^{-4}$ for a wide range of neutral scalar mass, which is well beyond the reach of current searches for charged LFV~\cite{BhupalDev:2018vpr}. Actually, the LFV signals induced by a neutral scalar are quite general in BSM scenarios~\cite{Dev:2017ftk}, e.g. in supersymmetric models with leptonic R-parity violation~\cite{Barbier:2004ez}, mirror models~\cite{Hung:2006ap,Bu:2008fx,Chang:2016ave}, and two-Higgs doublet models~\cite{Branco:2011iw,Crivellin:2015hha}, in addition to the LRSM~\cite{BhupalDev:2018vpr,Maiezza:2016ybz}.

\begin{figure}[!t]
  \centering
  \includegraphics[width=0.6\textwidth]{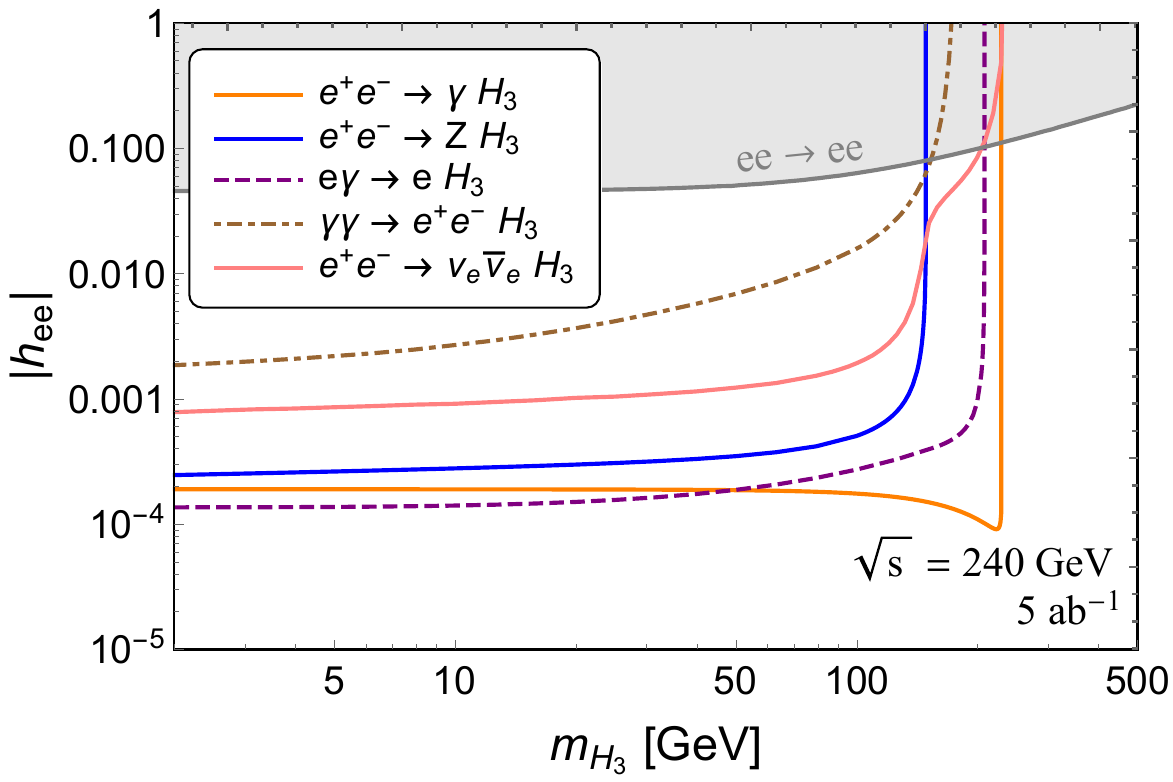} 
  \includegraphics[width=0.6\textwidth]{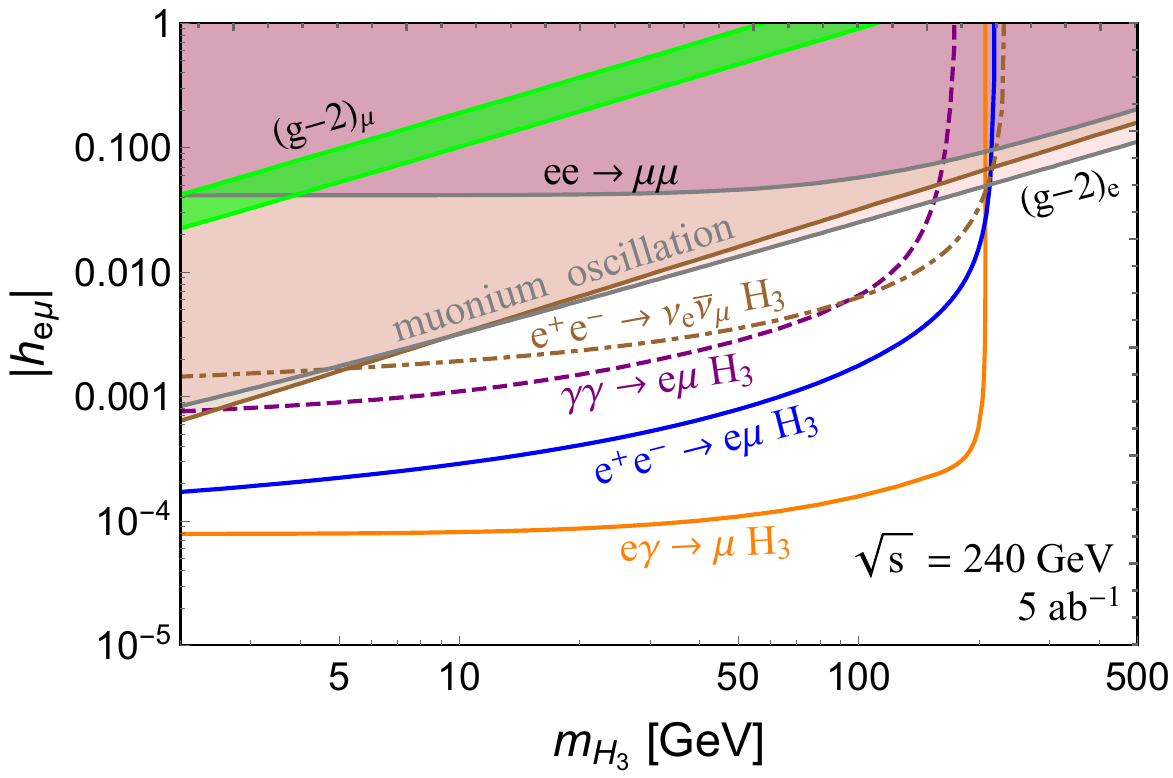}
  \caption{Prospects of the couplings $|h_{ee}|$ (top) and $|h_{e\mu}|$ (bottom) from the on-shell production of $H_3$ at CEPC (240 GeV and 5 ab$^{-1}$), in the channels of $e^+ e^- \to (\gamma/Z) H_3$, $e\gamma \to \ell H_3$, $e^+ e^-,\, \gamma\gamma \to \ell_\alpha^\pm \ell_\beta^\mp H_3$ and $e^+ e^- \to \nu \bar\nu H_3$. The shaded regions are excluded by current limits, while the yellow band corresponds to the muon $g-2$ discrepancy at $2\sigma$ C.L. Figure from Ref.~\cite{BhupalDev:2018vpr}. }
  \label{fig:H3:prospect1}
\end{figure}

In the LRSM, the neutral scalar $H_3$ can be produced at lepton colliders from its scalar coupling with the doubly-charged scalar $H^{\pm\pm}$,  Yukawa couplings to the RHNs and charged leptons, the 1-loop coupling to photons, and its mixing with the SM Higgs $h$~\cite{BhupalDev:2018vpr}. Then one can estimate the prospects of all the independent couplings $h_{\alpha\beta}$ at future lepton colliders. For illustration purpose, the prospects of $|h_{ee}|$ and $|h_{e\mu}|$ at the CEPC 240 GeV with an integrated luminosity of 5 ab$^{-1}$ are shown in the top and bottom panels of Fig.~\ref{fig:H3:prospect1}, respectively. The sensitivities of other couplings can be found in Ref.~\cite{BhupalDev:2018vpr}. The SM backgrounds are expected to be small, in particular for the LFV processes~\cite{Dev:2017ftk}. For simplicity, we have turned on only one of the couplings $h_{\alpha\beta}$ and set all others irrelevant to be zero. Neglecting the mixing of $H_3$ with the SM Higgs, the loop decay $H_3 \to \gamma\gamma$ and the decay $H_3 \to \nu\bar\nu$ suppressed by the heavy-light neutrino mixing $V_{\nu N}^4$, the neutral scalar $H_3$ decays predominantly into a pair of leptons, i.e. $H_3 \to \ell_\alpha^\pm \ell_\beta^\mp$. To be concrete, we assume a minimum number of 10 (30) for the signals with (without) LFV, and adopt an efficiency factor of $60\%$ for the tau lepton~\cite{Baer:2013cma}. In the process $e^+ e^- \to Z H_3$, only the visible decay products of $Z$ are taken into account. 
All the amplitudes for the on-shell production of $H_3$ depend linearly on the couplings $h_{\alpha\beta}$, 
thus free of the constraints from the rare LFV decays such as $\mu \to eee$ and $\tau \to e \gamma$ which depend quadratically on the Yukawa couplings $|h^\dagger h|$. 
The shaded regions are excluded by the muonium oscillation, electron $g-2$, muon $g-2$ (excluded by the theoretical-experimental discrepancy at the $5\sigma$ CL)~\cite{ParticleDataGroup:2024cfk} and the LEP $e^+e^- \to \ell^+ \ell^-$ data~\cite{DELPHI:2005wxt}. The yellow band in the bottom panel can explain the muon $g-2$ anomaly at the $2\sigma$ CL. 

\begin{figure}[!t]
  \centering
  \includegraphics[width=0.6\textwidth]{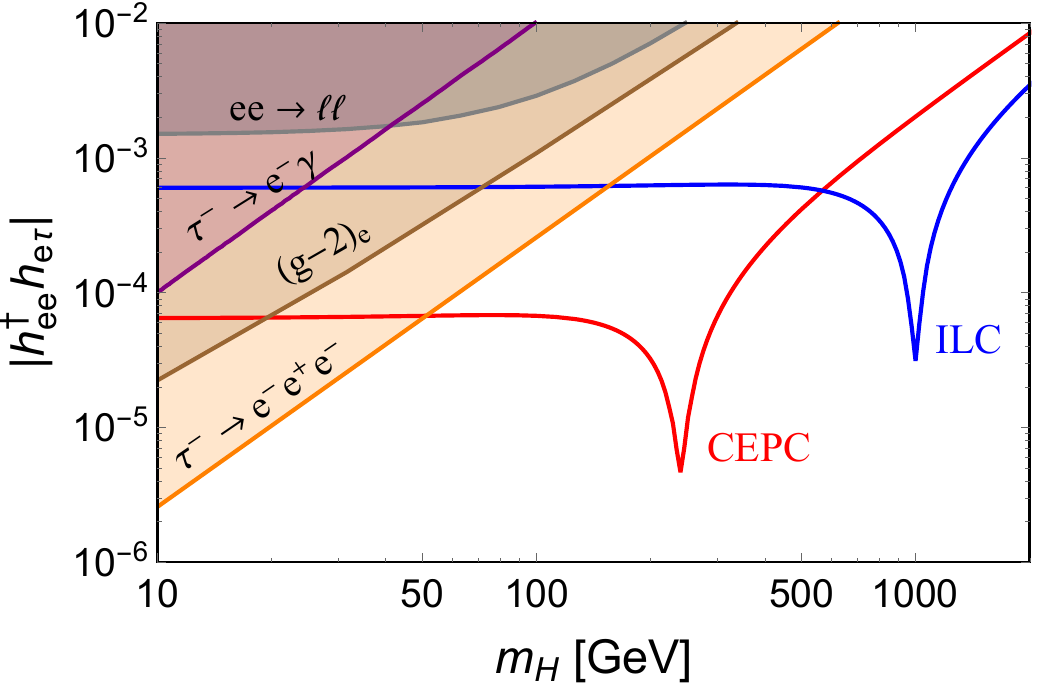} 
  \caption{Prospects of $|h^\dagger_{ee} h_{e\tau}|$ from searches of $e^+ e^- \to e^\pm \tau^\mp,\, \mu^\pm \tau^\mp$ at CEPC (red, $\sqrt{s} = 240$ GeV, ${\cal L} = 5$ ab$^{-1}$) and ILC (blue, 1 TeV and 1 ab$^{-1}$). The shaded regions are excluded by current limits. Figure from Ref.~\cite{Dev:2017ftk}.  }
  \label{fig:H3:prospect2}
\end{figure}

The neutral scalar $H$ may also mediate off-shell processes, e.g. $e^+ e^- \to \ell_\alpha^+ \ell_\beta^-$. For the illustration purpose, the sensitivities of $|h_{ee}^\dagger h_{e\tau}|$ are shown in Fig.~\ref{fig:H3:prospect2}. The red and blues lines are for the prospects at CEPC 240 GeV with an integrated luminosity of 5 ab$^{-1}$ and ILC 1 TeV with 1 ab$^{-1}$, respectively. With a nominal luminosity of 20 ab$^{-1}$ at CEPC, the corresponding sensitivities can be improved by a factor of $2$.
Also shown are the constraints from the rare lepton decays $\tau^- \to e^+ e^- e^-$, $\tau^- \to e^- \gamma$, electron $g-2$~\cite{ParticleDataGroup:2024cfk}, and the LEP $e^+ e^- \to \ell^+ \ell^-$ data~\cite{DELPHI:2005wxt}. More prospects of the couplings $|h^\dagger h|$ can be found in Refs.~\cite{Dev:2017ftk,BhupalDev:2018vpr}.

\begin{figure}[t!]
  \centering
  \includegraphics[width=0.6\textwidth]{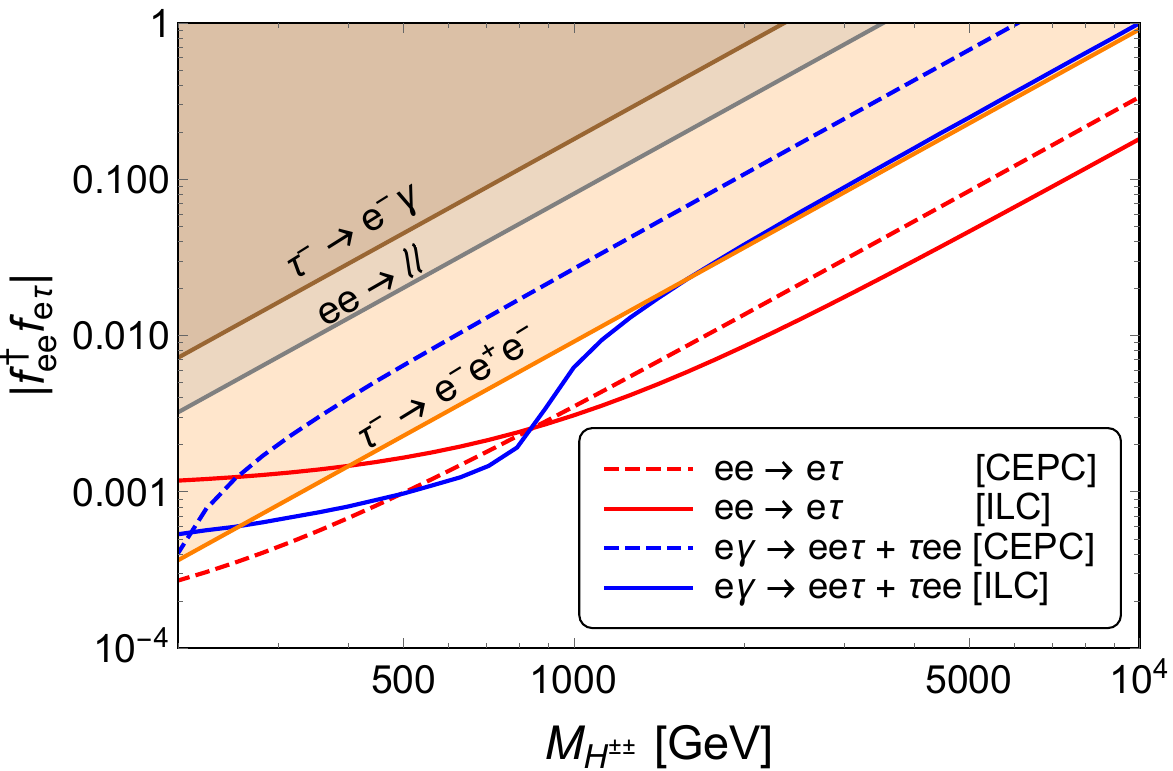}
  \caption{Prospects of the Yukawa couplings $|f_{ee}^\dagger f_{e\tau}|$ for the doubly-charged scalar $H^{\pm\pm}$ production via the $e^+ e^- \to e\tau$ (red) and $e^- \gamma \to e^+ e^- \tau^- + \tau^+ e^- e^-$ (blue) processes, at CEPC with $\sqrt{s} = 240$ GeV and an integrated luminosity of 5 ab$^{-1}$ (dashed) and ILC with $\sqrt{s} = 1$ TeV and 1 ab$^{-1}$ (solid). The shaded regions are excluded by current limits. Figure from Ref.~\cite{BhupalDev:2018vpr}. 
  }
  \label{fig:prospect:Hpp:2}
\end{figure}

The Yukawa couplings of the doubly-charged scalar $H^{\pm\pm}$ to the charged leptons might also be flavor-violating, which is directly correlated to the heavy RHN masses and mixings in the LRSM. With a combination of the pair, single and off-shell production of $H^{\pm\pm}$ like $e^+e^- \to H^{++} H^{--},\, H^{\pm\pm} e^\mp \mu^\mp,\, \mu^\pm \tau^\mp$, the Yukawa couplings can be probed up to $10^{-3}$ at future lepton colliders, which is allowed by current lepton flavor data in a large region of parameter space. As an explicit example, the prospects of $|f_{ee}^\dagger f_{e\tau}|$ in the the doubly-charged scalar induced processes $e^+ e^- \to e^\pm \tau^\mp$ and $e^- \gamma \to e^+ e^-\tau^- + \tau^+ e^- e^-$ are shown in Fig.~\ref{fig:prospect:Hpp:2}, as function of the doubly-charged scalar mass $M_{H^{\pm\pm}}$. The dashed lines are for the CEPC prospects, while the solid ones are the ILC sensitivities. The red and blue lines are for the $e^+ e^-$ and $e^- \gamma$ processes, respectively. The nominal luminosity of CEPC is 20 ab$^{-1}$ at 240 GeV, and as a result the corresponding sensitivities in Fig.~\ref{fig:prospect:Hpp:2} will be improved by a factor of $2$. The shaded regions are excluded by the rare tauon decays $\tau \to e \gamma$, $\tau \to eee$ and the LEP $ee \to \ell\ell$ data~\cite{DELPHI:2005wxt}. More prospects of other couplings of the doubly-charged scalar can be found in Ref.~\cite{BhupalDev:2018vpr}. As demonstrated in Figs.~\ref{fig:H3:prospect2} and \ref{fig:prospect:Hpp:2}, for both the neutral and doubly-charged cases, the scalar masses could be probed up to the few-TeV range in the off-shell channel. As a comparison, the center-of-mass energy at ILC is higher but the luminosity is relatively lower; therefore the CEPC and ILC are largely complementary in probing the LFV couplings of the neutral scalar. The LFV process can also be searched for at the high-energy muon and hadron colliders, e.g. in the processes of $\mu^+ \mu^-,\, q \bar{q} \to \ell_\alpha^\pm \ell_\beta^\mp + H_3$, which is expected to be largely complementary to the searches at the $e^+ e^-$ colliders.

The type-II seesaw mechanism with an isospin-triplet scalar $\Delta_L$ provides one of the most compelling explanations for the observed smallness of neutrino masses~\cite{Konetschny:1977bn,Magg:1980ut,Schechter:1980gr,Cheng:1980qt,Mohapatra:1980yp,Lazarides:1980nt}. The triplet contains a doubly-charged component $H_L^{\pm\pm}$, which decays predominantly to either same-sign dileptons or to a pair of $W$ bosons, depending on the size of the triplet vacuum expectation value. However, there exists a range of Yukawa couplings $f_L$ of the triplet to the charged leptons, wherein a relatively light $H_L^{\pm\pm}$ tends to be long-lived, giving rise to distinct displaced vertex signatures at the high-energy colliders~\cite{FileviezPerez:2008jbu,Melfo:2011nx,Kanemura:2014goa,BhupalDev:2018tox}. We find that the displaced vertex signals from the leptonic decays $H_L^{\pm\pm} \to \ell_\alpha^\pm \ell_\beta^\pm$ could probe a broad parameter space with $10^{-10} \lesssim |f_L| \lesssim 10^{-6}$ and 45.6 GeV $< M_{H_L^{\pm\pm}} \lesssim 200$ GeV at the high-luminosity LHC. Similar sensitivity can also be achieved at a future 1 TeV $e^+e^-$ collider. The mass reach can be extended to about 500 GeV at a future 100 TeV proton-proton collider. Similar conclusions apply for the right-handed triplet $H_R^{\pm\pm}$ in the TeV-scale LRSMs, which provide a natural embedding of the type-II seesaw. More details can be found in Ref.~\cite{BhupalDev:2018tox}. However, limited by the relatively low center-of-mass energy, it is expected that the CEPC 240 GeV can only probe much smaller parameter space of the doubly-charged scalar.

\subsection{Connection to Leptogenesis and Dark Matter}
\label{subsec:neutrino:DM}

Apart from the mysterious neutrino mass problem, there exists other well established evidence beyond the SM, e.g. baryon asymmetry of the universe~(BAU), DM, etc. An attractive solution can accommodate the explanations of all three problems in one unified model, the Neutrino Minimal Standard Model~($\nu$MSM)~\cite{Asaka:2005an,Asaka:2005pn}. In this model, three generations of RHNs $N_{1,2,3}$,  are added to the SM particle contents. These RHNs are all SM singlets, only interacts with the SM components via the active-sterile mixing.
Among them, the lightest one $N_1$ has tiny Yukawa couplings, thus tiny masses, can be the DM candidate~\cite{Dodelson:1993je,Shi:1998km,Abazajian:2001nj,Asaka:2006nq,Laine:2008pg}. The two heavier particles $N_{2,3}$ are responsible for generating the observed active neutrino masses via the aforementioned type-I seesaw mechanism. They also possess very similar masses, closing to the EW scale,
which can generate the asymmetry either via $CP$-violating RHN oscillations, or resonantly enhanced $CP$ asymmetry in RHN decay. Hence, the observed BAU can be explained by leptogenesis via neutrino oscillations~\cite{Akhmedov:1998qx, Asaka:2005pn} and resonant leptogenesis~\cite{Liu:1993tg,Flanz:1994yx,Flanz:1996fb,Covi:1996wh,Covi:1996fm,Pilaftsis:1997jf,Pilaftsis:1997dr,Pilaftsis:1998pd,Buchmuller:1997yu,Pilaftsis:2003gt,Pilaftsis:2005rv}.

\begin{figure}[!t]
\centering
\includegraphics[width=0.8\textwidth]{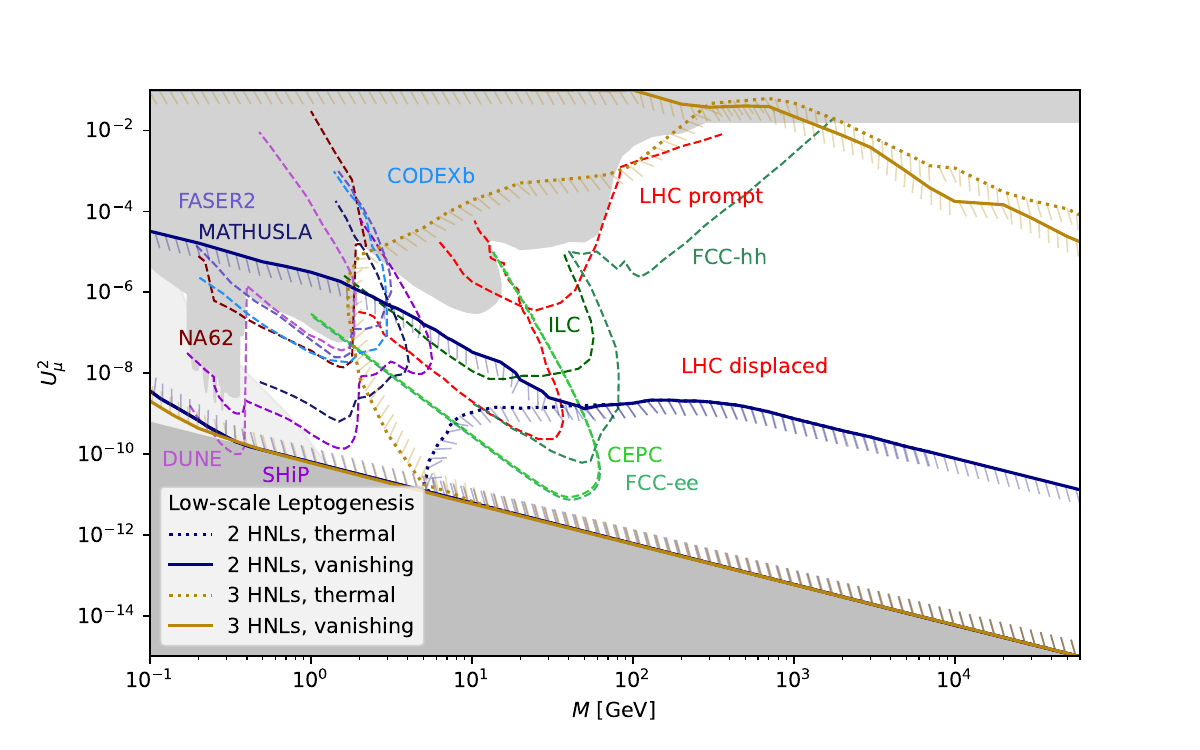}
\caption{
Theoretical predictions of the HNL mass $M$ and the ranges of $U_\mu^2$ in the framework of leptogenesis, with two  or three HNL flavours in the case of NO, for both vanishing and thermal initial HNL abundances. 
The shaded region in gray is excluded by existing limits,
complemented by the updated BBN bounds (light gray) and the lower bound from the seesaw mechanism (darker gray)~\cite{Klaric:2020phc,Klaric:2021cpi,Drewes:2021nqr}.
The colored lines indicate prospects at existing and future experiments.
Adapted from Ref.~\cite{Abdullahi:2022jlv}. 
}
\label{fig:lep}
\end{figure}

In the $\nu$MSM, the BAU is generated by ``low scale'' leptogenesis, since the mass scale of RHNs is below $10^9$ GeV, which is the Davidson-Ibrra bound implied by the `vanilla' leptogenesis~\cite{Davidson:2002qv}. The ``low scale'' leptogenesis includes leptogenesis via neutrino oscillations during freeze-in of the RHNs, and resonant leptogenesis during freeze-out~\cite{Klaric:2020phc,Klaric:2021cpi}. The two mechanisms can be united by a unique set of quantum kinetic equations, as described in Refs.~\cite{Klaric:2020phc,Klaric:2021cpi}. The viable parameter space of the model satisfying both the neutrino masses and BAU problems is thoroughly studied, with the summary shown in Fig.~\ref{fig:lep}~\cite{Drewes:2021nqr, Abdullahi:2022jlv}. 
The two \cite{Klaric:2020phc,Klaric:2021cpi} and three  \cite{Drewes:2021nqr} RHN scenarios in the case of normal ordering (NO) of neutrino masses, for both vanishing and thermal initial HNL abundances, are included.
The shaded region in gray is excluded by past experiments~\cite{CHARM:1985nku,Abela:1981nf,Yamazaki:1984sj,E949:2014gsn,Bernardi:1987ek,NuTeV:1999kej,Vaitaitis:2000vc,CMS:2018iaf,DELPHI:1996qcc,ATLAS:2019kpx,CMS:2022fut},
complemented by the updated BBN bounds in light gray from Refs.~\cite{Sabti:2020yrt,Boyarsky:2020dzc} and the lower bound from the seesaw mechanism in darker gray. See Refs.~\cite{Klaric:2020phc,Klaric:2021cpi,Drewes:2021nqr} for details.
    The various colored lines indicate existing~\cite{Izaguirre:2015pga,Das:2017gke,Pascoli:2018heg,Drewes:2019fou,Drewes:2018gkc} and future~\cite{Ballett:2019bgd,FASER:2018eoc,SHiP:2018xqw,Gorbunov:2020rjx,Curtin:2018mvb,Aielli:2019ivi,Antusch:2017pkq,Antusch:2016ejd,Blondel:2022qqo} experiments that will be able to probe the low-scale leptogenesis parameter space. 
As indicated in the figure, the sensitivity of the CEPC, shown in light green, can be sensitive to the parameter space where both seesaw and leptogenesis mechanisms are successful, no matter the number of RHNs and the initial condition.

The DM relic density $\Omega_{\rm DM}$ can also be explained by accounting the lightest right-handed neutrinos $N_1$ as the DM candidate. In the $\nu$MSM, $\Omega_{\rm DM}$ is only produced by mixing with active neutrinos~\footnote{It can also be produced by other mechanisms if the $\nu$MSM is extended, e.g. with Higgs inflation~\cite{Bezrukov:2007ep}. }~\cite{Dodelson:1993je,Shi:1998km,Abazajian:2001nj,Asaka:2006rw,Laine:2008pg}. Sufficient production is generated if large lepton symmetry is generated at the low temperature of $\mathcal{O}(200)$  MeV~\cite{Shi:1998km,Abazajian:2001nj,Laine:2008pg,Venumadhav:2015pla,Ghiglieri:2015jua,Ghiglieri:2019kbw,Ghiglieri:2020ulj,Bodeker:2020hbo}. To successfully reproduce $\Omega_{\rm DM}$, $M_{N_1} \sim \mathcal{O}(1)$ keV and $M_{N_2} \approx M_{N_3} \gtrsim \mathcal{O}(1)$ GeV~\cite{Canetti:2012vf}. 
More extended models, e.g. the LRSM, can also accommodate the origin of BAU and DM~\cite{Mohapatra:1974gc}. The discussions of the connection between them in such models can be found in Refs.~\cite{Nemevsek:2012cd, Nemevsek:2023yjl}. Such models have already been searched by $\ell$+MET signatures at LHC~\cite{Nemevsek:2023yjl}, and can be tested at CEPC, for example via $e^+ e^- \rightarrow N N$ processes mediated by the $W_R$ or $Z^\prime$ boson. If $M_N$ $\gtrsim \mathcal{O}(1)$ GeV, same sign dilepton plus multiple jets signatures are studied in Ref.~\cite{Biswal:2017nfl}, and displaced vertex signatures in Ref.~\cite{Urquia-Calderon:2023dkf},  and can be further extended by searching for monophoton signatures if the $N$ is even more stable, when
$M_{N} \sim \mathcal{O}(1)$ keV to explain the DM. 

\subsection{Summary}

\begin{figure}[!t]
\centering
\includegraphics[width=0.8\textwidth]{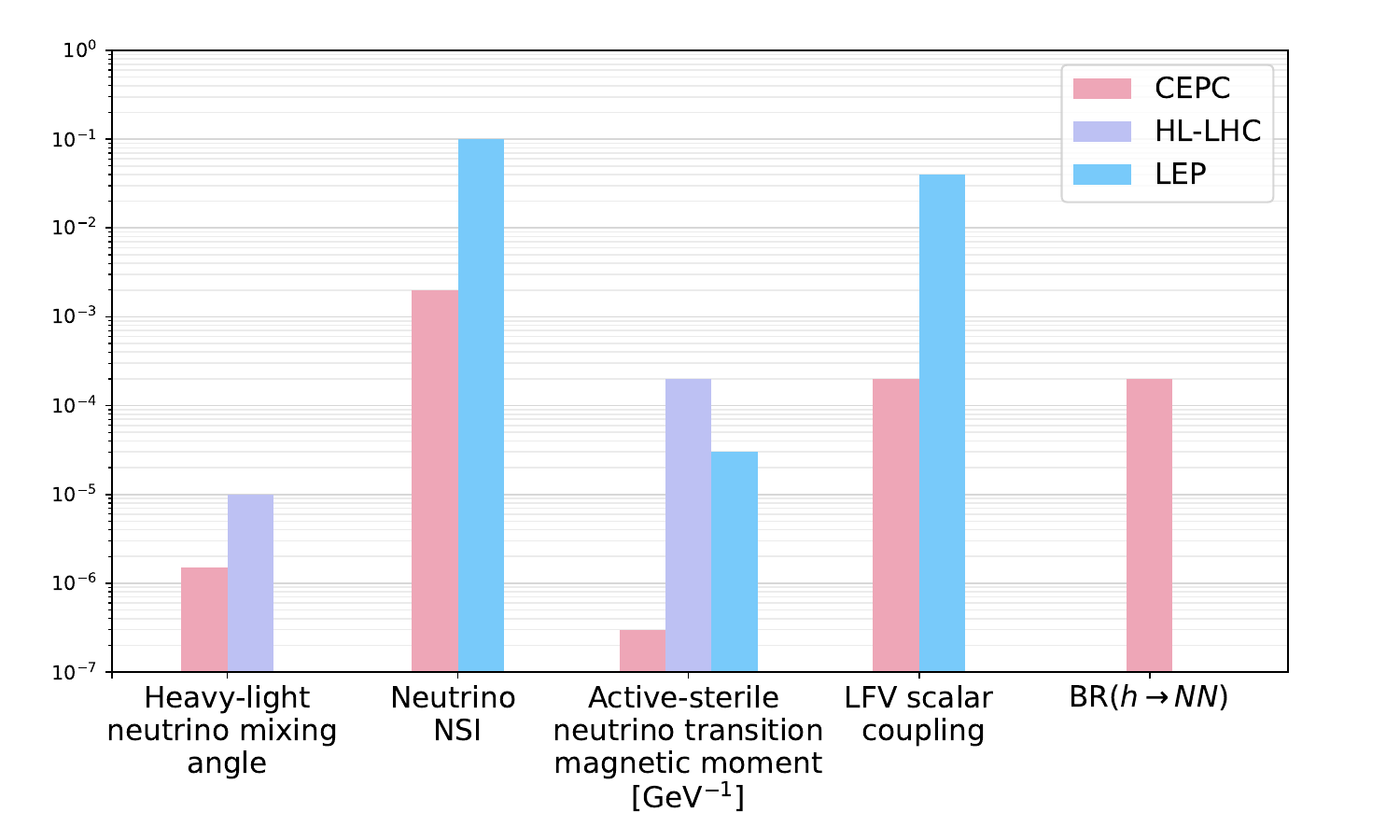}
\caption{Prospects of neutrino physics observables at the CEPC, in comparison with the LEP limits and LHC prospects. See text for more details.
}
\label{fig:neutrino}
\end{figure}

In this sections, we have examined the prospects of CEPC to some benchmark scenarios relevant to neutrino physics, in particular for the heavy neutrinos, non-standard neutrino interactions, active-sterile neutrino transition magnetic moments, the neutral and doubly-charged scalars in seesaw models, and the connection to leptogenesis and DM. The CEPC sensitivities of some representative neutrino physics relevant observables are presented in Fig.~\ref{fig:neutrino}. For illustration purpose we have chosen the heavy-light neutrino mixing angle from Fig.~\ref{fig:heavyneutrino:summary}, neutrino NSI from Fig.~\ref{fig:cepceelr}, active-sterile neutrino transition magnetic moment (in unit of GeV$^{-1}$) from Fig.~\ref{fig:total}, the LFV scalar coupling from Fig.~\ref{fig:H3:prospect1}, and the branching fraction BR($h\to NN$) from Fig.~\ref{fig:hNNlimits}. The comparison of CEPC with the existing LEP limits and the future prospects at the LHC are also shown in this figure; see the subsections above and figures therein for more details. In general, neutrino-relevant signals may behave like rare processes, e.g. the LFV or LNV events or displaced vertices.
Benefiting from the high luminosity and clean environment, the neutrino relevant sensitivities can be improved at the CEPC by one or two orders of magnitude, as exemplified in Fig.~\ref{fig:neutrino}.

\clearpage
\section{More Exotics}
\label{sec:exotica}

As a vast number of NP scenarios involve the exotic couplings to the SM electroweak and lepton sectors, it is generally expected that exotic searches can benefit from the high luminosity at CEPC's $Z$-pole and Higgs factory runs. High precision on the $Z$, $h$ widths and good reconstruction of the decay products offer powerful tests of exotic processes, including lepton number/flavor violation, sterile states, ALPs and many others. The CEPC's low hadronic activity level helps to minimize the contamination from hadronic initial state radiation, hence enhances the potential to accurately identify signals that involve relatively soft leptons, photons and jets. This section lists selected exotic physics studies that can potentially benefit from CEPC, with a partial focus on axion-like particles. Admittedly, many exotic studies still lack dedicated and quantitative analysis, and we list several possibilities for potential interest, such as electromagnetic form factors for hadrons and unstable leptons, exotic lepton mass and flavor models, etc. In addition, there is rising interest in spin-related kinematical observables, such as transverse spin~\cite{Wen:2023xxc,Wen:2024nff} and quantum entanglement~\cite{Aguilar-Saavedra:2022mpg,Barr:2021zcp,Bi:2023uop,Du:2024sly}. We expect these studies will provide more diversified avenues for the CEPC's physics science potential.

\subsection{Axion-like particles} %
\label{subsec:moreExotics-ALP}

As a relaxed solution to the ``strong-CP" problem, the Peccei-Quinn mechanism predicts the existence of the QCD axion~\cite{Peccei:1977hh,Weinberg:1977ma,Wilczek:1977pj}, which develops a coupling with gauge bosons at one-loop level. The characteristic Chern-Simons term $aF\tilde{F}/f_a$ leads to the generalization toward ALPs, which can arise in many NP scenarios containing the breaking of a global $U(1)$ symmetry~\cite{Svrcek:2006,Arvanitaki:2010,Cicoli:2012,Arias:2012}. The prospects for discovering ALPs via a light-by-light scattering at FCC-ee and CEPC have been extensively investigated. At future lepton colliders, promising sensitivities to the effective ALP-photon coupling $g_{a\gamma\gamma}$ can be derived for $m_a \alt 10$ GeV~\cite{Zhang:2021sio}. Here we list the projected limits from several recent ALP studies for the Higgs factory and higher-energy runs at the CEPC.

\begin{figure}[th!]
	\includegraphics[width=0.8\textwidth]{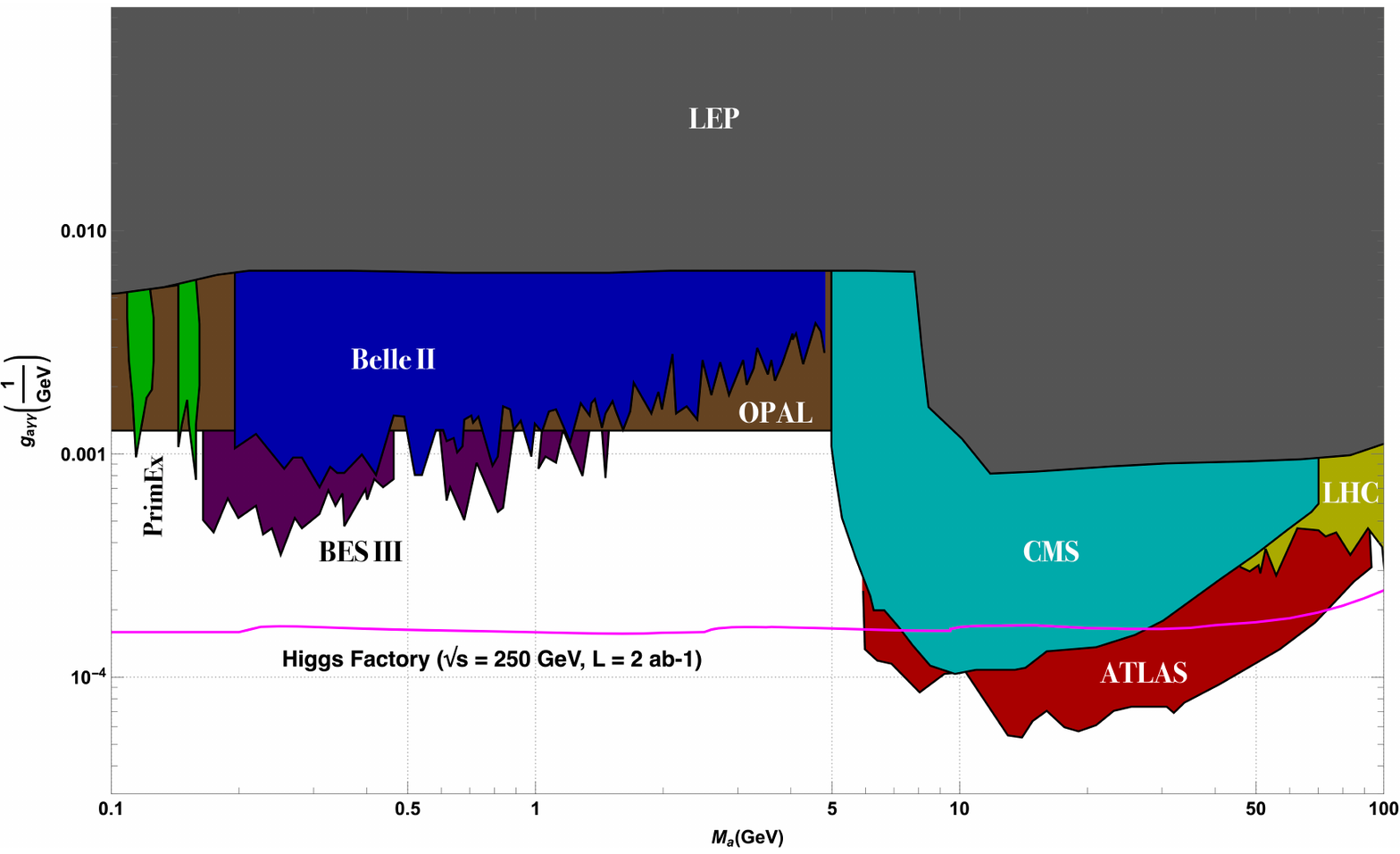}
	\caption{\small\label{fig:cheung_fig1}
Summary plot of the sensitivity to $g_{a\gamma\gamma}$ that can be achieved at $e^+e^-$ collider with $\sqrt{s}=250$ GeV and an integrated luminosity 2 ab$^{-1}$. Existing constraints are also shown for comparison, including  PrimEx~\cite{Aloni:2019ruo}, BES III~\cite{BESIII:2022rzz}, Belle II~\cite{Belle-II:2020jti}, LEP~\cite{Jaeckel:2015jla}, OPAL~\cite{Knapen:2016moh}, Pb-Pb at LHC~\cite{Knapen:2016moh}, CMS~\cite{CMS:2018erd} and ATLAS~\cite{ATLAS:2020hii}.}
\end{figure}

The ALP couplings to the SM electroweak gauge bosons read
\begin{equation}
  \mathcal{L}=
     -C_{BB}\frac{a}{f_a}B_{\mu\nu}\Tilde{B}^{\mu\nu}
     -C_{WW}\frac{a}{f_a}W^i_{\mu\nu}\Tilde{W}^{\mu\nu,i}.
\end{equation}
where $f_a$ is the ALP's decay constant. After electroweak symmetry breaking, the neutral fields $B,\,W^3$ will be rotated into the mass eigenstates $\gamma,\, Z$, and the conventional ALP couplings to $\gamma\gamma,\,WW,\,ZZ,\, Z\gamma$ are given, respectively, by 
\begin{align}
\begin{aligned}
g_{a\gamma\gamma}&=\frac{4}{f_a}(C_{BB}c_w^2+C_{WW}s_w^2), & \qquad
g_{aWW}&=\frac{4}{f_a}C_{WW}\,, \\
g_{aZZ}&=\frac{4}{f_a}(C_{BB}s_w^2+C_{WW}c_w^2), & \qquad
g_{aZ\gamma}&=\frac{8}{f_a}s_wc_w(C_{WW}-C_{BB}) \,.
\end{aligned}
\end{align}

\begin{figure}[t]
    \centering
    \includegraphics[width=0.7\textwidth]{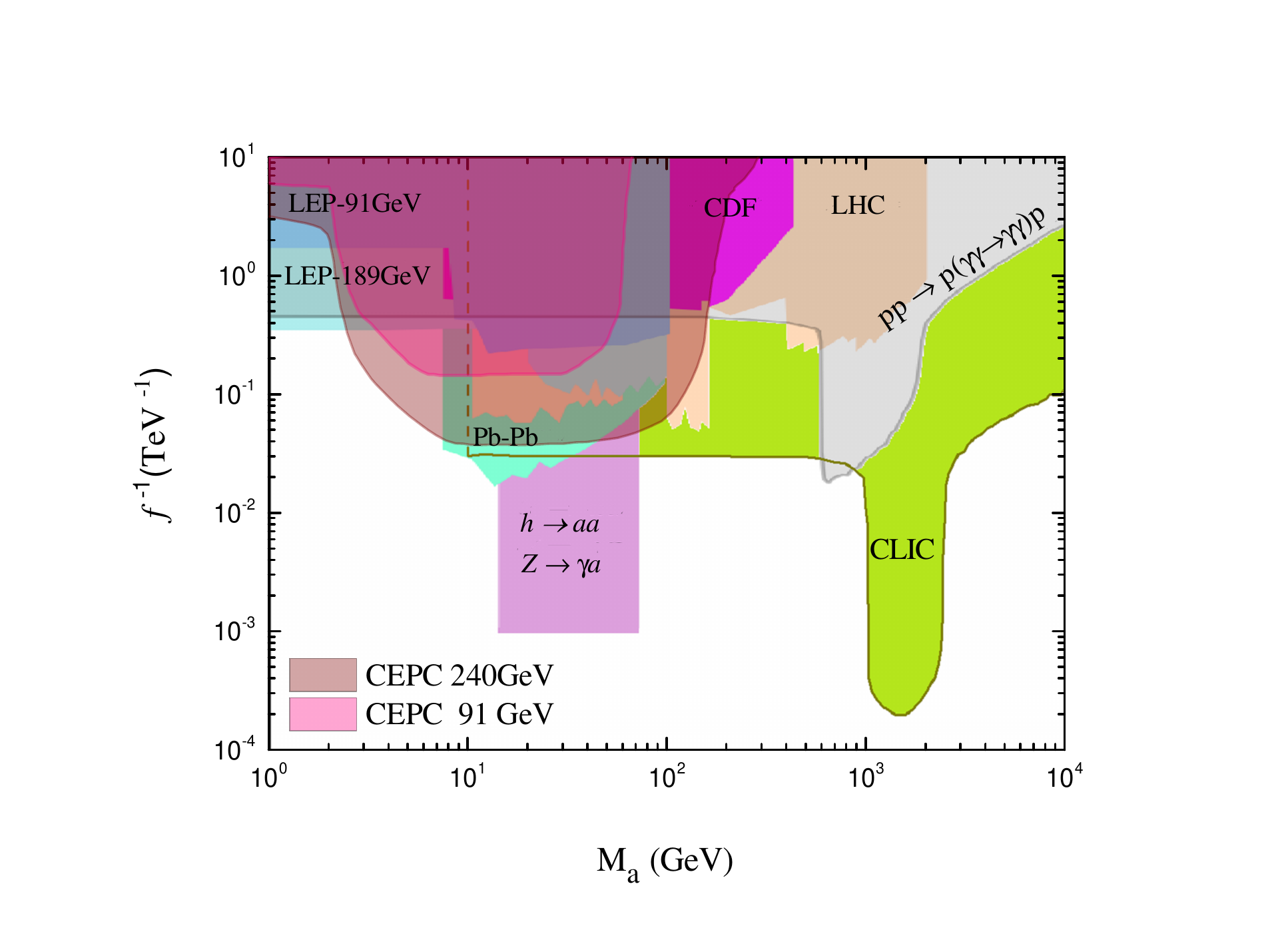}
    \includegraphics[width=0.49\textwidth]{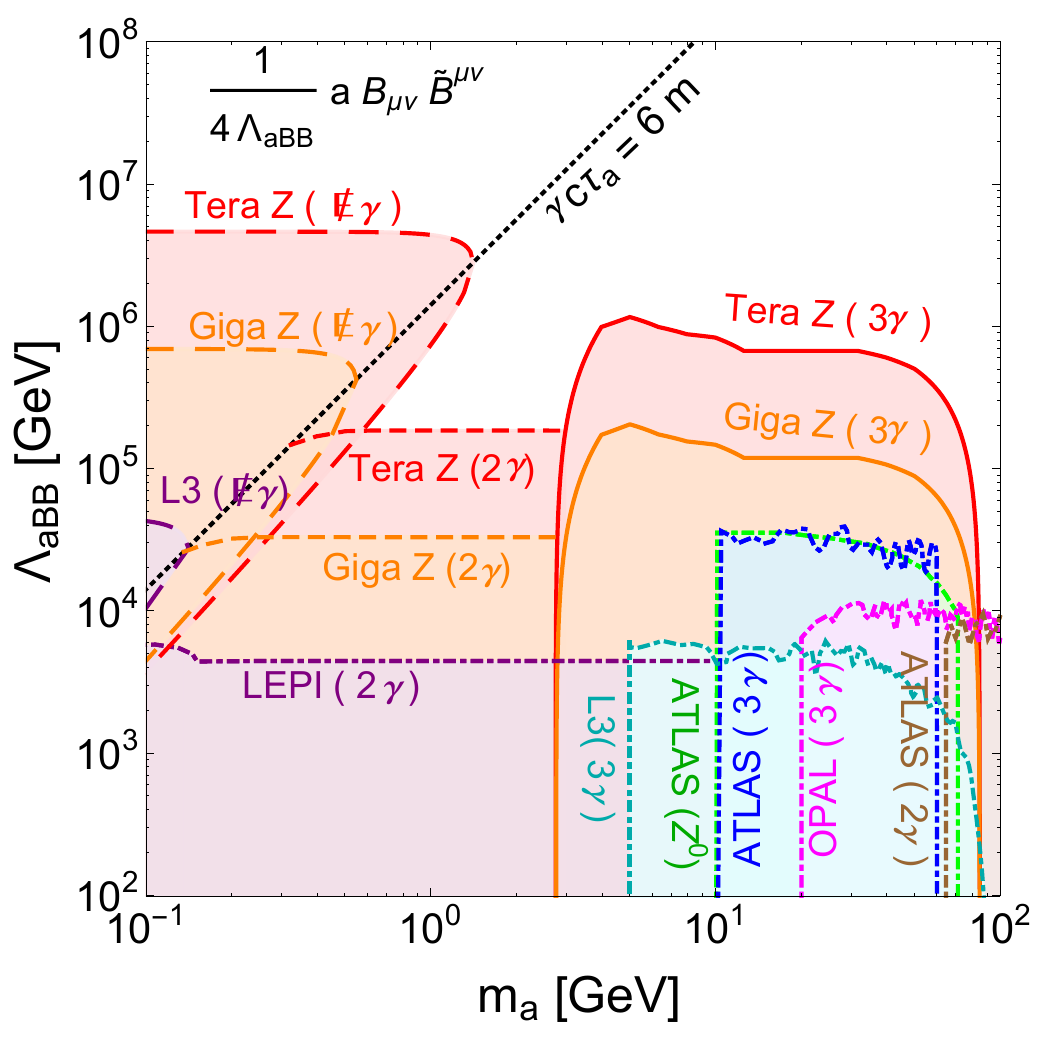}
    \caption{Top panel: 95 C.L. sensitivity regions on the ALP coupling $g_{a\gamma\gamma}$ as a function of $M_a$ for the process $e^+e^- \rightarrow \gamma\gamma e^{+}e^{-}$ at the 91 and 240 GeV runs of the CEPC. Bottom panel: sensitivity to the axion coupling to the SM $U(1)_Y$ field at the $Z$-factory. See text for references. }
    \label{fig:yangshuo}
\end{figure}

Ref.~\cite{Cheung:2023nzg} employed the ALP production processes $e^+ e^- \to f^+ f^- a$, where $f = e, \mu, \nu$, and devised a set of selection cuts to improve the signal-background ratio. The ALP is emitted by the gauge boson in the internal line of the process. The emitted ALP subsequently decays via $a\to \gamma\gamma$. 
Fig.~\ref{fig:cheung_fig1} illustrates the CEPC sensitivity reach for $\sqrt{s} = 250$ GeV with an integrated luminosity of $L = 2\,{\rm ab}^{-1}$, and the sensitivity can scale up by the square root of the luminosity for 20 ${\rm ab}^{-1}$. The upcoming Higgs factories can improve the sensitivity from the current constraints down to $2 \times 10^{-4}\,{\rm GeV}^{-1}$ for $m_a = 0.1 - 6$ GeV. See Ref.~\cite{Cheung:2023nzg} for further details of the study.

Ref.~\cite{Zhang:2021sio} investigated a similar light-by-light scattering $e^+e^-\rightarrow \gamma \gamma e^+e^-$ induced by ALP exchange at the CEPC, and derived the production cross-section and expected CEPC sensitivity reach for $\sqrt{s}=$91/240~GeV runs. The projected limits are shown in the upper panel of Fig.~\ref{fig:yangshuo}. Existing limits including LHC pb-pb~\cite{Knapen:2016moh}, LHC diphoton~\cite{Baldenegro:2018hng} and CLIC~\cite{Inan:2020aal} are shown for comparison. Also at the $Z$ pole, Ref.~\cite{Liu:2017zdh} investigated the axion coupling to the SM $U(1)_Y$ hypercharge field in the process of $Z$ boson decaying to $a\gamma$, with the axion subsequently decaying into two photons. The projected sensitivities are labeled as $3\gamma$ in the lower panel of Fig.~\ref{fig:yangshuo}, where the $3\gamma$ and $Z$ pole limits are given in terms of $g_{aBB}$, the ALP coupling to the SM $U(1)_Y$ gauge field. For comparison, the lower panel also includes the limits from ATLAS 3$\gamma$~\cite{ATLAS:2015rsn},$2\gamma$~\cite{ATLAS:2014jdv}, L3\cite{L3:1994shn,L3:1997exg} and OPAL\cite{OPAL:2002vhf}, etc.

\medskip

\begin{figure*}[ht!]
 \begin{center} 
 \includegraphics[width=1\textwidth]{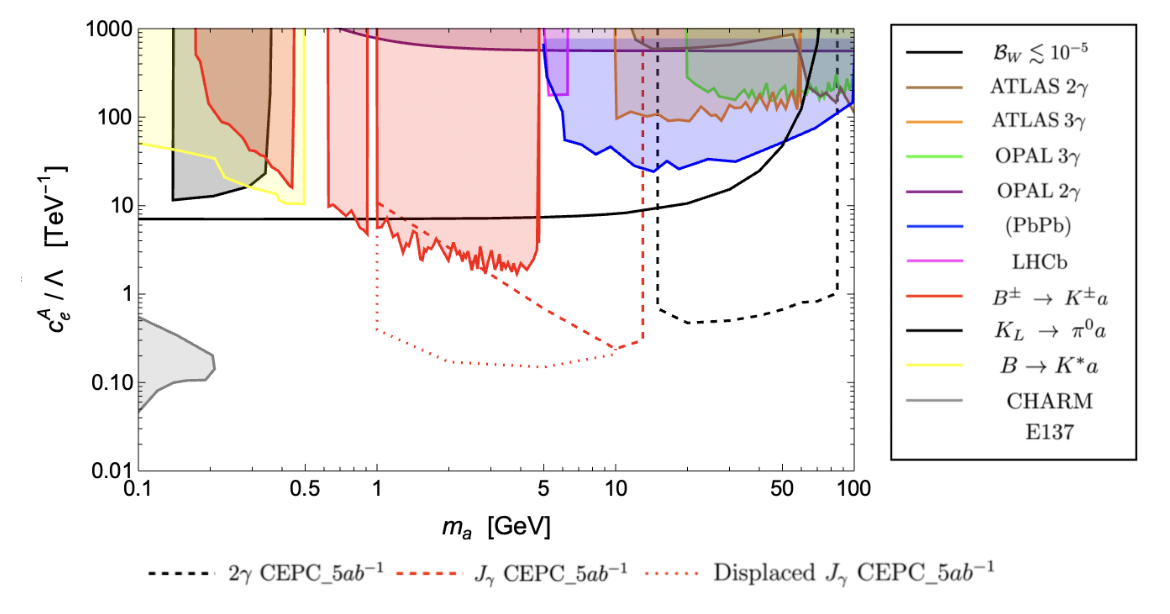}
 \end{center} 
\caption{Future bounds on the coupling $c^A_e/\Lambda$ of $e$ALPs from CEPC with $L = 5 {\rm ~ab}^{-1}$ within $95\%$ C.L. or equivalently $\ge 10$ survival events for background-free cases (dashed for the $e$ALP prompt decay and dotted for $e$ALP as a long-lived particle) as well as existing bounds (shaded). "$2\gamma$" and "$J_{\gamma}$" denote two distinct signatures at CEPC. The ${\cal B} (W^{\pm}\rightarrow\ell^{\pm}\nu a) < 10^{-5}$ limit~\cite{Altmannshofer:2022ckw} (black solid) and other collider bounds are presented for comparison, 
including: ATLAS $2\gamma$~\cite{ATLAS:2014jdv,Jaeckel:2015jla,Knapen:2016moh} (brown bulk), ATLAS $3\gamma$~\cite{ATLAS:2015rsn,Knapen:2016moh} (orange bulk), OPAL $3\gamma$~\cite{OPAL:2002vhf,Knapen:2016moh} (green bulk), OPAL $2\gamma$~\cite{OPAL:2002vhf,Knapen:2016moh}, ATLAS/CMS (PbPb)~\cite{dEnterria:2021ljz} (blue bulk) and LHCb~\cite{Benson:2018vya,CidVidal:2018blh} (magenta bulk). For light $e$ALPs, $B^{\pm}\rightarrow K^{\pm}a\rightarrow K^{\pm}(\gamma\gamma)$ from BaBar~\cite{BaBar:2021ich} (red bulk), $K_L\rightarrow\pi^0 a\rightarrow\pi^0 (e^+ e^-)$ from KTeV~\cite{KTeV:2003sls} (black bulk) and $B\rightarrow K^{\ast}a\rightarrow K^{\ast}(e^+ e^-)$ from LHCb~\cite{LHCb:2015ycz,Altmannshofer:2022ckw} (yellow bulk) are involved. Finally, the bounds from CHARM~\cite{CHARM:1985anb,Altmannshofer:2022ckw} and SLAC E137~\cite{Bjorken:1988as} (gray bulk) are also included.
}
\label{fig:summary_Lu}
\end{figure*} 

In light of an electroweak-violating scenario, Ref.~\cite{Altmannshofer:2022ckw} studied a four-point interaction denoted as $W$-$\ell$-$\nu$-$a$, a coupling that does not depend on the electron mass. This channel provides an opportunity to explore electrophilic ALPs ($e$ALPs) at the GeV scale. In this work, a novel $t$-channel process was investigated: $e^{+}e^{-}\rightarrow\nu_e a\overline{\nu}_e$, which involves the $W$-$\ell$-$\nu$-$a$ four-point interaction with effective coupling $\epsilon_e^A/\Lambda$. This process exhibits significant energy enhancement behaviors in its cross-sections as collision energy increases~\cite{Lu:2022zbe}. For GeV-scale $e$ALPs, their primary decay mode involves a photon pair, induced by the chiral anomaly, rather than an electron-positron pair. Consequently, the characteristic signal signature of this $t$-channel process consists of a photon pair accompanied by missing energy. Depending on the mass and decay width of the $e$ALPs, the final state can manifest as either two isolated photons ($2\gamma$), a photon-jet ($J_{\gamma}$), or a displaced $J_{\gamma}$. The analysis indicated that the potential future bounds on the coupling $c^A_e/\Lambda$ can be as stringent as $0.1-1.0$ TeV$^{-1}$ for $1$ GeV $\leqslant m_a\lesssim M_W$, at the CEPC with $L = 5~{\rm ab}^{-1}$ and $\sqrt{s}=240$ GeV. These constraints are depicted in Fig.~\ref{fig:summary_Lu}.

\subsection{Emergent Hadron Mass} 

It is common to regard the Higgs boson (HB) as the origin of mass within the SM of particle physics. Certainly, the Higgs mechanism is a mechanism that contributes to our understanding of the origin of mass of subatomic particles. Such Higgs couplings with the SM fermions 
produce the electron mass, $m_{e}=0.511\,$MeV, and the quark current masses, amongst them the light $u$ (up) and $d$ (down) quarks: $m_u \approx 4m_e \approx 2.2 \,$MeV, $m_d \approx 2 m_u$.
These particles combine to form the hydrogen atom, the most abundant element in the Universe, whose mass is 939\,MeV. Somehow one electron, two $u$ quarks and one $d$ quark, with a total Higgs-generated mass of $\sim 13 m_e \approx 6.6\,$MeV, combine to form an object whose mass is 140-times greater. Plainly, Nature must have another very effective mass-generating mechanism, which is now identified as emergent hadron mass (EHM) \cite{Roberts:2021nhw, Ding:2022ows}.

\begin{figure*}[h]
\includegraphics[width=0.49\textwidth]{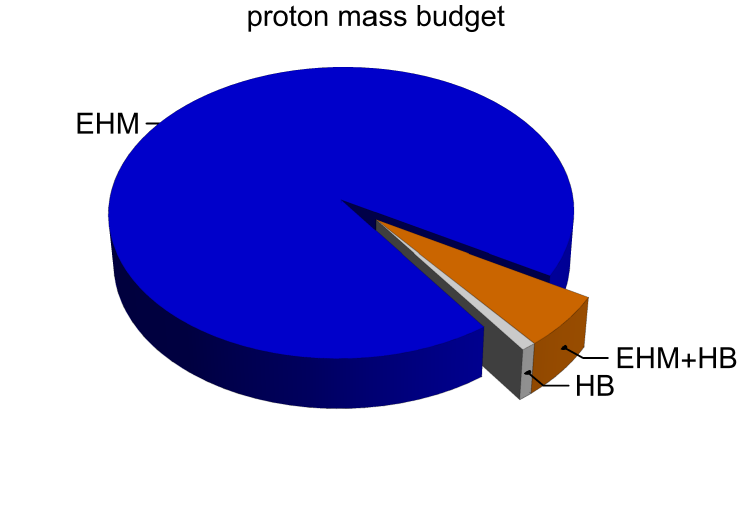} 
\includegraphics[width=0.49\textwidth]{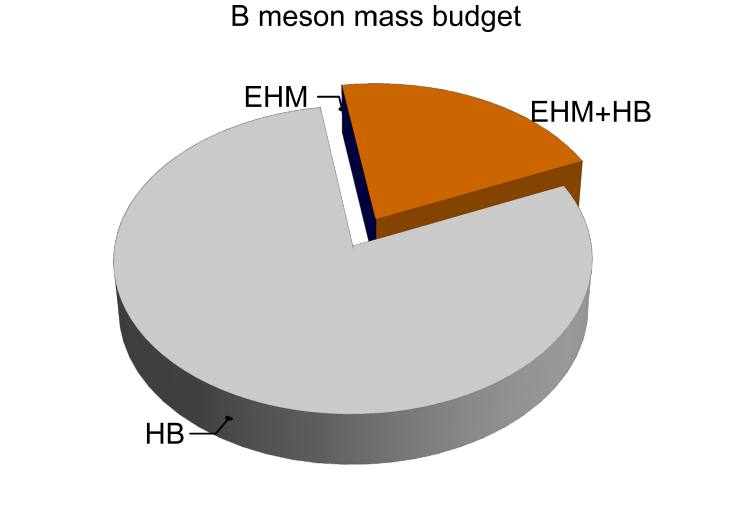}
\caption{\label{protonmass}
Poincar\'e-invariant decompositions of hadron masses:
(\textsf{A}) proton;
(\textsf{B}) $B$-meson.
EHM is the source of 94\% of the proton mass; by itself, the HB accounts for just 1\%; and the remaining 5\% is generated by constructive EHM+HB interference.
In stark contrast, EHM alone produces none of the $B$-meson mass.  Instead, the HB is responsible for 79\%; yet, there is a sizeable EHM+HB interference term.
See Refs.\,\cite{Roberts:2021nhw, Ding:2022ows} for details.
 }
\end{figure*}

Detailed pictures of the proton and $B$-meson mass budgets are drawn in Fig.\,\ref{protonmass}.  There are striking contrasts between the breakdowns into EHM, EHM+HB, and HB contributions. Modern science must discover and explain the source of these remarkable differences.

Contemporary theory explains EHM as the consequence of the dynamical generation of a gluon mass scale in QCD \cite{Binosi:2022djx, Ferreira:2023fva}.  This is \emph{mass from nothing}: the SM massless gluon parton becomes a massive quasiparticle owing to self-interactions.  The existence of such a mass entails that the QCD running coupling has a stable infrared completion, remaining finite at all energy scales, from the deep ultraviolet into the far infrared \cite{Cui:2019dwv, Deur:2023dzc}.  Together, these phenomena explain the character of mass in the matter sector of strong interactions \cite{Roberts:2021nhw, Ding:2022ows}.  Such extraordinary predictions require empirical verification.

An open road toward validation is provided by the study of semileptonic weak-interaction transitions between heavy and light hadrons.  In fact, heavy pseudoscalar meson to light pseudoscalar meson transitions serve to probe the relative impacts of the strength of EHM+HB interference in the initial and final states, whereas heavy-pseudoscalar to light-vector meson transitions overlap systems in which HB mass is dominant with those whose mass owes almost entirely to EHM.  Both classes of transitions, therefore, and analogues involving baryons, present excellent opportunities for exposing the character of EHM and its interference with HB effects in order to identify the source of visible mass and its impact on physical observables.  These cases are of heightened interest, of course, because the transitions have long been used to place constraints on the values of the elements of the CKM matrix, which parameterizes quark flavour mixing in the SM.  Furthermore, confronting measurements of transitions with different leptons in the final state with sound theoretical predictions can shine bright light onto the question of LFU. Searches for violations of CKM matrix unitarity and/or LFU are principal tools in the hunt for physics beyond the SM.
Modern theory is capable of delivering robust predictions for all hadron structure factors necessary for the sound SM prediction of such transitions \cite{Yao:2021pdy, Yao:2021pyf}.

Studying the evolution of hadron properties with quark current mass, \emph{i.e}., the strength of HB couplings into QCD, provides a clear window onto constructive interference between Nature's two sources of mass. This is a new feature of flavour physics, which adds enormously to its role in searching for physics beyond the SM. The CEPC will deliver copious numbers of hadrons containing heavy quarks. Exploiting this capacity, the CEPC can play a role in exposing the origin and character of mass.

\subsection{Lepton form factors}

The $e^+e^-$ collisions at CEPC offer high luminosity in photon-mediated processes, providing a unique opportunity to measure photon-lepton interactions. Leptons' effective electromagnetic vertices have long been a popular topic with high-energy lepton collisions, as measurements of the lepton-photon coupling above pair-production threshold or at the weak scale can be interpreted as probes of BSM theory that can modify such form factors.   

\subsubsection{General remarks on $\mu$/$e$ $g-2$} 

Muon/electron $g-2$ measurements can serve as important probes for NP beyond the SM. It has been known for a long time that the SM prediction
of the muon anomalous magnetic moment $a_\mu\equiv (g -2 )_\mu/ 2$ has subtle deviations
from the experimental values. Combining the recent reported FNAL muon $g-2$ measurement with the previous
BNL+FNAL results, the updated world-averaged experimental value~\cite{Muong-2:2023cdq} of $a_\mu$
has a $5.0\sigma$ deviation from the SM prediction provided by the 2020 White Paper from the Muon $g-2$ Theory Initiative \cite{Aoyama:2020ynm}.  Besides, the SM prediction of the electron $g-2$
also has a $2.4\sigma/1.6\sigma$ deviation using ${^{133}Cs}/{^{87}Rb}$~\cite{Hanneke:2008tm,Morel:2020dww} atoms experimental data with negative/positive central value. 
However, this picture is complicated by recent lattice results for the hadronic vacuum polarisation (HVP) contributions to the anomalous magnetic moment. The only lattice result with a comparable uncertainty comes from the Budapest-Marseille-Wuppertal (BMW) collaboration \cite{Borsanyi:2020mff,Boccaletti:2024guq} and it disagrees significantly with the data-driven estimate. Using the lattice value instead would reduce the deviation to only $0.9\sigma$. 
This result is also supported by recent lattice comparisons to the BMW result~\cite{Parrino:2025afq,FermilabLattice:2024yho,Bazavov:2024eou,RBC:2018dos}. A comparison of different lattice results and the data-driven approach~\cite{Wittig:2023pcl} gives a $3.8 \sigma$ tension between the data-driven estimate and the lattice QCD estimates. As a result, the discrepancy between the data-driven estimates and the lattice QCD calculations still needs to be settled before claiming NP beyond the SM\footnote{A newly released white paper from the Muon g-2 Theory Initiative uses only the lattice results for HVP contributions \cite{Aliberti:2025beg} in their prediction for the SM.}. However, NP explanations of the deviation between the theory white paper prediction and the measured value also make predictions that can be tested at future colliders \cite{Athron:2021iuf}, such as the CEPC.

\begin{figure}[!tp]
 \begin{center} 
   \includegraphics[width=0.6\textwidth]{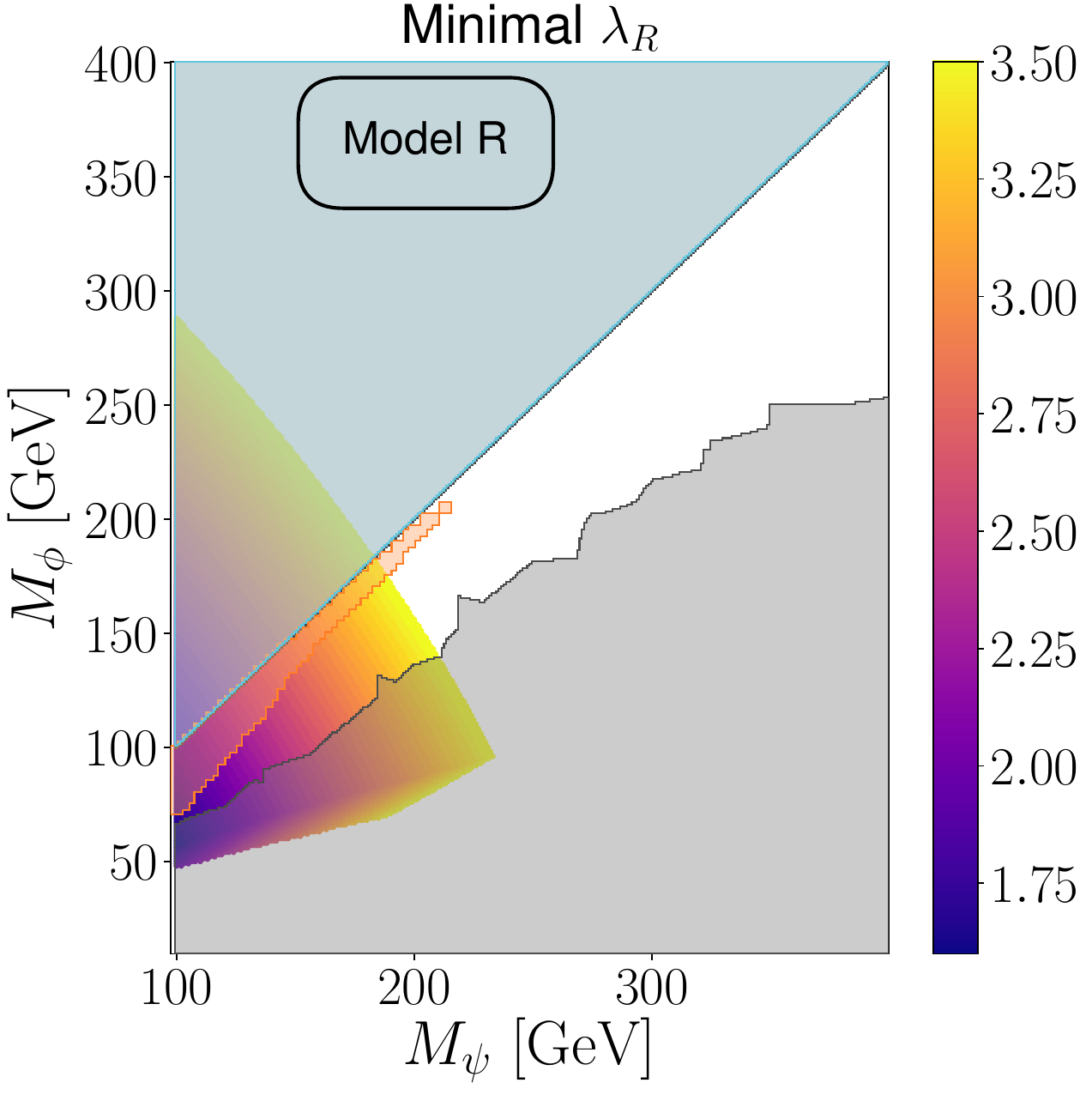}
   \end{center}
	\caption{A simplified BSM scenario with a scalar and a fermion. The color-coding shows the minimal coupling to right-handed muons to explain the muon $g-2$.  Outside of this colored region, the muon $g-2$ cannot be explained within $1\sigma$. The shaded grey region shows the LHC exclusions~\cite{GAMBIT:2018gjo} and the shaded orange region is excluded by compressed spectra. The shaded blue region (top left) is excluded due to a charged stable particle.}
	\label{fig:muong2}
\end{figure} 

The generic NP contributions to muon g-2 is expected to scale as
\begin{equation}
\Delta a_\mu^{\textrm{BSM}} \approx  C_{\textrm{BSM}} \frac{m_\mu^2}{M^2_{\textrm{BSM}}}, \label{Eq:amuBSM} 
\end{equation} 
where $M_\text{BSM}$ is the mass of the NP particles in the loop and  $C_{\textrm{BSM}}$ is a loop-suppressed coefficient. This means that without some special enhancement, the mass scale of the NP should be $\lesssim 200-300$ GeV for perturbative NP explanations. This is illustrated for a simple model with a new scalar and a new fermion in Fig.~\ref{fig:muong2}, where the colour contours show the minimum value of the coupling to right-handed muons required to explain the muon $g-2$, and anything outside that region cannot explain the muon $g-2$ within $1\sigma$.

LHC has already excluded many such scenarios. However, there can be gaps in the exclusion from compressed spectra (as shown in Fig.\ \ref{fig:muong2}) and it has been shown, for example, that constraints on light new electroweakinos are not very robust~\cite{GAMBIT:2018gjo}. Therefore, 
it still opens up the possibility of directly producing the states at future colliders. On the other hand, an elegant solution to the tension between the LHC and the muon $g-2$ comes in the form of a chirality flipping enhancement; see Ref.~\cite{Athron:2021iuf} for a recent review.  In Eq. \ref{Eq:amuBSM}, one factor of $m_\mu$ appears because the dipole operator flips the fermion chirality, and thus a muon mass insertion exists on one of the external legs.  If this chirality flip can instead be done inside the new loop diagram from NP, the muon mass will then be replaced with a mass parameter from the NP that can be much larger.  This chirality flipping enhancement is automatically present in SUSY extensions as well as in non-SUSY models like scalar and vector leptoquarks, which alleviates this tension and makes it easier to construct simultaneous explanations of dark matter and muon $g-2$.  
NP contribution to muon $g-2$ typically also implies large corrections to the self-energy of the muon. This leads to a fine-tuning in the muon mass if the new particle masses are much heavier than ${\mathcal O}(1\,\mathrm{TeV})$ \cite{Athron:2021iuf}. Furthermore, this also implies an enhanced Yukawa coupling, which means that precision measurements of $h\rightarrow \mu^+\mu^-$ at future colliders such as the CEPC could either exclude these explanations or give rise to a discovery-level deviation from the SM~\cite{Crivellin:2020tsz}. 

\subsubsection{$\mu/e$ dipole moments in SUSY} 

As a popular BSM scenario, SUSY contributes to muon $g-2$ mostly via the chargino-sneutrino and the neutralino-smuon loops, which always need light electroweakinos and sleptons to explain the anomaly. However, such requirements potentially have tensions with the observed 125 GeV Higgs mass and LHC exclusion bounds, which in general prefer heavy colored sparticles. 
Weak-scale phenomenological MSSM needs intricate parameter regions to survive the current LHC, dark matter and Higgs mass bounds, and can give a sizable contribution to muon/electron $g-2$ at the same time. 

Studies of the gluino-SUGRA, the anomaly-mediated SUSY breaking (AMSB) models~\cite{Wang:2017vxj,Ning:2017dng,Du:2017str} and the gauge/Yukawa mediated SUSY breaking models~\cite{Du:2022pbp} indicate that it is challenging yet possible to explain both $e$ and $\mu$ anomalies in a unified SUSY framework. This is because without any flavor violation in the lepton sector, NP contributions to the lepton $g-2$ are in general scaled with the corresponding lepton mass-square. Such a scaling relation can still explain the electron $g-2$ anomaly in a $2\sigma$ range for a positive central value of the electron $g-2$ experimental data when the muon $g-2$ anomaly is explained at $1\sigma$.
  
Generalized gravity mediation models always adopt various universal boundary conditions at the GUT scale. 
Given the stringent constraints on the first two generation squarks by LHC and the stop masses by the 125 GeV Higgs, the mSUGRA slepton masses cannot be light at the EW scale with universal sfermion mass inputs at the GUT scale. Thus, it is challenging to explain the muon $g-2$ anomaly within the framework of GUT-scale constrained SUSY, especially the mSUGRA~\cite{Wang:2021bcx}. 

Gluino-SUGRA ($\tilde{g}$SUGRA)~\cite{Akula:2013ioa} is an economical extension of mSUGRA, and it is a special case of non-universal gaugino mass realization at GUT scale. 
In $\tilde{g}$SUGRA the gluino mass can be much heavier than other gauginos and sfermions at the unification scale, hence the gaugino mass ratios at the EW scale will no longer constrain the electroweakino masses for a heavy gluino mass. 
The sleptons, which carry no color charge, will stay light. So, the RGE evolution will split the squark masses from slepton masses at the electroweak scale, 
which is needed for the muon $g-2$.  Ref.~\cite{Li:2021pnt} showed that with $M_1=M_2$ at the GUT scale and a viable
bino-like dark matter, $\tilde{g}$SUGRA can explain the muon $g-2$ anomaly at $1\sigma$ level and
be consistent with the updated LHC constraints. 
In this case, a light stau is needed for co-annihilation, 
and it is possible that the mild mass splittings between the first two generations of sleptons and $\tilde{\chi}_1^0$
will lead to energetic lepton final states so that they can be tested at CEPC. It is also possible to connect the $g-2$ explanation to neutrino masses~\cite{Li:2022zap}.

\subsubsection{$\tau$ weak-electric dipole moments} 

Electric dipole moments (EDM) and weak electric dipole moments (WDM) of fundamental fermions are important targets of experimental searches for NP, in particular for $\mathrm{CP}$-violation beyond the Kobayashi-Maskawa mechanism.
Any experimental observation of a nonzero value of an EDM ($d$) and/or WDM ($d^{w}$) for a lepton would be a smoking-gun evidence of non-SM sources of $\mathrm{CP}$-violation, because the (loop-induced) SM contributions to these quantities for leptons are extremely small~\cite{Bernreuther:1988jr,Booth:1993af,Yamaguchi:2020dsy}.
While $d_{\tau}$ can be probed via $e^+ e^- \rightarrow \tau^+ \tau^-$ at energies much lower than the $Z$-boson mass, such as at $\Upsilon(4{\rm S})$ in Belle~\cite{Belle:2002nla,Belle-II:2022cgf} and at $\psi(2{\rm S})$ in low-energy $e^+e^-$ colliders~\cite{He:2025ewk}, the optimal measurement of $d_{\tau}^{w}$ should be the $\tau$-pair production at the $Z$ resonance.

To date, the best results are from LEP by measuring the transverse and normal $\tau$-lepton polarizations~\cite{Bernabeu:1993er,Stiegler:1992sy},
which gave the limits on the real and imaginary parts~\cite{Stahl:2000aq,ALEPH:2002kbp,Lohmann:2005im}:
\begin{eqnarray}
\mathrm{Re}[d_{\tau}^{w}] &=& (-0.65 \pm 1.49) \times 10^{-18} \, e\,\mathrm{cm}, \nonumber \\
\mathrm{Im}[d_{\tau}^{w}] &=& (0.04 \pm 0.38) \times 10^{-17}\, e\,\mathrm{cm}.
\end{eqnarray}
The large amount of $\tau^+ \tau^-$ pairs during the planned Tera-$Z$ mode of CEPC, along with an improved $\tau$-reconstruction efficiency, will be able to test $d_{\tau}^{w}$ to a precision significantly higher than existing bounds.

Ref.~\cite{Bernreuther:2021elu} performed an exploratory study of the potential of CEPC for the measurements 
of $d_{\tau}^{w}$, using both simple and optimal $\mathrm{CP}$-violation observables~\cite{Atwood:1991ka,Davier:1992nw,Diehl:1993br}, albeit with a particular emphasis on the latter due to its clear advantage over the former. 
In this work, $e^+ e^- \rightarrow \tau^+ \tau^-$ is considered exactly at the $Z$ resonance with the leading-order SM couplings. 
$d_{\tau}^{w}$ is included through an effective $Z \tau\tau$ vertex followed by $\tau$ decays, taking into account all spin-correlation effects.
Assuming $1.38\times 10^{11}$ $\tau$-pairs collected at the $Z$ resonance, we obtain the 1~s.d.~statistical uncertainties in $\delta \mathrm{Re} [d^{w}_{\tau}]$ and $\delta \mathrm{Im} [d^{w}_{\tau}]$, using both the simple observables $T_{33},\,\hat{T}_{33},\, Q_{33},\, \hat{Q}_{33}$ where only one-prong decays of $\tau^{\pm}$ are included, and the optimal observables $O_{R},\, O_{I}$ that use the purely semi-hadronic decays of $\tau^{\pm}$. The analysis results are given in Table~\ref{sens_cepc} (see Ref.~\cite{Bernreuther:2021elu} for more details).%

\begin{table}[h!]
\begin{center}
\begin{tabular}{|c c c| c c c|}
\hline 
\multicolumn{3}{|c|}{$\delta \mathrm{Re} [d^{w}_{\tau}] \,[e\,\mathrm{cm}]$}   & \multicolumn{3}{c|}{$\delta \mathrm{Im} [d^{w}_{\tau}] \,  [e\,\mathrm{cm}]$}  \\ \hline
$\langle T_{33} \rangle$  &$\langle \hat{T}_{33} \rangle$    &$\langle O_{R} \rangle$	   &$\langle Q_{33} \rangle$ 	&$\langle \hat{Q}_{33}\rangle$	&$\langle O_{I} \rangle$	\\\hline
$3.4 \times10^{-21}$					&$3.4 \times10^{-21}$ 			& $1.4 \times10^{-21}$             &$3.2 \times10^{-19}$		       &$4.0 \times10^{-20}$        	&$2.1 \times10^{-21}$		\\\hline
\end{tabular}
\caption{Ideal 1~s.d.~statistical errors on $\mathrm{Re} [d^{w}_{\tau}]$ and $\mathrm{Im} [d^{w}_{\tau}]$~\cite{Bernreuther:2021elu}.}\label{sens_cepc}
\end{center}
\end{table}

These numbers show that the CEPC sensitivity on $d_\tau^{w}$ can reach the level of $10^{-21}\, e\, \mathrm{cm}$ using $O_R$ and $O_I$, far better than the current best bounds~\cite{Stahl:2000aq,Lohmann:2005im} quoted above. 
Note $e~$cm$ =5\times 10^{13}~e~$GeV$^{-1}$ in natural units, a sensitivity to $10^{-21}\, e~$cm is equivalent to an effective dipole moment of $d\sim 6\times 10^{-6}$ GeV$^{-1}$, and this indicate for a two orders of magnitude improvement over existing LEP limits.
In perspective, a more refined analysis will take SM radiative corrections into account, such as $Z\gamma$-interference etc., on top of the exploratory study above.

\subsection{Spin entanglement}
\label{subsect:spin_entanglement}
Recently, studies on quantum entanglement in the high-energy regime have gained significant attention. At colliders, reconstruction of the final-state particle helicity state offers an observation window on spin entanglement at energies much higher than those at optics laboratories. As a benchmark of spin entanglement, the test of the Bell inequality is of primary interest. It delivers a direct justification if Quantum Mechanics (QM) is a complete local theory and shows the contradiction of the local hidden variable theory (LHVT) with QM~\cite{Einstein:1935rr,Bohm:1957zz,Bell:1964kc}.

As the Higgs boson is the only spin-0 elementary particle in the SM, it offers a natural spin-singlet state to test the LHVT.
The decay of the Higgs boson into two spin-1/2 particles provides an ideal system
to reveal quantum entanglement and Bell inequality violation at high energies.
In Ref.~\cite{Ma:2023yvd}, it is proposed to test the Bell inequality through the Higgsstrahlung process $e^+e^-\to Zh$ at the CEPC.
Two realistic methods of testing Bell inequality, i.e., T\"{o}rnqvist's method~\cite{Tornqvist:1980af} and the Clauser, Horne, Shimony and Holt (CHSH) inequality~\cite{Clauser:1969ny}, are studied in terms of the polarization correlation in decay chain $h\to \tau^+\tau^-\to \pi^+\bar{\nu}_\tau \pi^- \nu_\tau$. We use the method of impact parameters for the reconstruction of $\tau$ lepton in our detector-level simulation and consider both the hadronic and leptonic decay modes of $Z$ boson.

\begin{figure}[t]
\centering
\includegraphics[width=0.7\textwidth]{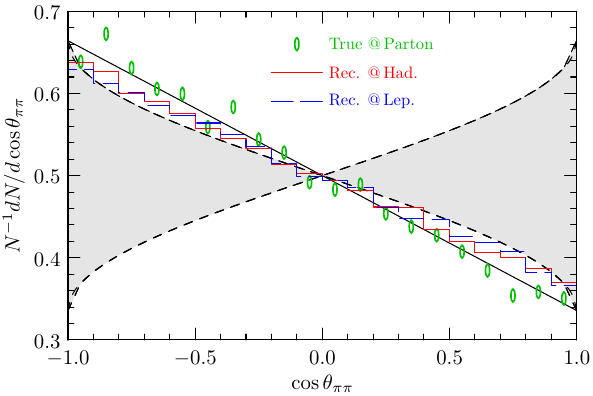}
\caption{Reconstructed distributions of $\cos\theta_{\pi\pi}$ for T\"{o}rnqvist's test of Bell inequality with $h\rightarrow \tau^+\tau^-$ at the CEPC Higgs factory mode. $1\sigma$ test sensitivity is projected at the CEPC, represented by the red and blue histograms for hadronic and leptonic decay modes of $Z$ boson, respectively. The gray-fitted region
is the phase space consistent with the classical prediction. 
}
\label{fig:Tornqvist}
\end{figure}

The experimental sensitivity of CEPC for the T\"{o}rnqvist's approach is studied by defining the following asymmetric observable:
\begin{equation}
\mathcal{A} = \frac{N(\cos\theta_{\pi\pi} < 0 ) - N(\cos\theta_{\pi\pi} > 0 ) }{ N(\cos\theta_{\pi\pi} < 0 ) + N(\cos\theta_{\pi\pi} > 0 ) }\,,
\end{equation}
where $\cos\theta_{\pi\pi}=\vec{p}_{\pi^-}\cdot\vec{p}_{\pi^+}/(|\vec{p}_{\pi^-}||\vec{p}_{\pi^+}|)$ in the Higgs rest frame. Fig.~\ref{fig:Tornqvist} shows the distribution of $\cos\theta_{\pi\pi}$ and the LHVT holds between the two dashed lines. The analytical prediction of the observable gives an upper bound $\mathcal{A}=0.119$ in the LHVT. From the simulation results of SM expectation, we obtain $\mathcal{A}=0.133\pm 0.269$ for $Z\to \ell\ell$ channel and
$\mathcal{A}=0.137\pm 0.1$ for $Z\to jj$ channel, respectively,
as listed in Table~\ref{tab:asybell}. Smaller uncertainties can be obtained with $\mathcal{A}=0.133\pm 0.142$ or
$\mathcal{A}=0.137\pm 0.053$ for an updated luminosity $\mathcal{L}=20~{\rm ab}^{-1}$.
In the CHSH approach, the LHVT supports the fact that the sum of the two largest eigenvalues (denoted by $m_1+m_2$) of the matrix $U=C^TC$ with $C$ being the spin correlation matrix is not larger than one. It turns out that both channels lead to $m_1+m_2>1$, as listed in Table~\ref{tab:asybell}.
For both the T\"{o}rnqvist's and CHSH approaches, the Bell inequality can be tested below $1\sigma$ level at the CEPC.
It is expected that the sensitivity can be further improved by using sophisticated jet reconstruction methods and an enhanced $\tau$-jet identification efficiency.

\begin{table}[h]
\renewcommand\arraystretch{1.22}
\begin{center}
\begin{tabular}{ccccc}
\hline
\makebox[0.15\textwidth]{Channels}  & \makebox[0.08\textwidth]{Observable} & 
\makebox[0.1\textwidth]{LHVT} &
\makebox[0.2\textwidth]{CEPC @\,5.6$~{\rm ab}^{-1}$} &
\makebox[0.2\textwidth]{CEPC @\,20$~{\rm ab}^{-1}$}
\\ \hline
\multirow{2}{*}{ $Z\to\ell\ell$} & $\mathcal{A}$
& $\le 0.119$ & $0.133 \pm 0.269$ & $0.133 \pm 0.142$
\\ 
 & $m_1+m_2$
 & $\le1$ & $1.04 \pm 0.921$ &
$1.04 \pm 0.481$
\\ \hline
\multirow{2}{*}{ $Z\to jj$} & $\mathcal{A}$
& $\le 0.119$ & $0.137 \pm 0.1$ & $0.137 \pm 0.053$
\\ 
 & $m_1+m_2$
& $\le1$ & $1.05 \pm 0.355$ & $1.05 \pm 0.188$
 \\
\hline
\end{tabular}
\end{center}
\caption{The results of observables testing the Bell inequality in T\"{o}rnqvist's method and the CHSH approach. The experimental predictions
are given for the CEPC with colliding energy $\sqrt{s}=240~{\rm GeV}$ and total luminosities $5.6~{\rm ab}^{-1}$ and $20~{\rm ab}^{-1}$.
}
\label{tab:asybell}
\end{table}

Particle decay involving more than two spins are also common at colliders. Recent studies on spin entanglement have brought forth entanglement construction in multi-body decays of the SM bosons. For instance,
Ref.~\cite{Morales:2024jhj} has studied quantum entanglement properties of the rare Higgs boson three-body decays into $\gamma$ and dileptons within the SM ($h\rightarrow l^+l^-\gamma$), with electroweak 1-loop corrections included. Novel observables for these three-body decays are presented for the analysis of three lepton families, with each family analyzed separately since they lead to different experimental channels and the 1-loop contribution dominates different energy regimes in each case. It offers a unique opportunity to examine quantum correlations between the spin degrees of freedom arising at next-to-leading-order in perturbation theory for 3-qubit systems.

\begin{figure}[t]
    \centering
\begin{tabular}{ccc}
\includegraphics[width=0.42\textwidth]{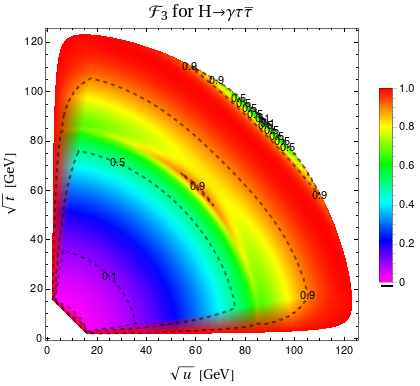}
\includegraphics[width=0.42\textwidth]{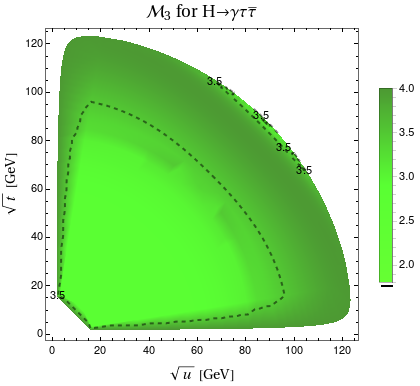} 
\end{tabular}
\caption{Dalitz plot representation for genuine multipartite entanglement (left) and Bell non-locality (right) quantifiers for the SM Higgs boson's $h\rightarrow \tau^+\tau^-\gamma$ decay. The Mandelstam variables $t$ and $u$ correspond to the subsystems photon-lepton and photon-anitlepton.}
\label{fig:moneyplot_morales}
\end{figure}

Based on the concurrence and Bell operator definitions for tripartite systems, the study was to identify regions of the phase-space where the final particles are entangled after the Higgs boson decay and to determine the feasibility of testing non-locality under these kinematical configurations.
For the three-body final state, various entanglement measures were computed, including one-to-other and one-to-one concurrences, the conditions for genuine entanglement of 3-qubit systems using the concurrence vector, the area of the concurrence triangle ($\mF_3$) and the three-tangle measure.
Regarding the Bell non-locality, both Mermin ($\mM_3$) and Svetlichny operators for 3-qubit systems were computed. Moreover, post-decay entanglement and auto-distillation phenomena for a dilepton invariant mass close to the $Z$-pole mass were analyzed.

The final state photon, lepton, and antilepton result in entanglement since $\mF_3$ is non-vanishing, as can be seen in the left panel of Fig.~\ref{fig:moneyplot_morales}. This also holds by considering the one-to-one and one-to-other concurrences of the subsystems among them. 
The amount of entanglement depends on the final-state kinematical configuration and maximally entangled subsystems appear in certain regions of the phase space (red regions where $\mF_3\sim1$).
Concerning the Bell non-locality (right panel), predictions incompatible with local realism ($2\leq\mM_3\leq4$) were obtained in the whole phase space, except for a few particular configurations, suggesting that $H\to l^+l^-\gamma$ could serve as a potential laboratory test of the Bell inequality.
On the other hand, CP-violating interactions in the Yukawa sector are suppressed by lepton masses, thus less powerful for such kind of NP. Furthermore, a natural multipartite extension is to consider the four-fermion Higgs decays, constituting a 4-qubit system.

\subsection{Exotic lepton mass models } 

Even more exotic models involving the lepton sector can benefit from the CEPC. For example, to a certain level of precision the SM's charged lepton masses seem to satisfy the curious formula:
\begin{equation}
K = \frac{m_e + m_\mu + m_\tau}{\left ( \sqrt{m_e} + \sqrt{m_\mu} + \sqrt{m_\tau} \right)^2}
  = \frac{2}{3},
\end{equation}
proposed by Y. Koide in early 1980's~\cite{Koide:1982si, Koide:1982ax, Koide:1983qe}, and it exhibits consistency with experimental data. The character $K$ calculated from the PDG 2022 data of charged lepton masses~\cite{ParticleDataGroup:2022pth} is $K = (2/3) \times (0.999991 \pm 0.000011)$, within $10^{-5}$ precision and within one sigma error. On the other hand, taking $K = 2/3$ as an input and using the measured electron and muon masses with high precision, the tau mass is predicted to be $m_\tau = 1776.969027 \pm 0.000036 \ \text{MeV} / c^2$.  It agrees with the PDG 2022 data $m_\tau = 1776.86 \pm 0.12 \ \text{MeV} / c^2$ and the Belle II 2023 result~\cite{Belle-II:2023izd} $m_\tau = 1777.09 \pm 0.08 \pm 0.11 \ \text{MeV} / c^2$ within one sigma error. 


\begin{figure}[t]
  \centering
  \includegraphics[width=0.7\linewidth]{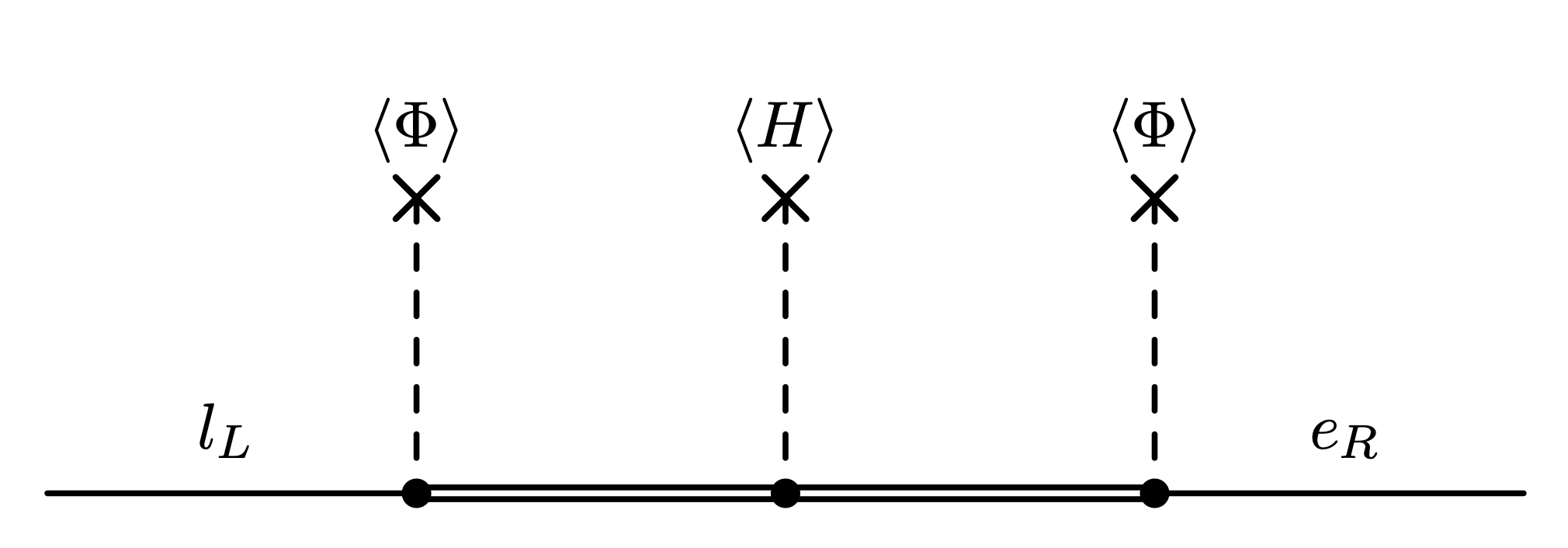}
  \caption{A Feynman diagram which generates the charged lepton mass matrix.  Plain lines represent the Standard Model charged leptons.  Double lines represent heavy fermions.  Dashed lines represent Higgs tadpoles as labeled.}\label{Koide:fig1}
\end{figure}

Proposals to explain such flavor mass pattern include the Froggatt-Nielsen model~\cite{Froggatt:1978nt}, the seesaw-type model~\cite{Koide:1989jq} and the supersymmetric Yukawaon model~\cite{Koide:2008tr}.  The key idea is to express the charged lepton mass matrix as $M \propto \langle \Phi \Phi \rangle$, where $\Phi$ is a Hermitian nonet scalar field in the $\mathbf{3} \otimes \mathbf{3}^* = \mathbf{8} \oplus \mathbf{1}$ representation of the $\mathrm{SU(3)}$ flavor symmetry.  In the Froggatt-Nielsen model or the seesaw-type model, the charged leptons couple to new fermions with a heavy mass $m_F$ through Yukawa couplings involving $\Phi$.  Diagrams similar to Fig.~\ref{Koide:fig1} generates a seesaw type mass matrix $M \propto \langle H \Phi \Phi \rangle / m_F^2$.  In the Yukawaon model, the SM Yukawa coupling terms for leptons are replaced by the dimension-five operators:
\begin{equation}
\mathcal{L}^{(5)} = - \frac{y_0}{\Lambda} \bar l_L Y H e_R + \text{h.c.},
\end{equation}
where $Y$ is a flavor nonet scalar field called the Yukawaon.  In a supersymmetric model with the superpotential
\begin{equation}
W = \mu \operatorname{Tr} \left ( Y A \right ) + \lambda \operatorname{Tr} \left ( \Phi \Phi A \right ),
\end{equation}
where $A$ is another flavor nonet scalar field, the F-term equations for $A$ gives $\langle Y \rangle \propto \langle \Phi \Phi \rangle$ and thus $M \propto \langle \Phi \Phi \rangle$ after electroweak symmetry breaking. Any of these models gives $K = \operatorname{Tr} \langle \Phi \Phi \rangle / (\operatorname{Tr} \langle \Phi \rangle)^2$.  Then a superpotential for $\Phi$ is built through symmetry considerations~\cite{Koide:2018gdm, Liang:2020oni}, and the F-term equations may set the vacuum expectation value of $\Phi$, which leads to $K = 2 / 3$. 
While precision lepton mass measurement is not a main target at the CEPC, the underlying physical models can be tested in leptonic channels of Higgs decay measurements at future colliders as well as from extra scalar/Higgs searches.

\begin{figure}[t]
    \centering
    \includegraphics[width=0.85\linewidth]{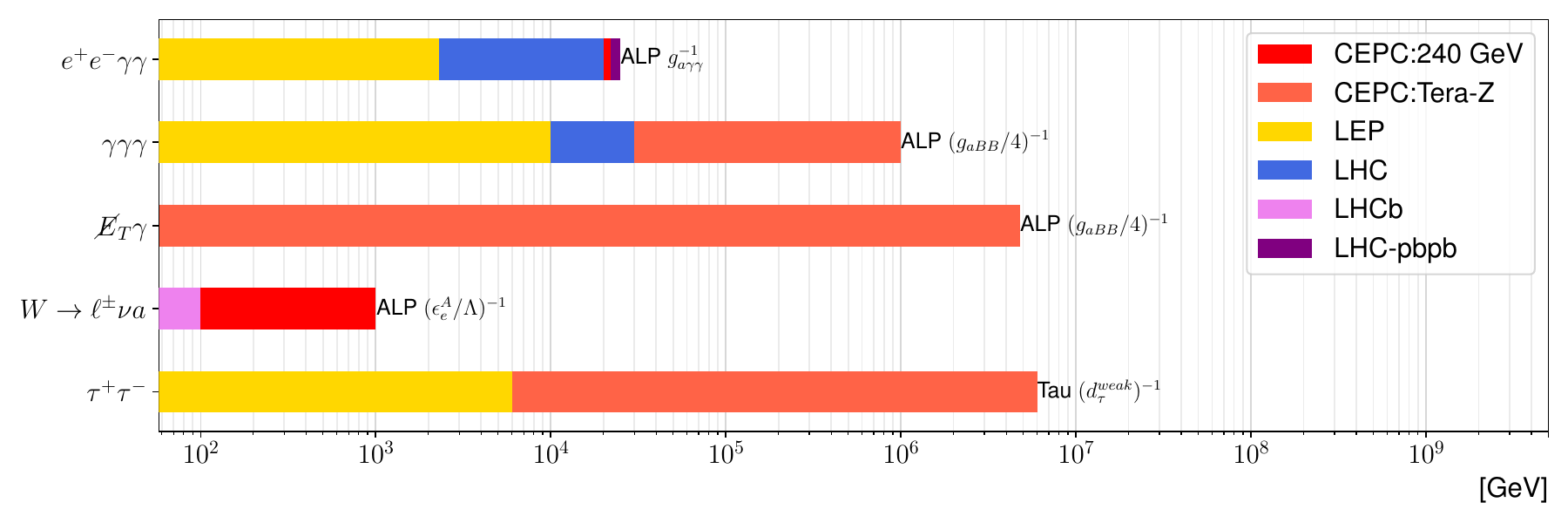}
    \caption{Energy reach in representative exotic search channels at the CEPC. Note the maximal reach may apply to different parameter regions between experiments.}
    \label{fig:exotic_summary_chart}
\end{figure}

\subsection{Summary}

For a brief summary, this section includes a number of CEPC-related exotic physics studies in recent literature. Qualitative results on the projected energy scale sensitivities from dedicated investigation focus on well-motivated topics such as the ALP, $\tau$ lepton form factors, Bell inequality tests, etc. 
The summary bar-chart Fig.~\ref{fig:exotic_summary_chart} illustrates the major sensitivity channels, and the quantitative results from these studies are listed in Table~\ref{tab:exotic_summary}.

\begin{table}[h]
    \centering
    \begin{tabular}{|c|c|c|c|}
    \hline
    \hline
       ~Quantity~  & ~Channel ~& ~Sensitivity scale (GeV)~& ~CEPC Run~\\
    \hline
        ALP $g_{a\gamma\gamma}^{-1}$ & $e^+e^-\gamma\gamma$ & $6.7\times 10^3$ ~~\cite{Zhang:2021sio}& Tera-$Z$ \\
        & $e^+e^-\gamma\gamma$ & $2.2\times 10^4$ ~~\cite{Zhang:2021sio}& 240 GeV \\
        & $\bar{f}fa$ & $6.5\times 10^3$ ~~\cite{Zhang:2021sio}& 250 GeV \\
        ALP $(g_{aBB}/4)^{-1}$ & $3\gamma$ & $10^6$~~\cite{Liu:2017zdh} & Tera-$Z$ \\
        & $\slashed{E}_T\gamma$ & $4.8\times 10^6$ ~~\cite{Liu:2017zdh} & Tera-$Z$ \\
      ALP $(\epsilon^A_e/\Lambda)^{-1}$  & $W\rightarrow \ell^{\pm}\nu a$ & $10^3$ ~~\cite{Altmannshofer:2022ckw}& 240 GeV \\
        Tau $(d_\tau^{weak})^{-1}$ & $\tau^+\tau^-$ & $6\times 10^6$ ~~\cite{Bernreuther:2021elu} & Tera-$Z$ \\
    \hline
    ~Bell Inequality~ & $Z,h\rightarrow \tau^+\tau^-$ & 1$\sigma$ ~~\cite{Ma:2023yvd} & 240 GeV \\
    \hline
    \hline
    \end{tabular}
    \caption{Projected energy scale sensitivities via exotic searches at the CEPC.}
    \label{tab:exotic_summary}
\end{table}

\clearpage
\section{Global Fits}
\label{sec:global git}

Global fits are an essential tool when it comes to obtaining a thorough understanding of a new physics model. They offer a comprehensive analysis by considering a wide range of experimental data. With global fits, we can extract the maximum amount of information possible from these datasets. 
The primary advantage of global fits is their ability to evaluate and compare the validity of different models. By exploring a variety of model parameters, they can identify the range of values that are most likely or have the highest posterior probability. This, in turn, helps us comprehend the implications and predictions of the models for future searches and experiments. 

In this section, we will discuss the latest research findings regarding the impact of the CEPC on the global fit analysis for SMEFT, 2HDMs, and various SUSY models.


\subsection{SMEFT global fits}

\textit{The SMEFT framework}---As has been previously addressed, the null result at the LHC since the Higgs discovery indicates that the SM is possibly a low-energy effective theory of some UV completed theories at a scale $\Lambda\gg \Lambda_{\rm EW}$, with $\Lambda_{\rm EW}=246\rm\,GeV$ the electroweak scale. This large energy gap then naturally renders the EFT framework ideal for a model-independent study on new physics based on the decoupling theorem\,\cite{Appelquist:1974tg}. In this section, we focus on the SMEFT framework based on the same ${\rm SU(3)}_c\otimes{\rm SU(2)}_L\otimes{\rm U(1)}_Y$ \textit{local} gauge symmetry as respected by the SM while relaxing the accidental \textit{global} symmetries in the SM. Generically, the SMEFT can be obtained by extending the SM Lagrangian as
\eqal{
\mathcal{L} = \mathcal{L}_{\rm SM} + \sum_{n=5}^\infty \sum_i \frac{\delta c_i}{\Lambda^{n-4}} \mathcal{O}_i^{(n)},
}
where $\mathcal{O}^{(n)}$ represents operators with mass dimension $n$, the index $i$ corresponds to the sum over the operator basis at dimension $n$, and $\delta c_i$ is the associated Wilson coefficient. Clearly, contributions from the SMEFT operators will be generically suppressed by powers of $p^2/\Lambda^2$, with $p^2$ the momentum transfer and satisfying $p^2\ll \Lambda^2$. Therefore, the dominant contribution will be coming from the dimension-5 operators, which are also known as the Weinberg operators\,\cite{Weinberg:1979sa}. These operators can induce non-vanishing neutrino masses after the electroweak spontaneous symmetry breaking (EWSSB), but will be otherwise irrelevant for the discussion in this section. With that, we focus on the dimension-6 operators, where a basis, known as the \textit{Warsaw basis}, was provided in\cite{Buchmuller:1985jz,Grzadkowski:2010es}. For phenomenological studies, a relatively convenient basis, known as the \textit{Higgs basis}, was proposed in\,\cite{Falkowski:2001958} in the broken phase, which is related to the \textit{Warsaw basis} by a linear transformation. The advantage of the \textit{Higgs basis} is that different physics becomes disentangled, thus reducing the number of operators in specific scenarios such as the Higgs, the electroweak, and the four-fermion induced physics as we will discuss in the following.

\textit{Methodology}---For this global analysis, we will only keep EFT corrections to the SM predictions at $\mathcal{O}(1/\Lambda^2)$, \textit{i.e.}, the interference between the SM and the SMEFT since pure EFT results will be further suppressed by $p^2/\Lambda^2$. Then under this approximation, one can generically parameterize
\eqal{
O_i &\, = O_i^{\rm SM} + \delta \vec{c}_i \cdot \vec O_{i,\rm SMEFT},\label{eq:eftlo}
}
for any observable $O_i$, with $\vec O_{i,\rm SMEFT}$ the collection of operator contributions and $O_i^{\rm SM}$ its SM prediction. The global fit is then carried out based on the $\chi^2$ constructed as
\eqal{
\chi^2 = \sum_{i,j} \left[ O_i - O_i^{\rm exp} \right] \sigma_{ij}^{-2} \left[ O_j - O_j^{\rm exp} \right],
}
where $O_i^{\rm exp}$ is the experimental measurement of $O_i$, and $\sigma_{ij}^{-2} = \left[\delta O_i \rho_{ij} \delta O_j\right]^{-1}$ with $\delta O_i$ the experimental uncertainties and $\rho_{ij}$ the experimental correlation between $O_{i,j}$. Note that under the linear SMEFT correction approximation as adopted here, $\chi^2$ is a quadratic function in terms of the $\delta c_i$'s, thus the global minimum can be analytically obtained from ${\partial\chi^2}/{\partial \vec\delta c_i}=0$. On the other hand, this leading order approximation in eq.\,\eqref{eq:eftlo} also enables the optimal observable approach\,\cite{Diehl:1993br}, which utilizes the full information of the process from the differential distributions and can also provide the best statistical reach. For this reason, optimal observables are used for this global analysis when applicable, $e^+e^-\to W^+W^-$ for instance. It is also worthy of mentioning, from a recent machine learning study in\,\cite{Chai:2024zyl}, that while this optimal observable approach can lead to satisfying results at the parton level, some bias in data interpretation will be introduced at the the detector level.

\begin{figure}[th!]
\centering
\includegraphics[width=0.9\textwidth]{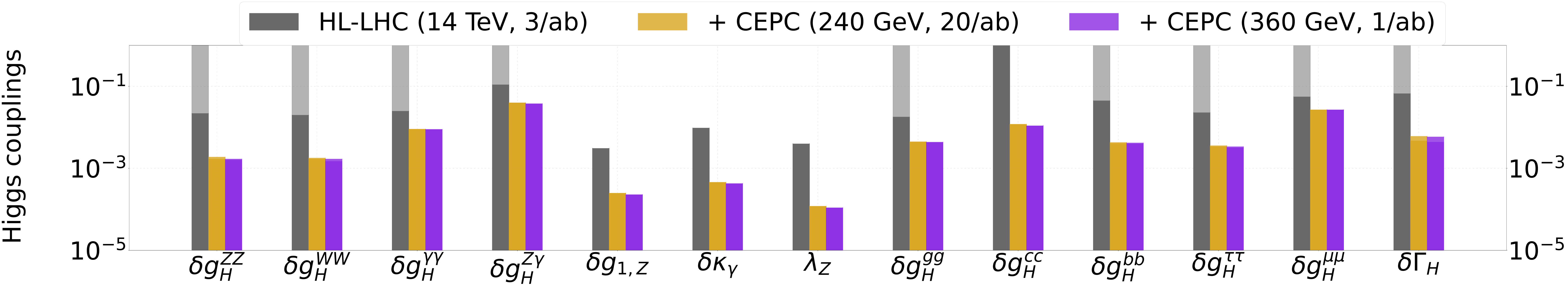}\caption{$1\sigma$ relative uncertainties on the SMEFT operators as shown along the horizontal axis from a global analysis of the Higgs couplings. The light bars are obtained by taking the total Higgs decay width $\Gamma_H$ as a free parameter, and the dark bars those with a constrained $\Gamma_H$.}\label{fig:smeft-higgs}
\end{figure}

\begin{figure}[th!]
 \centering
 \includegraphics[width=0.9\textwidth]{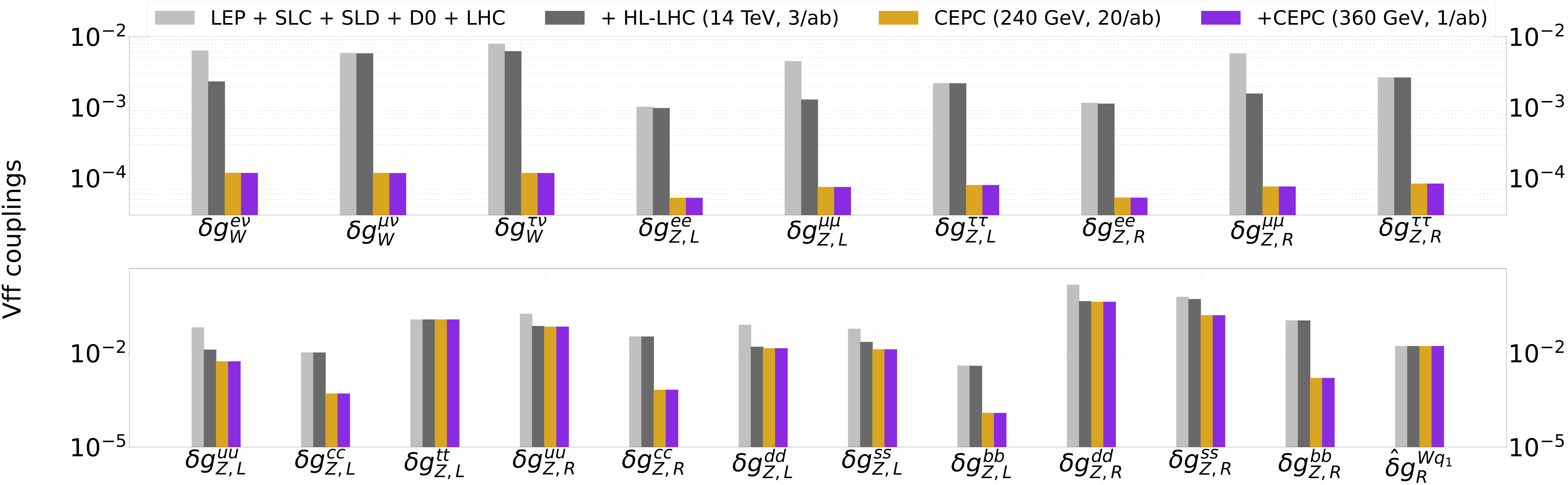}\\
 \includegraphics[width=0.9\textwidth]{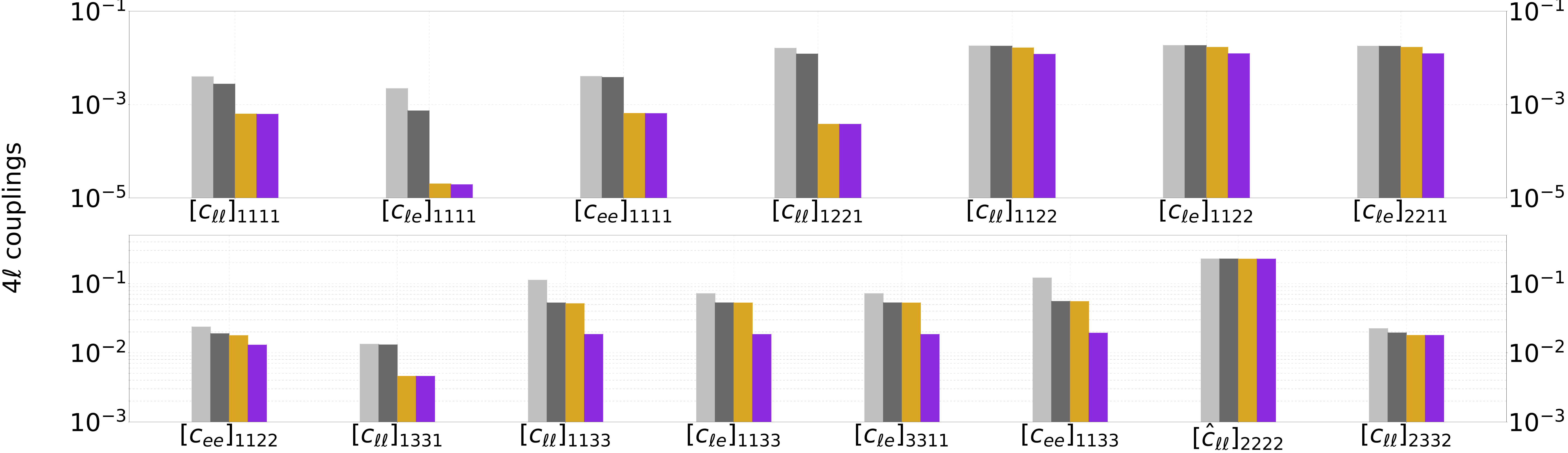}\\
 \includegraphics[width=0.9\textwidth]{figures/SMEFT-4f-2l2q.pdf}\\
 \caption{Same as Fig.~\ref{fig:smeft-higgs} but for the electroweak and 4-fermion operators.}
 \label{fig:smeft-4f}
\end{figure}

\begin{figure}[th!]
 \centering
 \includegraphics[width=0.9\textwidth]{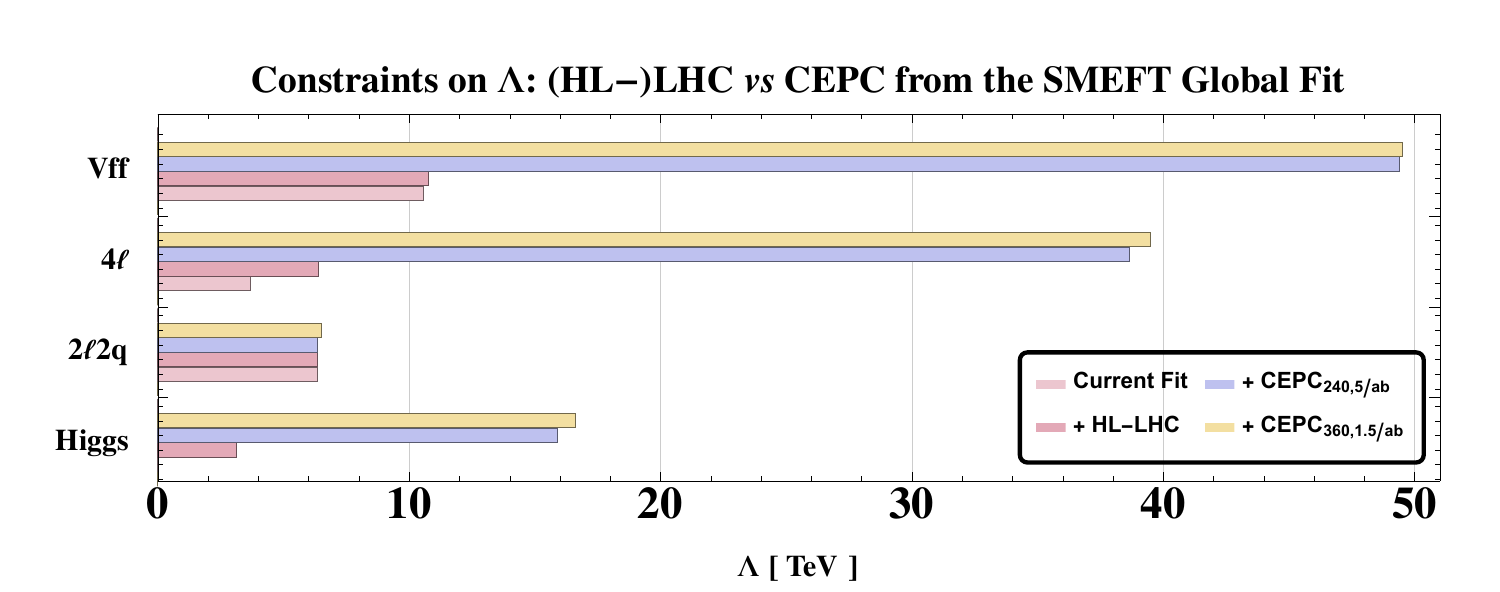}\\
 \caption{Sensitivity to the cutoff scale $\Lambda$ from the SMEFT global fit at the 95\% CL. The $y$ axis shows the optimal reach from electroweak physics (``Vff''), pure leptonic four-fermion operator induced physics (``$4\ell$''), semileptonic four-fermion operator induced physics (``$2\ell2q$''), and Higgs physics (``Higgs'').}
 \label{fig:smeft-sum}
\end{figure}

\begin{figure}[thb!]
 \centering
 \includegraphics[width=0.9\textwidth]{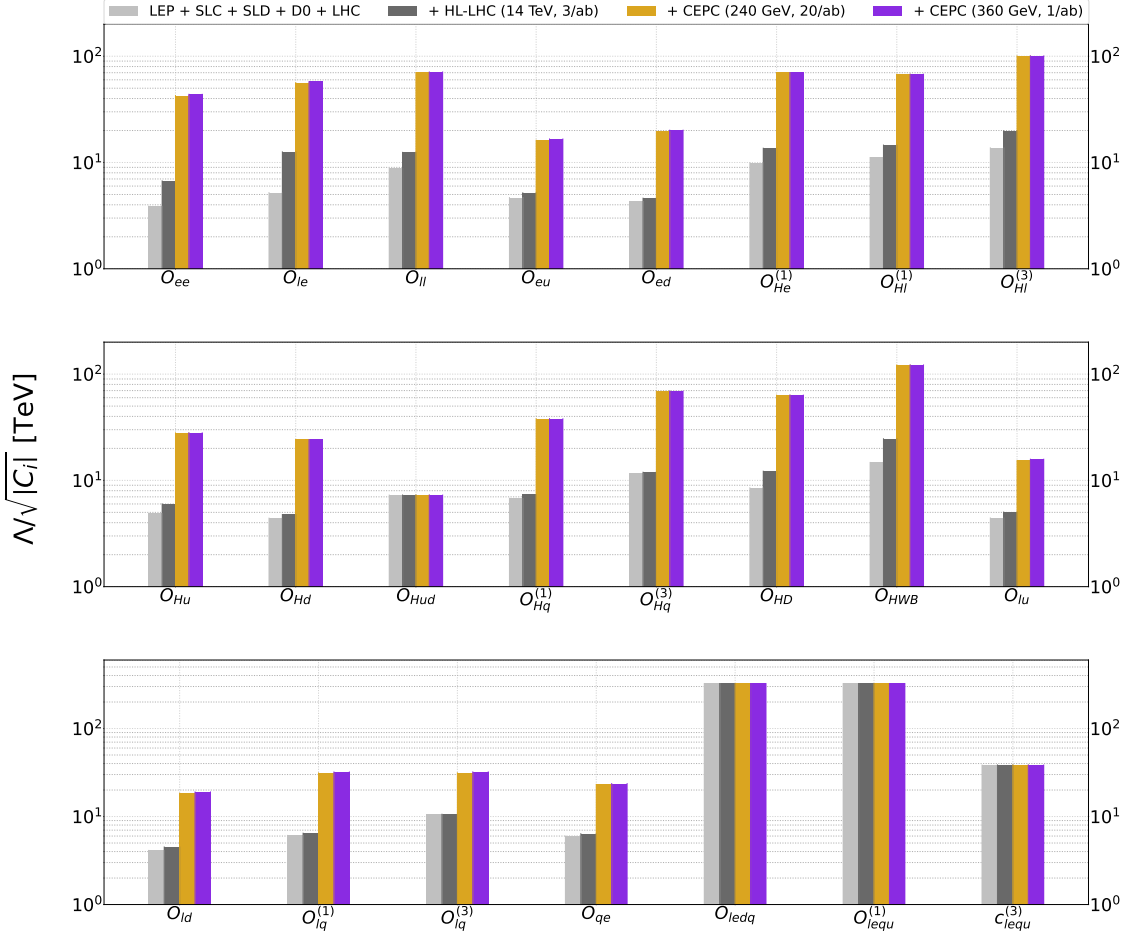}
 \caption{Lower bounds on $\Lambda/\sqrt{|C_i|}$ at the 95\% 
 CL as presented in the Warsaw basis, assuming flavor universality and one operator at a time.}
 \label{fig:cutoffwarsaw}
\end{figure}

\textit{Results}---As a Higgs factory, CEPC is expected to improve significantly the SMEFT global analysis due to its high energy and luminosity. Ref.\,\cite{deBlas:2022ofj} performed a detailed global study on electroweak and Higgs physics, semi-leptonic and pure leptonic 4-fermion operators using the latest CEPC projections. The results are reproduced and shown in figures\,\ref{fig:smeft-higgs}-\ref{fig:smeft-4f} by working in the most general flavor scenario. The central values of the Wilson coefficients in each plot are assumed to be aligned with the SM predictions, and only the $1\sigma$ relative uncertainties are shown for the LHC, HL-LHC, and CEPC with the center of mass energy $\sqrt s = 240\,(360){\rm\,GeV}$ and the integrated luminosity ${\cal L} = 20\,(1){\rm\,ab^{-1}}$. Fig.~\ref{fig:smeft-higgs} clearly shows that CEPC can improve the Higgs couplings by a factor of a few, or even orders of magnitude as can be seen, for example, for the triple gauge couplings $\delta g_{1,Z}$, $\delta \kappa_\gamma$, and $\lambda_Z$. The sensitivity reach of CEPC to leptonic electroweak vertices can be generically reduced down below the unprecedented $10^{-4}$ level as shown in the first row of Fig.~\ref{fig:smeft-4f}, thanks to the high-luminosity of CEPC and radiative return to the $Z$ pole from initial state radiation. The corresponding sensitivity to the hadronic electroweak vertices can also be improved by a few, or even an order of magnitude better as for $\delta g_{Z,L/R}^{bb}$ for instance. For the 4-fermion operators, CEPC can generically reduce the current uncertainties, as shown explicitly in the last five rows of figures\,\ref{fig:smeft-4f}. In particular, for the semi-leptonic operators, the global fit results with the inclusion of CEPC can be improved by $\mathcal{O}(10\sim10^2)$ for the 2nd and 3rd generation quarks due to tagging efficiencies of heavy quarks at CEPC. These stronger constraints as expected for CEPC will on the other hand also impose new challenges on the theoretical side, for example, theoretical uncertainties on the effective number of relative species during neutrino decoupling in the early Universe\,\cite{Du:2021idh,Du:2023upj}. These current theoretical errors will have to be further reduced at the CEPC era to match the precision of CEPC. In summary, CEPC can dramatically increase the sensitivity to Higgs, electroweak, and 4-fermion operators thanks to its high energy and luminosity, and as a result, enhancing its ability in discovering new physics at the 10$\sim$70\,TeV scale that could show up in either the Higgs, electroweak, or the 4-fermion sector as seen in the summary plot of Fig.~\ref{fig:smeft-sum}. For reference, the lower bounds on the new physics scale $\Lambda/\sqrt{|C_i|}$ are also shown in Fig.~\ref{fig:cutoffwarsaw} by switching to the Warsaw basis\,\cite{Grzadkowski:2010es} and at the 95\% CL. To obtain these results, we assume flavor universality and one operator at a time. In this figure, the light gray bar represents the constraints from the current data with the inclusion of $A_{FB}$ for neutral Drell-Yan processes at the LHC\,\cite{ATLAS:2018gqq}, and the gray bar that with the inclusion of $A_{FB}$ at the HL-LHC\,\cite{CMS:2017vxj}. The last two bars correspond to the sensitivity of CEPC with $\sqrt s= 250$\,GeV and 360\,GeV, respectively. In general, we find it feasible for CEPC to improve the new physics scale by a factor of 3$\sim$10 except for the $O_{Hud}$, $O_{ledq}$, $O_{lequ}^{(1)}$ and $O_{lequ}^{(3)}$ operators due to missing projections from the CKM unitarity and flavor physics. Finally, we comment on the global analysis of bosonic CP violating SMEFT operators at dimension 6, which will be important to understand the origin of matter-antimatter asymmetry of our Universe. In\,\cite{deBlas:2022ofj}, it was shown that CEPC could provide a complementary probe compared with that from the (HL-)LHC, and the new physics scale can be constrained to be above $\sim2.5$\,TeV. Furthermore, recent analyses in\,\cite{Ellis:2025ghl,Liu:2024tcz} based on the angular distribution of $e^+e^-\to Z \gamma$ with subsequent hadronic and leptonic $Z$ decay found that CEPC could even constrain the CP-violating dimension-8 SMEFT operators to be around $\sim1$\,TeV, assuming one operator at a time and a center of mass energy at 250\,GeV and an integrated luminosity of 5\,${\rm ab}^{-1}$. This highlights the opportunity of studying CP violation at the CEPC.

\subsection{2HDM global fits} 

While all the indications from the current particle physics measurements seem to confirm the validity of the Standard Model (SM) up to the electroweak scale of a few hundreds GeV, and the observed Higgs boson is SM-like, there are compelling arguments, both from theoretical and observational points of view, in favor of the existence of new physics beyond the SM (BSM). As such, searching for new Higgs bosons would be of high priority since they are present in many extensions of theories beyond the SM.  One of the most straightforward, but well-motivated extensions is the two Higgs doublet model (2HDM) \cite{Branco:2011iw}, . There are five massive spin-zero states in the spectrum ($h, H^0,A^0,H^\pm$) after the electroweak symmetry breaking.

Complementary to the direct searches, precision measurements of SM parameters and the Higgs properties could lead to relevant insights on new physics.
High precision achieved at future Higgs factories with about $10^6$ Higgs bosons, and possible $Z$ pole measurements with $10^{10} - 10^{12}\ Z$ bosons~\cite{CEPC-SPPCStudyGroup:2015csa,Gomez-Ceballos:2013zzn,Baer:2013cma,ALEPH:2005ab} would hopefully shed light on the new physics associated with the electroweak sector. To take advantage of these precisions~\cite{Gu:2017ckc,Chen:2018shg}, we make a global fit to explore their abilities of detecting new particles and constraining model parameter space.

\begin{figure}[t]
\begin{center}
\includegraphics[width=0.435\textwidth]{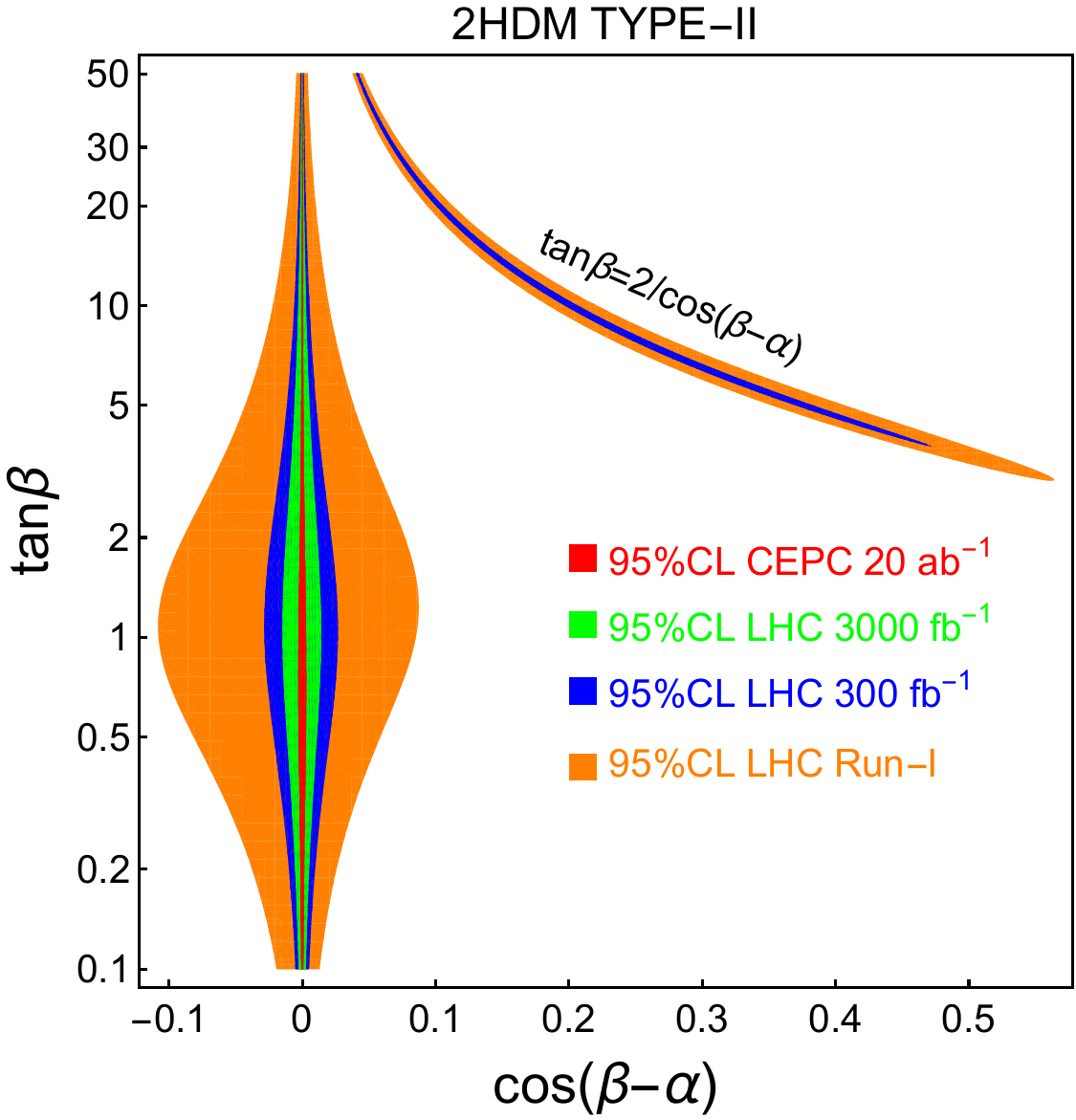}
\includegraphics[width=0.49\textwidth]{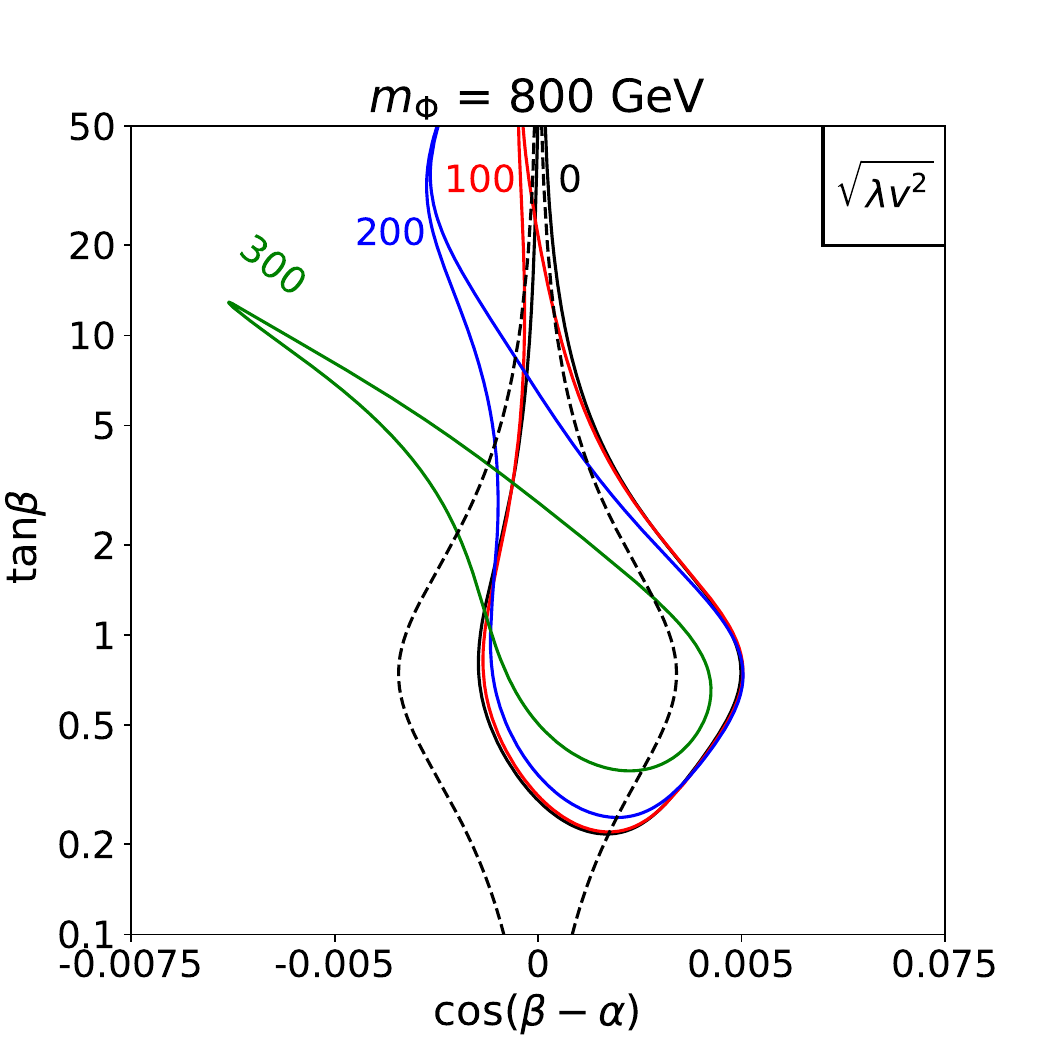}
\caption{The allowed region in the plane of $\cos(\beta-\alpha)$-$\tan \beta$ at 95\% C.L. for the Type-II of 2HDM, given LHC and CEPC Higgs precision measurements at tree level (left) and loop level under CEPC (right). For the tree-left global fit, the special ``arm" regions for the Type-II is the wrong-sign Yukawa region. The right panel shows the parameter space varying the value $\sqrt{\lambda v^2}$ with $m_A=m_H=m_{H^\pm}=m_\Phi=800$ GeV.  The tree-level only global fit results are shown by the dashed black lines for comparison. More details are shown in Ref.~\cite{Gu:2017ckc}.}
\label{fig:cepc-tree}
\end{center}
\end{figure}

There is a plethora of articles in the literature to study the effects of the heavy Higgs states on the Higgs couplings in Models with extended Higgs sector~\cite{Branco:2011iw,Gu:2017ckc,Chen:2018shg}. In 2HDM,
identifying the light CP-even Higgs $h$ to be the experimentally observed 125 GeV Higgs, the couplings of $h$ to the SM fermions and gauge bosons receive two contributions: tree-level values, which are controlled by the mixing angles $\alpha$ of the CP-even Higgses and $\tan\beta$, ratios of the vacuum expectation values of two Higgses: $\tan\beta=v_2/v_1$, and loop contributions with heavy Higgses running in the loop. 

With a global fit to the Higgs rate measurements at the LHC as well as the CEPC, assuming that no deviation to the SM values is observed at future measurements,  the 95\% C.L. region in the $\cos(\beta-\alpha)$ vs. $\tan\beta$ plane for various types of 2HDM (depending on how the two Higgs doublets are coupled to the quark and lepton sectors) are shown in~\autoref{fig:cepc-tree} for tree-level only effects. $\cos(\beta-\alpha)$ in all four types are tightly constrained at both small and large values of $\tan\beta$, except for Type-I (Ref.~\cite{Gu:2017ckc}), in which constraints are relaxed at large $\tan\beta$ due to suppressed Yukawa couplings.

To fully explore the Higgs factory potential in search for new physics beyond the SM, both the tree-level deviation and loop corrections need to be considered. The right panel of \autoref{fig:cepc-tree} shows the 95\% C.L. global fit results to all CEPC Higgs rate measurements in the Type-II 2HDM parameter space, including both tree level and loop corrections.  Degenerate Heavy Higgs masses $m_A=m_H=m_{H^\pm}=m_\Phi$ are assumed such that $Z$-pole precision measurements are automatically satisfied.  Black, red, blue and green curves are for model parameter $\sqrt{\lambda v^2}=\sqrt{m_\Phi^2-m_{12}^2/s_\beta c_\beta}=0$, 100, 200, and 300 GeV, respectively.  The tree-level only global fit results are shown by the dashed black lines for comparison.  $|\cos(\beta-\alpha)|$ is typically constrained to be less than about 0.008 for $\tan\beta\sim 1$. For smaller and larger values of $\tan\beta$, the allowed range of $\cos(\beta-\alpha)$ is greatly reduced.  Loop effects from heavy Higgses tilt the value of $\cos(\beta-\alpha)$ towards negative, especially in the large $\tan\beta$ region.
\begin{figure}[tb]
\begin{center}
 \includegraphics[width=1\textwidth]{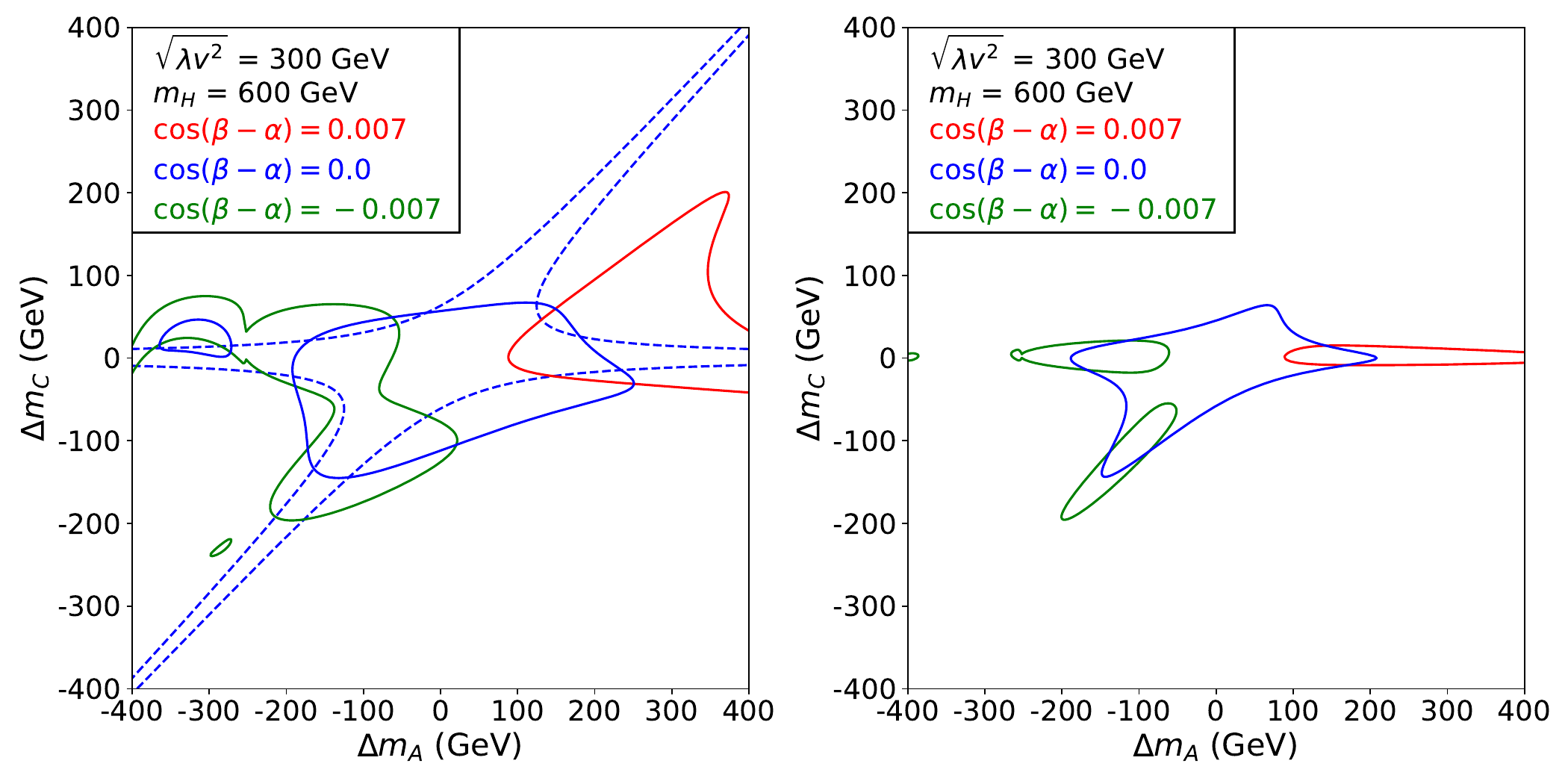}
  \caption{Three-parameter fitting 95\% C.L. range of $\Delta m_A$ - $\Delta m_C$ plane, focusing on the $\cos(\beta-\alpha)$ dependence (given by different colored lines), for Higgs and $Z$-pole precision constraints individually (left panel),  and combined constraints (right panel) in the Type-II 2HDM.  More details are shown in Ref.~\cite{Chen:2018shg}.  }
\label{fig:da-dc-cos}
\end{center}
\end{figure}

Going beyond the degenerate mass case, both the Higgs and $Z$-pole precision variables are sensitive to the mass splittings between the charged Higgs and the neutral ones.
~\autoref{fig:da-dc-cos} shows the 95\% C.L. range of   $\Delta m_A=m_A-m_H$ vs. $\Delta m_C=m_{H^\pm}-m_H$  plane, for Higgs and $Z$-pole precision constraints individually in (left panel),  and combined constraints (right panel), with  $m_H=600$ GeV  and $\sqrt{\lambda v^2}= 300$ GeV.   For the Higgs precision fit, the alignment limit (blue curve) leads to both $\Delta m_A$ and $\Delta m_C$ around 0 within a few hundred GeV range.   Even for small deviation away from the alignment limit,  $\Delta m_A$ is constrained to be positive for $\cos(\beta-\alpha)= 0.007$, and negative for  $\cos(\beta-\alpha)= -0.007$.   The $Z$ pole precision measurements (shown in region enclosed by blue dashed curves) constrain either $\Delta m_C\sim 0$ or $\Delta m_C \sim \Delta m_A$, equivalent to  $m_{H^\pm} \sim m_{H,A}$.  
Combining both the Higgs and $Z$ pole precisions (right panel), the range of $\Delta m_{A,C}$ are further constrained to a narrower range.   The expected accuracies at the $Z$-pole and at a Higgs factory are quite complementary in constraining heavy Higgs mass splittings.

In this section, we 
presented the results for the impacts of the precision measurements of the SM parameters at the proposed $Z$-factories and Higgs factories on the extended Higgs sector of 2HDM. For the tree-level 2HDM, $|\cos(\beta-\alpha)|$ can be restricted 0.008. When including the loop effects, CEPC precision can give lower bound on non-SM Higgs masses, as well as their splitting.
Combining the Higgs and Z-pole precisions, the typical heavy Higgs mass splitting is constrained to be less than about 200 GeV.

\subsection{SUSY global fits} 
\begin{figure}[th!]
 \centering
 \includegraphics[width=0.9\textwidth]{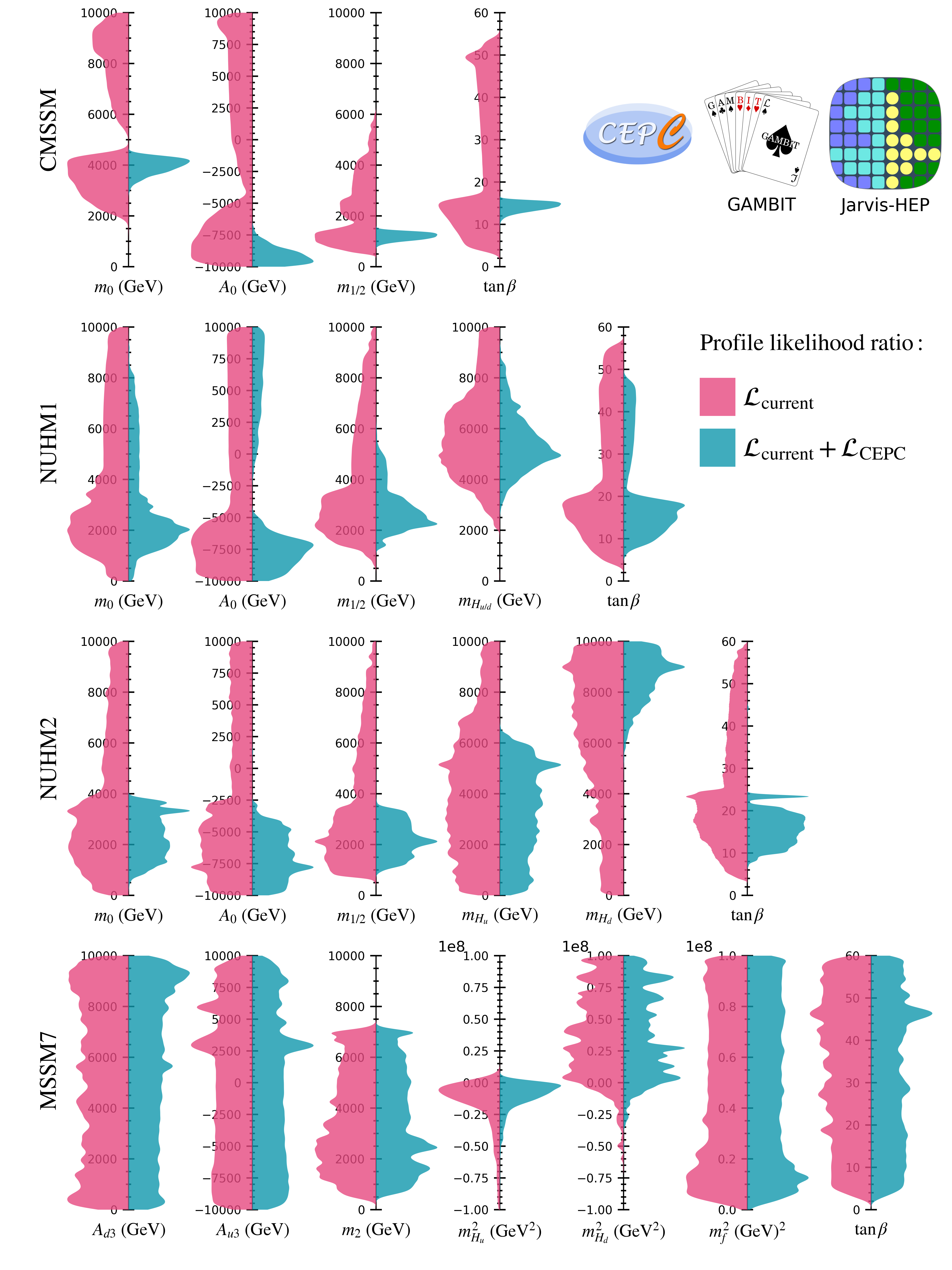}
 \caption{One-dimensional profiled likelihood ratio for the global fit of the CMSSM, NUHM1, NUHM2, and MSSM7 models, using the present experimental data (left parts) and considering additional CEPC measurements (right parts). }
 \label{fig:gambit}
\end{figure}

It is shown in Section~\ref{sec:SUSY} that the direct searches for sparticles at electron-positron colliders are restricted by collision energy. However, the high precision measurements of the Higgs and electroweak (EW) sector can significantly affect the global fit of SUSY models. Ref.~\cite{Li:2020glc} shown that conducting a global fit solely based on precise Higgs measurements at future Higgs factories could potentially raise the lower bound of the SUSY scale above TeV, for small values of $\tan\beta$. 

Ref~\cite{Athron:2022uzz} performed several comprehensive global fits by combining Higgs measurements at CEPC with existing experimental data,  using data provided by the GAMBIT community~\cite{GAMBIT:2017yxo,GAMBIT:2017snp,GAMBIT:2017zdo}, for four supersymmetric models:
\begin{itemize}
    \item CMSSM~(Constrained Minimal Supersymmetric Standard Model). Inspired by scenarios where SUSY breaking is transmitted through supergravity interactions, the soft mass parameters at the Grand Unified Theory (GUT) scale are set to a universal scalar mass $m_0$, a universal gaugino mass $m_{1/2}$ and a universal trilinear couping $A_0$. The Higgs sector has two remaining free parameters defined at the scale $m_Z$: the ratio of the vacuum expectation values of the two Higgs doublets, $\tan\beta = v_u/v_d$ and the sign of $\mu$.
    \item NUHM1~(Non-Universal Higgs Mass 1). The GUT-scale constraint on the soft scalar Higgs masses is relaxed by introducing an additional free parameter $m_H$ . The soft Higgs masses $m_{H_u}$ and $m_{H_d}$ are not set equal to $m_0$, but instead obey the relation $m_{H_u}=m_{H_d}=m_{H}$ at the GUT scale. 
    \item NUHM2~(Non-Universal Higgs Mass 2). The constraint on the soft Higgs masses is further relaxed so that $m_{H_u}$ and $m_{H_d}$ become independent, real, dimension-one parameters at the GUT scale.
    \item MSSM7~(seven-dimensional phenomenological MSSM). All the input parameters are defined at an energy near the electroweak scale. Inspired by GUT scale gaugino mass universality, the gaugino masses satisfy $3/5 \cos^2\theta_W M_1 = \sin^2\theta_W M_2 = \alpha/\alpha_s M_3$. 
    All entries in $A_u$, $A_d$ and $A_e$ are assumed to be zero except for $(A_u)_{33} = A_{u_3}$ and $(A_d)_{33}=A_{d_3}$. 
    All of the off-diagonal entries in $m_Q^2$, $m_u^2$, $m_d^2$, $m_L^2$ and $m_e^2$ to be zero, so as to suppress flavour-changing neutral currents. By setting all remaining mass matrix entries to a universal squared sfermion mass $m_{\tilde{f}}^2$, the final list of free parameters contains $M_2$, $A_{u_3}$, $A_{d_3}$, $m_{\tilde{f}}^2$, $m_{H_u}^2$, $m_{H_d}^2$ and $\tan\beta$ (plus the input scale $Q$ and the sign of $\mu$). 
\end{itemize}
Beside precise Higgs measurements at future Higgs factories, the likelihood functions include a number of direct and indirect dark matter searches, a large collection of electroweak precision and flavour observables, direct searches for supersymmetry at the LEP and Runs I and II of the LHC, and constraints from Higgs observables. 

Fig.~\ref{fig:gambit} shows the profile likelihoods with and without the additional likelihood for the Higgs measurements at CEPC. 
Here, the central values of measurements at the CEPC are assumed to be the same as those predicted by the best-fit point of each model, because GAMBIT employed advanced sampling methods, resulting in a majority of samples being clustered around the best-fit point. The theoretical uncertainties utilized in the likelihood functions are scaled to be 0.2 times smaller than the current theoretical uncertainties of the SM Higgs. It is evident that a significant portion of the parameter region favored by the current constraints is disallowed when considering the precise Higgs measurements obtained from the CEPC. The preferred regions of the parameter space undergo a noticeable reduction in size, converging closer to the best-fit point. Consequently, the additional measurements from the CEPC hold the potential to differentiate between various dark matter annihilation mechanisms present in the models, as well as provide insights into the signs of the $\mu$ parameter. Comparing the results across different models, the constraints placed on the model parameters tend to be weaker in models characterized by a larger number of input parameters,  i.e., looser correlations between the model parameters. For 19-parameter phenomenological MSSM, Ref.~\cite{Arbey:2021jdh} shows that  future the $e^+e^-$ collider can test up to $10\%\sim12\%$ of the samples obtained from a flat scan that have not been excluded by current LHC direct SUSY searches and flavor physics data.

In conclusion, future Higgs factories equipped with high-precision Higgs coupling measurements have the potential to greatly enhance our comprehension of the parameter space and mass spectrum in the MSSM. They offer valuable complementary information to dark matter searches and EW precision measurements. By providing precise data on Higgs couplings, these Higgs factories can contribute substantially to furthering our understanding of fundamental physics and refining our knowledge of the MSSM.

\clearpage
\section{Conclusion}

This document presents the Beyond the Standard Model physics potential of the Circular Electron-Positron Collider. Operating as a powerful probe of new physics, the CEPC will conduct dedicated runs at the Higgs, $Z$, and $W$ production thresholds, with upgrade capabilities for $\bar{t}t$ threshold operation. We systematically investigate BSM scenarios where the CEPC can provide significant advancements, organized into the following categories: exotic Higgs, $W$, $Z$, and top decays; electroweak phase transition and gravitational wave signatures; dark matter and dark sector studies; long-lived particle searches; supersymmetry; flavor physics; neutrino physics; exotic models and global fits.

\begin{figure}[t]
\centering
\includegraphics[width=1.0\textwidth]{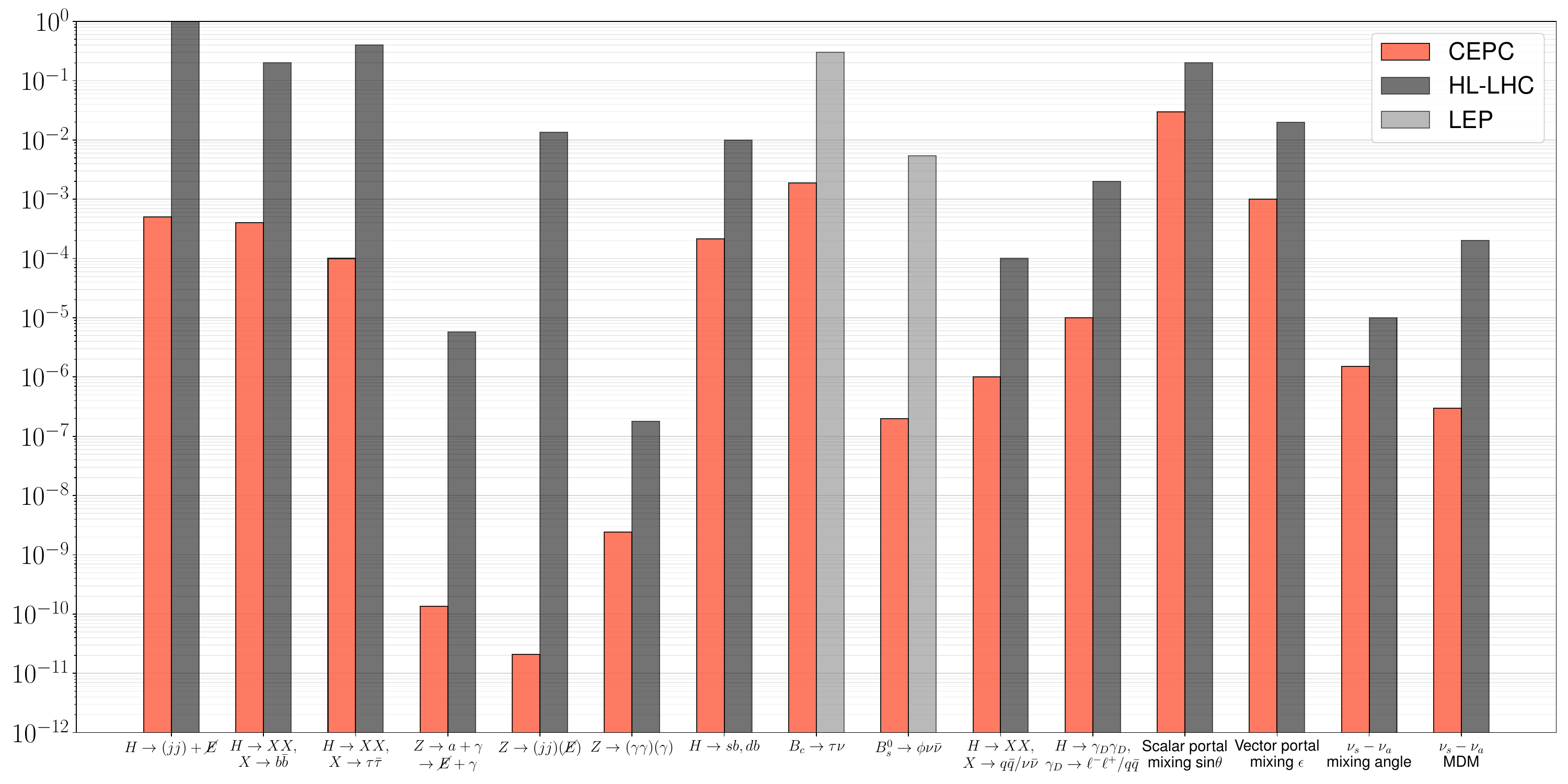}
\caption{Projected sensitivities of the CEPC and HL-LHC for various new physics scenarios, including exotic Higgs (first three categories) and  $Z$-boson decays (categories 4-6), flavor-violating Higgs and $B$-meson decays (categories 7-9), long-lived particle searches (categories 10-11), scalar and vector portal mixing (categories 12-13), as well as sterile-active neutrino mixing and magnetic dipole moment interactions (last two categories). For flavor-violating $B$-meson decay channels, there are existing constraints from LEP only.}
\label{fig:final-example-sen-LHC-CEPC}
\end{figure}

Given the CEPC's primary focus on precision Higgs physics, it serves as an ideal facility to investigate BSM scenarios mediated through Higgs portal operators and higher-dimensional Higgs-related operators. These operators can modify SM Higgs couplings and induce exotic Higgs decay channels. With its planned integrated luminosity of $20~\text{ab}^{-1}$, the CEPC will achieve unprecedented sensitivity to exotic Higgs decays, constraining branching ratios to: (i) $\sim 0.1\%$ for fully invisible decays, (ii) $0.03\%$--$0.2\%$ for semi-invisible decays, and (iii) $0.03\%$--$0.6\%$ for decays to dark sector particles that subsequently decay to SM final states.
Fig.~\ref{fig:final-example-sen-LHC-CEPC} presents a comprehensive sensitivity comparison between CEPC and HL-LHC for: semi-invisible decays $H \to jj+\slashed{E}$, dark sector decays $H \to XX$ with $X \to b\bar{b}$ or $X \to \tau^+\tau^-$. The cleaner lepton collider environment at CEPC provides 2--3 orders of magnitude improvement in fully visible channels, while its superior missing energy reconstruction enables 2--4 orders of magnitude better sensitivity for fully invisible and semi-invisible channels compared to HL-LHC.

Another key objective of the CEPC is conducting precision measurements of the $Z$ gauge boson at the $Z$-pole, which will substantially improve upon the electroweak precision achieved by LEP. Dark sector particles may carry effective electroweak charges, potentially generating exotic $Z$ decays through either off-shell mediation or on-shell production of the dark sector particles. These exotic decays can be categorized according to their final states: (i) decays containing invisible particles, (ii) light resonances producing SM particle pairs, or (iii) non-resonant three-body decays.
Fig.~\ref{fig:final-example-sen-LHC-CEPC} compares the sensitivity of CEPC and HL-LHC for: semi-invisible decays $Z \to \slashed{E} + \gamma$ and $Z \to jj + \slashed{E}$, and dark sector decays $Z \to a + \gamma$ with $a \to \gamma\gamma$. The Tera-$Z$ operation can probe branching ratios in the range $[10^{-7}, 10^{-11}]$ for semi-visible channels and $[10^{-6}, 10^{-9}]$ for fully visible channels. Additionally, measurements of fully invisible $Z$ decays will constrain the effective number of relativistic neutrino species ($N_{\rm eff}$) with a precision of approximately $10^{-3}$, which can probe $Z$ boson couplings to dark matter and sterile neutrinos.

The study of electroweak phase transitions and gravitational waves is highly pertinent to the CEPC. BSM physics could induce a First-Order Electroweak  Phase Transition, significantly altering the Higgs potential, as evidenced by deviations in the Higgs self-coupling ($h^3$) and its couplings to $Z$ bosons. Higgs precision measurements of the $hZ$ cross-section and the SM Higgs couplings, together with searches for exotic Higgs decays into lighter dark scalars can effectively probe the Standard Model Effective Field Theory in the context of FOEWPT and can cover almost the entire FOEWPT parameter space. CEPC is particularly well-suited for studying the extra scalar decays into $\tau^+\tau^-$ and $b\bar{b}$ channels, offering a significant advantage over the HL-LHC. 
CEPC provides a competitive method to explore the FOEWPT of the early Universe, which could be cross-checked by gravitational wave experiments.
In fact, the FOEWPT is in contradiction to the SM prediction of a cross-over in the early Universe. 
New particles need to be introduced to alter the thermal history of the early Universe. 
These new particles are naturally associated with the Higgs sector.
They could cast sizable deviations in Higgs boson properties that could be observed at the CEPC. Quantitative studies show that the CEPC could cover almost the entire phase space of those models, providing decisive tests of the FOEWPT scenario.

The CEPC offers exceptional sensitivity for probing light dark matter and dark sector particles that interact with the Standard Model electroweak sector. These particles may couple to the SM through scalar, fermion, or vector portals, with many such models testable via exotic Higgs and $Z$ decays. 
Additionally, the CEPC can directly produce dark sector particles through Drell-Yan processes and DM via associated production channels. Compared to HL-LHC, these channels enable up to an order-of-magnitude improvement in sensitivity to mixing angles or coupling strengths for scalar, fermion, and kinetic mixing portals, as well as 1-2 orders improvement in coupling strength sensitivity for DM with electric or magnetic dipole moments as shown in  
Fig.~\ref{fig:final-example-sen-LHC-CEPC}. 
For canonical scalar or fermion dark matter with Higgs-portal mediation to the SM, the CEPC provides excellent sensitivity at low masses region ($\lesssim 10$ GeV), complements direct-detection coverage at higher masses region ($\gtrsim  10$ GeV), and extends reach into the neutrino fog for both scalar/fermion DM scenarios beyond the HL-LHC capacity, as shown in Fig.~\ref{fig:Dark-Sector-Collider-DMDD}. The clean experimental environment and capability for full final-state reconstruction (including missing energy) are crucial for CEPC to deliver competitive sensitivity for dark sector models.

BSM particles may appear as long-lived particles in collider experiments, and the CEPC's sensitivity with additional far detectors has been extensively studied. Benefiting from its high integrated luminosity for $Z$, Higgs, and $W$ boson production, the CEPC offers exceptional competitiveness for investigating LLP decays originating from these bosons.
For instance, in cases of exotic Higgs decays to long-lived dark scalar or dark photon pairs that subsequently decay into quarks or leptons, the CEPC demonstrates sensitivities two orders of magnitude superior to the HL-LHC, as illustrated in Fig.~\ref{fig:final-example-sen-LHC-CEPC}. This advantage stems from the CEPC's cleaner experimental environment and enhanced sensitivity to quark and tau final states compared to the HL-LHC.
Furthermore, the CEPC provides superior probing capability for long-lived staus (superpartners of tau leptons) decaying into gravitinos and tau leptons compared to the HL-LHC. Consequently, LLP studies at the CEPC will deliver complementary measurements to those obtained at the HL-LHC.

Supersymmetry remains one of the most compelling beyond-Standard-Model scenarios. Although the CEPC operates at a lower center-of-mass energy than the HL-LHC, it maintains strong sensitivity to light electroweakinos and sleptons with masses up to half its collision energy. This mass reach can be further extended when one of the produced electroweakinos or sleptons in a pair-production process is permitted to be off-shell.
A particularly promising channel is the search for heavier selectrons via light bino pair production through t-channel selectron exchange. This process effectively bypasses conventional energy constraints, extending the selectron mass reach up to 4.5 TeV at the CEPC despite its lower center-of-mass energy. The low-background environment and excellent detection capabilities for soft final-state particles at CEPC provide significant advantages, especially for probing scenarios with compressed mass spectra where sensitivity at hadron colliders is typically reduced.

Flavor physics presents significant potential for uncovering new physics, and the CEPC holds great promise in this area due to the high statistics of $Z$ gauge boson events, a cleaner experimental environment, and the ability to fully reconstruct events with missing energy. The high integrated luminosity at the $Z$ pole allows sensitivity improvements of 2--4 orders of magnitude for exotic $Z$ decays into charged leptons that violate flavor, such as $Z \to \ell \ell'$. Additionally, the CEPC will allow for a highly precise examination of tau lepton charge lepton flavor violation decays, achieving sensitivity improvements of approximately two orders of magnitude. Sensitivity to rare $B$, $B_s$, and $B_c$ meson decays via FCNC processes can also be improved by 2–3 orders of magnitude, particularly in cases where the final state involves a $\tau$ lepton. In Fig.~\ref{fig:final-example-sen-LHC-CEPC}, we show the branching ratio sensitivity comparison for the flavor-violating Higgs decay $H \to sb,db$ and the $B$ meson decays $B_c \to \tau \nu$ and $B_s^0 \to \phi \nu\bar{\nu}$, where CEPC projected sensitivities are significantly better than HL-LHC and LEP experiments.

Neutrino masses and oscillations require BSM physics, and the CEPC can play a significant role in exploring these phenomena. The CEPC can complement existing studies by covering parameter spaces for sterile neutrinos and non-standard neutrino interactions. 
With its high integrated luminosity at the Z-pole and Higgs associated production, combined with a clean experimental environment, the CEPC can achieve superior sensitivity in detecting displaced sterile neutrinos from $Z$ and Higgs decays, surpassing the capabilities of the HL-LHC. Additionally, with potential main detector setups and far detector options, the CEPC could further enhance its reach. In Fig.~\ref{fig:final-example-sen-LHC-CEPC}, the transition dipole operator for sterile neutrinos, particularly for masses above the GeV scale, and the active neutrino mixing can be the most effectively probed at CEPC, providing sensitivity improvements by three and one orders of magnitude respectively compared to HL-LHC.

For exotic new physics models, the CEPC offers unique capabilities to constrain ALPs, particularly through their anomalous couplings to Standard Model gauge bosons. The production of ALPs from Higgs and $Z$ bosons at the CEPC provides complementary coverage to that of the HL-LHC and heavy ion colliders, with an advantage in the low-mass region due to the lower production thresholds and cleaner experimental environment. Additionally, the precise measurement of the electromagnetic form factor of charged leptons, especially tau leptons, is a crucial test of the Standard Model and a probe for new physics.

Global fits provide a comprehensive statistical framework for evaluating and comparing BSM models by simultaneously analyzing multiple experimental constraints to determine the most viable parameter ranges and model predictions. In this white paper, the global fit analysis has been performed for SMEFT, 2HDM, and various SUSY models for CEPC. The CEPC offers an order-of-magnitude improvement by 1--3 orders on Higgs coupling SMEFT operators and a 1--2 order improvement on electroweak vector-fermion couplings, especially for leptons and second/third-generation quarks, compared to HL-LHC. For 2HDM, the precision measurement of the Higgs and Z-pole at CEPC can constrain the Higgs parameter $|\cos(\beta-\alpha)|$ to less than 0.008, and the typical mass splitting of heavy Higgs to less than about 200 GeV. For small $\tan\beta$, the SUSY scale can be raised to above 1 TeV using the precision data of Higgs and electroweak measurements.

In summary, CEPC could determine the Higgs boson properties and explore Beyond the Standard Model physics with unprecedented sensitivity.
As a future Higgs factory, it will produce millions of Higgs bosons in a clean experimental environment, enabling detailed studies of Higgs properties that could reveal deviations from Standard Model predictions. 
These deviations may point to new physics, such as hidden couplings to dark matter, mixing with dark sector particles, or mechanisms explaining neutrino masses. The CEPC will also investigate the Higgs potential, which is crucial for understanding electroweak symmetry breaking and the possibility of a first-order phase transition in the early universe, a key requirement for explaining matter-antimatter asymmetry.  
In addition to Higgs studies, CEPC will excel in probing light dark matter and dark sector particles, which are challenging to detect at hadron colliders like the LHC. Its high-precision measurements of Higgs and Z bosons could uncover rare decays or signals of long-lived particles, offering insights into theories like supersymmetry or hidden dark forces. Its ability to reconstruct soft particles and missing energy makes it uniquely sensitive to compressed spectra and weakly interacting scenarios. 

\begin{figure}[t]
\centering
\includegraphics[width=1.0\textwidth]{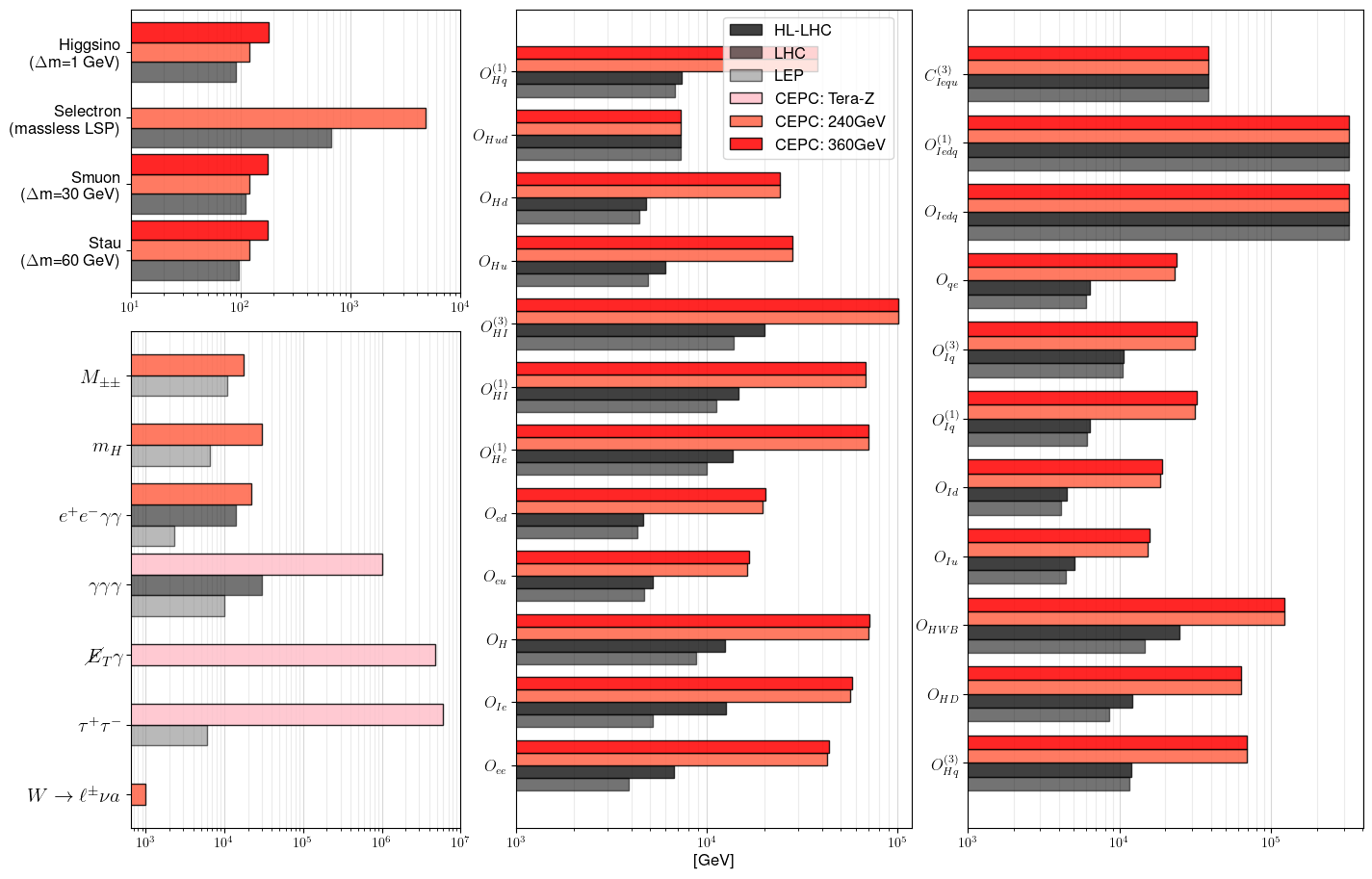}
\caption{The typical energy scale of New Physics that could be probed at the CEPC. The left column corresponds to a specialized New Physics Model of SUSY, 
and exotic New Physics signal searches, including Axion-like particles, and scalar particles in seesaw Models. 
The middle and right columns correspond to New Physics search via EFT.
}
\label{fig:NP_Summary}
\end{figure}

The new physics discovery power could also be expressed in the explorable energy range. Using the SMEFT framework and considering the high precision measurements on EW and Higgs sectors, the CEPC could explore the New Physics with an energy scale of 10--100 TeV, which is roughly one order of magnitude higher than that of the HL-LHC. CEPC is sensitive to the SUSY Particles with mass scale range from the beam energy up to 5 TeV (for the SUSY partner of the electron could be detected via $t$-channel exchanges). CEPC is sensitive to the New Physics at an energy scale much higher than the center-of-mass energy, for example, it could explore new physics with an energy scale of $1\times10^{4}$ TeV through the measurements of $\tau$ weak-electric dipole moments. 
More details could be found in the relevant sessions.

In conclusion, the CEPC stands as a versatile and powerful facility for probing new physics beyond the Standard Model. Its unparalleled precision in Higgs coupling measurements, electroweak parameter determinations, and exotic decay searches enables sensitivity to phenomena inaccessible at hadron colliders. The CEPC's clean collision environment and high luminosity make it exceptionally suited for investigating dark matter candidates, long-lived particles, and supersymmetric scenarios through both direct searches and precision deviations. Furthermore, its capability to explore the Higgs potential's structure provides critical insights into electroweak phase transition dynamics and potential gravitational wave signatures, bridging particle physics with cosmology.

Looking ahead, the CEPC serves as a vital complement to future 100 TeV proton-proton colliders like the SPPC. While the SPPC will extend the energy frontier to directly produce heavy new states, the CEPC's precision measurements will provide essential guidance by constraining model parameters and identifying promising discovery channels. This synergy creates a powerful multi-scale approach—the CEPC's electroweak-scale precision and the SPPC's ultra-high-energy reach together form a complete strategy for uncovering new physics. By establishing fundamental benchmarks and revealing subtle deviations, the CEPC will not only advance our understanding of the Higgs sector but also lay the groundwork for transformative discoveries at next-generation hadron colliders.

\newpage
\section*{Acknowledgements}

We sincerely thank Professor Tianchi Zhao for his invaluable support and suggestions.
This work was supported in part by the Natural Science Foundation of China (NSFC) under Grant Nos. 11905162, 12035008, 12075097, 12075123, 12090060, 12090064, 12105248, 12135006, 12175039, 12205227, 12205312, 12205387, 12205171, 12235008, 12321005, 12235001, 12305094, 12305115, 12335005, 12375091, 12375094, 12375096, 11975129, 12375194, 12447167, 12475094, 12475101, 12475106, 12475111, 12425506, 12375101, 12405119, 12405101, 12505121, 12135007, 12175218, 12075213, 12335005, 12175243, 12533001, 12125503, 12305103, 12505120, 12575099, 12505122, 12342502, 12575106, 12147214, W2432006, and W2441004.
Support is also acknowledged from 
the National Key R\&D Program of China under Grant No. 2024YFA1610603,
the China Postdoctoral Science Foundation under Grant Nos. 2023M732255, 2025M773403, and GZC20231613, 
the Natural Science Foundation of Jiangsu Province under Grant No. BK20210201, 
the Natural Science Foundation of Henan (Distinguished Young Scholars of Henan Province) under Grant No. 242300421046, 
the Natural Science Foundation of Sichuan Province under Grant No. 2025ZNSFSC0880, 
the Guangdong Major Project of Basic and Applied Basic Research under Grant No. 2020B0301030008, 
the Department of Science and Technology of Shandong province under Grant Nos. tsqn202312052 and 2024HWYQ-005,
the Startup Research Fund of Henan Academy of Sciences under Project No. 20251820001,
the Excellent Young Talents Program of Wuhan University of Technology under Grant No. 40122102,
the research program of the Wuhan University of Technology under Grant Nos. 3120625397 and 2020IB024,
the Fundamental Research Funds for the Central Universities under Grant Nos. JZ2023HGTB0222 and WUT: 2022IVA052, 
the Excellent Scholar Project of Southeast University (Class A), 
the Big Data Computing Center of Southeast University, 
National Science and Technology Council, the Ministry of Education (Higher Education Sprout Project NTU-114L104022-1), and the National Center for Theoretical Sciences of Taiwan, and Vietnam National Foundation for Science and Technology Development (NAFOSTED) under grant number 103.01-2023.50,
the Research Office of the University of the Witwatersrand and South African Department of Science and Innovation through the SA-CERN program, 
the self-determined research funds of Central China Normal University from the colleges’ basic research and operation of MOE under Grant No. CCNU24AI003, 
the SJTU Double First Class start-up fund (WF220442604), 
the Innovation Capability Support Program of Shaanxi under Program No. 2021KJXX-47,
the Slovenian Research Agency under the research core funding No. P1-0035 and the research grants J1-3013 and N1-0253,
CONICET and ANPCyT under project PICT-2021-00374,
Higher Education Sprout Project NTU-114L104022-1, KIAS Individual Grants (PG086002) at the Korea Institute for Advanced Study, FAPESP Grant No. 2021/09547-9, the Slovenian Research Agency under the research core funding No. P1-0035 and research grants J1-3013 and N1-0253, the bilateral project Proteus PR-12696 / Projet 50194VC.

\newpage
\section*{Glossary}
\noindent\textbf{ATLAS} A Toroidal LHC Apparatus \\
\textbf{ALP} Axion-like Particle \\
\textbf{BNL} Brookhaven national laboratory \\
\textbf{BSM} Beyond the Standard Model \\
\textbf{CDR} Conceptual Design Report \\
\textbf{CEPC} Circular Electron-Positron Collider \\
\textbf{cFLV} charged Lepton Flavour Violation\\
\textbf{CHARM} CERN-Hamburg-Amsterdam-Rome-Moscow collaboration \\
\textbf{CHSH} Clauser, Horne, Shimony and Holt \\
\textbf{CKM} Cabibbo–Kobayashi–Maskawa \\
\textbf{CMSSM} Constrained Minimal Supersymmetric Standard Model \\
\textbf{CP} Charge Parity \\
\textbf{DM} Dark Matter \\
\textbf{DS} Dark Sector \\
\textbf{EDM} Electric Dipole Moments \\
\textbf{EFT} Effective Field Theory \\
\textbf{EHM} Emergent Hadron Mass \\
\textbf{EW} Electroweak \\
\textbf{EWKino} Electroweakino \\
\textbf{EWPT} Electroweak Phase Transition \\
\textbf{FCC} Future circular collider \\
\textbf{FCMC} Flavor changing Neutral Current \\
\textbf{FD} Far Detector\\
\textbf{FNAL} Fermi national accelerator laboratory \\
\textbf{FOEWPT} First Order Electroweak Phase Transition \\
\textbf{HB} Higgs boson \\
\textbf{HVP} Hadronic vaccum polarisation \\
\textbf{GUT} Grand Unification Theory \\
\textbf{GW} Gravitational Waves\\
\textbf{LEP} Large electron-positron collider \\
\textbf{LFU} Lepton Flavour Unitarity \\
\textbf{LFV} Lepton Flavour Violation \\
\textbf{LHC} Large hadron collider \\
\textbf{LHVT} Local hidden variable theory \\
\textbf{LLP} Long-Lived Particle \\
\textbf{MC} Monte Carlo \\
\textbf{MD} Main Detector \\
\textbf{MSSM} Minimal Supersymmetric Standard Model \\
\textbf{ND} Near Detector\\
\textbf{NP} New Physics \\
\textbf{NSI} Non-Standard Interaction \\
\textbf{NUHM} Non-Universal Higgs Mass \\
\textbf{OPAL} Omni-Purpose Apparatus at LEP \\
\textbf{PDG} Particle data group \\
\textbf{QED} Quantum ElectroDynamics \\
\textbf{QCD} Quantum ChromoDynamics \\
\textbf{QM} Quantum Mechanics \\
\textbf{SM} the Standard Model \\
\textbf{SMEFT} the Standard Model Effective Field Theory\\
\textbf{SUSY} Supersymmetry \\
\textbf{SUGRA} Supersymmetric gravity \\
\textbf{2HDM} Two-Higgs-Doublet Model \\
\textbf{VLL} Vector-like Lepton \\
\textbf{WDM} Weak Dipole Moment  \\
\textbf{WIMP} Weakly Interacting Massive Particle  \\

\newpage

\bibliographystyle{utphys}
\bibliography{references,refs}

\providecommand{\href}[2]{#2}\begingroup\raggedright\begin{thebibliography}{10}

\bibitem{CMS:2023yay}
{\scshape CMS} collaboration, \emph{{Search for a standard model-like Higgs
  boson in the mass range between 70 and 110$~\mathrm{GeV}$ in the diphoton
  final state in proton-proton collisions at $\sqrt{s}=13~\mathrm{TeV}$}}, .

\bibitem{ATLAS:2023jzc}
{\scshape ATLAS} collaboration, \emph{{Search for diphoton resonances in the 66
  to 110 GeV mass range using 140 fb$^{-1}$ of 13 TeV $pp$ collisions collected
  with the ATLAS detector}}, .

\bibitem{Biekotter:2023oen}
T.~Biek\"otter, S.~Heinemeyer and G.~Weiglein, \emph{{95.4~GeV diphoton excess
  at ATLAS and CMS}},
  \href{https://doi.org/10.1103/PhysRevD.109.035005}{\emph{Phys. Rev. D}
  {\bfseries 109} (2024) 035005}
  [\href{https://arxiv.org/abs/2306.03889}{{\ttfamily 2306.03889}}].

\bibitem{Barate:2003sz}
{\scshape LEP Working Group for Higgs boson searches, ALEPH, DELPHI, L3, OPAL}
  collaboration, \emph{{Search for the standard model Higgs boson at LEP}},
  \href{https://doi.org/10.1016/S0370-2693(03)00614-2}{\emph{Phys. Lett. B}
  {\bfseries 565} (2003) 61}
  [\href{https://arxiv.org/abs/hep-ex/0306033}{{\ttfamily hep-ex/0306033}}].

\bibitem{Cao:2016uwt}
J.~Cao, X.~Guo, Y.~He, P.~Wu and Y.~Zhang, \emph{{Diphoton signal of the light
  Higgs boson in natural NMSSM}},
  \href{https://doi.org/10.1103/PhysRevD.95.116001}{\emph{Phys. Rev. D}
  {\bfseries 95} (2017) 116001}
  [\href{https://arxiv.org/abs/1612.08522}{{\ttfamily 1612.08522}}].

\bibitem{Azatov:2012bz}
A.~Azatov, R.~Contino and J.~Galloway, \emph{{Model-Independent Bounds on a
  Light Higgs}}, \href{https://doi.org/10.1007/JHEP04(2012)127}{\emph{JHEP}
  {\bfseries 04} (2012) 127} [\href{https://arxiv.org/abs/1202.3415}{{\ttfamily
  1202.3415}}].

\bibitem{Biekotter:2023jld}
T.~Biek\"otter, S.~Heinemeyer and G.~Weiglein, \emph{{The CMS di-photon excess
  at 95 GeV in view of the LHC Run 2 results}},
  \href{https://doi.org/10.1016/j.physletb.2023.138217}{\emph{Phys. Lett. B}
  {\bfseries 846} (2023) 138217}
  [\href{https://arxiv.org/abs/2303.12018}{{\ttfamily 2303.12018}}].

\bibitem{Biekotter:2021ovi}
T.~Biek\"otter and M.O.~Olea-Romacho, \emph{{Reconciling Higgs physics and
  pseudo-Nambu-Goldstone dark matter in the S2HDM using a genetic algorithm}},
  \href{https://doi.org/10.1007/JHEP10(2021)215}{\emph{JHEP} {\bfseries 10}
  (2021) 215} [\href{https://arxiv.org/abs/2108.10864}{{\ttfamily
  2108.10864}}].

\bibitem{Drechsel:2018mgd}
P.~Drechsel, G.~Moortgat-Pick and G.~Weiglein, \emph{{Prospects for direct
  searches for light Higgs bosons at the ILC with 250 GeV}},
  \href{https://doi.org/10.1140/epjc/s10052-020-08438-1}{\emph{Eur. Phys. J. C}
  {\bfseries 80} (2020) 922}
  [\href{https://arxiv.org/abs/1801.09662}{{\ttfamily 1801.09662}}].

\bibitem{Cepeda:2019klc}
M.~Cepeda et~al., \emph{{Report from Working Group 2}: {Higgs Physics at the
  HL-LHC and HE-LHC}},
  \href{https://doi.org/10.23731/CYRM-2019-007.221}{\emph{CERN Yellow Rep.
  Monogr.} {\bfseries 7} (2019) 221}
  [\href{https://arxiv.org/abs/1902.00134}{{\ttfamily 1902.00134}}].

\bibitem{Bambade:2019fyw}
P.~Bambade et~al., \emph{{The International Linear Collider: A Global
  Project}},  \href{https://arxiv.org/abs/1903.01629}{{\ttfamily 1903.01629}}.

\bibitem{Heinemeyer:2021msz}
S.~Heinemeyer, C.~Li, F.~Lika, G.~Moortgat-Pick and S.~Paasch,
  \emph{{Phenomenology of a 96~GeV Higgs boson in the 2HDM with an additional
  singlet}}, \href{https://doi.org/10.1103/PhysRevD.106.075003}{\emph{Phys.
  Rev. D} {\bfseries 106} (2022) 075003}
  [\href{https://arxiv.org/abs/2112.11958}{{\ttfamily 2112.11958}}].

\end{thebibliography}\endgroup


\providecommand{\href}[2]{#2}\begingroup\raggedright\begin{thebibliography}{100}

\bibitem{Dawson:2022zbb}
S.~Dawson {\em et al.}, ``{Report of the Topical Group on Higgs Physics for Snowmass 2021: The Case for Precision Higgs Physics},'' in {\em {Snowmass 2021}}.
\newblock 9, 2022.
\newblock \href{http://arxiv.org/abs/2209.07510}{{\tt arXiv:2209.07510 [hep-ph]}}.

\bibitem{ILCInternationalDevelopmentTeam:2022izu}
{\bf ILC International Development Team} Collaboration, I.~Adachi {\em et al.}, ``{The International Linear Collider: Report to Snowmass 2021},'' \href{http://arxiv.org/abs/2203.07622}{{\tt arXiv:2203.07622 [physics.acc-ph]}}.

\bibitem{CLIC:2018fvx}
{\bf CLIC} Collaboration, J.~de~Blas {\em et al.}, ``{The CLIC Potential for New Physics},'' \href{http://arxiv.org/abs/1812.02093}{{\tt arXiv:1812.02093 [hep-ph]}}.

\bibitem{FCC:2018evy}
{\bf FCC} Collaboration, A.~Abada {\em et al.}, ``{FCC-ee: The Lepton Collider}: {Future Circular Collider Conceptual Design Report Volume 2},'' \href{http://dx.doi.org/10.1140/epjst/e2019-900045-4}{{\em Eur. Phys. J. ST} {\bf 228} (2019) no.~2, 261--623}.

\bibitem{FCC:2025lpp}
{\bf FCC} Collaboration, M.~Benedikt {\em et al.}, ``{Future Circular Collider Feasibility Study Report: Volume 1, Physics, Experiments, Detectors},'' \href{http://arxiv.org/abs/2505.00272}{{\tt arXiv:2505.00272 [hep-ex]}}.

\bibitem{FCC:2025uan}
{\bf FCC} Collaboration, M.~Benedikt {\em et al.}, ``{Future Circular Collider Feasibility Study Report: Volume 2, Accelerators, Technical Infrastructure and Safety},'' \href{http://arxiv.org/abs/2505.00274}{{\tt arXiv:2505.00274 [physics.acc-ph]}}.

\bibitem{CEPC-SPPCStudyGroup:2015csa}
M.~Ahmad {\em et al.}, ``{CEPC-SPPC Preliminary Conceptual Design Report. 1. Physics and Detector},''
\newblock 3, 2015.
\newblock \href{http://arxiv.org/abs/IHEP-CEPC-DR-2015-01, IHEP-TH-2015-01, IHEP-EP-2015-01}{{\tt IHEP-CEPC-DR-2015-01, IHEP-TH-2015-01, IHEP-EP-2015-01}}.

\bibitem{CEPC-SPPCStudyGroup:2015esa}
``{CEPC-SPPC Preliminary Conceptual Design Report. 2. Accelerator},''
\newblock 1, 2015.
\newblock \href{http://arxiv.org/abs/IHEP-CEPC-DR-2015-01, IHEP-AC-2015-01}{{\tt IHEP-CEPC-DR-2015-01, IHEP-AC-2015-01}}.

\bibitem{deBlas:2022aow}
J.~de~Blas, J.~Gu, and Z.~Liu, ``{Higgs Precision at a 125 GeV Muon Collider},'' \href{http://arxiv.org/abs/2203.04324}{{\tt arXiv:2203.04324 [hep-ph]}}.

\bibitem{Bai:2021rdg}
M.~Bai {\em et al.}, ``{C$^3$: A ''Cool'' Route to the Higgs Boson and Beyond},'' in {\em {2022 Snowmass Summer Study}}.
\newblock 10, 2021.
\newblock \href{http://arxiv.org/abs/2110.15800}{{\tt arXiv:2110.15800 [hep-ex]}}.

\bibitem{Litvinenko:2022qbd}
V.~N. Litvinenko, N.~Bachhawat, M.~Chamizo-Llatas, Y.~Jing, F.~M\'eot, I.~Petrushina, and T.~Roser, ``{The ReLiC: Recycling Linear $e^+e^-$ Collider},'' \href{http://arxiv.org/abs/2203.06476}{{\tt arXiv:2203.06476 [hep-ex]}}.

\bibitem{Litvinenko:2022mrt}
V.~N. Litvinenko, N.~Bachhawat, M.~Chamizo-Llatas, F.~Meot, and T.~Roser, ``{CERC - Circular $e^+e^-$ Collider using Energy-Recovery Linac},'' in {\em {2022 Snowmass Summer Study}}.
\newblock 3, 2022.
\newblock \href{http://arxiv.org/abs/2203.07358}{{\tt arXiv:2203.07358 [physics.acc-ph]}}.

\bibitem{CEPCStudyGroup:2018ghi}
{\bf CEPC Study Group} Collaboration, M.~Dong {\em et al.}, ``{CEPC Conceptual Design Report: Volume 2 - Physics \& Detector},'' \href{http://arxiv.org/abs/1811.10545}{{\tt arXiv:1811.10545 [hep-ex]}}.

\bibitem{CEPCStudyGroup:2023quu}
{\bf CEPC Study Group} Collaboration, W.~Abdallah {\em et al.}, ``{CEPC Technical Design Report: Accelerator},'' \href{http://dx.doi.org/10.1007/s41605-024-00463-y}{{\em Radiat. Detect. Technol. Methods} {\bf 8} (2024) no.~1, 1--1105}, \href{http://arxiv.org/abs/2312.14363}{{\tt arXiv:2312.14363 [physics.acc-ph]}}.

\bibitem{Tang:2022fzs}
J.~Tang, Y.~Zhang, Q.~Xu, J.~Gao, X.~Lou, and Y.~Wang, ``{Snowmass 2021 White Paper AF4 - SPPC},'' in {\em {2022 Snowmass Summer Study}}.
\newblock 3, 2022.
\newblock \href{http://arxiv.org/abs/2203.07987}{{\tt arXiv:2203.07987 [hep-ex]}}.

\bibitem{Gao:2022lew}
{\bf CEPC Accelerator Study Group} Collaboration, J.~Gao, ``{Snowmass2021 White Paper AF3-CEPC},'' \href{http://arxiv.org/abs/2203.09451}{{\tt arXiv:2203.09451 [physics.acc-ph]}}.

\bibitem{Vcb_Hao_Talk}
H.~Liang, Y.~Zhu, and M.~Ruan, ``{\it Measurement of $V_{cb}$ from $WW\to\mu\nu qq$ at the CEPC}.'' \href{https://indico.ihep.ac.cn/event/19614/contributions/133214/attachments/68533/81829/Measurement%20of%20Vcb%20from%20W%E2%86%92%CE%BC%CE%BDqq%203-24.pdf}{\it talk at cepc flavor physics discussion}, 2023.

\bibitem{Zhao:2018jiq}
H.~Zhao, Y.-F. Zhu, C.-D. Fu, D.~Yu, and M.-Q. Ruan, ``{The Higgs Signatures at the CEPC CDR Baseline},'' \href{http://dx.doi.org/10.1088/1674-1137/43/2/023001}{{\em Chin. Phys. C} {\bf 43} (2019) no.~2, 023001}, \href{http://arxiv.org/abs/1806.04992}{{\tt arXiv:1806.04992 [hep-ex]}}.

\bibitem{Yu:2020bxh}
D.~Yu, M.~Ruan, V.~Boudry, H.~Videau, J.-C. Brient, Z.~Wu, Q.~Ouyang, Y.~Xu, and X.~Chen, ``{The measurement of the $H\rightarrow \tau \tau $ signal strength in the future $e^{+}e^{-}$ Higgs factories},'' \href{http://dx.doi.org/10.1140/epjc/s10052-019-7557-y}{{\em Eur. Phys. J. C} {\bf 80} (2020) no.~1, 7}.

\bibitem{Marshall:2013bda}
J.~S. Marshall and M.~A. Thomson, ``{Pandora Particle Flow Algorithm},'' in {\em {International Conference on Calorimetry for the High Energy Frontier}}, pp.~305--315.
\newblock 2013.
\newblock \href{http://arxiv.org/abs/1308.4537}{{\tt arXiv:1308.4537 [physics.ins-det]}}.

\bibitem{Ruan:2013rkk}
M.~Ruan and H.~Videau, ``{Arbor, a new approach of the Particle Flow Algorithm},'' in {\em {International Conference on Calorimetry for the High Energy Frontier}}, pp.~316--324.
\newblock 2013.
\newblock \href{http://arxiv.org/abs/1403.4784}{{\tt arXiv:1403.4784 [physics.ins-det]}}.

\bibitem{refTDR_Mingshui_talk}
M.~Chen, ``{\it Detector Baseline Concept and Performance}.'' \href{https://indico.ihep.ac.cn/event/25539/contributions/183751/attachments/89614/116133/20250414-v2.pdf}{\it talk at cepc international detector review committee meeting}, 2025.

\bibitem{IDEA_Snowmass_2021}
F.~Bedeschi, M.~Caccia, R.~Ferrari, P.~Giacomelli, F.~Grancagnolo, P.~Azzi, and S.~Braibant, ``{IDEA detector Letter of Intent}.'' \url{https://www.snowmass21.org/docs/files/summaries/EF/SNOWMASS21-EF1_EF4-IF3_IF6-096.pdf}, 2021.

\bibitem{Zhu:2022hyy}
Y.~Zhu, S.~Chen, H.~Cui, and M.~Ruan, ``{Requirement analysis for dE/dx measurement and PID performance at the CEPC baseline detector},'' \href{http://dx.doi.org/10.1016/j.nima.2022.167835}{{\em Nucl. Instrum. Meth. A} {\bf 1047} (2023)  167835}, \href{http://arxiv.org/abs/2209.14486}{{\tt arXiv:2209.14486 [physics.ins-det]}}.

\bibitem{refTDR_Jianchun_talk}
J.~Wang, ``{\it TDR of A Reference CEPC Detector}.'' \href{https://indico.ihep.ac.cn/event/25539/contributions/183763/attachments/89610/116147/250414_WJC.pdf}{\it talk at cepc international detector review committee meeting}, 2025.

\bibitem{Hu:2023dbm}
P.~Hu, Y.~Wang, D.~Du, Z.~Hua, S.~Qian, C.~Fu, Y.~Liu, M.~Ruan, J.~Wang, and Y.~Wang, ``{GSHCAL at future e+e\ensuremath{-} Higgs factories},'' \href{http://dx.doi.org/10.1016/j.nima.2023.168944}{{\em Nucl. Instrum. Meth. A} {\bf 1059} (2024)  168944}.

\bibitem{She:2023puo}
X.~She {\em et al.}, ``{Development of Time Projection Chamber prototype integrated with UV laser tracks for the future circular e$^{+}$e$^{-}$ collider},'' \href{http://dx.doi.org/10.1088/1748-0221/18/07/C07015}{{\em JINST} {\bf 18} (2023) no.~07, C07015}.

\bibitem{Cuna:2021sho}
F.~Cuna, N.~De~Filippis, F.~Grancagnolo, and G.~F. Tassielli, ``{Simulation of particle identification with the cluster counting technique},'' in {\em {International Workshop on Future Linear Colliders}}.
\newblock 5, 2021.
\newblock \href{http://arxiv.org/abs/2105.07064}{{\tt arXiv:2105.07064 [physics.ins-det]}}.

\bibitem{4th_vtx_talk}
Z.~Liang and Y.~Zhang, ``{\it CEPC vertex Detector}.'' \href{https://indico.ihep.ac.cn/event/22089/contributions/168824/attachments/83396/106033/CEPC-vertex_20241024_upload.pdf}{\it talk at 2024 international workshop of cepc}, 2024.

\bibitem{Wang:2024eji}
Y.~Wang, H.~Liang, Y.~Zhu, Y.~Che, X.~Xia, H.~Qu, C.~Zhou, X.~Zhuang, and M.~Ruan, ``{One-to-one correspondence reconstruction at the electron-positron Higgs factory},'' \href{http://dx.doi.org/10.1016/j.cpc.2025.109661}{{\em Comput. Phys. Commun.} {\bf 314} (2025)  109661}, \href{http://arxiv.org/abs/2411.06939}{{\tt arXiv:2411.06939 [hep-ex]}}.

\bibitem{Liang:2023wpt}
H.~Liang, Y.~Zhu, Y.~Wang, Y.~Che, C.~Zhou, H.~Qu, and M.~Ruan, ``{Jet-Origin Identification and Its Application at an Electron-Positron Higgs Factory},'' \href{http://dx.doi.org/10.1103/PhysRevLett.132.221802}{{\em Phys. Rev. Lett.} {\bf 132} (2024) no.~22, 221802}, \href{http://arxiv.org/abs/2310.03440}{{\tt arXiv:2310.03440 [hep-ex]}}.

\bibitem{Qu:2019gqs}
H.~Qu and L.~Gouskos, ``{ParticleNet: Jet Tagging via Particle Clouds},'' \href{http://dx.doi.org/10.1103/PhysRevD.101.056019}{{\em Phys. Rev. D} {\bf 101} (2020) no.~5, 056019}, \href{http://arxiv.org/abs/1902.08570}{{\tt arXiv:1902.08570 [hep-ph]}}.

\bibitem{Li_2022}
G.~Li, L.~Liao, X.~Lou, P.~Shen, W.~Song, S.~Wang, and Z.~Zhang, ``Classify the higgs decays with the pfn and particlenet at electron–positron colliders*,'' \href{http://dx.doi.org/10.1088/1674-1137/ac7f21}{{\em Chinese Physics C} {\bf 46} (2022) no.~11, 113001}. \url{https://dx.doi.org/10.1088/1674-1137/ac7f21}.

\bibitem{Liu:2016zki}
Z.~Liu, L.-T. Wang, and H.~Zhang, ``{Exotic decays of the 125 GeV Higgs boson at future $e^+e^-$ lepton colliders},'' \href{http://dx.doi.org/10.1088/1674-1137/41/6/063102}{{\em Chin. Phys. C} {\bf 41} (2017) no.~6, 063102}, \href{http://arxiv.org/abs/1612.09284}{{\tt arXiv:1612.09284 [hep-ph]}}.

\bibitem{Curtin:2013fra}
D.~Curtin {\em et al.}, ``{Exotic decays of the 125 GeV Higgs boson},'' \href{http://dx.doi.org/10.1103/PhysRevD.90.075004}{{\em Phys. Rev. D} {\bf 90} (2014) no.~7, 075004}, \href{http://arxiv.org/abs/1312.4992}{{\tt arXiv:1312.4992 [hep-ph]}}.

\bibitem{deFlorian:2016spz}
{\bf LHC Higgs Cross Section Working Group} Collaboration, D.~de~Florian {\em et al.}, ``{Handbook of LHC Higgs Cross Sections: 4. Deciphering the Nature of the Higgs Sector},''
\href{http://arxiv.org/abs/1610.07922}{{\tt arXiv:1610.07922 [hep-ph]}}.

\bibitem{Cepeda:2021rql}
M.~Cepeda, S.~Gori, V.~M. Outschoorn, and J.~Shelton, ``{Exotic Higgs Decays},'' \href{http://arxiv.org/abs/2111.12751}{{\tt arXiv:2111.12751 [hep-ph]}}.

\bibitem{Shelton:2021xwo}
J.~Shelton and D.~Xu, ``{Exotic Higgs Decays to Four Taus at Future Electron-Positron Colliders},''
\newblock 10, 2021.
\newblock \href{http://arxiv.org/abs/2110.13225}{{\tt arXiv:2110.13225 [hep-ph]}}.

\bibitem{Kato:2022}
Y.~Kato, ``Search for higgs decaying to exotic scalers at the ilc.'' Presentation at the 2021 international workshop on the high energy circular electron-positron collider (cepc), nov. 8-12, 2021, \url{https://indico.ihep.ac.cn/event/14938}.

\bibitem{CMS:2019rsy}
{\bf CMS} Collaboration, ``{Projection of searches for exotic Higgs boson decays to light pseudoscalars for the High-Luminosity LHC},''. \url{https://cds.cern.ch/record/2655340}.

\bibitem{Wang:2020tap}
K.~Wang and J.~Zhu, ``{A Novel Scenario in the Semi-constrained NMSSM},'' \href{http://dx.doi.org/10.1007/JHEP06(2020)078}{{\em JHEP} {\bf 06} (2020)  078}, \href{http://arxiv.org/abs/2002.05554}{{\tt arXiv:2002.05554 [hep-ph]}}.

\bibitem{Wang:2020dtb}
K.~Wang and J.~Zhu, ``{Funnel annihilations of light dark matter and the invisible decay of the Higgs boson},'' \href{http://dx.doi.org/10.1103/PhysRevD.101.095028}{{\em Phys. Rev. D} {\bf 101} (2020) no.~9, 095028}, \href{http://arxiv.org/abs/2003.01662}{{\tt arXiv:2003.01662 [hep-ph]}}.

\bibitem{Wang:2020xta}
K.~Wang, J.~Zhu, and Q.~Jie, ``{Higgsino Asymmetry and Direct-Detection Constraints of Light Dark Matter in the NMSSM with Non-Universal Higgs Masses},'' \href{http://dx.doi.org/10.1088/1674-1137/abe03c}{{\em Chin. Phys. C} {\bf 45} (2021) no.~4, 041003}, \href{http://arxiv.org/abs/2011.12848}{{\tt arXiv:2011.12848 [hep-ph]}}.

\bibitem{Ma:2020mjz}
S.~Ma, K.~Wang, and J.~Zhu, ``{Higgs decay to light (pseudo)scalars in the semi-constrained NMSSM},'' \href{http://dx.doi.org/10.1088/1674-1137/abce4f}{{\em Chin. Phys. C} {\bf 45} (2021) no.~2, 023113}, \href{http://arxiv.org/abs/2006.03527}{{\tt arXiv:2006.03527 [hep-ph]}}.

\bibitem{Bai:2013iqa}
Y.~Bai and J.~Berger, ``{Fermion Portal Dark Matter},'' \href{http://dx.doi.org/10.1007/JHEP11(2013)171}{{\em JHEP} {\bf 11} (2013)  171}, \href{http://arxiv.org/abs/1308.0612}{{\tt arXiv:1308.0612 [hep-ph]}}.

\bibitem{Liu:2021mhn}
J.~Liu, X.-P. Wang, and K.-P. Xie, ``{Searching for lepton portal dark matter with colliders and gravitational waves},'' \href{http://dx.doi.org/10.1007/JHEP06(2021)149}{{\em JHEP} {\bf 06} (2021)  149}, \href{http://arxiv.org/abs/2104.06421}{{\tt arXiv:2104.06421 [hep-ph]}}.

\bibitem{Chacko:2020zze}
Z.~Chacko, P.~J. Fox, R.~Harnik, and Z.~Liu, ``{Neutrino Masses from Low Scale Partial Compositeness},'' \href{http://arxiv.org/abs/2012.01443}{{\tt arXiv:2012.01443 [hep-ph]}}.

\bibitem{Craig:2015pha}
N.~Craig, A.~Katz, M.~Strassler, and R.~Sundrum, ``{Naturalness in the Dark at the LHC},'' \href{http://dx.doi.org/10.1007/JHEP07(2015)105}{{\em JHEP} {\bf 07} (2015)  105}, \href{http://arxiv.org/abs/1501.05310}{{\tt arXiv:1501.05310 [hep-ph]}}.

\bibitem{Curtin:2015fna}
D.~Curtin and C.~B. Verhaaren, ``{Discovering Uncolored Naturalness in Exotic Higgs Decays},'' \href{http://dx.doi.org/10.1007/JHEP12(2015)072}{{\em JHEP} {\bf 12} (2015)  072}, \href{http://arxiv.org/abs/1506.06141}{{\tt arXiv:1506.06141 [hep-ph]}}.

\bibitem{Liu:2018wte}
J.~Liu, Z.~Liu, and L.-T. Wang, ``{Enhancing Long-Lived Particles Searches at the LHC with Precision Timing Information},'' \href{http://dx.doi.org/10.1103/PhysRevLett.122.131801}{{\em Phys. Rev. Lett.} {\bf 122} (2019) no.~13, 131801}, \href{http://arxiv.org/abs/1805.05957}{{\tt arXiv:1805.05957 [hep-ph]}}.

\bibitem{Alipour-fard:2018mre}
S.~Alipour-Fard, N.~Craig, S.~Gori, S.~Koren, and D.~Redigolo, ``{The second Higgs at the lifetime frontier},'' \href{http://dx.doi.org/10.1007/JHEP07(2020)029}{{\em JHEP} {\bf 07} (2020)  029}, \href{http://arxiv.org/abs/1812.09315}{{\tt arXiv:1812.09315 [hep-ph]}}.

\bibitem{Liu:2020vur}
J.~Liu, Z.~Liu, L.-T. Wang, and X.-P. Wang, ``{Enhancing Sensitivities to Long-lived Particles with High Granularity Calorimeters at the LHC},'' \href{http://dx.doi.org/10.1007/JHEP11(2020)066}{{\em JHEP} {\bf 11} (2020)  066}, \href{http://arxiv.org/abs/2005.10836}{{\tt arXiv:2005.10836 [hep-ph]}}.

\bibitem{Carena:2022yvx}
M.~Carena, J.~Kozaczuk, Z.~Liu, T.~Ou, M.~J. Ramsey-Musolf, J.~Shelton, Y.~Wang, and K.-P. Xie, ``{Probing the Electroweak Phase Transition with Exotic Higgs Decays},'' in {\em {2022 Snowmass Summer Study}}.
\newblock 3, 2022.
\newblock \href{http://arxiv.org/abs/2203.08206}{{\tt arXiv:2203.08206 [hep-ph]}}.

\bibitem{Wang:2022dkz}
Z.~Wang, X.~Zhu, E.~E. Khoda, S.-C. Hsu, N.~Konstantinidis, K.~Li, S.~Li, M.~J. Ramsey-Musolf, Y.~Wu, and Y.~E. Zhang, ``{Study of Electroweak Phase Transition in Exotic Higgs Decays at the CEPC},'' in {\em {2022 Snowmass Summer Study}}.
\newblock 3, 2022.
\newblock \href{http://arxiv.org/abs/2203.10184}{{\tt arXiv:2203.10184 [hep-ex]}}.

\bibitem{Jung:2021tym}
S.~Jung, Z.~Liu, L.-T. Wang, and K.-P. Xie, ``{Probing Higgs exotic decay at the LHC with machine learning},'' \href{http://arxiv.org/abs/2109.03294}{{\tt arXiv:2109.03294 [hep-ph]}}.

\bibitem{Arkani-Hamed:2016rle}
N.~Arkani-Hamed, T.~Cohen, R.~T. D'Agnolo, A.~Hook, H.~D. Kim, and D.~Pinner, ``{Solving the Hierarchy Problem at Reheating with a Large Number of Degrees of Freedom},'' \href{http://dx.doi.org/10.1103/PhysRevLett.117.251801}{{\em Phys. Rev. Lett.} {\bf 117} (2016) no.~25, 251801}, \href{http://arxiv.org/abs/1607.06821}{{\tt arXiv:1607.06821 [hep-ph]}}.

\bibitem{Arkani-Hamed:2020yna}
N.~Arkani-Hamed, R.~Tito~D'agnolo, and H.~D. Kim, ``{The Weak Scale as a Trigger},'' \href{http://arxiv.org/abs/2012.04652}{{\tt arXiv:2012.04652 [hep-ph]}}.

\bibitem{Meade:2018saz}
P.~Meade and H.~Ramani, ``{Unrestored Electroweak Symmetry},'' \href{http://dx.doi.org/10.1103/PhysRevLett.122.041802}{{\em Phys. Rev. Lett.} {\bf 122} (2019) no.~4, 041802}, \href{http://arxiv.org/abs/1807.07578}{{\tt arXiv:1807.07578 [hep-ph]}}.

\bibitem{Baldes:2018nel}
I.~Baldes and G.~Servant, ``{High scale electroweak phase transition: baryogenesis and symmetry non-restoration},'' \href{http://dx.doi.org/10.1007/JHEP10(2018)053}{{\em JHEP} {\bf 10} (2018)  053}, \href{http://arxiv.org/abs/1807.08770}{{\tt arXiv:1807.08770 [hep-ph]}}.

\bibitem{Glioti:2018roy}
A.~Glioti, R.~Rattazzi, and L.~Vecchi, ``{Electroweak Baryogenesis above the Electroweak Scale},'' \href{http://dx.doi.org/10.1007/JHEP04(2019)027}{{\em JHEP} {\bf 04} (2019)  027}, \href{http://arxiv.org/abs/1811.11740}{{\tt arXiv:1811.11740 [hep-ph]}}.

\bibitem{Matsedonskyi:2020mlz}
O.~Matsedonskyi and G.~Servant, ``{High-Temperature Electroweak Symmetry Non-Restoration from New Fermions and Implications for Baryogenesis},'' \href{http://dx.doi.org/10.1007/JHEP09(2020)012}{{\em JHEP} {\bf 09} (2020)  012}, \href{http://arxiv.org/abs/2002.05174}{{\tt arXiv:2002.05174 [hep-ph]}}.

\bibitem{Liu:2017zdh}
J.~Liu, L.-T. Wang, X.-P. Wang, and W.~Xue, ``{Exposing the dark sector with future Z factories},'' \href{http://dx.doi.org/10.1103/PhysRevD.97.095044}{{\em Phys. Rev.} {\bf D97} (2018) no.~9, 095044},
\href{http://arxiv.org/abs/1712.07237}{{\tt arXiv:1712.07237 [hep-ph]}}.

\bibitem{Abada:2019lih}
{\bf FCC} Collaboration, A.~Abada {\em et al.}, ``{FCC Physics Opportunities}: {Future Circular Collider Conceptual Design Report Volume 1},'' \href{http://dx.doi.org/10.1140/epjc/s10052-019-6904-3}{{\em Eur. Phys. J. C} {\bf 79} (2019) no.~6, 474}.

\bibitem{Drees:1996pk}
M.~Drees, M.~M. Nojiri, D.~P. Roy, and Y.~Yamada, ``{Light Higgsino dark matter},'' \href{http://dx.doi.org/10.1103/PhysRevD.64.039901}{{\em Phys. Rev. D} {\bf 56} (1997)  276--290}, \href{http://arxiv.org/abs/hep-ph/9701219}{{\tt arXiv:hep-ph/9701219}}. [Erratum: Phys.Rev.D 64, 039901 (2001)].

\bibitem{Djouadi:2001kba}
A.~Djouadi, M.~Drees, P.~Fileviez~Perez, and M.~Muhlleitner, ``{Loop induced Higgs and Z boson couplings to neutralinos and implications for collider and dark matter searches},'' \href{http://dx.doi.org/10.1103/PhysRevD.65.075016}{{\em Phys. Rev. D} {\bf 65} (2002)  075016}, \href{http://arxiv.org/abs/hep-ph/0109283}{{\tt arXiv:hep-ph/0109283}}.

\bibitem{Eberl:2001vb}
H.~Eberl, M.~Kincel, W.~Majerotto, and Y.~Yamada, ``{One loop corrections to neutral Higgs boson decays into neutralinos},'' \href{http://dx.doi.org/10.1016/S0550-3213(02)00008-1}{{\em Nucl. Phys. B} {\bf 625} (2002)  372--388}, \href{http://arxiv.org/abs/hep-ph/0111303}{{\tt arXiv:hep-ph/0111303}}.

\bibitem{Berlin:2015njh}
A.~Berlin, D.~S. Robertson, M.~P. Solon, and K.~M. Zurek, ``{Bino variations: Effective field theory methods for dark matter direct detection},'' \href{http://dx.doi.org/10.1103/PhysRevD.93.095008}{{\em Phys. Rev. D} {\bf 93} (2016) no.~9, 095008}, \href{http://arxiv.org/abs/1511.05964}{{\tt arXiv:1511.05964 [hep-ph]}}.

\bibitem{ATLAS:2020cjb}
{\bf ATLAS} Collaboration, ``{Search for invisible Higgs boson decays with vector boson fusion signatures with the ATLAS detector using an integrated luminosity of 139 fb$^{-1}$},''. \url{https://cds.cern.ch/record/2715446}.

\bibitem{Bernaciak:2014pna}
C.~Bernaciak, T.~Plehn, P.~Schichtel, and J.~Tattersall, ``{Spying an invisible Higgs boson},'' \href{http://dx.doi.org/10.1103/PhysRevD.91.035024}{{\em Phys. Rev.} {\bf D91} (2015)  035024},
\href{http://arxiv.org/abs/1411.7699}{{\tt arXiv:1411.7699 [hep-ph]}}.

\bibitem{Zhang:2024bld}
Y.~Zhang, C.~Mo, X.~Chen, B.~Li, H.~Chen, J.~Hu, and L.~Li, ``{Search for Long-lived Particles at Future Lepton Colliders Using Deep Learning Techniques},'' \href{http://arxiv.org/abs/2401.05094}{{\tt arXiv:2401.05094 [hep-ex]}}.

\bibitem{Alipour-Fard:2018lsf}
S.~Alipour-Fard, N.~Craig, M.~Jiang, and S.~Koren, ``{Long Live the Higgs Factory: Higgs Decays to Long-Lived Particles at Future Lepton Colliders},'' \href{http://dx.doi.org/10.1088/1674-1137/43/5/053101}{{\em Chin. Phys. C} {\bf 43} (2019) no.~5, 053101}, \href{http://arxiv.org/abs/1812.05588}{{\tt arXiv:1812.05588 [hep-ph]}}.

\bibitem{Cheung:2019qdr}
K.~Cheung and Z.~S. Wang, ``{Probing Long-lived Particles at Higgs Factories},'' \href{http://dx.doi.org/10.1103/PhysRevD.101.035003}{{\em Phys. Rev. D} {\bf 101} (2020) no.~3, 035003}, \href{http://arxiv.org/abs/1911.08721}{{\tt arXiv:1911.08721 [hep-ph]}}.

\bibitem{Jeanty:2022cwr}
L.~Jeanty, L.~Nosler, and C.~Potter, ``{Sensitivity to decays of long-lived dark photons at the ILC},'' \href{http://arxiv.org/abs/2203.08347}{{\tt arXiv:2203.08347 [hep-ph]}}.

\bibitem{Wang:2019orr}
Z.~S. Wang and K.~Wang, ``{Long-lived light neutralinos at future $Z-$factories},'' \href{http://dx.doi.org/10.1103/PhysRevD.101.115018}{{\em Phys. Rev. D} {\bf 101} (2020) no.~11, 115018}, \href{http://arxiv.org/abs/1904.10661}{{\tt arXiv:1904.10661 [hep-ph]}}.

\bibitem{Calibbi:2022izs}
L.~Calibbi, Z.~Huang, S.~Qin, Y.~Yang, and X.~Yin, ``{Testing axion couplings to leptons in Z decays at future e+e- colliders},'' \href{http://dx.doi.org/10.1103/PhysRevD.108.015002}{{\em Phys. Rev. D} {\bf 108} (2023) no.~1, 015002}, \href{http://arxiv.org/abs/2212.02818}{{\tt arXiv:2212.02818 [hep-ph]}}.

\bibitem{Wang:2019xvx}
Z.~S. Wang and K.~Wang, ``{Physics with far detectors at future lepton colliders},'' \href{http://dx.doi.org/10.1103/PhysRevD.101.075046}{{\em Phys. Rev. D} {\bf 101} (2020) no.~7, 075046}, \href{http://arxiv.org/abs/1911.06576}{{\tt arXiv:1911.06576 [hep-ph]}}.

\bibitem{Cheng:2019yai}
H.-C. Cheng, L.~Li, E.~Salvioni, and C.~B. Verhaaren, ``{Light Hidden Mesons through the Z Portal},'' \href{http://dx.doi.org/10.1007/JHEP11(2019)031}{{\em JHEP} {\bf 11} (2019)  031}, \href{http://arxiv.org/abs/1906.02198}{{\tt arXiv:1906.02198 [hep-ph]}}.

\bibitem{Cheng:2021kjg}
H.-C. Cheng, L.~Li, and E.~Salvioni, ``{A theory of dark pions},'' \href{http://dx.doi.org/10.1007/JHEP01(2022)122}{{\em JHEP} {\bf 01} (2022)  122}, \href{http://arxiv.org/abs/2110.10691}{{\tt arXiv:2110.10691 [hep-ph]}}.

\bibitem{CMS:2024yhz}
{\bf CMS} Collaboration, A.~Hayrapetyan {\em et al.}, ``{Search for a standard model-like Higgs boson in the mass range between 70 and 110 GeV in the diphoton final state in proton-proton collisions at s=13TeV},'' \href{http://dx.doi.org/10.1016/j.physletb.2024.139067}{{\em Phys. Lett. B} {\bf 860} (2025)  139067}, \href{http://arxiv.org/abs/2405.18149}{{\tt arXiv:2405.18149 [hep-ex]}}.

\bibitem{ATLAS:2023jzc}
{\bf ATLAS} Collaboration, ``{Search for diphoton resonances in the 66 to 110 GeV mass range using 140 fb$^{-1}$ of 13 TeV $pp$ collisions collected with the ATLAS detector},''. \url{https://cds.cern.ch/record/2862024}.

\bibitem{Biekotter:2023oen}
T.~Biek\"otter, S.~Heinemeyer, and G.~Weiglein, ``{95.4~GeV diphoton excess at ATLAS and CMS},'' \href{http://dx.doi.org/10.1103/PhysRevD.109.035005}{{\em Phys. Rev. D} {\bf 109} (2024) no.~3, 035005}, \href{http://arxiv.org/abs/2306.03889}{{\tt arXiv:2306.03889 [hep-ph]}}.

\bibitem{Barate:2003sz}
{\bf LEP Working Group for Higgs boson searches, ALEPH, DELPHI, L3, OPAL} Collaboration, R.~Barate {\em et al.}, ``{Search for the standard model Higgs boson at LEP},'' \href{http://dx.doi.org/10.1016/S0370-2693(03)00614-2}{{\em Phys. Lett. B} {\bf 565} (2003)  61--75}, \href{http://arxiv.org/abs/hep-ex/0306033}{{\tt arXiv:hep-ex/0306033}}.

\bibitem{Cao:2016uwt}
J.~Cao, X.~Guo, Y.~He, P.~Wu, and Y.~Zhang, ``{Diphoton signal of the light Higgs boson in natural NMSSM},'' \href{http://dx.doi.org/10.1103/PhysRevD.95.116001}{{\em Phys. Rev. D} {\bf 95} (2017) no.~11, 116001}, \href{http://arxiv.org/abs/1612.08522}{{\tt arXiv:1612.08522 [hep-ph]}}.

\bibitem{Azatov:2012bz}
A.~Azatov, R.~Contino, and J.~Galloway, ``{Model-Independent Bounds on a Light Higgs},'' \href{http://dx.doi.org/10.1007/JHEP04(2012)127}{{\em JHEP} {\bf 04} (2012)  127}, \href{http://arxiv.org/abs/1202.3415}{{\tt arXiv:1202.3415 [hep-ph]}}. [Erratum: JHEP 04, 140 (2013)].

\bibitem{Biekotter:2023jld}
T.~Biek\"otter, S.~Heinemeyer, and G.~Weiglein, ``{The CMS di-photon excess at 95 GeV in view of the LHC Run 2 results},'' \href{http://dx.doi.org/10.1016/j.physletb.2023.138217}{{\em Phys. Lett. B} {\bf 846} (2023)  138217}, \href{http://arxiv.org/abs/2303.12018}{{\tt arXiv:2303.12018 [hep-ph]}}.

\bibitem{Biekotter:2021ovi}
T.~Biek\"otter and M.~O. Olea-Romacho, ``{Reconciling Higgs physics and pseudo-Nambu-Goldstone dark matter in the S2HDM using a genetic algorithm},'' \href{http://dx.doi.org/10.1007/JHEP10(2021)215}{{\em JHEP} {\bf 10} (2021)  215}, \href{http://arxiv.org/abs/2108.10864}{{\tt arXiv:2108.10864 [hep-ph]}}.

\bibitem{Drechsel:2018mgd}
P.~Drechsel, G.~Moortgat-Pick, and G.~Weiglein, ``{Prospects for direct searches for light Higgs bosons at the ILC with 250 GeV},'' \href{http://dx.doi.org/10.1140/epjc/s10052-020-08438-1}{{\em Eur. Phys. J. C} {\bf 80} (2020) no.~10, 922}, \href{http://arxiv.org/abs/1801.09662}{{\tt arXiv:1801.09662 [hep-ph]}}.

\bibitem{Cepeda:2019klc}
M.~Cepeda {\em et al.}, {\em {Report from Working Group 2}: {Higgs Physics at the HL-LHC and HE-LHC}}, vol.~7, \href{http://dx.doi.org/10.23731/CYRM-2019-007.221}{pp.~221--584}.
\newblock 12, 2019.
\newblock \href{http://arxiv.org/abs/1902.00134}{{\tt arXiv:1902.00134 [hep-ph]}}.

\bibitem{Bambade:2019fyw}
P.~Bambade {\em et al.}, ``{The International Linear Collider: A Global Project},'' \href{http://arxiv.org/abs/1903.01629}{{\tt arXiv:1903.01629 [hep-ex]}}.

\bibitem{Heinemeyer:2021msz}
S.~Heinemeyer, C.~Li, F.~Lika, G.~Moortgat-Pick, and S.~Paasch, ``{Phenomenology of a 96~GeV Higgs boson in the 2HDM with an additional singlet},'' \href{http://dx.doi.org/10.1103/PhysRevD.106.075003}{{\em Phys. Rev. D} {\bf 106} (2022) no.~7, 075003}, \href{http://arxiv.org/abs/2112.11958}{{\tt arXiv:2112.11958 [hep-ph]}}.

\bibitem{Kong:2014jwa}
K.~Kong, H.-S. Lee, and M.~Park, ``{Dark decay of the top quark},'' \href{http://dx.doi.org/10.1103/PhysRevD.89.074007}{{\em Phys. Rev. D} {\bf 89} (2014) no.~7, 074007}, \href{http://arxiv.org/abs/1401.5020}{{\tt arXiv:1401.5020 [hep-ph]}}.

\bibitem{Li:2022iav}
Z.~Li, X.~Sun, Y.~Fang, G.~Li, S.~Xin, S.~Wang, Y.~Wang, Y.~Zhang, H.~Zhang, and Z.~Liang, ``{Top quark mass measurements at the $t\bar{t}$ threshold with CEPC},'' \href{http://dx.doi.org/10.1140/epjc/s10052-023-11421-1}{{\em Eur. Phys. J. C} {\bf 83} (2023) no.~4, 269}, \href{http://arxiv.org/abs/2207.12177}{{\tt arXiv:2207.12177 [hep-ex]}}. [Erratum: Eur.Phys.J.C 83, 501 (2023)].

\bibitem{Dolan:1973qd}
L.~Dolan and R.~Jackiw, ``{Symmetry Behavior at Finite Temperature},'' \href{http://dx.doi.org/10.1103/PhysRevD.9.3320}{{\em Phys. Rev. D} {\bf 9} (1974)  3320--3341}.

\bibitem{Kajantie:1996qd}
K.~Kajantie, M.~Laine, K.~Rummukainen, and M.~E. Shaposhnikov, ``{A Nonperturbative analysis of the finite T phase transition in SU(2) x U(1) electroweak theory},'' \href{http://dx.doi.org/10.1016/S0550-3213(97)00164-8}{{\em Nucl. Phys. B} {\bf 493} (1997)  413--438}, \href{http://arxiv.org/abs/hep-lat/9612006}{{\tt arXiv:hep-lat/9612006}}.

\bibitem{Rummukainen:1998as}
K.~Rummukainen, M.~Tsypin, K.~Kajantie, M.~Laine, and M.~E. Shaposhnikov, ``{The Universality class of the electroweak theory},'' \href{http://dx.doi.org/10.1016/S0550-3213(98)00494-5}{{\em Nucl. Phys. B} {\bf 532} (1998)  283--314}, \href{http://arxiv.org/abs/hep-lat/9805013}{{\tt arXiv:hep-lat/9805013}}.

\bibitem{Laine:1998jb}
M.~Laine and K.~Rummukainen, ``{What's new with the electroweak phase transition?},'' \href{http://dx.doi.org/10.1016/S0920-5632(99)85017-8}{{\em Nucl. Phys. B Proc. Suppl.} {\bf 73} (1999)  180--185}, \href{http://arxiv.org/abs/hep-lat/9809045}{{\tt arXiv:hep-lat/9809045}}.

\bibitem{Dine:1992wr}
M.~Dine, R.~G. Leigh, P.~Y. Huet, A.~D. Linde, and D.~A. Linde, ``{Towards the theory of the electroweak phase transition},'' \href{http://dx.doi.org/10.1103/PhysRevD.46.550}{{\em Phys. Rev. D} {\bf 46} (1992)  550--571}, \href{http://arxiv.org/abs/hep-ph/9203203}{{\tt arXiv:hep-ph/9203203}}.

\bibitem{Morrissey:2012db}
D.~E. Morrissey and M.~J. Ramsey-Musolf, ``{Electroweak baryogenesis},'' \href{http://dx.doi.org/10.1088/1367-2630/14/12/125003}{{\em New J. Phys.} {\bf 14} (2012)  125003},
\href{http://arxiv.org/abs/1206.2942}{{\tt arXiv:1206.2942 [hep-ph]}}.

\bibitem{Joyce:1994fu}
M.~Joyce, T.~Prokopec, and N.~Turok, ``{Electroweak baryogenesis from a classical force},'' \href{http://dx.doi.org/10.1103/PhysRevLett.75.1695}{{\em Phys. Rev. Lett.} {\bf 75} (1995)  1695--1698}, \href{http://arxiv.org/abs/hep-ph/9408339}{{\tt arXiv:hep-ph/9408339}}. [Erratum: Phys.Rev.Lett. 75, 3375 (1995)].

\bibitem{Fromme:2006wx}
L.~Fromme and S.~J. Huber, ``{Top transport in electroweak baryogenesis},'' \href{http://dx.doi.org/10.1088/1126-6708/2007/03/049}{{\em JHEP} {\bf 03} (2007)  049}, \href{http://arxiv.org/abs/hep-ph/0604159}{{\tt arXiv:hep-ph/0604159}}.

\bibitem{Jiang:2015cwa}
M.~Jiang, L.~Bian, W.~Huang, and J.~Shu, ``{Impact of a complex singlet: Electroweak baryogenesis and dark matter},'' \href{http://dx.doi.org/10.1103/PhysRevD.93.065032}{{\em Phys. Rev. D} {\bf 93} (2016) no.~6, 065032}, \href{http://arxiv.org/abs/1502.07574}{{\tt arXiv:1502.07574 [hep-ph]}}.

\bibitem{Chiang:2017nmu}
C.-W. Chiang, M.~J. Ramsey-Musolf, and E.~Senaha, ``{Standard Model with a Complex Scalar Singlet: Cosmological Implications and Theoretical Considerations},'' \href{http://dx.doi.org/10.1103/PhysRevD.97.015005}{{\em Phys. Rev. D} {\bf 97} (2018) no.~1, 015005}, \href{http://arxiv.org/abs/1707.09960}{{\tt arXiv:1707.09960 [hep-ph]}}.

\bibitem{Xie:2020wzn}
K.-P. Xie, ``{Lepton-mediated electroweak baryogenesis, gravitational waves and the $4\tau$ final state at the collider},'' \href{http://dx.doi.org/10.1007/JHEP02(2021)090}{{\em JHEP} {\bf 02} (2021)  090}, \href{http://arxiv.org/abs/2011.04821}{{\tt arXiv:2011.04821 [hep-ph]}}. [Erratum: JHEP 8, 052 (2022)].

\bibitem{Bian:2019kmg}
L.~Bian, Y.~Wu, and K.-P. Xie, ``{Electroweak phase transition with composite Higgs models: calculability, gravitational waves and collider searches},'' \href{http://dx.doi.org/10.1007/JHEP12(2019)028}{{\em JHEP} {\bf 12} (2019)  028}, \href{http://arxiv.org/abs/1909.02014}{{\tt arXiv:1909.02014 [hep-ph]}}.

\bibitem{Xie:2020bkl}
K.-P. Xie, L.~Bian, and Y.~Wu, ``{Electroweak baryogenesis and gravitational waves in a composite Higgs model with high dimensional fermion representations},'' \href{http://dx.doi.org/10.1007/JHEP12(2020)047}{{\em JHEP} {\bf 12} (2020)  047}, \href{http://arxiv.org/abs/2005.13552}{{\tt arXiv:2005.13552 [hep-ph]}}.

\bibitem{Bell:2020gug}
N.~F. Bell, M.~J. Dolan, L.~S. Friedrich, M.~J. Ramsey-Musolf, and R.~R. Volkas, ``{Two-Step Electroweak Symmetry-Breaking: Theory Meets Experiment},'' \href{http://dx.doi.org/10.1007/JHEP05(2020)050}{{\em JHEP} {\bf 05} (2020)  050}, \href{http://arxiv.org/abs/2001.05335}{{\tt arXiv:2001.05335 [hep-ph]}}.

\bibitem{Bian:2017wfv}
L.~Bian, H.-K. Guo, and J.~Shu, ``{Gravitational Waves, baryon asymmetry of the universe and electric dipole moment in the CP-violating NMSSM},'' \href{http://dx.doi.org/10.1088/1674-1137/42/9/093106}{{\em Chin. Phys. C} {\bf 42} (2018) no.~9, 093106}, \href{http://arxiv.org/abs/1704.02488}{{\tt arXiv:1704.02488 [hep-ph]}}. [Erratum: Chin.Phys.C 43, 129101 (2019)].

\bibitem{Huang:2018aja}
F.~P. Huang, Z.~Qian, and M.~Zhang, ``{Exploring dynamical CP violation induced baryogenesis by gravitational waves and colliders},'' \href{http://dx.doi.org/10.1103/PhysRevD.98.015014}{{\em Phys. Rev. D} {\bf 98} (2018) no.~1, 015014}, \href{http://arxiv.org/abs/1804.06813}{{\tt arXiv:1804.06813 [hep-ph]}}.

\bibitem{Kanemura:2023juv}
S.~Kanemura and Y.~Mura, ``{Electroweak baryogenesis via top-charm mixing},'' \href{http://dx.doi.org/10.1007/JHEP09(2023)153}{{\em JHEP} {\bf 09} (2023)  153}, \href{http://arxiv.org/abs/2303.11252}{{\tt arXiv:2303.11252 [hep-ph]}}.

\bibitem{Enomoto:2024jyc}
K.~Enomoto, S.~Kanemura, and S.~Taniguchi, ``{The electric dipole moment in a model for neutrino mass, dark matter and baryon asymmetry of the Universe},'' \href{http://arxiv.org/abs/2403.13613}{{\tt arXiv:2403.13613 [hep-ph]}}.

\bibitem{Huang:2025cqi}
F.~P. Huang, ``{The First Particles},'' \href{http://arxiv.org/abs/2501.15543}{{\tt arXiv:2501.15543 [hep-ph]}}.

\bibitem{Weir:2017wfa}
D.~J. Weir, ``{Gravitational waves from a first order electroweak phase transition: a brief review},'' \href{http://dx.doi.org/10.1098/rsta.2017.0126}{{\em Phil. Trans. Roy. Soc. Lond. A} {\bf 376} (2018) no.~2114, 20170126}, \href{http://arxiv.org/abs/1705.01783}{{\tt arXiv:1705.01783 [hep-ph]}}. [Erratum: Phil.Trans.Roy.Soc.Lond.A 381, 20230212 (2023)].

\bibitem{Mazumdar:2018dfl}
A.~Mazumdar and G.~White, ``{Review of cosmic phase transitions: their significance and experimental signatures},'' \href{http://dx.doi.org/10.1088/1361-6633/ab1f55}{{\em Rept. Prog. Phys.} {\bf 82} (2019) no.~7, 076901}, \href{http://arxiv.org/abs/1811.01948}{{\tt arXiv:1811.01948 [hep-ph]}}.

\bibitem{Athron:2023xlk}
P.~Athron, C.~Bal\'azs, A.~Fowlie, L.~Morris, and L.~Wu, ``{Cosmological phase transitions: from perturbative particle physics to gravitational waves},'' \href{http://arxiv.org/abs/2305.02357}{{\tt arXiv:2305.02357 [hep-ph]}}.

\bibitem{Caldwell:2022qsj}
R.~Caldwell {\em et al.}, ``{Detection of early-universe gravitational-wave signatures and fundamental physics},'' \href{http://dx.doi.org/10.1007/s10714-022-03027-x}{{\em Gen. Rel. Grav.} {\bf 54} (2022) no.~12, 156}, \href{http://arxiv.org/abs/2203.07972}{{\tt arXiv:2203.07972 [gr-qc]}}.

\bibitem{Guo:2021qcq}
H.-K. Guo, K.~Sinha, D.~Vagie, and G.~White, ``{The benefits of diligence: how precise are predicted gravitational wave spectra in models with phase transitions?},'' \href{http://dx.doi.org/10.1007/JHEP06(2021)164}{{\em JHEP} {\bf 06} (2021)  164}, \href{http://arxiv.org/abs/2103.06933}{{\tt arXiv:2103.06933 [hep-ph]}}.

\bibitem{Grojean:2006bp}
C.~Grojean and G.~Servant, ``{Gravitational Waves from Phase Transitions at the Electroweak Scale and Beyond},'' \href{http://dx.doi.org/10.1103/PhysRevD.75.043507}{{\em Phys. Rev. D} {\bf 75} (2007)  043507}, \href{http://arxiv.org/abs/hep-ph/0607107}{{\tt arXiv:hep-ph/0607107}}.

\bibitem{Caprini:2015zlo}
C.~Caprini {\em et al.}, ``{Science with the space-based interferometer eLISA. II: Gravitational waves from cosmological phase transitions},'' \href{http://dx.doi.org/10.1088/1475-7516/2016/04/001}{{\em JCAP} {\bf 1604} (2016) no.~04, 001},
\href{http://arxiv.org/abs/1512.06239}{{\tt arXiv:1512.06239 [astro-ph.CO]}}.

\bibitem{Caprini:2019egz}
C.~Caprini {\em et al.}, ``{Detecting gravitational waves from cosmological phase transitions with LISA: an update},'' \href{http://dx.doi.org/10.1088/1475-7516/2020/03/024}{{\em JCAP} {\bf 03} (2020)  024}, \href{http://arxiv.org/abs/1910.13125}{{\tt arXiv:1910.13125 [astro-ph.CO]}}.

\bibitem{LISA:2017pwj}
{\bf LISA} Collaboration, P.~Amaro-Seoane {\em et al.}, ``{Laser Interferometer Space Antenna},'' \href{http://arxiv.org/abs/1702.00786}{{\tt arXiv:1702.00786 [astro-ph.IM]}}.

\bibitem{TianQin:2015yph}
{\bf TianQin} Collaboration, J.~Luo {\em et al.}, ``{TianQin: a space-borne gravitational wave detector},'' \href{http://dx.doi.org/10.1088/0264-9381/33/3/035010}{{\em Class. Quant. Grav.} {\bf 33} (2016) no.~3, 035010}, \href{http://arxiv.org/abs/1512.02076}{{\tt arXiv:1512.02076 [astro-ph.IM]}}.

\bibitem{TianQin:2020hid}
{\bf TianQin} Collaboration, J.~Mei {\em et al.}, ``{The TianQin project: current progress on science and technology},'' \href{http://dx.doi.org/10.1093/ptep/ptaa114}{{\em PTEP} {\bf 2021} (2021) no.~5, 05A107}, \href{http://arxiv.org/abs/2008.10332}{{\tt arXiv:2008.10332 [gr-qc]}}.

\bibitem{Hu:2017mde}
W.-R. Hu and Y.-L. Wu, ``{The Taiji Program in Space for gravitational wave physics and the nature of gravity},'' \href{http://dx.doi.org/10.1093/nsr/nwx116}{{\em Natl. Sci. Rev.} {\bf 4} (2017) no.~5, 685--686}.

\bibitem{Ruan:2018tsw}
W.-H. Ruan, Z.-K. Guo, R.-G. Cai, and Y.-Z. Zhang, ``{Taiji program: Gravitational-wave sources},'' \href{http://dx.doi.org/10.1142/S0217751X2050075X}{{\em Int. J. Mod. Phys. A} {\bf 35} (2020) no.~17, 2050075}, \href{http://arxiv.org/abs/1807.09495}{{\tt arXiv:1807.09495 [gr-qc]}}.

\bibitem{Baker:2016xzo}
M.~J. Baker and J.~Kopp, ``{Dark Matter Decay between Phase Transitions at the Weak Scale},'' \href{http://dx.doi.org/10.1103/PhysRevLett.119.061801}{{\em Phys. Rev. Lett.} {\bf 119} (2017) no.~6, 061801}, \href{http://arxiv.org/abs/1608.07578}{{\tt arXiv:1608.07578 [hep-ph]}}.

\bibitem{Baker:2017zwx}
M.~J. Baker, M.~Breitbach, J.~Kopp, and L.~Mittnacht, ``{Dynamic Freeze-In: Impact of Thermal Masses and Cosmological Phase Transitions on Dark Matter Production},'' \href{http://dx.doi.org/10.1007/JHEP03(2018)114}{{\em JHEP} {\bf 03} (2018)  114}, \href{http://arxiv.org/abs/1712.03962}{{\tt arXiv:1712.03962 [hep-ph]}}.

\bibitem{Wong:2023qon}
X.-R. Wong and K.-P. Xie, ``{Freeze-in of WIMP dark matter},'' \href{http://dx.doi.org/10.1103/PhysRevD.108.055035}{{\em Phys. Rev. D} {\bf 108} (2023) no.~5, 055035}, \href{http://arxiv.org/abs/2304.00908}{{\tt arXiv:2304.00908 [hep-ph]}}.

\bibitem{Hambye:2018qjv}
T.~Hambye, A.~Strumia, and D.~Teresi, ``{Super-cool Dark Matter},'' \href{http://dx.doi.org/10.1007/JHEP08(2018)188}{{\em JHEP} {\bf 08} (2018)  188}, \href{http://arxiv.org/abs/1805.01473}{{\tt arXiv:1805.01473 [hep-ph]}}.

\bibitem{Azatov:2021ifm}
A.~Azatov, M.~Vanvlasselaer, and W.~Yin, ``{Dark Matter production from relativistic bubble walls},'' \href{http://dx.doi.org/10.1007/JHEP03(2021)288}{{\em JHEP} {\bf 03} (2021)  288}, \href{http://arxiv.org/abs/2101.05721}{{\tt arXiv:2101.05721 [hep-ph]}}.

\bibitem{Azatov:2021irb}
A.~Azatov, M.~Vanvlasselaer, and W.~Yin, ``{Baryogenesis via relativistic bubble walls},'' \href{http://dx.doi.org/10.1007/JHEP10(2021)043}{{\em JHEP} {\bf 10} (2021)  043}, \href{http://arxiv.org/abs/2106.14913}{{\tt arXiv:2106.14913 [hep-ph]}}.

\bibitem{Ai:2024ikj}
W.-Y. Ai, M.~Fairbairn, K.~Mimasu, and T.~You, ``{Non-thermal production of heavy vector dark matter from relativistic bubble walls},'' \href{http://arxiv.org/abs/2406.20051}{{\tt arXiv:2406.20051 [hep-ph]}}.

\bibitem{Baldes:2021vyz}
I.~Baldes, S.~Blasi, A.~Mariotti, A.~Sevrin, and K.~Turbang, ``{Baryogenesis via relativistic bubble expansion},'' \href{http://dx.doi.org/10.1103/PhysRevD.104.115029}{{\em Phys. Rev. D} {\bf 104} (2021) no.~11, 115029}, \href{http://arxiv.org/abs/2106.15602}{{\tt arXiv:2106.15602 [hep-ph]}}.

\bibitem{Huang:2022vkf}
P.~Huang and K.-P. Xie, ``{Leptogenesis triggered by a first-order phase transition},'' \href{http://dx.doi.org/10.1007/JHEP09(2022)052}{{\em JHEP} {\bf 09} (2022)  052}, \href{http://arxiv.org/abs/2206.04691}{{\tt arXiv:2206.04691 [hep-ph]}}.

\bibitem{Chun:2023ezg}
E.~J. Chun, T.~P. Dutka, T.~H. Jung, X.~Nagels, and M.~Vanvlasselaer, ``{Bubble-assisted leptogenesis},'' \href{http://dx.doi.org/10.1007/JHEP09(2023)164}{{\em JHEP} {\bf 09} (2023)  164}, \href{http://arxiv.org/abs/2305.10759}{{\tt arXiv:2305.10759 [hep-ph]}}.

\bibitem{Baker:2019ndr}
M.~J. Baker, J.~Kopp, and A.~J. Long, ``{Filtered Dark Matter at a First Order Phase Transition},'' \href{http://dx.doi.org/10.1103/PhysRevLett.125.151102}{{\em Phys. Rev. Lett.} {\bf 125} (2020) no.~15, 151102}, \href{http://arxiv.org/abs/1912.02830}{{\tt arXiv:1912.02830 [hep-ph]}}.

\bibitem{Chway:2019kft}
D.~Chway, T.~H. Jung, and C.~S. Shin, ``{Dark matter filtering-out effect during a first-order phase transition},'' \href{http://dx.doi.org/10.1103/PhysRevD.101.095019}{{\em Phys. Rev. D} {\bf 101} (2020) no.~9, 095019}, \href{http://arxiv.org/abs/1912.04238}{{\tt arXiv:1912.04238 [hep-ph]}}.

\bibitem{Chao:2020adk}
W.~Chao, X.-F. Li, and L.~Wang, ``{Filtered pseudo-scalar dark matter and gravitational waves from first order phase transition},'' \href{http://dx.doi.org/10.1088/1475-7516/2021/06/038}{{\em JCAP} {\bf 06} (2021)  038}, \href{http://arxiv.org/abs/2012.15113}{{\tt arXiv:2012.15113 [hep-ph]}}.

\bibitem{Jiang:2023nkj}
S.~Jiang, F.~P. Huang, and C.~S. Li, ``{Hydrodynamic effects on the filtered dark matter produced by a first-order phase transition},'' \href{http://dx.doi.org/10.1103/PhysRevD.108.063508}{{\em Phys. Rev. D} {\bf 108} (2023) no.~6, 063508}, \href{http://arxiv.org/abs/2305.02218}{{\tt arXiv:2305.02218 [hep-ph]}}.

\bibitem{Huang:2017kzu}
F.~P. Huang and C.~S. Li, ``{Probing the baryogenesis and dark matter relaxed in phase transition by gravitational waves and colliders},'' \href{http://dx.doi.org/10.1103/PhysRevD.96.095028}{{\em Phys. Rev. D} {\bf 96} (2017) no.~9, 095028}, \href{http://arxiv.org/abs/1709.09691}{{\tt arXiv:1709.09691 [hep-ph]}}.

\bibitem{Bai:2018vik}
Y.~Bai and A.~J. Long, ``{Six Flavor Quark Matter},'' \href{http://dx.doi.org/10.1007/JHEP06(2018)072}{{\em JHEP} {\bf 06} (2018)  072}, \href{http://arxiv.org/abs/1804.10249}{{\tt arXiv:1804.10249 [hep-ph]}}.

\bibitem{Hong:2020est}
J.-P. Hong, S.~Jung, and K.-P. Xie, ``{Fermi-ball dark matter from a first-order phase transition},'' \href{http://dx.doi.org/10.1103/PhysRevD.102.075028}{{\em Phys. Rev. D} {\bf 102} (2020) no.~7, 075028}, \href{http://arxiv.org/abs/2008.04430}{{\tt arXiv:2008.04430 [hep-ph]}}.

\bibitem{Baker:2021nyl}
M.~J. Baker, M.~Breitbach, J.~Kopp, and L.~Mittnacht, ``{Primordial Black Holes from First-Order Cosmological Phase Transitions},'' \href{http://arxiv.org/abs/2105.07481}{{\tt arXiv:2105.07481 [astro-ph.CO]}}.

\bibitem{Kawana:2021tde}
K.~Kawana and K.-P. Xie, ``{Primordial black holes from a cosmic phase transition: The collapse of Fermi-balls},'' \href{http://dx.doi.org/10.1016/j.physletb.2021.136791}{{\em Phys. Lett. B} {\bf 824} (2022)  136791}, \href{http://arxiv.org/abs/2106.00111}{{\tt arXiv:2106.00111 [astro-ph.CO]}}.

\bibitem{Huang:2022him}
P.~Huang and K.-P. Xie, ``{Primordial black holes from an electroweak phase transition},'' \href{http://dx.doi.org/10.1103/PhysRevD.105.115033}{{\em Phys. Rev. D} {\bf 105} (2022) no.~11, 115033}, \href{http://arxiv.org/abs/2201.07243}{{\tt arXiv:2201.07243 [hep-ph]}}.

\bibitem{Kawana:2022lba}
K.~Kawana, P.~Lu, and K.-P. Xie, ``{First-order phase transition and fate of false vacuum remnants},'' \href{http://dx.doi.org/10.1088/1475-7516/2022/10/030}{{\em JCAP} {\bf 10} (2022)  030}, \href{http://arxiv.org/abs/2206.09923}{{\tt arXiv:2206.09923 [astro-ph.CO]}}.

\bibitem{Lu:2022paj}
P.~Lu, K.~Kawana, and K.-P. Xie, ``{Old phase remnants in first-order phase transitions},'' \href{http://dx.doi.org/10.1103/PhysRevD.105.123503}{{\em Phys. Rev. D} {\bf 105} (2022) no.~12, 123503}, \href{http://arxiv.org/abs/2202.03439}{{\tt arXiv:2202.03439 [astro-ph.CO]}}.

\bibitem{Lu:2022jnp}
P.~Lu, K.~Kawana, and A.~Kusenko, ``{Late-forming primordial black holes: Beyond the CMB era},'' \href{http://dx.doi.org/10.1103/PhysRevD.107.103037}{{\em Phys. Rev. D} {\bf 107} (2023) no.~10, 103037}, \href{http://arxiv.org/abs/2210.16462}{{\tt arXiv:2210.16462 [astro-ph.CO]}}.

\bibitem{Kim:2023ixo}
T.~Kim, P.~Lu, D.~Marfatia, and V.~Takhistov, ``{Regurgitated Dark Matter},'' \href{http://arxiv.org/abs/2309.05703}{{\tt arXiv:2309.05703 [hep-ph]}}.

\bibitem{Xie:2024mxr}
K.-P. Xie, ``{Revisiting the fermion-field nontopological solitons},'' \href{http://arxiv.org/abs/2405.01227}{{\tt arXiv:2405.01227 [hep-ph]}}.

\bibitem{Cline:2022xhx}
J.~M. Cline, B.~Laurent, S.~Raby, and J.-S. Roux, ``{PeV-scale leptogenesis, gravitational waves, and black holes from a SUSY-breaking phase transition},'' \href{http://dx.doi.org/10.1103/PhysRevD.107.095011}{{\em Phys. Rev. D} {\bf 107} (2023) no.~9, 095011}, \href{http://arxiv.org/abs/2211.00422}{{\tt arXiv:2211.00422 [hep-ph]}}.

\bibitem{Gehrman:2023qjn}
T.~C. Gehrman, B.~Shams Es~Haghi, K.~Sinha, and T.~Xu, ``{Recycled dark matter},'' \href{http://dx.doi.org/10.1088/1475-7516/2024/03/044}{{\em JCAP} {\bf 03} (2024)  044}, \href{http://arxiv.org/abs/2310.08526}{{\tt arXiv:2310.08526 [hep-ph]}}.

\bibitem{Liu:2021svg}
J.~Liu, L.~Bian, R.-G. Cai, Z.-K. Guo, and S.-J. Wang, ``{Primordial black hole production during first-order phase transitions},'' \href{http://dx.doi.org/10.1103/PhysRevD.105.L021303}{{\em Phys. Rev. D} {\bf 105} (2022) no.~2, L021303}, \href{http://arxiv.org/abs/2106.05637}{{\tt arXiv:2106.05637 [astro-ph.CO]}}.

\bibitem{Hashino:2021qoq}
K.~Hashino, S.~Kanemura, and T.~Takahashi, ``{Primordial black holes as a probe of strongly first-order electroweak phase transition},'' \href{http://dx.doi.org/10.1016/j.physletb.2022.137261}{{\em Phys. Lett. B} {\bf 833} (2022)  137261}, \href{http://arxiv.org/abs/2111.13099}{{\tt arXiv:2111.13099 [hep-ph]}}.

\bibitem{Kawana:2022olo}
K.~Kawana, T.~Kim, and P.~Lu, ``{PBH formation from overdensities in delayed vacuum transitions},'' \href{http://dx.doi.org/10.1103/PhysRevD.108.103531}{{\em Phys. Rev. D} {\bf 108} (2023) no.~10, 103531}, \href{http://arxiv.org/abs/2212.14037}{{\tt arXiv:2212.14037 [astro-ph.CO]}}.

\bibitem{Hashino:2022tcs}
K.~Hashino, S.~Kanemura, T.~Takahashi, and M.~Tanaka, ``{Probing first-order electroweak phase transition via primordial black holes in the effective field theory},'' \href{http://dx.doi.org/10.1016/j.physletb.2023.137688}{{\em Phys. Lett. B} {\bf 838} (2023)  137688}, \href{http://arxiv.org/abs/2211.16225}{{\tt arXiv:2211.16225 [hep-ph]}}.

\bibitem{Lewicki:2023ioy}
M.~Lewicki, P.~Toczek, and V.~Vaskonen, ``{Primordial black holes from strong first-order phase transitions},'' \href{http://dx.doi.org/10.1007/JHEP09(2023)092}{{\em JHEP} {\bf 09} (2023)  092}, \href{http://arxiv.org/abs/2305.04924}{{\tt arXiv:2305.04924 [astro-ph.CO]}}.

\bibitem{Gouttenoire:2023naa}
Y.~Gouttenoire and T.~Volansky, ``{Primordial Black Holes from Supercooled Phase Transitions},'' \href{http://arxiv.org/abs/2305.04942}{{\tt arXiv:2305.04942 [hep-ph]}}.

\bibitem{Kanemura:2024pae}
S.~Kanemura, M.~Tanaka, and K.-P. Xie, ``{Primordial black holes from slow phase transitions: a model-building perspective},'' \href{http://dx.doi.org/10.1007/JHEP06(2024)036}{{\em JHEP} {\bf 06} (2024)  036}, \href{http://arxiv.org/abs/2404.00646}{{\tt arXiv:2404.00646 [hep-ph]}}.

\bibitem{Cai:2024nln}
R.-G. Cai, Y.-S. Hao, and S.-J. Wang, ``{Primordial black holes and curvature perturbations from false-vacuum islands},'' \href{http://arxiv.org/abs/2404.06506}{{\tt arXiv:2404.06506 [astro-ph.CO]}}.

\bibitem{Arteaga:2024vde}
M.~Arteaga, A.~Ghoshal, and A.~Strumia, ``{Gravitational waves and black holes from the phase transition in models of dynamical symmetry breaking},'' \href{http://arxiv.org/abs/2409.04545}{{\tt arXiv:2409.04545 [hep-ph]}}.

\bibitem{Ai:2024cka}
W.-Y. Ai, L.~Heurtier, and T.~H. Jung, ``{Primordial black holes from an aborted phase transition},'' \href{http://arxiv.org/abs/2409.02175}{{\tt arXiv:2409.02175 [astro-ph.CO]}}.

\bibitem{Goncalves:2024vkj}
D.~Gon\c{c}alves, A.~Kaladharan, and Y.~Wu, ``{Primordial Black Holes from First-Order Phase Transition in the xSM},'' \href{http://arxiv.org/abs/2406.07622}{{\tt arXiv:2406.07622 [hep-ph]}}.

\bibitem{Linde:1981zj}
A.~D. Linde, ``{Decay of the False Vacuum at Finite Temperature},'' \href{http://dx.doi.org/10.1016/0550-3213(83)90072-X}{{\em Nucl. Phys. B} {\bf 216} (1983)  421}. [Erratum: Nucl.Phys.B 223, 544 (1983)].

\bibitem{Ramsey-Musolf:2019lsf}
M.~J. Ramsey-Musolf, ``{The electroweak phase transition: a collider target},'' \href{http://dx.doi.org/10.1007/JHEP09(2020)179}{{\em JHEP} {\bf 09} (2020)  179}, \href{http://arxiv.org/abs/1912.07189}{{\tt arXiv:1912.07189 [hep-ph]}}.

\bibitem{Chung:2012vg}
D.~J.~H. Chung, A.~J. Long, and L.-T. Wang, ``{125 GeV Higgs boson and electroweak phase transition model classes},'' \href{http://dx.doi.org/10.1103/PhysRevD.87.023509}{{\em Phys. Rev.} {\bf D87} (2013) no.~2, 023509},
\href{http://arxiv.org/abs/1209.1819}{{\tt arXiv:1209.1819 [hep-ph]}}.

\bibitem{Pietroni:1992in}
M.~Pietroni, ``{The Electroweak phase transition in a nonminimal supersymmetric model},'' \href{http://dx.doi.org/10.1016/0550-3213(93)90635-3}{{\em Nucl. Phys.} {\bf B402} (1993)  27--45},
\href{http://arxiv.org/abs/hep-ph/9207227}{{\tt arXiv:hep-ph/9207227 [hep-ph]}}.

\bibitem{Choi:1993cv}
J.~Choi and R.~R. Volkas, ``{Real Higgs singlet and the electroweak phase transition in the Standard Model},'' \href{http://dx.doi.org/10.1016/0370-2693(93)91013-D}{{\em Phys. Lett. B} {\bf 317} (1993)  385--391}, \href{http://arxiv.org/abs/hep-ph/9308234}{{\tt arXiv:hep-ph/9308234}}.

\bibitem{Espinosa1993}
J.~R. Espinosa and M.~Quiros, ``{The Electroweak phase transition with a singlet},'' \href{http://dx.doi.org/10.1016/0370-2693(93)91111-Y}{{\em Phys. Lett.} {\bf B305} (1993)  98--105},
\href{http://arxiv.org/abs/hep-ph/9301285}{{\tt arXiv:hep-ph/9301285 [hep-ph]}}.

\bibitem{Benson:1993qx}
K.~E.~C. Benson, ``{Avoiding baryon washout in the extended Standard Model},'' \href{http://dx.doi.org/10.1103/PhysRevD.48.2456}{{\em Phys. Rev. D} {\bf 48} (1993)  2456--2461}.

\bibitem{Ham:2004cf}
S.~W. Ham, Y.~S. Jeong, and S.~K. Oh, ``{Electroweak phase transition in an extension of the standard model with a real Higgs singlet},'' \href{http://dx.doi.org/10.1088/0954-3899/31/8/017}{{\em J. Phys. G} {\bf 31} (2005) no.~8, 857--871}, \href{http://arxiv.org/abs/hep-ph/0411352}{{\tt arXiv:hep-ph/0411352}}.

\bibitem{Profumo:2007wc}
S.~Profumo, M.~J. Ramsey-Musolf, and G.~Shaughnessy, ``{Singlet Higgs phenomenology and the electroweak phase transition},'' \href{http://dx.doi.org/10.1088/1126-6708/2007/08/010}{{\em JHEP} {\bf 08} (2007)  010},
\href{http://arxiv.org/abs/0705.2425}{{\tt arXiv:0705.2425 [hep-ph]}}.

\bibitem{Espinosa:2011ax}
J.~R. Espinosa, T.~Konstandin, and F.~Riva, ``{Strong Electroweak Phase Transitions in the Standard Model with a Singlet},'' \href{http://dx.doi.org/10.1016/j.nuclphysb.2011.09.010}{{\em Nucl. Phys.} {\bf B854} (2012)  592--630},
\href{http://arxiv.org/abs/1107.5441}{{\tt arXiv:1107.5441 [hep-ph]}}.

\bibitem{Alves:2018jsw}
A.~Alves, T.~Ghosh, H.-K. Guo, K.~Sinha, and D.~Vagie, ``{Collider and Gravitational Wave Complementarity in Exploring the Singlet Extension of the Standard Model},'' \href{http://dx.doi.org/10.1007/JHEP04(2019)052}{{\em JHEP} {\bf 04} (2019)  052}, \href{http://arxiv.org/abs/1812.09333}{{\tt arXiv:1812.09333 [hep-ph]}}.

\bibitem{Alves:2018oct}
A.~Alves, T.~Ghosh, H.-K. Guo, and K.~Sinha, ``{Resonant Di-Higgs Production at Gravitational Wave Benchmarks: A Collider Study using Machine Learning},'' \href{http://dx.doi.org/10.1007/JHEP12(2018)070}{{\em JHEP} {\bf 12} (2018)  070}, \href{http://arxiv.org/abs/1808.08974}{{\tt arXiv:1808.08974 [hep-ph]}}.

\bibitem{Alves:2019igs}
A.~Alves, D.~Gon\c{c}alves, T.~Ghosh, H.-K. Guo, and K.~Sinha, ``{Di-Higgs Production in the $4b$ Channel and Gravitational Wave Complementarity},'' \href{http://dx.doi.org/10.1007/JHEP03(2020)053}{{\em JHEP} {\bf 03} (2020)  053}, \href{http://arxiv.org/abs/1909.05268}{{\tt arXiv:1909.05268 [hep-ph]}}.

\bibitem{Alves:2020bpi}
A.~Alves, D.~Gon\c{c}alves, T.~Ghosh, H.-K. Guo, and K.~Sinha, ``{Di-Higgs Blind Spots in Gravitational Wave Signals},'' \href{http://dx.doi.org/10.1016/j.physletb.2021.136377}{{\em Phys. Lett. B} {\bf 818} (2021)  136377}, \href{http://arxiv.org/abs/2007.15654}{{\tt arXiv:2007.15654 [hep-ph]}}.

\bibitem{Liu:2021jyc}
W.~Liu and K.-P. Xie, ``{Probing electroweak phase transition with multi-TeV muon colliders and gravitational waves},'' \href{http://dx.doi.org/10.1007/JHEP04(2021)015}{{\em JHEP} {\bf 04} (2021)  015}, \href{http://arxiv.org/abs/2101.10469}{{\tt arXiv:2101.10469 [hep-ph]}}.

\bibitem{Ramsey-Musolf:2024ykk}
M.~J. Ramsey-Musolf, T.~V.~I. Tenkanen, and V.~Q. Tran, ``{Refining Gravitational Wave and Collider Physics Dialogue via Singlet Scalar Extension},'' \href{http://arxiv.org/abs/2409.17554}{{\tt arXiv:2409.17554 [hep-ph]}}.

\bibitem{Bian:2019bsn}
L.~Bian, H.-K. Guo, Y.~Wu, and R.~Zhou, ``{Gravitational wave and collider searches for electroweak symmetry breaking patterns},'' \href{http://dx.doi.org/10.1103/PhysRevD.101.035011}{{\em Phys. Rev. D} {\bf 101} (2020) no.~3, 035011}, \href{http://arxiv.org/abs/1906.11664}{{\tt arXiv:1906.11664 [hep-ph]}}.

\bibitem{Zhou:2018zli}
R.~Zhou, W.~Cheng, X.~Deng, L.~Bian, and Y.~Wu, ``{Electroweak phase transition and Higgs phenomenology in the Georgi-Machacek model},''
\href{http://arxiv.org/abs/1812.06217}{{\tt arXiv:1812.06217 [hep-ph]}}.

\bibitem{Huang:2017rzf}
F.~P. Huang and J.-H. Yu, ``{Exploring inert dark matter blind spots with gravitational wave signatures},'' \href{http://dx.doi.org/10.1103/PhysRevD.98.095022}{{\em Phys. Rev. D} {\bf 98} (2018) no.~9, 095022}, \href{http://arxiv.org/abs/1704.04201}{{\tt arXiv:1704.04201 [hep-ph]}}.

\bibitem{Wang:2020wrk}
Y.~Wang, C.~S. Li, and F.~P. Huang, ``{Complementary probe of dark matter blind spots by lepton colliders and gravitational waves},'' \href{http://dx.doi.org/10.1103/PhysRevD.104.053004}{{\em Phys. Rev. D} {\bf 104} (2021) no.~5, 053004}, \href{http://arxiv.org/abs/2012.03920}{{\tt arXiv:2012.03920 [hep-ph]}}.

\bibitem{Ghosh:2020ipy}
T.~Ghosh, H.-K. Guo, T.~Han, and H.~Liu, ``{Electroweak phase transition with an SU(2) dark sector},'' \href{http://dx.doi.org/10.1007/JHEP07(2021)045}{{\em JHEP} {\bf 07} (2021)  045}, \href{http://arxiv.org/abs/2012.09758}{{\tt arXiv:2012.09758 [hep-ph]}}.

\bibitem{Noble:2007kk}
A.~Noble and M.~Perelstein, ``{Higgs self-coupling as a probe of electroweak phase transition},'' \href{http://dx.doi.org/10.1103/PhysRevD.78.063518}{{\em Phys. Rev. D} {\bf 78} (2008)  063518}, \href{http://arxiv.org/abs/0711.3018}{{\tt arXiv:0711.3018 [hep-ph]}}.

\bibitem{Profumo:2014opa}
S.~Profumo, M.~J. Ramsey-Musolf, C.~L. Wainwright, and P.~Winslow, ``{Singlet-catalyzed electroweak phase transitions and precision Higgs boson studies},'' \href{http://dx.doi.org/10.1103/PhysRevD.91.035018}{{\em Phys. Rev. D} {\bf 91} (2015) no.~3, 035018}, \href{http://arxiv.org/abs/1407.5342}{{\tt arXiv:1407.5342 [hep-ph]}}.

\bibitem{Huang:2015izx}
F.~P. Huang, P.-H. Gu, P.-F. Yin, Z.-H. Yu, and X.~Zhang, ``{Testing the electroweak phase transition and electroweak baryogenesis at the LHC and a circular electron-positron collider},'' \href{http://dx.doi.org/10.1103/PhysRevD.93.103515}{{\em Phys. Rev. D} {\bf 93} (2016) no.~10, 103515}, \href{http://arxiv.org/abs/1511.03969}{{\tt arXiv:1511.03969 [hep-ph]}}.

\bibitem{Huang:2016cjm}
P.~Huang, A.~J. Long, and L.-T. Wang, ``{Probing the Electroweak Phase Transition with Higgs Factories and Gravitational Waves},'' \href{http://dx.doi.org/10.1103/PhysRevD.94.075008}{{\em Phys. Rev. D} {\bf 94} (2016) no.~7, 075008}, \href{http://arxiv.org/abs/1608.06619}{{\tt arXiv:1608.06619 [hep-ph]}}.

\bibitem{Cao:2017oez}
Q.-H. Cao, F.~P. Huang, K.-P. Xie, and X.~Zhang, ``{Testing the electroweak phase transition in scalar extension models at lepton colliders},'' \href{http://dx.doi.org/10.1088/1674-1137/42/2/023103}{{\em Chin. Phys.} {\bf C42} (2018) no.~2, 023103},
\href{http://arxiv.org/abs/1708.04737}{{\tt arXiv:1708.04737 [hep-ph]}}.

\bibitem{Chen:2019ebq}
N.~Chen, T.~Li, Y.~Wu, and L.~Bian, ``{Complementarity of the future $e^+ e^-$ colliders and gravitational waves in the probe of complex singlet extension to the standard model},'' \href{http://dx.doi.org/10.1103/PhysRevD.101.075047}{{\em Phys. Rev. D} {\bf 101} (2020) no.~7, 075047}, \href{http://arxiv.org/abs/1911.05579}{{\tt arXiv:1911.05579 [hep-ph]}}.

\bibitem{Su:2020pjw}
W.~Su, A.~G. Williams, and M.~Zhang, ``{Strong first order electroweak phase transition in 2HDM confronting future Z \& Higgs factories},'' \href{http://arxiv.org/abs/2011.04540}{{\tt arXiv:2011.04540 [hep-ph]}}.

\bibitem{Song:2022xts}
H.~Song, W.~Su, and M.~Zhang, ``{Electroweak phase transition in 2HDM under Higgs, Z-pole, and W precision measurements},'' \href{http://dx.doi.org/10.1007/JHEP10(2022)048}{{\em JHEP} {\bf 10} (2022)  048}, \href{http://arxiv.org/abs/2204.05085}{{\tt arXiv:2204.05085 [hep-ph]}}.

\bibitem{CEPCPhysicsStudyGroup:2022uwl}
{\bf CEPC Physics Study Group} Collaboration, H.~Cheng {\em et al.}, ``{The Physics potential of the CEPC. Prepared for the US Snowmass Community Planning Exercise (Snowmass 2021)},'' in {\em {Snowmass 2021}}.
\newblock 5, 2022.
\newblock \href{http://arxiv.org/abs/2205.08553}{{\tt arXiv:2205.08553 [hep-ph]}}.

\bibitem{Niemi:2024axp}
L.~Niemi, M.~J. Ramsey-Musolf, and G.~Xia, ``{Nonperturbative study of the electroweak phase transition in the real scalar singlet extended standard model},'' \href{http://dx.doi.org/10.1103/PhysRevD.110.115016}{{\em Phys. Rev. D} {\bf 110} (2024) no.~11, 115016}, \href{http://arxiv.org/abs/2405.01191}{{\tt arXiv:2405.01191 [hep-ph]}}.

\bibitem{Niemi:2018asa}
L.~Niemi, H.~H. Patel, M.~J. Ramsey-Musolf, T.~V. Tenkanen, and D.~J. Weir, ``{Electroweak phase transition in the real triplet extension of the SM: Dimensional reduction},'' \href{http://dx.doi.org/10.1103/PhysRevD.100.035002}{{\em Phys. Rev. D} {\bf 100} (2019) no.~3, 035002}, \href{http://arxiv.org/abs/1802.10500}{{\tt arXiv:1802.10500 [hep-ph]}}.

\bibitem{Niemi:2020hto}
L.~Niemi, M.~J. Ramsey-Musolf, T.~V.~I. Tenkanen, and D.~J. Weir, ``{Thermodynamics of a Two-Step Electroweak Phase Transition},'' \href{http://dx.doi.org/10.1103/PhysRevLett.126.171802}{{\em Phys. Rev. Lett.} {\bf 126} (2021) no.~17, 171802}, \href{http://arxiv.org/abs/2005.11332}{{\tt arXiv:2005.11332 [hep-ph]}}.

\bibitem{Ramsey-Musolf:2021ldh}
M.~J. Ramsey-Musolf, J.-H. Yu, and J.~Zhou, ``{Probing extended scalar sectors with precision e$^{+}$e$^{−}$\textrightarrow{} Zh and Higgs diphoton studies},'' \href{http://dx.doi.org/10.1007/JHEP10(2021)155}{{\em JHEP} {\bf 10} (2021)  155}, \href{http://arxiv.org/abs/2104.10709}{{\tt arXiv:2104.10709 [hep-ph]}}.

\bibitem{Huang:2016odd}
F.~P. Huang, Y.~Wan, D.-G. Wang, Y.-F. Cai, and X.~Zhang, ``{Hearing the echoes of electroweak baryogenesis with gravitational wave detectors},'' \href{http://dx.doi.org/10.1103/PhysRevD.94.041702}{{\em Phys. Rev. D} {\bf 94} (2016) no.~4, 041702}, \href{http://arxiv.org/abs/1601.01640}{{\tt arXiv:1601.01640 [hep-ph]}}.

\bibitem{Gong:2016jys}
Y.~Gong, Z.~Li, X.~Xu, L.~L. Yang, and X.~Zhao, ``{Mixed QCD-EW corrections for Higgs boson production at $e^+e^-$ colliders},''
\href{http://arxiv.org/abs/1609.03955}{{\tt arXiv:1609.03955 [hep-ph]}}.

\bibitem{Sun:2016bel}
Q.-F. Sun, F.~Feng, Y.~Jia, and W.-L. Sang, ``{Mixed electroweak-QCD corrections to $e^+e^-\to HZ$ at Higgs factories},''
\href{http://arxiv.org/abs/1609.03995}{{\tt arXiv:1609.03995 [hep-ph]}}.

\bibitem{Carena:2019une}
M.~Carena, Z.~Liu, and Y.~Wang, ``{Electroweak phase transition with spontaneous $Z_{2}$-breaking},'' \href{http://dx.doi.org/10.1007/JHEP08(2020)107}{{\em JHEP} {\bf 08} (2020)  107}, \href{http://arxiv.org/abs/1911.10206}{{\tt arXiv:1911.10206 [hep-ph]}}.

\bibitem{Kozaczuk:2019pet}
J.~Kozaczuk, M.~J. Ramsey-Musolf, and J.~Shelton, ``{Exotic Higgs boson decays and the electroweak phase transition},'' \href{http://dx.doi.org/10.1103/PhysRevD.101.115035}{{\em Phys. Rev. D} {\bf 101} (2020) no.~11, 115035}, \href{http://arxiv.org/abs/1911.10210}{{\tt arXiv:1911.10210 [hep-ph]}}.

\bibitem{Gershtein:2020mwi}
Y.~Gershtein, S.~Knapen, and D.~Redigolo, ``{Probing naturally light singlets with a displaced vertex trigger},'' \href{http://dx.doi.org/10.1016/j.physletb.2021.136758}{{\em Phys. Lett. B} {\bf 823} (2021)  136758}, \href{http://arxiv.org/abs/2012.07864}{{\tt arXiv:2012.07864 [hep-ph]}}.

\bibitem{deBlas:2019rxi}
J.~de~Blas {\em et al.}, ``{Higgs Boson Studies at Future Particle Colliders},'' \href{http://dx.doi.org/10.1007/JHEP01(2020)139}{{\em JHEP} {\bf 01} (2020)  139}, \href{http://arxiv.org/abs/1905.03764}{{\tt arXiv:1905.03764 [hep-ph]}}.

\bibitem{Wang:2023zys}
Z.~Wang, X.~Zhu, E.~E. Khoda, S.-C. Hsu, N.~Konstantinidis, K.~Li, S.~Li, M.~J. Ramsey-Musolf, Y.~Wu, and Y.~E. Zhang, ``{Probing Electroweak Phase Transition at CEPC via Exotic Higgs Decays with 4b Final States},'' \href{http://dx.doi.org/10.31526/lhep.2023.436}{{\em LHEP} {\bf 2023} (2023)  436}.

\bibitem{Liu:2022nvk}
W.~Liu, A.~Yang, and H.~Sun, ``{Shedding light on the electroweak phase transition from exotic Higgs boson decays at the lifetime frontiers},'' \href{http://dx.doi.org/10.1103/PhysRevD.105.115040}{{\em Phys. Rev. D} {\bf 105} (2022) no.~11, 115040}, \href{http://arxiv.org/abs/2205.08205}{{\tt arXiv:2205.08205 [hep-ph]}}.

\bibitem{Cermeno:2022rni}
M.~Cerme\~no, C.~Degrande, and L.~Mantani, ``{Signatures of leptophilic t-channel dark matter from active galactic nuclei},'' \href{http://dx.doi.org/10.1103/PhysRevD.105.083019}{{\em Phys. Rev. D} {\bf 105} (2022) no.~8, 083019}, \href{http://arxiv.org/abs/2201.07247}{{\tt arXiv:2201.07247 [hep-ph]}}.

\bibitem{Jueid:2020yfj}
A.~Jueid, S.~Nasri, and R.~Soualah, ``{Searching for GeV-scale Majorana Dark Matter: inter spem et metum},'' \href{http://dx.doi.org/10.1007/JHEP04(2021)012}{{\em JHEP} {\bf 04} (2021)  012}, \href{http://arxiv.org/abs/2006.01348}{{\tt arXiv:2006.01348 [hep-ph]}}.

\bibitem{Jueid:2021wla}
A.~Jueid and S.~Nasri, ``{Phenomenology of Minimal Leptophilic Dark Matter Models at Linear Colliders},'' in {\em {International Workshop on Future Linear Colliders}}.
\newblock 5, 2021.
\newblock \href{http://arxiv.org/abs/2105.02921}{{\tt arXiv:2105.02921 [hep-ph]}}.

\bibitem{Chen:2017bff}
X.~Chen and Y.~Wu, ``{Search for CP violation effects in the $h\to \tau\tau$ decay with $e^+e^-$ colliders},'' \href{http://dx.doi.org/10.1140/epjc/s10052-017-5258-y}{{\em Eur. Phys. J. C} {\bf 77} (2017) no.~10, 697}, \href{http://arxiv.org/abs/1703.04855}{{\tt arXiv:1703.04855 [hep-ph]}}.

\bibitem{Chen:2017nxp}
X.~Chen and Y.~Wu, ``{Probing the CP-Violation effects in the $h\tau\tau$ coupling at the LHC},'' \href{http://dx.doi.org/10.1016/j.physletb.2019.01.038}{{\em Phys. Lett. B} {\bf 790} (2019)  332--338}, \href{http://arxiv.org/abs/1708.02882}{{\tt arXiv:1708.02882 [hep-ph]}}.

\bibitem{Ge:2020mcl}
S.-F. Ge, G.~Li, P.~Pasquini, and M.~J. Ramsey-Musolf, ``{CP-violating Higgs Di-tau Decays: Baryogenesis and Higgs Factories},'' \href{http://dx.doi.org/10.1103/PhysRevD.103.095027}{{\em Phys. Rev. D} {\bf 103} (2021) no.~9, 095027}, \href{http://arxiv.org/abs/2012.13922}{{\tt arXiv:2012.13922 [hep-ph]}}.

\bibitem{Alonso-Gonzalez:2021jsa}
J.~Alonso-Gonz\'alez, L.~Merlo, and S.~Pokorski, ``{A new bound on CP violation in the \ensuremath{\tau} lepton Yukawa coupling and electroweak baryogenesis},'' \href{http://dx.doi.org/10.1007/JHEP06(2021)166}{{\em JHEP} {\bf 06} (2021)  166}, \href{http://arxiv.org/abs/2103.16569}{{\tt arXiv:2103.16569 [hep-ph]}}.

\bibitem{Li:2023bxy}
S.-P. Li and K.-P. Xie, ``{Collider test of nano-Hertz gravitational waves from pulsar timing arrays},'' \href{http://dx.doi.org/10.1103/PhysRevD.108.055018}{{\em Phys. Rev. D} {\bf 108} (2023) no.~5, 055018}, \href{http://arxiv.org/abs/2307.01086}{{\tt arXiv:2307.01086 [hep-ph]}}.

\bibitem{Bertone:2016nfn}
G.~Bertone and D.~Hooper, ``{History of dark matter},'' \href{http://dx.doi.org/10.1103/RevModPhys.90.045002}{{\em Rev. Mod. Phys.} {\bf 90} (2018) no.~4, 045002}, \href{http://arxiv.org/abs/1605.04909}{{\tt arXiv:1605.04909 [astro-ph.CO]}}.

\bibitem{Roszkowski:2017nbc}
L.~Roszkowski, E.~M. Sessolo, and S.~Trojanowski, ``{WIMP dark matter candidates and searches status and future prospects},'' \href{http://dx.doi.org/10.1088/1361-6633/aab913}{{\em Rept. Prog. Phys.} {\bf 81} (2018) no.~6, 066201}, \href{http://arxiv.org/abs/1707.06277}{{\tt arXiv:1707.06277 [hep-ph]}}.

\bibitem{Cirelli:2024ssz}
M.~Cirelli, A.~Strumia, and J.~Zupan, ``{Dark Matter},'' \href{http://arxiv.org/abs/2406.01705}{{\tt arXiv:2406.01705 [hep-ph]}}.

\bibitem{Boveia:2018yeb}
A.~Boveia and C.~Doglioni, ``{Dark Matter Searches at Colliders},'' \href{http://dx.doi.org/10.1146/annurev-nucl-101917-021008}{{\em Ann. Rev. Nucl. Part. Sci.} {\bf 68} (2018)  429--459}, \href{http://arxiv.org/abs/1810.12238}{{\tt arXiv:1810.12238 [hep-ex]}}.

\bibitem{Pospelov:2007mp}
M.~Pospelov, A.~Ritz, and M.~B. Voloshin, ``{Secluded WIMP Dark Matter},'' \href{http://dx.doi.org/10.1016/j.physletb.2008.02.052}{{\em Phys. Lett. B} {\bf 662} (2008)  53--61}, \href{http://arxiv.org/abs/0711.4866}{{\tt arXiv:0711.4866 [hep-ph]}}.

\bibitem{LHCNewPhysicsWorkingGroup:2011mji}
{\bf LHC New Physics Working Group} Collaboration, D.~Alves, ``{Simplified Models for LHC New Physics Searches},'' \href{http://dx.doi.org/10.1088/0954-3899/39/10/105005}{{\em J. Phys. G} {\bf 39} (2012)  105005}, \href{http://arxiv.org/abs/1105.2838}{{\tt arXiv:1105.2838 [hep-ph]}}.

\bibitem{Liu:2017lpo}
J.~Liu, X.-P. Wang, and F.~Yu, ``{A Tale of Two Portals: Testing Light, Hidden New Physics at Future $e^+ e^-$ Colliders},'' \href{http://dx.doi.org/10.1007/JHEP06(2017)077}{{\em JHEP} {\bf 06} (2017)  077}, \href{http://arxiv.org/abs/1704.00730}{{\tt arXiv:1704.00730 [hep-ph]}}.

\bibitem{Zhang:2021orr}
M.~Zhang, ``{Leptophilic composite asymmetric dark matter and its detection},'' \href{http://dx.doi.org/10.1103/PhysRevD.104.055008}{{\em Phys. Rev. D} {\bf 104} (2021) no.~5, 055008}, \href{http://arxiv.org/abs/2104.06988}{{\tt arXiv:2104.06988 [hep-ph]}}.

\bibitem{Cao:2023smj}
Q.-H. Cao, J.~Guo, J.~Liu, Y.~Luo, and X.-P. Wang, ``{Long-lived searches of vectorlike lepton and its accompanying scalar at colliders},'' \href{http://dx.doi.org/10.1103/PhysRevD.110.015029}{{\em Phys. Rev. D} {\bf 110} (2024) no.~1, 015029}, \href{http://arxiv.org/abs/2311.12934}{{\tt arXiv:2311.12934 [hep-ph]}}.

\bibitem{Liu:2019ogn}
Z.~Liu, Y.-H. Xu, and Y.~Zhang, ``{Probing dark matter particles at CEPC},'' \href{http://dx.doi.org/10.1007/JHEP06(2019)009}{{\em JHEP} {\bf 06} (2019)  009}, \href{http://arxiv.org/abs/1903.12114}{{\tt arXiv:1903.12114 [hep-ph]}}.

\bibitem{Zhang:2022ijx}
Y.~Zhang, M.~Song, and L.~Chen, ``{Dark states with electromagnetic form factors at electron colliders},'' \href{http://dx.doi.org/10.1103/PhysRevD.107.055023}{{\em Phys. Rev. D} {\bf 107} (2023) no.~5, 055023}, \href{http://arxiv.org/abs/2208.08142}{{\tt arXiv:2208.08142 [hep-ph]}}.

\bibitem{Kundu:2021cmo}
S.~Kundu, A.~Guha, P.~K. Das, and P.~S.~B. Dev, ``{A model-independent analysis of leptophilic dark matter at future electron-positron colliders in the mono-photon and mono-Z channels},'' \href{http://arxiv.org/abs/2110.06903}{{\tt arXiv:2110.06903 [hep-ph]}}.

\bibitem{Ge:2023wye}
S.-F. Ge, K.~Ma, X.-D. Ma, and J.~Sheng, ``{Associated Production of Neutrino and Dark Fermion at Future Lepton Colliders},'' \href{http://arxiv.org/abs/2306.00657}{{\tt arXiv:2306.00657 [hep-ph]}}.

\bibitem{Kumar:2013iva}
J.~Kumar and D.~Marfatia, ``{Matrix element analyses of dark matter scattering and annihilation},'' \href{http://dx.doi.org/10.1103/PhysRevD.88.014035}{{\em Phys. Rev. D} {\bf 88} (2013) no.~1, 014035}, \href{http://arxiv.org/abs/1305.1611}{{\tt arXiv:1305.1611 [hep-ph]}}.

\bibitem{Bai:2014osa}
Y.~Bai and J.~Berger, ``{Lepton Portal Dark Matter},'' \href{http://dx.doi.org/10.1007/JHEP08(2014)153}{{\em JHEP} {\bf 08} (2014)  153}, \href{http://arxiv.org/abs/1402.6696}{{\tt arXiv:1402.6696 [hep-ph]}}.

\bibitem{Akrami:2018vks}
{\bf Planck} Collaboration, N.~Aghanim {\em et al.}, ``{Planck 2018 results. I. Overview and the cosmological legacy of Planck},'' \href{http://dx.doi.org/10.1051/0004-6361/201833880}{{\em Astron. Astrophys.} {\bf 641} (2020)  A1}, \href{http://arxiv.org/abs/1807.06205}{{\tt arXiv:1807.06205 [astro-ph.CO]}}.

\bibitem{Ade:2015xua}
{\bf Planck} Collaboration, P.~A.~R. Ade {\em et al.}, ``{Planck 2015 results. XIII. Cosmological parameters},'' \href{http://dx.doi.org/10.1051/0004-6361/201525830}{{\em Astron. Astrophys.} {\bf 594} (2016)  A13},
\href{http://arxiv.org/abs/1502.01589}{{\tt arXiv:1502.01589 [astro-ph.CO]}}.

\bibitem{Kaplan:2009ag}
D.~E. Kaplan, M.~A. Luty, and K.~M. Zurek, ``{Asymmetric Dark Matter},'' \href{http://dx.doi.org/10.1103/PhysRevD.79.115016}{{\em Phys. Rev. D} {\bf 79} (2009)  115016}, \href{http://arxiv.org/abs/0901.4117}{{\tt arXiv:0901.4117 [hep-ph]}}.

\bibitem{Petraki:2013wwa}
K.~Petraki and R.~R. Volkas, ``{Review of asymmetric dark matter},'' \href{http://dx.doi.org/10.1142/S0217751X13300287}{{\em Int. J. Mod. Phys. A} {\bf 28} (2013)  1330028}, \href{http://arxiv.org/abs/1305.4939}{{\tt arXiv:1305.4939 [hep-ph]}}.

\bibitem{Zurek:2013wia}
K.~M. Zurek, ``{Asymmetric Dark Matter: Theories, Signatures, and Constraints},'' \href{http://dx.doi.org/10.1016/j.physrep.2013.12.001}{{\em Phys. Rept.} {\bf 537} (2014)  91--121}, \href{http://arxiv.org/abs/1308.0338}{{\tt arXiv:1308.0338 [hep-ph]}}.

\bibitem{Bai:2013xga}
Y.~Bai and P.~Schwaller, ``{Scale of dark QCD},'' \href{http://dx.doi.org/10.1103/PhysRevD.89.063522}{{\em Phys. Rev. D} {\bf 89} (2014) no.~6, 063522}, \href{http://arxiv.org/abs/1306.4676}{{\tt arXiv:1306.4676 [hep-ph]}}.

\bibitem{ATLAS:2016jza}
{\bf ATLAS} Collaboration, ``{Search for long-lived neutral particles decaying into displaced lepton jets in proton--proton collisions at $\sqrt{s}$ = 13 TeV with the ATLAS detector},''. \url{https://cds.cern.ch/record/2206083}.

\bibitem{Davoudiasl:2013aya}
H.~Davoudiasl, H.-S. Lee, I.~Lewis, and W.~J. Marciano, ``{Higgs Decays as a Window into the Dark Sector},'' \href{http://dx.doi.org/10.1103/PhysRevD.88.015022}{{\em Phys. Rev. D} {\bf 88} (2013) no.~1, 015022}, \href{http://arxiv.org/abs/1304.4935}{{\tt arXiv:1304.4935 [hep-ph]}}.

\bibitem{Davoudiasl:2012ag}
H.~Davoudiasl, H.-S. Lee, and W.~J. Marciano, ``{'Dark' Z implications for Parity Violation, Rare Meson Decays, and Higgs Physics},'' \href{http://dx.doi.org/10.1103/PhysRevD.85.115019}{{\em Phys. Rev. D} {\bf 85} (2012)  115019}, \href{http://arxiv.org/abs/1203.2947}{{\tt arXiv:1203.2947 [hep-ph]}}.

\bibitem{Curtin:2014cca}
D.~Curtin, R.~Essig, S.~Gori, and J.~Shelton, ``{Illuminating Dark Photons with High-Energy Colliders},'' \href{http://dx.doi.org/10.1007/JHEP02(2015)157}{{\em JHEP} {\bf 02} (2015)  157}, \href{http://arxiv.org/abs/1412.0018}{{\tt arXiv:1412.0018 [hep-ph]}}.

\bibitem{Giffin:2020jtl}
P.~Giffin, I.~M. Lewis, and Y.-J. Zheng, ``{Higgs production in association with a dark-Z at future electron positron colliders},'' \href{http://dx.doi.org/10.1088/1361-6471/ac38c1}{{\em J. Phys. G} {\bf 49} (2022) no.~1, 015003}, \href{http://arxiv.org/abs/2012.13404}{{\tt arXiv:2012.13404 [hep-ph]}}.

\bibitem{Davidson:2000hf}
S.~Davidson, S.~Hannestad, and G.~Raffelt, ``{Updated bounds on millicharged particles},'' \href{http://dx.doi.org/10.1088/1126-6708/2000/05/003}{{\em JHEP} {\bf 05} (2000)  003},
\href{http://arxiv.org/abs/hep-ph/0001179}{{\tt arXiv:hep-ph/0001179 [hep-ph]}}.

\bibitem{Soper:2014ska}
D.~E. Soper, M.~Spannowsky, C.~J. Wallace, and T.~M.~P. Tait, ``{Scattering of Dark Particles with Light Mediators},'' \href{http://dx.doi.org/10.1103/PhysRevD.90.115005}{{\em Phys. Rev.} {\bf D90} (2014) no.~11, 115005},
\href{http://arxiv.org/abs/1407.2623}{{\tt arXiv:1407.2623 [hep-ph]}}.

\bibitem{Magill:2018tbb}
G.~Magill, R.~Plestid, M.~Pospelov, and Y.-D. Tsai, ``{Millicharged particles in neutrino experiments},'' \href{http://dx.doi.org/10.1103/PhysRevLett.122.071801}{{\em Phys. Rev. Lett.} {\bf 122} (2019) no.~7, 071801},
\href{http://arxiv.org/abs/1806.03310}{{\tt arXiv:1806.03310 [hep-ph]}}.

\bibitem{Liu:2018jdi}
Z.~Liu and Y.~Zhang, ``{Probing millicharge at BESIII via monophoton searches},'' \href{http://dx.doi.org/10.1103/PhysRevD.99.015004}{{\em Phys. Rev.} {\bf D99} (2019) no.~1, 015004},
\href{http://arxiv.org/abs/1808.00983}{{\tt arXiv:1808.00983 [hep-ph]}}.

\bibitem{Chu:2018qrm}
X.~Chu, J.~Pradler, and L.~Semmelrock, ``{Light dark states with electromagnetic form factors},'' \href{http://dx.doi.org/10.1103/PhysRevD.99.015040}{{\em Phys. Rev. D} {\bf 99} (2019) no.~1, 015040}, \href{http://arxiv.org/abs/1811.04095}{{\tt arXiv:1811.04095 [hep-ph]}}.

\bibitem{Arina:2020mxo}
C.~Arina, A.~Cheek, K.~Mimasu, and L.~Pagani, ``{Light and Darkness: consistently coupling dark matter to photons via effective operators},'' \href{http://dx.doi.org/10.1140/epjc/s10052-021-09010-1}{{\em Eur. Phys. J. C} {\bf 81} (2021) no.~3, 223}, \href{http://arxiv.org/abs/2005.12789}{{\tt arXiv:2005.12789 [hep-ph]}}.

\bibitem{Chu:2020ysb}
X.~Chu, J.-L. Kuo, and J.~Pradler, ``{Dark sector-photon interactions in proton-beam experiments},'' \href{http://dx.doi.org/10.1103/PhysRevD.101.075035}{{\em Phys. Rev. D} {\bf 101} (2020) no.~7, 075035}, \href{http://arxiv.org/abs/2001.06042}{{\tt arXiv:2001.06042 [hep-ph]}}.

\bibitem{Chu:2019rok}
X.~Chu, J.-L. Kuo, J.~Pradler, and L.~Semmelrock, ``{Stellar probes of dark sector-photon interactions},'' \href{http://dx.doi.org/10.1103/PhysRevD.100.083002}{{\em Phys. Rev. D} {\bf 100} (2019) no.~8, 083002}, \href{http://arxiv.org/abs/1908.00553}{{\tt arXiv:1908.00553 [hep-ph]}}.

\bibitem{Krauss:2002px}
L.~M. Krauss, S.~Nasri, and M.~Trodden, ``{A Model for neutrino masses and dark matter},'' \href{http://dx.doi.org/10.1103/PhysRevD.67.085002}{{\em Phys. Rev. D} {\bf 67} (2003)  085002}, \href{http://arxiv.org/abs/hep-ph/0210389}{{\tt arXiv:hep-ph/0210389}}.

\bibitem{Baltz:2002we}
E.~A. Baltz and L.~Bergstrom, ``{Detection of leptonic dark matter},'' \href{http://dx.doi.org/10.1103/PhysRevD.67.043516}{{\em Phys. Rev. D} {\bf 67} (2003)  043516}, \href{http://arxiv.org/abs/hep-ph/0211325}{{\tt arXiv:hep-ph/0211325}}.

\bibitem{Ma:2006km}
E.~Ma, ``{Verifiable radiative seesaw mechanism of neutrino mass and dark matter},'' \href{http://dx.doi.org/10.1103/PhysRevD.73.077301}{{\em Phys. Rev. D} {\bf 73} (2006)  077301}, \href{http://arxiv.org/abs/hep-ph/0601225}{{\tt arXiv:hep-ph/0601225}}.

\bibitem{Hambye:2006zn}
T.~Hambye, K.~Kannike, E.~Ma, and M.~Raidal, ``{Emanations of Dark Matter: Muon Anomalous Magnetic Moment, Radiative Neutrino Mass, and Novel Leptogenesis at the TeV Scale},'' \href{http://dx.doi.org/10.1103/PhysRevD.75.095003}{{\em Phys. Rev. D} {\bf 75} (2007)  095003}, \href{http://arxiv.org/abs/hep-ph/0609228}{{\tt arXiv:hep-ph/0609228}}.

\bibitem{Bernabei:2007gr}
R.~Bernabei {\em et al.}, ``{Investigating electron interacting dark matter},'' \href{http://dx.doi.org/10.1103/PhysRevD.77.023506}{{\em Phys. Rev. D} {\bf 77} (2008)  023506}, \href{http://arxiv.org/abs/0712.0562}{{\tt arXiv:0712.0562 [astro-ph]}}.

\bibitem{Cirelli:2008pk}
M.~Cirelli, M.~Kadastik, M.~Raidal, and A.~Strumia, ``{Model-independent implications of the e+-, anti-proton cosmic ray spectra on properties of Dark Matter},'' \href{http://dx.doi.org/10.1016/j.nuclphysb.2008.11.031}{{\em Nucl. Phys. B} {\bf 813} (2009)  1--21}, \href{http://arxiv.org/abs/0809.2409}{{\tt arXiv:0809.2409 [hep-ph]}}. [Addendum: Nucl.Phys.B 873, 530--533 (2013)].

\bibitem{Chen:2008dh}
C.-R. Chen and F.~Takahashi, ``{Cosmic rays from Leptonic Dark Matter},'' \href{http://dx.doi.org/10.1088/1475-7516/2009/02/004}{{\em JCAP} {\bf 02} (2009)  004}, \href{http://arxiv.org/abs/0810.4110}{{\tt arXiv:0810.4110 [hep-ph]}}.

\bibitem{Bi:2009md}
X.-J. Bi, P.-H. Gu, T.~Li, and X.~Zhang, ``{ATIC and PAMELA Results on Cosmic e+- Excesses and Neutrino Masses},'' \href{http://dx.doi.org/10.1088/1126-6708/2009/04/103}{{\em JHEP} {\bf 04} (2009)  103}, \href{http://arxiv.org/abs/0901.0176}{{\tt arXiv:0901.0176 [hep-ph]}}.

\bibitem{Ibarra:2009bm}
A.~Ibarra, A.~Ringwald, D.~Tran, and C.~Weniger, ``{Cosmic Rays from Leptophilic Dark Matter Decay via Kinetic Mixing},'' \href{http://dx.doi.org/10.1088/1475-7516/2009/08/017}{{\em JCAP} {\bf 08} (2009)  017}, \href{http://arxiv.org/abs/0903.3625}{{\tt arXiv:0903.3625 [hep-ph]}}.

\bibitem{Dev:2013hka}
P.~S.~B. Dev, D.~K. Ghosh, N.~Okada, and I.~Saha, ``{Neutrino Mass and Dark Matter in light of recent AMS-02 results},'' \href{http://dx.doi.org/10.1103/PhysRevD.89.095001}{{\em Phys. Rev. D} {\bf 89} (2014)  095001}, \href{http://arxiv.org/abs/1307.6204}{{\tt arXiv:1307.6204 [hep-ph]}}.

\bibitem{Chang:2014tea}
S.~Chang, R.~Edezhath, J.~Hutchinson, and M.~Luty, ``{Leptophilic Effective WIMPs},'' \href{http://dx.doi.org/10.1103/PhysRevD.90.015011}{{\em Phys. Rev. D} {\bf 90} (2014) no.~1, 015011}, \href{http://arxiv.org/abs/1402.7358}{{\tt arXiv:1402.7358 [hep-ph]}}.

\bibitem{Agrawal:2014ufa}
P.~Agrawal, Z.~Chacko, and C.~B. Verhaaren, ``{Leptophilic Dark Matter and the Anomalous Magnetic Moment of the Muon},'' \href{http://dx.doi.org/10.1007/JHEP08(2014)147}{{\em JHEP} {\bf 08} (2014)  147}, \href{http://arxiv.org/abs/1402.7369}{{\tt arXiv:1402.7369 [hep-ph]}}.

\bibitem{Bell:2014tta}
N.~F. Bell, Y.~Cai, R.~K. Leane, and A.~D. Medina, ``{Leptophilic dark matter with $Z'$ interactions},'' \href{http://dx.doi.org/10.1103/PhysRevD.90.035027}{{\em Phys. Rev. D} {\bf 90} (2014) no.~3, 035027}, \href{http://arxiv.org/abs/1407.3001}{{\tt arXiv:1407.3001 [hep-ph]}}.

\bibitem{Freitas:2014jla}
A.~Freitas and S.~Westhoff, ``{Leptophilic Dark Matter in Lepton Interactions at LEP and ILC},'' \href{http://dx.doi.org/10.1007/JHEP10(2014)116}{{\em JHEP} {\bf 10} (2014)  116}, \href{http://arxiv.org/abs/1408.1959}{{\tt arXiv:1408.1959 [hep-ph]}}.

\bibitem{Cao:2014cda}
Q.-H. Cao, C.-R. Chen, and T.~Gong, ``{Leptophilic dark matter confronts AMS-02 cosmic-ray positron flux},'' \href{http://dx.doi.org/10.1016/j.cjph.2016.11.006}{{\em Chin. J. Phys.} {\bf 55} (2017)  10--15}, \href{http://arxiv.org/abs/1409.7317}{{\tt arXiv:1409.7317 [hep-ph]}}.

\bibitem{Lu:2016ups}
B.-Q. Lu and H.-S. Zong, ``{Leptophilic dark matter in Galactic Center excess},'' \href{http://dx.doi.org/10.1103/PhysRevD.93.083504}{{\em Phys. Rev. D} {\bf 93} (2016) no.~8, 083504}. [Addendum: Phys.Rev.D 93, 089910 (2016)].

\bibitem{Duan:2017pkq}
G.~H. Duan, L.~Feng, F.~Wang, L.~Wu, J.~M. Yang, and R.~Zheng, ``{Simplified TeV leptophilic dark matter in light of DAMPE data},'' \href{http://dx.doi.org/10.1007/JHEP02(2018)107}{{\em JHEP} {\bf 02} (2018)  107}, \href{http://arxiv.org/abs/1711.11012}{{\tt arXiv:1711.11012 [hep-ph]}}.

\bibitem{Madge:2018gfl}
E.~Madge and P.~Schwaller, ``{Leptophilic dark matter from gauged lepton number: Phenomenology and gravitational wave signatures},'' \href{http://dx.doi.org/10.1007/JHEP02(2019)048}{{\em JHEP} {\bf 02} (2019)  048}, \href{http://arxiv.org/abs/1809.09110}{{\tt arXiv:1809.09110 [hep-ph]}}.

\bibitem{Junius:2019dci}
S.~Junius, L.~Lopez-Honorez, and A.~Mariotti, ``{A feeble window on leptophilic dark matter},'' \href{http://dx.doi.org/10.1007/JHEP07(2019)136}{{\em JHEP} {\bf 07} (2019)  136}, \href{http://arxiv.org/abs/1904.07513}{{\tt arXiv:1904.07513 [hep-ph]}}.

\bibitem{Ghosh:2020fdc}
S.~Ghosh, A.~Dutta~Banik, E.~J. Chun, and D.~Majumdar, ``{Leptophilic-portal Dark Matter in the Light of AMS-02 positron excess},'' \href{http://arxiv.org/abs/2003.07675}{{\tt arXiv:2003.07675 [hep-ph]}}.

\bibitem{Chakraborti:2020zxt}
S.~Chakraborti and R.~Islam, ``{Implications of dark sector mixing on leptophilic scalar dark matter},'' \href{http://dx.doi.org/10.1007/JHEP03(2021)032}{{\em JHEP} {\bf 03} (2021)  032}, \href{http://arxiv.org/abs/2007.13719}{{\tt arXiv:2007.13719 [hep-ph]}}.

\bibitem{Horigome:2021qof}
S.-I. Horigome, T.~Katayose, S.~Matsumoto, and I.~Saha, ``{Leptophilic fermion WIMP: Role of future lepton colliders},'' \href{http://dx.doi.org/10.1103/PhysRevD.104.055001}{{\em Phys. Rev. D} {\bf 104} (2021) no.~5, 055001}, \href{http://arxiv.org/abs/2102.08645}{{\tt arXiv:2102.08645 [hep-ph]}}.

\bibitem{Abi:2021ojo}
{\bf Muon g-2} Collaboration, B.~Abi {\em et al.}, ``{Measurement of the Positive Muon Anomalous Magnetic Moment to 0.46 ppm},'' \href{http://dx.doi.org/10.1103/PhysRevLett.126.141801}{{\em Phys. Rev. Lett.} {\bf 126} (2021) no.~14, 141801}, \href{http://arxiv.org/abs/2104.03281}{{\tt arXiv:2104.03281 [hep-ex]}}.

\bibitem{Bernabei:2020mon}
{\bf DAMA/LIBRA} Collaboration, R.~Bernabei {\em et al.}, ``{The DAMA project: Achievements, implications and perspectives},'' \href{http://dx.doi.org/10.1016/j.ppnp.2020.103810}{{\em Prog. Part. Nucl. Phys.} {\bf 114} (2020)  103810}.

\bibitem{Abdollahi:2017nat}
{\bf Fermi-LAT} Collaboration, S.~Abdollahi {\em et al.}, ``{Cosmic-ray electron-positron spectrum from 7 GeV to 2 TeV with the Fermi Large Area Telescope},'' \href{http://dx.doi.org/10.1103/PhysRevD.95.082007}{{\em Phys. Rev. D} {\bf 95} (2017) no.~8, 082007}, \href{http://arxiv.org/abs/1704.07195}{{\tt arXiv:1704.07195 [astro-ph.HE]}}.

\bibitem{DAMPE:2017fbg}
{\bf DAMPE} Collaboration, G.~Ambrosi {\em et al.}, ``{Direct detection of a break in the teraelectronvolt cosmic-ray spectrum of electrons and positrons},'' \href{http://dx.doi.org/10.1038/nature24475}{{\em Nature} {\bf 552} (2017)  63--66}, \href{http://arxiv.org/abs/1711.10981}{{\tt arXiv:1711.10981 [astro-ph.HE]}}.

\bibitem{Adriani:2018ktz}
{\bf CALET} Collaboration, O.~Adriani {\em et al.}, ``{Extended Measurement of the Cosmic-Ray Electron and Positron Spectrum from 11 GeV to 4.8 TeV with the Calorimetric Electron Telescope on the International Space Station},'' \href{http://dx.doi.org/10.1103/PhysRevLett.120.261102}{{\em Phys. Rev. Lett.} {\bf 120} (2018) no.~26, 261102}, \href{http://arxiv.org/abs/1806.09728}{{\tt arXiv:1806.09728 [astro-ph.HE]}}.

\bibitem{AMS:2021nhj}
{\bf AMS} Collaboration, M.~Aguilar {\em et al.}, ``{The Alpha Magnetic Spectrometer (AMS) on the international space station: Part II \textemdash{} Results from the first seven years},'' \href{http://dx.doi.org/10.1016/j.physrep.2020.09.003}{{\em Phys. Rept.} {\bf 894} (2021)  1--116}.

\bibitem{TheFermi-LAT:2015kwa}
{\bf Fermi-LAT} Collaboration, M.~Ajello {\em et al.}, ``{Fermi-LAT Observations of High-Energy $\gamma$-Ray Emission Toward the Galactic Center},'' \href{http://dx.doi.org/10.3847/0004-637X/819/1/44}{{\em Astrophys. J.} {\bf 819} (2016) no.~1, 44}, \href{http://arxiv.org/abs/1511.02938}{{\tt arXiv:1511.02938 [astro-ph.HE]}}.

\bibitem{XENON:2020rca}
{\bf XENON} Collaboration, E.~Aprile {\em et al.}, ``{Excess electronic recoil events in XENON1T},'' \href{http://dx.doi.org/10.1103/PhysRevD.102.072004}{{\em Phys. Rev. D} {\bf 102} (2020) no.~7, 072004}, \href{http://arxiv.org/abs/2006.09721}{{\tt arXiv:2006.09721 [hep-ex]}}.

\bibitem{XENON100:2015tol}
{\bf XENON100} Collaboration, E.~Aprile {\em et al.}, ``{Exclusion of Leptophilic Dark Matter Models using XENON100 Electronic Recoil Data},'' \href{http://dx.doi.org/10.1126/science.aab2069}{{\em Science} {\bf 349} (2015) no.~6250, 851--854}, \href{http://arxiv.org/abs/1507.07747}{{\tt arXiv:1507.07747 [astro-ph.CO]}}.

\bibitem{XENON:2019gfn}
{\bf XENON} Collaboration, E.~Aprile {\em et al.}, ``{Light Dark Matter Search with Ionization Signals in XENON1T},'' \href{http://dx.doi.org/10.1103/PhysRevLett.123.251801}{{\em Phys. Rev. Lett.} {\bf 123} (2019) no.~25, 251801}, \href{http://arxiv.org/abs/1907.11485}{{\tt arXiv:1907.11485 [hep-ex]}}.

\bibitem{LZ:2021xov}
{\bf LZ} Collaboration, D.~S. Akerib {\em et al.}, ``{Projected sensitivities of the LUX-ZEPLIN (LZ) experiment to new physics via low-energy electron recoils},'' \href{http://arxiv.org/abs/2102.11740}{{\tt arXiv:2102.11740 [hep-ex]}}.

\bibitem{Chen:2018vkr}
C.-Y. Chen, J.~Kozaczuk, and Y.-M. Zhong, ``{Exploring leptophilic dark matter with NA64-$\mu$},'' \href{http://dx.doi.org/10.1007/JHEP10(2018)154}{{\em JHEP} {\bf 10} (2018)  154}, \href{http://arxiv.org/abs/1807.03790}{{\tt arXiv:1807.03790 [hep-ph]}}.

\bibitem{Marsicano:2018vin}
L.~Marsicano, M.~Battaglieri, A.~Celentano, R.~De~Vita, and Y.-M. Zhong, ``{Probing Leptophilic Dark Sectors at Electron Beam-Dump Facilities},'' \href{http://dx.doi.org/10.1103/PhysRevD.98.115022}{{\em Phys. Rev. D} {\bf 98} (2018) no.~11, 115022}, \href{http://arxiv.org/abs/1812.03829}{{\tt arXiv:1812.03829 [hep-ex]}}.

\bibitem{DELPHI:2003dlq}
{\bf DELPHI} Collaboration, J.~Abdallah {\em et al.}, ``{Photon events with missing energy in e+ e- collisions at s**(1/2) = 130-GeV to 209-GeV},'' \href{http://dx.doi.org/10.1140/epjc/s2004-02051-8}{{\em Eur. Phys. J. C} {\bf 38} (2005)  395--411}, \href{http://arxiv.org/abs/hep-ex/0406019}{{\tt arXiv:hep-ex/0406019}}.

\bibitem{Birkedal:2004xn}
A.~Birkedal, K.~Matchev, and M.~Perelstein, ``{Dark matter at colliders: A Model independent approach},'' \href{http://dx.doi.org/10.1103/PhysRevD.70.077701}{{\em Phys. Rev. D} {\bf 70} (2004)  077701}, \href{http://arxiv.org/abs/hep-ph/0403004}{{\tt arXiv:hep-ph/0403004}}.

\bibitem{Fox:2008kb}
P.~J. Fox and E.~Poppitz, ``{Leptophilic Dark Matter},'' \href{http://dx.doi.org/10.1103/PhysRevD.79.083528}{{\em Phys. Rev. D} {\bf 79} (2009)  083528}, \href{http://arxiv.org/abs/0811.0399}{{\tt arXiv:0811.0399 [hep-ph]}}.

\bibitem{Konar:2009ae}
P.~Konar, K.~Kong, K.~T. Matchev, and M.~Perelstein, ``{Shedding Light on the Dark Sector with Direct WIMP Production},'' \href{http://dx.doi.org/10.1088/1367-2630/11/10/105004}{{\em New J. Phys.} {\bf 11} (2009)  105004}, \href{http://arxiv.org/abs/0902.2000}{{\tt arXiv:0902.2000 [hep-ph]}}.

\bibitem{Fox:2011fx}
P.~J. Fox, R.~Harnik, J.~Kopp, and Y.~Tsai, ``{LEP Shines Light on Dark Matter},'' \href{http://dx.doi.org/10.1103/PhysRevD.84.014028}{{\em Phys. Rev. D} {\bf 84} (2011)  014028}, \href{http://arxiv.org/abs/1103.0240}{{\tt arXiv:1103.0240 [hep-ph]}}.

\bibitem{Bartels:2012ex}
C.~Bartels, M.~Berggren, and J.~List, ``{Characterising WIMPs at a future $e^+e^-$ Linear Collider},'' \href{http://dx.doi.org/10.1140/epjc/s10052-012-2213-9}{{\em Eur. Phys. J. C} {\bf 72} (2012)  2213}, \href{http://arxiv.org/abs/1206.6639}{{\tt arXiv:1206.6639 [hep-ex]}}.

\bibitem{Dreiner:2012xm}
H.~Dreiner, M.~Huck, M.~Kr\"amer, D.~Schmeier, and J.~Tattersall, ``{Illuminating Dark Matter at the ILC},'' \href{http://dx.doi.org/10.1103/PhysRevD.87.075015}{{\em Phys. Rev. D} {\bf 87} (2013) no.~7, 075015}, \href{http://arxiv.org/abs/1211.2254}{{\tt arXiv:1211.2254 [hep-ph]}}.

\bibitem{Chae:2012bq}
Y.~J. Chae and M.~Perelstein, ``{Dark Matter Search at a Linear Collider: Effective Operator Approach},'' \href{http://dx.doi.org/10.1007/JHEP05(2013)138}{{\em JHEP} {\bf 05} (2013)  138}, \href{http://arxiv.org/abs/1211.4008}{{\tt arXiv:1211.4008 [hep-ph]}}.

\bibitem{Habermehl:2020njb}
M.~Habermehl, M.~Berggren, and J.~List, ``{WIMP Dark Matter at the International Linear Collider},'' \href{http://dx.doi.org/10.1103/PhysRevD.101.075053}{{\em Phys. Rev. D} {\bf 101} (2020) no.~7, 075053}, \href{http://arxiv.org/abs/2001.03011}{{\tt arXiv:2001.03011 [hep-ex]}}.

\bibitem{Kalinowski:2021tyr}
J.~Kalinowski, W.~Kotlarski, K.~Mekala, P.~Sopicki, and A.~F. Zarnecki, ``{Sensitivity of future $e^+e^-$ colliders to processes of dark matter production with light mediator exchange},'' \href{http://arxiv.org/abs/2107.11194}{{\tt arXiv:2107.11194 [hep-ph]}}.

\bibitem{Barman:2021hhg}
B.~Barman, S.~Bhattacharya, S.~Girmohanta, and S.~Jahedi, ``{Catch 'em all: Effective Leptophilic WIMPs at the $e^+\,e^-$ Collider},'' \href{http://arxiv.org/abs/2109.10936}{{\tt arXiv:2109.10936 [hep-ph]}}.

\bibitem{Wan:2014rhl}
N.~Wan, M.~Song, G.~Li, W.-G. Ma, R.-Y. Zhang, and J.-Y. Guo, ``{Searching for dark matter via mono-$Z$ boson production at the ILC},'' \href{http://dx.doi.org/10.1140/epjc/s10052-014-3219-2}{{\em Eur. Phys. J. C} {\bf 74} (2014) no.~12, 3219}, \href{http://arxiv.org/abs/1403.7921}{{\tt arXiv:1403.7921 [hep-ph]}}.

\bibitem{Yu:2014ula}
Z.-H. Yu, X.-J. Bi, Q.-S. Yan, and P.-F. Yin, ``{Dark matter searches in the mono-$Z$ channel at high energy $e^+e^-$ colliders},'' \href{http://dx.doi.org/10.1103/PhysRevD.90.055010}{{\em Phys. Rev. D} {\bf 90} (2014) no.~5, 055010}, \href{http://arxiv.org/abs/1404.6990}{{\tt arXiv:1404.6990 [hep-ph]}}.

\bibitem{Dutta:2017ljq}
S.~Dutta, D.~Sachdeva, and B.~Rawat, ``{Signals of Leptophilic Dark Matter at the ILC},'' \href{http://dx.doi.org/10.1140/epjc/s10052-017-5188-8}{{\em Eur. Phys. J. C} {\bf 77} (2017) no.~9, 639}, \href{http://arxiv.org/abs/1704.03994}{{\tt arXiv:1704.03994 [hep-ph]}}.

\bibitem{Grzadkowski:2020frj}
B.~Grzadkowski, M.~Iglicki, K.~Mekala, and A.~F. Zarnecki, ``{Dark-matter-spin effects at future $e^{+} e^{-}$ colliders},'' \href{http://dx.doi.org/10.1007/JHEP08(2020)052}{{\em JHEP} {\bf 08} (2020)  052}, \href{http://arxiv.org/abs/2003.06719}{{\tt arXiv:2003.06719 [hep-ph]}}.

\bibitem{Kopp:2009et}
J.~Kopp, V.~Niro, T.~Schwetz, and J.~Zupan, ``{DAMA/LIBRA and leptonically interacting Dark Matter},'' \href{http://dx.doi.org/10.1103/PhysRevD.80.083502}{{\em Phys. Rev. D} {\bf 80} (2009)  083502}, \href{http://arxiv.org/abs/0907.3159}{{\tt arXiv:0907.3159 [hep-ph]}}.

\bibitem{Beltran:2010ww}
M.~Beltran, D.~Hooper, E.~W. Kolb, Z.~A.~C. Krusberg, and T.~M.~P. Tait, ``{Maverick dark matter at colliders},'' \href{http://dx.doi.org/10.1007/JHEP09(2010)037}{{\em JHEP} {\bf 09} (2010)  037}, \href{http://arxiv.org/abs/1002.4137}{{\tt arXiv:1002.4137 [hep-ph]}}.

\bibitem{Goodman:2010yf}
J.~Goodman, M.~Ibe, A.~Rajaraman, W.~Shepherd, T.~M.~P. Tait, and H.-B. Yu, ``{Constraints on Light Majorana dark Matter from Colliders},'' \href{http://dx.doi.org/10.1016/j.physletb.2010.11.009}{{\em Phys. Lett. B} {\bf 695} (2011)  185--188}, \href{http://arxiv.org/abs/1005.1286}{{\tt arXiv:1005.1286 [hep-ph]}}.

\bibitem{Bai:2010hh}
Y.~Bai, P.~J. Fox, and R.~Harnik, ``{The Tevatron at the Frontier of Dark Matter Direct Detection},'' \href{http://dx.doi.org/10.1007/JHEP12(2010)048}{{\em JHEP} {\bf 12} (2010)  048}, \href{http://arxiv.org/abs/1005.3797}{{\tt arXiv:1005.3797 [hep-ph]}}.

\bibitem{Goodman:2010ku}
J.~Goodman, M.~Ibe, A.~Rajaraman, W.~Shepherd, T.~M.~P. Tait, and H.-B. Yu, ``{Constraints on Dark Matter from Colliders},'' \href{http://dx.doi.org/10.1103/PhysRevD.82.116010}{{\em Phys. Rev. D} {\bf 82} (2010)  116010}, \href{http://arxiv.org/abs/1008.1783}{{\tt arXiv:1008.1783 [hep-ph]}}.

\bibitem{Fox:2011pm}
P.~J. Fox, R.~Harnik, J.~Kopp, and Y.~Tsai, ``{Missing Energy Signatures of Dark Matter at the LHC},'' \href{http://dx.doi.org/10.1103/PhysRevD.85.056011}{{\em Phys. Rev. D} {\bf 85} (2012)  056011}, \href{http://arxiv.org/abs/1109.4398}{{\tt arXiv:1109.4398 [hep-ph]}}.

\bibitem{Rajaraman:2011wf}
A.~Rajaraman, W.~Shepherd, T.~M.~P. Tait, and A.~M. Wijangco, ``{LHC Bounds on Interactions of Dark Matter},'' \href{http://dx.doi.org/10.1103/PhysRevD.84.095013}{{\em Phys. Rev. D} {\bf 84} (2011)  095013}, \href{http://arxiv.org/abs/1108.1196}{{\tt arXiv:1108.1196 [hep-ph]}}.

\bibitem{Kahlhoefer:2017dnp}
F.~Kahlhoefer, ``{Review of LHC Dark Matter Searches},'' \href{http://dx.doi.org/10.1142/S0217751X1730006X}{{\em Int. J. Mod. Phys. A} {\bf 32} (2017) no.~13, 1730006}, \href{http://arxiv.org/abs/1702.02430}{{\tt arXiv:1702.02430 [hep-ph]}}.

\bibitem{Penning:2017tmb}
B.~Penning, ``{The pursuit of dark matter at colliders\textemdash{}an overview},'' \href{http://dx.doi.org/10.1088/1361-6471/aabea7}{{\em J. Phys. G} {\bf 45} (2018) no.~6, 063001}, \href{http://arxiv.org/abs/1712.01391}{{\tt arXiv:1712.01391 [hep-ex]}}.

\bibitem{CMS:2017zts}
{\bf CMS} Collaboration, A.~M. Sirunyan {\em et al.}, ``{Search for new physics in final states with an energetic jet or a hadronically decaying $W$ or $Z$ boson and transverse momentum imbalance at $\sqrt{s}=13\text{ }\text{ }\mathrm{TeV}$},'' \href{http://dx.doi.org/10.1103/PhysRevD.97.092005}{{\em Phys. Rev. D} {\bf 97} (2018) no.~9, 092005}, \href{http://arxiv.org/abs/1712.02345}{{\tt arXiv:1712.02345 [hep-ex]}}.

\bibitem{ATLAS:2021kxv}
{\bf ATLAS} Collaboration, G.~Aad {\em et al.}, ``{Search for new phenomena in events with an energetic jet and missing transverse momentum in $pp$ collisions at $\sqrt {s}$ =13 TeV with the ATLAS detector},'' \href{http://dx.doi.org/10.1103/PhysRevD.103.112006}{{\em Phys. Rev. D} {\bf 103} (2021) no.~11, 112006}, \href{http://arxiv.org/abs/2102.10874}{{\tt arXiv:2102.10874 [hep-ex]}}.

\bibitem{Matsumoto:2016hbs}
S.~Matsumoto, S.~Mukhopadhyay, and Y.-L.~S. Tsai, ``{Effective Theory of WIMP Dark Matter supplemented by Simplified Models: Singlet-like Majorana fermion case},'' \href{http://dx.doi.org/10.1103/PhysRevD.94.065034}{{\em Phys. Rev. D} {\bf 94} (2016) no.~6, 065034}, \href{http://arxiv.org/abs/1604.02230}{{\tt arXiv:1604.02230 [hep-ph]}}.

\bibitem{XENON:2018voc}
{\bf XENON} Collaboration, E.~Aprile {\em et al.}, ``{Dark Matter Search Results from a One Ton-Year Exposure of XENON1T},'' \href{http://dx.doi.org/10.1103/PhysRevLett.121.111302}{{\em Phys. Rev. Lett.} {\bf 121} (2018) no.~11, 111302}, \href{http://arxiv.org/abs/1805.12562}{{\tt arXiv:1805.12562 [astro-ph.CO]}}.

\bibitem{PandaX-4T:2021bab}
{\bf PandaX-4T} Collaboration, Y.~Meng {\em et al.}, ``{Dark Matter Search Results from the PandaX-4T Commissioning Run},'' \href{http://arxiv.org/abs/2107.13438}{{\tt arXiv:2107.13438 [hep-ex]}}.

\bibitem{Leane:2018kjk}
R.~K. Leane, T.~R. Slatyer, J.~F. Beacom, and K.~C.~Y. Ng, ``{GeV-scale thermal WIMPs: Not even slightly ruled out},'' \href{http://dx.doi.org/10.1103/PhysRevD.98.023016}{{\em Phys. Rev. D} {\bf 98} (2018) no.~2, 023016}, \href{http://arxiv.org/abs/1805.10305}{{\tt arXiv:1805.10305 [hep-ph]}}.

\bibitem{John:2021ugy}
I.~John and T.~Linden, ``{Cosmic-Ray Positrons Strongly Constrain Leptophilic Dark Matter},'' \href{http://arxiv.org/abs/2107.10261}{{\tt arXiv:2107.10261 [astro-ph.HE]}}.

\bibitem{Guha:2018mli}
A.~Guha, P.~S.~B. Dev, and P.~K. Das, ``{Model-independent Astrophysical Constraints on Leptophilic Dark Matter in the Framework of Tsallis Statistics},'' \href{http://dx.doi.org/10.1088/1475-7516/2019/02/032}{{\em JCAP} {\bf 02} (2019)  032}, \href{http://arxiv.org/abs/1810.00399}{{\tt arXiv:1810.00399 [hep-ph]}}.

\bibitem{XENON:2019rxp}
{\bf XENON} Collaboration, E.~Aprile {\em et al.}, ``{Constraining the spin-dependent WIMP-nucleon cross sections with XENON1T},'' \href{http://dx.doi.org/10.1103/PhysRevLett.122.141301}{{\em Phys. Rev. Lett.} {\bf 122} (2019) no.~14, 141301}, \href{http://arxiv.org/abs/1902.03234}{{\tt arXiv:1902.03234 [astro-ph.CO]}}.

\bibitem{Liu:2017drf}
J.~Liu, X.~Chen, and X.~Ji, ``{Current status of direct dark matter detection experiments},'' \href{http://dx.doi.org/10.1038/nphys4039}{{\em Nature Phys.} {\bf 13} (2017) no.~3, 212--216}, \href{http://arxiv.org/abs/1709.00688}{{\tt arXiv:1709.00688 [astro-ph.CO]}}.

\bibitem{Schumann:2019eaa}
M.~Schumann, ``{Direct Detection of WIMP Dark Matter: Concepts and Status},'' \href{http://dx.doi.org/10.1088/1361-6471/ab2ea5}{{\em J. Phys. G} {\bf 46} (2019) no.~10, 103003}, \href{http://arxiv.org/abs/1903.03026}{{\tt arXiv:1903.03026 [astro-ph.CO]}}.

\bibitem{Billard:2021uyg}
J.~Billard {\em et al.}, ``{Direct detection of dark matter\textemdash{}APPEC committee report*},'' \href{http://dx.doi.org/10.1088/1361-6633/ac5754}{{\em Rept. Prog. Phys.} {\bf 85} (2022) no.~5, 056201}, \href{http://arxiv.org/abs/2104.07634}{{\tt arXiv:2104.07634 [hep-ex]}}.

\bibitem{Gaskins:2016cha}
J.~M. Gaskins, ``{A review of indirect searches for particle dark matter},'' \href{http://dx.doi.org/10.1080/00107514.2016.1175160}{{\em Contemp. Phys.} {\bf 57} (2016) no.~4, 496--525}, \href{http://arxiv.org/abs/1604.00014}{{\tt arXiv:1604.00014 [astro-ph.HE]}}.

\bibitem{Leane:2020liq}
R.~K. Leane, ``{Indirect Detection of Dark Matter in the Galaxy},'' in {\em {3rd World Summit on Exploring the Dark Side of the Universe}}, pp.~203--228.
\newblock 2020.
\newblock \href{http://arxiv.org/abs/2006.00513}{{\tt arXiv:2006.00513 [hep-ph]}}.

\bibitem{Slatyer:2021qgc}
T.~R. Slatyer, ``{Les Houches Lectures on Indirect Detection of Dark Matter},'' \href{http://dx.doi.org/10.21468/SciPostPhysLectNotes.53}{{\em SciPost Phys. Lect. Notes} {\bf 53} (2022)  1}, \href{http://arxiv.org/abs/2109.02696}{{\tt arXiv:2109.02696 [hep-ph]}}.

\bibitem{Gori:2022vri}
S.~Gori {\em et al.}, ``{Dark Sector Physics at High-Intensity Experiments},'' \href{http://arxiv.org/abs/2209.04671}{{\tt arXiv:2209.04671 [hep-ph]}}.

\bibitem{Lagouri:2022ier}
T.~Lagouri, ``{Review on Higgs Hidden\textendash{}Dark Sector Physics at High-Energy Colliders},'' \href{http://dx.doi.org/10.3390/sym14071299}{{\em Symmetry} {\bf 14} (2022) no.~7, 1299}.

\bibitem{Ge:2022ius}
S.-F. Ge, X.-G. He, X.-D. Ma, and J.~Sheng, ``{Revisiting the fermionic dark matter absorption on electron target},'' \href{http://dx.doi.org/10.1007/JHEP05(2022)191}{{\em JHEP} {\bf 05} (2022)  191}, \href{http://arxiv.org/abs/2201.11497}{{\tt arXiv:2201.11497 [hep-ph]}}.

\bibitem{Cao:2014ita}
J.~Cao, Z.~Heng, D.~Li, L.~Shang, and P.~Wu, ``{Higgs-strahlung production process $e^{+} e^{-} \to Zh$ at the future Higgs factory in the Minimal Dilaton Model},'' \href{http://dx.doi.org/10.1007/JHEP08(2014)138}{{\em JHEP} {\bf 08} (2014)  138}, \href{http://arxiv.org/abs/1405.4489}{{\tt arXiv:1405.4489 [hep-ph]}}.

\bibitem{Fedderke:2015txa}
M.~A. Fedderke, T.~Lin, and L.-T. Wang, ``{Probing the fermionic Higgs portal at lepton colliders},'' \href{http://dx.doi.org/10.1007/JHEP04(2016)160}{{\em JHEP} {\bf 04} (2016)  160}, \href{http://arxiv.org/abs/1506.05465}{{\tt arXiv:1506.05465 [hep-ph]}}.

\bibitem{Cao:2016qgc}
Q.-H. Cao, Y.~Li, B.~Yan, Y.~Zhang, and Z.~Zhang, ``{Probing dark particles indirectly at the CEPC},'' \href{http://dx.doi.org/10.1016/j.nuclphysb.2016.05.010}{{\em Nucl. Phys. B} {\bf 909} (2016)  197--217}, \href{http://arxiv.org/abs/1604.07536}{{\tt arXiv:1604.07536 [hep-ph]}}.

\bibitem{Cai:2016sjz}
C.~Cai, Z.-H. Yu, and H.-H. Zhang, ``{CEPC Precision of Electroweak Oblique Parameters and Weakly Interacting Dark Matter: the Fermionic Case},'' \href{http://dx.doi.org/10.1016/j.nuclphysb.2017.05.015}{{\em Nucl. Phys. B} {\bf 921} (2017)  181--210}, \href{http://arxiv.org/abs/1611.02186}{{\tt arXiv:1611.02186 [hep-ph]}}.

\bibitem{Liu:2017msv}
N.~Liu and L.~Wu, ``{An indirect probe of the higgsino world at the CEPC},'' \href{http://dx.doi.org/10.1140/epjc/s10052-017-5443-z}{{\em Eur. Phys. J. C} {\bf 77} (2017) no.~12, 868}, \href{http://arxiv.org/abs/1705.02534}{{\tt arXiv:1705.02534 [hep-ph]}}.

\bibitem{Cai:2017wdu}
C.~Cai, Z.-H. Yu, and H.-H. Zhang, ``{CEPC Precision of Electroweak Oblique Parameters and Weakly Interacting Dark Matter: the Scalar Case},'' \href{http://dx.doi.org/10.1016/j.nuclphysb.2017.09.007}{{\em Nucl. Phys. B} {\bf 924} (2017)  128--152}, \href{http://arxiv.org/abs/1705.07921}{{\tt arXiv:1705.07921 [hep-ph]}}.

\bibitem{Xiang:2017yfs}
Q.-F. Xiang, X.-J. Bi, P.-F. Yin, and Z.-H. Yu, ``{Exploring Fermionic Dark Matter via Higgs Boson Precision Measurements at the Circular Electron Positron Collider},'' \href{http://dx.doi.org/10.1103/PhysRevD.97.055004}{{\em Phys. Rev. D} {\bf 97} (2018) no.~5, 055004}, \href{http://arxiv.org/abs/1707.03094}{{\tt arXiv:1707.03094 [hep-ph]}}.

\bibitem{Wang:2017sxx}
J.-W. Wang, X.-J. Bi, Q.-F. Xiang, P.-F. Yin, and Z.-H. Yu, ``{Exploring triplet-quadruplet fermionic dark matter at the LHC and future colliders},'' \href{http://dx.doi.org/10.1103/PhysRevD.97.035021}{{\em Phys. Rev. D} {\bf 97} (2018) no.~3, 035021}, \href{http://arxiv.org/abs/1711.05622}{{\tt arXiv:1711.05622 [hep-ph]}}.

\bibitem{Gao:2021jip}
L.-Q. Gao, X.-J. Bi, J.-W. Wang, Q.-F. Xiang, and P.-F. Yin, ``{Exploring fermionic multiplet dark matter through precision measurements at the CEPC *},'' \href{http://dx.doi.org/10.1088/1674-1137/ac7547}{{\em Chin. Phys. C} {\bf 46} (2022) no.~9, 093112}, \href{http://arxiv.org/abs/2112.02519}{{\tt arXiv:2112.02519 [hep-ph]}}.

\bibitem{Mahbubani:2005pt}
R.~Mahbubani and L.~Senatore, ``{The Minimal model for dark matter and unification},'' \href{http://dx.doi.org/10.1103/PhysRevD.73.043510}{{\em Phys. Rev. D} {\bf 73} (2006)  043510}, \href{http://arxiv.org/abs/hep-ph/0510064}{{\tt arXiv:hep-ph/0510064}}.

\bibitem{Cohen:2011ec}
T.~Cohen, J.~Kearney, A.~Pierce, and D.~Tucker-Smith, ``{Singlet-Doublet Dark Matter},'' \href{http://dx.doi.org/10.1103/PhysRevD.85.075003}{{\em Phys. Rev. D} {\bf 85} (2012)  075003}, \href{http://arxiv.org/abs/1109.2604}{{\tt arXiv:1109.2604 [hep-ph]}}.

\bibitem{Dedes:2014hga}
A.~Dedes and D.~Karamitros, ``{Doublet-Triplet Fermionic Dark Matter},'' \href{http://dx.doi.org/10.1103/PhysRevD.89.115002}{{\em Phys. Rev. D} {\bf 89} (2014) no.~11, 115002}, \href{http://arxiv.org/abs/1403.7744}{{\tt arXiv:1403.7744 [hep-ph]}}.

\bibitem{Planck:2018vyg}
{\bf Planck} Collaboration, N.~Aghanim {\em et al.}, ``{Planck 2018 results. VI. Cosmological parameters},'' \href{http://dx.doi.org/10.1051/0004-6361/201833910}{{\em Astron. Astrophys.} {\bf 641} (2020)  A6}, \href{http://arxiv.org/abs/1807.06209}{{\tt arXiv:1807.06209 [astro-ph.CO]}}. [Erratum: Astron.Astrophys. 652, C4 (2021)].

\bibitem{LZtalk}
{\bf LZ} Collaboration, S.~Haselschwardt, ``Status of the lux-zeplin dark matter experiment.'' \url{https://indico.uchicago.edu/event/427/contributions/1325/}.

\bibitem{Cheung:2012qy}
C.~Cheung, L.~J. Hall, D.~Pinner, and J.~T. Ruderman, ``{Prospects and Blind Spots for Neutralino Dark Matter},'' \href{http://dx.doi.org/10.1007/JHEP05(2013)100}{{\em JHEP} {\bf 05} (2013)  100}, \href{http://arxiv.org/abs/1211.4873}{{\tt arXiv:1211.4873 [hep-ph]}}.

\bibitem{Huang:2014xua}
P.~Huang and C.~E.~M. Wagner, ``{Blind Spots for neutralino Dark Matter in the MSSM with an intermediate $m_A$},'' \href{http://dx.doi.org/10.1103/PhysRevD.90.015018}{{\em Phys. Rev. D} {\bf 90} (2014) no.~1, 015018}, \href{http://arxiv.org/abs/1404.0392}{{\tt arXiv:1404.0392 [hep-ph]}}.

\bibitem{ATLAS:2023tkt}
{\bf ATLAS} Collaboration, G.~Aad {\em et al.}, ``{Combination of searches for invisible decays of the Higgs boson using 139 fb{\ensuremath{-}}1 of proton-proton collision data at s=13 TeV collected with the ATLAS experiment},'' \href{http://dx.doi.org/10.1016/j.physletb.2023.137963}{{\em Phys. Lett. B} {\bf 842} (2023)  137963}, \href{http://arxiv.org/abs/2301.10731}{{\tt arXiv:2301.10731 [hep-ex]}}.

\bibitem{DarkSide-50:2022qzh}
{\bf DarkSide-50} Collaboration, P.~Agnes {\em et al.}, ``{Search for low-mass dark matter WIMPs with 12~ton-day exposure of DarkSide-50},'' \href{http://dx.doi.org/10.1103/PhysRevD.107.063001}{{\em Phys. Rev. D} {\bf 107} (2023) no.~6, 063001}, \href{http://arxiv.org/abs/2207.11966}{{\tt arXiv:2207.11966 [hep-ex]}}.

\bibitem{LZ:2024zvo}
{\bf LZ} Collaboration, J.~Aalbers {\em et al.}, ``{Dark Matter Search Results from 4.2{\,}{\,}Tonne-Years of Exposure of the LUX-ZEPLIN (LZ) Experiment},'' \href{http://dx.doi.org/10.1103/4dyc-z8zf}{{\em Phys. Rev. Lett.} {\bf 135} (2025) no.~1, 011802}, \href{http://arxiv.org/abs/2410.17036}{{\tt arXiv:2410.17036 [hep-ex]}}.

\bibitem{PandaX:2022aac}
{\bf PandaX} Collaboration, W.~Ma {\em et al.}, ``{Search for Solar B8 Neutrinos in the PandaX-4T Experiment Using Neutrino-Nucleus Coherent Scattering},'' \href{http://dx.doi.org/10.1103/PhysRevLett.130.021802}{{\em Phys. Rev. Lett.} {\bf 130} (2023) no.~2, 021802}, \href{http://arxiv.org/abs/2207.04883}{{\tt arXiv:2207.04883 [hep-ex]}}.

\bibitem{PandaX:2024qfu}
{\bf PandaX} Collaboration, Z.~Bo {\em et al.}, ``{Dark Matter Search Results from 1.54{\,}{\,}Tonne{\textperiodcentered}Year Exposure of PandaX-4T},'' \href{http://dx.doi.org/10.1103/PhysRevLett.134.011805}{{\em Phys. Rev. Lett.} {\bf 134} (2025) no.~1, 011805}, \href{http://arxiv.org/abs/2408.00664}{{\tt arXiv:2408.00664 [hep-ex]}}.

\bibitem{XENON:2024hup}
{\bf XENON} Collaboration, E.~Aprile {\em et al.}, ``{First Search for Light Dark Matter in the Neutrino Fog with XENONnT},'' \href{http://dx.doi.org/10.1103/PhysRevLett.134.111802}{{\em Phys. Rev. Lett.} {\bf 134} (2025) no.~11, 111802}, \href{http://arxiv.org/abs/2409.17868}{{\tt arXiv:2409.17868 [hep-ex]}}.

\bibitem{XENON:2025vwd}
{\bf XENON} Collaboration, E.~Aprile {\em et al.}, ``{WIMP Dark Matter Search using a 3.1 tonne $\times$ year Exposure of the XENONnT Experiment},'' \href{http://arxiv.org/abs/2502.18005}{{\tt arXiv:2502.18005 [hep-ex]}}.

\bibitem{McDonald:2024osu}
A.~B. McDonald, ``{Dark matter detection with liquid argon},'' \href{http://dx.doi.org/10.1016/j.nuclphysb.2024.116436}{{\em Nucl. Phys. B} {\bf 1003} (2024)  116436}.

\bibitem{CDEX:2023vvc}
{\bf CDEX} Collaboration, X.~P. Geng {\em et al.}, ``{Projected WIMP sensitivity of the CDEX-50 dark matter experiment},'' \href{http://dx.doi.org/10.1088/1475-7516/2024/07/009}{{\em JCAP} {\bf 07} (2024)  009}, \href{http://arxiv.org/abs/2309.01843}{{\tt arXiv:2309.01843 [hep-ex]}}.

\bibitem{Baudis:2024jnk}
L.~Baudis, ``{DARWIN/XLZD: A future xenon observatory for dark matter and other rare interactions},'' \href{http://dx.doi.org/10.1016/j.nuclphysb.2024.116473}{{\em Nucl. Phys. B} {\bf 1003} (2024)  116473}, \href{http://arxiv.org/abs/2404.19524}{{\tt arXiv:2404.19524 [astro-ph.IM]}}.

\bibitem{PANDA-X:2024dlo}
{\bf PANDA-X, PandaX} Collaboration, A.~Abdukerim {\em et al.}, ``{PandaX-xT{\textemdash}A deep underground multi-ten-tonne liquid xenon observatory},'' \href{http://dx.doi.org/10.1007/s11433-024-2539-y}{{\em Sci. China Phys. Mech. Astron.} {\bf 68} (2025) no.~2, 221011}, \href{http://arxiv.org/abs/2402.03596}{{\tt arXiv:2402.03596 [hep-ex]}}.

\bibitem{OHare:2021utq}
C.~A.~J. O'Hare, ``{New Definition of the Neutrino Floor for Direct Dark Matter Searches},'' \href{http://dx.doi.org/10.1103/PhysRevLett.127.251802}{{\em Phys. Rev. Lett.} {\bf 127} (2021) no.~25, 251802}, \href{http://arxiv.org/abs/2109.03116}{{\tt arXiv:2109.03116 [hep-ph]}}.

\bibitem{Essig:2013lka}
R.~Essig {\em et al.}, ``{Working Group Report: New Light Weakly Coupled Particles},'' in {\em {Proceedings, Community Summer Study 2013: Snowmass on the Mississippi (CSS2013): Minneapolis, MN, USA, July 29-August 6, 2013}}.
\newblock 2013.
\newblock \href{http://arxiv.org/abs/1311.0029}{{\tt arXiv:1311.0029 [hep-ph]}}.
\newblock
\url{https://inspirehep.net/record/1263039/files/arXiv:1311.0029.pdf}.
\newblock

\bibitem{Alekhin:2015byh}
S.~Alekhin {\em et al.}, ``{A facility to Search for Hidden Particles at the CERN SPS: the SHiP physics case},'' \href{http://dx.doi.org/10.1088/0034-4885/79/12/124201}{{\em Rept. Prog. Phys.} {\bf 79} (2016) no.~12, 124201}, \href{http://arxiv.org/abs/1504.04855}{{\tt arXiv:1504.04855 [hep-ph]}}.

\bibitem{Beacham:2019nyx}
J.~Beacham {\em et al.}, ``{Physics Beyond Colliders at CERN: Beyond the Standard Model Working Group Report},'' \href{http://dx.doi.org/10.1088/1361-6471/ab4cd2}{{\em J. Phys.} {\bf G47} (2020) no.~1, 010501},
\href{http://arxiv.org/abs/1901.09966}{{\tt arXiv:1901.09966 [hep-ex]}}.

\bibitem{Curtin:2018mvb}
D.~Curtin {\em et al.}, ``{Long-Lived Particles at the Energy Frontier: The MATHUSLA Physics Case},'' \href{http://dx.doi.org/10.1088/1361-6633/ab28d6}{{\em Rept. Prog. Phys.} {\bf 82} (2019) no.~11, 116201},
\href{http://arxiv.org/abs/1806.07396}{{\tt arXiv:1806.07396 [hep-ph]}}.

\bibitem{Antel:2023hkf}
C.~Antel {\em et al.}, ``{Feebly Interacting Particles: FIPs 2022 workshop report},'' in {\em {Workshop on Feebly-Interacting Particles}}.
\newblock 5, 2023.
\newblock \href{http://arxiv.org/abs/2305.01715}{{\tt arXiv:2305.01715 [hep-ph]}}.

\bibitem{ATLAS:2018nud}
{\bf ATLAS} Collaboration, M.~Aaboud {\em et al.}, ``{Search for photonic signatures of gauge-mediated supersymmetry in 13 TeV $pp$ collisions with the ATLAS detector},'' \href{http://dx.doi.org/10.1103/PhysRevD.97.092006}{{\em Phys. Rev. D} {\bf 97} (2018) no.~9, 092006}, \href{http://arxiv.org/abs/1802.03158}{{\tt arXiv:1802.03158 [hep-ex]}}.

\bibitem{CMS:2017brl}
{\bf CMS} Collaboration, A.~M. Sirunyan {\em et al.}, ``{Search for gauge-mediated supersymmetry in events with at least one photon and missing transverse momentum in pp collisions at $\sqrt{s} = $ 13 TeV},'' \href{http://dx.doi.org/10.1016/j.physletb.2018.02.045}{{\em Phys. Lett. B} {\bf 780} (2018)  118--143}, \href{http://arxiv.org/abs/1711.08008}{{\tt arXiv:1711.08008 [hep-ex]}}.

\bibitem{CMS:2019vzo}
{\bf CMS} Collaboration, A.~M. Sirunyan {\em et al.}, ``{Search for supersymmetry in final states with photons and missing transverse momentum in proton-proton collisions at 13 TeV},'' \href{http://dx.doi.org/10.1007/JHEP06(2019)143}{{\em JHEP} {\bf 06} (2019)  143}, \href{http://arxiv.org/abs/1903.07070}{{\tt arXiv:1903.07070 [hep-ex]}}.

\bibitem{CMS:2019zmd}
{\bf CMS} Collaboration, T.~C. Collaboration {\em et al.}, ``{Search for supersymmetry in proton-proton collisions at 13 TeV in final states with jets and missing transverse momentum},'' \href{http://dx.doi.org/10.1007/JHEP10(2019)244}{{\em JHEP} {\bf 10} (2019)  244}, \href{http://arxiv.org/abs/1908.04722}{{\tt arXiv:1908.04722 [hep-ex]}}.

\bibitem{ATLAS:2020xgt}
{\bf ATLAS} Collaboration, G.~Aad {\em et al.}, ``{Search for new phenomena in final states with large jet multiplicities and missing transverse momentum using $ \sqrt{s} $ = 13 TeV proton-proton collisions recorded by ATLAS in Run 2 of the LHC},'' \href{http://dx.doi.org/10.1007/JHEP10(2020)062}{{\em JHEP} {\bf 10} (2020)  062}, \href{http://arxiv.org/abs/2008.06032}{{\tt arXiv:2008.06032 [hep-ex]}}.

\bibitem{Lee:2018pag}
L.~Lee, C.~Ohm, A.~Soffer, and T.-T. Yu, ``{Collider Searches for Long-Lived Particles Beyond the Standard Model},''
\href{http://arxiv.org/abs/1810.12602}{{\tt arXiv:1810.12602 [hep-ph]}}.

\bibitem{Alimena:2019zri}
J.~Alimena {\em et al.}, ``{Searching for Long-Lived Particles beyond the Standard Model at the Large Hadron Collider},'' \href{http://dx.doi.org/10.1088/1361-6471/ab4574}{{\em J. Phys. G} {\bf 47} (2019) no.~9, 090501},
\href{http://arxiv.org/abs/1903.04497}{{\tt arXiv:1903.04497 [hep-ex]}}.

\bibitem{DeRoeck:2019yvg}
A.~De~Roeck, ``{Searching for long-lived particles at the Large Hadron Collider and beyond},'' \href{http://dx.doi.org/10.1098/rsta.2019.0047}{{\em Phil. Trans. Roy. Soc. Lond. A} {\bf 377} (2019) no.~2161, 20190047}.

\bibitem{Alimena:2021mdu}
D.~Acosta {\em et al.}, ``{Review of opportunities for new long-lived particle triggers in Run 3 of the Large Hadron Collider},'' \href{http://arxiv.org/abs/2110.14675}{{\tt arXiv:2110.14675 [hep-ex]}}.

\bibitem{Knapen:2022afb}
S.~Knapen and S.~Lowette, ``{A guide to hunting long-lived particles at the LHC},'' \href{http://arxiv.org/abs/2212.03883}{{\tt arXiv:2212.03883 [hep-ph]}}.

\bibitem{ATLAS:2012yve}
{\bf ATLAS} Collaboration, G.~Aad {\em et al.}, ``{Observation of a new particle in the search for the Standard Model Higgs boson with the ATLAS detector at the LHC},'' \href{http://dx.doi.org/10.1016/j.physletb.2012.08.020}{{\em Phys. Lett. B} {\bf 716} (2012)  1--29}, \href{http://arxiv.org/abs/1207.7214}{{\tt arXiv:1207.7214 [hep-ex]}}.

\bibitem{CMS:2012qbp}
{\bf CMS} Collaboration, S.~Chatrchyan {\em et al.}, ``{Observation of a New Boson at a Mass of 125 GeV with the CMS Experiment at the LHC},'' \href{http://dx.doi.org/10.1016/j.physletb.2012.08.021}{{\em Phys. Lett. B} {\bf 716} (2012)  30--61}, \href{http://arxiv.org/abs/1207.7235}{{\tt arXiv:1207.7235 [hep-ex]}}.

\bibitem{CEPCStudyGroup:2018rmc}
{\bf CEPC Study Group} Collaboration, ``{CEPC Conceptual Design Report: Volume 1 - Accelerator},'' \href{http://arxiv.org/abs/1809.00285}{{\tt arXiv:1809.00285 [physics.acc-ph]}}.

\bibitem{An:2018dwb}
F.~An {\em et al.}, ``{Precision Higgs physics at the CEPC},'' \href{http://dx.doi.org/10.1088/1674-1137/43/4/043002}{{\em Chin. Phys. C} {\bf 43} (2019) no.~4, 043002}, \href{http://arxiv.org/abs/1810.09037}{{\tt arXiv:1810.09037 [hep-ex]}}.

\bibitem{CEPCAcceleratorStudyGroup:2019myu}
{\bf CEPC Accelerator Study Group} Collaboration, ``{CEPC Input to the ESPP 2018 -Accelerator},'' \href{http://arxiv.org/abs/1901.03169}{{\tt arXiv:1901.03169 [physics.acc-ph]}}.

\bibitem{Phinney:2007gp}
G.~Aarons {\em et al.}, ``{ILC Reference Design Report Volume 3 - Accelerator},'' \href{http://arxiv.org/abs/0712.2361}{{\tt arXiv:0712.2361 [physics.acc-ph]}}.

\bibitem{Behnke:2013xla}
``{The International Linear Collider Technical Design Report - Volume 1: Executive Summary},'' \href{http://arxiv.org/abs/1306.6327}{{\tt arXiv:1306.6327 [physics.acc-ph]}}.

\bibitem{Baer:2013cma}
H.~Baer, T.~Barklow, K.~Fujii, Y.~Gao, A.~Hoang, S.~Kanemura, J.~List, H.~E. Logan, A.~Nomerotski, M.~Perelstein, {\em et al.}, ``{The International Linear Collider Technical Design Report - Volume 2: Physics},''
\href{http://arxiv.org/abs/1306.6352}{{\tt arXiv:1306.6352 [hep-ph]}}.

\bibitem{Behnke:2013lya}
H.~Abramowicz {\em et al.}, ``{The International Linear Collider Technical Design Report - Volume 4: Detectors},'' \href{http://arxiv.org/abs/1306.6329}{{\tt arXiv:1306.6329 [physics.ins-det]}}.

\bibitem{Fujii:2017vwa}
K.~Fujii {\em et al.}, ``{Physics Case for the 250 GeV Stage of the International Linear Collider},'' \href{http://arxiv.org/abs/1710.07621}{{\tt arXiv:1710.07621 [hep-ex]}}.

\bibitem{FCC:2018byv}
{\bf FCC} Collaboration, A.~Abada {\em et al.}, ``{FCC Physics Opportunities}: {Future Circular Collider Conceptual Design Report Volume 1},'' \href{http://dx.doi.org/10.1140/epjc/s10052-019-6904-3}{{\em Eur. Phys. J. C} {\bf 79} (2019) no.~6, 474}.

\bibitem{FCC:2018vvp}
{\bf FCC} Collaboration, A.~Abada {\em et al.}, ``{FCC-hh: The Hadron Collider}: {Future Circular Collider Conceptual Design Report Volume 3},'' \href{http://dx.doi.org/10.1140/epjst/e2019-900087-0}{{\em Eur. Phys. J. ST} {\bf 228} (2019) no.~4, 755--1107}.

\bibitem{FCC:2018bvk}
{\bf FCC} Collaboration, A.~Abada {\em et al.}, ``{HE-LHC: The High-Energy Large Hadron Collider}: {Future Circular Collider Conceptual Design Report Volume 4},'' \href{http://dx.doi.org/10.1140/epjst/e2019-900088-6}{{\em Eur. Phys. J. ST} {\bf 228} (2019) no.~5, 1109--1382}.

\bibitem{Linssen:2012hp}
L.~Linssen, A.~Miyamoto, M.~Stanitzki, and H.~Weerts, ``{Physics and Detectors at CLIC: CLIC Conceptual Design Report},''
\href{http://arxiv.org/abs/1202.5940}{{\tt arXiv:1202.5940 [physics.ins-det]}}.

\bibitem{Klamka:2021cjt}
{\bf CLICdp} Collaboration, J.~Klamka, ``{The CLIC potential for new physics},'' \href{http://dx.doi.org/10.22323/1.398.0714}{{\em PoS} {\bf EPS-HEP2021} (2022)  714}, \href{http://arxiv.org/abs/2111.04787}{{\tt arXiv:2111.04787 [hep-ex]}}.

\bibitem{Blondel:2022qqo}
A.~Blondel {\em et al.}, ``{Searches for long-lived particles at the future FCC-ee},'' \href{http://dx.doi.org/10.3389/fphy.2022.967881}{{\em Front. in Phys.} {\bf 10} (2022)  967881}, \href{http://arxiv.org/abs/2203.05502}{{\tt arXiv:2203.05502 [hep-ex]}}.

\bibitem{Lu:2024fxs}
Y.~Lu, Y.-n. Mao, K.~Wang, and Z.~S. Wang, ``{LAYCAST: LAYered CAvern Surface Tracker at future electron-positron colliders},'' \href{http://arxiv.org/abs/2406.05770}{{\tt arXiv:2406.05770 [hep-ph]}}.

\bibitem{Tian:2022rsi}
M.~Tian, K.~Wang, and Z.~S. Wang, ``{Search for long-lived axions with far detectors at future lepton colliders},'' \href{http://arxiv.org/abs/2201.08960}{{\tt arXiv:2201.08960 [hep-ph]}}.

\bibitem{Kucharczyk:2022pie}
M.~Kucharczyk and M.~Goncerz, ``{Search for exotic decays of the Higgs boson into long-lived particles with jet pairs in the final state at CLIC},'' \href{http://dx.doi.org/10.1007/JHEP03(2023)131}{{\em JHEP} {\bf 03} (2023)  131}, \href{http://arxiv.org/abs/2212.04147}{{\tt arXiv:2212.04147 [hep-ex]}}.

\bibitem{Fuchs:2020cmm}
E.~Fuchs, O.~Matsedonskyi, I.~Savoray, and M.~Schlaffer, ``{Collider searches for scalar singlets across lifetimes},'' \href{http://dx.doi.org/10.1007/JHEP04(2021)019}{{\em JHEP} {\bf 04} (2021)  019}, \href{http://arxiv.org/abs/2008.12773}{{\tt arXiv:2008.12773 [hep-ph]}}.

\bibitem{Strassler:2006ri}
M.~J. Strassler and K.~M. Zurek, ``{Discovering the Higgs through highly-displaced vertices},'' \href{http://dx.doi.org/10.1016/j.physletb.2008.02.008}{{\em Phys. Lett. B} {\bf 661} (2008)  263--267}, \href{http://arxiv.org/abs/hep-ph/0605193}{{\tt arXiv:hep-ph/0605193}}.

\bibitem{Helo:2018qej}
J.~C. Helo, M.~Hirsch, and Z.~S. Wang, ``{Heavy neutral fermions at the high-luminosity LHC},'' \href{http://dx.doi.org/10.1007/JHEP07(2018)056}{{\em JHEP} {\bf 07} (2018)  056},
\href{http://arxiv.org/abs/1803.02212}{{\tt arXiv:1803.02212 [hep-ph]}}.

\bibitem{Dercks:2018wum}
D.~Dercks, H.~K. Dreiner, M.~Hirsch, and Z.~S. Wang, ``{Long-Lived Fermions at AL3X},'' \href{http://dx.doi.org/10.1103/PhysRevD.99.055020}{{\em Phys. Rev.} {\bf D99} (2019) no.~5, 055020},
\href{http://arxiv.org/abs/1811.01995}{{\tt arXiv:1811.01995 [hep-ph]}}.

\bibitem{Kao:2009fg}
Y.~Kao and T.~Takeuchi, ``{Single-Coupling Bounds on R-parity violating Supersymmetry, an update},''
\href{http://arxiv.org/abs/0910.4980}{{\tt arXiv:0910.4980 [hep-ph]}}.

\bibitem{ATLAS:2008xda}
{\bf ATLAS} Collaboration, G.~Aad {\em et al.}, ``{The ATLAS Experiment at the CERN Large Hadron Collider},'' \href{http://dx.doi.org/10.1088/1748-0221/3/08/S08003}{{\em JINST} {\bf 3} (2008)  S08003}.

\bibitem{Gligorov:2018vkc}
V.~V. Gligorov, S.~Knapen, B.~Nachman, M.~Papucci, and D.~J. Robinson, ``{Leveraging the ALICE/L3 cavern for long-lived particle searches},'' \href{http://dx.doi.org/10.1103/PhysRevD.99.015023}{{\em Phys. Rev.} {\bf D99} (2019) no.~1, 015023},
\href{http://arxiv.org/abs/1810.03636}{{\tt arXiv:1810.03636 [hep-ph]}}.

\bibitem{Gligorov:2017nwh}
V.~V. Gligorov, S.~Knapen, M.~Papucci, and D.~J. Robinson, ``{Searching for Long-lived Particles: A Compact Detector for Exotics at LHCb},'' \href{http://dx.doi.org/10.1103/PhysRevD.97.015023}{{\em Phys. Rev.} {\bf D97} (2018) no.~1, 015023},
\href{http://arxiv.org/abs/1708.09395}{{\tt arXiv:1708.09395 [hep-ph]}}.

\bibitem{Feng:2017uoz}
J.~L. Feng, I.~Galon, F.~Kling, and S.~Trojanowski, ``{ForwArd Search ExpeRiment at the LHC},'' \href{http://dx.doi.org/10.1103/PhysRevD.97.035001}{{\em Phys. Rev. D} {\bf 97} (2018) no.~3, 035001}, \href{http://arxiv.org/abs/1708.09389}{{\tt arXiv:1708.09389 [hep-ph]}}.

\bibitem{Croon:2020lrf}
D.~Croon, G.~Elor, R.~K. Leane, and S.~D. McDermott, ``{Supernova Muons: New Constraints on $Z$' Bosons, Axions and ALPs},'' \href{http://dx.doi.org/10.1007/JHEP01(2021)107}{{\em JHEP} {\bf 01} (2021)  107}, \href{http://arxiv.org/abs/2006.13942}{{\tt arXiv:2006.13942 [hep-ph]}}.

\bibitem{Heisig:2012px}
J.~Heisig, ``{Long-lived charged sleptons at the ILC/CLIC},'' in {\em {3rd Linear Collider Forum}}, pp.~398--404.
\newblock DESY, Hamburg, 11, 2012.
\newblock \href{http://arxiv.org/abs/1211.2195}{{\tt arXiv:1211.2195 [hep-ph]}}.

\bibitem{Martyn:2007mj}
H.-U. Martyn, ``{Detection of long-lived staus and gravitinos at the ILC},'' {\em eConf} {\bf C0705302} (2007)  SUS03, \href{http://arxiv.org/abs/0709.1030}{{\tt arXiv:0709.1030 [hep-ph]}}.

\bibitem{Ibarra:2006sz}
A.~Ibarra and S.~Roy, ``{Lepton flavour violation in future linear colliders in the long-lived stau NLSP scenario},'' \href{http://dx.doi.org/10.1088/1126-6708/2007/05/059}{{\em JHEP} {\bf 05} (2007)  059}, \href{http://arxiv.org/abs/hep-ph/0606116}{{\tt arXiv:hep-ph/0606116}}.

\bibitem{SNDatLHC:2022}
{SND@LHC Collaboration}, ``{SND@LHC – Scattering and Neutrino Detector at the LHC}.'' \url{http://www.ship-korea.com/SND.html}, 2022.
\newblock Accessed: 2025-05-20. Describes detector layout and physics goals.

\bibitem{SHiP:2020sos}
{\bf SHiP} Collaboration, C.~Ahdida {\em et al.}, ``{SND@LHC},'' \href{http://arxiv.org/abs/2002.08722}{{\tt arXiv:2002.08722 [physics.ins-det]}}.

\bibitem{SNDLHC:2022ihg}
{\bf SND@LHC} Collaboration, G.~Acampora {\em et al.}, ``{SND@LHC: The Scattering and Neutrino Detector at the LHC},'' \href{http://arxiv.org/abs/2210.02784}{{\tt arXiv:2210.02784 [hep-ex]}}.

\bibitem{FASER:2019dxq}
{\bf FASER} Collaboration, H.~Abreu {\em et al.}, ``{Detecting and Studying High-Energy Collider Neutrinos with FASER at the LHC},'' \href{http://dx.doi.org/10.1140/epjc/s10052-020-7631-5}{{\em Eur. Phys. J. C} {\bf 80} (2020) no.~1, 61}, \href{http://arxiv.org/abs/1908.02310}{{\tt arXiv:1908.02310 [hep-ex]}}.

\bibitem{FASER:2020gpr}
{\bf FASER} Collaboration, H.~Abreu {\em et al.}, ``{Technical Proposal: FASERnu},'' \href{http://arxiv.org/abs/2001.03073}{{\tt arXiv:2001.03073 [physics.ins-det]}}.

\bibitem{Feng:2022inv}
J.~L. Feng {\em et al.}, ``{The Forward Physics Facility at the High-Luminosity LHC},'' \href{http://dx.doi.org/10.1088/1361-6471/ac865e}{{\em J. Phys. G} {\bf 50} (2023) no.~3, 030501}, \href{http://arxiv.org/abs/2203.05090}{{\tt arXiv:2203.05090 [hep-ex]}}.

\bibitem{MammenAbraham:2022xoc}
R.~Mammen~Abraham {\em et al.}, ``{Tau neutrinos in the next decade: from GeV to EeV},'' \href{http://dx.doi.org/10.1088/1361-6471/ac89d2}{{\em J. Phys. G} {\bf 49} (2022) no.~11, 110501}, \href{http://arxiv.org/abs/2203.05591}{{\tt arXiv:2203.05591 [hep-ph]}}.

\bibitem{MammenAbraham:2020hex}
R.~Mammen~Abraham {\em et al.}, ``{Forward Physics Facility - Snowmass 2021 Letter of Interest},''. \url{https://www.osti.gov/biblio/1865358}.

\bibitem{Anchordoqui:2021ghd}
L.~A. Anchordoqui {\em et al.}, ``{The Forward Physics Facility: Sites, experiments, and physics potential},'' \href{http://dx.doi.org/10.1016/j.physrep.2022.04.004}{{\em Phys. Rept.} {\bf 968} (2022)  1--50}, \href{http://arxiv.org/abs/2109.10905}{{\tt arXiv:2109.10905 [hep-ph]}}.

\bibitem{Batell:2021blf}
B.~Batell, J.~L. Feng, and S.~Trojanowski, ``{Detecting Dark Matter with Far-Forward Emulsion and Liquid Argon Detectors at the LHC},'' \href{http://dx.doi.org/10.1103/PhysRevD.103.075023}{{\em Phys. Rev. D} {\bf 103} (2021) no.~7, 075023}, \href{http://arxiv.org/abs/2101.10338}{{\tt arXiv:2101.10338 [hep-ph]}}.

\bibitem{FASER:2019aik}
{\bf FASER} Collaboration, A.~Ariga {\em et al.}, ``{FASER: ForwArd Search ExpeRiment at the LHC},'' \href{http://arxiv.org/abs/1901.04468}{{\tt arXiv:1901.04468 [hep-ex]}}.

\bibitem{FASER:2022hcn}
{\bf FASER} Collaboration, H.~Abreu {\em et al.}, ``{The FASER Detector},'' \href{http://arxiv.org/abs/2207.11427}{{\tt arXiv:2207.11427 [physics.ins-det]}}.

\bibitem{Chou:2016lxi}
J.~P. Chou, D.~Curtin, and H.~J. Lubatti, ``{New Detectors to Explore the Lifetime Frontier},'' \href{http://dx.doi.org/10.1016/j.physletb.2017.01.043}{{\em Phys. Lett.} {\bf B767} (2017)  29--36},
\href{http://arxiv.org/abs/1606.06298}{{\tt arXiv:1606.06298 [hep-ph]}}.

\bibitem{MATHUSLA:2020uve}
{\bf MATHUSLA} Collaboration, C.~Alpigiani {\em et al.}, ``{An Update to the Letter of Intent for MATHUSLA: Search for Long-Lived Particles at the HL-LHC},'' \href{http://arxiv.org/abs/2009.01693}{{\tt arXiv:2009.01693 [physics.ins-det]}}.

\bibitem{Bauer:2019vqk}
M.~Bauer, O.~Brandt, L.~Lee, and C.~Ohm, ``{ANUBIS: Proposal to search for long-lived neutral particles in CERN service shafts},'' \href{http://arxiv.org/abs/1909.13022}{{\tt arXiv:1909.13022 [physics.ins-det]}}.

\bibitem{Cerci:2021nlb}
S.~Cerci {\em et al.}, ``{FACET: A new long-lived particle detector in the very forward region of the CMS experiment},'' \href{http://dx.doi.org/10.1007/JHEP06(2022)110}{{\em JHEP} {\bf 06} (2022)  110}, \href{http://arxiv.org/abs/2201.00019}{{\tt arXiv:2201.00019 [hep-ex]}}.

\bibitem{Aielli:2019ivi}
G.~Aielli {\em et al.}, ``{Expression of Interest for the CODEX-b Detector},''
\href{http://arxiv.org/abs/1911.00481}{{\tt arXiv:1911.00481 [hep-ex]}}.

\bibitem{Pinfold:2019nqj}
J.~L. Pinfold, ``{The MoEDAL Experiment at the LHC\textemdash{}A Progress Report},'' \href{http://dx.doi.org/10.3390/universe5020047}{{\em Universe} {\bf 5} (2019) no.~2, 47}.

\bibitem{Pinfold:2019zwp}
J.~L. Pinfold, ``{The MoEDAL experiment: a new light on the high-energy frontier},'' \href{http://dx.doi.org/10.1098/rsta.2019.0382}{{\em Phil. Trans. Roy. Soc. Lond. A} {\bf 377} (2019) no.~2161, 20190382}.

\bibitem{DeVries:2020jbs}
J.~De~Vries, H.~K. Dreiner, J.~Y. G\"unther, Z.~S. Wang, and G.~Zhou, ``{Long-lived Sterile Neutrinos at the LHC in Effective Field Theory},'' \href{http://dx.doi.org/10.1007/JHEP03(2021)148}{{\em JHEP} {\bf 03} (2021)  148}, \href{http://arxiv.org/abs/2010.07305}{{\tt arXiv:2010.07305 [hep-ph]}}.

\bibitem{Mao:2023zzk}
Y.-n. Mao, K.~Wang, and Z.~S. Wang, ``{Can we discover lepton number violation with the LHC far detectors?},'' \href{http://dx.doi.org/10.1103/PhysRevD.108.095025}{{\em Phys. Rev. D} {\bf 108} (2023) no.~9, 095025}, \href{http://arxiv.org/abs/2305.03908}{{\tt arXiv:2305.03908 [hep-ph]}}.

\bibitem{Chrzaszcz:2020emg}
M.~Chrzaszcz, M.~Drewes, and J.~Hajer, ``{HECATE: A long-lived particle detector concept for the FCC-ee or CEPC},'' \href{http://dx.doi.org/10.1140/epjc/s10052-021-09253-y}{{\em Eur. Phys. J. C} {\bf 81} (2021) no.~6, 546}, \href{http://arxiv.org/abs/2011.01005}{{\tt arXiv:2011.01005 [hep-ph]}}.

\bibitem{Schafer:2022shi}
R.~Sch\"afer, F.~Tillinger, and S.~Westhoff, ``{Near or far detectors? A case study for long-lived particle searches at electron-positron colliders},'' \href{http://dx.doi.org/10.1103/PhysRevD.107.076022}{{\em Phys. Rev. D} {\bf 107} (2023) no.~7, 076022}, \href{http://arxiv.org/abs/2202.11714}{{\tt arXiv:2202.11714 [hep-ph]}}.

\bibitem{Sakaki:2020mqb}
Y.~Sakaki and D.~Ueda, ``{Searching for new light particles at the international linear collider main beam dump},'' \href{http://dx.doi.org/10.1103/PhysRevD.103.035024}{{\em Phys. Rev. D} {\bf 103} (2021) no.~3, 035024}, \href{http://arxiv.org/abs/2009.13790}{{\tt arXiv:2009.13790 [hep-ph]}}.

\bibitem{Satyamurthy:2012zz}
P.~Satyamurthy {\em et al.}, ``{Design of an 18-MW vortex flow water beam dump for 500-GeV electrons/positrons of an international linear collider},'' \href{http://dx.doi.org/10.1016/j.nima.2012.01.075}{{\em Nucl. Instrum. Meth. A} {\bf 679} (2012)  67--81}.

\bibitem{Dolan:2017osp}
M.~J. Dolan, T.~Ferber, C.~Hearty, F.~Kahlhoefer, and K.~Schmidt-Hoberg, ``{Revised constraints and Belle II sensitivity for visible and invisible axion-like particles},'' \href{http://dx.doi.org/10.1007/JHEP12(2017)094}{{\em JHEP} {\bf 12} (2017)  094},
\href{http://arxiv.org/abs/1709.00009}{{\tt arXiv:1709.00009 [hep-ph]}}.

\bibitem{Jaeckel:2017tud}
J.~Jaeckel, P.~C. Malta, and J.~Redondo, ``{Decay photons from the axionlike particles burst of type II supernovae},'' \href{http://dx.doi.org/10.1103/PhysRevD.98.055032}{{\em Phys. Rev. D} {\bf 98} (2018) no.~5, 055032}, \href{http://arxiv.org/abs/1702.02964}{{\tt arXiv:1702.02964 [hep-ph]}}.

\bibitem{Cadamuro:2011fd}
D.~Cadamuro and J.~Redondo, ``{Cosmological bounds on pseudo Nambu-Goldstone bosons},'' \href{http://dx.doi.org/10.1088/1475-7516/2012/02/032}{{\em JCAP} {\bf 02} (2012)  032}, \href{http://arxiv.org/abs/1110.2895}{{\tt arXiv:1110.2895 [hep-ph]}}.

\bibitem{Dobrich:2015jyk}
B.~Doebrich, J.~Jaeckel, F.~Kahlhoefer, A.~Ringwald, and K.~Schmidt-Hoberg, ``{ALPtraum: ALP production in proton beam dump experiments},'' \href{http://dx.doi.org/10.1007/JHEP02(2016)018}{{\em JHEP} {\bf 02} (2016)  018},
\href{http://arxiv.org/abs/1512.03069}{{\tt arXiv:1512.03069 [hep-ph]}}.

\bibitem{Dobrich:2019dxc}
B.~Döbrich, J.~Jaeckel, and T.~Spadaro, ``{Light in the beam dump. Axion-Like Particle production from decay photons in proton beam-dumps},'' \href{http://dx.doi.org/10.1007/JHEP05(2019)213}{{\em JHEP} {\bf 05} (2019)  213},
\href{http://arxiv.org/abs/1904.02091}{{\tt arXiv:1904.02091 [hep-ph]}}.

\bibitem{Chen:2017awl}
C.-Y. Chen, M.~Pospelov, and Y.-M. Zhong, ``{Muon Beam Experiments to Probe the Dark Sector},'' \href{http://dx.doi.org/10.1103/PhysRevD.95.115005}{{\em Phys. Rev. D} {\bf 95} (2017) no.~11, 115005}, \href{http://arxiv.org/abs/1701.07437}{{\tt arXiv:1701.07437 [hep-ph]}}.

\bibitem{Asai:2021xtg}
K.~Asai, T.~Moroi, and A.~Niki, ``{Leptophilic Gauge Bosons at ILC Beam Dump Experiment},'' \href{http://dx.doi.org/10.1016/j.physletb.2021.136374}{{\em Phys. Lett. B} {\bf 818} (2021)  136374}, \href{http://arxiv.org/abs/2104.00888}{{\tt arXiv:2104.00888 [hep-ph]}}.

\bibitem{Asai:2022zxw}
K.~Asai, A.~Das, J.~Li, T.~Nomura, and O.~Seto, ``{Chiral Z' in FASER, FASER2, DUNE, and ILC beam dump experiments},'' \href{http://dx.doi.org/10.1103/PhysRevD.106.095033}{{\em Phys. Rev. D} {\bf 106} (2022) no.~9, 095033}, \href{http://arxiv.org/abs/2206.12676}{{\tt arXiv:2206.12676 [hep-ph]}}.

\bibitem{ATLAS:2022hsp}
{\bf ATLAS} Collaboration, ``{Snowmass White Paper Contribution: Physics with the Phase-2 ATLAS and CMS Detectors},''. \url{https://cds.cern.ch/record/2805993}.

\bibitem{Kling:2021fwx}
F.~Kling and S.~Trojanowski, ``{Forward experiment sensitivity estimator for the LHC and future hadron colliders},'' \href{http://dx.doi.org/10.1103/PhysRevD.104.035012}{{\em Phys. Rev. D} {\bf 104} (2021) no.~3, 035012}, \href{http://arxiv.org/abs/2105.07077}{{\tt arXiv:2105.07077 [hep-ph]}}.

\bibitem{Domingo:2023dew}
F.~Domingo, J.~G\"unther, J.~S. Kim, and Z.~S. Wang, ``{A C++ program for estimating detector sensitivities to long-lived particles: Displaced Decay Counter},'' \href{http://arxiv.org/abs/2308.07371}{{\tt arXiv:2308.07371 [hep-ph]}}.

\bibitem{Ovchynnikov:2023cry}
M.~Ovchynnikov, J.-L. Tastet, O.~Mikulenko, and K.~Bondarenko, ``{Sensitivities to feebly interacting particles: Public and unified calculations},'' \href{http://dx.doi.org/10.1103/PhysRevD.108.075028}{{\em Phys. Rev. D} {\bf 108} (2023) no.~7, 075028}, \href{http://arxiv.org/abs/2305.13383}{{\tt arXiv:2305.13383 [hep-ph]}}.

\bibitem{Delahaye:2019omf}
J.~P. Delahaye, M.~Diemoz, K.~Long, B.~Mansouli\'e, N.~Pastrone, L.~Rivkin, D.~Schulte, A.~Skrinsky, and A.~Wulzer, ``{Muon Colliders},'' \href{http://arxiv.org/abs/1901.06150}{{\tt arXiv:1901.06150 [physics.acc-ph]}}.

\bibitem{Long:2020wfp}
K.~Long, D.~Lucchesi, M.~Palmer, N.~Pastrone, D.~Schulte, and V.~Shiltsev, ``{Muon colliders to expand frontiers of particle physics},'' \href{http://arxiv.org/abs/2007.15684}{{\tt arXiv:2007.15684 [physics.acc-ph]}}.

\bibitem{AlAli:2021let}
H.~Al~Ali {\em et al.}, ``{The Muon Smasher's Guide},'' \href{http://arxiv.org/abs/2103.14043}{{\tt arXiv:2103.14043 [hep-ph]}}.

\bibitem{Accettura:2023ked}
C.~Accettura {\em et al.}, ``{Towards a muon collider},'' \href{http://dx.doi.org/10.1140/epjc/s10052-023-11889-x}{{\em Eur. Phys. J. C} {\bf 83} (2023) no.~9, 864}, \href{http://arxiv.org/abs/2303.08533}{{\tt arXiv:2303.08533 [physics.acc-ph]}}.

\bibitem{MuonCollider:2022xlm}
{\bf Muon Collider} Collaboration, J.~de~Blas {\em et al.}, ``{The physics case of a 3 TeV muon collider stage},'' \href{http://arxiv.org/abs/2203.07261}{{\tt arXiv:2203.07261 [hep-ph]}}.

\bibitem{Lu:2020dkx}
M.~Lu, A.~M. Levin, C.~Li, A.~Agapitos, Q.~Li, F.~Meng, S.~Qian, J.~Xiao, and T.~Yang, ``{The physics case for an electron-muon collider},'' \href{http://dx.doi.org/10.1155/2021/6693618}{{\em Adv. High Energy Phys.} {\bf 2021} (2021)  6693618}, \href{http://arxiv.org/abs/2010.15144}{{\tt arXiv:2010.15144 [hep-ph]}}.

\bibitem{Bouzas:2021sif}
A.~O. Bouzas and F.~Larios, ``{Two-to-Two Processes at an Electron-Muon Collider},'' \href{http://dx.doi.org/10.1155/2022/3603613}{{\em Adv. High Energy Phys.} {\bf 2022} (2022)  3603613}, \href{http://arxiv.org/abs/2109.02769}{{\tt arXiv:2109.02769 [hep-ph]}}.

\bibitem{Dainton:2006wd}
J.~B. Dainton, M.~Klein, P.~Newman, E.~Perez, and F.~Willeke, ``{Deep inelastic electron-nucleon scattering at the LHC},'' \href{http://dx.doi.org/10.1088/1748-0221/1/10/P10001}{{\em JINST} {\bf 1} (2006)  P10001}, \href{http://arxiv.org/abs/hep-ex/0603016}{{\tt arXiv:hep-ex/0603016}}.

\bibitem{LHeCStudyGroup:2012zhm}
{\bf LHeC Study Group} Collaboration, J.~L. Abelleira~Fernandez {\em et al.}, ``{A Large Hadron Electron Collider at CERN: Report on the Physics and Design Concepts for Machine and Detector},'' \href{http://dx.doi.org/10.1088/0954-3899/39/7/075001}{{\em J. Phys. G} {\bf 39} (2012)  075001}, \href{http://arxiv.org/abs/1206.2913}{{\tt arXiv:1206.2913 [physics.acc-ph]}}.

\bibitem{Klein:2018rhq}
M.~Klein, {\em {Future Deep Inelastic Scattering with the LHeC}}, \href{http://dx.doi.org/10.1142/9789813238053_0015}{pp.~303--347}.
\newblock 2019.
\newblock \href{http://arxiv.org/abs/1802.04317}{{\tt arXiv:1802.04317 [hep-ph]}}.

\bibitem{Cheung:2021iev}
K.~Cheung and Z.~S. Wang, ``{Physics potential of a muon-proton collider},'' \href{http://dx.doi.org/10.1103/PhysRevD.103.116009}{{\em Phys. Rev. D} {\bf 103} (2021)  116009}, \href{http://arxiv.org/abs/2101.10476}{{\tt arXiv:2101.10476 [hep-ph]}}.

\bibitem{Caliskan:2017meb}
A.~Caliskan, S.~O. Kara, and A.~Ozansoy, ``{Excited muon searches at the FCC based muon-hadron colliders},'' \href{http://dx.doi.org/10.1155/2017/1540243}{{\em Adv. High Energy Phys.} {\bf 2017} (2017)  1540243}, \href{http://arxiv.org/abs/1701.03426}{{\tt arXiv:1701.03426 [hep-ph]}}.

\bibitem{Acar:2017eli}
Y.~C. Acar, U.~Kaya, and B.~B. Oner, ``{Resonant production of color octet muons at Future Circular Collider-based muon-proton colliders},'' \href{http://dx.doi.org/10.1088/1674-1137/42/8/083108}{{\em Chin. Phys. C} {\bf 42} (2018) no.~8, 083108}, \href{http://arxiv.org/abs/1703.04030}{{\tt arXiv:1703.04030 [hep-ph]}}.

\bibitem{Aad:2020sgw}
{\bf ATLAS} Collaboration, G.~Aad {\em et al.}, ``{Search for a scalar partner of the top quark in the all-hadronic $t{\bar{t}}$ plus missing transverse momentum final state at $\sqrt{s}=13$ TeV with the ATLAS detector},'' \href{http://dx.doi.org/10.1140/epjc/s10052-020-8102-8}{{\em Eur. Phys. J. C} {\bf 80} (2020) no.~8, 737}, \href{http://arxiv.org/abs/2004.14060}{{\tt arXiv:2004.14060 [hep-ex]}}.

\bibitem{Aad:2019vvi}
{\bf ATLAS} Collaboration, G.~Aad {\em et al.}, ``{Search for chargino-neutralino production with mass splittings near the electroweak scale in three-lepton final states in $\sqrt {s}$=13 TeV $pp$ collisions with the ATLAS detector},'' \href{http://dx.doi.org/10.1103/PhysRevD.101.072001}{{\em Phys. Rev. D} {\bf 101} (2020) no.~7, 072001}, \href{http://arxiv.org/abs/1912.08479}{{\tt arXiv:1912.08479 [hep-ex]}}.

\bibitem{Aad:2019qnd}
{\bf ATLAS} Collaboration, G.~Aad {\em et al.}, ``{Searches for electroweak production of supersymmetric particles with compressed mass spectra in $\sqrt{s}=$ 13 TeV $pp$ collisions with the ATLAS detector},'' \href{http://dx.doi.org/10.1103/PhysRevD.101.052005}{{\em Phys. Rev. D} {\bf 101} (2020) no.~5, 052005}, \href{http://arxiv.org/abs/1911.12606}{{\tt arXiv:1911.12606 [hep-ex]}}.

\bibitem{Han:2013usa}
C.~Han, A.~Kobakhidze, N.~Liu, A.~Saavedra, L.~Wu, and J.~M. Yang, ``{Probing Light Higgsinos in Natural SUSY from Monojet Signals at the LHC},'' \href{http://dx.doi.org/10.1007/JHEP02(2014)049}{{\em JHEP} {\bf 02} (2014)  049}, \href{http://arxiv.org/abs/1310.4274}{{\tt arXiv:1310.4274 [hep-ph]}}.

\bibitem{Han:2014xoa}
C.~Han, L.~Wu, J.~M. Yang, M.~Zhang, and Y.~Zhang, ``{New approach for detecting a compressed bino/wino at the LHC},'' \href{http://dx.doi.org/10.1103/PhysRevD.91.055030}{{\em Phys. Rev. D} {\bf 91} (2015)  055030}, \href{http://arxiv.org/abs/1409.4533}{{\tt arXiv:1409.4533 [hep-ph]}}.

\bibitem{Kobakhidze:2016mdx}
A.~Kobakhidze, M.~Talia, and L.~Wu, ``{Probing the MSSM explanation of the muon g-2 anomaly in dark matter experiments and at a 100 TeV $pp$ collider},'' \href{http://dx.doi.org/10.1103/PhysRevD.95.055023}{{\em Phys. Rev. D} {\bf 95} (2017) no.~5, 055023}, \href{http://arxiv.org/abs/1608.03641}{{\tt arXiv:1608.03641 [hep-ph]}}.

\bibitem{Han:2016gvr}
C.~Han, K.-i. Hikasa, L.~Wu, J.~M. Yang, and Y.~Zhang, ``{Status of CMSSM in light of current LHC Run-2 and LUX data},'' \href{http://dx.doi.org/10.1016/j.physletb.2017.04.026}{{\em Phys. Lett. B} {\bf 769} (2017)  470--476}, \href{http://arxiv.org/abs/1612.02296}{{\tt arXiv:1612.02296 [hep-ph]}}.

\bibitem{Abdughani:2017dqs}
M.~Abdughani, L.~Wu, and J.~M. Yang, ``{Status and prospects of light bino\textendash{}higgsino dark matter in natural SUSY},'' \href{http://dx.doi.org/10.1140/epjc/s10052-017-5485-2}{{\em Eur. Phys. J. C} {\bf 78} (2018) no.~1, 4}, \href{http://arxiv.org/abs/1705.09164}{{\tt arXiv:1705.09164 [hep-ph]}}.

\bibitem{Ren:2017ymm}
J.~Ren, L.~Wu, J.~M. Yang, and J.~Zhao, ``{Exploring supersymmetry with machine learning},'' \href{http://dx.doi.org/10.1016/j.nuclphysb.2019.114613}{{\em Nucl. Phys. B} {\bf 943} (2019)  114613}, \href{http://arxiv.org/abs/1708.06615}{{\tt arXiv:1708.06615 [hep-ph]}}.

\bibitem{Duan:2017ucw}
G.~H. Duan, W.~Wang, L.~Wu, J.~M. Yang, and J.~Zhao, ``{Probing GeV-scale MSSM neutralino dark matter in collider and direct detection experiments},'' \href{http://dx.doi.org/10.1016/j.physletb.2018.01.030}{{\em Phys. Lett. B} {\bf 778} (2018)  296--302}, \href{http://arxiv.org/abs/1711.03893}{{\tt arXiv:1711.03893 [hep-ph]}}.

\bibitem{Duan:2018rls}
G.~H. Duan, K.-I. Hikasa, J.~Ren, L.~Wu, and J.~M. Yang, ``{Probing bino-wino coannihilation dark matter below the neutrino floor at the LHC},'' \href{http://dx.doi.org/10.1103/PhysRevD.98.015010}{{\em Phys. Rev. D} {\bf 98} (2018) no.~1, 015010}, \href{http://arxiv.org/abs/1804.05238}{{\tt arXiv:1804.05238 [hep-ph]}}.

\bibitem{Abdughani:2019wss}
M.~Abdughani and L.~Wu, ``{On the coverage of neutralino dark matter in coannihilations at the upgraded LHC},'' \href{http://dx.doi.org/10.1140/epjc/s10052-020-7793-1}{{\em Eur. Phys. J. C} {\bf 80} (2020) no.~3, 233}, \href{http://arxiv.org/abs/1908.11350}{{\tt arXiv:1908.11350 [hep-ph]}}.

\bibitem{Abdughani:2019wai}
M.~Abdughani, K.-I. Hikasa, L.~Wu, J.~M. Yang, and J.~Zhao, ``{Testing electroweak SUSY for muon $g$ \ensuremath{-} 2 and dark matter at the LHC and beyond},'' \href{http://dx.doi.org/10.1007/JHEP11(2019)095}{{\em JHEP} {\bf 11} (2019)  095}, \href{http://arxiv.org/abs/1909.07792}{{\tt arXiv:1909.07792 [hep-ph]}}.

\bibitem{Gu:2020ozv}
Y.~Gu, M.~Khlopov, L.~Wu, J.~M. Yang, and B.~Zhu, ``{Light gravitino dark matter: LHC searches and the Hubble tension},'' \href{http://dx.doi.org/10.1103/PhysRevD.102.115005}{{\em Phys. Rev. D} {\bf 102} (2020) no.~11, 115005}, \href{http://arxiv.org/abs/2006.09906}{{\tt arXiv:2006.09906 [hep-ph]}}.

\bibitem{Abdughani:2021pdc}
M.~Abdughani, Y.-Z. Fan, L.~Feng, Y.-L.~S. Tsai, L.~Wu, and Q.~Yuan, ``{A common origin of muon g-2 anomaly, Galaxy Center GeV excess and AMS-02 anti-proton excess in the NMSSM},'' \href{http://dx.doi.org/10.1016/j.scib.2021.07.029}{{\em Sci. Bull.} {\bf 66} (2021)  2170--2174}, \href{http://arxiv.org/abs/2104.03274}{{\tt arXiv:2104.03274 [hep-ph]}}.

\bibitem{Wang:2021bcx}
F.~Wang, L.~Wu, Y.~Xiao, J.~M. Yang, and Y.~Zhang, ``{GUT-scale constrained SUSY in light of new muon g-2 measurement},'' \href{http://dx.doi.org/10.1016/j.nuclphysb.2021.115486}{{\em Nucl. Phys. B} {\bf 970} (2021)  115486}, \href{http://arxiv.org/abs/2104.03262}{{\tt arXiv:2104.03262 [hep-ph]}}.

\bibitem{Gu:2021lni}
Y.~Gu, L.~Wu, and B.~Zhu, ``{Axion dark radiation: Hubble tension and the Hyper-Kamiokande neutrino experiment},'' \href{http://dx.doi.org/10.1103/PhysRevD.105.095008}{{\em Phys. Rev. D} {\bf 105} (2022) no.~9, 095008}, \href{http://arxiv.org/abs/2105.07232}{{\tt arXiv:2105.07232 [hep-ph]}}.

\bibitem{Lv:2022pme}
H.~Lv, D.~Wang, and L.~Wu, ``{Deep learning jet images as a probe of light Higgsino dark matter at the LHC},'' \href{http://dx.doi.org/10.1103/PhysRevD.106.055008}{{\em Phys. Rev. D} {\bf 106} (2022) no.~5, 055008}, \href{http://arxiv.org/abs/2203.14569}{{\tt arXiv:2203.14569 [hep-ph]}}.

\bibitem{Flambaum:2022lkc}
V.~V. Flambaum, X.~Liu, I.~Samsonov, L.~Wu, and B.~Zhu, ``{Probing supersymmetry breaking scale with atomic clocks},'' \href{http://dx.doi.org/10.1016/j.nuclphysb.2023.116260}{{\em Nucl. Phys. B} {\bf 993} (2023)  116260}, \href{http://arxiv.org/abs/2209.03231}{{\tt arXiv:2209.03231 [hep-ph]}}.

\bibitem{Hu:2014eia}
S.~L. Hu, N.~Liu, J.~Ren, and L.~Wu, ``{Revisiting Associated Production of 125 GeV Higgs Boson with a Photon at a Higgs Factory},'' \href{http://dx.doi.org/10.1088/0954-3899/41/12/125004}{{\em J. Phys. G} {\bf 41} (2014) no.~12, 125004}, \href{http://arxiv.org/abs/1402.3050}{{\tt arXiv:1402.3050 [hep-ph]}}.

\bibitem{Cao:2014rma}
J.~Cao, C.~Han, J.~Ren, L.~Wu, J.~M. Yang, and Y.~Zhang, ``{SUSY effects in Higgs productions at high energy $e^+e^-$ colliders},'' \href{http://dx.doi.org/10.1088/1674-1137/40/11/113104}{{\em Chin. Phys. C} {\bf 40} (2016) no.~11, 113104}, \href{http://arxiv.org/abs/1410.1018}{{\tt arXiv:1410.1018 [hep-ph]}}.

\bibitem{Duan:2018cgb}
G.~H. Duan, C.~Han, B.~Peng, L.~Wu, and J.~M. Yang, ``{Vacuum stability in stau-neutralino coannihilation in MSSM},'' \href{http://dx.doi.org/10.1016/j.physletb.2018.12.001}{{\em Phys. Lett. B} {\bf 788} (2019)  475--479}, \href{http://arxiv.org/abs/1809.10061}{{\tt arXiv:1809.10061 [hep-ph]}}.

\bibitem{Han:2020exx}
C.~Han, M.~L. L\'opez-Ib\'a\~nez, A.~Melis, O.~Vives, L.~Wu, and J.~M. Yang, ``{LFV and (g-2) in non-universal SUSY models with light higgsinos},'' \href{http://dx.doi.org/10.1007/JHEP05(2020)102}{{\em JHEP} {\bf 05} (2020)  102}, \href{http://arxiv.org/abs/2003.06187}{{\tt arXiv:2003.06187 [hep-ph]}}.

\bibitem{Athron:2022uzz}
P.~Athron, C.~Balazs, A.~Fowlie, H.~Lv, W.~Su, L.~Wu, J.~M. Yang, and Y.~Zhang, ``{Global fits of SUSY at future Higgs factories},'' \href{http://dx.doi.org/10.1103/PhysRevD.105.115029}{{\em Phys. Rev. D} {\bf 105} (2022) no.~11, 115029}, \href{http://arxiv.org/abs/2203.04828}{{\tt arXiv:2203.04828 [hep-ph]}}.

\bibitem{Leggett:2014mza}
T.~Leggett, T.~Li, J.~A. Maxin, D.~V. Nanopoulos, and J.~W. Walker, ``{No Naturalness or Fine-tuning Problems from No-Scale Supergravity},'' \href{http://arxiv.org/abs/1403.3099}{{\tt arXiv:1403.3099 [hep-ph]}}.

\bibitem{Leggett:2014hha}
T.~Leggett, T.~Li, J.~A. Maxin, D.~V. Nanopoulos, and J.~W. Walker, ``{Confronting Electroweak Fine-tuning with No-Scale Supergravity},'' \href{http://dx.doi.org/10.1016/j.physletb.2014.11.023}{{\em Phys. Lett. B} {\bf 740} (2015)  66--72}, \href{http://arxiv.org/abs/1408.4459}{{\tt arXiv:1408.4459 [hep-ph]}}.

\bibitem{Du:2015una}
G.~Du, T.~Li, D.~V. Nanopoulos, and S.~Raza, ``{Super-Natural MSSM},'' \href{http://dx.doi.org/10.1103/PhysRevD.92.025038}{{\em Phys. Rev. D} {\bf 92} (2015) no.~2, 025038}, \href{http://arxiv.org/abs/1502.06893}{{\tt arXiv:1502.06893 [hep-ph]}}.

\bibitem{Li:2015dil}
T.~Li, S.~Raza, and X.-C. Wang, ``{Supernatural supersymmetry and its classic example: M-theory inspired NMSSM},'' \href{http://dx.doi.org/10.1103/PhysRevD.93.115014}{{\em Phys. Rev. D} {\bf 93} (2016) no.~11, 115014}, \href{http://arxiv.org/abs/1510.06851}{{\tt arXiv:1510.06851 [hep-ph]}}.

\bibitem{Yuan:2022ewk}
J.-R. Yuan, H.-J. Cheng, and X.-A. Zhuang, ``{Prospects for chargino pair production at the CEPC},'' \href{http://dx.doi.org/10.1088/1674-1137/ac2f93}{{\em Chinese Physics C} {\bf 46} (2022) no.~1, 013104}.

\bibitem{Pagels:1981ke}
H.~Pagels and J.~R. Primack, ``{Supersymmetry, Cosmology and New TeV Physics},'' \href{http://dx.doi.org/10.1103/PhysRevLett.48.223}{{\em Phys. Rev. Lett.} {\bf 48} (1982)  223}.

\bibitem{Chen:2021omv}
J.~Chen, C.~Han, J.~M. Yang, and M.~Zhang, ``{Probing a bino NLSP at lepton colliders},'' \href{http://dx.doi.org/10.1103/PhysRevD.104.015009}{{\em Phys. Rev. D} {\bf 104} (2021) no.~1, 015009}, \href{http://arxiv.org/abs/2101.12131}{{\tt arXiv:2101.12131 [hep-ph]}}.

\bibitem{Yuan:2022ykg}
J.~Yuan, H.~Cheng, and X.~Zhuang, ``{Prospects for slepton pair production in the future $e^-e^+$ Higgs factories},'' \href{http://arxiv.org/abs/2203.10580}{{\tt arXiv:2203.10580 [hep-ex]}}.

\bibitem{Lyu:2025slp}
J.-R. Lyu~Feng, Yuan, H.-J. Cheng, and X.-A. Zhuang, ``{Potential search for direct slepton pair production in sqrt{s} = 360 GeV at CEPC},'' \href{http://arxiv.org/abs/2501.03600}{{\tt arXiv:2501.03600 [hep-ex]}}.

\bibitem{Ahmed:2022ude}
W.~Ahmed, I.~Khan, T.~Li, S.~Raza, and W.~Zhang, ``{Probing Relatively Heavier Right-Handed Selectron at the CEPC, $\rm\bf {FCC_{ee}}$ and ILC},'' \href{http://arxiv.org/abs/2202.11011}{{\tt arXiv:2202.11011 [hep-ex]}}.

\bibitem{Yang:2022qga}
J.~M. Yang, Y.~Zhang, P.~Zhu, and R.~Zhu, ``{Reconstructing masses for semi-invisibly decaying particles pair-produced at lepton colliders},'' \href{http://arxiv.org/abs/2211.08132}{{\tt arXiv:2211.08132 [hep-ph]}}.

\bibitem{Li:2023pab}
T.~Li, J.~A. Maxin, D.~V. Nanopoulos, and X.~Yin, ``The right-handed slepton bulk region for dark matter in generalized no-scale $\mathcal{F}$-$su(5)$ with effective super-natural supersymmetry,'' \href{http://arxiv.org/abs/2310.03622}{{\tt arXiv:2310.03622 [hep-ph]}}.

\bibitem{Barr:1981qv}
S.~M. Barr, ``{A New Symmetry Breaking Pattern for SO(10) and Proton Decay},'' \href{http://dx.doi.org/10.1016/0370-2693(82)90966-2}{{\em Phys. Lett. B} {\bf 112} (1982)  219--222}.

\bibitem{Jiang:2006hf}
J.~Jiang, T.~Li, and D.~V. Nanopoulos, ``{Testable Flipped SU(5) x U(1)(X) Models},'' \href{http://dx.doi.org/10.1016/j.nuclphysb.2007.02.025}{{\em Nucl. Phys. B} {\bf 772} (2007)  49--66}, \href{http://arxiv.org/abs/hep-ph/0610054}{{\tt arXiv:hep-ph/0610054}}.

\bibitem{Jiang:2008xrg}
J.~Jiang, T.~Li, D.~V. Nanopoulos, and D.~Xie, ``{F-SU(5)},'' \href{http://dx.doi.org/10.1016/j.physletb.2009.05.012}{{\em Phys. Lett. B} {\bf 677} (2009)  322--325}, \href{http://arxiv.org/abs/0811.2807}{{\tt arXiv:0811.2807 [hep-th]}}.

\bibitem{Jiang:2009za}
J.~Jiang, T.~Li, D.~V. Nanopoulos, and D.~Xie, ``{Flipped SU(5) x U(1)(X) Models from F-Theory},'' \href{http://dx.doi.org/10.1016/j.nuclphysb.2009.12.028}{{\em Nucl. Phys. B} {\bf 830} (2010)  195--220}, \href{http://arxiv.org/abs/0905.3394}{{\tt arXiv:0905.3394 [hep-th]}}.

\bibitem{Lopez:1992kg}
J.~L. Lopez, D.~V. Nanopoulos, and K.-j. Yuan, ``{The Search for a realistic flipped SU(5) string model},'' \href{http://dx.doi.org/10.1016/0550-3213(93)90513-O}{{\em Nucl. Phys. B} {\bf 399} (1993)  654--690}, \href{http://arxiv.org/abs/hep-th/9203025}{{\tt arXiv:hep-th/9203025}}.

\bibitem{CEPCFlavorWP}
X.~Ai {\em et al.}, ``{Flavor Physics at CEPC: a General Perspective},'' \href{http://arxiv.org/abs/2412.19743}{{\tt arXiv:2412.19743 [hep-ex]}}.

\bibitem{Zheng:2020ult}
T.~Zheng, J.~Xu, L.~Cao, D.~Yu, W.~Wang, S.~Prell, Y.-K.~E. Cheung, and M.~Ruan, ``{Analysis of $B_c \to \tau\nu_\tau$ at CEPC},'' \href{http://dx.doi.org/10.1088/1674-1137/abcf1f}{{\em Chin. Phys. C} {\bf 45} (2021) no.~2, 023001}, \href{http://arxiv.org/abs/2007.08234}{{\tt arXiv:2007.08234 [hep-ex]}}.

\bibitem{Li:2022tlo}
X.~Li, M.~Ruan, and M.~Zhao, ``{Prospect for measurement of CP-violation phase $\phi_s$ study in the $B_s\rightarrow J/\Psi\phi$ channel at future $Z$ factory},'' \href{http://arxiv.org/abs/2205.10565}{{\tt arXiv:2205.10565 [hep-ex]}}.

\bibitem{Aleksan:2021gii}
R.~Aleksan, L.~Oliver, and E.~Perez, ``{CP violation and determination of the $bs$ ''flat'' unitarity triangle at FCCee},'' \href{http://arxiv.org/abs/2107.02002}{{\tt arXiv:2107.02002 [hep-ph]}}.

\bibitem{Aleksan:2021fbx}
R.~Aleksan, L.~Oliver, and E.~Perez, ``{Study of CP violation in $B^\pm$ decays to $\overline{D^0}(D^0) K^\pm$ at FCCee},'' \href{http://arxiv.org/abs/2107.05311}{{\tt arXiv:2107.05311 [hep-ph]}}.

\bibitem{Amhis:2021cfy}
Y.~Amhis, C.~Helsens, D.~Hill, and O.~Sumensari, ``{Prospects for $B_{c}^+\to \tau^+ \nu_\tau$ at FCC-ee},'' \href{http://arxiv.org/abs/2105.13330}{{\tt arXiv:2105.13330 [hep-ex]}}.

\bibitem{Kamenik:2017ghi}
J.~F. Kamenik, S.~Monteil, A.~Semkiv, and L.~V. Silva, ``{Lepton polarization asymmetries in rare semi-tauonic $ b \rightarrow s $ exclusive decays at FCC-$ee$},'' \href{http://dx.doi.org/10.1140/epjc/s10052-017-5272-0}{{\em Eur. Phys. J.} {\bf C77} (2017) no.~10, 701},
\href{http://arxiv.org/abs/1705.11106}{{\tt arXiv:1705.11106 [hep-ph]}}.

\bibitem{Li:2020bvr}
L.~Li and T.~Liu, ``{$b\to s\tau^+\tau^-$ Physics at Future $Z$ Factories},'' \href{http://arxiv.org/abs/2012.00665}{{\tt arXiv:2012.00665 [hep-ph]}}.

\bibitem{Monteil:2021ith}
S.~Monteil and G.~Wilkinson, ``{Heavy-quark opportunities and challenges at FCC-ee},'' \href{http://arxiv.org/abs/2106.01259}{{\tt arXiv:2106.01259 [hep-ex]}}.

\bibitem{Chrzaszcz:2021nuk}
M.~Chrzaszcz, R.~G. Suarez, and S.~Monteil, ``{Hunt for rare processes and long-lived particles at FCC-ee},'' \href{http://dx.doi.org/10.1140/epjp/s13360-021-01961-4}{{\em Eur. Phys. J. Plus} {\bf 136} (2021) no.~10, 1056}, \href{http://arxiv.org/abs/2106.15459}{{\tt arXiv:2106.15459 [hep-ex]}}.

\bibitem{Dam:2018rfz}
M.~Dam, ``{Tau-lepton Physics at the FCC-ee circular e$^+$e$^-$ Collider},'' \href{http://dx.doi.org/10.21468/SciPostPhysProc.1.041}{{\em SciPost Phys. Proc.} {\bf 1} (2019)  041}, \href{http://arxiv.org/abs/1811.09408}{{\tt arXiv:1811.09408 [hep-ex]}}.

\bibitem{Qin:2017aju}
Q.~Qin, Q.~Li, C.-D. L\"u, F.-S. Yu, and S.-H. Zhou, ``{Charged lepton flavor violating Higgs decays at future $e^+e^-$ colliders},'' \href{http://dx.doi.org/10.1140/epjc/s10052-018-6298-7}{{\em Eur. Phys. J. C} {\bf 78} (2018) no.~10, 835}, \href{http://arxiv.org/abs/1711.07243}{{\tt arXiv:1711.07243 [hep-ph]}}.

\bibitem{Li:2018cod}
T.~Li and M.~A. Schmidt, ``{Sensitivity of future lepton colliders to the search for charged lepton flavor violation},'' \href{http://dx.doi.org/10.1103/PhysRevD.99.055038}{{\em Phys. Rev.} {\bf D99} (2019) no.~5, 055038},
\href{http://arxiv.org/abs/1809.07924}{{\tt arXiv:1809.07924 [hep-ph]}}.

\bibitem{Calibbi:2021pyh}
L.~Calibbi, X.~Marcano, and J.~Roy, ``{Z lepton flavour violation as a probe for new physics at future $e^+e^-$ colliders},'' \href{http://dx.doi.org/10.1140/epjc/s10052-021-09777-3}{{\em Eur. Phys. J. C} {\bf 81} (2021) no.~12, 1054}, \href{http://arxiv.org/abs/2107.10273}{{\tt arXiv:2107.10273 [hep-ph]}}.

\bibitem{Altmannshofer:2023tsa}
W.~Altmannshofer, P.~Munbodh, and T.~Oh, ``{Probing lepton flavor violation at Circular Electron-Positron Colliders},'' \href{http://dx.doi.org/10.1007/JHEP08(2023)026}{{\em JHEP} {\bf 08} (2023)  026}, \href{http://arxiv.org/abs/2305.03869}{{\tt arXiv:2305.03869 [hep-ph]}}.

\bibitem{Glashow:1970gm}
S.~L. Glashow, J.~Iliopoulos, and L.~Maiani, ``{Weak Interactions with Lepton-Hadron Symmetry},'' \href{http://dx.doi.org/10.1103/PhysRevD.2.1285}{{\em Phys. Rev. D} {\bf 2} (1970)  1285--1292}.

\bibitem{Cheng:2024hvq}
H.-C. Cheng, X.-H. Jiang, L.~Li, and E.~Salvioni, ``{Dark showers from Z-dark Z' mixing},'' \href{http://arxiv.org/abs/2401.08785}{{\tt arXiv:2401.08785 [hep-ph]}}.

\bibitem{Cheng:2024aco}
H.-C. Cheng, X.-H. Jiang, and L.~Li, ``{Phenomenology of electroweak portal dark showers: high energy direct probes and low energy complementarity},'' \href{http://dx.doi.org/10.1007/JHEP01(2025)149}{{\em JHEP} {\bf 01} (2025)  149}, \href{http://arxiv.org/abs/2408.13304}{{\tt arXiv:2408.13304 [hep-ph]}}.

\bibitem{Fukuda:1998mi}
{\bf Super-Kamiokande} Collaboration, Y.~Fukuda {\em et al.}, ``{Evidence for oscillation of atmospheric neutrinos},'' \href{http://dx.doi.org/10.1103/PhysRevLett.81.1562}{{\em Phys. Rev. Lett.} {\bf 81} (1998)  1562--1567}, \href{http://arxiv.org/abs/hep-ex/9807003}{{\tt arXiv:hep-ex/9807003}}.

\bibitem{Ahmad:2002jz}
{\bf SNO} Collaboration, Q.~R. Ahmad {\em et al.}, ``{Direct evidence for neutrino flavor transformation from neutral current interactions in the Sudbury Neutrino Observatory},'' \href{http://dx.doi.org/10.1103/PhysRevLett.89.011301}{{\em Phys. Rev. Lett.} {\bf 89} (2002)  011301}, \href{http://arxiv.org/abs/nucl-ex/0204008}{{\tt arXiv:nucl-ex/0204008}}.

\bibitem{FASER:2023zcr}
{\bf FASER} Collaboration, H.~Abreu {\em et al.}, ``{First Direct Observation of Collider Neutrinos with FASER at the LHC},'' \href{http://dx.doi.org/10.1103/PhysRevLett.131.031801}{{\em Phys. Rev. Lett.} {\bf 131} (2023) no.~3, 031801}, \href{http://arxiv.org/abs/2303.14185}{{\tt arXiv:2303.14185 [hep-ex]}}.

\bibitem{Deppisch:2015qwa}
F.~F. Deppisch, P.~S. Bhupal~Dev, and A.~Pilaftsis, ``{Neutrinos and Collider Physics},'' \href{http://dx.doi.org/10.1088/1367-2630/17/7/075019}{{\em New J. Phys.} {\bf 17} (2015) no.~7, 075019},
\href{http://arxiv.org/abs/1502.06541}{{\tt arXiv:1502.06541 [hep-ph]}}.

\bibitem{Cai:2017mow}
Y.~Cai, T.~Han, T.~Li, and R.~Ruiz, ``{Lepton Number Violation: Seesaw Models and Their Collider Tests},'' \href{http://dx.doi.org/10.3389/fphy.2018.00040}{{\em Front.in Phys.} {\bf 6} (2018)  40},
\href{http://arxiv.org/abs/1711.02180}{{\tt arXiv:1711.02180 [hep-ph]}}.

\bibitem{Keung:1983uu}
W.-Y. Keung and G.~Senjanovic, ``{Majorana Neutrinos and the Production of the Right-handed Charged Gauge Boson},'' \href{http://dx.doi.org/10.1103/PhysRevLett.50.1427}{{\em Phys. Rev. Lett.} {\bf 50} (1983)  1427}.

\bibitem{Weinberg:1979sa}
S.~Weinberg, ``{Baryon and Lepton Nonconserving Processes},'' \href{http://dx.doi.org/10.1103/PhysRevLett.43.1566}{{\em Phys. Rev. Lett.} {\bf 43} (1979)  1566--1570}.

\bibitem{Minkowski:1977sc}
P.~Minkowski, ``{$\mu \to e\gamma$ at a Rate of One Out of $10^{9}$ Muon Decays?},''
\href{http://dx.doi.org/10.1016/0370-2693(77)90435-X}{{\em Phys. Lett.} {\bf 67B} (1977)  421--428}.

\bibitem{Mohapatra:1979ia}
R.~N. Mohapatra and G.~Senjanovic, ``{Neutrino Mass and Spontaneous Parity Nonconservation},'' \href{http://dx.doi.org/10.1103/PhysRevLett.44.912}{{\em Phys. Rev. Lett.} {\bf 44} (1980)  912}.
[,231(1979)].

\bibitem{Yanagida:1979as}
T.~Yanagida, ``{Horizontal gauge symmetry and masses of neutrinos},'' {\em Conf. Proc. C} {\bf 7902131} (1979)  95--99. \url{https://inspirehep.net/literature/143150}.

\bibitem{GellMann:1980vs}
M.~Gell-Mann, P.~Ramond, and R.~Slansky, ``{Complex Spinors and Unified Theories},'' {\em Conf. Proc.} {\bf C790927} (1979)  315--321,
\href{http://arxiv.org/abs/1306.4669}{{\tt arXiv:1306.4669 [hep-th]}}.

\bibitem{Glashow:1979nm}
S.~L. Glashow, ``{The Future of Elementary Particle Physics},''
\href{http://dx.doi.org/10.1007/978-1-4684-7197-7_15}{{\em NATO Sci. Ser. B} {\bf 61} (1980)  687}.

\bibitem{Schechter:1980gr}
J.~Schechter and J.~W.~F. Valle, ``{Neutrino Masses in SU(2) x U(1) Theories},''
\href{http://dx.doi.org/10.1103/PhysRevD.22.2227}{{\em Phys. Rev.} {\bf D22} (1980)  2227}.

\bibitem{Davidson:1979wr}
A.~Davidson, M.~Koca, and K.~C. Wali, ``{U(1) as the Minimal Horizontal Gauge Symmetry},'' \href{http://dx.doi.org/10.1103/PhysRevLett.43.92}{{\em Phys. Rev. Lett.} {\bf 43} (1979)  92}.

\bibitem{Marshak:1979fm}
R.~E. Marshak and R.~N. Mohapatra, ``{Quark - Lepton Symmetry and B-L as the U(1) Generator of the Electroweak Symmetry Group},'' \href{http://dx.doi.org/10.1016/0370-2693(80)90436-0}{{\em Phys. Lett. B} {\bf 91} (1980)  222--224}.

\bibitem{Buchmuller:1991ce}
W.~Buchmuller, C.~Greub, and P.~Minkowski, ``{Neutrino masses, neutral vector bosons and the scale of B-L breaking},'' \href{http://dx.doi.org/10.1016/0370-2693(91)90952-M}{{\em Phys. Lett. B} {\bf 267} (1991)  395--399}.

\bibitem{Mohapatra:1974hk}
R.~N. Mohapatra and J.~C. Pati, ``{Left-Right Gauge Symmetry and an Isoconjugate Model of CP Violation},'' \href{http://dx.doi.org/10.1103/PhysRevD.11.566}{{\em Phys. Rev. D} {\bf 11} (1975)  566--571}.

\bibitem{Mohapatra:1974gc}
R.~N. Mohapatra and J.~C. Pati, ``{A Natural Left-Right Symmetry},'' \href{http://dx.doi.org/10.1103/PhysRevD.11.2558}{{\em Phys. Rev. D} {\bf 11} (1975)  2558}.

\bibitem{Senjanovic:1975rk}
G.~Senjanovic and R.~N. Mohapatra, ``{Exact Left-Right Symmetry and Spontaneous Violation of Parity},'' \href{http://dx.doi.org/10.1103/PhysRevD.12.1502}{{\em Phys. Rev. D} {\bf 12} (1975)  1502}.

\bibitem{Fritzsch:1974nn}
H.~Fritzsch and P.~Minkowski, ``{Unified Interactions of Leptons and Hadrons},'' \href{http://dx.doi.org/10.1016/0003-4916(75)90211-0}{{\em Annals Phys.} {\bf 93} (1975)  193--266}.

\bibitem{Witten:1979nr}
E.~Witten, ``{Neutrino Masses in the Minimal O(10) Theory},'' \href{http://dx.doi.org/10.1016/0370-2693(80)90666-8}{{\em Phys. Lett. B} {\bf 91} (1980)  81--84}.

\bibitem{Konetschny:1977bn}
W.~Konetschny and W.~Kummer, ``{Nonconservation of Total Lepton Number with Scalar Bosons},''
\href{http://dx.doi.org/10.1016/0370-2693(77)90407-5}{{\em Phys. Lett.} {\bf 70B} (1977)  433--435}.

\bibitem{Magg:1980ut}
M.~Magg and C.~Wetterich, ``{Neutrino Mass Problem and Gauge Hierarchy},''
\href{http://dx.doi.org/10.1016/0370-2693(80)90825-4}{{\em Phys. Lett.} {\bf 94B} (1980)  61--64}.

\bibitem{Cheng:1980qt}
T.~P. Cheng and L.-F. Li, ``{Neutrino Masses, Mixings and Oscillations in SU(2) x U(1) Models of Electroweak Interactions},''
\href{http://dx.doi.org/10.1103/PhysRevD.22.2860}{{\em Phys. Rev.} {\bf D22} (1980)  2860}.

\bibitem{Mohapatra:1980yp}
R.~N. Mohapatra and G.~Senjanovic, ``{Neutrino Masses and Mixings in Gauge Models with Spontaneous Parity Violation},''
\href{http://dx.doi.org/10.1103/PhysRevD.23.165}{{\em Phys. Rev.} {\bf D23} (1981)  165}.

\bibitem{Lazarides:1980nt}
G.~Lazarides, Q.~Shafi, and C.~Wetterich, ``{Proton Lifetime and Fermion Masses in an SO(10) Model},''
\href{http://dx.doi.org/10.1016/0550-3213(81)90354-0}{{\em Nucl. Phys.} {\bf B181} (1981)  287--300}.

\bibitem{Foot:1988aq}
R.~Foot, H.~Lew, X.~G. He, and G.~C. Joshi, ``{Seesaw Neutrino Masses Induced by a Triplet of Leptons},''
\href{http://dx.doi.org/10.1007/BF01415558}{{\em Z. Phys.} {\bf C44} (1989)  441}.

\bibitem{Ma:2002pf}
E.~Ma and D.~P. Roy, ``{Heavy triplet leptons and new gauge boson},'' \href{http://dx.doi.org/10.1016/S0550-3213(02)00815-5}{{\em Nucl. Phys. B} {\bf 644} (2002)  290--302}, \href{http://arxiv.org/abs/hep-ph/0206150}{{\tt arXiv:hep-ph/0206150}}.

\bibitem{Zee:1980ai}
A.~Zee, ``{A Theory of Lepton Number Violation, Neutrino Majorana Mass, and Oscillation},'' \href{http://dx.doi.org/10.1016/0370-2693(80)90349-4}{{\em Phys. Lett. B} {\bf 93} (1980)  389}. [Erratum: Phys.Lett.B 95, 461 (1980)].

\bibitem{Zee:1985id}
A.~Zee, ``{Quantum Numbers of Majorana Neutrino Masses},'' \href{http://dx.doi.org/10.1016/0550-3213(86)90475-X}{{\em Nucl. Phys. B} {\bf 264} (1986)  99--110}.

\bibitem{Babu:1988ki}
K.~S. Babu, ``{Model of 'Calculable' Majorana Neutrino Masses},'' \href{http://dx.doi.org/10.1016/0370-2693(88)91584-5}{{\em Phys. Lett. B} {\bf 203} (1988)  132--136}.

\bibitem{Cai:2017jrq}
Y.~Cai, J.~Herrero-Garc\'\i{}a, M.~A. Schmidt, A.~Vicente, and R.~R. Volkas, ``{From the trees to the forest: a review of radiative neutrino mass models},'' \href{http://dx.doi.org/10.3389/fphy.2017.00063}{{\em Front. in Phys.} {\bf 5} (2017)  63}, \href{http://arxiv.org/abs/1706.08524}{{\tt arXiv:1706.08524 [hep-ph]}}.

\bibitem{Dev:2018sel}
P.~S.~B. Dev, M.~J. Ramsey-Musolf, and Y.~Zhang, ``{Doubly-Charged Scalars in the Type-II Seesaw Mechanism: Fundamental Symmetry Tests and High-Energy Searches},'' \href{http://dx.doi.org/10.1103/PhysRevD.98.055013}{{\em Phys. Rev. D} {\bf 98} (2018) no.~5, 055013}, \href{http://arxiv.org/abs/1806.08499}{{\tt arXiv:1806.08499 [hep-ph]}}.

\bibitem{Cirigliano:2004tc}
V.~Cirigliano, A.~Kurylov, M.~J. Ramsey-Musolf, and P.~Vogel, ``{Neutrinoless double beta decay and lepton flavor violation},'' \href{http://dx.doi.org/10.1103/PhysRevLett.93.231802}{{\em Phys. Rev. Lett.} {\bf 93} (2004)  231802}, \href{http://arxiv.org/abs/hep-ph/0406199}{{\tt arXiv:hep-ph/0406199}}.

\bibitem{Abada:2007ux}
A.~Abada, C.~Biggio, F.~Bonnet, M.~B. Gavela, and T.~Hambye, ``{Low energy effects of neutrino masses},'' \href{http://dx.doi.org/10.1088/1126-6708/2007/12/061}{{\em JHEP} {\bf 12} (2007)  061}, \href{http://arxiv.org/abs/0707.4058}{{\tt arXiv:0707.4058 [hep-ph]}}.

\bibitem{Tello:2010am}
V.~Tello, M.~Nemevsek, F.~Nesti, G.~Senjanovic, and F.~Vissani, ``{Left-Right Symmetry: from LHC to Neutrinoless Double Beta Decay},'' \href{http://dx.doi.org/10.1103/PhysRevLett.106.151801}{{\em Phys. Rev. Lett.} {\bf 106} (2011)  151801}, \href{http://arxiv.org/abs/1011.3522}{{\tt arXiv:1011.3522 [hep-ph]}}.

\bibitem{Chakrabortty:2012mh}
J.~Chakrabortty, H.~Z. Devi, S.~Goswami, and S.~Patra, ``{Neutrinoless double-$\beta$ decay in TeV scale Left-Right symmetric models},'' \href{http://dx.doi.org/10.1007/JHEP08(2012)008}{{\em JHEP} {\bf 08} (2012)  008}, \href{http://arxiv.org/abs/1204.2527}{{\tt arXiv:1204.2527 [hep-ph]}}.

\bibitem{Barry:2013xxa}
J.~Barry and W.~Rodejohann, ``{Lepton number and flavour violation in TeV-scale left-right symmetric theories with large left-right mixing},'' \href{http://dx.doi.org/10.1007/JHEP09(2013)153}{{\em JHEP} {\bf 09} (2013)  153}, \href{http://arxiv.org/abs/1303.6324}{{\tt arXiv:1303.6324 [hep-ph]}}.

\bibitem{BhupalDev:2014qbx}
P.~S. Bhupal~Dev, S.~Goswami, and M.~Mitra, ``{TeV Scale Left-Right Symmetry and Large Mixing Effects in Neutrinoless Double Beta Decay},'' \href{http://dx.doi.org/10.1103/PhysRevD.91.113004}{{\em Phys. Rev. D} {\bf 91} (2015) no.~11, 113004}, \href{http://arxiv.org/abs/1405.1399}{{\tt arXiv:1405.1399 [hep-ph]}}.

\bibitem{Awasthi:2015ota}
R.~L. Awasthi, P.~S.~B. Dev, and M.~Mitra, ``{Implications of the Diboson Excess for Neutrinoless Double Beta Decay and Lepton Flavor Violation in TeV Scale Left Right Symmetric Model},'' \href{http://dx.doi.org/10.1103/PhysRevD.93.011701}{{\em Phys. Rev. D} {\bf 93} (2016) no.~1, 011701}, \href{http://arxiv.org/abs/1509.05387}{{\tt arXiv:1509.05387 [hep-ph]}}.

\bibitem{Bambhaniya:2015ipg}
G.~Bambhaniya, P.~S.~B. Dev, S.~Goswami, and M.~Mitra, ``{The Scalar Triplet Contribution to Lepton Flavour Violation and Neutrinoless Double Beta Decay in Left-Right Symmetric Model},'' \href{http://dx.doi.org/10.1007/JHEP04(2016)046}{{\em JHEP} {\bf 04} (2016)  046}, \href{http://arxiv.org/abs/1512.00440}{{\tt arXiv:1512.00440 [hep-ph]}}.

\bibitem{Borah:2016iqd}
D.~Borah and A.~Dasgupta, ``{Charged lepton flavour violcxmation and neutrinoless double beta decay in left-right symmetric models with type I+II seesaw},'' \href{http://dx.doi.org/10.1007/JHEP07(2016)022}{{\em JHEP} {\bf 07} (2016)  022}, \href{http://arxiv.org/abs/1606.00378}{{\tt arXiv:1606.00378 [hep-ph]}}.

\bibitem{Li:2020flq}
G.~Li, M.~Ramsey-Musolf, and J.~C. Vasquez, ``{Left-Right Symmetry and Leading Contributions to Neutrinoless Double Beta Decay},'' \href{http://dx.doi.org/10.1103/PhysRevLett.126.151801}{{\em Phys. Rev. Lett.} {\bf 126} (2021) no.~15, 151801}, \href{http://arxiv.org/abs/2009.01257}{{\tt arXiv:2009.01257 [hep-ph]}}.

\bibitem{Li:2022cuq}
G.~Li, M.~J. Ramsey-Musolf, and J.~C. Vasquez, ``{Unraveling the left-right mixing using 0\ensuremath{\nu}\ensuremath{\beta}\ensuremath{\beta} decay and collider probes},'' \href{http://dx.doi.org/10.1103/PhysRevD.105.115021}{{\em Phys. Rev. D} {\bf 105} (2022) no.~11, 115021}, \href{http://arxiv.org/abs/2202.01789}{{\tt arXiv:2202.01789 [hep-ph]}}.

\bibitem{deVries:2022nyh}
J.~de~Vries, G.~Li, M.~J. Ramsey-Musolf, and J.~C. Vasquez, ``{Light sterile neutrinos, left-right symmetry, and 0\ensuremath{\nu}\ensuremath{\beta}\ensuremath{\beta} decay},'' \href{http://dx.doi.org/10.1007/JHEP11(2022)056}{{\em JHEP} {\bf 11} (2022)  056}, \href{http://arxiv.org/abs/2209.03031}{{\tt arXiv:2209.03031 [hep-ph]}}.

\bibitem{Li:2024djp}
G.~Li, M.~J. Ramsey-Musolf, S.~Urrutia~Quiroga, and J.~C. Vasquez, ``{Dissecting Lepton Number Violation in the Left-Right Symmetric Model: $0\nu\beta\beta$ decay, M\o{}ller scattering, and collider searches},'' \href{http://arxiv.org/abs/2408.06306}{{\tt arXiv:2408.06306 [hep-ph]}}.

\bibitem{Antusch:2016vyf}
S.~Antusch, E.~Cazzato, and O.~Fischer, ``{Displaced vertex searches for sterile neutrinos at future lepton colliders},'' \href{http://dx.doi.org/10.1007/JHEP12(2016)007}{{\em JHEP} {\bf 12} (2016)  007}, \href{http://arxiv.org/abs/1604.02420}{{\tt arXiv:1604.02420 [hep-ph]}}.

\bibitem{neutrinolimits}
``Sterile neutrino constraints.'' \url{https://www.hep.ucl.ac.uk/~pbolton/plots.html}.

\bibitem{Abdullahi:2022jlv}
A.~M. Abdullahi {\em et al.}, ``{The present and future status of heavy neutral leptons},'' \href{http://dx.doi.org/10.1088/1361-6471/ac98f9}{{\em J. Phys. G} {\bf 50} (2023) no.~2, 020501}, \href{http://arxiv.org/abs/2203.08039}{{\tt arXiv:2203.08039 [hep-ph]}}.

\bibitem{Gao:2021one}
Y.~Gao and K.~Wang, ``{Heavy Neutrino Searches via Same-sign Lepton Pairs at the Higgs Factory},'' \href{http://arxiv.org/abs/2102.12826}{{\tt arXiv:2102.12826 [hep-ph]}}.

\bibitem{Deppisch:2018eth}
F.~F. Deppisch, W.~Liu, and M.~Mitra, ``{Long-lived Heavy Neutrinos from Higgs Decays},'' \href{http://dx.doi.org/10.1007/JHEP08(2018)181}{{\em JHEP} {\bf 08} (2018)  181}, \href{http://arxiv.org/abs/1804.04075}{{\tt arXiv:1804.04075 [hep-ph]}}.

\bibitem{Das:2022rbl}
A.~Das, S.~Mandal, T.~Nomura, and S.~Shil, ``{Heavy Majorana neutrino pair production from Z' at hadron and lepton colliders},'' \href{http://dx.doi.org/10.1103/PhysRevD.105.095031}{{\em Phys. Rev. D} {\bf 105} (2022) no.~9, 095031}, \href{http://arxiv.org/abs/2202.13358}{{\tt arXiv:2202.13358 [hep-ph]}}.

\bibitem{Nemevsek:2016enw}
M.~Nemev\v{s}ek, F.~Nesti, and J.~C. Vasquez, ``{Majorana Higgses at colliders},'' \href{http://dx.doi.org/10.1007/JHEP04(2017)114}{{\em JHEP} {\bf 04} (2017)  114}, \href{http://arxiv.org/abs/1612.06840}{{\tt arXiv:1612.06840 [hep-ph]}}.

\bibitem{Liao:2021rjw}
J.~Liao and Y.~Zhang, ``{Constraining nonstandard neutrino interactions at electron colliders},'' \href{http://dx.doi.org/10.1103/PhysRevD.104.035043}{{\em Phys. Rev. D} {\bf 104} (2021) no.~3, 035043}, \href{http://arxiv.org/abs/2105.11215}{{\tt arXiv:2105.11215 [hep-ph]}}.

\bibitem{Zhang:2023nxy}
Y.~Zhang and W.~Liu, ``{Probing active-sterile neutrino transition magnetic moments at LEP and CEPC},'' \href{http://dx.doi.org/10.1103/PhysRevD.107.095031}{{\em Phys. Rev. D} {\bf 107} (2023) no.~9, 095031}, \href{http://arxiv.org/abs/2301.06050}{{\tt arXiv:2301.06050 [hep-ph]}}.

\bibitem{Dev:2017ftk}
P.~S.~B. Dev, R.~N. Mohapatra, and Y.~Zhang, ``{Lepton Flavor Violation Induced by a Neutral Scalar at Future Lepton Colliders},'' \href{http://dx.doi.org/10.1103/PhysRevLett.120.221804}{{\em Phys. Rev. Lett.} {\bf 120} (2018) no.~22, 221804}, \href{http://arxiv.org/abs/1711.08430}{{\tt arXiv:1711.08430 [hep-ph]}}.

\bibitem{BhupalDev:2018vpr}
P.~S. Bhupal~Dev, R.~N. Mohapatra, and Y.~Zhang, ``{Probing TeV scale origin of neutrino mass at future lepton colliders via neutral and doubly-charged scalars},'' \href{http://dx.doi.org/10.1103/PhysRevD.98.075028}{{\em Phys. Rev. D} {\bf 98} (2018) no.~7, 075028}, \href{http://arxiv.org/abs/1803.11167}{{\tt arXiv:1803.11167 [hep-ph]}}.

\bibitem{Antusch:2015mia}
S.~Antusch and O.~Fischer, ``{Testing sterile neutrino extensions of the Standard Model at future lepton colliders},'' \href{http://dx.doi.org/10.1007/JHEP05(2015)053}{{\em JHEP} {\bf 05} (2015)  053}, \href{http://arxiv.org/abs/1502.05915}{{\tt arXiv:1502.05915 [hep-ph]}}.

\bibitem{Barducci:2022hll}
D.~Barducci and E.~Bertuzzo, ``{The see-saw portal at future Higgs factories: the role of dimension six operators},'' \href{http://dx.doi.org/10.1007/JHEP06(2022)077}{{\em JHEP} {\bf 06} (2022)  077}, \href{http://arxiv.org/abs/2201.11754}{{\tt arXiv:2201.11754 [hep-ph]}}.

\bibitem{Rygaard:2022qms}
L.~Rygaard, ``Long-lived heavy neutral leptons at the fcc-ee,'' master's thesis, Uppsala University, 2022.
\newblock \url{https://urn.kb.se/resolve?urn=urn:nbn:se:uu:diva-479595}.

\bibitem{Drewes:2022rsk}
M.~Drewes, ``{Distinguishing Dirac and Majorana Heavy Neutrinos at Lepton Colliders},'' \href{http://dx.doi.org/10.22323/1.414.0608}{{\em PoS} {\bf ICHEP2022} (2022)  608}, \href{http://arxiv.org/abs/2210.17110}{{\tt arXiv:2210.17110 [hep-ph]}}.

\bibitem{Antusch:2022ceb}
S.~Antusch, J.~Hajer, and J.~Rosskopp, ``{Simulating lepton number violation induced by heavy neutrino-antineutrino oscillations at colliders},'' \href{http://dx.doi.org/10.1007/JHEP03(2023)110}{{\em JHEP} {\bf 03} (2023)  110}, \href{http://arxiv.org/abs/2210.10738}{{\tt arXiv:2210.10738 [hep-ph]}}.

\bibitem{Antusch:2023nqd}
S.~Antusch, J.~Hajer, and J.~Rosskopp, ``{Decoherence effects on lepton number violation from heavy neutrino-antineutrino oscillations},'' \href{http://dx.doi.org/10.1007/JHEP11(2023)235}{{\em JHEP} {\bf 11} (2023)  235}, \href{http://arxiv.org/abs/2307.06208}{{\tt arXiv:2307.06208 [hep-ph]}}.

\bibitem{Antusch:2023jsa}
S.~Antusch, J.~Hajer, and B.~M.~S. Oliveira, ``{Heavy neutrino-antineutrino oscillations at the FCC-ee},'' \href{http://dx.doi.org/10.1007/JHEP10(2023)129}{{\em JHEP} {\bf 10} (2023)  129}, \href{http://arxiv.org/abs/2308.07297}{{\tt arXiv:2308.07297 [hep-ph]}}.

\bibitem{Antusch:2024otj}
S.~Antusch, J.~Hajer, and B.~M.~S. Oliveira, ``{Discovering heavy neutrino-antineutrino oscillations at the Z-pole},'' \href{http://dx.doi.org/10.1007/JHEP11(2024)102}{{\em JHEP} {\bf 11} (2024)  102}, \href{http://arxiv.org/abs/2408.01389}{{\tt arXiv:2408.01389 [hep-ph]}}.

\bibitem{Ajmal:2024kwi}
S.~Ajmal, P.~Azzi, S.~Giappichini, M.~Klute, O.~Panella, M.~Presilla, and X.~Zuo, ``{Searching for type I seesaw mechanism in a two Heavy Neutral Leptons scenario at FCC-ee},'' \href{http://arxiv.org/abs/2410.03615}{{\tt arXiv:2410.03615 [hep-ph]}}.

\bibitem{Bellagamba:2025xpd}
L.~Bellagamba, G.~Polesello, and N.~Valle, ``{Searches for Heavy Neutral Leptons at FCC-ee in final states including a muon},'' \href{http://arxiv.org/abs/2503.19464}{{\tt arXiv:2503.19464 [hep-ex]}}.

\bibitem{Ovchynnikov:2023wgg}
M.~Ovchynnikov and J.-Y. Zhu, ``{Search for the dipole portal of heavy neutral leptons at future colliders},'' \href{http://dx.doi.org/10.1007/JHEP07(2023)039}{{\em JHEP} {\bf 07} (2023)  039}, \href{http://arxiv.org/abs/2301.08592}{{\tt arXiv:2301.08592 [hep-ph]}}.

\bibitem{Antonello:2020tzq}
{\bf RD-FA} Collaboration, M.~Antonello, ``{IDEA: A detector concept for future leptonic colliders},'' \href{http://dx.doi.org/10.1393/ncc/i2020-20027-2}{{\em Nuovo Cim. C} {\bf 43} (2020) no.~2-3, 27}.

\bibitem{SHiP:2018xqw}
{\bf SHiP} Collaboration, C.~Ahdida {\em et al.}, ``{Sensitivity of the SHiP experiment to Heavy Neutral Leptons},'' \href{http://dx.doi.org/10.1007/JHEP04(2019)077}{{\em JHEP} {\bf 04} (2019)  077}, \href{http://arxiv.org/abs/1811.00930}{{\tt arXiv:1811.00930 [hep-ph]}}.

\bibitem{Ballett:2019bgd}
P.~Ballett, T.~Boschi, and S.~Pascoli, ``{Heavy Neutral Leptons from low-scale seesaws at the DUNE Near Detector},'' \href{http://dx.doi.org/10.1007/JHEP03(2020)111}{{\em JHEP} {\bf 03} (2020)  111}, \href{http://arxiv.org/abs/1905.00284}{{\tt arXiv:1905.00284 [hep-ph]}}.

\bibitem{CHARM:1985nku}
{\bf CHARM} Collaboration, F.~Bergsma {\em et al.}, ``{A Search for Decays of Heavy Neutrinos in the Mass Range 0.5-{GeV} to 2.8-{GeV}},'' \href{http://dx.doi.org/10.1016/0370-2693(86)91601-1}{{\em Phys. Lett. B} {\bf 166} (1986)  473--478}.

\bibitem{CMS:2018iaf}
{\bf CMS} Collaboration, A.~M. Sirunyan {\em et al.}, ``{Search for heavy neutral leptons in events with three charged leptons in proton-proton collisions at $\sqrt{s} =$ 13 TeV},'' \href{http://dx.doi.org/10.1103/PhysRevLett.120.221801}{{\em Phys. Rev. Lett.} {\bf 120} (2018) no.~22, 221801}, \href{http://arxiv.org/abs/1802.02965}{{\tt arXiv:1802.02965 [hep-ex]}}.

\bibitem{ATLAS:2019kpx}
{\bf ATLAS} Collaboration, G.~Aad {\em et al.}, ``{Search for heavy neutral leptons in decays of $W$ bosons produced in 13 TeV $pp$ collisions using prompt and displaced signatures with the ATLAS detector},'' \href{http://dx.doi.org/10.1007/JHEP10(2019)265}{{\em JHEP} {\bf 10} (2019)  265}, \href{http://arxiv.org/abs/1905.09787}{{\tt arXiv:1905.09787 [hep-ex]}}.

\bibitem{DELPHI:1996qcc}
{\bf DELPHI} Collaboration, P.~Abreu {\em et al.}, ``{Search for neutral heavy leptons produced in Z decays},'' \href{http://dx.doi.org/10.1007/s002880050370}{{\em Z. Phys. C} {\bf 74} (1997)  57--71}. [Erratum: Z.Phys.C 75, 580 (1997)].

\bibitem{E949:2014gsn}
{\bf E949} Collaboration, A.~V. Artamonov {\em et al.}, ``{Search for heavy neutrinos in $K^+\to\mu^+\nu_H$ decays},'' \href{http://dx.doi.org/10.1103/PhysRevD.91.052001}{{\em Phys. Rev. D} {\bf 91} (2015) no.~5, 052001}, \href{http://arxiv.org/abs/1411.3963}{{\tt arXiv:1411.3963 [hep-ex]}}. [Erratum: Phys.Rev.D 91, 059903 (2015)].

\bibitem{LHCb:2016inz}
{\bf LHCb} Collaboration, R.~Aaij {\em et al.}, ``{Search for massive long-lived particles decaying semileptonically in the LHCb detector},'' \href{http://dx.doi.org/10.1140/epjc/s10052-017-4744-6}{{\em Eur. Phys. J. C} {\bf 77} (2017) no.~4, 224}, \href{http://arxiv.org/abs/1612.00945}{{\tt arXiv:1612.00945 [hep-ex]}}.

\bibitem{Antusch:2017hhu}
S.~Antusch, E.~Cazzato, and O.~Fischer, ``{Sterile neutrino searches via displaced vertices at LHCb},'' \href{http://dx.doi.org/10.1016/j.physletb.2017.09.057}{{\em Phys. Lett. B} {\bf 774} (2017)  114--118}, \href{http://arxiv.org/abs/1706.05990}{{\tt arXiv:1706.05990 [hep-ph]}}.

\bibitem{NuTeV:1999kej}
{\bf NuTeV, E815} Collaboration, A.~Vaitaitis {\em et al.}, ``{Search for neutral heavy leptons in a high-energy neutrino beam},'' \href{http://dx.doi.org/10.1103/PhysRevLett.83.4943}{{\em Phys. Rev. Lett.} {\bf 83} (1999)  4943--4946}, \href{http://arxiv.org/abs/hep-ex/9908011}{{\tt arXiv:hep-ex/9908011}}.

\bibitem{Bernardi:1987ek}
G.~Bernardi {\em et al.}, ``{Further Limits On Heavy Neutrino Couplings},''
\href{http://dx.doi.org/10.1016/0370-2693(88)90563-1}{{\em Phys. Lett.} {\bf B203} (1988)  332--334}.

\bibitem{Boyarsky:2020dzc}
A.~Boyarsky, M.~Ovchynnikov, O.~Ruchayskiy, and V.~Syvolap, ``{Improved big bang nucleosynthesis constraints on heavy neutral leptons},'' \href{http://dx.doi.org/10.1103/PhysRevD.104.023517}{{\em Phys. Rev. D} {\bf 104} (2021) no.~2, 023517}, \href{http://arxiv.org/abs/2008.00749}{{\tt arXiv:2008.00749 [hep-ph]}}.

\bibitem{Asaka:2005an}
T.~Asaka, S.~Blanchet, and M.~Shaposhnikov, ``{The nuMSM, dark matter and neutrino masses},'' \href{http://dx.doi.org/10.1016/j.physletb.2005.09.070}{{\em Phys. Lett. B} {\bf 631} (2005)  151--156}, \href{http://arxiv.org/abs/hep-ph/0503065}{{\tt arXiv:hep-ph/0503065}}.

\bibitem{Asaka:2005pn}
T.~Asaka and M.~Shaposhnikov, ``{The $\nu$MSM, dark matter and baryon asymmetry of the universe},'' \href{http://dx.doi.org/10.1016/j.physletb.2005.06.020}{{\em Phys. Lett. B} {\bf 620} (2005)  17--26}, \href{http://arxiv.org/abs/hep-ph/0505013}{{\tt arXiv:hep-ph/0505013}}.

\bibitem{Liao:2017jiz}
W.~Liao and X.-H. Wu, ``{Signature of heavy sterile neutrinos at CEPC},'' \href{http://dx.doi.org/10.1103/PhysRevD.97.055005}{{\em Phys. Rev.} {\bf D97} (2018) no.~5, 055005},
\href{http://arxiv.org/abs/1710.09266}{{\tt arXiv:1710.09266 [hep-ph]}}.

\bibitem{Ding:2019tqq}
J.-N. Ding, Q.~Qin, and F.-S. Yu, ``{Heavy neutrino searches at future $Z$-factories},'' \href{http://dx.doi.org/10.1140/epjc/s10052-019-7277-3}{{\em Eur. Phys. J. C} {\bf 79} (2019) no.~9, 766}, \href{http://arxiv.org/abs/1903.02570}{{\tt arXiv:1903.02570 [hep-ph]}}.

\bibitem{Blondel:2021mss}
A.~Blondel, A.~de~Gouv\^ea, and B.~Kayser, ``{Z-boson decays into Majorana or Dirac heavy neutrinos},'' \href{http://dx.doi.org/10.1103/PhysRevD.104.055027}{{\em Phys. Rev. D} {\bf 104} (2021) no.~5, 055027}, \href{http://arxiv.org/abs/2105.06576}{{\tt arXiv:2105.06576 [hep-ph]}}.

\bibitem{Shoemaker:2010fg}
I.~M. Shoemaker, K.~Petraki, and A.~Kusenko, ``{Collider signatures of sterile neutrinos in models with a gauge-singlet Higgs},'' \href{http://dx.doi.org/10.1007/JHEP09(2010)060}{{\em JHEP} {\bf 09} (2010)  060},
\href{http://arxiv.org/abs/1006.5458}{{\tt arXiv:1006.5458 [hep-ph]}}.

\bibitem{Robens:2022cun}
T.~Robens, ``{Constraining Extended Scalar Sectors at~Current and~Future Colliders\textemdash{}An Update},'' \href{http://dx.doi.org/10.1007/978-3-031-30459-0_13}{{\em Springer Proc. Phys.} {\bf 292} (2023)  141--152}, \href{http://arxiv.org/abs/2209.15544}{{\tt arXiv:2209.15544 [hep-ph]}}.

\bibitem{Aihara:2009ad}
``{SiD Letter of Intent},'' \href{http://arxiv.org/abs/0911.0006}{{\tt arXiv:0911.0006 [physics.ins-det]}}.

\bibitem{Gu:2017ckc}
J.~Gu, H.~Li, Z.~Liu, S.~Su, and W.~Su, ``{Learning from Higgs Physics at Future Higgs Factories},'' \href{http://dx.doi.org/10.1007/JHEP12(2017)153}{{\em JHEP} {\bf 12} (2017)  153}, \href{http://arxiv.org/abs/1709.06103}{{\tt arXiv:1709.06103 [hep-ph]}}.

\bibitem{Das:2021esm}
A.~Das, P.~S.~B. Dev, Y.~Hosotani, and S.~Mandal, ``{Probing the minimal U(1)X model at future electron-positron colliders via fermion pair-production channels},'' \href{http://dx.doi.org/10.1103/PhysRevD.105.115030}{{\em Phys. Rev. D} {\bf 105} (2022) no.~11, 115030}, \href{http://arxiv.org/abs/2104.10902}{{\tt arXiv:2104.10902 [hep-ph]}}.

\bibitem{Das:2017flq}
A.~Das, N.~Okada, and D.~Raut, ``{Enhanced pair production of heavy Majorana neutrinos at the LHC},'' \href{http://dx.doi.org/10.1103/PhysRevD.97.115023}{{\em Phys. Rev. D} {\bf 97} (2018) no.~11, 115023}, \href{http://arxiv.org/abs/1710.03377}{{\tt arXiv:1710.03377 [hep-ph]}}.

\bibitem{Pati:1974yy}
J.~C. Pati and A.~Salam, ``{Lepton Number as the Fourth Color},'' \href{http://dx.doi.org/10.1103/PhysRevD.10.275}{{\em Phys. Rev. D} {\bf 10} (1974)  275--289}. [Erratum: Phys.Rev.D 11, 703--703 (1975)].

\bibitem{Djouadi:2005gi}
A.~Djouadi, ``{The Anatomy of electro-weak symmetry breaking. I: The Higgs boson in the standard model},'' \href{http://dx.doi.org/10.1016/j.physrep.2007.10.004}{{\em Phys. Rept.} {\bf 457} (2008)  1--216}, \href{http://arxiv.org/abs/hep-ph/0503172}{{\tt arXiv:hep-ph/0503172}}.

\bibitem{Gomez-Ceballos:2013zzn}
{\bf TLEP Design Study Working Group} Collaboration, M.~Bicer {\em et al.}, ``{First Look at the Physics Case of TLEP},'' \href{http://dx.doi.org/10.1007/JHEP01(2014)164}{{\em JHEP} {\bf 01} (2014)  164},
\href{http://arxiv.org/abs/1308.6176}{{\tt arXiv:1308.6176 [hep-ex]}}.

\bibitem{Barranco:2007ej}
J.~Barranco, O.~G. Miranda, C.~A. Moura, and J.~W.~F. Valle, ``{Constraining non-standard neutrino-electron interactions},'' \href{http://dx.doi.org/10.1103/PhysRevD.77.093014}{{\em Phys. Rev. D} {\bf 77} (2008)  093014}, \href{http://arxiv.org/abs/0711.0698}{{\tt arXiv:0711.0698 [hep-ph]}}.

\bibitem{Magill:2018jla}
G.~Magill, R.~Plestid, M.~Pospelov, and Y.-D. Tsai, ``{Dipole Portal to Heavy Neutral Leptons},'' \href{http://dx.doi.org/10.1103/PhysRevD.98.115015}{{\em Phys. Rev. D} {\bf 98} (2018) no.~11, 115015}, \href{http://arxiv.org/abs/1803.03262}{{\tt arXiv:1803.03262 [hep-ph]}}.

\bibitem{OPAL:1994kgw}
{\bf OPAL} Collaboration, R.~Akers {\em et al.}, ``{Measurement of single photon production in e+ e- collisions near the Z0 resonance},'' \href{http://dx.doi.org/10.1007/BF01571303}{{\em Z. Phys. C} {\bf 65} (1995)  47--66}.

\bibitem{ATLAS:2020uiq}
{\bf ATLAS} Collaboration, G.~Aad {\em et al.}, ``{Search for dark matter in association with an energetic photon in $pp$ collisions at $\sqrt{s}$ = 13 TeV with the ATLAS detector},'' \href{http://dx.doi.org/10.1007/JHEP02(2021)226}{{\em JHEP} {\bf 02} (2021)  226}, \href{http://arxiv.org/abs/2011.05259}{{\tt arXiv:2011.05259 [hep-ex]}}.

\bibitem{CMS:2018fon}
{\bf CMS} Collaboration, A.~M. Sirunyan {\em et al.}, ``{Search for supersymmetry in events with a photon, a lepton, and missing transverse momentum in proton-proton collisions at $\sqrt{s} =$ 13 TeV},'' \href{http://dx.doi.org/10.1007/JHEP01(2019)154}{{\em JHEP} {\bf 01} (2019)  154}, \href{http://arxiv.org/abs/1812.04066}{{\tt arXiv:1812.04066 [hep-ex]}}.

\bibitem{Barbier:2004ez}
R.~Barbier {\em et al.}, ``{R-parity violating supersymmetry},'' \href{http://dx.doi.org/10.1016/j.physrep.2005.08.006}{{\em Phys. Rept.} {\bf 420} (2005)  1--202},
\href{http://arxiv.org/abs/hep-ph/0406039}{{\tt arXiv:hep-ph/0406039 [hep-ph]}}.

\bibitem{Hung:2006ap}
P.~Q. Hung, ``{A Model of electroweak-scale right-handed neutrino mass},'' \href{http://dx.doi.org/10.1016/j.physletb.2007.03.067}{{\em Phys. Lett. B} {\bf 649} (2007)  275--279}, \href{http://arxiv.org/abs/hep-ph/0612004}{{\tt arXiv:hep-ph/0612004}}.

\bibitem{Bu:2008fx}
J.-P. Bu, Y.~Liao, and J.-Y. Liu, ``{Lepton Flavor Violating Muon Decays in a Model of Electroweak-Scale Right-Handed Neutrinos},'' \href{http://dx.doi.org/10.1016/j.physletb.2008.05.059}{{\em Phys. Lett. B} {\bf 665} (2008)  39--43}, \href{http://arxiv.org/abs/0802.3241}{{\tt arXiv:0802.3241 [hep-ph]}}.

\bibitem{Chang:2016ave}
C.-F. Chang, C.-H.~V. Chang, C.~S. Nugroho, and T.-C. Yuan, ``{Lepton Flavor Violating Decays of Neutral Higgses in Extended Mirror Fermion Model},'' \href{http://dx.doi.org/10.1016/j.nuclphysb.2016.07.009}{{\em Nucl. Phys. B} {\bf 910} (2016)  293--308}, \href{http://arxiv.org/abs/1602.00680}{{\tt arXiv:1602.00680 [hep-ph]}}.

\bibitem{Branco:2011iw}
G.~C. Branco, P.~M. Ferreira, L.~Lavoura, M.~N. Rebelo, M.~Sher, and J.~P. Silva, ``{Theory and phenomenology of two-Higgs-doublet models},'' \href{http://dx.doi.org/10.1016/j.physrep.2012.02.002}{{\em Phys. Rept.} {\bf 516} (2012)  1--102},
\href{http://arxiv.org/abs/1106.0034}{{\tt arXiv:1106.0034 [hep-ph]}}.

\bibitem{Crivellin:2015hha}
A.~Crivellin, J.~Heeck, and P.~Stoffer, ``{A perturbed lepton-specific two-Higgs-doublet model facing experimental hints for physics beyond the Standard Model},'' \href{http://dx.doi.org/10.1103/PhysRevLett.116.081801}{{\em Phys. Rev. Lett.} {\bf 116} (2016) no.~8, 081801}, \href{http://arxiv.org/abs/1507.07567}{{\tt arXiv:1507.07567 [hep-ph]}}.

\bibitem{Maiezza:2016ybz}
A.~Maiezza, G.~Senjanovi\'c, and J.~C. Vasquez, ``{Higgs sector of the minimal left-right symmetric theory},'' \href{http://dx.doi.org/10.1103/PhysRevD.95.095004}{{\em Phys. Rev. D} {\bf 95} (2017) no.~9, 095004}, \href{http://arxiv.org/abs/1612.09146}{{\tt arXiv:1612.09146 [hep-ph]}}.

\bibitem{ParticleDataGroup:2024cfk}
{\bf Particle Data Group} Collaboration, S.~Navas {\em et al.}, ``{Review of particle physics},'' \href{http://dx.doi.org/10.1103/PhysRevD.110.030001}{{\em Phys. Rev. D} {\bf 110} (2024) no.~3, 030001}.

\bibitem{DELPHI:2005wxt}
{\bf DELPHI} Collaboration, J.~Abdallah {\em et al.}, ``{Measurement and interpretation of fermion-pair production at LEP energies above the Z resonance},'' \href{http://dx.doi.org/10.1140/epjc/s2005-02461-0}{{\em Eur. Phys. J. C} {\bf 45} (2006)  589--632}, \href{http://arxiv.org/abs/hep-ex/0512012}{{\tt arXiv:hep-ex/0512012}}.

\bibitem{FileviezPerez:2008jbu}
P.~Fileviez~Perez, T.~Han, G.-y. Huang, T.~Li, and K.~Wang, ``{Neutrino Masses and the CERN LHC: Testing Type II Seesaw},'' \href{http://dx.doi.org/10.1103/PhysRevD.78.015018}{{\em Phys. Rev. D} {\bf 78} (2008)  015018}, \href{http://arxiv.org/abs/0805.3536}{{\tt arXiv:0805.3536 [hep-ph]}}.

\bibitem{Melfo:2011nx}
A.~Melfo, M.~Nemevsek, F.~Nesti, G.~Senjanovic, and Y.~Zhang, ``{Type II Seesaw at LHC: The Roadmap},'' \href{http://dx.doi.org/10.1103/PhysRevD.85.055018}{{\em Phys. Rev. D} {\bf 85} (2012)  055018}, \href{http://arxiv.org/abs/1108.4416}{{\tt arXiv:1108.4416 [hep-ph]}}.

\bibitem{Kanemura:2014goa}
S.~Kanemura, M.~Kikuchi, K.~Yagyu, and H.~Yokoya, ``{Bounds on the mass of doubly-charged Higgs bosons in the same-sign diboson decay scenario},'' \href{http://dx.doi.org/10.1103/PhysRevD.90.115018}{{\em Phys. Rev. D} {\bf 90} (2014) no.~11, 115018}, \href{http://arxiv.org/abs/1407.6547}{{\tt arXiv:1407.6547 [hep-ph]}}.

\bibitem{BhupalDev:2018tox}
P.~S. Bhupal~Dev and Y.~Zhang, ``{Displaced vertex signatures of doubly charged scalars in the type-II seesaw and its left-right extensions},'' \href{http://dx.doi.org/10.1007/JHEP10(2018)199}{{\em JHEP} {\bf 10} (2018)  199}, \href{http://arxiv.org/abs/1808.00943}{{\tt arXiv:1808.00943 [hep-ph]}}.

\bibitem{Dodelson:1993je}
S.~Dodelson and L.~M. Widrow, ``{Sterile-neutrinos as dark matter},'' \href{http://dx.doi.org/10.1103/PhysRevLett.72.17}{{\em Phys. Rev. Lett.} {\bf 72} (1994)  17--20}, \href{http://arxiv.org/abs/hep-ph/9303287}{{\tt arXiv:hep-ph/9303287}}.

\bibitem{Shi:1998km}
X.-D. Shi and G.~M. Fuller, ``{A New dark matter candidate: Nonthermal sterile neutrinos},'' \href{http://dx.doi.org/10.1103/PhysRevLett.82.2832}{{\em Phys. Rev. Lett.} {\bf 82} (1999)  2832--2835}, \href{http://arxiv.org/abs/astro-ph/9810076}{{\tt arXiv:astro-ph/9810076}}.

\bibitem{Abazajian:2001nj}
K.~Abazajian, G.~M. Fuller, and M.~Patel, ``{Sterile neutrino hot, warm, and cold dark matter},'' \href{http://dx.doi.org/10.1103/PhysRevD.64.023501}{{\em Phys. Rev. D} {\bf 64} (2001)  023501}, \href{http://arxiv.org/abs/astro-ph/0101524}{{\tt arXiv:astro-ph/0101524}}.

\bibitem{Asaka:2006nq}
T.~Asaka, M.~Laine, and M.~Shaposhnikov, ``{Lightest sterile neutrino abundance within the nuMSM},'' \href{http://dx.doi.org/10.1088/1126-6708/2007/01/091}{{\em JHEP} {\bf 01} (2007)  091}, \href{http://arxiv.org/abs/hep-ph/0612182}{{\tt arXiv:hep-ph/0612182}}. [Erratum: JHEP 02, 028 (2015)].

\bibitem{Laine:2008pg}
M.~Laine and M.~Shaposhnikov, ``{Sterile neutrino dark matter as a consequence of nuMSM-induced lepton asymmetry},'' \href{http://dx.doi.org/10.1088/1475-7516/2008/06/031}{{\em JCAP} {\bf 06} (2008)  031}, \href{http://arxiv.org/abs/0804.4543}{{\tt arXiv:0804.4543 [hep-ph]}}.

\bibitem{Akhmedov:1998qx}
E.~K. Akhmedov, V.~A. Rubakov, and A.~Y. Smirnov, ``{Baryogenesis via neutrino oscillations},'' \href{http://dx.doi.org/10.1103/PhysRevLett.81.1359}{{\em Phys. Rev. Lett.} {\bf 81} (1998)  1359--1362}, \href{http://arxiv.org/abs/hep-ph/9803255}{{\tt arXiv:hep-ph/9803255}}.

\bibitem{Liu:1993tg}
J.~Liu and G.~Segre, ``{Reexamination of generation of baryon and lepton number asymmetries by heavy particle decay},'' \href{http://dx.doi.org/10.1103/PhysRevD.48.4609}{{\em Phys. Rev. D} {\bf 48} (1993)  4609--4612}, \href{http://arxiv.org/abs/hep-ph/9304241}{{\tt arXiv:hep-ph/9304241}}.

\bibitem{Flanz:1994yx}
M.~Flanz, E.~A. Paschos, and U.~Sarkar, ``{Baryogenesis from a lepton asymmetric universe},'' \href{http://dx.doi.org/10.1016/0370-2693(94)01555-Q}{{\em Phys. Lett. B} {\bf 345} (1995)  248--252}, \href{http://arxiv.org/abs/hep-ph/9411366}{{\tt arXiv:hep-ph/9411366}}. [Erratum: Phys.Lett.B 384, 487--487 (1996), Erratum: Phys.Lett.B 382, 447--447 (1996)].

\bibitem{Flanz:1996fb}
M.~Flanz, E.~A. Paschos, U.~Sarkar, and J.~Weiss, ``{Baryogenesis through mixing of heavy Majorana neutrinos},'' \href{http://dx.doi.org/10.1016/S0370-2693(96)01337-8}{{\em Phys. Lett. B} {\bf 389} (1996)  693--699}, \href{http://arxiv.org/abs/hep-ph/9607310}{{\tt arXiv:hep-ph/9607310}}.

\bibitem{Covi:1996wh}
L.~Covi, E.~Roulet, and F.~Vissani, ``{CP violating decays in leptogenesis scenarios},'' \href{http://dx.doi.org/10.1016/0370-2693(96)00817-9}{{\em Phys. Lett. B} {\bf 384} (1996)  169--174}, \href{http://arxiv.org/abs/hep-ph/9605319}{{\tt arXiv:hep-ph/9605319}}.

\bibitem{Covi:1996fm}
L.~Covi and E.~Roulet, ``{Baryogenesis from mixed particle decays},'' \href{http://dx.doi.org/10.1016/S0370-2693(97)00287-6}{{\em Phys. Lett. B} {\bf 399} (1997)  113--118}, \href{http://arxiv.org/abs/hep-ph/9611425}{{\tt arXiv:hep-ph/9611425}}.

\bibitem{Pilaftsis:1997jf}
A.~Pilaftsis, ``{CP violation and baryogenesis due to heavy Majorana neutrinos},'' \href{http://dx.doi.org/10.1103/PhysRevD.56.5431}{{\em Phys. Rev. D} {\bf 56} (1997)  5431--5451}, \href{http://arxiv.org/abs/hep-ph/9707235}{{\tt arXiv:hep-ph/9707235}}.

\bibitem{Pilaftsis:1997dr}
A.~Pilaftsis, ``{Resonant CP violation induced by particle mixing in transition amplitudes},'' \href{http://dx.doi.org/10.1016/S0550-3213(97)00469-0}{{\em Nucl. Phys. B} {\bf 504} (1997)  61--107}, \href{http://arxiv.org/abs/hep-ph/9702393}{{\tt arXiv:hep-ph/9702393}}.

\bibitem{Pilaftsis:1998pd}
A.~Pilaftsis, ``{Heavy Majorana neutrinos and baryogenesis},'' \href{http://dx.doi.org/10.1142/S0217751X99000932}{{\em Int. J. Mod. Phys. A} {\bf 14} (1999)  1811--1858}, \href{http://arxiv.org/abs/hep-ph/9812256}{{\tt arXiv:hep-ph/9812256}}.

\bibitem{Buchmuller:1997yu}
W.~Buchmuller and M.~Plumacher, ``{CP asymmetry in Majorana neutrino decays},'' \href{http://dx.doi.org/10.1016/S0370-2693(97)01548-7}{{\em Phys. Lett. B} {\bf 431} (1998)  354--362}, \href{http://arxiv.org/abs/hep-ph/9710460}{{\tt arXiv:hep-ph/9710460}}.

\bibitem{Pilaftsis:2003gt}
A.~Pilaftsis and T.~E.~J. Underwood, ``{Resonant leptogenesis},'' \href{http://dx.doi.org/10.1016/j.nuclphysb.2004.05.029}{{\em Nucl. Phys. B} {\bf 692} (2004)  303--345}, \href{http://arxiv.org/abs/hep-ph/0309342}{{\tt arXiv:hep-ph/0309342}}.

\bibitem{Pilaftsis:2005rv}
A.~Pilaftsis and T.~E.~J. Underwood, ``{Electroweak-scale resonant leptogenesis},'' \href{http://dx.doi.org/10.1103/PhysRevD.72.113001}{{\em Phys. Rev. D} {\bf 72} (2005)  113001}, \href{http://arxiv.org/abs/hep-ph/0506107}{{\tt arXiv:hep-ph/0506107}}.

\bibitem{Davidson:2002qv}
S.~Davidson and A.~Ibarra, ``{A Lower bound on the right-handed neutrino mass from leptogenesis},'' \href{http://dx.doi.org/10.1016/S0370-2693(02)01735-5}{{\em Phys. Lett. B} {\bf 535} (2002)  25--32}, \href{http://arxiv.org/abs/hep-ph/0202239}{{\tt arXiv:hep-ph/0202239}}.

\bibitem{Klaric:2020phc}
J.~Klari\'c, M.~Shaposhnikov, and I.~Timiryasov, ``{Uniting Low-Scale Leptogenesis Mechanisms},'' \href{http://dx.doi.org/10.1103/PhysRevLett.127.111802}{{\em Phys. Rev. Lett.} {\bf 127} (2021) no.~11, 111802}, \href{http://arxiv.org/abs/2008.13771}{{\tt arXiv:2008.13771 [hep-ph]}}.

\bibitem{Klaric:2021cpi}
J.~Klari\'c, M.~Shaposhnikov, and I.~Timiryasov, ``{Reconciling resonant leptogenesis and baryogenesis via neutrino oscillations},'' \href{http://dx.doi.org/10.1103/PhysRevD.104.055010}{{\em Phys. Rev. D} {\bf 104} (2021) no.~5, 055010}, \href{http://arxiv.org/abs/2103.16545}{{\tt arXiv:2103.16545 [hep-ph]}}.

\bibitem{Drewes:2021nqr}
M.~Drewes, Y.~Georis, and J.~Klari\'c, ``{Mapping the Viable Parameter Space for Testable Leptogenesis},'' \href{http://dx.doi.org/10.1103/PhysRevLett.128.051801}{{\em Phys. Rev. Lett.} {\bf 128} (2022) no.~5, 051801}, \href{http://arxiv.org/abs/2106.16226}{{\tt arXiv:2106.16226 [hep-ph]}}.

\bibitem{Abela:1981nf}
R.~Abela, M.~Daum, G.~H. Eaton, R.~Frosch, B.~Jost, P.~R. Kettle, and E.~Steiner, ``{Search for an Admixture of Heavy Neutrino in Pion Decay},'' \href{http://dx.doi.org/10.1016/0370-2693(81)90884-4}{{\em Phys. Lett. B} {\bf 105} (1981)  263--266}. [Erratum: Phys.Lett.B 106, 513 (1981)].

\bibitem{Yamazaki:1984sj}
T.~Yamazaki {\em et al.}, ``{Search for Heavy Neutrinos in Kaon Decay},'' {\em Conf. Proc. C} {\bf 840719} (1984)  262. \url{https://inspirehep.net/literature/211342}.

\bibitem{Vaitaitis:2000vc}
A.~G. Vaitaitis, \href{http://dx.doi.org/10.2172/1421441}{{\em {Search for neutral heavy leptons in a high-energy neutrino beam}}}.
\newblock PhD thesis, Columbia U., 2000.

\bibitem{CMS:2022fut}
{\bf CMS} Collaboration, A.~Tumasyan {\em et al.}, ``{Search for long-lived heavy neutral leptons with displaced vertices in proton-proton collisions at $ \sqrt{\mathrm{s}} $ =13 TeV},'' \href{http://dx.doi.org/10.1007/JHEP07(2022)081}{{\em JHEP} {\bf 07} (2022)  081}, \href{http://arxiv.org/abs/2201.05578}{{\tt arXiv:2201.05578 [hep-ex]}}.

\bibitem{Sabti:2020yrt}
N.~Sabti, A.~Magalich, and A.~Filimonova, ``{An Extended Analysis of Heavy Neutral Leptons during Big Bang Nucleosynthesis},'' \href{http://dx.doi.org/10.1088/1475-7516/2020/11/056}{{\em JCAP} {\bf 11} (2020)  056}, \href{http://arxiv.org/abs/2006.07387}{{\tt arXiv:2006.07387 [hep-ph]}}.

\bibitem{Izaguirre:2015pga}
E.~Izaguirre and B.~Shuve, ``{Multilepton and Lepton Jet Probes of Sub-Weak-Scale Right-Handed Neutrinos},'' \href{http://dx.doi.org/10.1103/PhysRevD.91.093010}{{\em Phys. Rev. D} {\bf 91} (2015) no.~9, 093010}, \href{http://arxiv.org/abs/1504.02470}{{\tt arXiv:1504.02470 [hep-ph]}}.

\bibitem{Das:2017gke}
A.~Das, P.~Konar, and A.~Thalapillil, ``{Jet substructure shedding light on heavy Majorana neutrinos at the LHC},'' \href{http://dx.doi.org/10.1007/JHEP02(2018)083}{{\em JHEP} {\bf 02} (2018)  083}, \href{http://arxiv.org/abs/1709.09712}{{\tt arXiv:1709.09712 [hep-ph]}}.

\bibitem{Pascoli:2018heg}
S.~Pascoli, R.~Ruiz, and C.~Weiland, ``{Heavy neutrinos with dynamic jet vetoes: multilepton searches at $ \sqrt{s}=14 $ , 27, and 100 TeV},'' \href{http://dx.doi.org/10.1007/JHEP06(2019)049}{{\em JHEP} {\bf 06} (2019)  049}, \href{http://arxiv.org/abs/1812.08750}{{\tt arXiv:1812.08750 [hep-ph]}}.

\bibitem{Drewes:2019fou}
M.~Drewes and J.~Hajer, ``{Heavy Neutrinos in displaced vertex searches at the LHC and HL-LHC},'' \href{http://dx.doi.org/10.1007/JHEP02(2020)070}{{\em JHEP} {\bf 20} (2019)  070},
\href{http://arxiv.org/abs/1903.06100}{{\tt arXiv:1903.06100 [hep-ph]}}.

\bibitem{Drewes:2018gkc}
M.~Drewes, J.~Hajer, J.~Klaric, and G.~Lanfranchi, ``{NA62 sensitivity to heavy neutral leptons in the low scale seesaw model},'' \href{http://dx.doi.org/10.1007/JHEP07(2018)105}{{\em JHEP} {\bf 07} (2018)  105}, \href{http://arxiv.org/abs/1801.04207}{{\tt arXiv:1801.04207 [hep-ph]}}.

\bibitem{FASER:2018eoc}
{\bf FASER} Collaboration, A.~Ariga {\em et al.}, ``{FASER\textquoteright{}s physics reach for long-lived particles},'' \href{http://dx.doi.org/10.1103/PhysRevD.99.095011}{{\em Phys. Rev. D} {\bf 99} (2019) no.~9, 095011}, \href{http://arxiv.org/abs/1811.12522}{{\tt arXiv:1811.12522 [hep-ph]}}.

\bibitem{Gorbunov:2020rjx}
D.~Gorbunov, I.~Krasnov, Y.~Kudenko, and S.~Suvorov, ``{Heavy Neutral Leptons from kaon decays in the SHiP experiment},'' \href{http://dx.doi.org/10.1016/j.physletb.2020.135817}{{\em Phys. Lett. B} {\bf 810} (2020)  135817}, \href{http://arxiv.org/abs/2004.07974}{{\tt arXiv:2004.07974 [hep-ph]}}.

\bibitem{Antusch:2017pkq}
S.~Antusch, E.~Cazzato, M.~Drewes, O.~Fischer, B.~Garbrecht, D.~Gueter, and J.~Klaric, ``{Probing Leptogenesis at Future Colliders},'' \href{http://dx.doi.org/10.1007/JHEP09(2018)124}{{\em JHEP} {\bf 09} (2018)  124}, \href{http://arxiv.org/abs/1710.03744}{{\tt arXiv:1710.03744 [hep-ph]}}.

\bibitem{Antusch:2016ejd}
S.~Antusch, E.~Cazzato, and O.~Fischer, ``{Sterile neutrino searches at future $e^-e^+$, $pp$, and $e^-p$ colliders},'' \href{http://dx.doi.org/10.1142/S0217751X17500786}{{\em Int. J. Mod. Phys. A} {\bf 32} (2017) no.~14, 1750078}, \href{http://arxiv.org/abs/1612.02728}{{\tt arXiv:1612.02728 [hep-ph]}}.

\bibitem{Bezrukov:2007ep}
F.~L. Bezrukov and M.~Shaposhnikov, ``{The Standard Model Higgs boson as the inflaton},'' \href{http://dx.doi.org/10.1016/j.physletb.2007.11.072}{{\em Phys. Lett. B} {\bf 659} (2008)  703--706}, \href{http://arxiv.org/abs/0710.3755}{{\tt arXiv:0710.3755 [hep-th]}}.

\bibitem{Asaka:2006rw}
T.~Asaka, M.~Laine, and M.~Shaposhnikov, ``{On the hadronic contribution to sterile neutrino production},'' \href{http://dx.doi.org/10.1088/1126-6708/2006/06/053}{{\em JHEP} {\bf 06} (2006)  053}, \href{http://arxiv.org/abs/hep-ph/0605209}{{\tt arXiv:hep-ph/0605209}}.

\bibitem{Venumadhav:2015pla}
T.~Venumadhav, F.-Y. Cyr-Racine, K.~N. Abazajian, and C.~M. Hirata, ``{Sterile neutrino dark matter: Weak interactions in the strong coupling epoch},'' \href{http://dx.doi.org/10.1103/PhysRevD.94.043515}{{\em Phys. Rev. D} {\bf 94} (2016) no.~4, 043515}, \href{http://arxiv.org/abs/1507.06655}{{\tt arXiv:1507.06655 [astro-ph.CO]}}.

\bibitem{Ghiglieri:2015jua}
J.~Ghiglieri and M.~Laine, ``{Improved determination of sterile neutrino dark matter spectrum},'' \href{http://dx.doi.org/10.1007/JHEP11(2015)171}{{\em JHEP} {\bf 11} (2015)  171}, \href{http://arxiv.org/abs/1506.06752}{{\tt arXiv:1506.06752 [hep-ph]}}.

\bibitem{Ghiglieri:2019kbw}
J.~Ghiglieri and M.~Laine, ``{Sterile neutrino dark matter via GeV-scale leptogenesis?},'' \href{http://dx.doi.org/10.1007/JHEP07(2019)078}{{\em JHEP} {\bf 07} (2019)  078}, \href{http://arxiv.org/abs/1905.08814}{{\tt arXiv:1905.08814 [hep-ph]}}.

\bibitem{Ghiglieri:2020ulj}
J.~Ghiglieri and M.~Laine, ``{Sterile neutrino dark matter via coinciding resonances},'' \href{http://dx.doi.org/10.1088/1475-7516/2020/07/012}{{\em JCAP} {\bf 07} (2020)  012}, \href{http://arxiv.org/abs/2004.10766}{{\tt arXiv:2004.10766 [hep-ph]}}.

\bibitem{Bodeker:2020hbo}
D.~Bodeker and A.~Klaus, ``{Sterile neutrino dark matter: Impact of active-neutrino opacities},'' \href{http://dx.doi.org/10.1007/JHEP07(2020)218}{{\em JHEP} {\bf 07} (2020)  218}, \href{http://arxiv.org/abs/2005.03039}{{\tt arXiv:2005.03039 [hep-ph]}}.

\bibitem{Canetti:2012vf}
L.~Canetti, M.~Drewes, and M.~Shaposhnikov, ``{Sterile Neutrinos as the Origin of Dark and Baryonic Matter},'' \href{http://dx.doi.org/10.1103/PhysRevLett.110.061801}{{\em Phys. Rev. Lett.} {\bf 110} (2013) no.~6, 061801}, \href{http://arxiv.org/abs/1204.3902}{{\tt arXiv:1204.3902 [hep-ph]}}.

\bibitem{Nemevsek:2012cd}
M.~Nemevsek, G.~Senjanovic, and Y.~Zhang, ``{Warm Dark Matter in Low Scale Left-Right Theory},'' \href{http://dx.doi.org/10.1088/1475-7516/2012/07/006}{{\em JCAP} {\bf 07} (2012)  006}, \href{http://arxiv.org/abs/1205.0844}{{\tt arXiv:1205.0844 [hep-ph]}}.

\bibitem{Nemevsek:2023yjl}
M.~Nemev\v{s}ek and Y.~Zhang, ``{Anatomy of diluted dark matter in the minimal left-right symmetric model},'' \href{http://dx.doi.org/10.1103/PhysRevD.109.056021}{{\em Phys. Rev. D} {\bf 109} (2024) no.~5, 056021}, \href{http://arxiv.org/abs/2312.00129}{{\tt arXiv:2312.00129 [hep-ph]}}.

\bibitem{Biswal:2017nfl}
S.~S. Biswal and P.~S.~B. Dev, ``{Probing left-right seesaw models using beam polarization at an $e^+e^-$ collider},'' \href{http://dx.doi.org/10.1103/PhysRevD.95.115031}{{\em Phys. Rev. D} {\bf 95} (2017) no.~11, 115031}, \href{http://arxiv.org/abs/1701.08751}{{\tt arXiv:1701.08751 [hep-ph]}}.

\bibitem{Urquia-Calderon:2023dkf}
K.~A. Urqu\'\i{}a-Calder\'on, ``{Long-lived heavy neutral leptons at lepton colliders as a probe of left-right-symmetric models},'' \href{http://dx.doi.org/10.1103/PhysRevD.109.055002}{{\em Phys. Rev. D} {\bf 109} (2024) no.~5, 055002}, \href{http://arxiv.org/abs/2310.17406}{{\tt arXiv:2310.17406 [hep-ph]}}.

\bibitem{Wen:2023xxc}
X.-K. Wen, B.~Yan, Z.~Yu, and C.~P. Yuan, ``{Single Transverse Spin Asymmetry as a New Probe of Standard-Model-Effective-Field-Theory Dipole Operators},'' \href{http://dx.doi.org/10.1103/PhysRevLett.131.241801}{{\em Phys. Rev. Lett.} {\bf 131} (2023) no.~24, 241801}, \href{http://arxiv.org/abs/2307.05236}{{\tt arXiv:2307.05236 [hep-ph]}}.

\bibitem{Wen:2024nff}
X.-K. Wen, B.~Yan, Z.~Yu, and C.~P. Yuan, ``{Transverse spin effects and light-quark dipole moments at lepton colliders},'' \href{http://arxiv.org/abs/2411.13845}{{\tt arXiv:2411.13845 [hep-ph]}}.

\bibitem{Aguilar-Saavedra:2022mpg}
J.~A. Aguilar-Saavedra, ``{Laboratory-frame tests of quantum entanglement in H\textrightarrow{}WW},'' \href{http://dx.doi.org/10.1103/PhysRevD.107.076016}{{\em Phys. Rev. D} {\bf 107} (2023) no.~7, 076016}, \href{http://arxiv.org/abs/2209.14033}{{\tt arXiv:2209.14033 [hep-ph]}}.

\bibitem{Barr:2021zcp}
A.~J. Barr, ``{Testing Bell inequalities in Higgs boson decays},'' \href{http://dx.doi.org/10.1016/j.physletb.2021.136866}{{\em Phys. Lett. B} {\bf 825} (2022)  136866}, \href{http://arxiv.org/abs/2106.01377}{{\tt arXiv:2106.01377 [hep-ph]}}.

\bibitem{Bi:2023uop}
Q.~Bi, Q.-H. Cao, K.~Cheng, and H.~Zhang, ``{New observables for testing Bell inequalities in $W$ boson pair production},'' \href{http://arxiv.org/abs/2307.14895}{{\tt arXiv:2307.14895 [hep-ph]}}.

\bibitem{Du:2024sly}
Y.~Du, X.-G. He, C.-W. Liu, and J.-P. Ma, ``{Impact of parity violation on quantum entanglement and Bell nonlocality},'' \href{http://arxiv.org/abs/2409.15418}{{\tt arXiv:2409.15418 [hep-ph]}}.

\bibitem{Peccei:1977hh}
R.~D. Peccei and H.~R. Quinn, ``{CP Conservation in the Presence of Instantons},'' \href{http://dx.doi.org/10.1103/PhysRevLett.38.1440}{{\em Phys. Rev. Lett.} {\bf 38} (1977)  1440--1443}.
[,328(1977)].

\bibitem{Weinberg:1977ma}
S.~Weinberg, ``{A New Light Boson?},''
\href{http://dx.doi.org/10.1103/PhysRevLett.40.223}{{\em Phys. Rev. Lett.} {\bf 40} (1978)  223--226}.

\bibitem{Wilczek:1977pj}
F.~Wilczek, ``{Problem of Strong $P$ and $T$ Invariance in the Presence of Instantons},''
\href{http://dx.doi.org/10.1103/PhysRevLett.40.279}{{\em Phys. Rev. Lett.} {\bf 40} (1978)  279--282}.

\bibitem{Svrcek:2006}
P.~Svrcek and E.~Witten {\em JHEP} {\bf 06} (2006)  051, \href{http://arxiv.org/abs/hep-th/0605206}{{\tt hep-th/0605206}}.

\bibitem{Arvanitaki:2010}
A.~Arvanitaki, S.~Dimopoulos, S.and~Dubovsky, N.~Kaloper, and J.~March-Russell {\em PRD} {\bf 81} (2010)  123530, \href{http://arxiv.org/abs/0905.4720}{{\tt arXiv:0905.4720 [hep-ph]}}.

\bibitem{Cicoli:2012}
M.~Cicoli, M.~Goodsell, and A.~Goodsell {\em JHEP} {\bf 10} (2012)  146, \href{http://arxiv.org/abs/1206.0819}{{\tt arXiv:1206.0819 [hep-ph]}}.

\bibitem{Arias:2012}
P.~Arias and et~al. {\em JCAP} {\bf 06} (2012)  013, \href{http://arxiv.org/abs/1201.5902}{{\tt arXiv:1201.5902 [hep-ph]}}.

\bibitem{Zhang:2021sio}
H.-Y. Zhang, C.-X. Yue, Y.-C. Guo, and S.~Yang, ``{Searching for axionlike particles at future electron-positron colliders},'' \href{http://dx.doi.org/10.1103/PhysRevD.104.096008}{{\em Phys. Rev. D} {\bf 104} (2021) no.~9, 096008}, \href{http://arxiv.org/abs/2103.05218}{{\tt arXiv:2103.05218 [hep-ph]}}.

\bibitem{Aloni:2019ruo}
D.~Aloni, C.~Fanelli, Y.~Soreq, and M.~Williams, ``{Photoproduction of Axionlike Particles},'' \href{http://dx.doi.org/10.1103/PhysRevLett.123.071801}{{\em Phys. Rev. Lett.} {\bf 123} (2019) no.~7, 071801},
\href{http://arxiv.org/abs/1903.03586}{{\tt arXiv:1903.03586 [hep-ph]}}.

\bibitem{BESIII:2022rzz}
{\bf BESIII} Collaboration, M.~Ablikim {\em et al.}, ``{Search for an axion-like particle in radiative J/\ensuremath{\psi} decays},'' \href{http://dx.doi.org/10.1016/j.physletb.2023.137698}{{\em Phys. Lett. B} {\bf 838} (2023)  137698}, \href{http://arxiv.org/abs/2211.12699}{{\tt arXiv:2211.12699 [hep-ex]}}.

\bibitem{Belle-II:2020jti}
{\bf Belle-II} Collaboration, F.~Abudin\'en {\em et al.}, ``{Search for Axion-Like Particles produced in $e^+e^-$ collisions at Belle II},'' \href{http://dx.doi.org/10.1103/PhysRevLett.125.161806}{{\em Phys. Rev. Lett.} {\bf 125} (2020) no.~16, 161806}, \href{http://arxiv.org/abs/2007.13071}{{\tt arXiv:2007.13071 [hep-ex]}}.

\bibitem{Jaeckel:2015jla}
J.~Jaeckel and M.~Spannowsky, ``{Probing MeV to 90 GeV axion-like particles with LEP and LHC},'' \href{http://dx.doi.org/10.1016/j.physletb.2015.12.037}{{\em Phys. Lett.} {\bf B753} (2016)  482--487},
\href{http://arxiv.org/abs/1509.00476}{{\tt arXiv:1509.00476 [hep-ph]}}.

\bibitem{Knapen:2016moh}
S.~Knapen, T.~Lin, H.~K. Lou, and T.~Melia, ``{Searching for Axionlike Particles with Ultraperipheral Heavy-Ion Collisions},'' \href{http://dx.doi.org/10.1103/PhysRevLett.118.171801}{{\em Phys. Rev. Lett.} {\bf 118} (2017) no.~17, 171801},
\href{http://arxiv.org/abs/1607.06083}{{\tt arXiv:1607.06083 [hep-ph]}}.

\bibitem{CMS:2018erd}
{\bf CMS} Collaboration, A.~M. Sirunyan {\em et al.}, ``{Evidence for light-by-light scattering and searches for axion-like particles in ultraperipheral PbPb collisions at $\sqrt{s_\mathrm{NN}} =$ 5.02 TeV},'' \href{http://dx.doi.org/10.1016/j.physletb.2019.134826}{{\em Phys. Lett. B} {\bf 797} (2019)  134826}, \href{http://arxiv.org/abs/1810.04602}{{\tt arXiv:1810.04602 [hep-ex]}}.

\bibitem{ATLAS:2020hii}
{\bf ATLAS} Collaboration, G.~Aad {\em et al.}, ``{Measurement of light-by-light scattering and search for axion-like particles with 2.2 nb$^{-1}$ of Pb+Pb data with the ATLAS detector},'' \href{http://dx.doi.org/10.1007/JHEP03(2021)243}{{\em JHEP} {\bf 03} (2021)  243}, \href{http://arxiv.org/abs/2008.05355}{{\tt arXiv:2008.05355 [hep-ex]}}. [Erratum: JHEP 11, 050 (2021)].

\bibitem{Cheung:2023nzg}
K.~Cheung and C.~J. Ouseph, ``{Axionlike particle search at Higgs factories},'' \href{http://dx.doi.org/10.1103/PhysRevD.108.035003}{{\em Phys. Rev. D} {\bf 108} (2023) no.~3, 035003}, \href{http://arxiv.org/abs/2303.16514}{{\tt arXiv:2303.16514 [hep-ph]}}.

\bibitem{Baldenegro:2018hng}
C.~Baldenegro, S.~Fichet, G.~von Gersdorff, and C.~Royon, ``{Searching for axion-like particles with proton tagging at the LHC},'' \href{http://dx.doi.org/10.1007/JHEP06(2018)131}{{\em JHEP} {\bf 06} (2018)  131}, \href{http://arxiv.org/abs/1803.10835}{{\tt arXiv:1803.10835 [hep-ph]}}.

\bibitem{Inan:2020aal}
S.~C. \.Inan and A.~V. Kisselev, ``{A search for axion-like particles in light-by-light scattering at the CLIC},'' \href{http://dx.doi.org/10.1007/JHEP06(2020)183}{{\em JHEP} {\bf 06} (2020)  183}, \href{http://arxiv.org/abs/2003.01978}{{\tt arXiv:2003.01978 [hep-ph]}}.

\bibitem{ATLAS:2015rsn}
{\bf ATLAS} Collaboration, G.~Aad {\em et al.}, ``{Search for new phenomena in events with at least three photons collected in $pp$ collisions at $\sqrt{s}$ = 8 TeV with the ATLAS detector},'' \href{http://dx.doi.org/10.1140/epjc/s10052-016-4034-8}{{\em Eur. Phys. J. C} {\bf 76} (2016) no.~4, 210}, \href{http://arxiv.org/abs/1509.05051}{{\tt arXiv:1509.05051 [hep-ex]}}.

\bibitem{ATLAS:2014jdv}
{\bf ATLAS} Collaboration, G.~Aad {\em et al.}, ``{Search for Scalar Diphoton Resonances in the Mass Range $65-600$ GeV with the ATLAS Detector in $pp$ Collision Data at $\sqrt{s}$ = 8 $TeV$},'' \href{http://dx.doi.org/10.1103/PhysRevLett.113.171801}{{\em Phys. Rev. Lett.} {\bf 113} (2014) no.~17, 171801}, \href{http://arxiv.org/abs/1407.6583}{{\tt arXiv:1407.6583 [hep-ex]}}.

\bibitem{L3:1994shn}
{\bf L3} Collaboration, M.~Acciarri {\em et al.}, ``{Search for anomalous Z --\ensuremath{>} gamma gamma gamma events at LEP},'' \href{http://dx.doi.org/10.1016/0370-2693(95)01612-T}{{\em Phys. Lett. B} {\bf 345} (1995)  609--616}.

\bibitem{L3:1997exg}
{\bf L3} Collaboration, M.~Acciarri {\em et al.}, ``{Search for new physics in energetic single photon production in $e^{+} e^{-}$ annihilation at the $Z$ resonance},'' \href{http://dx.doi.org/10.1016/S0370-2693(97)01003-4}{{\em Phys. Lett. B} {\bf 412} (1997)  201--209}.

\bibitem{OPAL:2002vhf}
{\bf OPAL} Collaboration, G.~Abbiendi {\em et al.}, ``{Multiphoton production in e+ e- collisions at s**(1/2) = 181-GeV to 209-GeV},'' \href{http://dx.doi.org/10.1140/epjc/s2002-01074-5}{{\em Eur. Phys. J. C} {\bf 26} (2003)  331--344}, \href{http://arxiv.org/abs/hep-ex/0210016}{{\tt arXiv:hep-ex/0210016}}.

\bibitem{Altmannshofer:2022ckw}
W.~Altmannshofer, J.~A. Dror, and S.~Gori, ``{New Opportunities for Detecting Axion-Lepton Interactions},'' \href{http://dx.doi.org/10.1103/PhysRevLett.130.241801}{{\em Phys. Rev. Lett.} {\bf 130} (2023) no.~24, 241801}, \href{http://arxiv.org/abs/2209.00665}{{\tt arXiv:2209.00665 [hep-ph]}}.

\bibitem{dEnterria:2021ljz}
D.~d'Enterria, ``{Collider constraints on axion-like particles},'' in {\em {Workshop on Feebly Interacting Particles}}.
\newblock 2, 2021.
\newblock \href{http://arxiv.org/abs/2102.08971}{{\tt arXiv:2102.08971 [hep-ex]}}.

\bibitem{Benson:2018vya}
S.~Benson and A.~Puig~Navarro, ``{Triggering $B_s^0 \to \gamma\gamma$ at LHCb},''.

\bibitem{CidVidal:2018blh}
X.~Cid~Vidal, A.~Mariotti, D.~Redigolo, F.~Sala, and K.~Tobioka, ``{New Axion Searches at Flavor Factories},'' \href{http://dx.doi.org/10.1007/JHEP01(2019)113}{{\em JHEP} {\bf 01} (2019)  113},
\href{http://arxiv.org/abs/1810.09452}{{\tt arXiv:1810.09452 [hep-ph]}}.

\bibitem{BaBar:2021ich}
{\bf BaBar} Collaboration, J.~P. Lees {\em et al.}, ``{Search for an Axionlike Particle in $B$ Meson Decays},'' \href{http://dx.doi.org/10.1103/PhysRevLett.128.131802}{{\em Phys. Rev. Lett.} {\bf 128} (2022) no.~13, 131802}, \href{http://arxiv.org/abs/2111.01800}{{\tt arXiv:2111.01800 [hep-ex]}}.

\bibitem{KTeV:2003sls}
{\bf KTeV} Collaboration, A.~Alavi-Harati {\em et al.}, ``{Search for the rare decay K(L) ---\ensuremath{>} pi0 e+ e-},'' \href{http://dx.doi.org/10.1103/PhysRevLett.93.021805}{{\em Phys. Rev. Lett.} {\bf 93} (2004)  021805}, \href{http://arxiv.org/abs/hep-ex/0309072}{{\tt arXiv:hep-ex/0309072}}.

\bibitem{LHCb:2015ycz}
{\bf LHCb} Collaboration, R.~Aaij {\em et al.}, ``{Angular analysis of the B0 to K*0 e+ e- decay in the low-q\^2 region},'' \href{http://dx.doi.org/10.1007/JHEP04(2015)064}{{\em JHEP} {\bf 04} (2015)  064}, \href{http://arxiv.org/abs/1501.03038}{{\tt arXiv:1501.03038 [hep-ex]}}.

\bibitem{CHARM:1985anb}
{\bf CHARM} Collaboration, F.~Bergsma {\em et al.}, ``{Search for Axion Like Particle Production in 400-{GeV} Proton - Copper Interactions},'' \href{http://dx.doi.org/10.1016/0370-2693(85)90400-9}{{\em Phys. Lett. B} {\bf 157} (1985)  458--462}.

\bibitem{Bjorken:1988as}
J.~D. Bjorken, S.~Ecklund, W.~R. Nelson, A.~Abashian, C.~Church, B.~Lu, L.~W. Mo, T.~A. Nunamaker, and P.~Rassmann, ``{Search for Neutral Metastable Penetrating Particles Produced in the SLAC Beam Dump},'' \href{http://dx.doi.org/10.1103/PhysRevD.38.3375}{{\em Phys. Rev. D} {\bf 38} (1988)  3375}.

\bibitem{Lu:2022zbe}
C.-T. Lu, ``{Lighting electroweak-violating ALP-lepton interactions at $e^+e^-$ and $ep$ colliders},'' \href{http://dx.doi.org/10.1103/PhysRevD.108.115029}{{\em Phys. Rev. D} {\bf 108} (2023) no.~11, 115029}, \href{http://arxiv.org/abs/2210.15648}{{\tt arXiv:2210.15648 [hep-ph]}}.

\bibitem{Roberts:2021nhw}
C.~D. Roberts, D.~G. Richards, T.~Horn, and L.~Chang, ``{Insights into the emergence of mass from studies of pion and kaon structure},'' \href{http://dx.doi.org/10.1016/j.ppnp.2021.103883}{{\em Prog. Part. Nucl. Phys.} {\bf 120} (2021)  103883}, \href{http://arxiv.org/abs/2102.01765}{{\tt arXiv:2102.01765 [hep-ph]}}.

\bibitem{Ding:2022ows}
M.~Ding, C.~D. Roberts, and S.~M. Schmidt, ``{Emergence of Hadron Mass and Structure},'' {\em Particles} {\bf 6} (2023)  57--120, \href{http://arxiv.org/abs/2211.07763}{{\tt arXiv:2211.07763 [hep-ph]}}.

\bibitem{Binosi:2022djx}
D.~Binosi, ``{Emergent Hadron Mass in Strong Dynamics},'' \href{http://dx.doi.org/10.1007/s00601-022-01740-6}{{\em Few Body Syst.} {\bf 63} (2022) no.~2, 42}, \href{http://arxiv.org/abs/2203.00942}{{\tt arXiv:2203.00942 [hep-ph]}}.

\bibitem{Ferreira:2023fva}
M.~N. Ferreira and J.~Papavassiliou, ``{Gauge Sector Dynamics in QCD},'' \href{http://dx.doi.org/10.3390/particles6010017}{{\em Particles} {\bf 6} (2023) no.~1, 312--363}, \href{http://arxiv.org/abs/2301.02314}{{\tt arXiv:2301.02314 [hep-ph]}}.

\bibitem{Cui:2019dwv}
Z.-F. Cui, J.-L. Zhang, D.~Binosi, F.~de~Soto, C.~Mezrag, J.~Papavassiliou, C.~D. Roberts, J.~Rodr\'\i{}guez-Quintero, J.~Segovia, and S.~Zafeiropoulos, ``{Effective charge from lattice QCD},'' \href{http://dx.doi.org/10.1088/1674-1137/44/8/083102}{{\em Chin. Phys. C} {\bf 44} (2020) no.~8, 083102}, \href{http://arxiv.org/abs/1912.08232}{{\tt arXiv:1912.08232 [hep-ph]}}.

\bibitem{Deur:2023dzc}
A.~Deur, S.~J. Brodsky, and C.~D. Roberts, ``{QCD Running Couplings and Effective Charges},'' \href{http://arxiv.org/abs/2303.00723}{{\tt arXiv:2303.00723 [hep-ph]}}.

\bibitem{Yao:2021pdy}
Z.-Q. Yao, D.~Binosi, Z.-F. Cui, and C.~D. Roberts, ``{Semileptonic transitions: $B(s)\textrightarrow{}\ensuremath{\pi}(K); Ds\textrightarrow{}K; D\textrightarrow{}\ensuremath{\pi},K; and K\textrightarrow{}\ensuremath{\pi}$},'' \href{http://dx.doi.org/10.1016/j.physletb.2021.136793}{{\em Phys. Lett. B} {\bf 824} (2022)  136793}, \href{http://arxiv.org/abs/2111.06473}{{\tt arXiv:2111.06473 [hep-ph]}}.

\bibitem{Yao:2021pyf}
Z.-Q. Yao, D.~Binosi, Z.-F. Cui, and C.~D. Roberts, ``{Semileptonic $B_c \to \eta_c, J/\psi$ transitions},'' \href{http://dx.doi.org/10.1016/j.physletb.2021.136344}{{\em Phys. Lett. B} {\bf 818} (2021)  136344}, \href{http://arxiv.org/abs/2104.10261}{{\tt arXiv:2104.10261 [hep-ph]}}.

\bibitem{Muong-2:2023cdq}
{\bf Muon g-2} Collaboration, D.~P. Aguillard {\em et al.}, ``{Measurement of the Positive Muon Anomalous Magnetic Moment to 0.20~ppm},'' \href{http://dx.doi.org/10.1103/PhysRevLett.131.161802}{{\em Phys. Rev. Lett.} {\bf 131} (2023) no.~16, 161802}, \href{http://arxiv.org/abs/2308.06230}{{\tt arXiv:2308.06230 [hep-ex]}}.

\bibitem{Aoyama:2020ynm}
T.~Aoyama {\em et al.}, ``{The anomalous magnetic moment of the muon in the Standard Model},'' \href{http://dx.doi.org/10.1016/j.physrep.2020.07.006}{{\em Phys. Rept.} {\bf 887} (2020)  1--166}, \href{http://arxiv.org/abs/2006.04822}{{\tt arXiv:2006.04822 [hep-ph]}}.

\bibitem{Hanneke:2008tm}
D.~Hanneke, S.~Fogwell, and G.~Gabrielse, ``{New Measurement of the Electron Magnetic Moment and the Fine Structure Constant},'' \href{http://dx.doi.org/10.1103/PhysRevLett.100.120801}{{\em Phys. Rev. Lett.} {\bf 100} (2008)  120801}, \href{http://arxiv.org/abs/0801.1134}{{\tt arXiv:0801.1134 [physics.atom-ph]}}.

\bibitem{Morel:2020dww}
L.~Morel, Z.~Yao, P.~Clad\'e, and S.~Guellati-Kh\'elifa, ``{Determination of the fine-structure constant with an accuracy of 81 parts per trillion},'' \href{http://dx.doi.org/10.1038/s41586-020-2964-7}{{\em Nature} {\bf 588} (2020) no.~7836, 61--65}.

\bibitem{Borsanyi:2020mff}
S.~Borsanyi {\em et al.}, ``{Leading hadronic contribution to the muon magnetic moment from lattice QCD},'' \href{http://dx.doi.org/10.1038/s41586-021-03418-1}{{\em Nature} {\bf 593} (2021) no.~7857, 51--55}, \href{http://arxiv.org/abs/2002.12347}{{\tt arXiv:2002.12347 [hep-lat]}}.

\bibitem{Boccaletti:2024guq}
A.~Boccaletti {\em et al.}, ``{High precision calculation of the hadronic vacuum polarisation contribution to the muon anomaly},'' \href{http://arxiv.org/abs/2407.10913}{{\tt arXiv:2407.10913 [hep-lat]}}.

\bibitem{Parrino:2025afq}
J.~Parrino, V.~Biloshytskyi, E.-H. Chao, H.~B. Meyer, and V.~Pascalutsa, ``{Computing the UV-finite electromagnetic corrections to the hadronic vacuum polarization in the muon $(g-2)$ from lattice QCD},'' \href{http://arxiv.org/abs/2501.03192}{{\tt arXiv:2501.03192 [hep-lat]}}.

\bibitem{FermilabLattice:2024yho}
{\bf Fermilab Lattice, HPQCD, MILC} Collaboration, A.~Bazavov {\em et al.}, ``{Hadronic vacuum polarization for the muon $g-2$ from lattice QCD: Complete short and intermediate windows},'' \href{http://arxiv.org/abs/2411.09656}{{\tt arXiv:2411.09656 [hep-lat]}}.

\bibitem{Bazavov:2024eou}
A.~Bazavov {\em et al.}, ``{Hadronic vacuum polarization for the muon $g-2$ from lattice QCD: Long-distance and full light-quark connected contribution},'' \href{http://arxiv.org/abs/2412.18491}{{\tt arXiv:2412.18491 [hep-lat]}}.

\bibitem{RBC:2018dos}
{\bf RBC, UKQCD} Collaboration, T.~Blum, P.~A. Boyle, V.~G\"ulpers, T.~Izubuchi, L.~Jin, C.~Jung, A.~J\"uttner, C.~Lehner, A.~Portelli, and J.~T. Tsang, ``{Calculation of the hadronic vacuum polarization contribution to the muon anomalous magnetic moment},'' \href{http://dx.doi.org/10.1103/PhysRevLett.121.022003}{{\em Phys. Rev. Lett.} {\bf 121} (2018) no.~2, 022003}, \href{http://arxiv.org/abs/1801.07224}{{\tt arXiv:1801.07224 [hep-lat]}}.

\bibitem{Wittig:2023pcl}
H.~Wittig, ``{Progress on $(g-2)_\mu$ from Lattice QCD},'' in {\em {57th Rencontres de Moriond on Electroweak Interactions and Unified Theories}}.
\newblock 6, 2023.
\newblock \href{http://arxiv.org/abs/2306.04165}{{\tt arXiv:2306.04165 [hep-ph]}}.

\bibitem{Aliberti:2025beg}
R.~Aliberti {\em et al.}, ``{The anomalous magnetic moment of the muon in the Standard Model: an update},'' \href{http://arxiv.org/abs/2505.21476}{{\tt arXiv:2505.21476 [hep-ph]}}.

\bibitem{Athron:2021iuf}
P.~Athron, C.~Bal\'azs, D.~H.~J. Jacob, W.~Kotlarski, D.~St\"ockinger, and H.~St\"ockinger-Kim, ``{New physics explanations of $a_{\mu}$ in light of the FNAL muon g \ensuremath{-} 2 measurement},'' \href{http://dx.doi.org/10.1007/JHEP09(2021)080}{{\em JHEP} {\bf 09} (2021)  080}, \href{http://arxiv.org/abs/2104.03691}{{\tt arXiv:2104.03691 [hep-ph]}}.

\bibitem{GAMBIT:2018gjo}
{\bf GAMBIT} Collaboration, P.~Athron {\em et al.}, ``{Combined collider constraints on neutralinos and charginos},'' \href{http://dx.doi.org/10.1140/epjc/s10052-019-6837-x}{{\em Eur. Phys. J. C} {\bf 79} (2019) no.~5, 395}, \href{http://arxiv.org/abs/1809.02097}{{\tt arXiv:1809.02097 [hep-ph]}}.

\bibitem{Crivellin:2020tsz}
A.~Crivellin, D.~Mueller, and F.~Saturnino, ``{Correlating h\textrightarrow{}\ensuremath{\mu}+\ensuremath{\mu}- to the Anomalous Magnetic Moment of the Muon via Leptoquarks},'' \href{http://dx.doi.org/10.1103/PhysRevLett.127.021801}{{\em Phys. Rev. Lett.} {\bf 127} (2021) no.~2, 021801}, \href{http://arxiv.org/abs/2008.02643}{{\tt arXiv:2008.02643 [hep-ph]}}.

\bibitem{Wang:2017vxj}
F.~Wang, W.~Wang, and J.~M. Yang, ``{Solving the muon g-2 anomaly in deflected anomaly mediated SUSY breaking with messenger-matter interactions},'' \href{http://dx.doi.org/10.1103/PhysRevD.96.075025}{{\em Phys. Rev. D} {\bf 96} (2017) no.~7, 075025}, \href{http://arxiv.org/abs/1703.10894}{{\tt arXiv:1703.10894 [hep-ph]}}.

\bibitem{Ning:2017dng}
X.~Ning and F.~Wang, ``{Solving the muon g-2 anomaly within the NMSSM from generalized deflected AMSB},'' \href{http://dx.doi.org/10.1007/JHEP08(2017)089}{{\em JHEP} {\bf 08} (2017)  089}, \href{http://arxiv.org/abs/1704.05079}{{\tt arXiv:1704.05079 [hep-ph]}}.

\bibitem{Du:2017str}
X.~Du and F.~Wang, ``{NMSSM From Alternative Deflection in Generalized Deflected Anomaly Mediated SUSY Breaking},'' \href{http://dx.doi.org/10.1140/epjc/s10052-018-5921-y}{{\em Eur. Phys. J. C} {\bf 78} (2018) no.~5, 431}, \href{http://arxiv.org/abs/1710.06105}{{\tt arXiv:1710.06105 [hep-ph]}}.

\bibitem{Du:2022pbp}
X.~K. Du, Z.~Li, F.~Wang, and Y.~K. Zhang, ``{The muon g \ensuremath{-} 2 anomaly in EOGM with adjoint messengers},'' \href{http://dx.doi.org/10.1016/j.nuclphysb.2023.116151}{{\em Nucl. Phys. B} {\bf 989} (2023)  116151}, \href{http://arxiv.org/abs/2204.04286}{{\tt arXiv:2204.04286 [hep-ph]}}.

\bibitem{Akula:2013ioa}
S.~Akula and P.~Nath, ``{Gluino-driven radiative breaking, Higgs boson mass, muon g-2, and the Higgs diphoton decay in supergravity unification},'' \href{http://dx.doi.org/10.1103/PhysRevD.87.115022}{{\em Phys. Rev. D} {\bf 87} (2013) no.~11, 115022}, \href{http://arxiv.org/abs/1304.5526}{{\tt arXiv:1304.5526 [hep-ph]}}.

\bibitem{Li:2021pnt}
Z.~Li, G.-L. Liu, F.~Wang, J.~M. Yang, and Y.~Zhang, ``{Gluino-SUGRA scenarios in light of FNAL muon g \textendash{} 2 anomaly},'' \href{http://dx.doi.org/10.1007/JHEP12(2021)219}{{\em JHEP} {\bf 12} (2021)  219}, \href{http://arxiv.org/abs/2106.04466}{{\tt arXiv:2106.04466 [hep-ph]}}.

\bibitem{Li:2022zap}
S.~Li, Z.~Li, F.~Wang, and J.~M. Yang, ``{Explanation of electron and muon g \ensuremath{-} 2 anomalies in AMSB},'' \href{http://dx.doi.org/10.1016/j.nuclphysb.2022.115927}{{\em Nucl. Phys. B} {\bf 983} (2022)  115927}, \href{http://arxiv.org/abs/2205.15153}{{\tt arXiv:2205.15153 [hep-ph]}}.

\bibitem{Bernreuther:1988jr}
W.~Bernreuther, U.~Low, J.~P. Ma, and O.~Nachtmann, ``{CP Violation and Z Boson Decays},'' \href{http://dx.doi.org/10.1007/BF02430617}{{\em Z. Phys. C} {\bf 43} (1989)  117}.

\bibitem{Booth:1993af}
M.~J. Booth, ``{The Electric dipole moment of the W and electron in the Standard Model},'' \href{http://arxiv.org/abs/hep-ph/9301293}{{\tt arXiv:hep-ph/9301293}}.

\bibitem{Yamaguchi:2020dsy}
Y.~Yamaguchi and N.~Yamanaka, ``{Quark level and hadronic contributions to the electric dipole moment of charged leptons in the standard model},'' \href{http://dx.doi.org/10.1103/PhysRevD.103.013001}{{\em Phys. Rev. D} {\bf 103} (2021) no.~1, 013001}, \href{http://arxiv.org/abs/2006.00281}{{\tt arXiv:2006.00281 [hep-ph]}}.

\bibitem{Belle:2002nla}
{\bf Belle} Collaboration, K.~Inami {\em et al.}, ``{Search for the electric dipole moment of the tau lepton},'' \href{http://dx.doi.org/10.1016/S0370-2693(02)02984-2}{{\em Phys. Lett. B} {\bf 551} (2003)  16--26}, \href{http://arxiv.org/abs/hep-ex/0210066}{{\tt arXiv:hep-ex/0210066}}.

\bibitem{Belle-II:2022cgf}
{\bf Belle-II} Collaboration, L.~Aggarwal {\em et al.}, ``{Snowmass White Paper: Belle II physics reach and plans for the next decade and beyond},'' \href{http://arxiv.org/abs/2207.06307}{{\tt arXiv:2207.06307 [hep-ex]}}.

\bibitem{He:2025ewk}
X.-G. He, C.-W. Liu, J.-P. Ma, C.~Yang, and Z.-Y. Zou, ``{Precise measurement of CP violating $\tau$ EDM through $e^+ e^- \to \gamma^*, \psi(2s) \to \tau^+ \tau^-$},'' \href{http://arxiv.org/abs/2501.06687}{{\tt arXiv:2501.06687 [hep-ph]}}.

\bibitem{Bernabeu:1993er}
J.~Bernabeu, G.~A. Gonzalez-Sprinberg, and J.~Vidal, ``{Normal and transverse single tau polarization at the Z peak},'' \href{http://dx.doi.org/10.1016/0370-2693(94)91209-2}{{\em Phys. Lett. B} {\bf 326} (1994)  168--174}.

\bibitem{Stiegler:1992sy}
U.~Stiegler, ``{On the magnetic moment in Z ---\ensuremath{>} tau tau},'' \href{http://dx.doi.org/10.1007/BF01474346}{{\em Z. Phys. C} {\bf 57} (1993)  511--514}.

\bibitem{Stahl:2000aq}
A.~Stahl, ``{Physics with tau leptons},'' \href{http://dx.doi.org/10.1007/BFb0109630}{{\em Springer Tracts Mod. Phys.} {\bf 160} (2000)  1--316}.

\bibitem{ALEPH:2002kbp}
{\bf ALEPH} Collaboration, A.~Heister {\em et al.}, ``{Search for anomalous weak dipole moments of the tau lepton},'' \href{http://dx.doi.org/10.1140/epjc/s2003-01286-1}{{\em Eur. Phys. J. C} {\bf 30} (2003)  291--304}, \href{http://arxiv.org/abs/hep-ex/0209066}{{\tt arXiv:hep-ex/0209066}}.

\bibitem{Lohmann:2005im}
W.~Lohmann, ``{Electromagnetic and weak moments of the tau-lepton},'' \href{http://dx.doi.org/10.1016/j.nuclphysbps.2005.02.017}{{\em Nucl. Phys. B Proc. Suppl.} {\bf 144} (2005)  122--127}, \href{http://arxiv.org/abs/hep-ex/0501065}{{\tt arXiv:hep-ex/0501065}}.

\bibitem{Bernreuther:2021elu}
W.~Bernreuther, L.~Chen, and O.~Nachtmann, ``{Electric dipole moment of the tau lepton revisited},'' \href{http://dx.doi.org/10.1103/PhysRevD.103.096011}{{\em Phys. Rev. D} {\bf 103} (2021) no.~9, 096011}, \href{http://arxiv.org/abs/2101.08071}{{\tt arXiv:2101.08071 [hep-ph]}}.

\bibitem{Atwood:1991ka}
D.~Atwood and A.~Soni, ``{Analysis for magnetic moment and electric dipole moment form-factors of the top quark via e+ e- ---\ensuremath{>} t anti-t},'' \href{http://dx.doi.org/10.1103/PhysRevD.45.2405}{{\em Phys. Rev. D} {\bf 45} (1992)  2405--2413}.

\bibitem{Davier:1992nw}
M.~Davier, L.~Duflot, F.~Le~Diberder, and A.~Rouge, ``{The Optimal method for the measurement of tau polarization},'' \href{http://dx.doi.org/10.1016/0370-2693(93)90101-M}{{\em Phys. Lett. B} {\bf 306} (1993)  411--417}.

\bibitem{Diehl:1993br}
M.~Diehl and O.~Nachtmann, ``{Optimal observables for the measurement of three gauge boson couplings in e+ e- ---\ensuremath{>} W+ W-},'' \href{http://dx.doi.org/10.1007/BF01555899}{{\em Z. Phys. C} {\bf 62} (1994)  397--412}.

\bibitem{Einstein:1935rr}
A.~Einstein, B.~Podolsky, and N.~Rosen, ``{Can quantum mechanical description of physical reality be considered complete?},'' \href{http://dx.doi.org/10.1103/PhysRev.47.777}{{\em Phys. Rev.} {\bf 47} (1935)  777--780}.

\bibitem{Bohm:1957zz}
D.~Bohm and Y.~Aharonov, ``{Discussion of Experimental Proof for the Paradox of Einstein, Rosen, and Podolsky},'' \href{http://dx.doi.org/10.1103/PhysRev.108.1070}{{\em Phys. Rev.} {\bf 108} (1957)  1070--1076}.

\bibitem{Bell:1964kc}
J.~S. Bell, ``{On the Einstein-Podolsky-Rosen paradox},'' \href{http://dx.doi.org/10.1103/PhysicsPhysiqueFizika.1.195}{{\em Physics Physique Fizika} {\bf 1} (1964)  195--200}.

\bibitem{Ma:2023yvd}
K.~Ma and T.~Li, ``{Testing Bell inequality through $h\to\tau\tau$ at CEPC},'' \href{http://dx.doi.org/https://doi.org/10.1088/1674-1137/ad62d8}{{\em Chinese Phys. C} (2024)  }.

\bibitem{Tornqvist:1980af}
N.~A. Tornqvist, ``{Suggestion for Einstein-podolsky-rosen Experiments Using Reactions Like $e^+ e^- \to \Lambda \bar{\Lambda} \to \pi^- p \pi^+ \bar{p}$},'' \href{http://dx.doi.org/10.1007/BF00715204}{{\em Found. Phys.} {\bf 11} (1981)  171--177}.

\bibitem{Clauser:1969ny}
J.~F. Clauser, M.~A. Horne, A.~Shimony, and R.~A. Holt, ``{Proposed experiment to test local hidden variable theories},'' \href{http://dx.doi.org/10.1103/PhysRevLett.23.880}{{\em Phys. Rev. Lett.} {\bf 23} (1969)  880--884}.

\bibitem{Morales:2024jhj}
R.~A. Morales, ``{Tripartite entanglement and Bell non-locality in loop-induced Higgs boson decays},'' \href{http://dx.doi.org/10.1140/epjc/s10052-024-12921-4}{{\em Eur. Phys. J. C} {\bf 84} (2024) no.~6, 581}, \href{http://arxiv.org/abs/2403.18023}{{\tt arXiv:2403.18023 [hep-ph]}}.

\bibitem{Koide:1982si}
Y.~Koide, ``{Fermion - Boson Two-body Model of Quarks and Leptons and Cabibbo Mixing},'' \href{http://dx.doi.org/10.1007/BF02817096}{{\em Lett. Nuovo Cim.} {\bf 34} (1982)  201}.

\bibitem{Koide:1982ax}
Y.~Koide, ``{A Fermion - Boson Composite Model of Quarks and Leptons},'' \href{http://dx.doi.org/10.1016/0370-2693(83)90644-5}{{\em Phys. Lett. B} {\bf 120} (1983)  161--165}.

\bibitem{Koide:1983qe}
Y.~Koide, ``{A New View of Quark and Lepton Mass Hierarchy},'' \href{http://dx.doi.org/10.1103/PhysRevD.28.252}{{\em Phys. Rev. D} {\bf 28} (1983)  252}.

\bibitem{ParticleDataGroup:2022pth}
{\bf Particle Data Group} Collaboration, R.~L. Workman {\em et al.}, ``{Review of Particle Physics},'' \href{http://dx.doi.org/10.1093/ptep/ptac097}{{\em PTEP} {\bf 2022} (2022)  083C01}.

\bibitem{Belle-II:2023izd}
{\bf Belle-II} Collaboration, I.~Adachi {\em et al.}, ``{Measurement of the \ensuremath{\tau}-lepton mass with the Belle II experiment},'' \href{http://dx.doi.org/10.1103/PhysRevD.108.032006}{{\em Phys. Rev. D} {\bf 108} (2023) no.~3, 032006}, \href{http://arxiv.org/abs/2305.19116}{{\tt arXiv:2305.19116 [hep-ex]}}.

\bibitem{Froggatt:1978nt}
C.~D. Froggatt and H.~B. Nielsen, ``{Hierarchy of Quark Masses, Cabibbo Angles and CP Violation},'' \href{http://dx.doi.org/10.1016/0550-3213(79)90316-X}{{\em Nucl. Phys. B} {\bf 147} (1979)  277--298}.

\bibitem{Koide:1989jq}
Y.~Koide, ``{Charged lepton mass sum rule from U(3) family Higgs potential model},'' \href{http://dx.doi.org/10.1142/S0217732390002663}{{\em Mod. Phys. Lett. A} {\bf 5} (1990)  2319--2324}.

\bibitem{Koide:2008tr}
Y.~Koide, ``{Charged Lepton Mass Relations in a Supersymmetric Yukawaon Model},'' \href{http://dx.doi.org/10.1103/PhysRevD.79.033009}{{\em Phys. Rev. D} {\bf 79} (2009)  033009}, \href{http://arxiv.org/abs/0811.3470}{{\tt arXiv:0811.3470 [hep-ph]}}.

\bibitem{Koide:2018gdm}
Y.~Koide and T.~Yamashita, ``{Charged Lepton Mass Relations in a SUSY Scenario},'' \href{http://dx.doi.org/10.1016/j.physletb.2018.10.058}{{\em Phys. Lett. B} {\bf 787} (2018)  171--174}, \href{http://arxiv.org/abs/1805.09533}{{\tt arXiv:1805.09533 [hep-ph]}}.

\bibitem{Liang:2020oni}
Z.~Liang and Z.~Sun, ``{A modified version of the Koide formula from flavor nonets in a scalar potential model and in a Yukawaon model},'' \href{http://dx.doi.org/10.1016/j.nuclphysb.2021.115546}{{\em Nucl. Phys. B} {\bf 972} (2021)  115546}, \href{http://arxiv.org/abs/2007.05878}{{\tt arXiv:2007.05878 [hep-ph]}}.

\bibitem{Appelquist:1974tg}
T.~Appelquist and J.~Carazzone, ``{Infrared Singularities and Massive Fields},'' \href{http://dx.doi.org/10.1103/PhysRevD.11.2856}{{\em Phys. Rev. D} {\bf 11} (1975)  2856}.

\bibitem{Buchmuller:1985jz}
W.~Buchmuller and D.~Wyler, ``{Effective Lagrangian Analysis of New Interactions and Flavor Conservation},''
\href{http://dx.doi.org/10.1016/0550-3213(86)90262-2}{{\em Nucl. Phys.} {\bf B268} (1986)  621--653}.

\bibitem{Grzadkowski:2010es}
B.~Grzadkowski, M.~Iskrzynski, M.~Misiak, and J.~Rosiek, ``{Dimension-Six Terms in the Standard Model Lagrangian},'' \href{http://dx.doi.org/10.1007/JHEP10(2010)085}{{\em JHEP} {\bf 10} (2010)  085},
\href{http://arxiv.org/abs/1008.4884}{{\tt arXiv:1008.4884 [hep-ph]}}.

\bibitem{Falkowski:2001958}
A.~Falkowski and A.~Falkowski, ``{Higgs Basis: Proposal for an EFT basis choice for LHC HXSWG},''. \url{https://cds.cern.ch/record/2001958}.

\bibitem{Chai:2024zyl}
S.~Chai, J.~Gu, and L.~Li, ``{From optimal observables to machine learning: an effective-field-theory analysis of $e^+e^-\to W^+W^-$ at future lepton colliders},'' \href{http://dx.doi.org/10.1007/JHEP05(2024)292}{{\em JHEP} {\bf 05} (2024)  292}, \href{http://arxiv.org/abs/2401.02474}{{\tt arXiv:2401.02474 [hep-ph]}}.

\bibitem{deBlas:2022ofj}
J.~de~Blas, Y.~Du, C.~Grojean, J.~Gu, V.~Miralles, M.~E. Peskin, J.~Tian, M.~Vos, and E.~Vryonidou, ``{Global SMEFT Fits at Future Colliders},'' in {\em {Snowmass 2021}}.
\newblock 6, 2022.
\newblock \href{http://arxiv.org/abs/2206.08326}{{\tt arXiv:2206.08326 [hep-ph]}}.

\bibitem{Du:2021idh}
Y.~Du and J.-H. Yu, ``{Neutrino non-standard interactions meet precision measurements of N$_{eff}$},'' \href{http://dx.doi.org/10.1007/JHEP05(2021)058}{{\em JHEP} {\bf 05} (2021)  058}, \href{http://arxiv.org/abs/2101.10475}{{\tt arXiv:2101.10475 [hep-ph]}}.

\bibitem{Du:2023upj}
Y.~Du, ``{Neff as a new physics probe in the precision era of cosmology},'' \href{http://dx.doi.org/10.1103/PhysRevD.110.055030}{{\em Phys. Rev. D} {\bf 110} (2024) no.~5, 055030}, \href{http://arxiv.org/abs/2310.10034}{{\tt arXiv:2310.10034 [hep-ph]}}.

\bibitem{ATLAS:2018gqq}
{\bf ATLAS} Collaboration, ``{Measurement of the effective leptonic weak mixing angle using electron and muon pairs from $Z$-boson decay in the ATLAS experiment at $\sqrt s = 8$ TeV},''.

\bibitem{CMS:2017vxj}
{\bf CMS} Collaboration, ``{A proposal for the measurement of the weak mixing angle at the HL-LHC},''.

\bibitem{Ellis:2025ghl}
J.~Ellis, H.-J. He, and R.-Q. Xiao, ``{Probing CP-Violating Neutral Triple Gauge Couplings at Electron-Positron Colliders},'' \href{http://arxiv.org/abs/2504.13135}{{\tt arXiv:2504.13135 [hep-ph]}}.

\bibitem{Liu:2024tcz}
D.~Liu, R.-Q. Xiao, S.~Li, J.~Ellis, H.-J. He, and R.~Yuan, ``{Probing Neutral Triple Gauge Couplings via $\boldsymbol{Z\gamma\,(\ell^+\ell^-\gamma)}$ Production at $\boldsymbol{e^+e^-}$ Colliders},'' \href{http://dx.doi.org/10.15302/frontphys.2025.015201}{{\em Front. Phys. (Beijing)} {\bf 20} (2025) no.~1, 015201}, \href{http://arxiv.org/abs/2404.15937}{{\tt arXiv:2404.15937 [hep-ph]}}.

\bibitem{ALEPH:2005ab}
{\bf ALEPH, DELPHI, L3, OPAL, SLD, LEP Electroweak Working Group, SLD Electroweak Group, SLD Heavy Flavour Group} Collaboration, S.~Schael {\em et al.}, ``{Precision electroweak measurements on the $Z$ resonance},'' \href{http://dx.doi.org/10.1016/j.physrep.2005.12.006}{{\em Phys. Rept.} {\bf 427} (2006)  257--454},
\href{http://arxiv.org/abs/hep-ex/0509008}{{\tt arXiv:hep-ex/0509008 [hep-ex]}}.

\bibitem{Chen:2018shg}
N.~Chen, T.~Han, S.~Su, W.~Su, and Y.~Wu, ``{Type-II 2HDM under the Precision Measurements at the $Z$-pole and a Higgs Factory},'' \href{http://dx.doi.org/10.1007/JHEP03(2019)023}{{\em JHEP} {\bf 03} (2019)  023}, \href{http://arxiv.org/abs/1808.02037}{{\tt arXiv:1808.02037 [hep-ph]}}.

\bibitem{Li:2020glc}
H.~Li, H.~Song, S.~Su, W.~Su, and J.~M. Yang, ``{MSSM at future Higgs factories},'' \href{http://dx.doi.org/10.1088/1674-1137/abe19b}{{\em Chin. Phys. C} {\bf 45} (2021) no.~4, 045106}, \href{http://arxiv.org/abs/2010.09782}{{\tt arXiv:2010.09782 [hep-ph]}}.

\bibitem{GAMBIT:2017yxo}
{\bf GAMBIT} Collaboration, P.~Athron {\em et al.}, ``{GAMBIT: The Global and Modular Beyond-the-Standard-Model Inference Tool},'' \href{http://dx.doi.org/10.1140/epjc/s10052-017-5321-8}{{\em Eur. Phys. J. C} {\bf 77} (2017) no.~11, 784}, \href{http://arxiv.org/abs/1705.07908}{{\tt arXiv:1705.07908 [hep-ph]}}. [Addendum: Eur.Phys.J.C 78, 98 (2018)].

\bibitem{GAMBIT:2017snp}
{\bf GAMBIT} Collaboration, P.~Athron {\em et al.}, ``{Global fits of GUT-scale SUSY models with GAMBIT},'' \href{http://dx.doi.org/10.1140/epjc/s10052-017-5167-0}{{\em Eur. Phys. J. C} {\bf 77} (2017) no.~12, 824}, \href{http://arxiv.org/abs/1705.07935}{{\tt arXiv:1705.07935 [hep-ph]}}.

\bibitem{GAMBIT:2017zdo}
{\bf GAMBIT} Collaboration, P.~Athron {\em et al.}, ``{A global fit of the MSSM with GAMBIT},'' \href{http://dx.doi.org/10.1140/epjc/s10052-017-5196-8}{{\em Eur. Phys. J. C} {\bf 77} (2017) no.~12, 879}, \href{http://arxiv.org/abs/1705.07917}{{\tt arXiv:1705.07917 [hep-ph]}}.

\bibitem{Arbey:2021jdh}
A.~Arbey, M.~Battaglia, A.~Djouadi, F.~Mahmoudi, M.~Muhlleitner, and M.~Spira, ``{Higgs boson properties and supersymmetry: Constraints and sensitivity from the LHC to an e+e- collider},'' \href{http://dx.doi.org/10.1103/PhysRevD.106.055002}{{\em Phys. Rev. D} {\bf 106} (2022) no.~5, 055002}, \href{http://arxiv.org/abs/2201.00070}{{\tt arXiv:2201.00070 [hep-ph]}}.

\end{thebibliography}\endgroup

\end{document}